%% file: main.tex
\documentclass[graybox,envcountchap,sectrefs]{svmono}

\usepackage{type1cm}         

\usepackage{makeidx}         
\usepackage{graphicx}        
\usepackage{multicol}        
\usepackage[bottom]{footmisc}

\usepackage{newtxtext}       %
\usepackage[varvw]{newtxmath}       

\usepackage[colorlinks,linkcolor=blue,anchorcolor=blue,citecolor=blue]{hyperref}

\usepackage{pdfpages}

\makeindex             

\graphicspath{{figures/}}

\begin{document}

\includepdfmerge{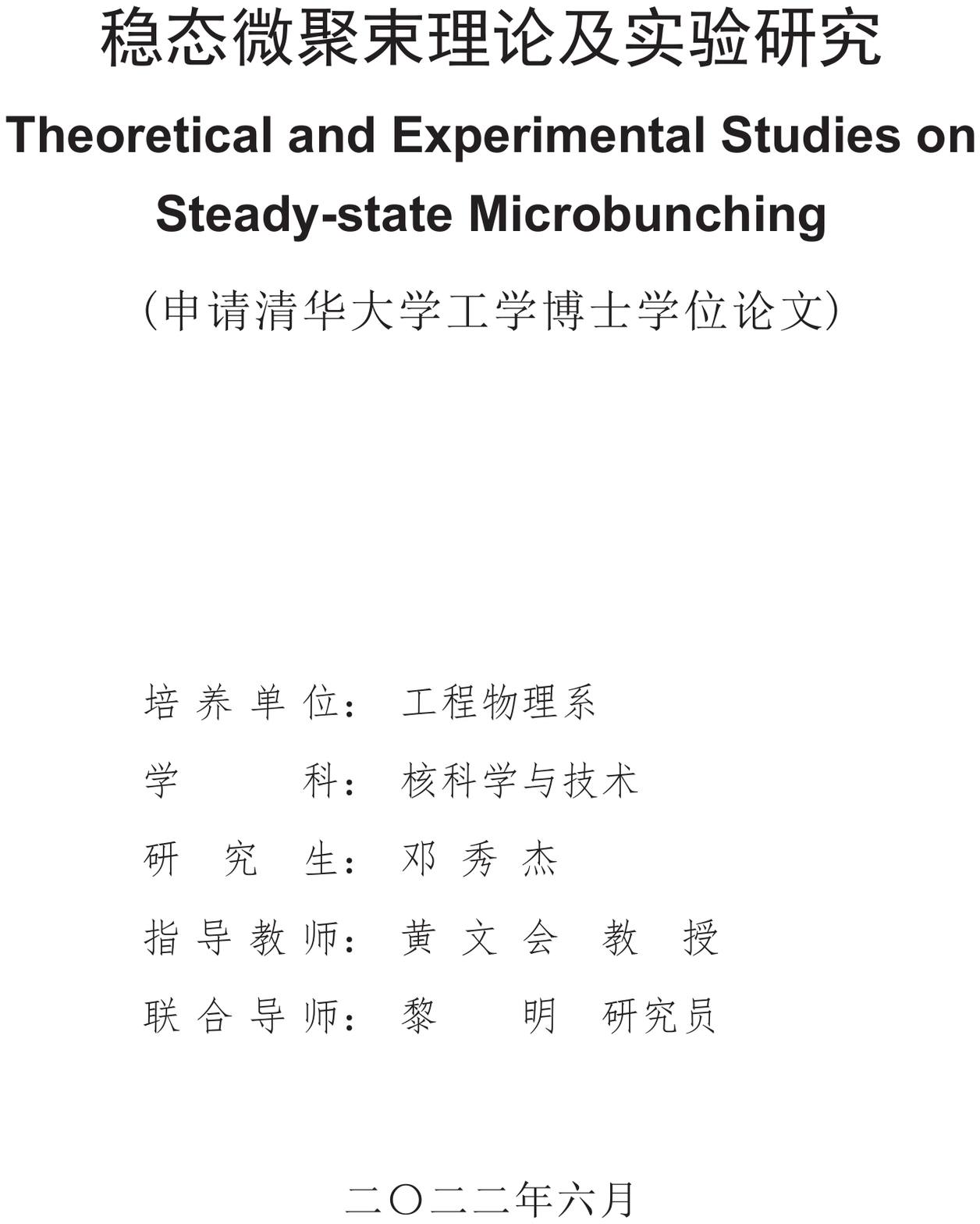}
\includepdfmerge{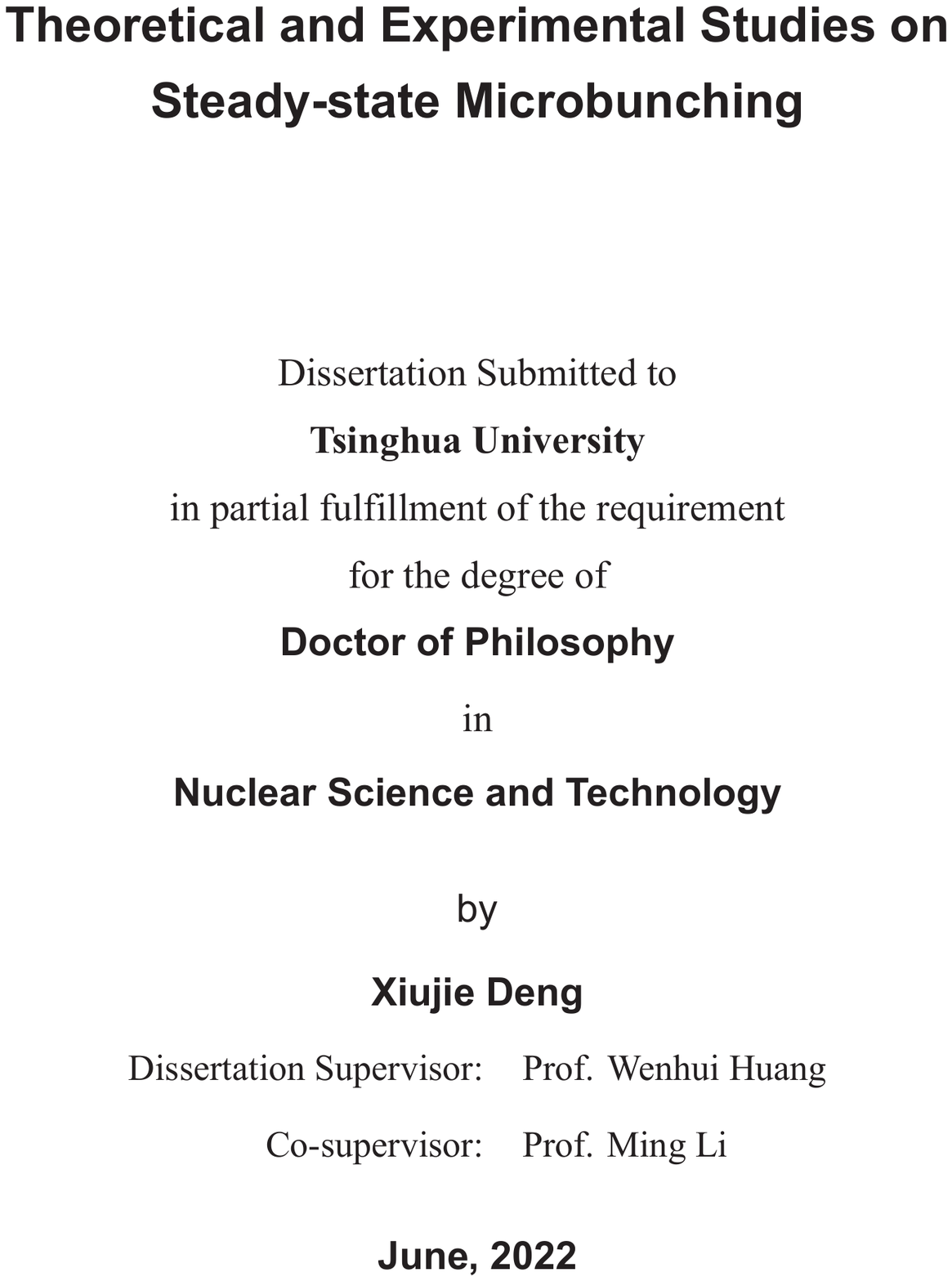}
\includepdfmerge{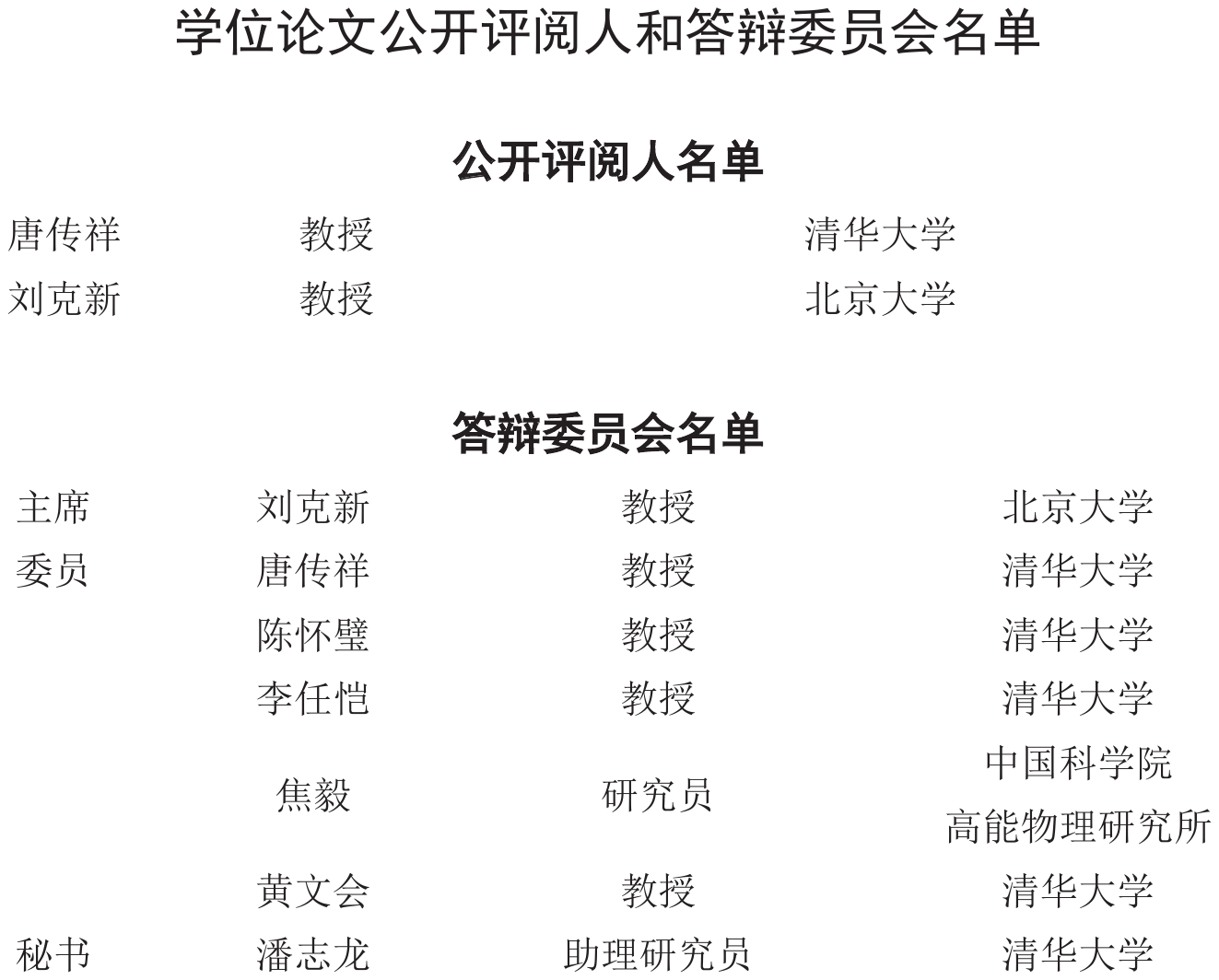}
\includepdfmerge{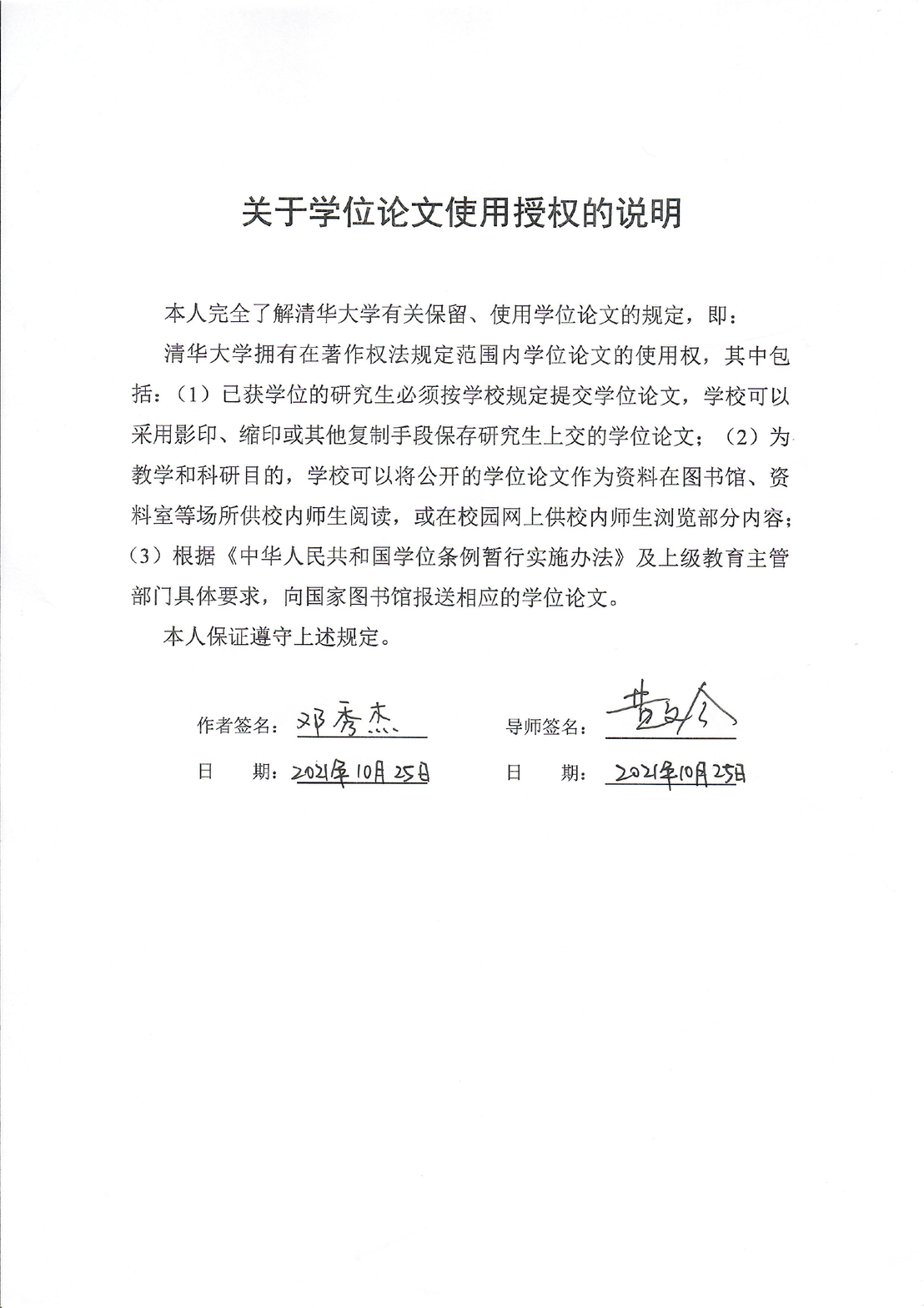}

\frontmatter

\includepdfmerge{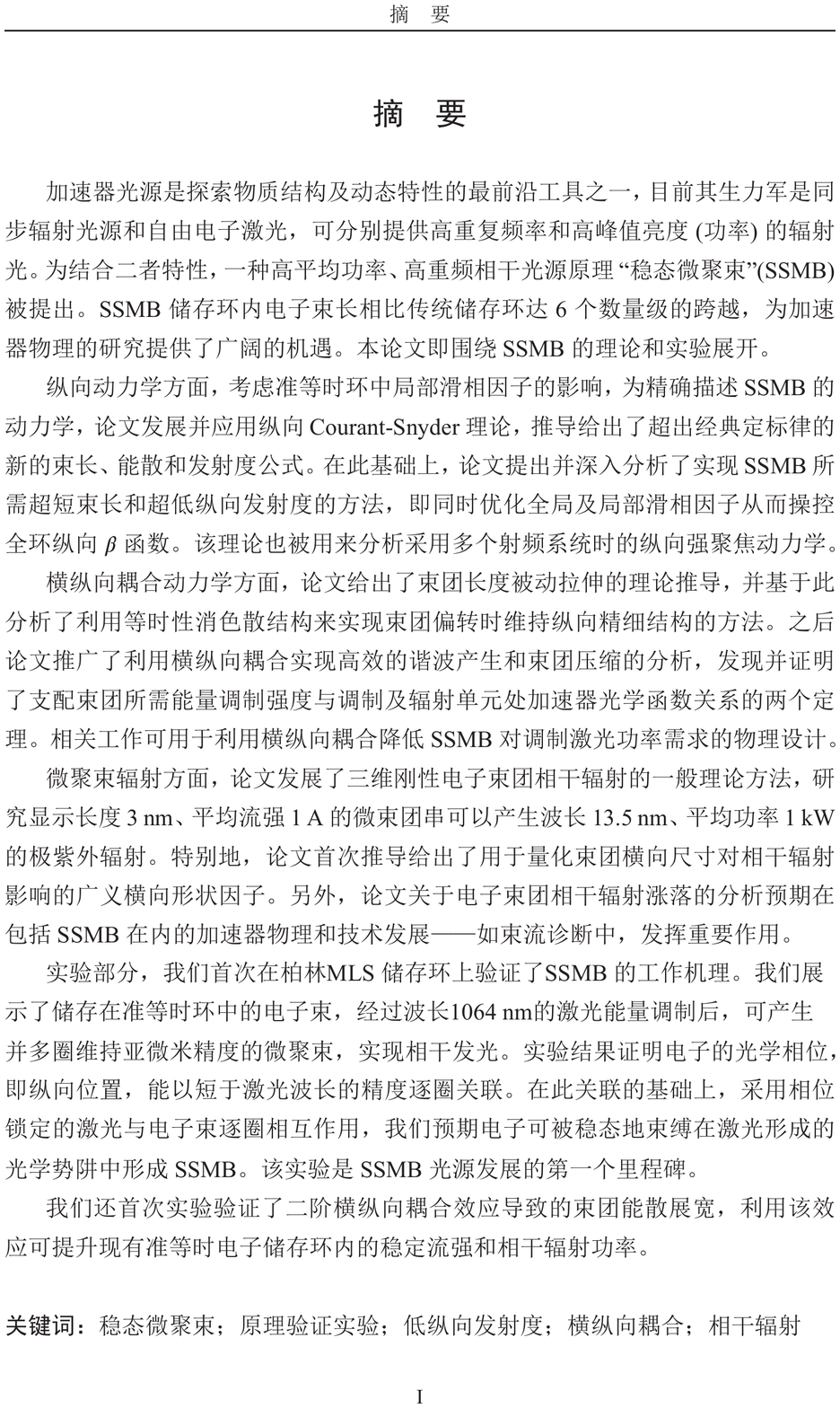}
\includepdfmerge{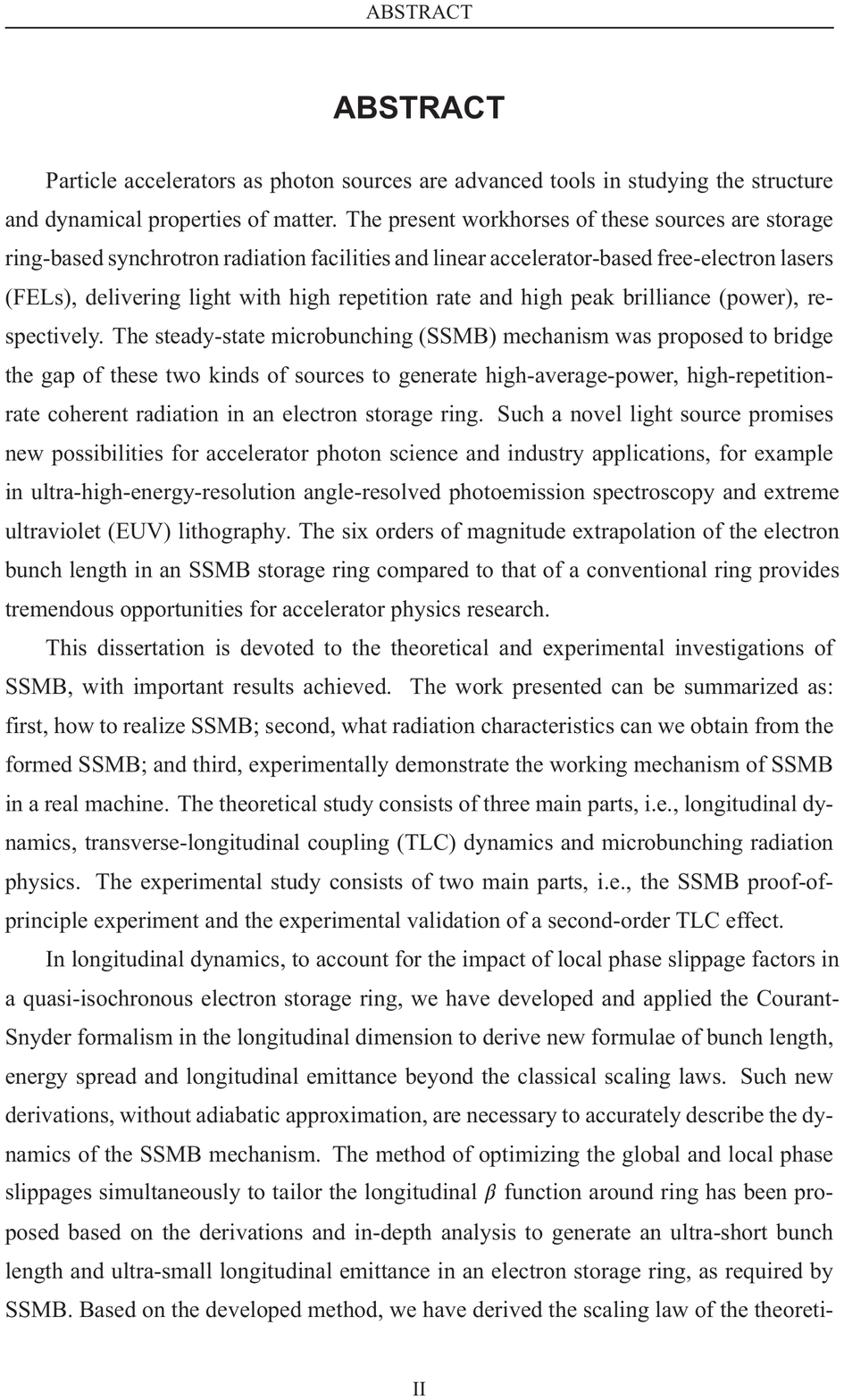}
\includepdfmerge{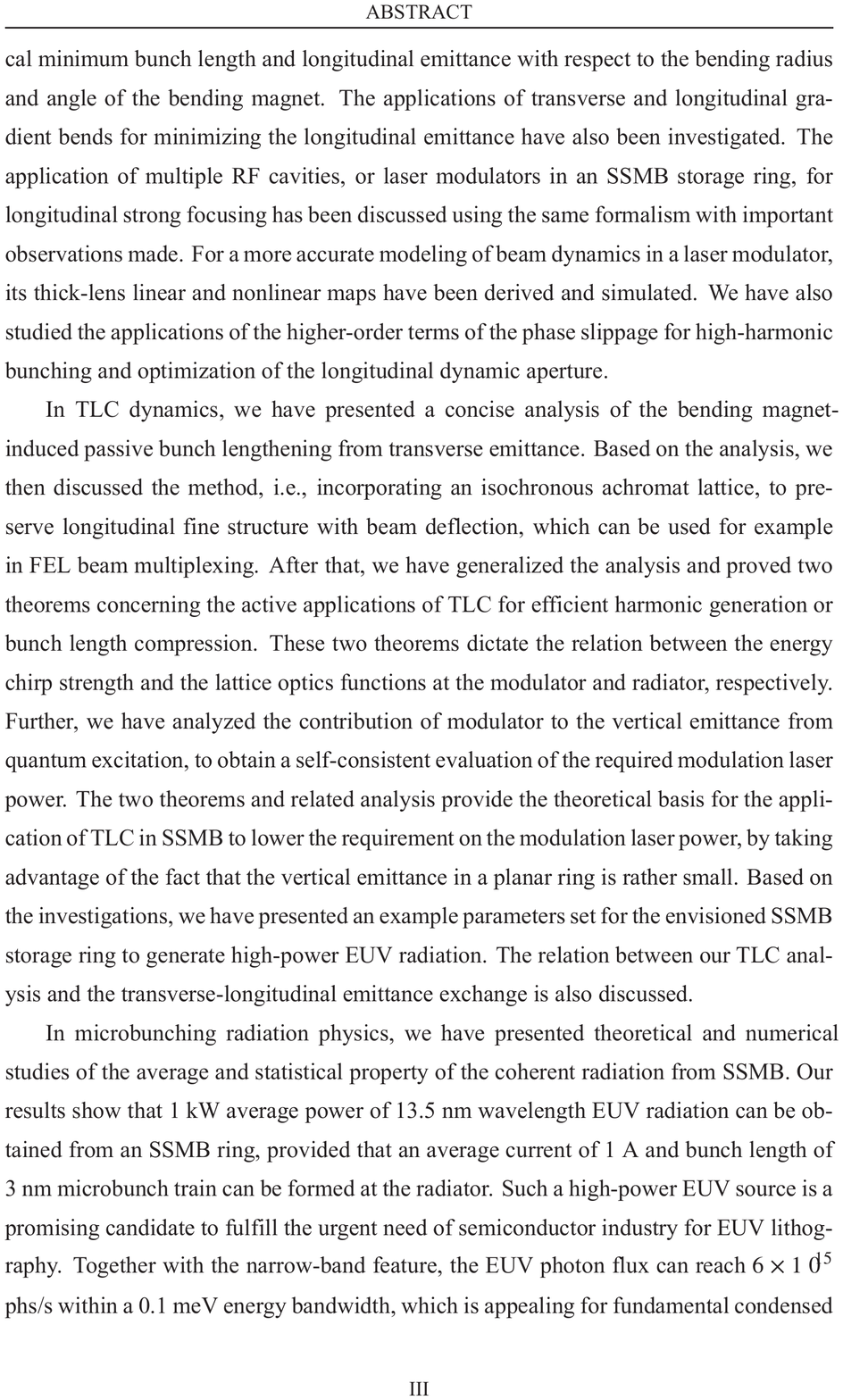}
\includepdfmerge{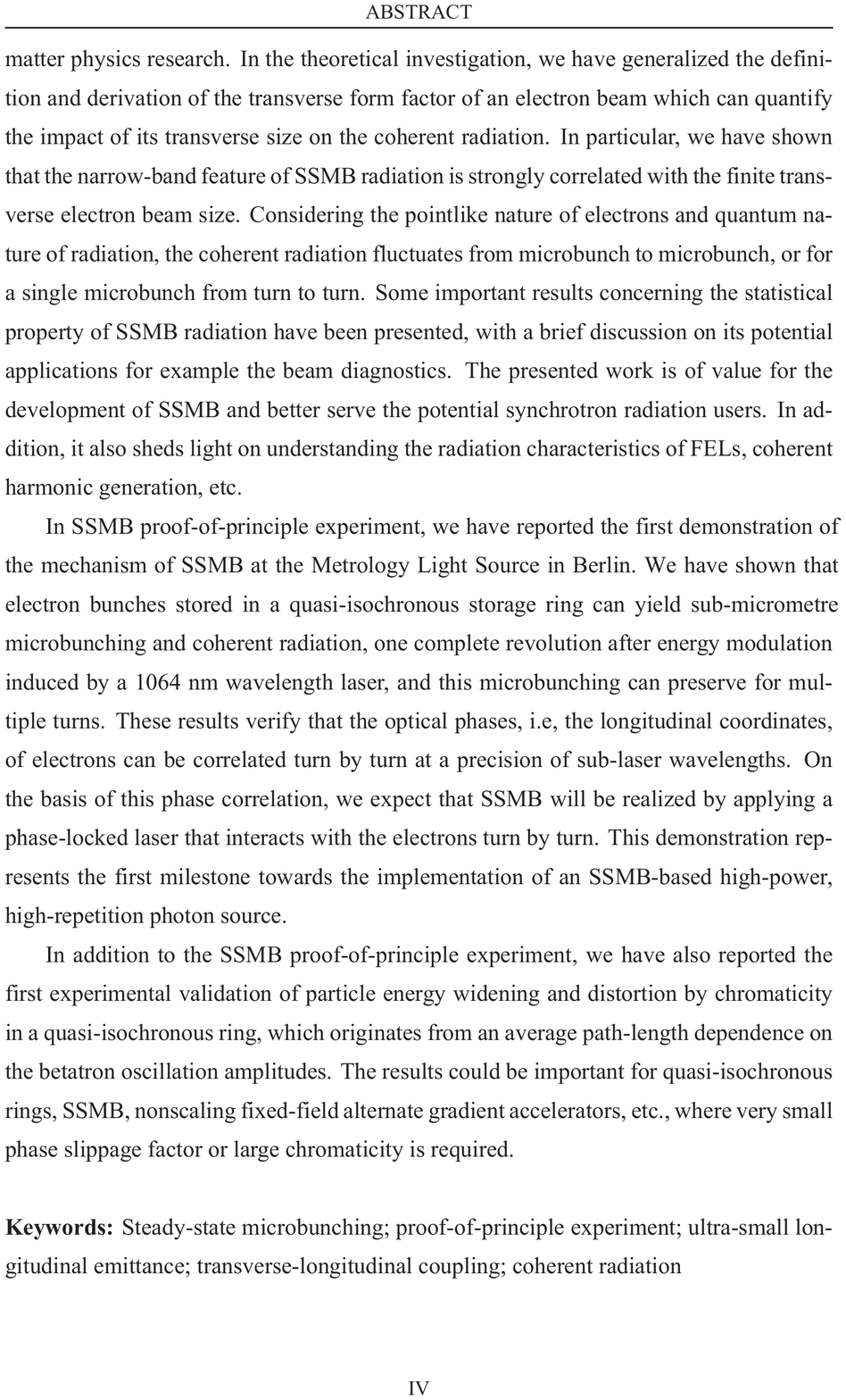}
\includepdfmerge{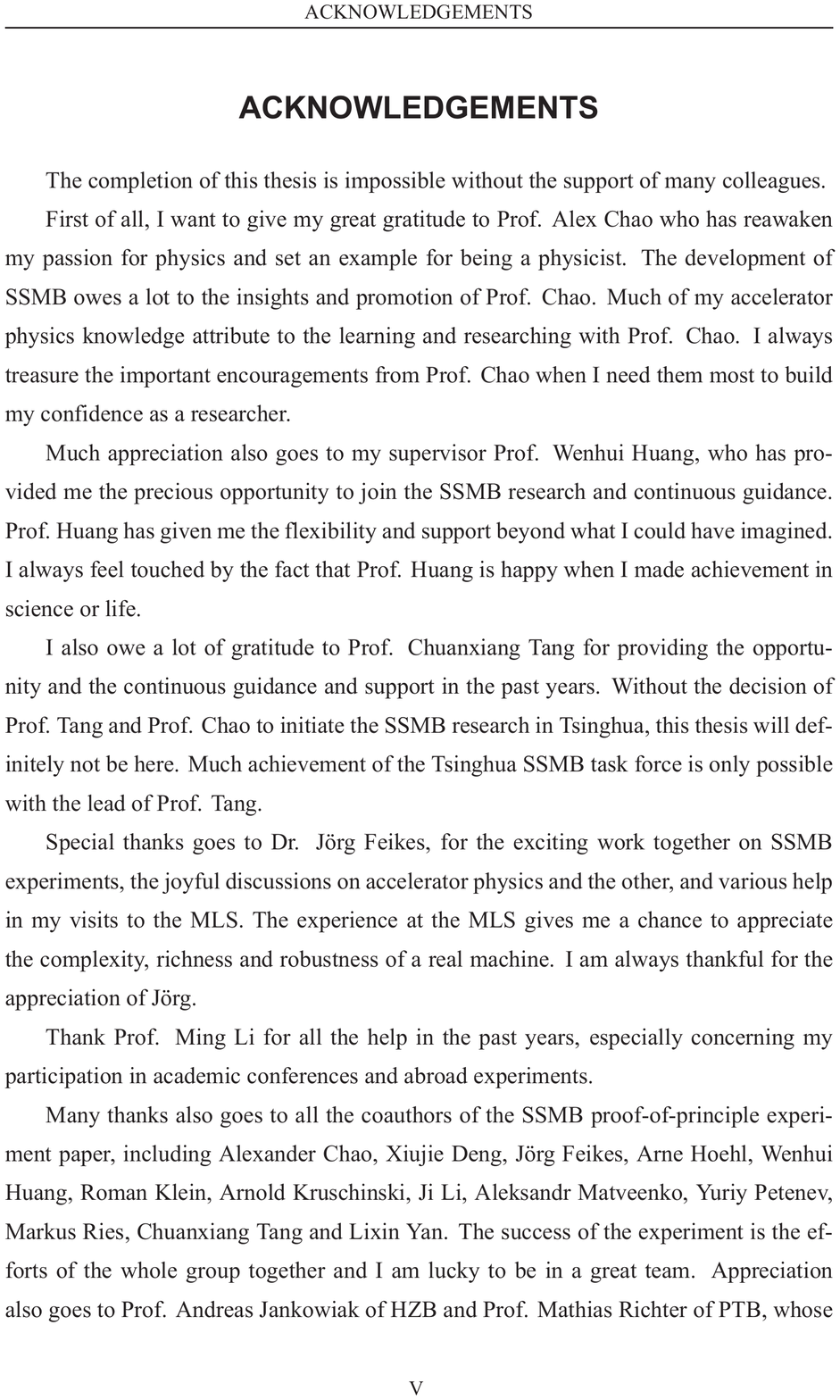}
\includepdfmerge{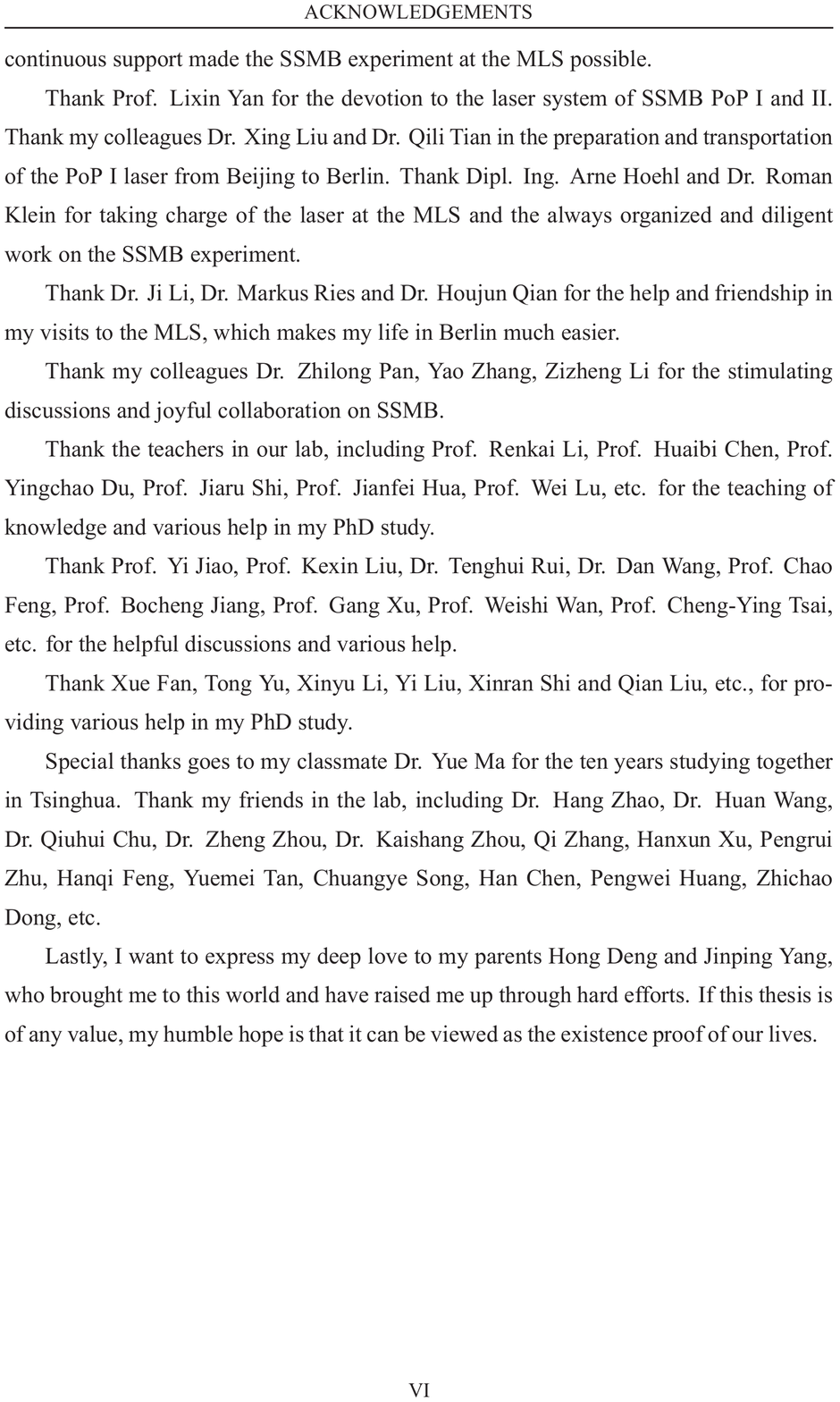}





{
	\hypersetup{linkcolor=black}
	\tableofcontents
}


\include{data/acronym}

\mainmatter
\include{data/chap01}

\include{data/chap02}

\include{data/chap03}
\include{data/chap04}

\include{data/chap05}
\include{data/chap06}

\bibliographystyle{ieeetr}
\bibliography{ref/refs}  

\appendix
\include{data/appendix}

\includepdfmerge{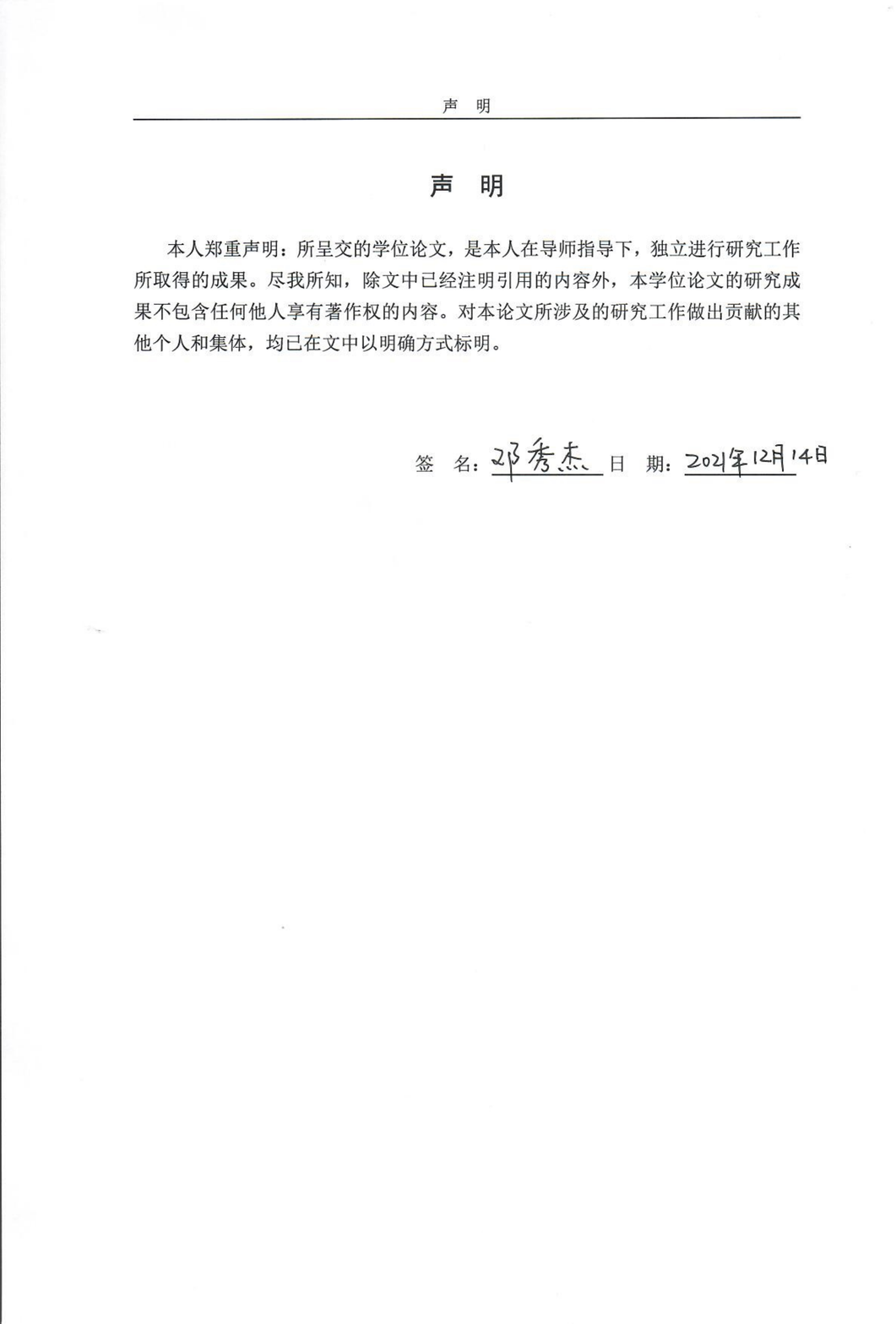}
\includepdfmerge{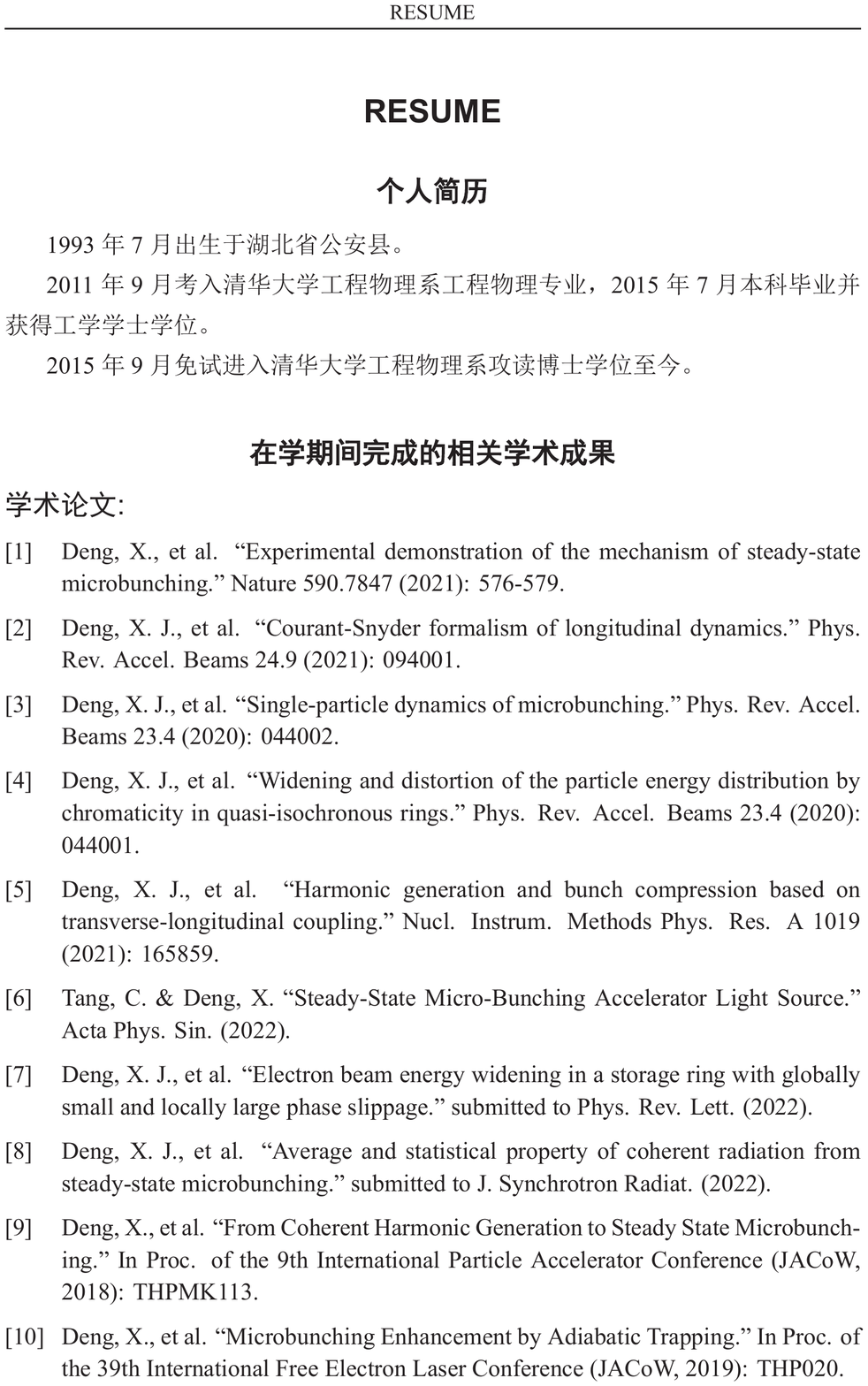}
\includepdfmerge{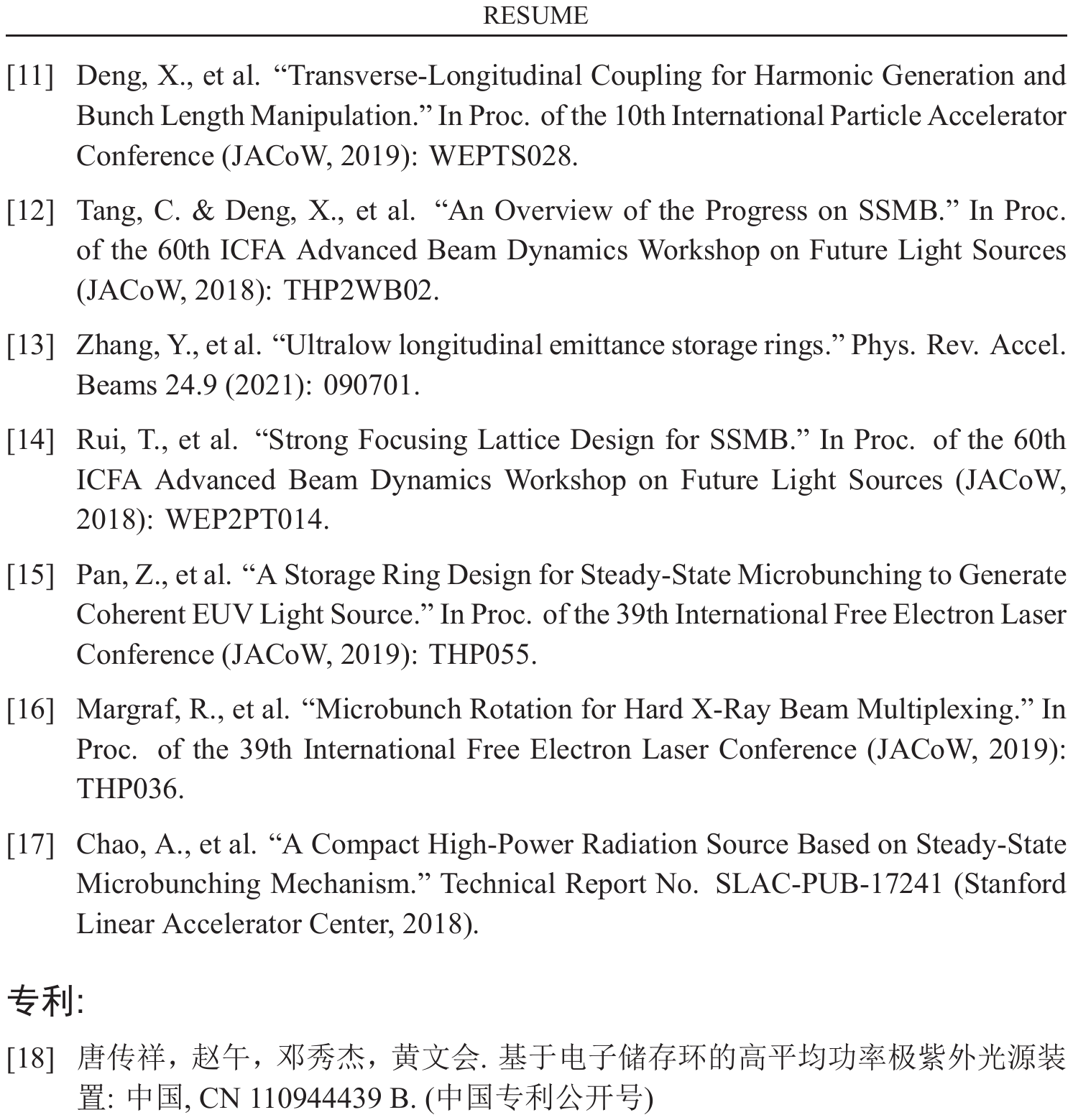}
\includepdfmerge{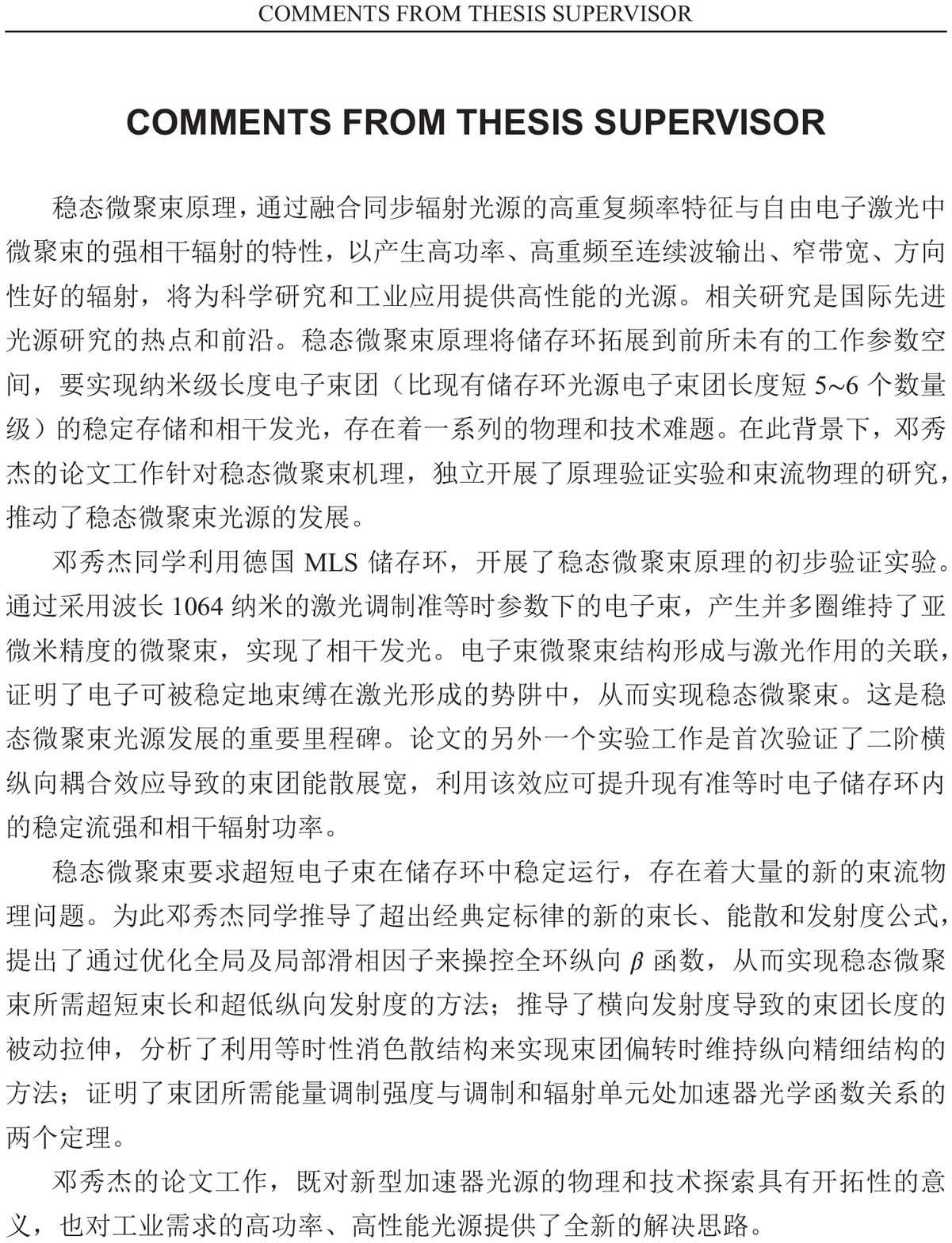}
\includepdfmerge{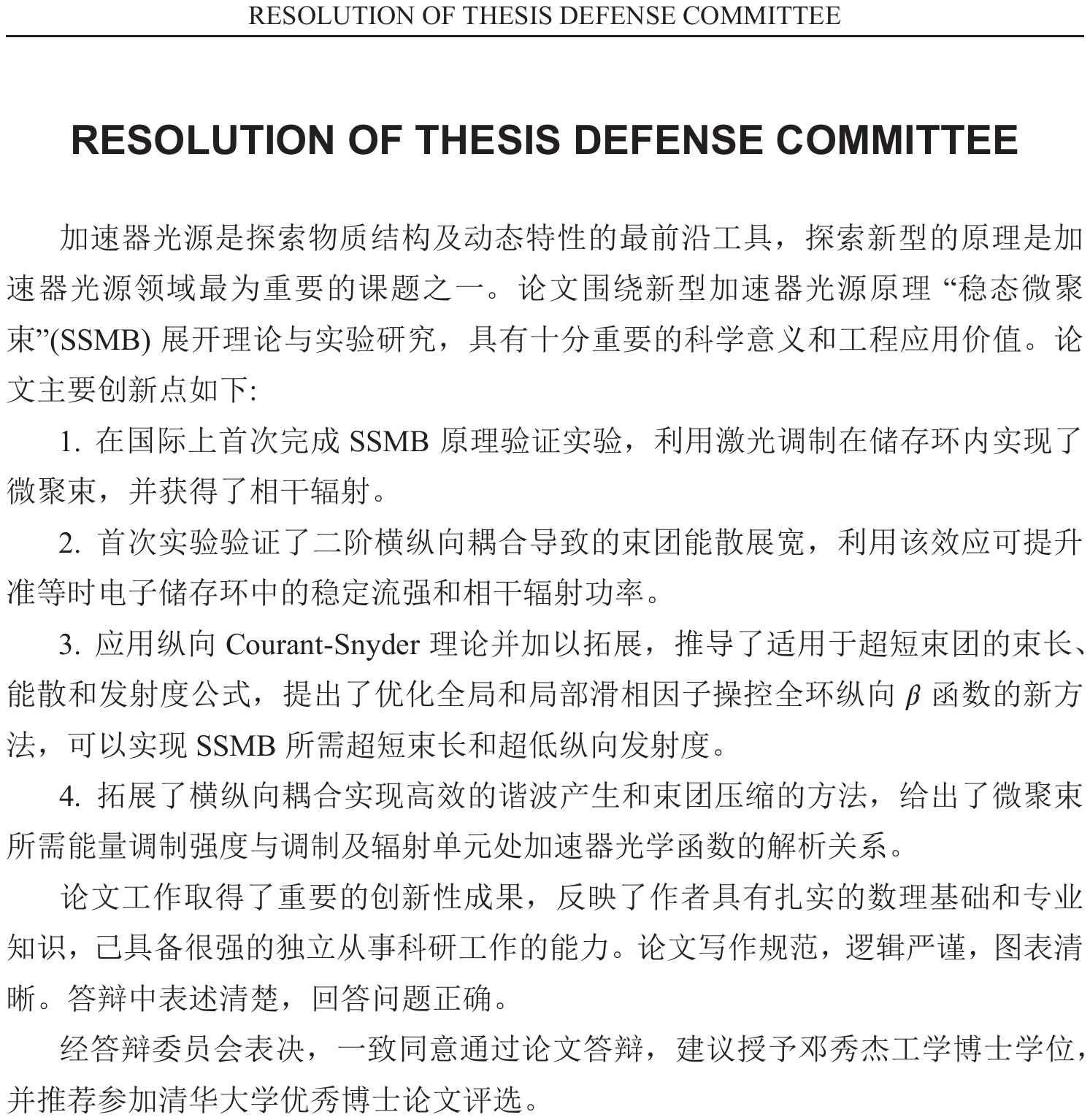}

%


\end{document}

%% file: data/acronym.tex
%
%

\extrachap{Acronyms}

%

\begin{description}[CABR]
 \item[ADM] Angular dispersion-induced microbunching
 \item[ARPES] Angle-resolved photoemission spectroscopy
 \item[BPM] Beam position monitor
 \item[CBS] Compton-backscattering
 \item[CHG] Coherent harmonic generation
 \item[CSR] Coherent synchrotron radiation
 \item[CW] Continuous-wave
 \item[DA] Dynamic aperture
 \item[DC] Direct current
 \item[EEHG] Echo-enabled harmonic generation 
 \item[EEX] Emittance exchange
 \item[ERL] Energy recovery linac
 \item[EUV] Extreme ultraviolet
 \item[FEL] Free-electron laser
 \item[FFAG] Fixed-field alternating-gradient
 \item[FWHM] Full-width at half-maximum
 \item[GHz] Gigahertz
 \item[HGHG] High-gain harmonic generation
 \item[HZB] Helmholtz-Zentrum Berlin
 \item[IBS] Intrabeam scattering
 \item[IOTA] Integrable Optics Test Accelerator
 \item[kHz] Kilohertz
 \item[LCLS] Linac Coherent Light Source
 \item[LGB] Longitudinal gradient bend
 \item[LSF] Longitudinal strong focusing
 \item[MA] Momentum aperture
 \item[meV] Milli-electronvolt
 \item[MeV] Mega-electronvolt
 \item[MHz] Megahertz
 \item[MLS] Metrology Light Source
 \item[OEC] Optical enhancement cavity
 \item[PEHG] Phase-merging enhanced harmonic generation
 \item[PoP] Proof-of-principle
 \item[PSD] Power spectral density
 \item[PTB] Physikalisch-Technische Bundesanstalt
 \item[RMS] Root mean square
 \item[RF] Radio frequency
 \item[SASE] Self-Amplified Spontaneous Emission
 \item[SFP] Stable fixed point
 \item[SLIM] Solution by linear matrices
 \item[SSMB] Steady-state microbunching
 \item[TEM] Transverse electromagnetic
 \item[TGB] Transverse gradient bend
 \item[THz] Terahertz
 \item[TLC] Transverse-longitudinal coupling
 \item[TME] Theoretical minimum emittance
 \item[UFP] Unstable fixed point
 \item[XFEL] X-ray free-electron laser
\end{description}

%% file: data/chap01.tex
\chapter{Introduction}
\label{cha:intro}

Particle accelerators as photon sources are advanced tools in investigating the structure and dynamical properties of matter, and have enabled advances in science and technology for more than half a century~\cite{chao2011reviews}. The present workhorses of these sources are storage ring-based synchrotron radiation facilities~\cite{elder1947radiation,tzu1948,schwinger1949classical,zhao2010storage} and linear accelerator-based free-electron lasers (FELs)~\cite{madey1971stimulated,kroll1978stimulated,kondratenko1980generating,bonifacio1984collective,emma2010first,huang2007review,pellegrini2016physics}. These two kinds of sources deliver light with high repetition rate and high peak brilliance (power), respectively. Some applications, however, do need high average power and high photon flux. Kilowatt extreme ultraviolet (EUV) light sources, for example, are urgently needed by the semiconductor industry for EUV lithography~\cite{bakshi2018euv}. Another example is that the ultra-high-energy-resolution angle-resolved photoemission spectroscopy (ARPES) requires that the initial photon flux before monochromator is high enough~\cite{damascelli2003angle,lv2019angle}. To realize high average power and photon flux, a high peak power or a high repetition rate alone is not sufficient. We need both of them simultaneously. 

The key of the high peak power of FELs lies in microbunching, which means the electrons are bunched or sub-bunched to a longitudinal dimension smaller than the radiation wavelength so that the electrons radiate in phase and thus cohere~\cite{schwinger1996radiation,nodvick1954suppression,gover2019superradiant}. The power of coherent radiation is proportional to the number of the radiating electrons squared, therefore can be orders of magnitude stronger than the equivalent incoherent radiation in which the power dependence on the electron number is linear. The Self-Amplified Spontaneous Emission (SASE) scheme~\cite{kondratenko1980generating,bonifacio1984collective} of microbunching making the high-gain FELs so powerful, however, is actually a collective beam instability which cannot sustain in a continuous manner and the  repetition rate of the radiation is limited by the driving source. There are now active efforts devoted to the high-repetition-rate FELs, for example the superconducting FEL~\cite{altarelli2007european,galayda2018lcls,zhao2018sclf}, X-ray FEL oscillator~\cite{kim2008proposal}, energy recovery linac (ERL)~\cite{nakamura2012review} and similar. However, the realization of a high-average-power, narrow-band, and continuous-wave (CW) short-wavelength light source remains a challenge.

A mechanism called steady-state microbunching (SSMB) was proposed in Ref.~\cite{ratner2010steady} to resolve this issue. The idea of SSMB is that by a phase-space manipulation of an electron beam, microbunching forms and stays in a steady state each time going through a radiator in a storage ring. The steady state here means a balance of  quantum excitation and radiation damping, a true equilibrium in the context of storage ring beam dynamics. Once realized, the strong coherent radiation from microbunching and the high repetition rate of a storage ring combine to make a high-average-power photon source. 

\begin{figure}[tb]
	\centering 
	\includegraphics[width=0.9\textwidth]{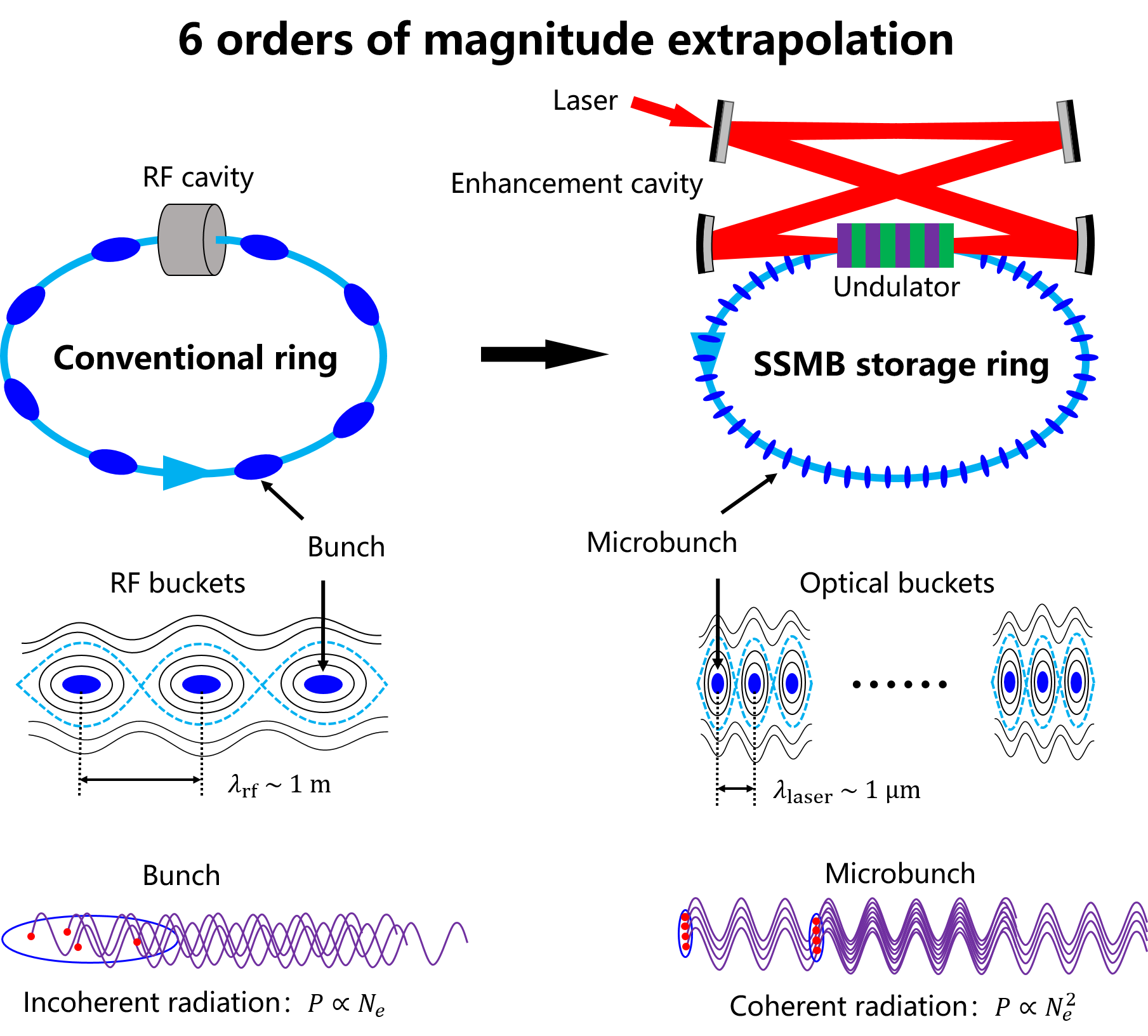}
	\caption{
		\label{fig:Chap1-SSMBSchematic} 
		Schematic layout of a conventional storage ring (left) and an SSMB storage ring (right). 
	}
\end{figure}

The schematic layout of an SSMB storage ring and its operating principle in comparison to a conventional storage ring are shown in Fig.~\ref{fig:Chap1-SSMBSchematic}.  SSMB replaces the  conventional bunching system in a storage ring, namely the radio frequency (RF) cavity, with a laser modulation system. As the wavelength of laser ($\sim\mu$m) is typically six orders of magnitude smaller than that of an RF wave ($\sim$ m), a much shorter bunch, i.e., microbunch, can thus be anticipated by invoking this replacement together with a dedicated magnetic lattice. To provide adequate and stable longitudinal focusing such that microbunches can be formed and sustained, SSMB requires a powerful phase-locked laser to interact with electrons on a turn-by-turn basis. The realization of such a laser system usually demands an optical enhancement cavity. A laser cannot effectively interact with the co-propagating electrons if the electrons go through a straight line, as the electric field of a laser is perpendicular to the  laser propagation direction. A modulator which bends the electron trajectory transversely is thus needed. The modulator is usually an undulator, which is a periodic structure of dipole magnets.   Note that to avoid the head-on collisions (Compton back-scattering) between the reflected laser and the electrons, a four-mirror optical cavity, instead of a two-mirror one, is chosen for the illustration in Fig.~\ref{fig:Chap1-SSMBSchematic}. 

The microbunching in SSMB is from the active longitudinal focusing provided by the laser modulator, just similar to the conventional RF bunching through phase stability principle \cite{veksler1944new,mcmillan1945synchrotron}. The radiation in SSMB, unlike that in an FEL, is a passive process and the radiator can be rather short, for example it can be a simple dipole magnet or a short undulator. The modulator undulator is also much shorter than the radiator undulator in a high-gain FEL. Therefore, there is no FEL mechanism invoked in the bunching or radiation process in SSMB. If there is some unavoidable FEL effects, it needs to be controlled within a safe region to not destroy the steady state micobunches.

Note that we have not presented explicitly the energy compensation system for SSMB in Fig.~\ref{fig:Chap1-SSMBSchematic}. The modulation laser in principle can be used to supplement the radiation energy loss of the electrons, just like the traditional RF, but this may not be a cost-effective choice. Besides, the electron beam current and output radiation power will also be limited by the incident laser power. Instead, one may just use a traditional RF cavity for the energy compensation. If a larger filling factor of the electron beam is desired, it could also be one or several induction linacs. In the present envisioned high-average-power SSMB photon source, induction linac is tentatively used as the energy compensation system and the filling factor of the electron beam in the storage ring can be rather large, for example larger than $50\%$. 

SSMB is a general concept with two key ingredients, i.e., microbunching and steady state. Several specific SSMB scenarios \cite{ratner2010steady,ratner2011reversible,jiao2011terahertz,chao2016high} have been proposed for different target radiation wavelengths since the first publication advancing the SSMB concept. One natural SSMB scheme, as just mentioned, is using a laser modulator to microbunch the electron beam in the optical microbuckets, similar to the RF-bunched bunches in conventional RF buckets, as shown in Fig.~\ref{fig:Chap1-SSMBSchematic}.
Such an SSMB setup can be used as a laser amplifier to amplify the modulation laser power. If the microbunch is much shorter than the modulation laser wavelength, it can also be used for high harmonic generation.


The above approach can realize a bunch length one to two orders of magnitude smaller than the modulation laser wavelength. To generate coherent radiation at even shorter wavelengths, for example, EUV wavelengths or soft X-ray, bunch length as short as nm level is needed. The  bunch length therefore needs to be compressed further. Two main scenarios are most promising for this goal. The first approach is called longitudinal strong focusing SSMB~\cite{chao2016high}. In this scheme, the longitudinal strong focusing principle is applied in the longitudinal dimension, not unlike its transverse counterpart \cite{christofilos1950focusing,courant1952strong} which is the basis for modern particle accelerators. In addition to beam dynamics in the longitudinal dimension alone, ultra-short bunches can also be obtained by implementing a transverse-longitudinal coupled lattice, since then there will be some freedom in projecting the three eigen emittances of the electron beam into different physical dimensions.  By taking advantage of the fact that the vertical emittance in a planar electron storage ring is rather small, the application of such transverse-longitudinal coupling schemes can help relax the requirement on the modulation laser power to realize extremely short electron bunches in SSMB. Such a scheme can be viewed as a generalized longitudinal strong focusing \cite{deng2021theorem}. In addition to strong focusing, the second lattice approach is called reversible seeding SSMB \cite{ratner2011reversible}. 
This approach produces microbunching and high laser harmonics similar to the harmonic generation schemes in FEL seeding techniques. However, to ensure that the microbunching process can repeat turn by turn, the modulation and microbunching processes are perfectly canceled by means of de-microbunching and reverse modulation following the radiator. Outside the reversible seeding module, the beam is identical to that in an ordinary ring. It can be an RF or laser-bunched beam or a coasting beam. This scheme relaxes the requirements on the overall lattice design outside the reversible seeding module, but the modulation cancellation requires high precision. 

The potential of SSMB as a new light source mechanism is tremendous, as can be viewed from two perspectives.
\begin{itemize}
	\item From the accelerator physics perspective: there is now active and important development on low-transverse-emittance or diffraction-limited storage rings~\cite{teng1984minimizing,eriksson2014diffraction}, which focus on the transverse dimension of the electron beam. SSMB, however, is trying to dig the potential of the longitudinal dimension of electron beam in storage rings. The large compactification of the bunching system wavelength (from m to $\mu$m) gives room for exciting development of accelerator physics.
	\item From the synchrotron radiation application perspective: once realized, SSMB can deliver photons with high average power, high repetition rate (megahertz (MHz) to CW) and narrow bandwidth, at wavelength ranging from terahertz (THz) to soft X-ray. Such a novel photon source could provide unprecedented opportunities
	for accelerator photon science and technological applications. For
	example, SSMB is promising for generating high-power EUV radiation
	for EUV lithography~\cite{bakshi2018euv}. Energy-tunable high-flux narrowband EUV
	photons are also highly desirable in condensed matter physics study, for example in the application of ultra-high-energy-resolution ARPES to study the energy gap distribution and electron states of magic-angle graphene in superconductivity \cite{cao2018unconventional}. Ultrahigh-power
	deep ultraviolet and infrared sources are potential research tools in
	atomic and molecular physics. Moreover, new nonlinear phenomena
	and dynamical properties of materials can be driven and studied
	by high-peak and average-power THz sources~\cite{carr2002high,cole2001coherent}. Besides high
	power, SSMB can also produce ultrashort (sub-femtosecond to attosecond)
	photon pulses and pulse trains with definite phase relations,
	which could be useful in attosecond physics investigations~\cite{krausz2009attosecond}.	
\end{itemize}   

To promote the SSMB research and in particular develop an EUV SSMB storage ring, a task force has been established at Tsinghua University~\cite{tang2018overview,chao2018SSMB,rui2018strong,pan2019storage,pan2020research,li2019lattice,deng2020single,deng2020widening,deng2021experimental,tang2020first,feikes2021progress,zhang2021ultralow,deng2021courant,deng2021theorem,tang2022steady,deng2022electron,deng2022average}, in collaboration with researchers from different institutes. A key progress of the collaboration is the recent success of the SSMB proof-of-principle (PoP) experiment conducted at the Metrology Light Source (MLS) in Berlin ~\cite{deng2021experimental,tang2020first,feikes2021progress}. The study of SSMB beam physics~\cite{deng2020single,deng2020widening,zhang2021ultralow,deng2021courant,deng2021theorem,deng2022electron}, radiation characteristics~\cite{deng2022average} and magnet lattice design~\cite{rui2018strong,pan2019storage,pan2020research,li2019lattice} for the EUV SSMB storage ring is also ongoing and encouraging progresses are being achieved. Both lattice approaches mentioned above, i.e., longitudinal strong focusing and reversible seeding, are being actively studied by the SSMB task force. A hybrid lattice scheme to incorporate the advantages of both approaches is also under investigation. Novel 6D phase-space manipulation schemes and interesting single-particle and collective effects can be envisioned and need to be investigated. The work presented in this dissertation is part of these active ongoing investigations.

\begin{figure}[tb]
	\centering 
	\includegraphics[width=0.8\textwidth]{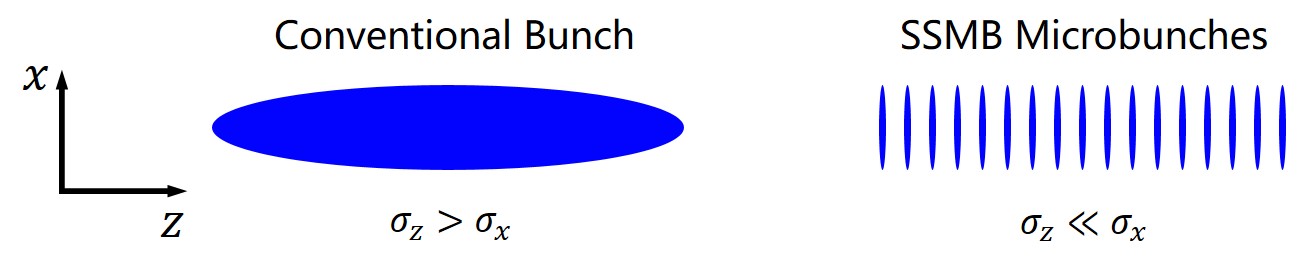}
	\caption{
		\label{fig:Chap1-SSMBMicrobunches} 
		Bunch in a conventional storage ring (left) and microbunches in an SSMB storage ring (right).  
	}
\end{figure}

Before the detailed description of the research work, it might be helpful for us to get a feeling about the qualitative difference of SSMB microbunches compared to conventional bunches in a storage ring. The bunch length in SSMB is orders of magnitude smaller than that in a conventional ring. The transverse size, however, is not so much different. The transverse size of an SSMB beam can thus be much larger than the microbunch length, while the contrary is true in a conventional ring, as shown in Fig.~\ref{fig:Chap1-SSMBMicrobunches}.  Therefore for SSMB, it can be anticipated that the longitudinal dynamics and the coupling of the transverse and longitudinal dimensions should be controlled in precision. Another remark is that in a conventional storage ring, the bunches in different RF buckets form a beam current with clear spikes separated from each other. As we will see later, in SSMB, the current spikes may overlap with each other and becomes less sharp in some places of the ring, as the bunch lengthening from the transverse emittance can easily be comparable or even longer than the modulation laser wavelength which is the separation length of microbunches. This bunch lengthening and current overlapping may have significant impact on collective effects, for example intrabeam scattering (IBS) and coherent synchrotorn radiation (CSR), of the microbunches in a storage ring. The fast variation of bunch length around the ring may make the adiabatic approximation usually adopted for the longitudinal dimension, in the study of single-particle and collective effects, break down. Besides, in an SSMB ring, the radiation of one microbunch can catch up with its frontier microbunches, while this usually does not happen in a conventional ring. Therefore new collective instability mechanism may arise from this new characteristic. The study of such new beam dynamics is important for SSMB. 

In this dissertation, we have conducted theoretical and experimental studies for SSMB, with important results achieved. The work presented can be summarized as: first, how to realize SSMB; second, what radiation characteristics can we obtain from the formed SSMB; and third, experimentally demonstrate the working mechanism of SSMB in a real machine.   In the theoretical part, we first conducted in-depth analysis of the single-particle effects vital for the formation and transport of microbunching in a storage ring. Chapter~\ref{cha:Longitudinal} is devoted to the longitudinal dynamics and Chap.~\ref{cha:TLC} focuses on the transverse-longitudinal coupling (TLC) dynamics. Chapter~\ref{cha:Radiation} is the theoretical and numerical investigations on the average and statistical characteristics of radiation generated from the formed microbunching. In the experimental part, we report our work on the first demonstration of the mechanism of SSMB in Chap.~\ref{cha:pop}. In addition to the SSMB PoP experiment, we have also reported the first experimental validation of energy widening induced by a second-order TLC  effect in Chap.~\ref{cha:TLC}.  Finally, in Chap.~\ref{cha:summary} we present a short summary of the dissertation and some outlooks for the future research.

The highlights of the dissertation contribution are:

1. The first proof-of-principle experiment of SSMB, which is the first milestone in the development of an SSMB light source. Challenging as it is, we have shown that SSMB is doable in a real machine with efforts;

2. The first experimental verification of beam energy widening arising from betatron motion, which can be used to boost the stable beam current and coherent radiation power of an electron storage ring working in quasi-isochronous regime; 

3. The development and application of the longitudinal Courant-Snyder formalism and derivation of new formulae of bunch length, energy spread and longitudinal emittance in an electron storage ring, beyond the classical scaling laws. Based on the derivation and analysis, the method of optimizing the global and local phase slippages simultaneously to tailor the longitudinal $\beta$ function around ring is proposed to generate the ultra-short bunch length and ultra-small longitudinal emittance, as required by SSMB;

4. An universal analysis and the proof of two theorems on transverse-longitudinal coupling-based bunch compression and harmonic generation schemes, which are of value for the promotion of such schemes for various applications, for example to lower the requirement on modulation laser power in SSMB by taking advantage of the fact that the vertical emittance in a planar electron storage ring is rather small;

5. A definition and derivation of the generalized transverse form factor of an electron beam, which can quantify the impact of electron beam transverse size on the coherent undulator radiation in an efficient way. An elegant formula for the coherent radiation power fluctuation, which is useful in investigating its potential applications for example in beam diagnostics.

%% file: data/chap02.tex
\chapter{Longitudinal Dynamics}
\label{cha:Longitudinal}

In this chapter, we investigate the single-particle longitudinal dynamics of SSMB. The motivation is to answer the question: how to realize the short bunch length and small longitudinal emittance in an electron storage ring, as required by SSMB? We start from the linear dynamics and then step into the nonlinear ones. For the longitudinal dynamics without coupling from the transverse dimension, what we can play are the momentum compaction and RF systems, for SSMB the laser modulators. The momentum compaction is a measure of particle energy dependence of the recirculation path length
\begin{equation}
\alpha=\frac{\Delta C/C_{0}}{\Delta E/E_{0}}=\frac{1}{C_{0}}\oint\frac{D_{x}(s)}{\rho(s)}ds,
\end{equation}
where $C_{0}$ is the ring circumference, $E_{0}$ is the particle energy, ${D}_{x}$ is the horizontal dispersion which is a measure
of the energy dependence of the horizontal position of the particle, $\rho$ is the bending radius. Note that the curvilinear (Frenet-Serret) coordinate system and the state vector ${\bf X}=(x,x',y,'y',z,\delta)^{T}$, with $^{T}$ representing the transpose, are used throughout this dissertation. Considering the energy-dependent velocity, the particle energy dependence of the revolution time can be quantifies by a parameter named phase slippage factor $\eta=\frac{\Delta T/T_{0}}{\Delta E/E_{0}}=\alpha-\frac{1}{\gamma^{2}}$, with $\gamma$ the relativistic factor. For linear dynamics, the phase slippage to the longitudinal dimension is like the drift space to the transverse dimension, while the RF kick in linear approximation to the longitudinal dimension is like the quadrupole to the transverse dimension. The difference is that the phase slippage can either be positive or negative, while the drift space can only have a positive physical length. As mentioned in Chap.~\ref{cha:intro}, the adiabatic approximation may not apply in an SSMB storage ring, as the bunch length can vary significantly around the ring. Courant-Snyder analysis can be invoked for the linear dynamics analysis beyond adiabatic approximation. Usually there is only one RF cavity in a storage ring, the longitudinal optics can be manipulated with more freedom with multiple RFs. For example, the strong focusing principle can be implemented in the longitudinal dimension to realize ultra-short bunch length and small longitudinal emittance, not unlike its transverse counterpart. For nonlinear dynamics, both the nonlinearity of the phase slippage and the sinusoidal modulation waveform can lead to complex and rich beam dynamics. In the following we do the investigation along this brief review.

\begin{figure}[tb] 
	\centering 
	\includegraphics[width=0.4\textwidth]{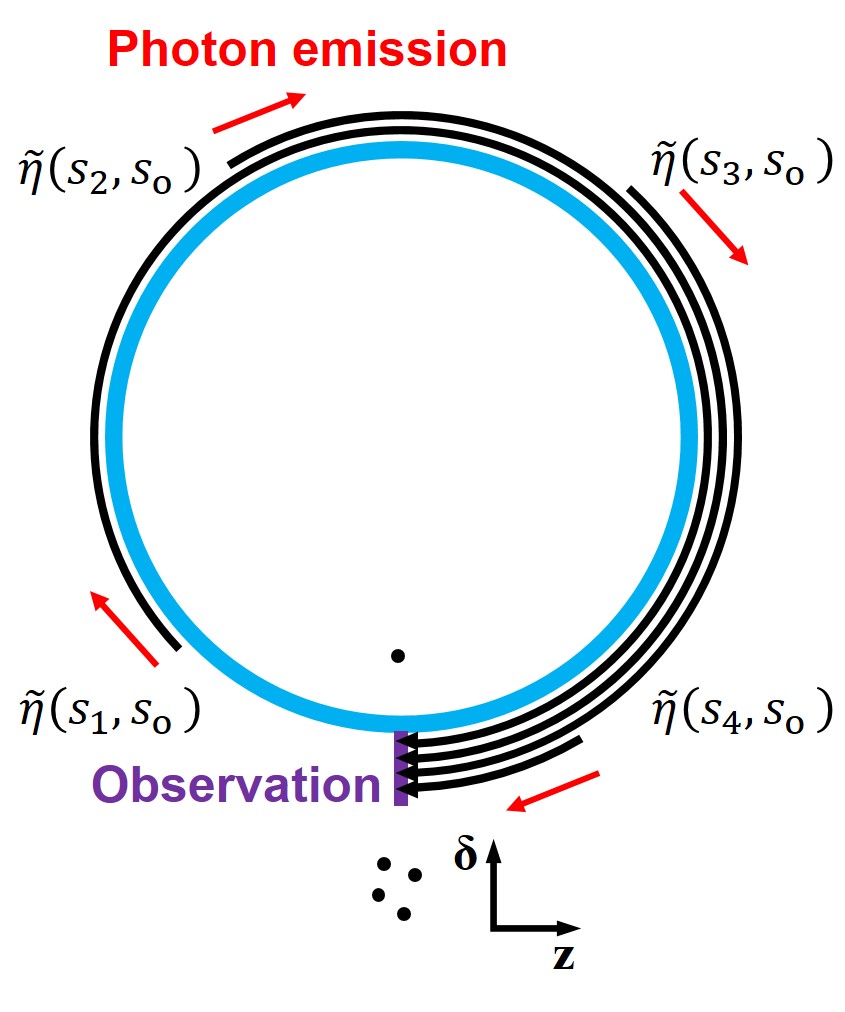}
	\caption{
		\label{fig:Chap2-PartialAlpha} 
		Physical picture of the partial phase slippage factors and quantum excitation. Particles undergo diffusion in both the particle energy and the longitudinal coordinate in each turn, giving rise to longitudinal emittance, as a result of quantum excitation.
	}
\end{figure}

\section{Linear Phase Slippage}\label{sec:PartialAlpha}

\subsection{Global and Local Phase Slippage}
SSMB means an ultrashort electron bunch in an equilibrium state. One successful method of realizing short bunches in an electron storage ring is the implementation of a quasi-isochronous (low-alpha) lattice, which means particles with different energies complete one revolution using almost the same time. The reason behind is the well-known $\sigma_{z}\propto\sqrt{|\eta|}$ scaling law of the ``zero-current" bunch length given by Sands \cite{sands1970physics}, in which $\eta=\alpha-\frac{1}{\gamma^{2}}$ is the global phase slippage factor of the ring as introduced just now. However, from single-particle dynamics perspective, there is a fundamental effect limiting the lowest bunch length realizable in an electron storage ring originating from the stochasticity of photon emission time. This effect is first theoretically investigated by Shoji et al. \cite{shoji1996longitudinal,shoji1998longitudinal}, and recently more accurately analyzed by us using the longitudinal Courant-Snyder formalism~\cite{deng2021courant,biscari2005bunch}.   This stochasticity results in a diffusion of the electron longitudinal coordinate $z$ even if the global phase slippage of the ring is zero as we cannot make all the local or partial phase slippages zero simultaneously. The partial phase slippage factor from the $j^{\text{th}}$ photon emission point $s_{j}$ to the observation point $s_{o}$ is defined as
\begin{equation}\label{eq:PartialAlpha}
\tilde{\eta}(s_{j},s_{o})=\frac{1}{C_{0}}\int_{s_{j}}^{s_{o}}\left(\frac{{D}_{x}(s)}{\rho(s)}-\frac{1}{\gamma^{2}}\right)ds.
\end{equation} 
The physical picture of the partial phase slippage and quantum excitation in both the particle energy and longitudinal coordinate, therefore the longitudinal emittance, is shown in Fig.~\ref{fig:Chap2-PartialAlpha}.

Due to this quantum diffusion, there exists a bunch length limit and the energy spread grows significantly when the bunch length is pushed close to the limit. 
This effect is of vital importance for SSMB and other ideas invoking ultra-short electron bunches or ultra-small longitudinal emittance in storage rings. 
While the treatment of this quantum excitation is intrinsically embedded in Chao's  solution by linear matrices (SLIM) formalism~\cite{chao1979evaluation}, which can give the equilibrium electron beam parameters in a storage ring straightforwardly by a numerical calculation, analytical formulas for this effect are still highly desired. The motivation is that these analytical results can help us better understand its physics and find the method to control this quantum diffusion to minimize the longitudinal emittance and bunch length, for example in an SSMB storage ring. In the following, we apply the longitudinal Courant-Snyder formalism in a planar $x$-$y$ uncoupled ring for the analysis of this effect, which can be viewed as an application of the SLIM formalism where concise analytical result can be obtained. Formulas for the equilibrium longitudinal emittance, energy spread and bunch length are derived, based on which some important points of this effect are analyzed. The application of multiple RF cavities, or laser modulators in an SSMB storage ring, for longitudinal strong focusing are discussed using the same formalism with important observations made.

\subsection{Courant-Snyder Formalism}\label{sec:CS}

The quantum diffusion in $z$ in a storage ring emphasizes the fact that we should consider the quantum excitation on the longitudinal emittance, rather than that on the particle energy alone. The key to understanding the effect is to change from the global viewpoint to a local one, i.e., the quantum excitation at different places around the ring contribute to the longitudinal emittance with different strengths, just like its transverse counterpart. For an accurate analysis of this effect, here we apply Chao's SLIM formalism \cite{chao1979evaluation,chao1981evaluation}.  
SLIM is an early effort to generalize the Courant-Snyder theory~\cite{courant1958theory} from 1D (2D phase space) to higher dimensions. The spirit of SLIM formalism is eigen analysis. It invokes $6\times6$ ($8\times8$ if spin considered) general transport matrices and applies to 3D (6D phase space) general coupled lattice without the assumption of longitudinal weak focusing. Concerning the evaluation of equilibrium beam parameters in an electron storage ring, SLIM can be viewed as a method of solving linear Fokker-Planck equation \cite{wang1945theory,chandrasekhar1943stochastic} without adopting the adiabatic approximation. Therefore, it can account for the variation of one-turn map around the ring. In other words, the contribution of diffusion and damping, for example the quantum excitation and radiation damping, to the eigen emittances depend on the local one-turn map.  

According to SLIM, the equilibrium eigen emittances and second moments of the beam in an electron storage ring are
\begin{equation}\label{eq:emittance}
\epsilon_{k}=\frac{C_{L}\gamma^{5}}{c\alpha_{k}}\oint \frac{|{\bf E}_{k5}(s)|^{2}}{|\rho(s)|^{3}}ds
\end{equation}
and
\begin{equation}\label{eq:SLIMsecondmoments}
\Sigma_{ij}(s)=\langle {\bf X}_{i}{\bf X}_{j}\rangle(s)=2\sum_{k=I, II, III}\epsilon_{k}\text{Re}[{\bf E}_{ki}(s){\bf E}_{kj}^{*}(s)],
\end{equation}
where  $c$ is the speed of light in free space, $\alpha_{k}$ are the damping constants of the three eigen modes, $C_{L}={55r_{e}\hbar}/{(48\sqrt{3}m_{e})}$, with $\hbar$ the reduced Planck's constant, $r_{e}$ the electron classical radius and $m_{e}$ the electron rest mass, ${\bf E}_{k}$ are the eigenvectors of the linear one-turn map ${\bf M}$, $^{*}$ denotes complex conjugate and Re[] means taking the real part of a complex number. The eigenvectors satisfy ${\bf E}_{k}^{*}={\bf E}_{-k}$, ${\bf E}_{j}^{T}{\bf S}{\bf E}_{i}=0$ unless $j=-i$, and the normalization condition
\begin{equation}
({\bf E}_{k}^{*})^{T}{\bf S}{\bf E}_{k}=\begin{cases}
&i,\ k=I,II,III,\\
&-i,\ k=-I,-II,-III.
\end{cases}
\end{equation}
This normalization condition is preserved as a function of $s$ due to the symplecticity of the transfer matrices.
The eigen emittances $\epsilon_{k}$ are defined as the positive eigenvalues of $i\Sigma{\bf S}$ where $i$ is the imaginary unit, $\Sigma=\langle {\bf X}{\bf X}^{T}\rangle$ is the second moments matrix and ${\bf S}$ is the symplectic form
\begin{equation}
{\bf S}=\left(\begin{matrix}
0&1&0&0&0&0\\
-1&0&0&0&0&0\\
0&0&0&1&0&0\\
0&0&-1&0&0&0\\
0&0&0&0&0&1\\
0&0&0&0&-1&0\\
\end{matrix}\right).
\end{equation}
The subscripts $_{I,II,III}$ represent the three eigen modes and in the planar uncoupled case correspond to the usual horizontal, vertical and longitudinal dimensions ${x,y,z}$. Note that a linear symplectic transport does not change the eigen-emittances of a particle beam~\cite{dragt1992general}, which can be seen from the following argument. Suppose a particle beam is transported by a linear symplectic matrix ${\bf T}$, then we have
\begin{equation}
i\Sigma_{\text{new}}{\bf S}=i{\bf T}\Sigma_{\text{old}}{\bf T}^{T}{\bf S}={\bf T}(i\Sigma_{\text{old}}{\bf S}){\bf T}^{-1},
\end{equation}
in which the last step has invoked the symplecticity of ${\bf T}$, i.e., ${\bf T}^{T}{\bf S}{\bf T}={\bf S}$.
Therefore, $i\Sigma_{\text{new}}{\bf S}$ is related to $i\Sigma_{\text{old}}{\bf S}$ by a similarity transform, thus having the same eigenvalues.


To simplify the discussion, here we only consider the horizontal and longitudinal dimensions and use the state vector 
${\bf X}=\left(x,x',
z,\  \delta
\right)^{T}
$. Under the assumptions that the ring is planar $x$-$y$ uncoupled and the RFs are placed at dispersion-free locations, which is the typical setup for present synchrotron sources, the betatron coordinate ${\bf X}_{\beta}={\bf B}{\bf X}$ can be introduced to parametrize the transfer matrix in a diagonal form, with the dispersion matrix given by
\begin{equation}
\bf{B}=\left(\begin{matrix}
1&0&0&-D_{x}\\
0&1&0&-D_{x}'\\
D_{x}'&-D_{x}&1&0\\
0&0&0&1\\
\end{matrix}\right).
\end{equation}
The one-turn map ${\bf M}$ of ${\bf X}$ is related to the one-turn map ${\bf M}_{\beta}$ of ${\bf X}_{\beta}$ by
\begin{equation}
{\bf M}={\bf B}^{-1}{\bf M}_{\beta}{\bf B},\ 
\end{equation}
with
\begin{equation}\label{eq:OTMCS1}
\begin{aligned}
{\bf M}_{\beta}&=\left(
\begin{matrix}
{\bf M}_{x\beta}&0\\
0&{\bf M}_{z\beta}
\end{matrix}
\right),\\
{\bf M}_{x,z\beta}&=\left(
\begin{matrix}
\cos\Phi_{x,z}+\alpha_{x,z}\sin\Phi_{x,z}&\beta_{x,z}\sin\Phi_{x,z}\\
-\gamma_{x,z}\sin\Phi_{x,z}&\cos\Phi_{x,z}-\alpha_{x,z}\sin\Phi_{x,z}
\end{matrix}
\right),\
\end{aligned}
\end{equation}
in which $\Phi_{x}=2\pi\nu_{x}$ and $\Phi_{z}=2\pi\nu_{s}$ are the betatron and synchrotron phase advance per turn.
The eigenvectors of ${\bf M}_{\beta}$ can be expressed using the Courant-Snyder functions as
\begin{equation}
{\bf v}_{x}=\frac{1}{\sqrt{2}}\left(\begin{matrix}
\sqrt{\beta_{x}}\\
\frac{i-\alpha_{x}}{\sqrt{\beta_{x}}}\\
0\\
0\\
\end{matrix}\right),\
{\bf v}_{z}=\frac{1}{\sqrt{2}}\left(\begin{matrix}
0\\
0\\
\sqrt{\beta_{z}}\\
\frac{i-\alpha_{z}}{\sqrt{\beta_{z}}}\\
\end{matrix}\right).
\end{equation}
Therefore, the eigenvectors of ${\bf M}$ are 
\begin{equation}\label{eq:PlanarEigenVectors}
\begin{aligned}
{\bf E}_{x}&={\bf B}^{-1}{\bf v}_{x}=\frac{1}{\sqrt{2}}\left(\begin{matrix}
\sqrt{\beta_{x}}\\
\frac{i-\alpha_{x}}{\sqrt{\beta_{x}}}\\
-\sqrt{\beta_{x}}D_{x}'+\frac{i-\alpha_{x}}{\sqrt{\beta_{x}}}D_{x}\\
0\\
\end{matrix}\right),\ {\bf E}_{z}={\bf B}^{-1}{\bf v}_{z}=\frac{1}{\sqrt{2}}\left(\begin{matrix}
\frac{i-\alpha_{z}}{\sqrt{\beta_{z}}}D_{x}\\
\frac{i-\alpha_{z}}{\sqrt{\beta_{z}}}D_{x}'\\
\sqrt{\beta_{z}}\\
\frac{i-\alpha_{z}}{\sqrt{\beta_{z}}}\\
\end{matrix}\right).
\end{aligned}
\end{equation}
According to SLIM,  the equilibrium horizontal and longitudinal emittance are then
\begin{equation}\label{eq:longitudinalEmittance}
\begin{aligned}
\epsilon_{x}&=\langle J_{x}\rangle=\frac{55}{96\sqrt{3}}\frac{\alpha_{F}{\lambdabar}_{e}^{2}\gamma^{5}}{\alpha_{\text{H}}}\oint \frac{\mathcal{H}_{x}(s)}{|\rho(s)|^{3}}ds,\\
\epsilon_{z}&=\langle J_{z}\rangle=\frac{55}{96\sqrt{3}}\frac{\alpha_{F}{\lambdabar}_{e}^{2}\gamma^{5}}{\alpha_{\text{L}}}\oint \frac{\beta_{z}(s)}{|\rho(s)|^{3}}ds,
\end{aligned}
\end{equation}
in which
\begin{equation}
\begin{aligned}
J_{x}&=\frac{\left(x-D_{x}\delta\right)^{2}+\left[\alpha_{x}(x-D_{x}\delta)+\beta_{x}(x'-D_{x}'\delta)\right]^{2}}{2\beta_{x}},\\
J_{z}&=\frac{\left(z-D_{x}'x-D_{x}x'\right)^{2}+\left[\alpha_{z}(z-D_{x}'x-D_{x}x')+\beta_{z}\delta\right]^{2}}{2\beta_{z}},\\
\end{aligned}
\end{equation} 
and $\langle\rangle$ here means particle ensemble average, $\alpha_{\text{H}}$ and $\alpha_{\text{L}}$ are the horizontal and longitudinal damping constants,  $\alpha_{F}=\frac{1}{137}$ is the fine structure constant, $\lambdabar_{e}={\lambda_{e}}/{2\pi}=386\ \text{fm}$  is the reduced Compton wavelength
of electron and $\mathcal{H}_{x}=\gamma_{z}{D}_{x}^{2}+2\alpha_{x}{D_{x}}{D_{x}'}+\beta_{x}{D_{x}'}^{2}=\frac{D_{x}^{2}+\left(\alpha_{x}D_{x}+\beta_{x}D_{x}'\right)^{2}}{\beta_{x}}$ is the horizontal chromatic function. 

After getting the eigen emittances, we can obtain the second moments of the beam according to Eq.~(\ref{eq:SLIMsecondmoments}) and (\ref{eq:PlanarEigenVectors}), 
%
more specifically,
\begin{equation}
\Sigma_{\beta}=\langle {\bf X}_{\beta}{\bf X}^{T}_{\beta}\rangle=\left(\begin{matrix}
\epsilon_{x}\beta_{x}&-\epsilon_{x}\alpha_{x}&0&0\\
-\epsilon_{x}\alpha_{x}&\epsilon_{x}{\gamma_{x}}&0&0\\
0&0&\epsilon_{z}\beta_{z}&-\epsilon_{z}\alpha_{z}\\
0&0&-\epsilon_{z}\alpha_{z}&\epsilon_{z}{\gamma_{z}}\\
\end{matrix}\right),
\end{equation}
and
\begin{equation}\label{eq:2ndmoments}
\begin{aligned}
\Sigma&={\bf B}^{-1} \Sigma_{\beta}\left({\bf B}^{-1}\right)^{T}=\left(\begin{matrix}
\Sigma_{\text{H}}&\Sigma_{\text{HL}}\\
\Sigma_{\text{HL}}^{T}&\Sigma_{\text{L}}\\
\end{matrix}\right),
\end{aligned}
\end{equation}
where 
\begin{equation}\label{eq:2ndmomentsSeparate}
\begin{aligned}
\Sigma_{\text{H}}&=\left(\begin{matrix}
\epsilon _x\beta _x+\epsilon _z \gamma _z  D_{x}^2  &  -\epsilon _x\alpha _x  + \epsilon _z\gamma _z D_{x}D_{x}'\\
-\epsilon _x\alpha _x+\epsilon _z\gamma _z D_{x}D_{x}'  &  \epsilon _x\gamma _x +\epsilon _z\gamma _zD_{x}'^2 \\
\end{matrix}\right),\\
\Sigma_{\text{HL}}&=\left(\begin{matrix}
- (\epsilon _x\alpha _x + \epsilon _z\alpha _z)D_{x} - \epsilon _x \beta _x D_{x}' &  \epsilon _z \gamma _z D_{x}\\
\epsilon _x \gamma _x D_{x} +( \epsilon _x\alpha _x-\epsilon _z\alpha _z )D_{x}'  & \epsilon _z\gamma _z D_{x}' \\
\end{matrix}\right),\\
\Sigma_{\text{L}}&=\left(\begin{matrix}
\epsilon_{z}\beta_{z}+\epsilon_{x}\mathcal{H}_{x} & - \epsilon _z \alpha _z \\
- \epsilon _z\alpha _z &  \epsilon _z \gamma _z \\
\end{matrix}\right).
\end{aligned}
\end{equation}
The distribution of a Gaussian beam is related to the second moments matrix of the beam according to
\begin{equation}
\begin{aligned}
\psi({\bf X})&=\frac{1}{(2\pi)^{2}\sqrt{\text{det} \Sigma}}\text{exp}\left(-\frac{1}{2}{\bf X}^{T} \Sigma^{-1}{\bf X}\right)=\frac{1}{(2\pi)^{2}\epsilon_{x}\epsilon_{z}}\text{exp}\left(-\frac{J_{x}}{\epsilon_{x}}-\frac{J_{z}}{\epsilon_{z}}\right).
\end{aligned}
\end{equation}

\subsection{Single RF}\label{sec:singleRF}
\subsubsection{Classical $\sqrt{|\eta|}$ scaling}
Now we first reproduce the classical $\sigma_{z}\propto\sqrt{|\eta|}$ scaling using this longitudinal Courant-Snyder parameterization. To simplify the discussion further, in this section and the following  we focus on the longitudinal dimension only and the state vector 
${\bf X}=\left(
z,\  \delta
\right)^{T}
$ is used. We treat first the case where there is only one RF placed at a dispersion-free location. In this case, the linear longitudinal one-turn map observed in the middle of the RF cavity is
\begin{equation}
\begin{aligned}
{\bf M}&=\left(\begin{matrix}
1&0\\
\frac{h}{2}&1
\end{matrix}\right)\left(\begin{matrix}
1&-\eta C_{0}\\
0&1
\end{matrix}\right)
\left(\begin{matrix}
1&0\\
\frac{h}{2}&1
\end{matrix}\right)=\left(\begin{matrix}
1-\frac{h}{2}\eta C_{0}&-\eta C_{0}\\
h-\left(\frac{h}{2}\right)^{2}\eta C_{0}&1-\frac{h}{2}\eta C_{0}
\end{matrix}\right),
\end{aligned}
\end{equation}
with
$
h={eV_{\text{RF}}k_{\text{RF}}\cos\phi_{\text{s}}}/{E_{0}}
$ quantifying the RF acceleration gradient, where $e$ is the elementary charge, $V_{\text{RF}}$ is the RF voltage, $k_{\text{RF}}=2\pi/\lambda_{\text{RF}}$ is the RF wavenumber, $\phi_{\text{s}}$ is the synchronous phase and $E_{0}=\gamma m_{e}c^2$ is the electron energy. The $R_{56}=-\eta C_{0}$, a measure for the dependence of $z$ on $\delta$, of the ring and the RF kick $h$ can be viewed as the longitudinal drift space and quadrupole, in correspondence to their transverse counterparts, respectively. Note however that as mentioned before, the $R_{56}$ can be either positive or negative, while the physical length of a drift space is always positive.

The linear stability requires that
\begin{equation}
\left|1-\frac{h}{2}\eta C_{0}\right|<1\Rightarrow0<h\eta C_{0}<4.
\end{equation}
For rings working in the longitudinal weak focusing regime, $|\nu_{s}|\ll1$, we then have
\begin{equation}
\begin{aligned}
&1-\frac{h}{2}\eta C_{0}=\cos\Phi_{z}\approx1-\frac{\Phi_{z}^{2}}{2}\Rightarrow\Phi_{z}\approx\begin{cases}
&-\sqrt{h\eta C_{0}}\ \text{if}\ \eta>0,\\
&\sqrt{h\eta C_{0}}\ \text{if}\ \eta<0.
\end{cases}
\end{aligned}
\end{equation}
Therefore the longitudinal beta function $\beta_{z}$ at the RF center is
\begin{equation}\label{eq:longitudinalbeta1}
\beta_{z\text{S}}=\frac{{\bf M}_{12}}{\sin\Phi_{z}}\approx\frac{-\eta C_{0}}{\Phi_{z}}\approx\sqrt{\eta C_{0}/h}.
\end{equation}
In this dissertation, we use the subscript $_{\text{S}}$ to denote the results which are the same with that obtained in Sands' classical analysis \cite{sands1970physics}, although the method used here to obtain these results is different from that of Sands. 
As $|\nu_{s}|\ll1$, therefore
\begin{equation}
\beta_{z\text{S}}\gg|-\eta C_{0}|.
\end{equation}
We will see later in Sec.~\ref{sec:multiRFs} that in  a longitudinal strong focusing ring, $|\nu_{s}|$ can be close to or even larger than 1, and $\beta_{z}$ then can be the same level of or smaller than $|-\eta C_{0}|$. 

Using this $\beta_{z\text{S}}$ to represent $\beta_{z}$ of the whole ring, we then get the longitudinal emittance obtained in Sands' analysis
\begin{equation}\label{eq:SandsEmittance}
\epsilon_{z\text{S}}=\frac{55}{96\sqrt{3}}\frac{\alpha_{F}{\lambdabar}_{e}^{2}\gamma^{5}\beta_{z\text{S}}}{\alpha_{\text{L}}}\oint \frac{1}{|\rho(s)|^{3}}ds.
\end{equation}
For a ring consisting of isomagnets which bend always inwards, $\rho$ is a positive constant and $\alpha_{\text{L}}=J_{s}{U_{0}}/{2E_{0}}=J_{s}{2\pi{\lambdabar}_{e}\alpha_{F}\gamma^3}/{3\rho}$, with $U_{0}$ the radiation energy loss per turn and $J_{s}$ the longitudinal damping partition number \cite{sands1970physics} and nominally $J_{s}\approx2$, we have
\begin{equation}\label{eq:classicalScaling}
\begin{aligned}
\epsilon_{z\text{S}}&\approx \frac{C_{q}}{J_{s}}\frac{\gamma^{2}}{\rho}\sqrt{\frac{\eta C_{0}}{h}}\approx\sigma_{\delta\text{S}}^{2}\beta_{z\text{S}},\\
\sigma_{z\text{S}}&=\sqrt{\epsilon_{z\text{S}}\beta_{z\text{S}}}\approx\sqrt{\frac{C_{q}}{J_{s}}\frac{\gamma^{2}}{\rho}}\sqrt{\frac{\eta C_{0}}{h}}\approx\sigma_{\delta\text{S}}\beta_{z\text{S}},\\
\sigma_{\delta\text{S}}&=\sqrt{\epsilon_{z\text{S}}\gamma_{z\text{S}}}\approx\sqrt{\epsilon_{z\text{S}}/\beta_{z\text{S}}}\approx\sqrt{\frac{C_{q}}{J_{s}}\frac{\gamma^{2}}{\rho}},
\end{aligned}
\end{equation}
where $C_{q}=\frac{55{\lambdabar}_{e}}{32\sqrt{3}}=3.8319\times10^{-13}$ m. Therefore, to generate short bunches in an electron storage ring, we need to implement a quasi-isochronous lattice, i.e., a small $\eta$, and a high RF gradient, i.e., a large $h$. 

\subsubsection{Beyond the classical $\sqrt{|\eta|}$ scaling}

Using a single $\beta_{z\text{S}}$ to represent that of the whole ring is valid in usual rings where the relative variation of $\beta_{z}$ is negligible and the electron distribution in the longitudinal phase space is always upright. But when $\eta$ is small, the partial phase slippage can be significantly larger than the global one and the variation of $\beta_{z}$ and beam orientation in the longitudinal phase space  around the ring can be significant, thus the classical $\sqrt{|\eta|}$ scaling fails. Now we use the longitudinal Courant-Snyder formalism to conduct an analysis of this. 
At a position $s_{j}$, the ring can be divided into three parts, with their transfer matrices given by
\begin{equation}
\begin{aligned}
{\bf T}(s_{j},s_{\text{RF}-})&=\left(\begin{matrix}
1&F(s_{j},s_{\text{RF}-})\\
0&1
\end{matrix}\right),\\
{\bf T}(s_{\text{RF}-},s_{\text{RF}+})&=\left(\begin{matrix}
1&0\\
h&1
\end{matrix}\right),\\
{\bf T}(s_{\text{RF}+},s_{j})&=\left(\begin{matrix}
1&F(s_{\text{RF}+},s_{j})\\
0&1
\end{matrix}\right),
\end{aligned}
\end{equation}
where
\begin{equation}\label{eq:Fdefinition}
F(s_{1},s_{2})=-\int_{s_{1}}^{s_{2}}\left(\frac{D_{x}(s)}{\rho(s)}-\frac{1}{\gamma^{2}}\right)ds=-\tilde{\eta}(s_{1},s_{2})C_{0}
\end{equation}
is the partial $R_{56}$ from $s_{1}$ to $s_{2}$, and 
\begin{equation}
F(s_{j},s_{\text{RF}-})+F(s_{\text{RF}+},s_{j})=-\eta C_{0}
\end{equation}
is the global $R_{56}$ of the ring, with ${+/-}$ denoting right after/before the corresponding location.
The one-turn map at $s_{j}$ is
\begin{equation}
\begin{aligned}
{\bf M}(s_{j})&={\bf T}(s_{\text{RF}+},s_{j}){\bf T}(s_{\text{RF}-},s_{\text{RF}+}){\bf T}(s_{j},s_{\text{RF}-})\\
&=\left(\begin{matrix}
1+F(s_{j},s_{\text{RF}-})h&-\eta C_{0}+F(s_{\text{RF}+},s_{j})F(s_{j},s_{\text{RF}-})h\\
h&1+F(s_{\text{RF}+},s_{j})h
\end{matrix}\right).
\end{aligned}
\end{equation}
Therefore,
\begin{equation}
\begin{aligned}
\beta_{z}(s_{j})&=\frac{{\bf M}_{12}(s_{j})}{\sin\Phi_{z}}=\frac{-\eta C_{0}+F(s_{\text{RF}+},s_{j})F(s_{j},s_{\text{RF}-})h}{\sin\Phi_{z}}.
\end{aligned}
\end{equation}
As can be seen, the first term in the numerator is the usual global phase slippage of the ring. The second term represents the impact of partial phase slippage on $\beta_{z}$. Sands' analysis means to drop the second term but keep only the first term. With both terms considered, the more accurate longitudinal emittance is then
\begin{equation}\label{eq:SLIMemittance}
\begin{aligned}
\epsilon_{z}&=\epsilon_{z\text{S}}\left(1+\frac{\left\langle F^{2}(s_{\text{RF}+},s_{j})\right\rangle+\eta C_{0}\left\langle F(s_{\text{RF}+},s_{j})\right\rangle }{\eta C_{0}/h}\right)\\
&\approx\epsilon_{z\text{S}}\left(1+\frac{\left\langle F^{2}(s_{\text{RF}+},s_{j})\right\rangle+\eta C_{0}\left\langle F(s_{\text{RF}+},s_{j})\right\rangle }{\beta_{z\text{S}}^{2}}\right),
\end{aligned}
\end{equation}
in which $\langle \rangle$ means the radiation-weighted average around the ring, defined as
\begin{equation}
\langle P\rangle=\frac{\oint\frac{P}{|\rho(s)|^{3}}ds}{\oint\frac{1}{|\rho(s)|^{3}}ds},
\end{equation}
i,e., the average actually takes place only at the bend-related elements. 
After getting the longitudinal emittance and Courant-Snyder functions, the bunch length and energy spread at $s_{j}$ can be obtained by
\begin{equation}\label{eq:BLES}
\begin{aligned}
\sigma_{z}(s_{j})&=\sqrt{\epsilon_{z}\beta_{z}(s_{j})},\\ \sigma_{\delta}(s_{j})&=\sqrt{\epsilon_{z}\gamma_{z}(s_{j})}\approx\sigma_{\delta\text{S}}\sqrt{\frac{\epsilon_{z}}{\epsilon_{z\text{S}}}}.
\end{aligned}
\end{equation}
with $\gamma_{z}(s_{j})=\frac{-{\bf M}_{21}(s_{j})}{\sin\Phi_{z}}$.
For the bunch length more accurately 
\begin{equation}\label{eq:BLESwithH}
\sigma_{z}(s_{j})=\sqrt{\epsilon_{z}\beta_{z}(s_{j})+\epsilon_{x}\mathcal{H}_{x}(s_{j})},\
\end{equation}
considering the coupling from the transverse emittance according to Eq.~(\ref{eq:2ndmomentsSeparate}). In a planar $x$-$y$ uncoupled ring which means there is only passive transverse-longitudinal coupling introduced by bending magnets,  the transverse emittance always lengthen the bunch at places where $\mathcal{H}_{x}\neq0$. 
But this conclusion is not generally true if we take advantage of transverse-longitudinal coupling actively for bunch compression when the transverse emittance is small, as we will see in Chap.~\ref{cha:TLC}. The reason is that there is flexibility in projecting the three eigen emittances of a beam into different physical dimensions as shown in Eq.~(\ref{eq:SLIMsecondmoments}), although their values cannot be changed in a linear symplectic lattice as we have explained.

As analyzed before and shown in Eq.~(\ref{eq:classicalScaling}), we know that in a longitudinal weak focusing ring $\epsilon_{z\text{S}}\propto\sqrt{\eta/h}$, but now according to Eq.~(\ref{eq:SLIMemittance}), 
there exists a specific $|\eta|$ where a minimum $\epsilon_{z}$ will be reached when adjusting $|\eta|$ by changing the dispersion function slightly with the overall pattern unchanged. Below this $|\eta|$, $\epsilon_{z}$ will grow with the decrease of $|\eta|$. The same argument applies also to the RF gradient $h$. Above a specific $|h|$, the emittance will grow, instead of decreasing, with the increase of $|h|$. 

Correspondingly, considering this unavoidable diffusion of longitudinal coordinate, there exists a bunch length limit
\begin{equation}
\sigma_{z,\text{limit}}=\sigma_{\delta\text{S}}\sqrt{\langle F^{2}(s_{\text{RF}+},s_{j})\rangle},
\end{equation}
and the energy spread grows significantly as the bunch length is pushed close to this limit. Note that this limit has little dependence on the RF acceleration gradient or the global phase slippage factor.  For usual rings, such bunch length limit is in a couple of 10 fs to about 100~fs level, while the typical bunch length in operation is 1~ps to 10 ps level. So this effect is negligible in almost all existing rings. However, with the accelerator physics and technology continue to advance, more ambitious goals of bunch length can be envisioned and realized in the future to benefit more from the electron beam. For example, SSMB aims at a bunch length of sub-micron or even nanometer, which corresponds to sub-fs in unit of time. The quantum diffusion investigated here then becomes the first fundamental issue we need to resolve. Therefore, it is of crucial importance to acquire a good knowledge of this effect.

\begin{figure}[tb] 
	\centering 
	\includegraphics[width=1\textwidth]{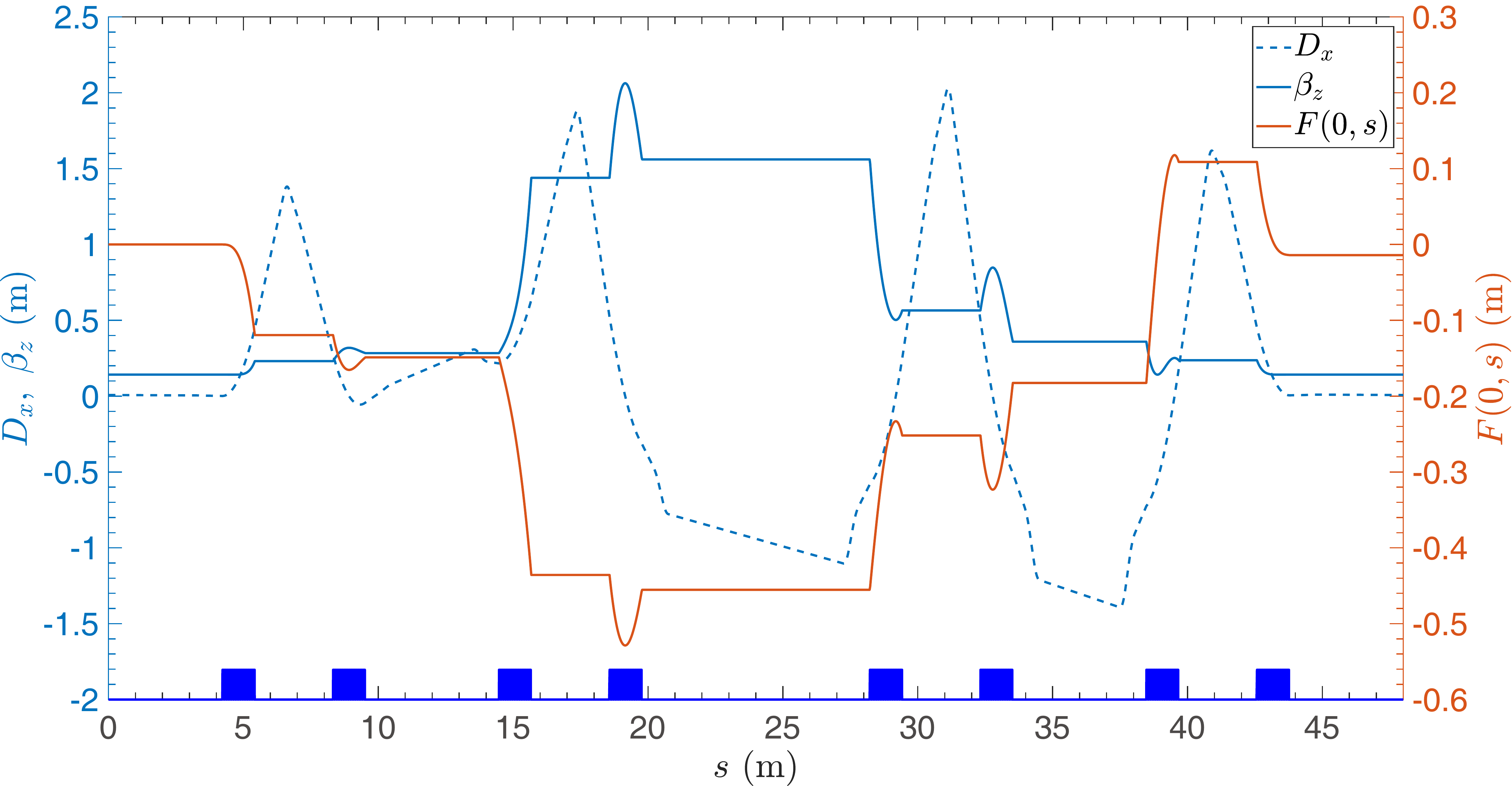}
	\caption{
		\label{fig:Chap2-LongitudinalBeta} 
		The horizontal dispersion $D_{x}$, longitudinal beta function $\beta_{z}$ and $F(0,s)$ of the MLS lattice used for analysis and simulation. In this plot, the zero-length RF is placed at $s_{\text{RF}}=0$ m and $V_{\text{RF}}=80$ MV is applied. The dipoles are shown at the bottom as blue rectangles. Each dipole has a length of 1.2 m and bends the electron trajectory for an angle of $\pi/4$. 
	}
\end{figure}

\begin{table}[tb]
	\caption{\label{tab:tab1}
		Parameters of the MLS lattice used in simulation.}
	\centering
	\begin{tabular}{lll}
		\hline
		Parameter & \multicolumn{1}{l}{\textrm{Value}}  & Description \\
		\hline
		$E_0$ & 1.2  GeV & Beam energy \\
		$C_0$ & 48  m & Ring circumference \\
		$\eta$  & $2.9\times10^{-4}$  & Phase slippage factor \\
		$f_{\text{RF}}$  & 500  MHz & RF frequency \\
		$V_{\text{RF}}$  & 2.5 - 160  MV & RF voltage \\	
		$U_{0}$  & 0.12  MeV & Radiation energy loss per turn \\
		$J_{s}$  & 1.94 & Longitudinal damping partition \\	
		$\tau_{\delta}$ & 1.6 ms & Longitudinal radiation damping time \\
		$\nu_{x}$ & 3.17 & Horizontal betatron tune \\
		$\nu_{s}$ & -0.016@80 MV RF & Synchrotron tune \\
		$h$  & 0.7  m$^{-1}$@80 MV RF & rf acceleration gradient \\
		$\sigma_{\delta\text{S}}$ & $8.4\times10^{-4}$ & Natural energy spread \\
		$-\eta C_{0}$ & -0.014 m & -\\
		$\langle F(s_{\text{RF}+},s_{j})\rangle$ & -0.19 m & -\\
		$\sqrt{\langle F^{2}(s_{\text{RF}+},s_{j})\rangle}$ & $0.26$ m & -\\
		$\sqrt{\langle F^{2}\rangle-\langle F\rangle^{2}}$ & $0.18$ m & -\\
		$\sigma_{z,\text{limit}}$ & 218 $\mu$m (727 fs) & Bunch length limit	\\
		\hline	
	\end{tabular}
\end{table}

\begin{figure}[tb] 
	\centering 
	\includegraphics[width=0.6\columnwidth]{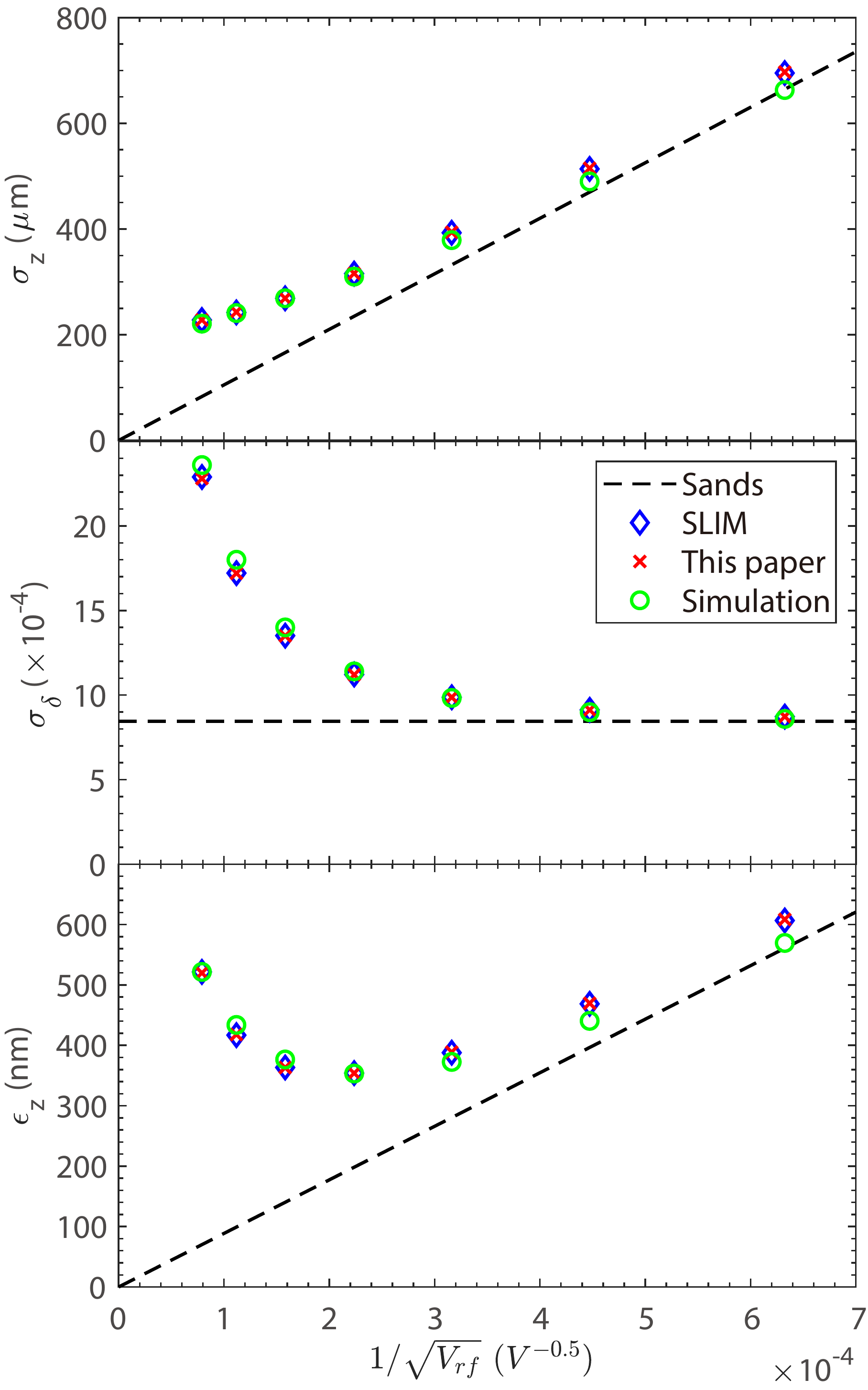}
	\caption{
		\label{fig:Chap2-SLIMvsSands} 
		Bunch length saturation and energy spread and longitudinal emittance divergence when the bunch length is pushed to the limit, in the simulation by increasing the RF voltage, given by quantum diffusion. The lattice optics and related parameters are shown in Fig.~\ref{fig:Chap2-LongitudinalBeta} and Tab.~\ref{tab:tab1}. In each simulation, 1000 electrons are tracked using the code ELEGANT \cite{borland2000elegant} for 50000 turns which correspond to about 5 longitudinal radiation damping times, and the averaged result of the last 10000 turns are used to get the $\sigma_{z}$, $\sigma_{\delta}$ and~$\epsilon_{z}$. The observation point is at the RF.
	}
\end{figure}

The above analysis has been confirmed by particle tracking simulation using a lattice of the Metrology Light Source (MLS) storage ring of Physikalisch-Technische Bundesanstalt (PTB) in Berlin~\cite{klein2008operation,feikes2011metrology,ries2014nonlinear}.The dispersion function pattern of the lattice used in the simulation is shown in Fig.~\ref{fig:Chap2-LongitudinalBeta}. Also shown are $\beta_{z}(s)$ (for the case of $V_{\text{RF}}=80$ MV) and $F(0,s)$ around the ring. Note that $\beta_{\text{z}}>0$ and varies significantly around the ring. The other related parameters for the simulation are given in Tab.~\ref{tab:tab1}. This lattice has a globally small and locally large phase slippage, which makes the bunch length limit arising from the quantum diffusion a large value (727 fs), and thus easier to see the influence of this effect. The beam energy used is higher than the actual operation value of the MLS and the RF voltages applied is also dramatically high. They are chosen here for the numerical example to present the difference of our derivation and the classical analysis of Sands in a more notable way. The simulation results along with the theoretical predictions from Sands' analysis and this dissertation are shown in Fig.~\ref{fig:Chap2-SLIMvsSands}. 
Also shown is the theoretical numerical calculation based on SLIM~\cite{chao1979evaluation}. As can be seen from Fig.~\ref{fig:Chap2-SLIMvsSands}, our result and SLIM agree well with the simulation in the cases of ultra-high RF voltages, while the predication of Sands deviates from the simulation.

\subsubsection{Discussions}

Now we make some more discussions. First, we notice that in Fig.~\ref{fig:Chap2-LongitudinalBeta}, $\beta_{z}$ changes only notably inside dipoles, which reflects the fact that the partial phase slippage, i.e., the effective longitudinal drift space, changes only significantly inside dipoles where $\rho\neq0$, which is consistent with the definition Eq.~(\ref{eq:Fdefinition}). We will see later in Sec.~\ref{sec:passiveLengthening} that the bunch lengthening from transverse emittance, quantified by the chromatic $\mathcal{H}$ function, also changes only inside dipoles. In other words, ignoring the contribution of $\frac{1}{\gamma^{2}}$ on $R_{56}$, the bunch length changes only inside the dipoles, since the longitudinal velocities of the relativistic particles in drift space are all close to the speed of light.

The minimum $\beta_{z}$ around the ring is realized at a place $s_{j}$ where $F(s_{\text{RF}+},s_{j})=F(s_{j},s_{\text{RF}-})=\frac{-\eta C_{0}}{2}$, and
\begin{equation}
\begin{aligned}
\beta_{z,\text{min}}=\left(1-\frac{h\eta C_{0}}{4}\right)\beta_{z\text{S}}.
\end{aligned}
\end{equation}
Furthermore, we remind the readers that in some cases, $\epsilon_{z}$ can actually be smaller than $\epsilon_{z\text{S}}$. The physical meaning is then the average $\beta_{z}$ at the dipole-related elements is smaller than that at the RF, for example if the dispersions all around the ring have the same sign. 
This however is usually not satisfied in a quasi-isochronous lattice, where the integration cancellation of positive and negative dispersions are needed to achieve a small global phase slippage, and the change of $F(s_{\text{RF}+},s_{j})$ around the ring can be significantly larger than the global $R_{56}=-\eta C_{0}$. Therefore, $\beta_{z}$ at other places can be significantly larger than that at the RF, and $\epsilon_{z}>\epsilon_{z\text{S}}$, just like the case shown in Fig.~\ref{fig:Chap2-LongitudinalBeta}. 

We point out that quantifying the impact of variation of $\beta_{z}$ around the ring on $\epsilon_{z}$ using partial alpha or partial phase slippage variance $\langle F^{2}\rangle-\langle F\rangle^2$, as that done in Ref.~\cite{shoji1996longitudinal} and also our previous publication Ref.~\cite{deng2020single}, is not generally accurate. Given the same dispersion function pattern, which means the same $\langle F^{2}\rangle-\langle F\rangle^2$ as it is independent of the observation point, a different longitudinal beta function pattern can be generated if the RF is placed at a different location, therefore resulting in a different longitudinal emittance according to Eq.~(\ref{eq:longitudinalEmittance}). 

We can also view this from another way. Changing the RF location means shifting  $F(s_{\text{RF}+},s_{j})$ up or down as a whole, which can also be done by artificially inserting a pair of $\pm r_{56}$ besides the RF.  We assume that there is no synchrotron radiation within these artificially inserted $\pm r_{56}$, and therefore these manipulations do not change the $R_{56}$ and also $\langle F^{2}\rangle-\langle F\rangle^2$ of the ring, but they do affect $\beta_{z}$ and therefore also $\epsilon_{z}$.
The dependence of the longitudinal emittance on the artificially inserted $r_{56}$ is parabolic. When $r_{56}=\langle F(s_{\text{RF}+},s_{j})\rangle+\frac{\eta C_{0}}{2},$
we arrive at the minimum longitudinal emittance
\begin{equation}\label{eq:SLIMemittanceMin}
\epsilon_{z,\text{min}}=\epsilon_{z\text{S}}\left(1+\frac{\langle F^{2}\rangle-\langle F\rangle^2-\left(\frac{\eta C_{0}}{2}\right)^{2}}{\eta C_{0}/h}\right).
\end{equation}
Equivalently in the case of shifting RF location, $\epsilon_{z,\text{min}}$ is reached when the RF is placed at a location such that $\langle F(s_{\text{RF}+},s_{j})\rangle=-\frac{\eta C_{0}}{2}$.
The maximum longitudinal emittance is realized when the RF is placed at a place such that $\left|\langle F(s_{\text{RF}+},s_{j})\rangle+\frac{\eta C_{0}}{2}\right|$ reaches the maximum possible value. 

\begin{figure}[tb] 
	\centering 
	\includegraphics[width=1\columnwidth]{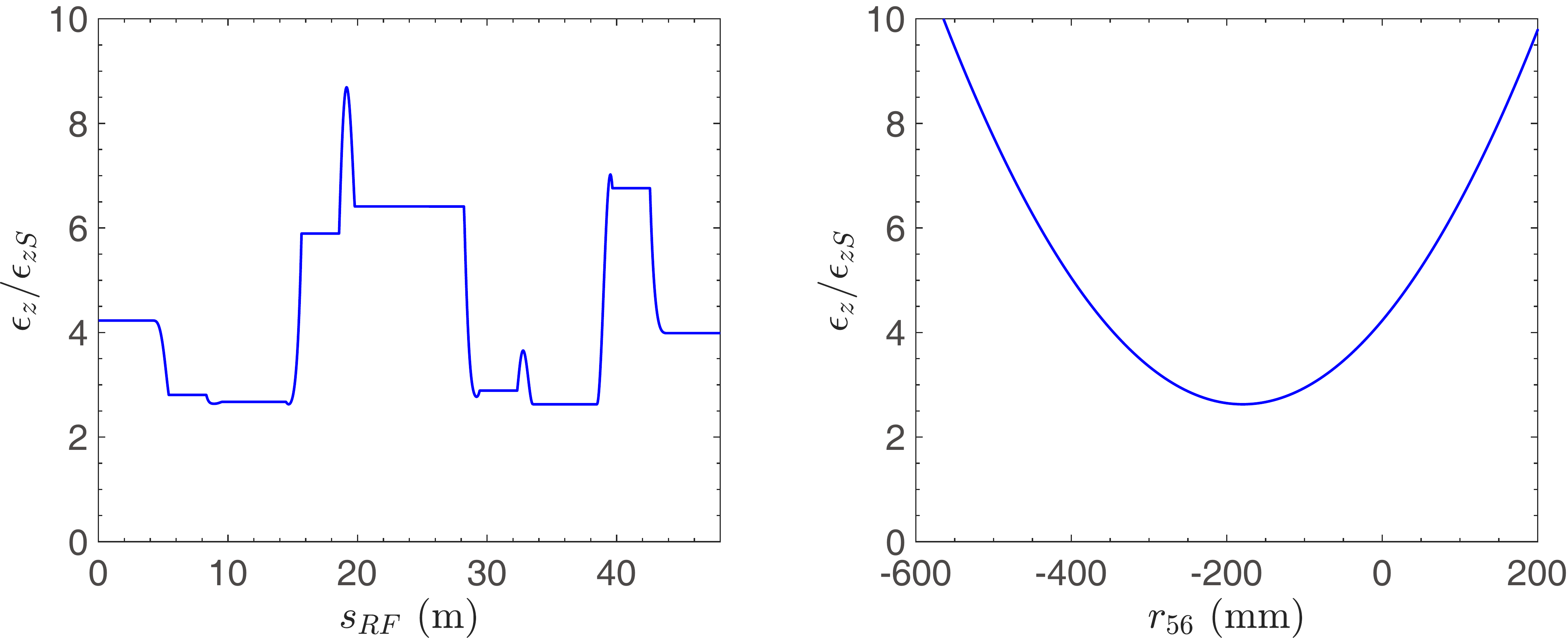}
	\caption{
		\label{fig:Chap2-EMittanceRatio} 
		The emittance ratio $\frac{\epsilon_{z}}{\epsilon_{z\text{S}}}$ as a function of the RF place $s_{\text{RF}}$ (left), and the $r_{56}$ artificially inserted in front of the RF whose location is fixed at $s_{\text{RF}}=0$ m (right). The lattice optics and related parameters are shown in Fig.~\ref{fig:Chap2-LongitudinalBeta} and Tab.~\ref{tab:tab1}. The RF voltage applied is $V_{\text{RF}}=80$~MV.  
	}
\end{figure}

To give the readers a more concrete feeling, here we present some calculations based on the lattice optics presented in Fig.~\ref{fig:Chap2-LongitudinalBeta} and other related parameters given in Tab.~\ref{tab:tab1}.   Figure~\ref{fig:Chap2-EMittanceRatio} shows the emittance ratio ${\epsilon_{z}}/{\epsilon_{z\text{S}}}$ as a function of the RF place $s_{\text{RF}}$, and the $\pm r_{56}$ artificially inserted besides the RF whose location is fixed at $s_{\text{RF}}=0$~m, respectively. An 80 MV RF voltage is applied in both calculations. As can be seen, indeed the place of the RF can have a crucial impact on the longitudinal emittance in this specific lattice optics and choice of RF voltage, and the impact of the artificially inserted $\pm r_{56}$ is also as expected. 

When the minimum longitudinal emittance is reached, the bunch length at the RF is 
\begin{equation}
\begin{aligned}
\sigma_{z,\text{min}}(s_{\text{RF}})
=\sigma_{\delta\text{S}}\sqrt{\frac{\eta C_{0}}{h}+\langle F^{2}\rangle-\langle F\rangle^2-\left(\frac{\eta C_{0}}{2}\right)^{2} }.\\
\end{aligned}
\end{equation}
In the case of ultra-small $\eta$,
\begin{equation}\label{eq:steadyBLLimit}
\sigma_{z,\text{min}}(s_{\text{RF}})\approx\sigma_{\delta\text{S}}\sqrt{\langle F^{2}\rangle-\langle F\rangle^2},
\end{equation}
and
\begin{equation}\label{eq:steadyEmittanceLimit}
\epsilon_{z,\text{min}}=\frac{\sigma_{z,\text{min}}^{2}(s_{\text{RF}})}{\beta_{z\text{S}}}\approx\frac{\epsilon_{z\text{S}}}{\sqrt{\eta C_{0}/h}}\left(\sqrt{\eta C_{0}/h}+\frac{\langle F^{2}\rangle-\langle F\rangle^2}{\sqrt{\eta C_{0}/h}}\right)\geq2\sigma_{\delta\text{S}}^{2}\sqrt{\langle F^{2}\rangle-\langle F\rangle^2}.
\end{equation}
The equality holds when
\begin{equation}
\beta_{z\text{S}}=\sqrt{\eta C_{0}/h}=\sqrt{\langle F^{2}\rangle-\langle F\rangle^2}.
\end{equation}
Therefore, the variance of $F$ can be viewed as a parameter to quantify the lowest possible contribution of this effect to the equilibrium bunch length at the RF and the longitudinal emittance with a dispersion function pattern given, if we can choose the location of the RF. However, in a real machine, the RF location is fixed, and Eqs.~(\ref{eq:SLIMemittance}), (\ref{eq:BLES}) and (\ref{eq:BLESwithH}) should be used instead. The analysis of shifting RF location or adding artificial $\pm r_{56}$ helps us better understand the physics, and we recognize the fact that the RF can actually not be placed inside dipoles and also that the Courant-Snyder functions lose their well-defined meaning if the RF is placed at a dispersive location.

A simplified derivation of the bunch length, energy spread and longitudinal emittance beyond the classical $\sqrt{|\eta|}$ scaling using Fokker-Planck equation can also be found in Appendix.~\ref{app:Fokker-Planck}. We remind the readers that the spirit of the simplified derivation is the same with SLIM.

\subsubsection{Campbell's theorem}

Now we explain in more detail why $\langle F^{2}\rangle-\langle F\rangle^2$ is not a generally accurate criterion in quantifying this quantum diffusion of longitudinal coordinate. The reason is that while the photon emission process is stochastic, the evolution of $F$ around the ring is deterministic. So 
the diffusion of $z$ each turn $d_{z}^{2}$ due to quantum excitation  is 
\begin{equation}\label{eq:zdiffusion}
d_{z}^{2}=\langle z^2\rangle-\langle z\rangle^2=\langle F^{2}\rangle\langle \mathcal{N}\rangle\left\langle \frac{u^{2}}{E_{0}^{2}}\right\rangle,
\end{equation}
instead of 
\begin{equation}
d_{z}^{2}=\langle z^2\rangle-\langle z\rangle^2=\left(\langle F^{2}\rangle-\langle F\rangle^2\right)\langle \mathcal{N}\rangle\left\langle \frac{u^{2}}{E_{0}^{2}}\right\rangle
\end{equation}
as that given in Ref.~\cite{shoji1996longitudinal},
where $F$ is calculated with the final observation location as the ending point, $\langle \mathcal{N}\rangle$ is the expected number of emitted photons, $u$ is the photon energy, $\langle u^2 \rangle$ and later also $\langle u \rangle$ mean the average is taken with respect to the photon energy spectrum. 

This result can be understood with the help of Campbell's theorem \cite{campbell1909study}.  From this theorem some expectation result for the Poisson point process follows. For example, for the application in synchrotron radiation, we have $\delta=-\sum_{i} \frac{u_{i}}{E_{0}}$, where $_{i}$ means the $i^{\text{th}}$ photon emission. Then according to Campbell's theorem we have
\begin{equation}\label{eq:CampbellStorageRing}
\begin{aligned}
\langle\delta\rangle&=-\langle \mathcal{N}\rangle\left\langle \frac{u}{E_{0}}\right\rangle=-T_{\text{dipole}}\dot{\mathcal{N}}\left\langle \frac{u}{E_{0}}\right\rangle,\\
\langle\delta^2\rangle-\langle\delta\rangle^2&=\langle \mathcal{N}\rangle\left\langle \frac{u^{2}}{E_{0}^{2}}\right\rangle=T_{\text{dipole}} \dot{\mathcal{N}}\left\langle \frac{u^{2}}{E_{0}^{2}}\right\rangle,
\end{aligned}
\end{equation}
where $\dot{\mathcal{N}}$ is the number of photons emitted per unit time in the dipoles and $T_{\text{dipole}}$ is the total time within dipoles. 
Equation~(\ref{eq:CampbellStorageRing}) is why $\dot{\mathcal{N}}\langle u^{2}\rangle$  appears so often in the calculation of energy spread, emittance, etc., in electron storage ring physics. 
Note that the relation in Eq.~(\ref{eq:CampbellStorageRing}) holds as long as the radiation is a Poisson point process. It is independent of whether $\langle u\rangle=0$ or not, and is also independent of the detailed spectrum of the photon energy. In other words, the key of a Poisson point process is the randomness in whether there is a kick or not, i.e, the kick number, and not in the randomness of the size of the kicks. The importance of this theorem for electron dynamics was first pointed out by Sands \cite{sands1955synchrotron}. A proof can be found in the article of Rice~\cite{rice1944mathematical} and a less rigorous but simpler one in the lecture note of Jowett~\cite{jowett1987introductory}. 

Now we can understand Eq.~(\ref{eq:zdiffusion}) like this. Suppose that the RF is the observation point. We divide the ring into many sections, and in each section $F(s_{j},s_{\text{RF}})$ does not change much. Then the change of electron longitudinal coordinate in one turn is $z=\sum_{j}z_{j}$, with $z_{j}=-\sum_{i}F(s_{ji},s_{\text{RF}})\frac{u_{ji}}{E_{0}}$ the contribution due to photon emissions within the section $j$. So according to Campbell's theorem, the variance of $z_{j}$ is 
\begin{equation}
\begin{aligned}
\text{Var}(z_{j})
&= F^2(s_{j},s_{\text{RF}}) t_j \dot{\mathcal{N}}\left\langle \frac{u^{2}}{E_{0}^{2}}\right\rangle,
\end{aligned}
\end{equation}
where $t_j$ is the time within the dipoles in section $j$. As the photon emissions in different sections are uncorrelated, then the variance of $z$ is the sum of variance of $z_{j}$
\begin{equation}\label{eq:diffusionZ}
\begin{aligned}
\langle z^2\rangle-\langle z\rangle^2
&=\frac{\sum_{j}\left[F^{2}(s_{j},s_{\text{RF}}) t_j\right]}{T_{\text{total}}} T_{\text{total}}\dot{\mathcal{N}}\left\langle \frac{u^{2}}{E_{0}^{2}}\right\rangle\\
&=\langle F^{2}(s_{j},s_{\text{RF}})\rangle\langle \mathcal{N}\rangle\left\langle \frac{u^{2}}{E_{0}^{2}}\right\rangle,
\end{aligned}
\end{equation}
in which $T_{\text{total}}=\sum_{j}t_{j}$ is the total time within the dipoles. So now we  obtain Eq.~(\ref{eq:zdiffusion}) following Campbell's theorem.
In fact the same argument can be applied to explain why $|E_{k5}(s)|^{2}$ in Eq.~(\ref{eq:emittance}), instead of its variance, are used in SLIM to calculate the increase of eigen emittances due to quantum excitation.

\subsection{Minimizing Longitudinal Emittance}\label{sec:miniLonemit}

It is clear that the quantum diffusion of $z$ needs to be carefully treated for the realization and long-term maintenance of ultrashort bunches or small longitudinal emittances in either a multipass device or a single-pass transport line with bending magnets and large dispersion. It can be seen from Eqs.~(\ref{eq:steadyBLLimit}), (\ref{eq:steadyEmittanceLimit}) and (\ref{eq:zdiffusion}) that the minimum contribution of this effect is determined by the beam energy and the variance of the partial phase slippage. A lower operating energy is preferred for suppressing this quantum diffusion. Note that the energy scaling laws for this effect are different in the one-turn (single-pass) and steady-state cases; for the single-pass case, the root-mean-square (RMS) diffusion of longitudinal coordinate $d_{z}\propto\gamma^{2.5}$, while for the steady-state case,  $\sigma_{z,\text{min}}\propto\gamma$, because the radiation damping time also depends on $\gamma$.  

\begin{figure}[tb] 
	\centering 
	\includegraphics[width=0.7\columnwidth]{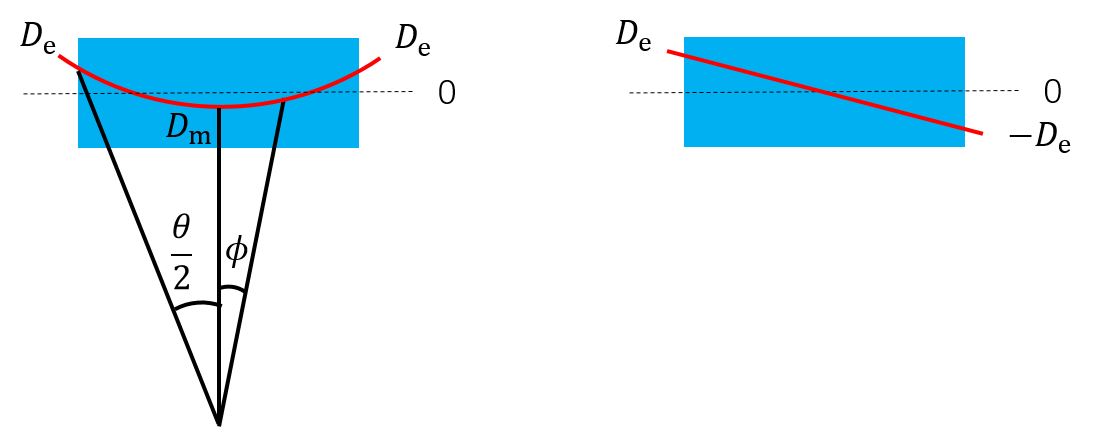}
	\caption{
		\label{fig:Chap2-Dispersion} 
		Two possible evolution of dispersion function inside a single bending magnet which satisfy the isochoronous condition.  Given the same bending angle of the dipole, the left one has a smaller variation of partial phase slippage compared to the right one,  and thus can result a smaller longitudinal emittance. The lattice design for the realization of the dispersion function pattern in the left can be found in Refs.~\cite{pan2019storage,pan2020research}.
	}
\end{figure}

In addition to ensuring a small global phase slippage, the variation in the partial phase slippage should also be well confined by means of dedicated lattice design, more specifically by the tailoring of horizontal dispersion function, to tailor the longitudinal beta function all around the ring. At the MLS, the small global phase slippage is usually achieved by means of an overall integration cancellation between the large positive and large negative horizontal dispersions at different dipoles \cite{feikes2011metrology,ries2014nonlinear}. Therefore, the partial phase slippage varies sharply within the dipoles, leading to a large partial phase slippage variation and significant quantum diffusion of $z$, as can be seen in Fig.~\ref{fig:Chap2-LongitudinalBeta} and the simulation result Fig.~\ref{fig:Chap2-SLIMvsSands}. To obtain small global and partial phase slippages simultaneously, such cancellation should be done as locally as possible, and the magnitudes of the dispersion at the dipoles should also be minimized, thus making the partial phase slippage vary as gently as possible. In other words, each partial component of the ring should be made as isochronous as possible. In this sense, the dispersion function pattern in the left of Fig.~\ref{fig:Chap2-Dispersion} is better than that right one in mitigating the longitudinal emittance~\cite{pan2020research}, for an isochronous bending magnet. This requirement on the dispersion is similar (but not identical) to that for low-transverse-emittance ring design. So the transverse emittance in an SSMB ring is usually also not very large.  

\subsubsection{Constant Bending Radius}
To give the readers a more concrete feeling of the above argument, here we present some quantitative analysis of the minimization of longitudinal emittance. As can seen in Eq.~(\ref{eq:longitudinalEmittance}), the longitudinal beta function $\beta_{z}$ with respect to the longitudinal dimension plays a role similar to that of the chromatic function $\mathcal{H}_{x}$ in the transverse dimension. 
As both the longitudinal and transverse emittances originate from quantum excitation, for a ring consisting of identical isochronous cells,  
the same scaling law of the theoretical minimum emittance (TME), i.e., $\epsilon_{x,z,\text{TME}}\propto\gamma^{2}\theta^{3}$, concerning the beam energy and bending angle of the dipole can be obtained. Note that the TME is independent of the bending radius. But we will show soon that the bunch length limit does depend on the bending radius. According to the scaling, a ring consisting of a larger number of isochronous cells, each with a smaller bending angle, can better minimize the emittance than a ring consisting of fewer cells with larger bending. Generally, it is easier to realize small emittance in a larger ring. Increasing the longitudinal damping partition number is also a method of lowering the longitudinal emittance. 

Here we present an analysis of the above scaling by evaluating $\sqrt{\langle F^{2}\rangle-\langle F\rangle^2}$ of a single constant bending-radius magnet whose dispersion function is symmetric with respect to the dipole middle point as shown in the left of Fig.~\ref{fig:Chap2-Dispersion}. To simplify the analysis, we start from the middle point of the dipole where the dispersion angle is zero $D'_{\text{m}}=0$, then the dispersion as a function of angle $\phi$ is
\begin{align}\label{eq:dispersion}
D(\phi)=D_{\text{m}}\cos{\phi}+\rho(1-\cos{\phi}),
\end{align}
where $D_{m}$ is the dispersion at the middle of the dipole.
As the case we are interested in usually has a large $\gamma$, ignoring the contribution of $\frac{1}{\gamma^{2}}$ on $F$, the condition of isochronicity of each dipole is
\begin{align}\label{eq:isochronicity}
\int_{0}^{\frac{\theta}{2}}D(\phi)d\phi=0,
\end{align}
with $\theta$ the bending angle of each dipole. Substituting Eq.~(\ref{eq:dispersion}) into Eq.~(\ref{eq:isochronicity}), we get 
\begin{equation}
\begin{aligned}
D_{\text{m}}&=\rho\left(1-\frac{\frac{\theta}{2}}{\sin{\frac{\theta}{2}}}\right)\approx-\frac{1}{24}\rho\theta^{2},\\
D_{\text{e}}&=\rho\left(1-\frac{\frac{\theta}{2}}{\tan{\frac{\theta}{2}}}\right)\approx\frac{1}{12}\rho\theta^{2},
\end{aligned}
\end{equation}
where $D_{\text{e}}$ is the dispersion at the entrance and exit of the dipole.
Now we can calculate $\sqrt{\langle F^{2}\rangle-\langle F\rangle^2}$ of the dipole as follows
\begin{equation}
\begin{aligned}
&F(\phi)=\int_{0}^{\phi}D(\beta)d\beta=\rho\left(\phi-\frac{\frac{\theta}{2}}{\sin\frac{\theta}{2}}\sin\phi\right),\\
&\langle F\rangle=\frac{1}{\frac{\theta}{2}}\int_{0}^{\frac{\theta}{2}} F(\phi)d\phi=\rho\left(\frac{\theta}{4}-\tan{\frac{\theta}{4}}\right)\approx-\frac{1}{192}\rho\theta^{3},\\
&\langle F^{2}\rangle=\frac{1}{\frac{\theta}{2}}\int_{0}^{\frac{\theta}{2}} F^{2}(\phi)d\phi=\frac{1}{6}\rho^{2}\left[2\left(-6+\left(\frac{\theta}{2}\right)^{2}\right)+9\frac{\frac{\theta}{2}}{\tan{\frac{\theta}{2}}}+3\frac{\left(\frac{\theta}{2}\right)^{2}}{\sin{\left(\frac{\theta}{2}\right)}^{2}}\right]\approx\frac{1}{30240}\rho^{2}\theta^{6},\\
&\sqrt{\langle F^{2}\rangle-\langle F\rangle^2}\approx\frac{\sqrt{2415}}{20160}\rho\theta^{3}.
\end{aligned}
\end{equation}
Note that the $F$ in this section is defined with the middle point of the bending magnet as the starting point. Therefore, for a ring consisting of such isochronous isomagnets (note that the global phase slippage of the ring is non-zero for a stable beam motion), we have
\begin{align}\label{eq:BLLimit}
\sigma_{z,\text{min}}\approx\sigma_{\delta\text{S}}\sqrt{\langle F^{2}\rangle-\langle F\rangle^2}=\frac{\sqrt{2415}}{20160}\sqrt{\frac{C_{q}}{J_{s}}}\sqrt{\rho}\gamma\theta^{3}\propto\sqrt{\rho}\gamma\theta^{3},
\end{align}
and 
\begin{equation}\label{eq:emittanceLimit}
\epsilon_{z,\text{min}}\approx2\sigma_{\delta\text{S}}^{2}\sqrt{\langle F^{2}\rangle-\langle F\rangle^2}=2\frac{\sqrt{2415}}{20160}{\frac{C_{q}}{J_{s}}}\gamma^{2}\theta^{3}\propto\gamma^{2}\theta^{3}.
\end{equation}
where $C_{q}=\frac{55{\lambdabar}_{e}}{32\sqrt{3}}=3.8319\times10^{-13}$ m, and $J_{s}$ is the longitudinal damping partition number and nominally $J_{s}\approx2$. A comprehensive analysis of minimizing longitudinal emittance in an electron storage ring can also be found in Ref.~\cite{zhang2021ultralow}.	

Now we can use Eqs.~(\ref{eq:BLLimit}) and (\ref{eq:emittanceLimit}) to do some evaluation, for example for the envisioned EUV SSMB. For coherent 13.5 nm EUV radiation generation, we need an electron bunch length around 3 nm. Since the energy spread will grow when we push the bunch length to the limit value $\sigma_{z,\text{min}}$, therefore, if we want to obtain a 3 nm bunch length without significant energy widening compared to the natural value, then the bunch length limit due to the quantum diffusion we have analyzed should be smaller than 3 nm.  Here we conduct some evaluation for the case of $\sigma_{z,\text{min}}=2$ nm. Suppose the beam energy is $E_{0}=400$ MeV and $J_{s}=2$, then to get a bunch length limit of 2~nm, the maximum acceptable $\sqrt{\rho}\theta^{3}$ is $2.4\times10^{-3}\ \text{m}^{\frac{1}{2}}$.
If $\rho=2\ \text{m}$, which means the bending magnet field strength is $B=0.67\ \text{T}$, 
the bending angle of each dipole is then
\begin{align}\label{eq:bendingangle}
\theta&=0.12\text{ rad}=6.8^{\circ}.
\end{align}
The bending magnet length in this case is 
\begin{align}
L_{\text{bend}}=\rho\theta=0.24\ \text{m}.
\end{align}
The corresponding dispersion at the entrance  (exit) and middle of the bending magnet are 
\begin{equation}
\begin{aligned}
D_{\text{e}}&=2.4\text{ mm},\\ D_{\text{m}}&=-1.2\text{ mm}.
\end{aligned}
\end{equation} 
The natural energy spread is
\begin{align}
\sigma_{\delta\text{S}}&=2.4\times10^{-4},
\end{align}
and the minimum possible emittance is 
\begin{equation}
\epsilon_{z,\text{min}}\approx0.98\ \text{pm}.
\end{equation}

In the above evaluation, the dispersion at the entrance and exit of the bending magnet $D_{\text{e}}$ is small. The potential challenges for this lattice design is to correct the three chromaticities $\xi_{x,y,z}=\frac{d\nu_{x,y,z}}{d\delta}$ and get a large enough dynamic aperture (DA) at the same time.  Due to the different scaling laws of $\sigma_{z,\text{min}}\propto\sqrt{\rho}\theta^{3}$
and $D_{\text{e}}\propto\rho\theta^{2}$, there is some flexibility in choosing the bending magnet parameters to increase $D_{\text{e}}$ without change of $\sigma_{z,\text{min}}$. For example if we want to increase $D_{\text{e}}$ by a factor of two and keep $\sigma_{z,\text{min}}$ unchanged at the same time, we can choose the new bending radius and angle of the bending magnet as 
\begin{equation}\label{eq:bendingangleNew}
\begin{aligned}
\rho_{\text{new}}&=2\sqrt{2}\rho_{\text{old}}=4\sqrt{2}\ \text{m},\\
\theta_{\text{new}}&=\frac{1}{\sqrt[4]{2}}\theta_{\text{old}}=0.1\text{ rad}=5.7^{\circ}.
\end{aligned}
\end{equation}
The new magntic field is $B_{\text{new}}=0.24\ \text{T}$ and the bending magnet length in this case is 
\begin{align}
L_{\text{bend,new}}=\rho_{\text{new}}\theta_{\text{new}}=0.57\ \text{m}.
\end{align}
The corresponding dispersion at the entrance (exit) and middle of the bending magnet are now
\begin{equation}
\begin{aligned}
D_{\text{e},\text{new}}&=2D_{\text{e},\text{old}}=4.8\text{ mm},\\ D_{\text{m},\text{new}}&=2D_{\text{m},\text{new}}=-2.4\text{ mm}.
\end{aligned}
\end{equation} 
The natural energy spread is now
\begin{align}
\sigma_{\delta\text{S},\text{new}}=1.4\times10^{-4},
\end{align}
and the minimum possible emittance is now
\begin{equation}
\epsilon_{z,\text{min},\text{new}}\approx 0.58\ \text{pm}.
\end{equation}

Equation~(\ref{eq:bendingangleNew})  means that in total $\frac{2\pi}{0.1}\approx63$ bending magnets are needed in the ring. If the length of each isochronous cell with a single bending magnet can be designed to be around $2.5$ m, then the circumference of the ring can be about $180$ m, considering the sections of laser modulation, radiation generation and the energy supply system, etc. However, there is one practical issue to realize 3 nm in this ring. If we want to realize CW output radiation to obtain a high average power, then it means a CW mode optical enhancement cavity is needed. At present, 1 MW stored laser power is what we can aim for. For a 1 MW power $1\ \mu$m wavelength laser, the energy chirp strength $|h|$ realizable is around $4000\ \text{m}^{-1}$ for a practical undulator modulator. Then according to Eq.~(\ref{eq:classicalScaling}),
\begin{equation}
|\eta C_{0}|\approx |h|\left(\frac{\sigma_{z\text{S}}}{\sigma_{\delta\text{S}}}\right)^{2}=4000\times\left(\frac{3\times10^{-9}}{1.4\times10^{-4}}\right)^{2}\ \text{m}=1.8\ \mu\text{m},
\end{equation}
which means the global phase slippage factor $\eta$ required is $\sim1\times10^{-8}$, and is too small at present for a real machine. To make the requirement on global phase slippage less demanding, we can aim for a bunch length of $\sim30$ nm in the ring, and invoke the bunch compression scheme in the insertion to realize 3 nm at the radiator. The phase slippage factor needed can then be $1\times10^{-6}$ level, which is realizable in a real machine. For the bunch compression, we may invoke the longitudinal strong focusing scheme to be introduced in Sec.\ref{sec:multiRFs}, or other novel bunch compression schemes as will be introduced in Chap.~\ref{cha:TLC} which can be viewed as a generalized longitudinal strong focusing.

\subsubsection{Transverse Gradient Bends}
The above analysis is for a constant bending radius. To minimize the longitudinal emittance further, transverse and longitudinal gradient bending magnets (TGB and LGB) can be invoked.  Below we conduct some calculations based on the similar dispersion configuration as shown in the left part of Fig.~\ref{fig:Chap2-Dispersion}, but this time using a TGB.
The Hill's equation for the dispersion is \cite{chao2020lectures} 
\begin{align}\label{eq:TGBDx}
\frac{d^{2}D(s)}{ds^{2}}+\left(\frac{1}{\rho(s)^{2}}-k(s)\right)D(s)=\frac{1}{\rho(s)}.
\end{align}
For simplicity, here we only investigate the case of a constant bending radius $\rho(s)=\rho$ and a constant transverse gradient $k(s)=k$. To simplify the writing, we denote 
\begin{align}
g=\frac{1}{\rho^{2}}-k.
\end{align}
If $g>0$, then the solution of Eq.~(\ref{eq:TGBDx}) is
\begin{equation}
\begin{aligned}
D(s)&=D_{\text{i}}\cos\left(\sqrt{g}s\right)+D'_{\text{i}}\frac{\sin\left(\sqrt{g}s\right)}{\sqrt{g}}+\frac{1}{g\rho}\left[1-\cos\left(\sqrt{g}s\right)\right],\\
D'(s)&=-D_{\text{i}}\sqrt{g}\sin\left(\sqrt{g}s\right)+D'_{\text{i}}\cos\left(\sqrt{g}s\right)+\frac{1}{\sqrt{g}\rho}\sin\left(\sqrt{g}s\right),
\end{aligned}
\end{equation}
where $D_{\text{i}}$ and $D'_{\text{i}}$ are the initial dispersion and dispersion angle at the origin $s=0$ m, respectively.
If $g<0$, then the solution of Eq.~(\ref{eq:TGBDx}) is
\begin{equation}
\begin{aligned}
D(s)&=D_{\text{i}}\cosh\left(\sqrt{|g|}s\right)+D'_{\text{i}}\frac{\sinh\left(\sqrt{|g|}s\right)}{\sqrt{|g|}}+\frac{1}{|g|\rho}\left[-1+\cosh\left(\sqrt{|g|}s\right)\right],\\
D'(s)&=-D_{\text{i}}\sqrt{|g|}\sinh\left(\sqrt{|g|}s\right)+D'_{\text{i}}\cosh\left(\sqrt{|g|}s\right)+\frac{1}{\sqrt{|g|}\rho}\sinh\left(\sqrt{|g|}s\right).
\end{aligned}
\end{equation}

\begin{figure}[tb] 
	\centering 
	\includegraphics[width=0.5\columnwidth]{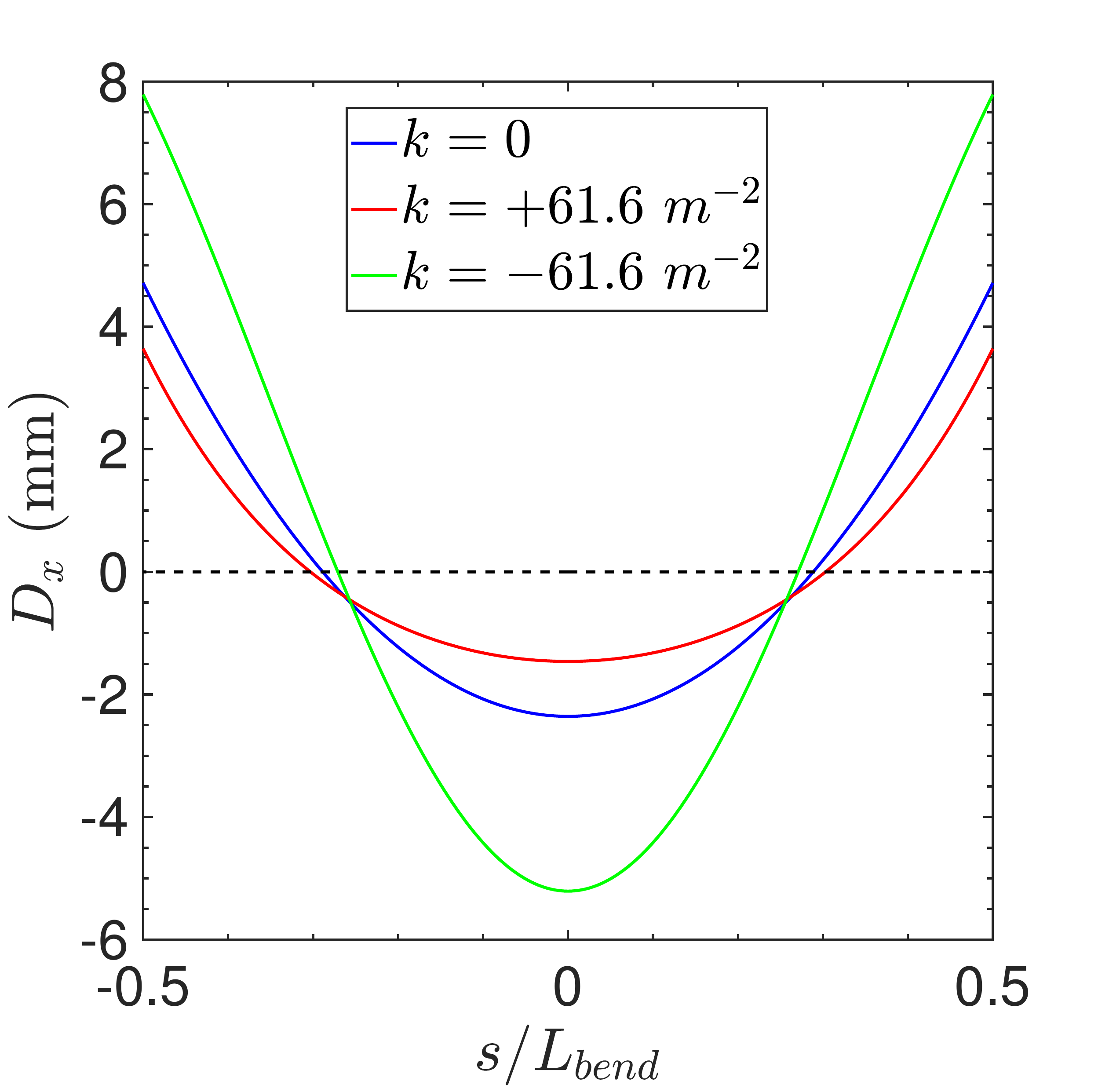}
	\caption{
		\label{fig:Chap2-DisTGB} 
		Dispersion function in an isochronous dipole with zero, positive and negative transverse gradients, respectively. The bending radius $\rho=4\sqrt{2}\ \text{m}$, total bending angle $\theta=0.1\text{ rad}=5.7^{\circ}$, and total length of the dipole $L_{\text{bend}}=\rho\theta=0.57$ m.
	}
\end{figure}

Below, we do the derivation for the case of $g>0$ and the results are similar when $g<0$. Similar to the previous calculations, we set the origin at the middle of the dipole where $D'_{\text{m}}=0$, the dispersion as a function of angle $\phi$ is then
\begin{align}\label{eq:TGBDxAlpha}
D(\phi)&=D_{\text{m}}\cos\left(\sqrt{g}\rho\phi\right)+\frac{1}{g\rho}\left[1-\cos\left(\sqrt{g}\rho\phi\right)\right].
\end{align} 
Substitute Eq.~(\ref{eq:TGBDxAlpha}) into the isochronicity condition Eq.~(\ref{eq:isochronicity}),  we get 
\begin{equation}
\begin{aligned}
D_{\text{m}}&=\frac{1}{g\rho}\left[1-\frac{\sqrt{g}\rho\frac{\theta}{2}}{\sin\left(\sqrt{g}\rho\frac{\theta}{2}\right)}\right]=-\frac{1}{24}\rho\theta^{2}\left(1+\frac{7}{240}g\rho^{2}\theta^{2}\right)+\mathcal{O}(g^{2}\rho^{5}\theta^{6}),\\
D_{\text{e}}&=D\left(\frac{\theta}{2}\right)=\frac{1}{g\rho}\left[1-\frac{\sqrt{g}\rho\frac{\theta}{2}}{\tan\left(\sqrt{g}\rho\frac{\theta}{2}\right)}\right]=\frac{1}{12}\rho\theta^{2}\left(1+\frac{1}{60}g\rho^{2}\theta^{2}\right)+\mathcal{O}(g^{2}\rho^{5}\theta^{6}),
\end{aligned}
\end{equation}
where $\mathcal{O}(x^{n})$ means terms of order $x^{n}$ and higher. The $\sqrt{\langle F^{2}\rangle-\langle F\rangle^2}$ in this case can be calculated as follows
\begin{equation}
\begin{aligned}
&F(\phi)=\int_{0}^{\phi}D(\phi')d\phi'=\frac{1}{g\rho}\left[\phi-\frac{\frac{\theta}{2}}{\sin\left(\sqrt{g}\rho\frac{\theta}{2}\right)}\sin\left(\sqrt{g}\rho\phi\right)\right],\\
&\langle F\rangle=\frac{1}{\frac{\theta}{2}}\int_{0}^{\frac{\theta}{2}} F(\phi)d\phi=\frac{1}{\sqrt{g}^{3}\rho^{2}}\left[\frac{\sqrt{g}\rho\theta}{4}-\tan\left(\frac{\sqrt{g}\rho\theta}{4}\right)\right]\\
&\ \ \ \ \ \ \ \ =-\frac{1}{192}\rho\theta^{3}\left(1+\frac{g\rho^{2}\theta^{2}}{40}\right)+\mathcal{O}(g^{2}\rho^{5}\theta^{7}),\\
&\langle F^{2}\rangle=\frac{1}{\frac{\theta}{2}}\int_{0}^{\frac{\theta}{2}} F^{2}(\phi)d\phi\\
&\ \ \ \ \ \ \   =\frac{1}{6g^3\rho^{4}}\left[-12+2g\rho^{2}\left(\frac{\theta}{2}\right)^{2}+9\frac{\sqrt{g}\rho\left(\frac{\theta}{2}\right)}{\tan\left(\sqrt{g}\rho\left(\frac{\theta}{2}\right)\right)}+3\frac{\left(\sqrt{g}\rho\left(\frac{\theta}{2}\right)a\right)^2}{\sin\left(\sqrt{g}\rho\left(\frac{\theta}{2}\right)\right)^2}\right]\\
&\ \ \ \ \ \ \  =\frac{1}{30240}\rho^{2}\theta^{6}\left(1+\frac{g\rho^{2}\theta^{2}}{20}\right)+\mathcal{O}(g^{2}\rho^{6}\theta^{10}),\\
&\sqrt{\langle F^{2}\rangle-\langle F\rangle^{2}}\approx\frac{\sqrt{2415}}{20160}\rho\theta^{3}\left(1+\frac{g\rho^{2}\theta^{2}}{40}\right)+\mathcal{O}(g^{2}\rho^{5}\theta^{7}). 
\end{aligned}
\end{equation}
For example, to reduce $\sqrt{\langle F^{2}\rangle-\langle F\rangle^{2}}$ by a factor of two compared to the case of no transverse gradient, we need
\begin{equation}
\frac{g\rho^{2}\theta^{2}}{40}=-\frac{1}{2}\Rightarrow g=-\frac{20}{L_{\text{bend}}^{2}}.
\end{equation}
For the case of the second example shown in last section, then
$k=\frac{1}{\rho^{2}} -g=\frac{1}{(4\sqrt{2})^{2}}+\frac{20}{0.57^{2}}\ \text{m}^{-2}=61.6\ \text{m}^{-2}$ which is a practical gradient. Figure~\ref{fig:Chap2-DisTGB} gives an example plot of the dispersion within an isochronous bending magnet with $k=0$ and $=\pm61.6\ \text{m}^{-2}$, respectively. As can be seen, a positive $k$ is favorable in suppressing the magnitude of dispersion and the partial phase slippage.

Besides the influence on $\sqrt{\langle F^{2}\rangle-\langle F\rangle^{2}}$, the transverse gradient may also influence the damping partition and hence has an influence on the bunch length and longitudinal emittance. For the specific case of a constant bending radius with a constant transverse gradient we are treating, we have
\begin{equation}\label{eq:radiationIntegrals}
\begin{aligned}
I_{2}&=\oint\left(\frac{1}{\rho_{x}^{2}}+\frac{1}{\rho_{y}^{2}}\right)ds=\frac{2\pi}{\rho},\\
I_{4x}&=\oint\frac{D_{x}}{\rho_{x}^{3}}(1+2\rho_{x}^{2}k)ds=0,\\
I_{4y}&=\oint\frac{D_{y}}{\rho_{y}^{3}}(1-2\rho_{y}^{2}k)ds=0,\\
J_{s}&=2+\frac{I_{4x}+I_{4y}}{I_{2}}=2,\\
\end{aligned}
\end{equation}
where $I_{2}$, $I_{4x,y}$ and $I_{3}$ in the following are the radiation integrals~\cite{sands1970physics}.
So a dipole with a constant bending radius and a constant transverse gradient is not very flexible in controlling the damping partition number, due to the constraint of isochronous condition. A varying transverse gradient may be helpful to minimize the longitudinal emittance, and optimization of the transverse gradient profile based on numerical method can be invoked.


\subsubsection{Longitudinal Gradient Bends}

Now we investigate the application of LGB to minimize longitudinal emittance. For simplicity, we assume that there is no quadrupole component in the LGB. Then inside the sector dipole, the dispersion function satisfy
\begin{equation}\label{eq:DFunctionInDipole}
D''(s)+\frac{D(s)}{\rho(s)^{2}}=\frac{1}{\rho(s)}.
\end{equation}
For simplicity, first we neglect the $\frac{D(s)}{\rho(s)^{2}}$ in Eq.~(\ref{eq:DFunctionInDipole}).  The condition for such simplification is $\frac{D(s)}{\rho(s)}\ll1$, which is satisfied in our case as the $\rho(s)$ is usually in meter level while $D(s)$ is in centimeter or millimeter level.
Considering the symmetry, we still set the middle point of the dipole as the starting point where $D_{\text{m}}'=0$. Then
\begin{equation}
D(\phi)=D_{\text{m}}+\int_{0}^{\phi}\phi'\rho(\phi')d\phi'.
\end{equation}
Ignoring the contribution of $\frac{1}{\gamma^{2}}$ on $F$, then
\begin{equation}
F(\phi)=\int_{0}^{\phi}D(\phi')d\phi'=D_{\text{m}}\phi+\int_{0}^{\phi}d\phi'\int_{0}^{\phi'}d\phi''\phi''\rho(\phi'').
\end{equation}
Suppose the total bending angle of a LGB is $\theta$, then the length of the bend is
\begin{equation}
L_{\text{bend}}=2\int_{0}^{\frac{\theta}{2}}\rho(\phi)d\phi.
\end{equation}
Requirement on zero $F$ (isochronousity) of the LGB is
\begin{equation}
F\left(\frac{\theta}{2}\right)=D_{\text{m}}\frac{\theta}{2}+\int_{0}^{\frac{\theta}{2}}d\phi'\int_{0}^{\phi'}d\phi''\phi''\rho(\phi'')=0.
\end{equation}
Therefore,
\begin{equation}
D_{\text{m}}=-\frac{\int_{0}^{\frac{\theta}{2}}d\phi'\int_{0}^{\phi'}d\phi''\phi''\rho(\phi'')}{\frac{\theta}{2}}.
\end{equation}
If $\rho(s)=\rho$ is a constant, then $D_{\text{m}}=-\frac{\rho\theta^{2}}{24}$, which is the result we analyzed before.

Suppose that the main ring consists of $\frac{2\pi}{\theta}$ isochronous LGBs, and use $\phi$ as the free variable, the problem is then framed as: keep the bending magnet length $L_{\text{bend}}$ and bending angle $\theta$ the same, find function $\rho(\phi)$ over $\phi\in[-\frac{\theta}{2},\frac{\theta}{2}]$ 
to minimize 
\begin{equation}\label{eq:TMEL}
\begin{aligned}
\Delta\epsilon_{z}[\theta,\rho(\phi),\beta_{z0},\alpha_{z0}]&=\frac{2\pi}{\theta}\frac{55}{96\sqrt{3}}\frac{\alpha_{F}{\lambdabar}_{e}^{2}\gamma^{5}}{\alpha_{\text{L}}}\int_{-\frac{\theta}{2}}^{\frac{\theta}{2}}\frac{\beta_{z}(\phi)}{|\rho(\phi)|^{2}}d\phi\\
&=\int_{-\frac{\theta}{2}}^{\frac{\theta}{2}}f[\theta,\rho(\phi),\beta_{z0},\alpha_{z0}]d\phi,
\end{aligned}
\end{equation}
in which
\begin{equation}
\begin{aligned}
\beta_{z}(\phi)=&\beta_{z0}-2\alpha_{z0}F(\phi)+\frac{1+\alpha_{z0}^2}{\beta_{z0}}F^{2}(\phi)
\end{aligned}
\end{equation}
and 
\begin{equation}
\alpha_{\text{L}}\approx\frac{U_{0}}{E_{0}}=\frac{2\pi}{\theta}\frac{2{\lambdabar}_{e}\alpha_{F}\gamma^{3}}{3}\int_{-\frac{\theta}{2}}^{\frac{\theta}{2}}\frac{1}{|\rho(\phi)|}d\phi.
\end{equation}
In other words we need to minimize
\begin{equation}
f(\rho(\phi))=\frac{\int_{-\frac{\theta}{2}}^{\frac{\theta}{2}}\frac{\beta_{z}(\phi)}{|\rho(\phi)|^{2}}d\phi}{\int_{-\frac{\theta}{2}}^{\frac{\theta}{2}}\frac{1}{|\rho(\phi)|}d\phi}.
\end{equation}
It seems that analytical results of $\rho(\phi)$ can only be obtained in some special cases.  In the general case, it is more straightforward to seek for numerical optimization using the formulation presented in this section. 

Note that $\rho(-\phi)=\rho(\phi)$ and $F(-\phi)=-F(\phi)$.  For simplicity, we assume that $\rho(\phi)>0$ and $\alpha_{z0}=0$, then $\beta_{z}(-\phi)=\beta_{z}(\phi)$, and
\begin{equation}
f(\rho(\phi))=\frac{\int_{0}^{\frac{\theta}{2}}\frac{\beta_{z}(\phi)}{\rho(\phi)^{2}}d\phi}{\int_{0}^{\frac{\theta}{2}}\frac{1}{\rho(\phi)}d\phi}.
\end{equation}
In order to see how effective LGB is in minimizing longitudinal emittance, we need to get the maximum ratio of the minimum emittance reachable by using normal bends and LGBs 
\begin{equation}
\mathcal{R}=\text{max}\left[\frac{\text{min}\left(\frac{\int_{0}^{\frac{\theta}{2}}\frac{\beta_{z,\text{NB}}(\phi)}{\rho_{\text{NB}}^{2}}d\phi}{\int_{0}^{\frac{\theta}{2}}\frac{1}{\rho_{\text{NB}}}d\phi}\right)}{\text{min}\left(\frac{\int_{0}^{\frac{\theta}{2}}\frac{\beta_{z,\text{LGB}}(\phi)}{\rho_{\text{LGB}}(\phi)^{2}}d\phi}{\int_{0}^{\frac{\theta}{2}}\frac{1}{\rho_{\text{LGB}}(\phi)}d\phi}\right)}\right].
\end{equation}
For a normal bend
\begin{equation}
\begin{aligned}
&\text{min}\left(\frac{\int_{0}^{\frac{\theta}{2}}\frac{\beta_{z,\text{NB}}(\phi)}{\rho_{\text{NB}}^{2}}d\phi}{\int_{0}^{\frac{\theta}{2}}\frac{1}{\rho_{\text{NB}}}d\phi}\right)\\
&=\text{min}\left(\frac{2}{\rho_{\text{NB}}\theta}\int_{0}^{\frac{\theta}{2}}\left\{\beta_{z0}+\frac{1}{\beta_{z0}}\left[\left(-\frac{1}{24} \rho_{\text{NB}}\theta ^2+\frac{1}{6} \rho_{\text{NB}}\phi ^2\right)\phi\right]^{2}\right\}d\phi\right)\\
&=\text{min}\left[\frac{2}{\rho_{\text{NB}}\theta}\left(\beta _{\text{z0}}\frac{\theta}{2}+\frac{  \rho_{\text{NB}}^2\theta ^7}{60480 \beta _{\text{z0}}} \right)\right]\\
&=2\sqrt{\frac{1}{30240}}\theta^{3},
\end{aligned}
\end{equation}
which is independent of $\rho_{\text{NB}}$ as expected. The minimum value is reached when $\beta_{\text{z0}}=\sqrt{\frac{1}{30240}}\rho_{\text{NB}}\theta^{3}$. Note that according to Eq.~(\ref{eq:TMEL}), the emittance is proportional to $\gamma^{2}$, therefore the $\epsilon_{z,\text{min}}\propto\gamma^{2}\theta^{3}$.

For a LGB, as mentioned it is not easy to obtain the general analytical result. Here we consider first the simple case of a polynomial function of $\rho_{\text{LGB}}(\phi)$ that for $\phi\in[0,\frac{\theta}{2}]$,
\begin{equation}
\rho_{\text{LGB}}(\phi)=a_{0}+a_{1}\phi+a_{2}\phi^{2}.
\end{equation} 
Note that for $\phi\in[-\frac{\theta}{2},0]$, then
\begin{equation}
\rho_{\text{LGB}}(\phi)=a_{0}-a_{1}\phi+a_{2}\phi^{2}.
\end{equation}
The length of a LGB is
\begin{equation}
L_{\text{bend}}=2\int_{0}^{\frac{\theta}{2}}\rho_{\text{LGB}}(\phi)d\phi=2\left(a_0 \frac{\theta}{2}+\frac{1}{2}a_1 \left(\frac{\theta}{2}\right) ^2+\frac{1}{3}a_2 \left(\frac{\theta}{2}\right) ^3\right).
\end{equation}
To make the bending magnet length the same, the bending radius of the normal magnet is $\rho_{\text{NB}}=a_{0}+\frac{1}{2}a_{1}\theta+\frac{1}{3}a_{2}\theta^{2}$.
The isochronousity of the LGB gives
\begin{equation}
\begin{aligned}
D_{\text{m}}&=-\frac{\int_{0}^{\theta}d\phi\int_{0}^{\phi}\phi'\rho_{\text{LGB}}(\phi')d\phi'}{\frac{\theta}{2}}\\
&=-\left(\frac{1}{6} a_0+\frac{1}{12}a_{1}\frac{\theta}{2}+\frac{1}{20}a_{2}\left(\frac{\theta}{2}\right)^{2}\right)\left(\frac{\theta}{2}\right) ^2.
\end{aligned}
\end{equation}
Then
\begin{equation}
\begin{aligned}
F(\phi)&=D_{\text{m}}\phi+\int_{0}^{\phi}d\phi'\int_{0}^{\phi'}\phi''\rho_{\text{LGB}}(\phi'')d\phi''\\
&=\left[-\left(\frac{1}{6} a_0+\frac{1}{12}a_{1}\frac{\theta}{2}+\frac{1}{20}a_{2}\left(\frac{\theta}{2}\right)^{2}\right)\left(\frac{\theta}{2}\right) ^2\right.\\
&\left.\ \ \ \ \ \ \ \ +\left(\frac{1}{6} a_0+\frac{1}{12}a_{1}\phi+\frac{1}{20}a_{2}\phi^{2}\right)\phi ^2\right]\phi,
\end{aligned}
\end{equation}
and
\begin{equation}
\begin{aligned}
\beta_{z}(\phi)&=\beta_{z0}+\frac{1}{\beta_{z0}}F^{2}(\phi)\\
&=\beta_{z0}+\frac{1}{\beta_{z0}}\left\{\left[-\left(\frac{1}{6} a_0+\frac{1}{12}a_{1}\frac{\theta}{2}+\frac{1}{20}a_{2}\left(\frac{\theta}{2}\right)^{2}\right)\left(\frac{\theta}{2}\right) ^2\right.\right.\\
&\left.\left.\ \ \ \  +\left(\frac{1}{6} a_0+\frac{1}{12}a_{1}\phi+\frac{1}{20}a_{2}\phi^{2}\right)\phi ^2\right]\phi\right\}^{2}.
\end{aligned}
\end{equation}
The complete expressions for $\int_{0}^{\frac{\theta}{2}}\frac{\beta_{z}(\phi)}{\rho_{\text{LGB}}(\phi)^{2}}d\phi$ is lengthy, although the integration is straightforward.
With the help of software {\it Mathematica}~\cite{Mathematica}, we have
\begin{equation}
\text{min}\left(\frac{\int_{0}^{\frac{\theta}{2}}\frac{\beta_{z,\text{LGB}}(\phi)}{\rho_{\text{LGB}}(\phi)^{2}}d\phi}{\int_{0}^{\frac{\theta}{2}}\frac{1}{\rho_{\text{LGB}}(\phi)}d\phi}\right)=2\sqrt{\frac{1}{30240}}\theta^{3}\left[1+\frac{7 a_1  }{128 a_0}\frac{\theta}{2}+\left(\frac{1487 a_1^2}{32768 a_0^2}+\frac{a_2}{15 a_0}\right) \left(\frac{\theta}{2}\right) ^2\right]+\mathcal{O}(\theta^{6}).
\end{equation}
Note the optimal $\beta_{\text{z0}}$ to reach the minimum emittance in the case of a LGB is different from that of a normal bend.
Then
\begin{equation}
\mathcal{R}=\text{max}\left[\frac{\text{min}\left(\frac{\int_{0}^{\theta}\frac{\beta_{z,\text{NB}}(\phi)}{\rho_{\text{NB}}^{2}}d\phi}{\int_{0}^{\theta}\frac{1}{\rho_{\text{NB}}}d\phi}\right)}{\text{min}\left(\frac{\int_{0}^{\theta}\frac{\beta_{z,\text{LGB}}(\phi)}{\rho_{\text{LGB}}(\phi)^{2}}d\phi}{\int_{0}^{\theta}\frac{1}{\rho_{\text{LGB}}(\phi)}d\phi}\right)}\right]\approx\frac{1}{\text{min}\left[1+\frac{7 a_1  }{128 a_0}\frac{\theta}{2}+\left(\frac{1487 a_1^2}{32768 a_0^2}+\frac{a_2}{15 a_0}\right) \left(\frac{\theta}{2}\right) ^2\right]}.
\end{equation}

If $a_{2}=0$, one can see that when $\frac{a_{1}\frac{\theta}{2}}{a_{0}}=-\frac{\frac{7}{128}}{2\frac{1487}{32768}}\approx-0.6$, the denominator of the above formula reaches the minimum and $\mathcal{R}=\frac{1}{0.98}$. Therefore, a linear longitudinal gradient is not very effective in minimizing the longitudinal emittance. If $a_{2}\neq0$, one can see that the condition of $\rho_{\text{LGB}}\left(\frac{\theta}{2}\right)>0$, will make $\frac{a_{2}\left(\frac{\theta}{2}\right)^{2}}{a_{0}}>-\frac{1}{15}\left(1-\frac{\frac{7}{128}}{2\frac{1487}{32768}}\right)$, and $\mathcal{R}<\frac{1}{0.96}$. So generally, a LGB with a polynomial function form of positive $\rho(\phi)$ is not very effective in minimizing the longitudinal emittance. There could be other form of $\rho(\phi)$ which is more effective than the polynomial function form. Here, however, we do not present the details of the other functional form.

\subsection{Multiple RFs}\label{sec:multiRFs}

\subsubsection{Analysis}
\begin{figure}[tb] 
	\centering 
	\includegraphics[width=0.7\columnwidth]{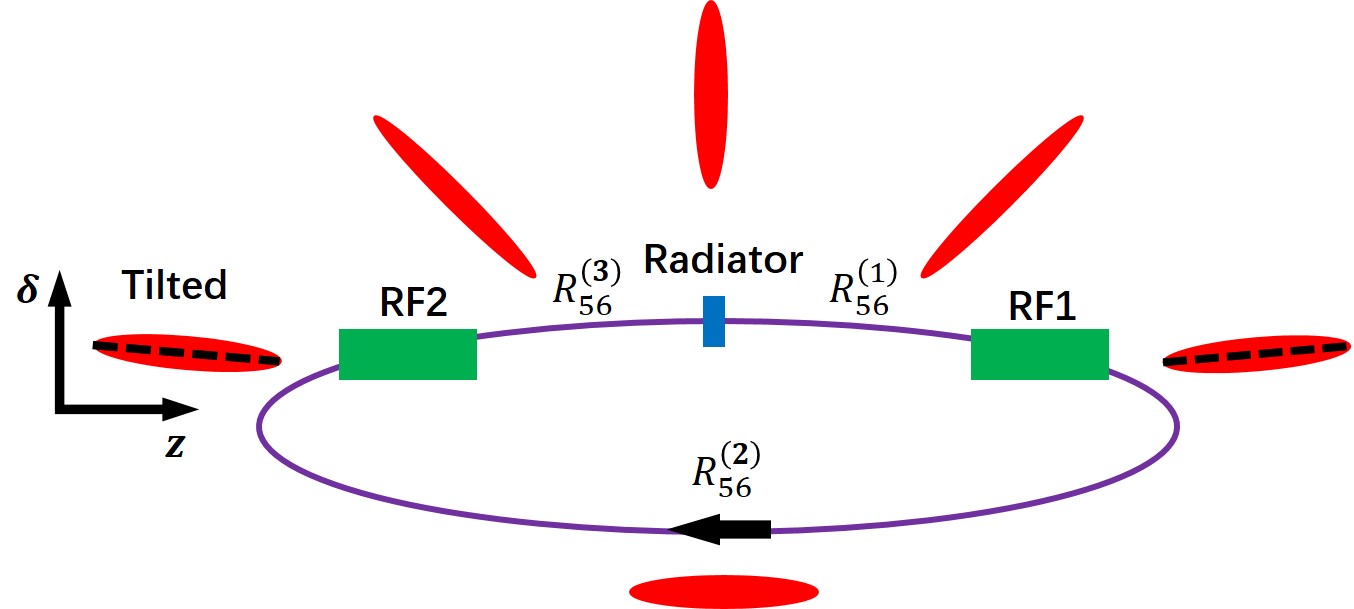}
	\caption{
		\label{fig:Chap2-TwoRFs} 
		Schematic layout of a storage ring using two RF systems for longitudinal strong focusing and an example beam distribution evolution in the longitudinal phase space. Note that the tilted
		angles of the beam distribution and bunch length ratios at different places do not strictly correspond to the parameters in Tab.~\ref{tab:tab2}, but only to present the qualitative characteristics.
	}
\end{figure}

The analysis in the above sections considers the case with only a single RF. When there are multiple RFs, for the longitudinal dynamics, it is similar to implement multiple quadrupoles in the transverse dimension, and the beam dynamics can have more possibilities. Longitudinal strong focusing scheme for example can be invoked~\cite{biscari2005bunch,chao2016high}, not unlike its transverse counterpart which almost all the storage rings today implement.  The linear beam dynamics with multiple RFs can be treated the same way as that with a single RF using SLIM. When all the RFs are placed at dispersion-free locations, the Courant-Snyder parametrization can be applied as analyzed in Sec.\ref{sec:CS}. Here we use a setup with two RFs as an example to show the scheme of manipulating $\beta_{z}$ around the ring. The schematic layout of the ring is shown in Fig.~\ref{fig:Chap2-TwoRFs}. The treatment of cases with more RFs is similar.

\begin{table}[tb]
	\centering
	\caption{\label{tab:tab2}
		An example parameters set corresponding to the setup shown in Fig.~\ref{fig:Chap2-TwoRFs}.}
	\begin{tabular}{cc}
		\hline
		Parameter & \multicolumn{1}{c}{\textrm{Value}}\\
		\hline
		$R_{56}^{(1)}$ & 15 $\mu$m \\
		$R_{56}^{(2)}$ & -100 $\mu$m \\
		$R_{56}^{(3)}$ & 15 $\mu$m \\
		$h$ & $-5\times10^{4}\ \text{m}^{-1}$ \\
		$\zeta_{1}$ & 0.25 \\
		$\zeta_{2}$ & 7 \\
		$\nu_{s}$ & 0.115  \\
		$\beta_{z}(s_{\text{rad}})$ & 1.9  $\mu$m\\
		$\beta_{z}(s_{\text{RF}})$ & 121  $\mu$m\\
		$\alpha_{z}(s_{\text{rad}})$ & 0  \\
		$\alpha_{z}(s_{\text{RF1-}})$ & -7.9  \\
		$\alpha_{z}(s_{\text{RF1+}})$ & -1.9  \\
		$\alpha_{z}(s_{\text{RF2-}})$ & 1.9  \\
		$\alpha_{z}(s_{\text{RF2+}})$ & 7.9  \\
		\hline
	\end{tabular}
\end{table}	

We divide the ring into five sections, i.e., three longitudinal drifts ($R_{56}$) and two longitudinal quadrupole kicks ($h$), with the transfer matrices given by
\begin{equation}
\begin{aligned}
&{\bf T}_{\text{D1}}=\left(\begin{matrix}
1&R_{56}^{(1)}\\
0&1
\end{matrix}\right),\ {\bf T}_{\text{RF1}}=\left(\begin{matrix}
1&0\\
h_{1}&1
\end{matrix}\right),\\
&{\bf T}_{\text{D2}}=\left(\begin{matrix}
1&R_{56}^{(2)}\\
0&1
\end{matrix}\right),\
{\bf T}_{\text{RF2}}=\left(\begin{matrix}
1&0\\
h_{2}&1
\end{matrix}\right),\\
& {\bf T}_{\text{D3}}=\left(\begin{matrix}
1&R_{56}^{(3)}\\
0&1
\end{matrix}\right).
\end{aligned}
\end{equation}
Then the one-turn map at the radiator center is
\begin{equation}
\begin{aligned}
{\bf M}_{\text{R}}&={\bf T}_{\text{D3}}{\bf T}_{\text{RF2}}{\bf T}_{\text{D2}}{\bf T}_{\text{RF1}}{\bf T}_{\text{D1}}.\\
\end{aligned}
\end{equation}
Linear stability requires that $|\text{Tr}\left({\bf M_{\text{R}}}\right)|<2$, where Tr() means the trace of.
For the generation of coherent radiation, we usually want the bunch length to reach its minimum at the radiator, then we need $\alpha_{z}=0$ for ${\bf M}_{\text{R}}$.

With the primary purpose to present the principle, instead of a detailed design, here for simplicity we only discuss one special case: $R_{56}^{(1)}=R_{56}^{(3)}$, $h_{1}=h_{2}=h$. 
The treatment of more general cases with different signs and magnitudes of $R_{56}^{(1)}$ and $R_{56}^{(3)}$ and $h_{1}$ and $h_{2}$ is similar, but the same-signed $R_{56}^{(1)}$ and $R_{56}^{(3)}$ might be easier for a real lattice to fulfill. For example if $R_{56}^{(1)},R_{56}^{(3)}>0$, a possible realization of them are chicanes. 

For the special case of $R_{56}^{(1)}=R_{56}^{(3)}$, $h_{1}=h_{2}=h$ and denote 
\begin{equation}
1+R_{56}^{(1)}h=\zeta_{1},\ 2+R_{56}^{(2)}h=\zeta_{2},
\end{equation}
we then have 
\begin{equation}
\begin{aligned}
{\bf M}_{\text{R}}=\left(\begin{matrix}
\zeta_{1}\zeta_{2}-1&\frac{\zeta_{1}^{2}\zeta_{2}-2\zeta_{1}}{h}\\
h\zeta_{2}&\zeta_{1}\zeta_{2}-1
\end{matrix}\right).
\end{aligned}
\end{equation}
The linear stability requires 
$
|\zeta_{1}\zeta_{2}-1|<1,
$
and the synchrotron tune is
\begin{equation}
\nu_{s}=
\begin{cases}
&\frac{1}{2\pi}\arccos\left[\zeta_{1}\zeta_{2}-1\right]\ \text{if}\ \frac{\zeta_{1}^{2}\zeta_{2}-2\zeta_{1}}{h} > 0, \\
&1-\frac{1}{2\pi}\arccos\left[\zeta_{1}\zeta_{2}-1\right]\ \text{if}\ \frac{\zeta_{1}^{2}\zeta_{2}-2\zeta_{1}}{h} < 0.
\end{cases}
\end{equation} 

Here we give one example parameter set with a stable linear motion as shown in Tab.~\ref{tab:tab2}. 
We recognize the fact that the large energy chirp strength $h$ used is demanding, for SSMB only pulsed laser can now reach such large value (100 MW level modulation laser power required), and here our primary goal is to present the principle based on which the interested readers can choose and optimize the parameters for their target applications. We will discuss in Chap.~\ref{cha:TLC} the application of transverse-longitudinal coupling scheme to lower the requirement on the modulation laser power. According to the longitudinal Courant-Snyder functions given in Tab.~\ref{tab:tab2} (note the values of $\beta_{z}$ and the signs of $\alpha_{z}$), the evolution of electron distribution in the longitudinal phase space around the ring (note the bunch lengths and orientations) is qualitatively shown in Fig.~\ref{fig:Chap2-TwoRFs}.

\subsubsection{Discussions}
Here we make several observations from the above analysis and numerical example, which we believe are important. First, $\beta_{z}$ in a longitudinal strong focusing ring is at the same level of or even smaller than the ring $|R_{56}=-\eta C_{0}|$, while in a longitudinal weak focusing ring $\beta_{z}\gg|-\eta C_{0}|$. Therefore, according to Eq.~(\ref{eq:longitudinalEmittance}), the equilibrium longitudinal emittance in a longitudinal strong focusing ring can be much smaller than that in a longitudinal weak focusing ring \cite{zhang2021ultralow}. Together with a smaller $\beta_{z}$, the bunch length can thus be much smaller than that in a longitudinal weak focusing ring. This is the reason behind the application of longitudinal strong focusing in SSMB to realize extreme short bunches~\cite{chao2016high,zhang2021ultralow}.

Second, $\beta_{z}$ changes significantly around the ring in the longitudinal strong focusing regime. Therefore, the bunch length and beam orientation in the longitudinal phase space vary greatly around the ring, as shown qualitatively in Fig.~\ref{fig:Chap2-TwoRFs}. This means the adiabatic approximation cannot be applied for the longitudinal dimension anymore.  Actually the adiabatic approximation also breaks down in the case corresponds to Fig.~\ref{fig:Chap2-LongitudinalBeta}, where the change of $\beta_{z}$ around the ring is significant although the total synchrotron phase advance per turn is small. Therefore, the global synchrotron tune is not a general criterion in the classification of a ring to be weak focusing or strong focusing. The evolution of $\beta_{z}$ is more relevant.  The argument is based on the fact that $R_{56}$ can be either positive or negative, therefore the local synchrotron phase advance can also both be positive and negative. While in the transverse dimension, the drift length and betatron phase advance are always be positive. 

The breakdown of adiabatic approximation can have crucial impacts on the study of both the single-particle and collective effects. For linear single-particle dynamics, the longitudinal and transverse dimensions should be treated the same way on equal footing and SLIM formalism can be invoked. The treatment of nonlinear single-particle dynamics is more subtle as the longitudinal dynamics now is strongly chaotic. 
For the collective effects, many classical treatments should be re-evaluated and some new formalism needs to be developed. For example, the Haissinski equation \cite{haissinski1973exact} for calculating the equilibrium beam distortion cannot be applied directly then. Also, to our knowledge, there is no discussion on CSR-induced microwave instability in a longitudinal strong focusing ring. The scaling law obtained in the longitudinal weak focusing \cite{bane2010threshold} cannot be applied directly. 3D CSR effects and also the impact of bunch lengthening from the transverse emittance on CSR needs more in-depth study. This is especially true for an SSMB ring, considering the fact that the beam width there is much larger than the microbunch length, while the contrary is true in a conventional ring. The contribution from horizontal emittance can easily dominate the bunch length at many places in an SSMB ring. This on the other hand, will be helpful to suppress unwanted CSR and may also be helpful in mitigating IBS, as extreme short bunches occur only at limited locations. The IBS in a longitudinal strong focusing ring, and a general coupled lattice, also deserves special attention. To our knowledge, the IBS formalism of Nash \cite{nash2006analytical} and that of Kubo and Oide~\cite{kubo2001intrabeam} can be applied for such purposes, as they are based on $6\times6$ general transport matrices.
An IBS formalism can be developed based on Chao's SLIM formalism \cite{chao1979evaluation}, in which eigen analysis has been invoked and applies to 3D general coupled lattice with longitudinal strong focusing.  


\subsection{Linear and Nonlinear Maps of a Laser Modulator}\label{eq:secLM}
In the previous discussions, we have approximated the function of a laser modulator by a thin-lens RF kick. This means that we have ignored the phase slippage or $R_{56}$ of the laser modulator itself. We need to know if this approximation is valid or under what circumstance can we use this approximation. 

Here we derive the phase slippage factor of the undulator first and then get the thick-lens transfer matrix of the laser modulator. The path length of an electron with a relative energy deviation of $\delta$ wiggling in a planar undulator is
\begin{equation}
\begin{aligned}
L(\delta)&=\int_{0}^{L_{u}}\sqrt{1+(x')^{2}}dz
\approx\int_{0}^{L_{u}}\left[1+\frac{1}{2}(x')^{2}\right]dz\\
&\approx\int_{0}^{L_{u}}\left[1+\frac{1}{2}\left(\frac{K}{\gamma}\cos(k_{u}z)\right)^{2}\right]dz\\
&\approx(1+\frac{1}{4}\frac{K^{2}}{\gamma^{2}})L_{u}
=\left[1+\frac{1}{4}\frac{K^{2}}{\gamma_{r}^{2}(1+\delta)^{2}}\right]L_{u}\\
&\approx(1-\frac{1}{2}\frac{K^{2}}{\gamma_{r}^{2}}\delta)L_{u},
\end{aligned}
\end{equation}
in which $k_{u}=2\pi/\lambda_{u}$ is the undulator wave number, $\gamma_{r}$ is the Lorentz factor corresponding to the resonant energy. The $R_{56}$ of an undulator is then 
\begin{equation}\label{eq:undulatorR56}
R_{56}=\frac{L_{u}-L(\delta)}{\delta}+\frac{L_{u}}{\gamma_{r}^{2}}=\frac{L_{u}(1+K^{2}/2)}{\gamma_{r}^{2}}=2N_{u}\lambda_{0},
\end{equation}
where $N_{u}$ is the number of undulator periods, $\lambda_{0}=\frac{1+K^{2}/2}{2\gamma_{r}^{2}}\lambda_{u}$ is the central wavelength of on-axis fundamental spontaneous radiation. As can be seen from Eq.~(\ref{eq:undulatorR56}), the undulator $R_{56}$ is twice the slippage length of the undulator radiation. Take an example parameter of the modulator for SSMB:
$\lambda_{0}=1\ \mu\text{m},\ N_{u}=50$, the $R_{56}$ of the modulator is then $R_{56}=100\ \mu\text{m}$, which is not a negligible value compared to the $R_{56}$ of the whole ring in the case of a quasi-isochronous ring. This $R_{56}$ of the undulator will also has an impact on the radiation if undulator is used as the radiator, especially in the longitudinal strong focusing scheme, where the energy spread at the radiator undulator is large.

As mentioned, the RF or laser modulator kick in linear approximation is like a longitudinal quadrupole and the $R_{56}$ of the laser modulator is like the length of this longitudinal quadrupole. If we represent a laser modulator as an RF kick followed by a dispersion $R_{56}=2N_{u}\lambda_{0}$, then the transfer matrix is 
\begin{equation}
\begin{aligned}
{\bf M}_{N=1}&=\left(\begin{matrix}
1&R_{56}\\
0&1
\end{matrix}\right)
\left(\begin{matrix}
1&0\\
h&1
\end{matrix}\right)=\left(\begin{matrix}
1+R_{56} h&R_{56}\\
h&1
\end{matrix}\right).
\end{aligned}
\end{equation}
A more symmetric form is ``$R_{56}/2+\text{RF}+R_{56}/2$"
\begin{equation}\label{eq:matrix3}
\begin{aligned}
{\bf M}_{N=1,\text{sym1}}&=\left(\begin{matrix}
1&R_{56}/2\\
0&1
\end{matrix}\right)
\left(\begin{matrix}
1&0\\
h&1
\end{matrix}\right)\left(\begin{matrix}
1&R_{56}/2\\
0&1
\end{matrix}\right)\\
&=\left(\begin{matrix}
1+R_{56} h/2&R_{56}+(R_{56}/2)^{2} h\\
h&1+R_{56} h/2
\end{matrix}\right),
\end{aligned}
\end{equation}
or ``$\text{RF}/2+R_{56}+\text{RF}/2$"
\begin{equation}\label{eq:matrix32}
\begin{aligned}
{\bf M}_{N=1,\text{sym2}}&=\left(\begin{matrix}
1&0\\
h/2&1
\end{matrix}\right)
\left(\begin{matrix}
1&R_{56}\\
0&1
\end{matrix}\right)\left(\begin{matrix}
1&0\\
h/2&1
\end{matrix}\right)\\
&=\left(\begin{matrix}
1+R_{56} h/2&R_{56}\\
h+(h/2)^{2}R_{56}&1+R_{56} h/2
\end{matrix}\right).
\end{aligned}
\end{equation}
If we treat it as $N$ equal tiny ``$\text{RF}/N+R_{56}/N$", then
\begin{equation}
\begin{aligned}
{\bf M}_{N}&=\left[\left(\begin{matrix}
1&R_{56}/N\\
0&1
\end{matrix}\right)
\left(\begin{matrix}
1&0\\
h/N&1
\end{matrix}\right)\right]^{N}=\left(\begin{matrix}
1+R_{56}h/N^{2}&R_{56}/N\\
h/N&1
\end{matrix}\right)^{N}.
\end{aligned}
\end{equation}
When $N\rightarrow\infty$, similar to the thick-lens quadrupole in the transverse dimension, we have
\begin{equation}\label{eq:thicklens}
\begin{aligned}
{\bf M}_{N\rightarrow\infty}&=\begin{cases}
&\left(\begin{matrix}
\cos\left(\sqrt{-R_{56}h}\right)&\frac{R_{56}\sin\left(\sqrt{-R_{56}h}\right)}{\sqrt{-R_{56}h}}\\
\frac{h\sin\left(\sqrt{-R_{56}h}\right)}{\sqrt{-R_{56}h}}&\cos\left(\sqrt{-R_{56}h}\right)
\end{matrix}\right),\ \text{if}\ R_{56}h<0,\\
&\left(\begin{matrix}
\cosh\left(\sqrt{R_{56}h}\right)&\frac{R_{56}\sinh\left(\sqrt{R_{56}h}\right)}{\sqrt{R_{56}h}}\\
\frac{h\sinh\left(\sqrt{R_{56}h}\right)}{\sqrt{R_{56}h}}&\cosh\left(\sqrt{R_{56}h}\right)
\end{matrix}\right),\ \text{if}\ R_{56}h>0.
\end{cases}
\end{aligned}
\end{equation}
A laser modulator is therefore like a thick-lens quadrupole in the longitudinal dimension.  A thin-lens approximation is applicable when $\sqrt{-R_{56}h}\ll1$.

Equation~(\ref{eq:thicklens}) can also be obtained using Courant-Snyder parameterization and De Moivre's formula. Following Courant-Snyder~\cite{courant1958theory}, a $2\times2$ symplectic matrix can be parameterized as
\begin{equation}
\begin{aligned}
{\bf T}&=\cos(\mu){\bf I}+\sin(\mu){\bf J},
\end{aligned}
\end{equation}
where
\begin{equation}
\begin{aligned}
{\bf I}&=\left(\begin{matrix}
1&0\\
0&1
\end{matrix}\right),\ {\bf J}=\left(\begin{matrix}
\alpha&\beta\\
-\gamma&-\alpha
\end{matrix}\right)
\end{aligned}
\end{equation}
and ${\bf J}^{2}=-{\bf I}$. The De Moivre's formula then gives
\begin{equation}
\begin{aligned}
{\bf T}^{N}&=\cos(N\mu){\bf I}+\sin(N\mu){\bf J}.
\end{aligned}
\end{equation}
For the laser modulator,  
\begin{equation}
\begin{aligned}
{\bf T}&=\left(\begin{matrix}
1+R_{56}h/N^{2}&R_{56}/N\\
h/N&1
\end{matrix}\right),
\end{aligned}
\end{equation}
we have
\begin{equation}
\mu=\arccos\left(1+\frac{R_{56}h}{2N^{2}}\right),\ \beta=\frac{R_{56}/N}{\sin(\mu)},\ \alpha=\frac{\frac{R_{56}h}{2N^{2}}}{\sin(\mu)},\ \gamma=-\frac{h/N}{\sin(\mu)}.
\end{equation}
Then 
\begin{equation}\label{eq:Nkicks}
\begin{aligned}
{\bf M}&={\bf T}^{N}=\cos(N\mu)\left(\begin{matrix}
1&0\\
0&1
\end{matrix}\right)+\sin(N\mu)\left(\begin{matrix}
\alpha&\beta\\
-\gamma&-\alpha
\end{matrix}\right)\\
&=\left(\begin{matrix}
\cos(N\mu)+\sin(N\mu)\frac{\frac{R_{56}h}{2N^{2}}}{\sin(\mu)}&\sin(N\mu)\frac{R_{56}/N}{\sin(\mu)}\\
\sin(N\mu)\frac{h/N}{\sin(\mu)}&\cos(N\mu)-\sin(N\mu)\frac{\frac{R_{56}h}{2N^{2}}}{\sin(\mu)}
\end{matrix}\right).
\end{aligned}
\end{equation}
Equation~(\ref{eq:thicklens}) is then the limit of Eq.~(\ref{eq:Nkicks}) when $N\rightarrow\infty\ \left(N\mu\rightarrow\sqrt{-R_{56}h}\right)$.

After discussing the linear map, now we take into account the fact that the modulation waveform is sinusoidal. In principle we can get the nonlinear thick-lens transfer map of the laser modulator using the technique of Lie algebra~\cite{dragt2011lie,chao2022special}.
Here we use the more straightforward method, i.e., to implement a symplectic kick map as below in the numerical code, to give the readers a picture,
\begin{equation}\label{eq:symkicks1}
\begin{aligned}
\text{for}&\ i=1:1:N_{u}\\
z&=z+\lambda_{0}\delta\\
\delta&=\delta+A_{i}\sin(k_{\text{L}}z)\\
z&=z+\lambda_{0}\delta\\
\text{end} & 
\end{aligned}
\end{equation}
in which $\lambda_{\text{L}}$ is the modulation laser wavelength.
In other words, we have split the undulator into $N_{u}$ small ``$\frac{1}{2}$dispersion + modulation + $\frac{1}{2}$dispersion". For a plane wave
$
A_{i}=\frac{A}{N_{u}}
$ with $A$ being the total energy modulation strength,
while for a Gaussian laser beam $A_{i}$ is a function of $i$.

\begin{figure}[tb] 
	\centering 
	\includegraphics[width=0.32\columnwidth]{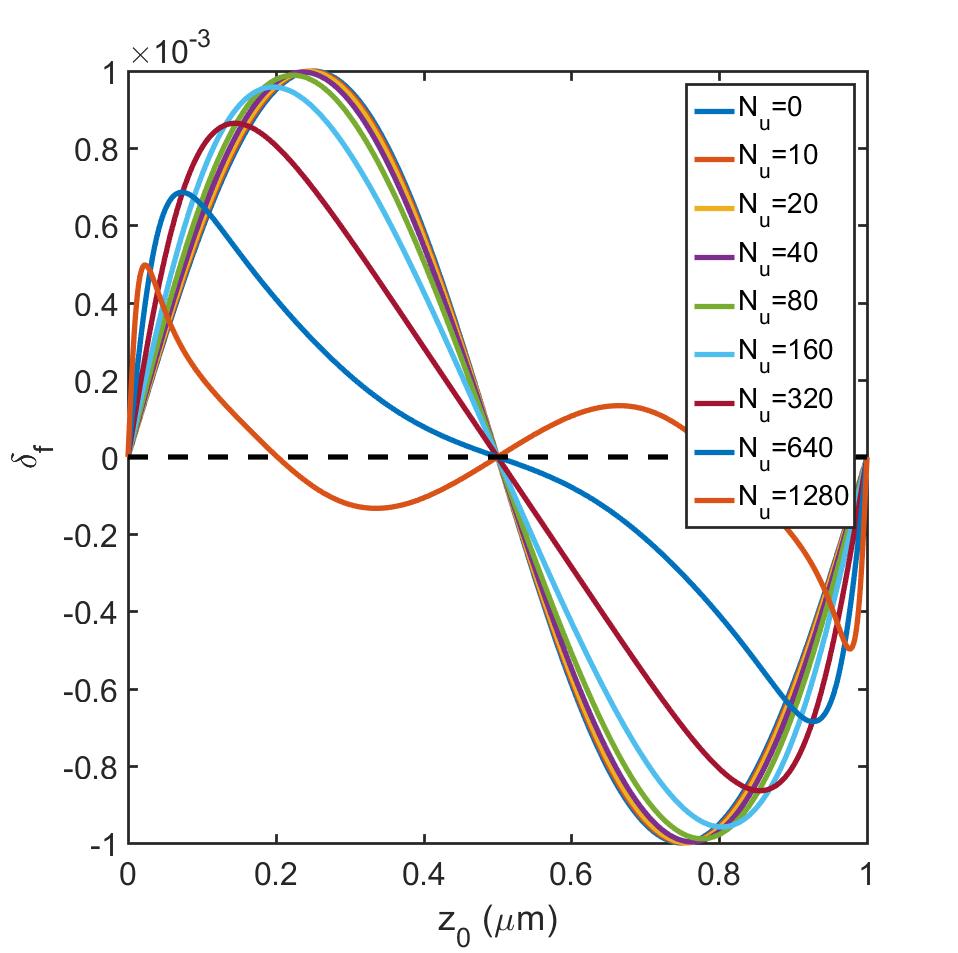}
	\includegraphics[width=0.32\columnwidth]{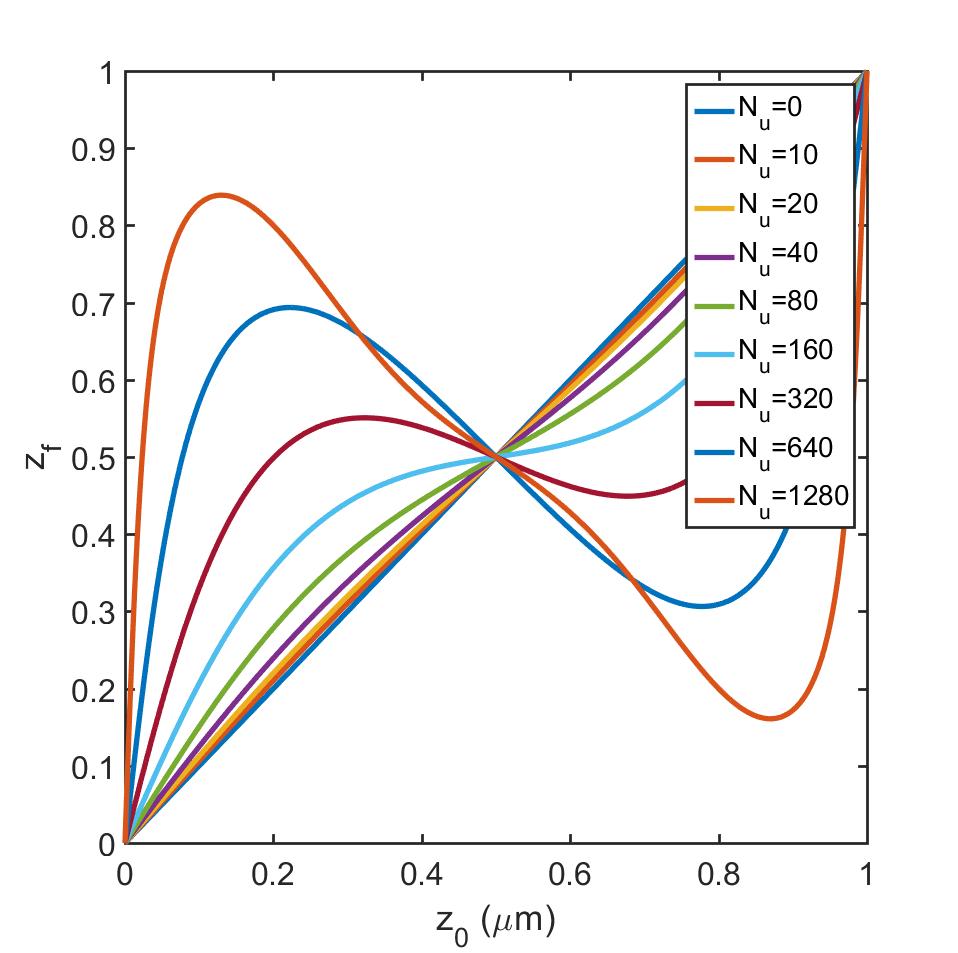}
	\includegraphics[width=0.32\columnwidth]{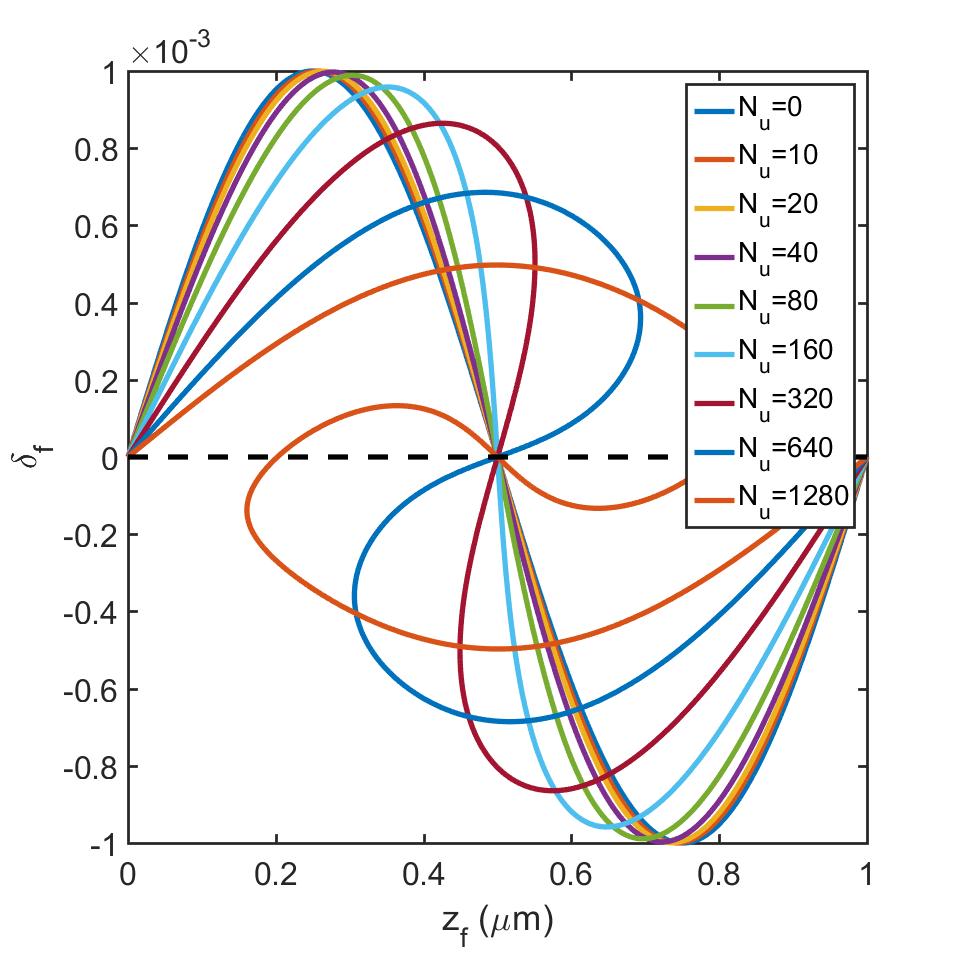}
	\caption{
		\label{fig:Chap2-ModulationWaveform} 
		%
		The single-pass modulation of a line beam in a laser modulator, with ``$\delta=0$ and $z\in[0,\lambda_{0}]$", as a function of $N_{u}$ (namely $R_{56}$) of the undulator. $(x,y)$ in the figures from left to right: ($z_{\text{entrance}}$, $\delta_{\text{exit}}$);  ($z_{\text{entrance}}$, $z_{\text{exit}}$);  ($z_{\text{exit}}$, $\delta_{\text{exit}}$). Parameters used: $\lambda_{\text{L}}=\lambda_{0}=1\ \mu{m},\ A=1\times10^{-3}$.
	}
\end{figure}

\begin{figure}[tb] 
	\centering 
	\includegraphics[width=0.32\columnwidth]{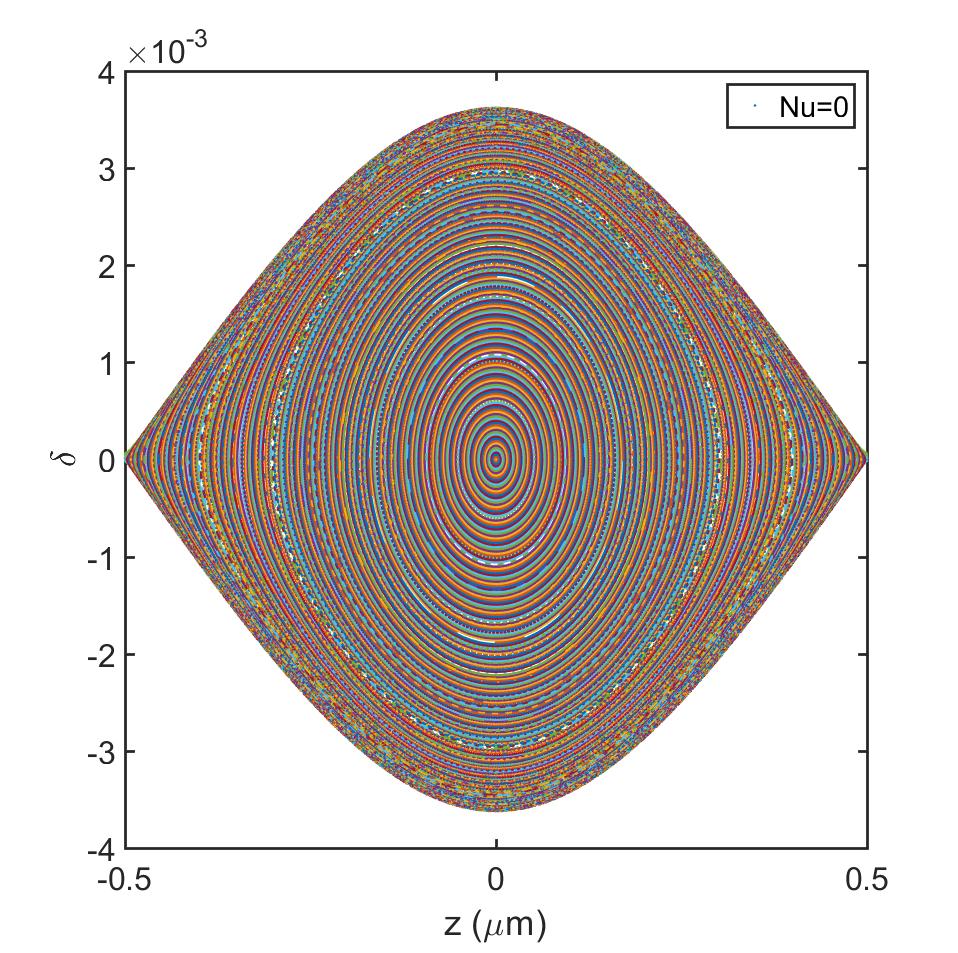}
	\includegraphics[width=0.32\columnwidth]{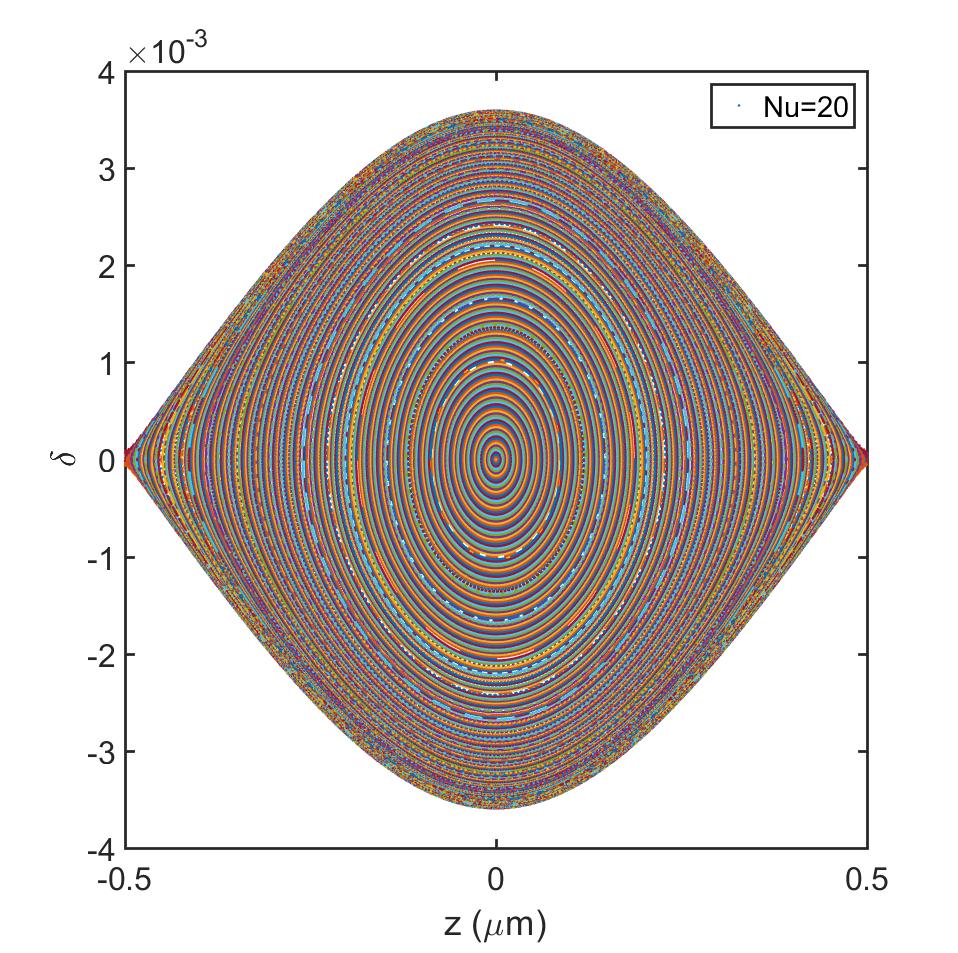}
	\includegraphics[width=0.32\columnwidth]{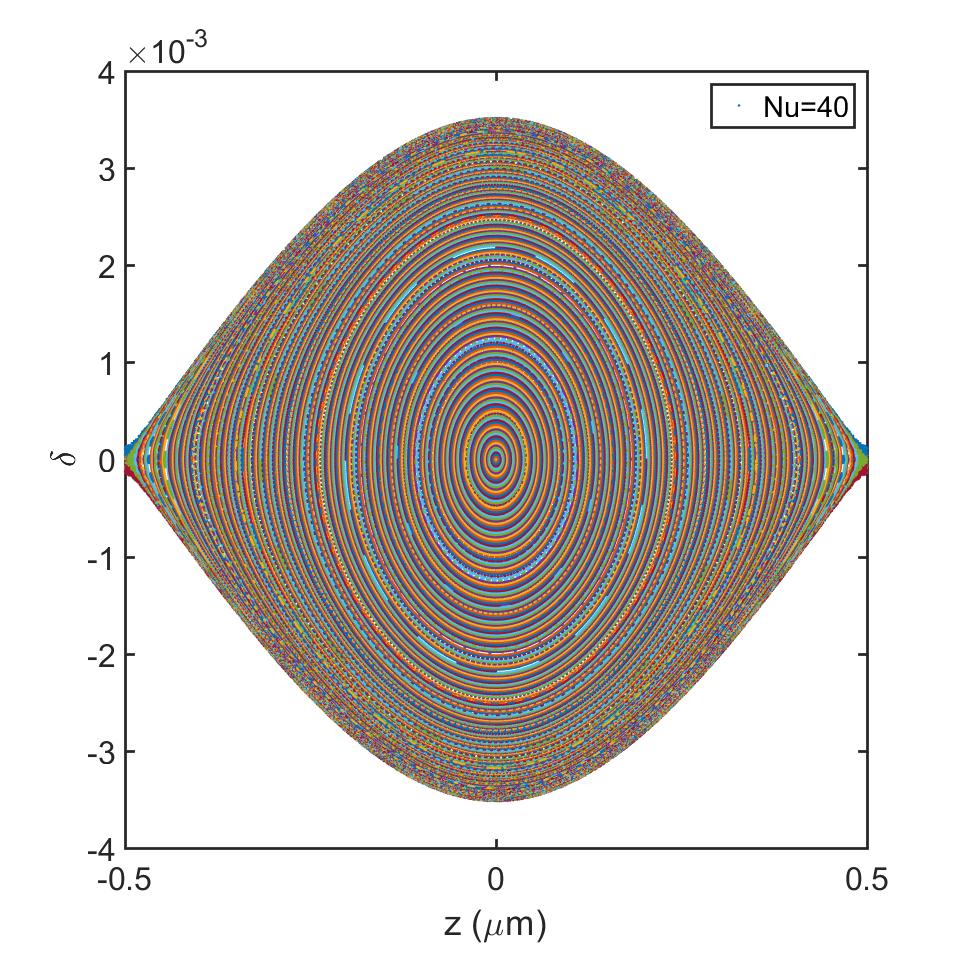}
	\includegraphics[width=0.32\columnwidth]{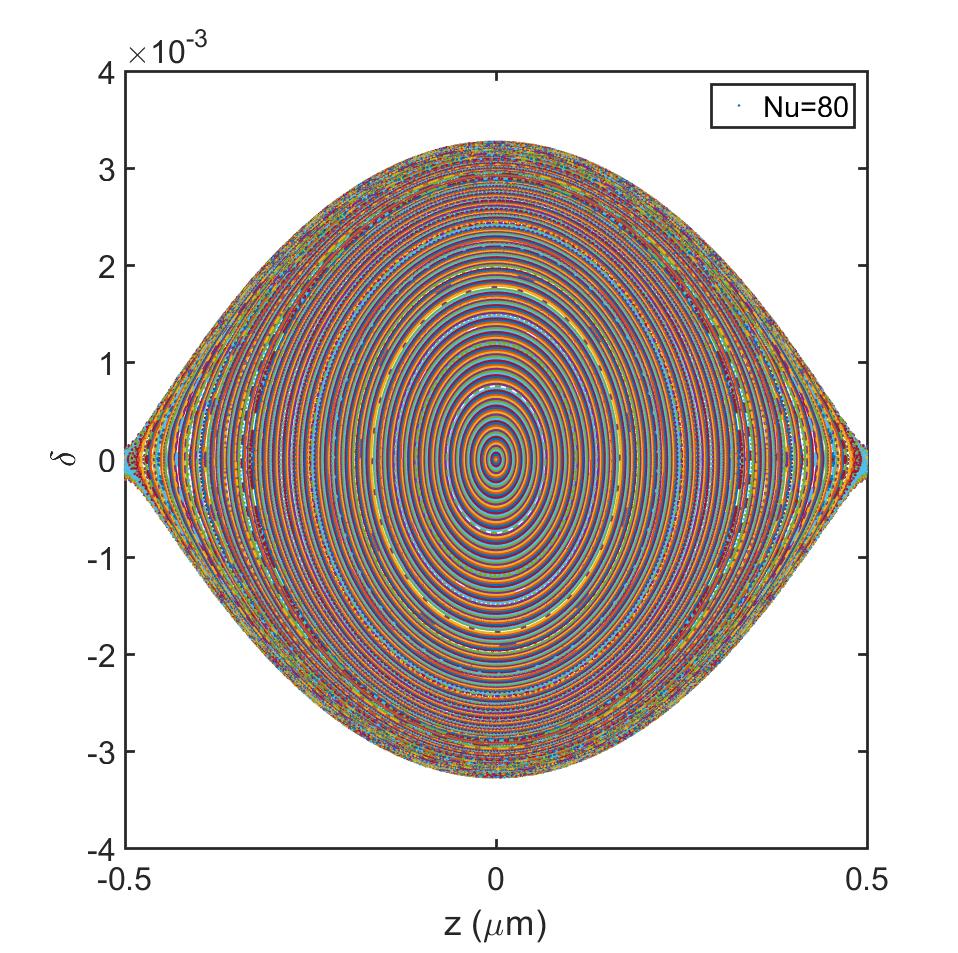}
	\includegraphics[width=0.32\columnwidth]{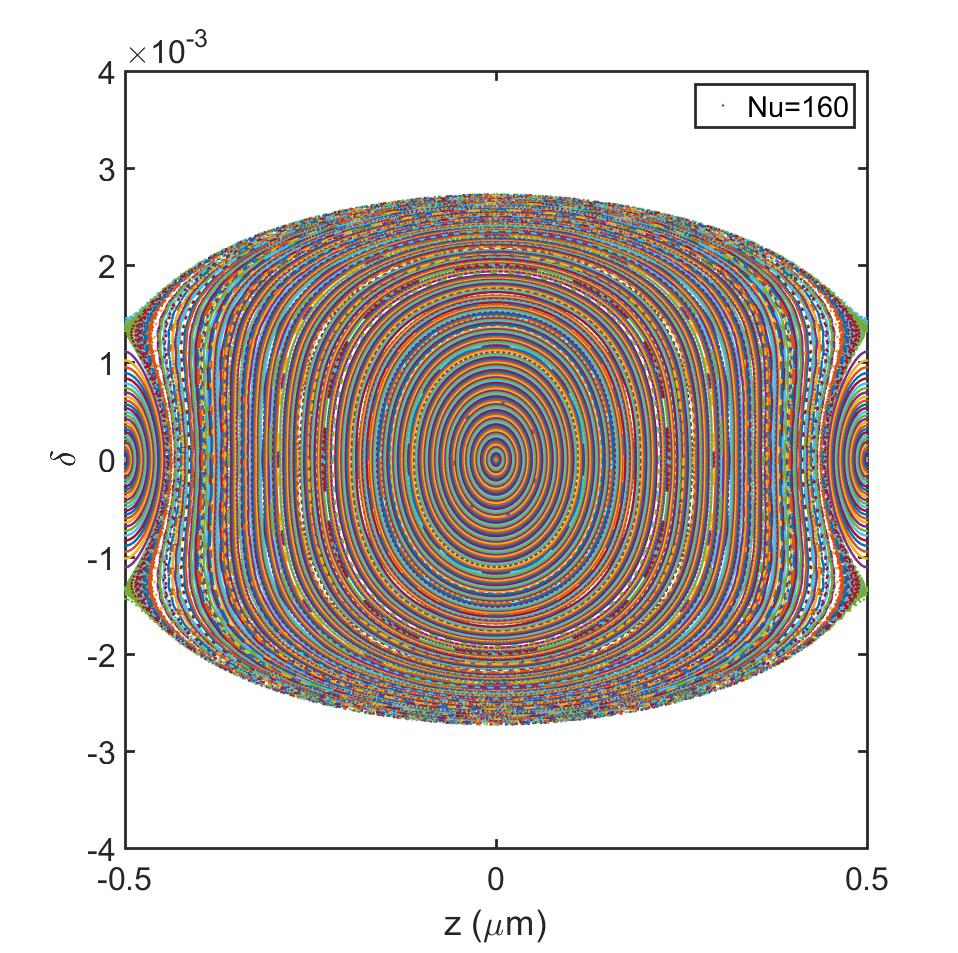}
	\includegraphics[width=0.32\columnwidth]{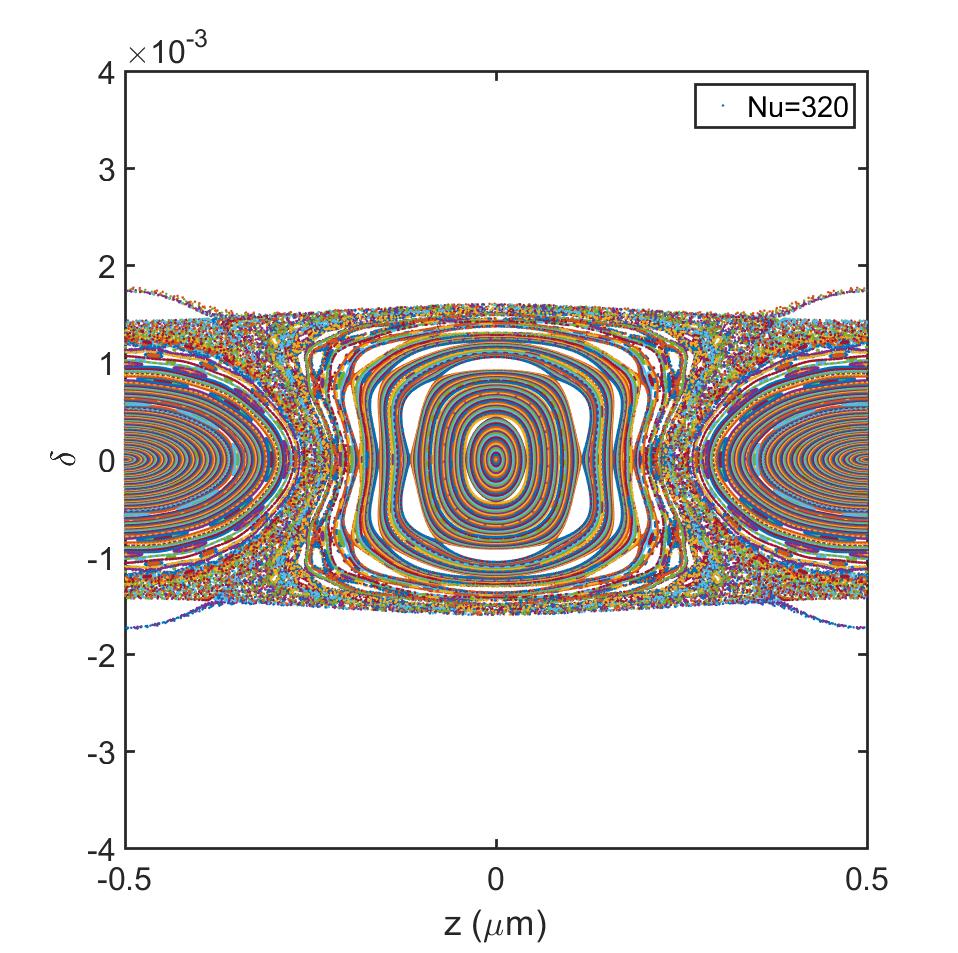}
	\caption{
		\label{fig:Chap2-ModulatorR56Bucket} 
		Longitudinal phase space opposite the modulator center with $N_{u}=0,20,40,80,160,320$. Parameters used: $
		\lambda_{\text{L}}=\lambda_{0}=1\ \mu{m},\ A=1\times10^{-3},\ C_{0}=100\ \text{m},\ \eta=5\times10^{-7}
		$ 
	}
\end{figure}

To simplify the analysis, in the example numerical simulation, we consider the case of a plane wave, i.e., $
A_{i}=\frac{A}{N_{u}}$. First, we want to see how the single-pass modulation waveform is like based on Eq.~(\ref{eq:symkicks1}).
We choose parameters $\lambda_{\text{L}}=\lambda_{0}=1\ \mu{m},\ A=1\times10^{-3}$.
The single-pass modulation of a line beam, with ``$\delta=0$ and $z\in[0,\lambda_{0}]$", as a function of $N_{u}$ (namely $R_{56}$) of the undulator is shown in Fig.~\ref{fig:Chap2-ModulationWaveform}. When changing $N_{u}$, we keep $A$ unchanged. As can be seen, the waveform deviates from {\it{sine}} wave when $N_{u}$ increases. The beam distribution in phase space at the undulator exit (right sub-figure of Fig.~\ref{fig:Chap2-ModulationWaveform}) is similar to beam de-coherence in an RF bucket. 

Now we consider the multi-pass cases, i.e., we consider the impact of modulator $R_{56}$ on the phase space bucket. But here we only do the simulation for the longitudinal weak focusing with a single laser modulator, as we only aim to give the readers a picture about such impact. 
We use parameters of
$
\lambda_{\text{L}}=\lambda_{0}=1\ \mu{m},\ A=1\times10^{-3},\ C_{0}=100\ \text{m},\ \eta=5\times10^{-7}
$ in the simulation, and choose to observe the beam opposite the modulator center where $\alpha_{z}=0$ with $N_{u}=0,20,40,80,160,320$, respectively.  Note that $\eta$ is the phase slippage factor of the whole ring, including the modulator. When changing the undulator priod number $N_{u}$, we keep $\eta$ a constant. The results are shown in Fig.~\ref{fig:Chap2-ModulatorR56Bucket}. It can be seen that the modulator $R_{56}$ only distorts the bucket slightly when $N_{u}$ is not very large. It is only when $N_{u}$ is as large, for example $N_{u}=160$, will it has a profound effect in the longitudinal weak focusing case. Its impact on longitudinal strong focusing is more subtle as the motion in longitudinal strong focusing is strongly chaotic if the nonlinear modulation waveform is taken into account. The study of such effect can refer more straightforwardly to numerical simulations. Besides, the undulator $R_{56}$ could also have a crucial impact on the coherent radiation induced collective instability in the laser modulator~\cite{tsai2021coherent,tsai2021longitudinal,tsai2021theoretical}. We also remind the readers that the $6\times6$ map of a laser modulator could be more complicated than what analyzed in this section. More in-depth work is ongoing.

%

\section{Nonlinear Phase Slippage}

After resolving the issue of quantum diffusion of $z$ by means of dedicated lattice design, we can then apply the low-alpha or low-phase slippage method to realize short bunches in SSMB. However, the phase slippage is actually a function of the particle energy 
\begin{equation}\label{eq:alpha}
\eta(\delta)=\eta_{0}+\eta_{1}\delta+\eta_{2}\delta^{2}+...\ 
\end{equation} 
When $\eta_{0}$ is sufficiently small, the higher-order terms in Eq.~(\ref{eq:alpha}) may become relevant or even dominant, and the beam dynamics can be significantly different from those in a linear-phase slippage state. Proper application of dedicated sextupoles and octupoles may be needed to control these higher-order terms.

The beam dynamics of the quasi-isochronous rings have been studied by many authors \cite{pellegrini1991quasi,feikes2011metrology,ries2014nonlinear}. Here, we wish to emphasize two further points that have not previously been well investigated and might be important in some cases, for example, in the SSMB proof-of-principle experiment to be introduced in Chap.~\ref{cha:pop} and the longitudinal dynamic aperture in SSMB.  

\subsection{For High-harmonic Bunching}\label{sec:NonlinearAlpha}

\begin{figure}[tb] 
	\centering
	\includegraphics[width=0.195\textwidth]{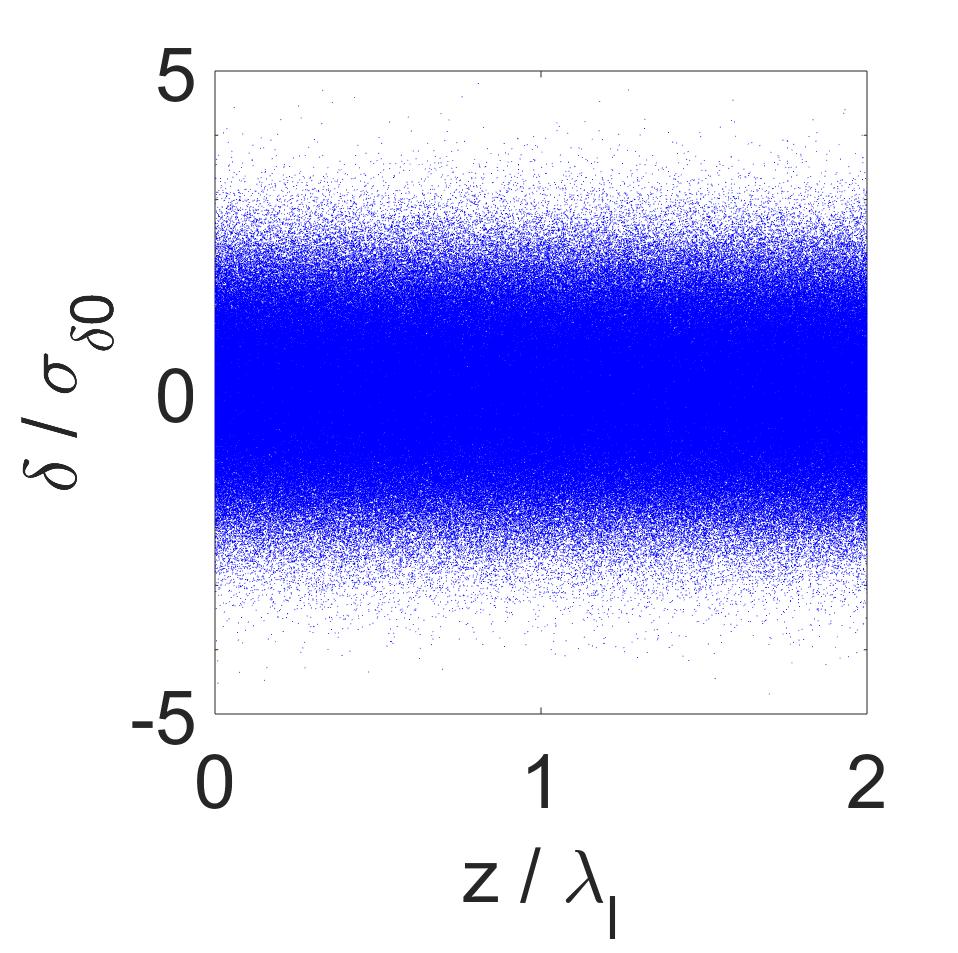}
	\includegraphics[width=0.195\textwidth]{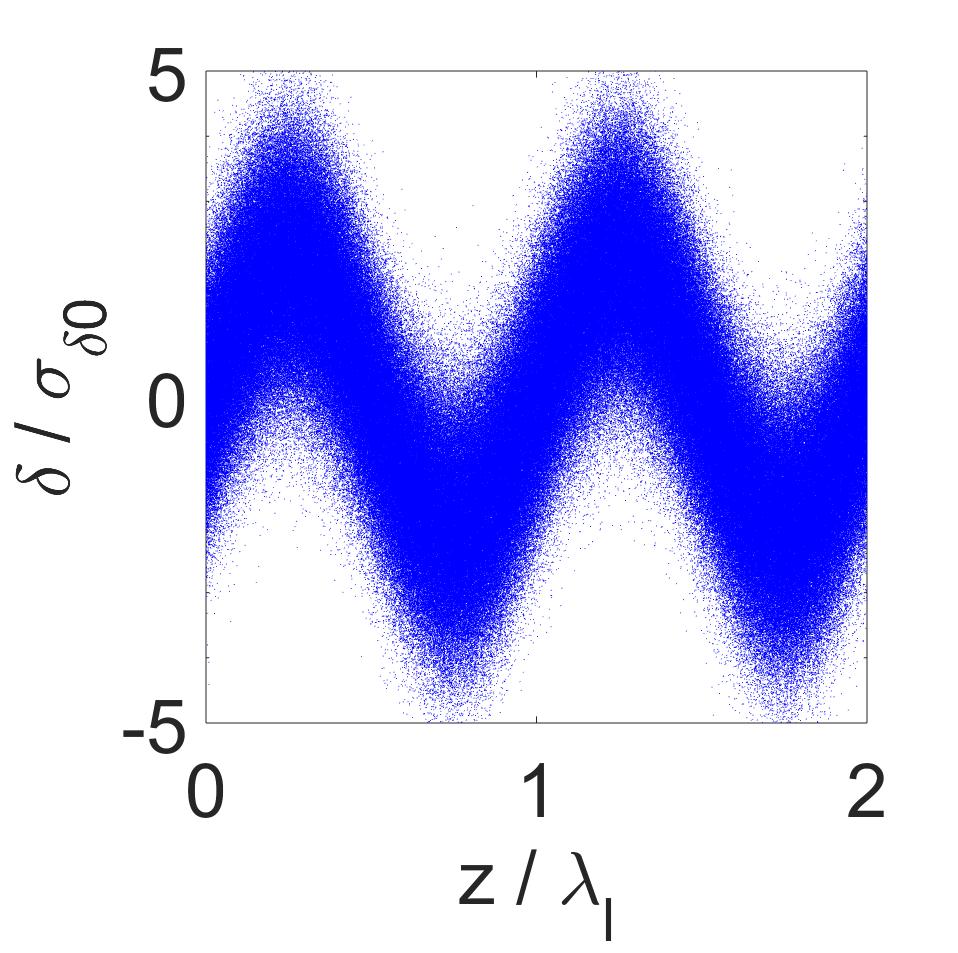}
	\includegraphics[width=0.195\textwidth]{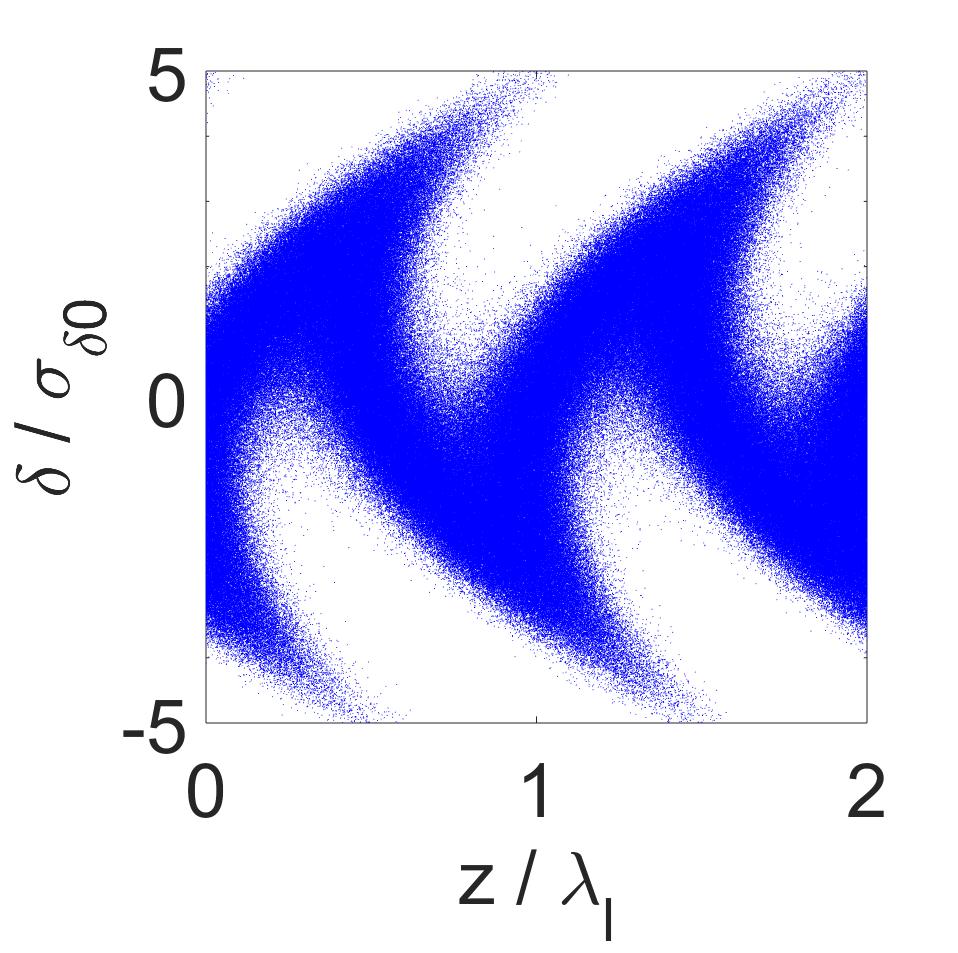}
	\includegraphics[width=0.195\textwidth]{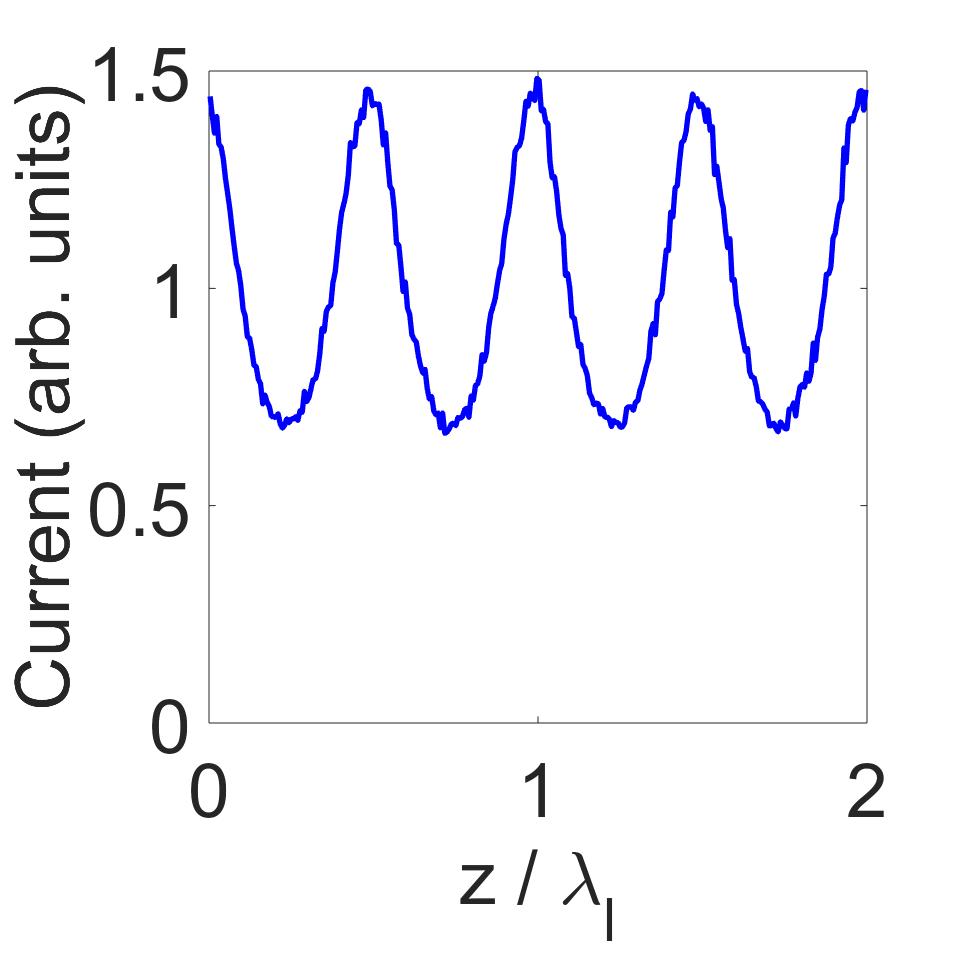}
	\includegraphics[width=0.195\textwidth]{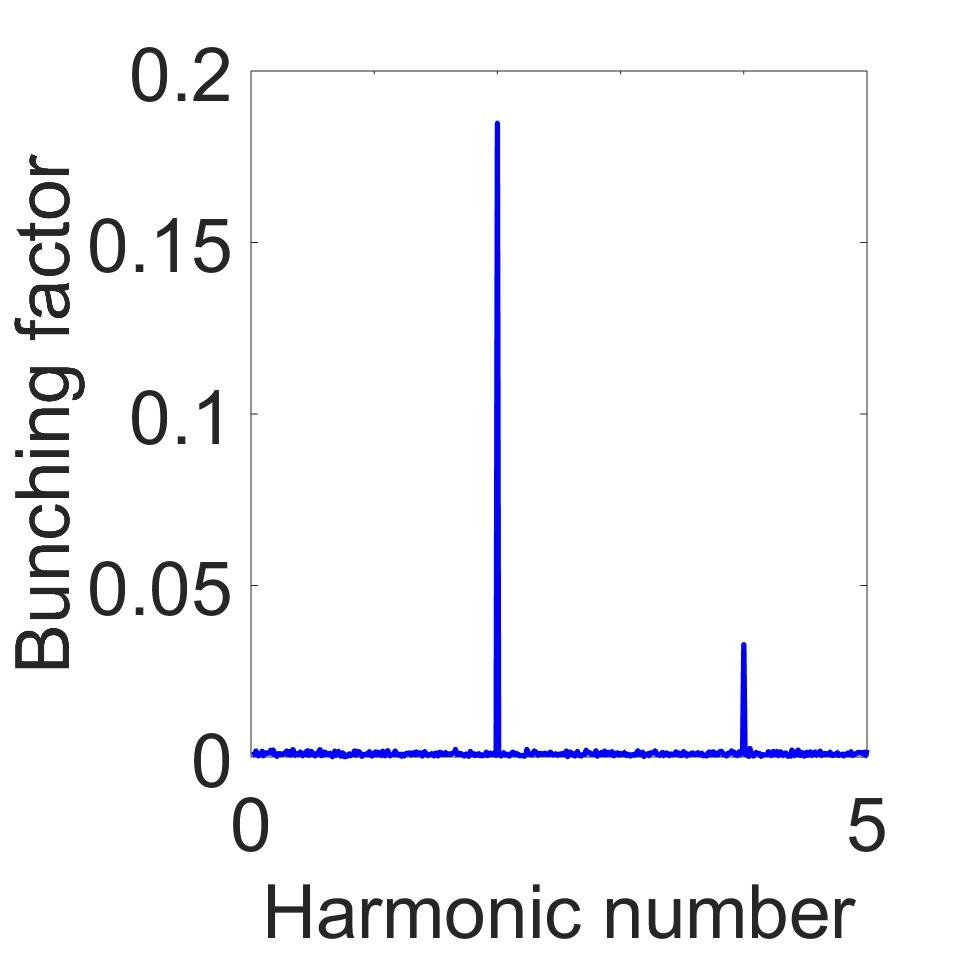}
	\caption{
		\label{fig:Chap2-NonlinearAlphaMicrobunching} 
		The beam evolution in longitudinal phase space, final current distribution and bunching factor, when $\eta(\delta)=\eta_{1}\delta$ is used for microbunching in CHG or HGHG, as modeled by Eq.~(\ref{eq:HGHG}).
	}
\end{figure}

For seeding techniques such as coherent harmonic generation (CHG) \cite{girard1984optical} and high-gain harmonic generation (HGHG) \cite{yu1991generation,yu2000high}, it seems that to date, linear phase slippage or $R_{56}$ has been applied for microbunching formation. Here, we wish to point out that one can actually take advantage of the nonlinearity of the phase slippage for high harmonic generation. Intuitively, this is because a sinusoidal energy modulation followed by a nonlinear dispersion can lead to a distorted current distribution, which, in some cases, can lead to large bunching at a specific harmonic number. Figure~\ref{fig:Chap2-NonlinearAlphaMicrobunching} shows an example simulation of using $\eta(\delta)=\eta_{1}\delta$ for microbunching. It can be seen that there is significant bunching in the second and fourth harmonics, while no bunching is produced in the fundamental and third harmonics. The reason can be found from the following derivation of the bunching factor.

The microbunching process in the case of a single energy modulation followed by a dispersion section, as that in CHG and HGHG, can be modeled as \cite{yu1991generation}	
\begin{align}\label{eq:HGHG}
\delta'&=\delta+A\sin(k_{\text{L}}z)\notag,\\
z'&=z-\eta(\delta')C_{0}\delta',
\end{align}
where $k_{\text{L}}=2\pi/\lambda_{\text{L}}$ is the wavenumber of the modulation laser,  $A$ is the electron energy modulation strength induced by the laser. The bunching factor at the wavenumber of $k$ is defined as
\begin{equation}\label{eq:BF}
b(k)=\left|\int_{-\infty}^{\infty}dze^{-ikz}f(z)\right|,\\
\end{equation}
where $f(z)$ is the normalized longitudinal density distribution of the electron beam satisfying $\int_{-\infty}^{\infty}dzf(z)=1$.
According to Liouville's theorem, we have $dzd\delta=dz'd\delta'$. Therefore, the bunching factor can be calculated in accordance with the initial distribution of the particles $f_{0}(z,\delta)$ as \cite{yu1991generation,stupakov2009using,xiang2009echo}

\begin{align}\label{eq:BF2}
b(k)&=\left|\int_{-\infty}^{\infty}\int_{-\infty}^{\infty}dzd\delta\ f_{0}(z,\delta)e^{-ikz'(z,\delta)}\right|.
\end{align}

First we consider the simple case of $\eta(\delta)\equiv\eta_{0}+\eta_{1}\delta$, then 
\begin{equation}
z'=z-\left[\eta_{0}+\eta_{1}\left(\delta+A\sin(k_{\text{L}}z)\right)\right]C_{0}\left[\delta+A\sin(k_{\text{L}}z)\right]
\end{equation}
and
\begin{equation}
\begin{aligned}\label{eq:bessel}
b(k)
&=\left|\int_{-\infty}^{\infty}\int_{-\infty}^{\infty}dzd\delta\ f_{0}(z,\delta) e^{ik\left(\eta_{0}C_{0}\delta+\eta_{1}C_{0}\delta^{2}+\frac{\eta_{1}C_{0}A^{2}}{2}\right)}\right. \\
&\left. \ \ \ \  e^{ik\left[-z+\left(\eta_{0}C_{0}A+2\eta_{1}C_{0}\delta A\right)\sin(k_{\text{L}}z)-\frac{\eta_{1}C_{0}A^{2}}{2}\cos(2k_{\text{L}}z)\right]}\right|.
\end{aligned}
\end{equation} 
Adopting the notation $Y=k\left(\eta_{0}C_{0}A+2\eta_{1}C_{0}\delta A\right),\ Z=-k\eta_{1}C_{0}A^{2}/{2}$, and using the mathematical identity
$
e^{ix\sin(k_{\text{L}}z)}=\sum_{n=-\infty}^{\infty}e^{ink_{\text{L}}z}J_{n}[x],
$	
we have
\begin{equation}
\begin{aligned}
&e^{-ikz+iY\sin(k_{\text{L}}z)+iZ\cos(2k_{\text{L}}z)}\\
&=\sum_{p=-\infty}^{\infty}\sum_{q=-\infty}^{\infty}J_{p}[Y]J_{q}[Z]\text{exp}\left(-i\left[(k-(p-2q)k_{\text{L}})z-q\frac{\pi}{2}\right]\right).
\end{aligned}
\end{equation}
If the initial beam is much longer than the laser wavelength, and considering that 
\begin{equation}
\frac{1}{2\pi}\int_{-\infty}^{\infty}e^{-i\omega t}d\omega=\delta(t),
\end{equation}
where $\delta(t)$ is the Dirac delta function, the bunching factor will not vanish only if 
$
k=(p-2q)k_{\text{L}}
$. The bunching factor at the $n^{\text{th}}$ harmonic of the modulation laser is then
\begin{align}\label{eq:NFNonlinearAlpha}
b_{n}&=\left|\int_{-\infty}^{\infty}d\delta\  e^{ink_{\text{L}}\left(\eta_{0}C_{0}\delta+\eta_{1}C_{0}\delta^{2}+\frac{\eta_{1}C_{0}A^{2}}{2}\right)} f_{0}(\delta)\sum_{m=-\infty}^{\infty}J_{n+2m}[Y]J_{m}[Z]\right|.
\end{align}	
Here we consider the simple case of an initial Gaussian energy distribution $f_{0}(\delta)=\frac{1}{\sqrt{2\pi}\sigma_{\delta}}\text{exp}\left(-\frac{\delta^{2}}{2\sigma_{\delta}^{2}}\right)$, where $\sigma_{\delta}$ is the initial RMS energy spread.

If $\eta_{1}=0$, then $Y=nk_{\text{L}}\eta_{0}C_{0}A$, $Z=0$, and $\sum_{m=-\infty}^{\infty}J_{n+2m}[Y]J_{m}[Z]=J_{n}[nk_{\text{L}}\eta_{0}C_{0}A]$, and we have
\begin{equation}\label{eq:HGHGBF}
b_{n}=\left|J_{n}[nk_{\text{L}}\eta_{0}C_{0}A]\right|\text{exp}\left[-\frac{\left(nk_{\text{L}}\eta_{0}C_{0}\sigma_{\delta}\right)^{2}}{2}\right],
\end{equation}
which is a familiar result for HGHG~\cite{yu1991generation} if we adopt the notation $R_{56}=-\eta_{0}C_{0}$. If $\eta_{0}=0$, then $Y=2nk_{\text{L}}\eta_{1}C_{0}\delta A$, $Z=-nk_{\text{L}}\eta_{1}C_{0}A^{2}/{2}$, meaning that we have
\begin{equation}
\begin{aligned}\label{eq:NFNonlinearAlpha1}
b_{n}=\frac{1}{\sqrt{2\pi}\sigma_{\delta}}\Bigg|&\int_{-\infty}^{\infty}d\delta\ \text{exp}\left[{ink_{\text{L}}\left(\eta_{1}C_{0}\delta^{2}+\frac{\eta_{1}C_{0}A^{2}}{2}\right)}\right]\\
&\text{exp}\left(-\frac{\delta^{2}}{2\sigma_{\delta}^{2}}\right)\sum_{m=-\infty}^{\infty}J_{n+2m}[Y]J_{m}[Z]\Bigg|.
\end{aligned}	
\end{equation}
The two exponential terms in the integral are even functions of $\delta$, while $J_{n+2m}[Y]J_{m}[Z]$ is an odd function of $\delta$ when $n$ is odd; thus, $b_{n}$ is nonzero only for even $n$. This is why bunching occurs only in the second and fourth harmonics but not in the fundamental and third harmonics when we use $\eta(\delta)=\eta_{1}\delta$ for microbunching, as shown in Fig.~\ref{fig:Chap2-NonlinearAlphaMicrobunching}.

Following the derivations and according to the relation 
\begin{gather}
\begin{align}
\cos^{n}(x)
&=\frac{1}{2^{n-1}}\sum_{m=(n+1)/2}^{n}\left(\begin{matrix}
n\\
m
\end{matrix}\right)\cos(2m-n)x,
\end{align}
\end{gather} 
it can be seen that the energy modulation at the fundamental frequency can be cast into $\left[i\times (n-2p)+j\times (n-2q)\right]^{\text{th}}$-harmonic bunching through the term $\eta_{n-1}\delta^{n}$ in the function of $\eta(\delta)$. For an odd $n$, bunching at all harmonic numbers are possibles, while for an even $n$, only bunching at the even harmonic numbers are possible. The optimal bunching condition for a specific harmonic requires the matching of $\eta(\delta)$ with the modulation strength. However, the analytical formula for the bunching factor will become increasingly involved with more higher-order terms of the phase slippage considered. Thus, it would be better to refer to numerical integration and simulation to calculate and optimize the bunching factor directly for a specific application case. For storage rings, another relevant point is that the distribution of the particle energy in the nonlinear phase slippage state may also have an impact on the high harmonic generation, and this phenomenon is also easier to be studied by means of numerical simulation.

The approach of applying a nonlinear phase slippage for high harmonic bunching can be considered to share some similarity with echo-enabled harmonic generation (EEHG) \cite{stupakov2009using,xiang2009echo}, in which the sinusoidal energy modulation and dispersion in the first stage can be viewed as the source of the distorted current distribution in the second stage of modulation and dispersion for microbunching. We have also noticed the work of Stupakov and Zolotorev on optimizing the nonlinearity of the dispersion to increase the bunching factor for EEHG~\cite{stupakov2011using}. Based on similar considerations, tricks can also be applied on the energy-modulation waveform side using different harmonics of the modulation laser, for example, forming a sawtooth waveform to boost bunching, as will be discussed in Chap.~\ref{cha:TLC}.

\subsection{For Longitudinal Dynamic Aperture}
Similar to the transverse dimension, there is a region in the longitudinal phase space outside of which particle motion is not bounded and will be lost in a ring. We refer this stable region as the longitudinal dynamic aperture. Here in this section, we want to show that, by properly tailoring the nonlinear phase slippage, the longitudinal dynamic aperture can be enlarged significantly compared to the case of a pure linear phase slippage. Only symplectic dynamics is considered in this discussion.
\subsubsection{Longitudinal Weak Focusing}
The longitudinal dynamics of a particle in a ring with a single RF can be modeled by the kick map
\begin{equation}\label{eq:standardMap}
\begin{cases}
&\delta_{n+1}=\delta_{n}+A[\sin(k_{\text{RF}}z_{n})-\sin\phi_{\text{s}}],\\
&z_{n+1}=z_{n}-\eta(\delta_{n+1})C_{0}\delta_{n+1},
\end{cases}
\end{equation}
where $A\sin\phi_{\text{s}}=U_{0}/E_{0}$ where $U_{0}$ is the radiation loss per turn.
For the case of longitudinal weak focusing, adiabatic approximation can be used and the kicik map can be appximated by differentiation and Hamiltonian formalism can be invoked for the analysis. Denote $\phi=k_{\text{RF}}z$, then the equation of motion is  
\begin{equation}\label{eq:eom2}
\begin{cases}
&\frac{d\phi}{dt}=-\frac{k_{\text{RF}}\eta(\delta_{n+1})C_{0}}{T_{0}}\delta=\frac{\partial{\mathcal{H}}}{\partial{\delta}},\\
&\frac{d\delta}{dt}=\frac{A}{T_{0}}(\sin\phi-\sin\phi_{\text{s}})=-\frac{\partial{\mathcal{H}}}{\partial{\phi}},\\
\end{cases}
\end{equation}
with $T_{0}$ being the revolution period of the particle in the ring.
For $\eta(\delta)=\eta_{0}+\eta_{1}\delta+\eta_{2}\delta^{2}$, the corresponding Hamiltonian is
\begin{equation}
\begin{aligned}
\mathcal{H}(\phi,\delta)&=-\omega_{\text{RF}}\left(\frac{1}{2}\eta_{0}\delta^{2}+\frac{1}{3}\eta_{1}\delta^{3}+\frac{1}{4}\eta_{2}\delta^{4}\right)\\
&\ \ \ \  +\frac{A}{T_{0}}\left[\cos\phi-\cos\phi_{\text{s}}+(\phi-\phi_{\text{s}})\sin\phi_{\text{s}}\right].
\end{aligned}
\end{equation}
In writing down the closed-form Hamiltonian, we have implicitly assumed that the motion is integrable, i.e., there is no chaos. But we need to keep in mind that the dynamics dictated by Eq.~(\ref{eq:standardMap}) is actually chaotic even with a linear phase slippage~\cite{chirikov1979universal}. But here we ignore this subtle point as the chaotic layer is very thin in the longitudinal weak focusing regime. We remind the readers that the chaotic dynamics, for example the bucket bifurcation, can actually also be applied for ultra-short bunch generation~\cite{jiao2011terahertz}.

To analyze the motion, we need to find the fixed points of the system
\begin{equation}
\begin{cases}
\frac{\partial{H}}{\partial{\phi}}&=0\\
\frac{\partial{H}}{\partial{\delta}}&=0
\end{cases}
\Longrightarrow
\begin{cases}
\sin\phi_{\text{s}}-\sin\phi&=0,\\
\delta\eta(\delta)&=0.
\end{cases}
\end{equation}
To determine whether a fixed point is stable or not, we need to check the trace of the Jacobian matrix around the fixed point. If $\eta(\delta)=\eta_{0}$, there are two sets of fixed points:
\begin{equation}
\begin{cases}
\text{SFP}:& (\phi_{\text{s}},0),\\ 
\text{UFP}:& (\pi-\phi_{\text{s}},0),
\end{cases}
\end{equation}
in which SFP stands for stable fixed point while UFP stands for unstable fixed point. If $\eta(\delta)=\eta_{0}+\eta_{1}\delta$, there are four sets of fixed points:
\begin{equation}
\begin{cases}
\text{SFP}:& (\phi_{\text{s}},0),\ \left(\pi-\phi_{\text{s}},-\frac{\eta_{0}}{\eta_{1}}\right), \\ 
\text{UFP}:& (\pi-\phi_{\text{s}},0), \left(\phi_{\text{s}},-\frac{\eta_{0}}{\eta_{1}}\right).\ 
\end{cases}
\end{equation}
If $\eta(\delta)=\eta_{0}+\eta_{1}\delta+\eta_{2}\delta^{2}$, there are six sets of fixed points 
\begin{equation}
\begin{cases}
\text{SFP}:& (\phi_{\text{s}},0),\ \left(\pi-\phi_{\text{s}},-\frac{\eta_{1}}{2\eta_{2}}+\sqrt{\left(\frac{\eta_{1}}{2\eta_{2}}\right)^{2}-\frac{\eta_{0}}{\eta_{2}}}\right),\ \left(\pi-\phi_{\text{s}},-\frac{\eta_{1}}{2\eta_{2}}-\sqrt{\left(\frac{\eta_{1}}{2\eta_{2}}\right)^{2}-\frac{\eta_{0}}{\eta_{2}}}\right), \\ 
\text{UFP}:& (\pi-\phi_{\text{s}},0),\ \left(\phi_{\text{s}},-\frac{\eta_{1}}{2\eta_{2}}+\sqrt{\left(\frac{\eta_{1}}{2\eta_{2}}\right)^{2}-\frac{\eta_{0}}{\eta_{2}}}\right),\ \left(\phi_{\text{s}},-\frac{\eta_{1}}{2\eta_{2}}-\sqrt{\left(\frac{\eta_{1}}{2\eta_{2}}\right)^{2}-\frac{\eta_{0}}{\eta_{2}}}\right).\ 
\end{cases}
\end{equation}

\begin{figure}[tb] 
	\centering
	\includegraphics[width=0.32\textwidth]{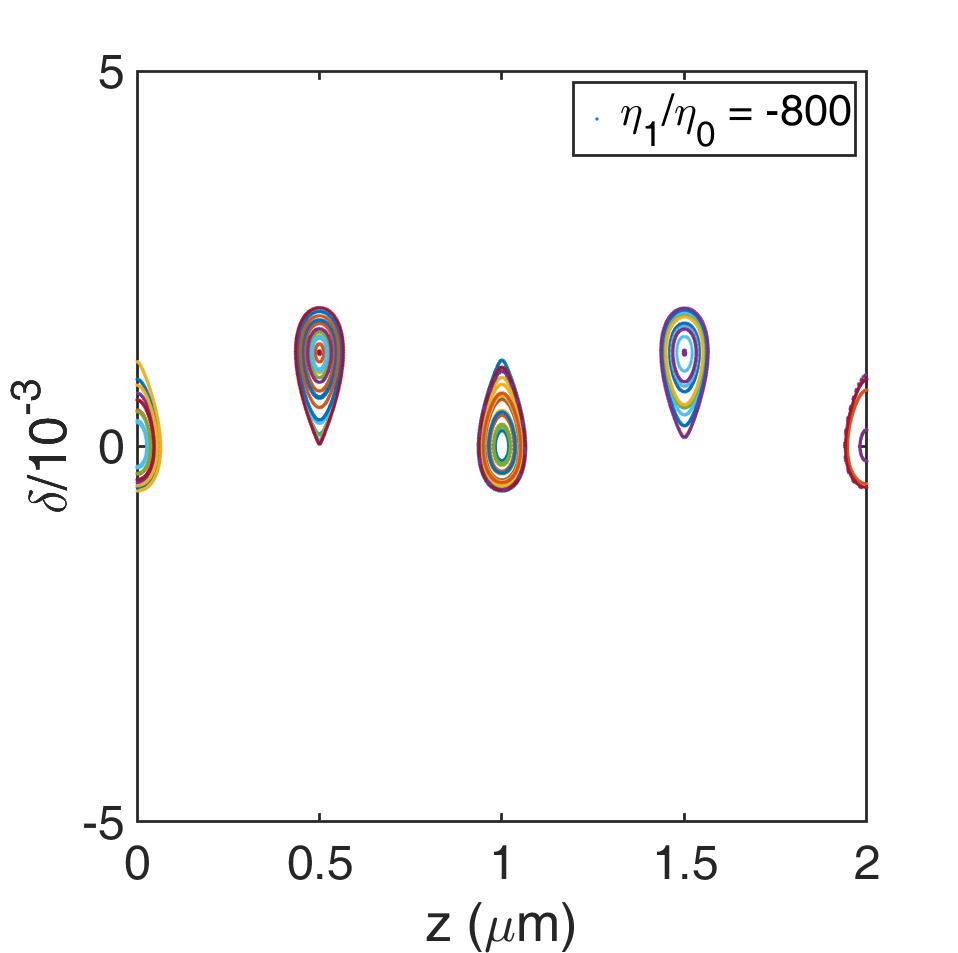}
	\includegraphics[width=0.32\textwidth]{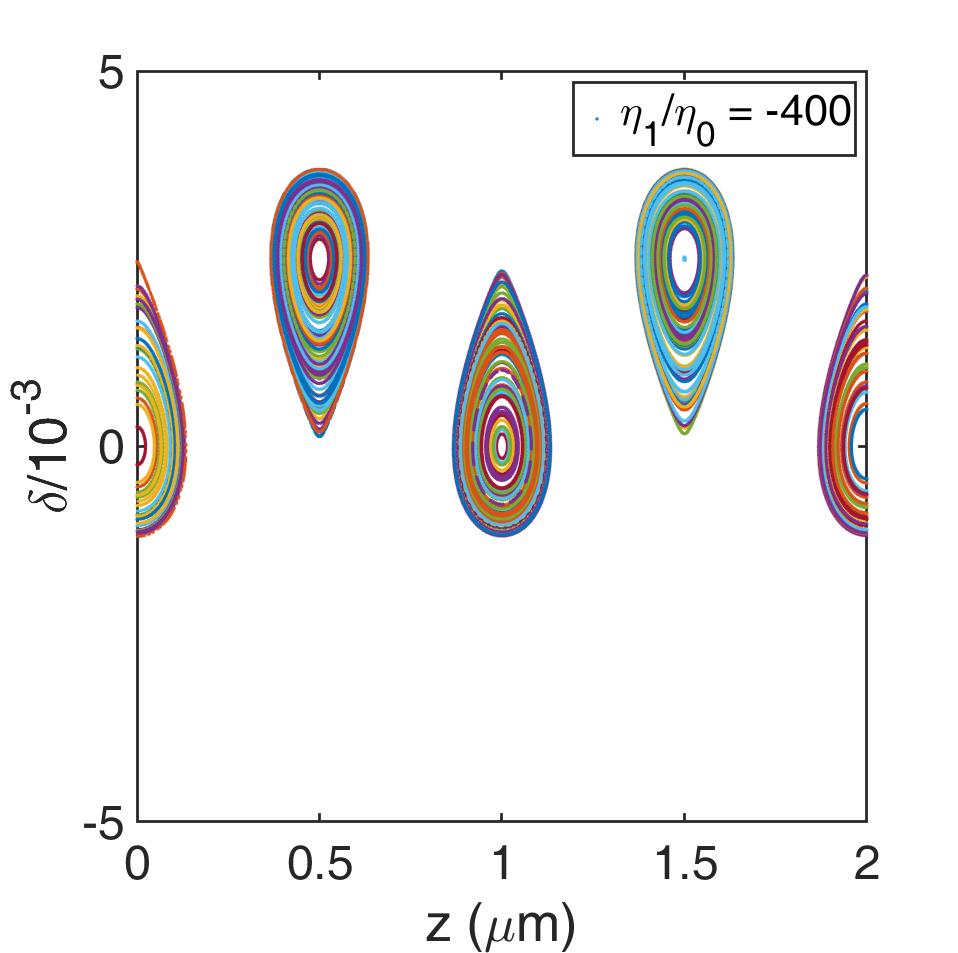}
	\includegraphics[width=0.32\textwidth]{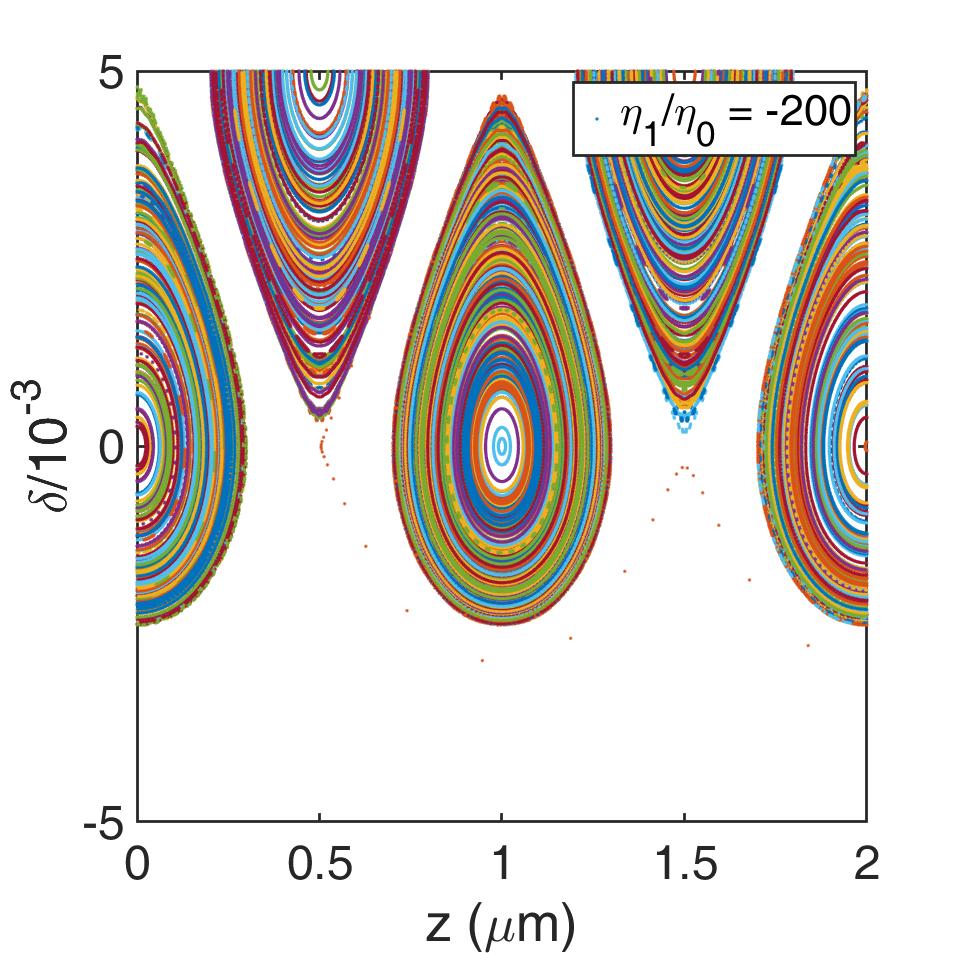}\\	
	\includegraphics[width=0.32\textwidth]{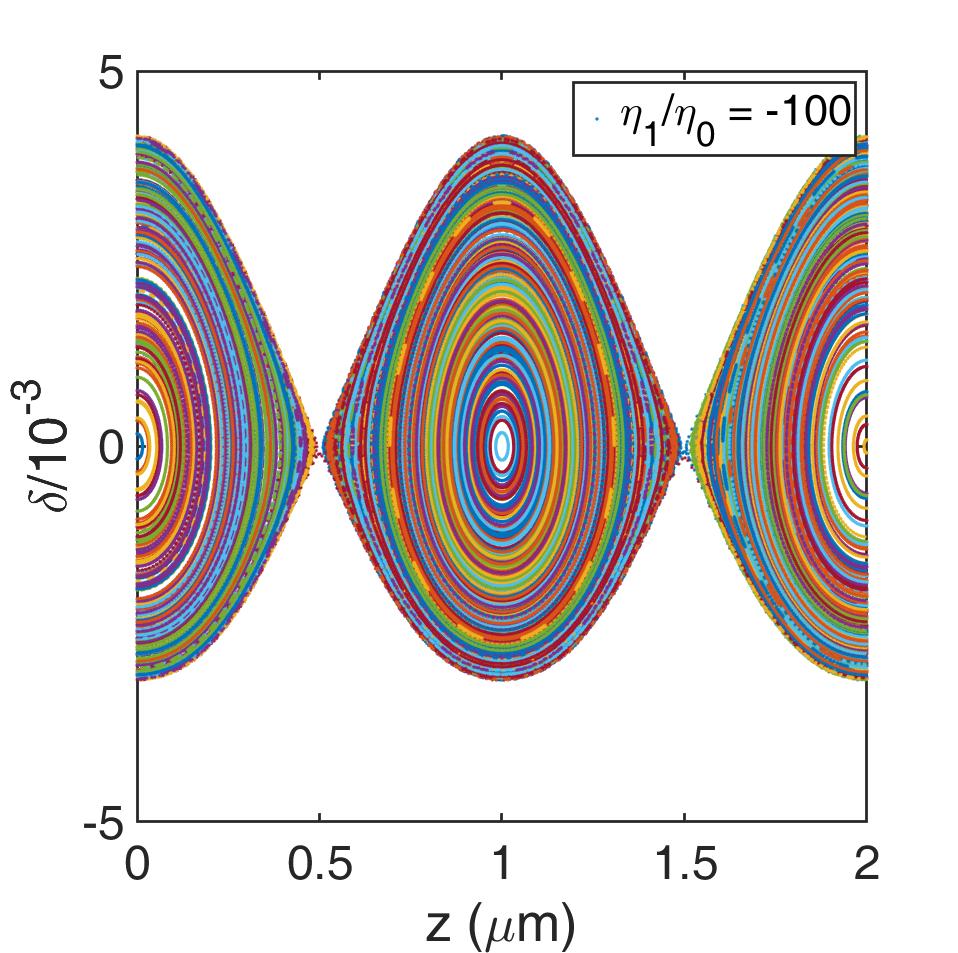}	
	\includegraphics[width=0.32\textwidth]{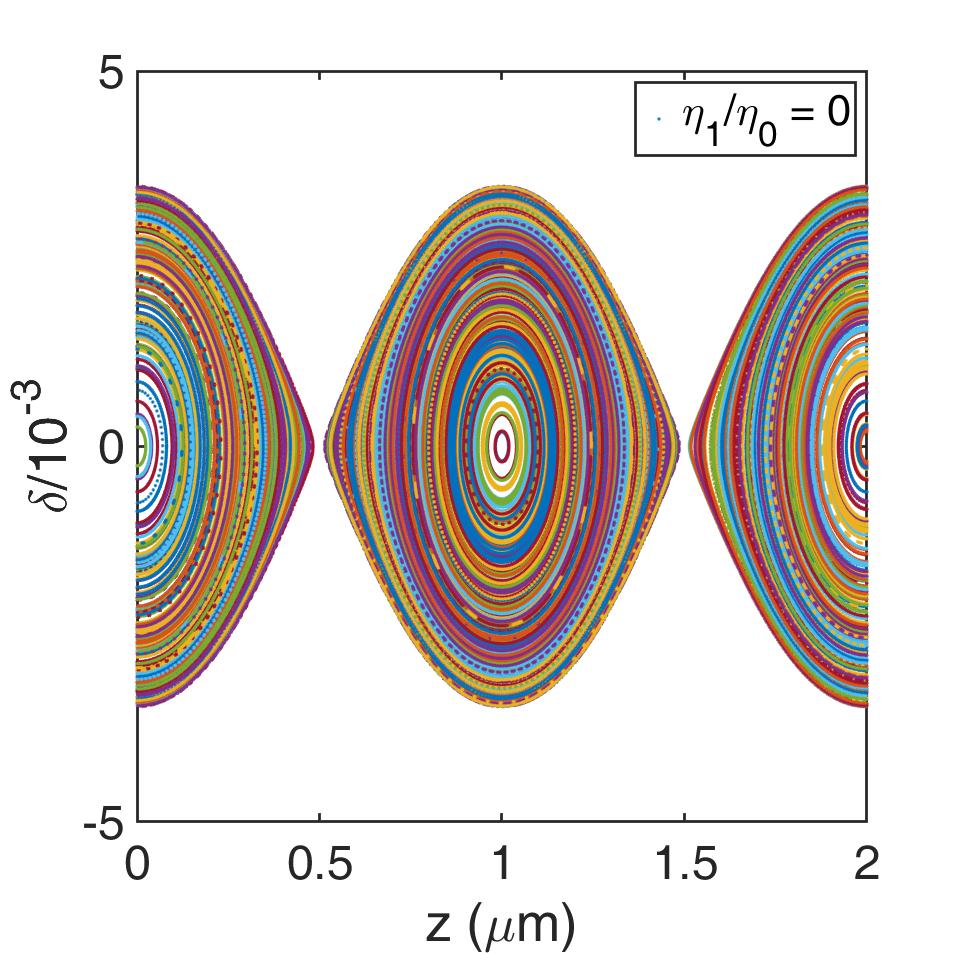}
	\includegraphics[width=0.32\textwidth]{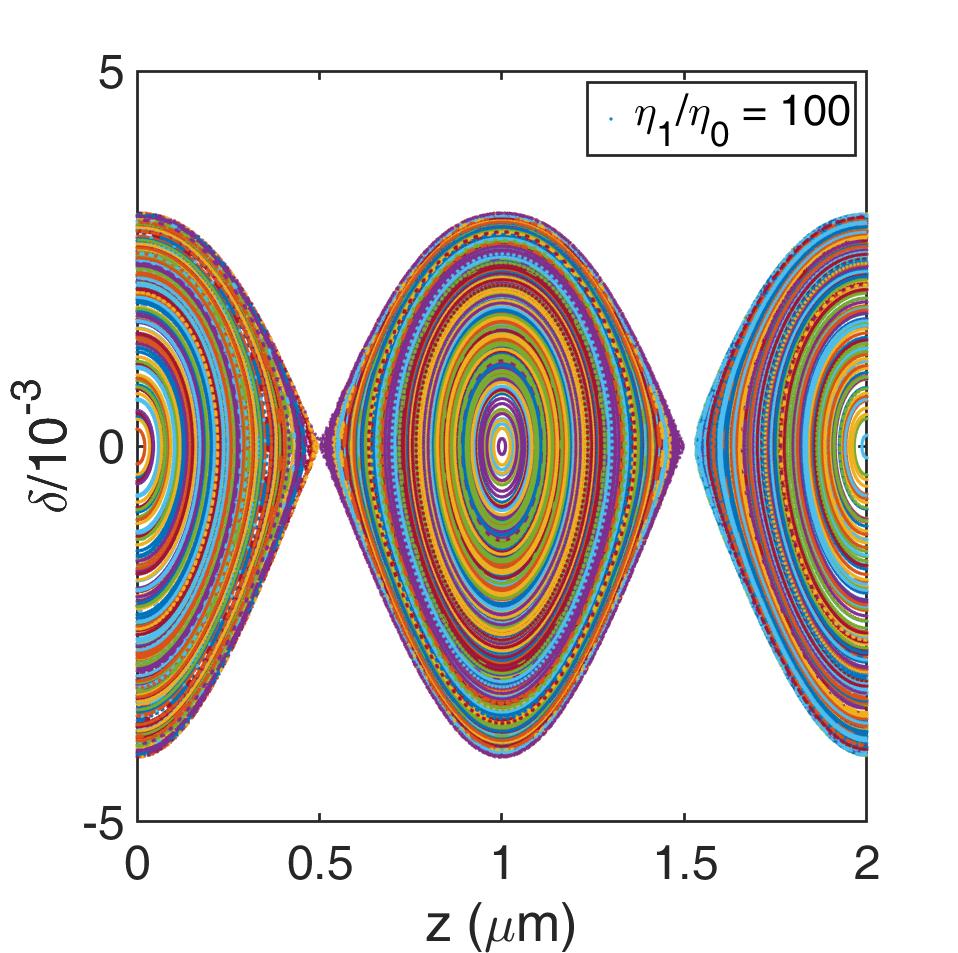}\\
	\includegraphics[width=0.32\textwidth]{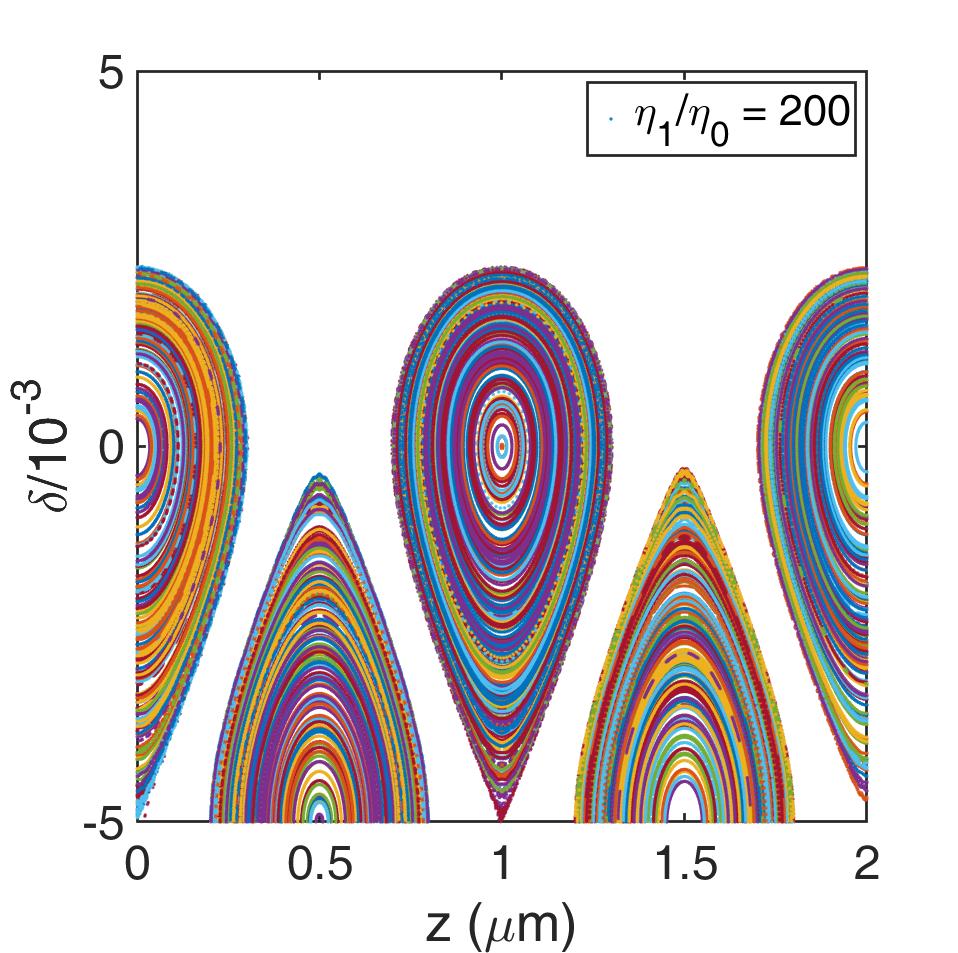}
	\includegraphics[width=0.32\textwidth]{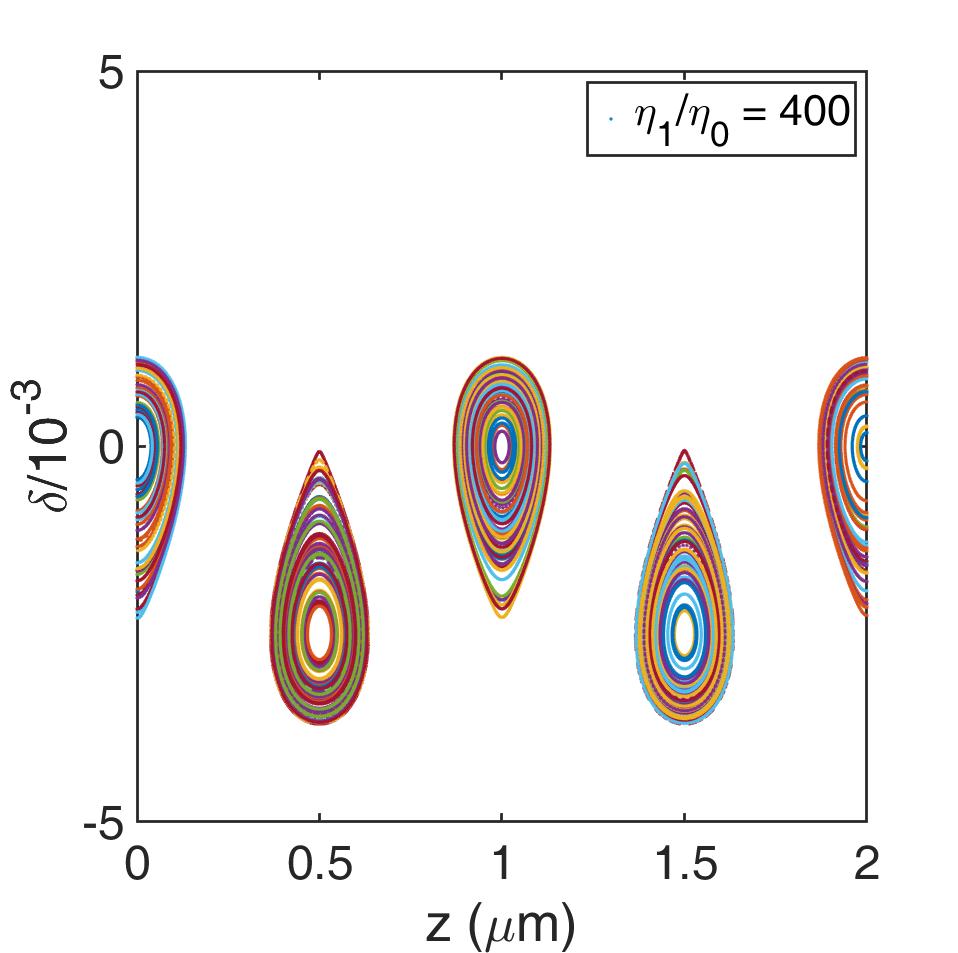}
	\includegraphics[width=0.32\textwidth]{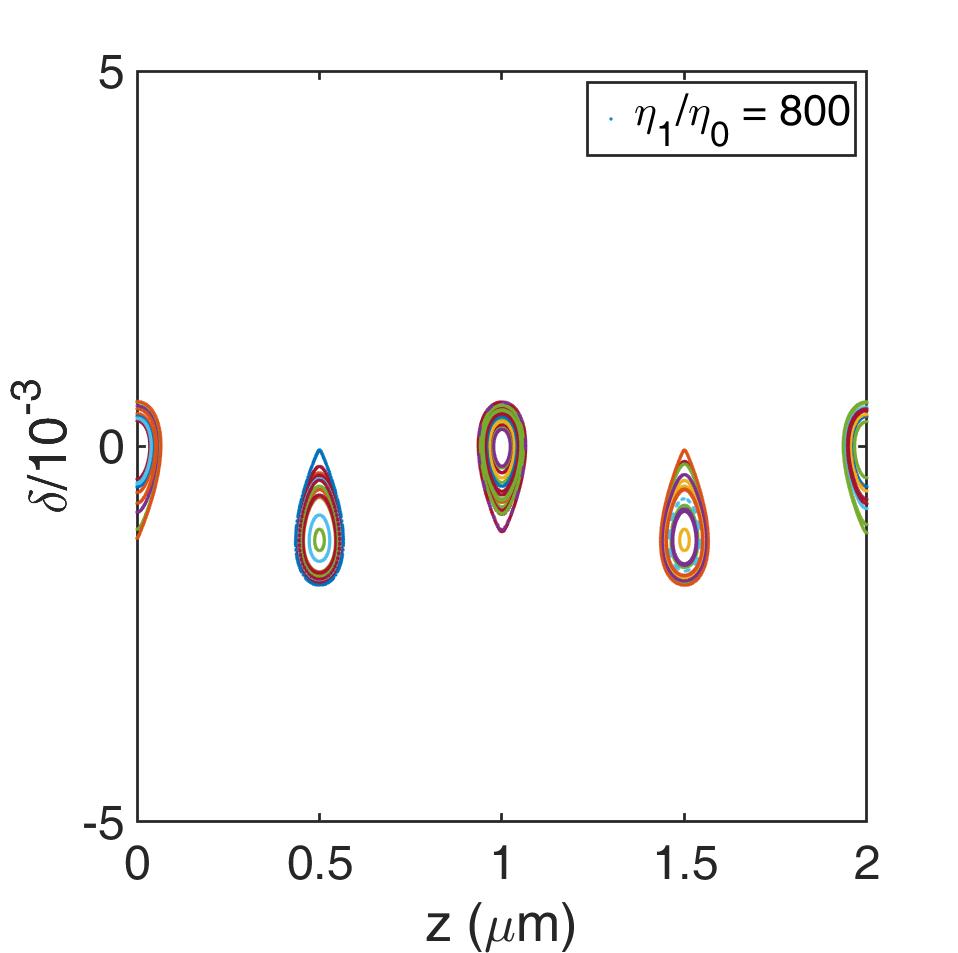}
	
	\caption{
		\label{fig:Chap2-LWFNonLinearAlpha1} 
		The impact of $\eta_{1}$ on the longitudinal phase space bucket in the longitudinal weak focusing regime. Simulation parameters: $\lambda_{\text{RF}}=1\ \mu$m, $A=1\times10^{-3}$,  $C_{0}=100$ m, $\eta_{0}=5\times10^{-7}$. 
	}
\end{figure} 

\begin{figure}[tb] 
	\centering
	\includegraphics[width=0.32\textwidth]{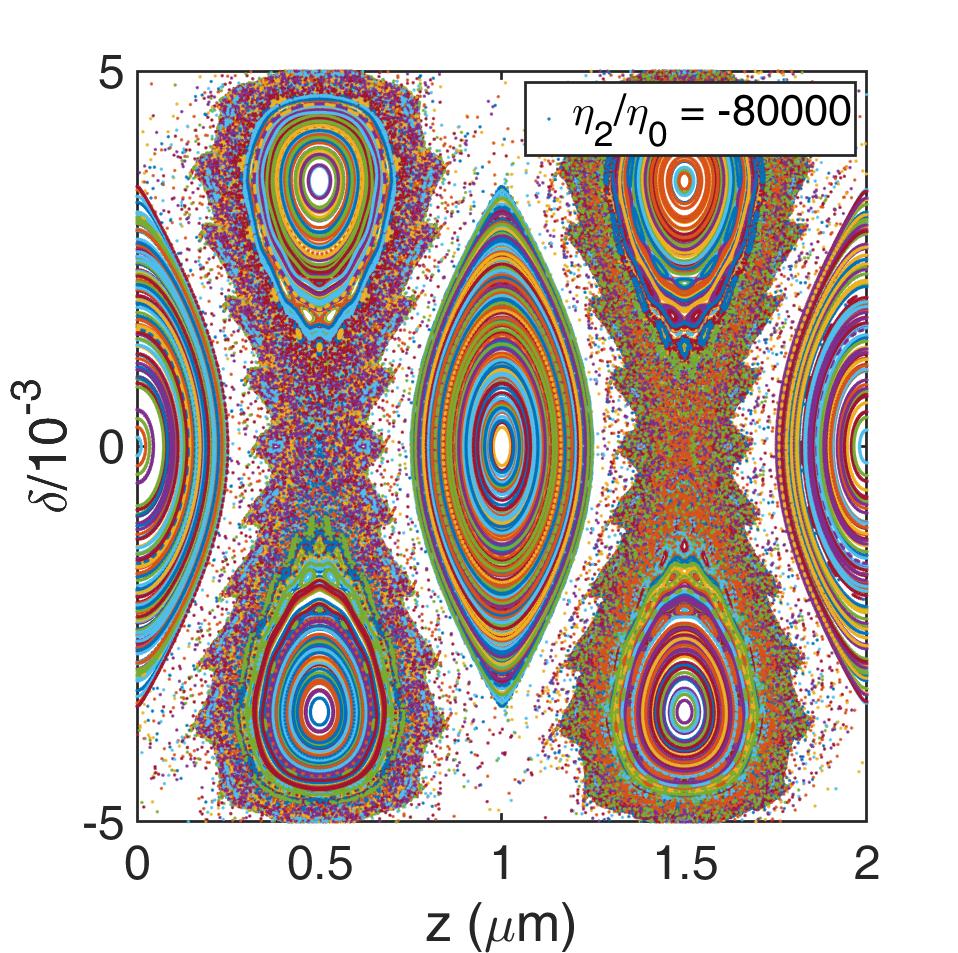}
	\includegraphics[width=0.32\textwidth]{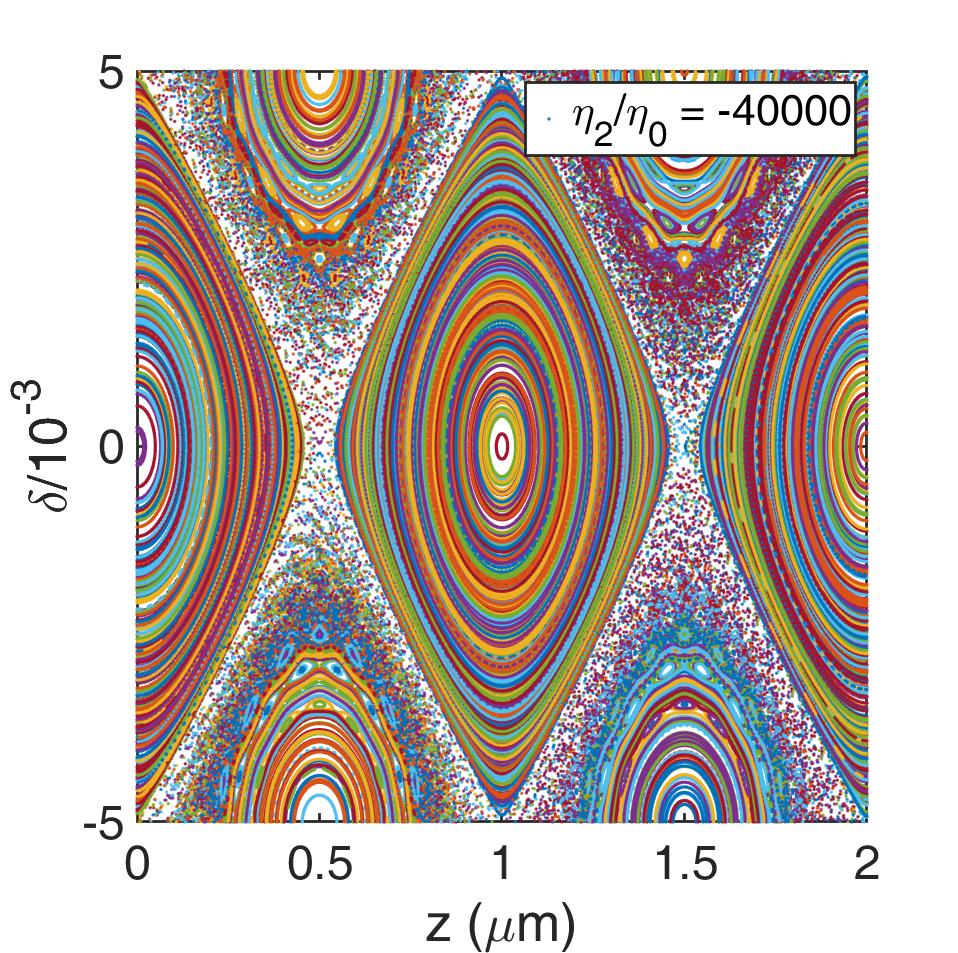}
	\includegraphics[width=0.32\textwidth]{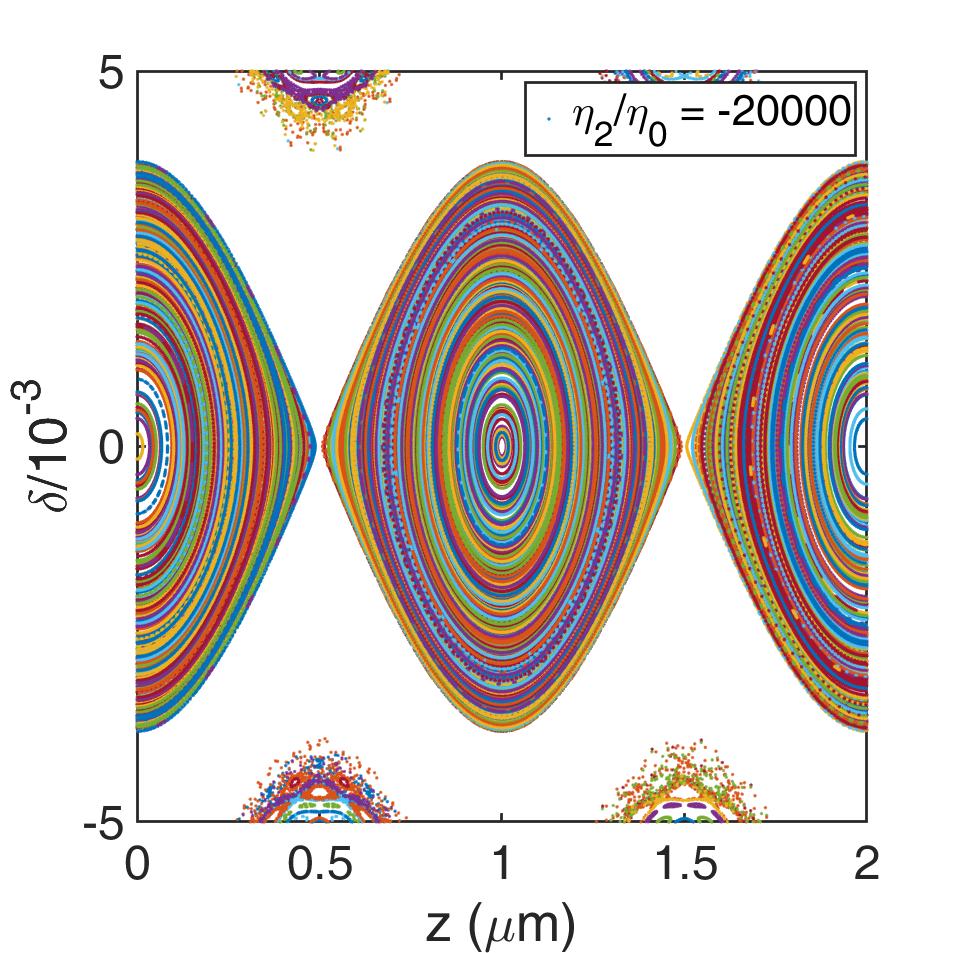}\\	
	\includegraphics[width=0.32\textwidth]{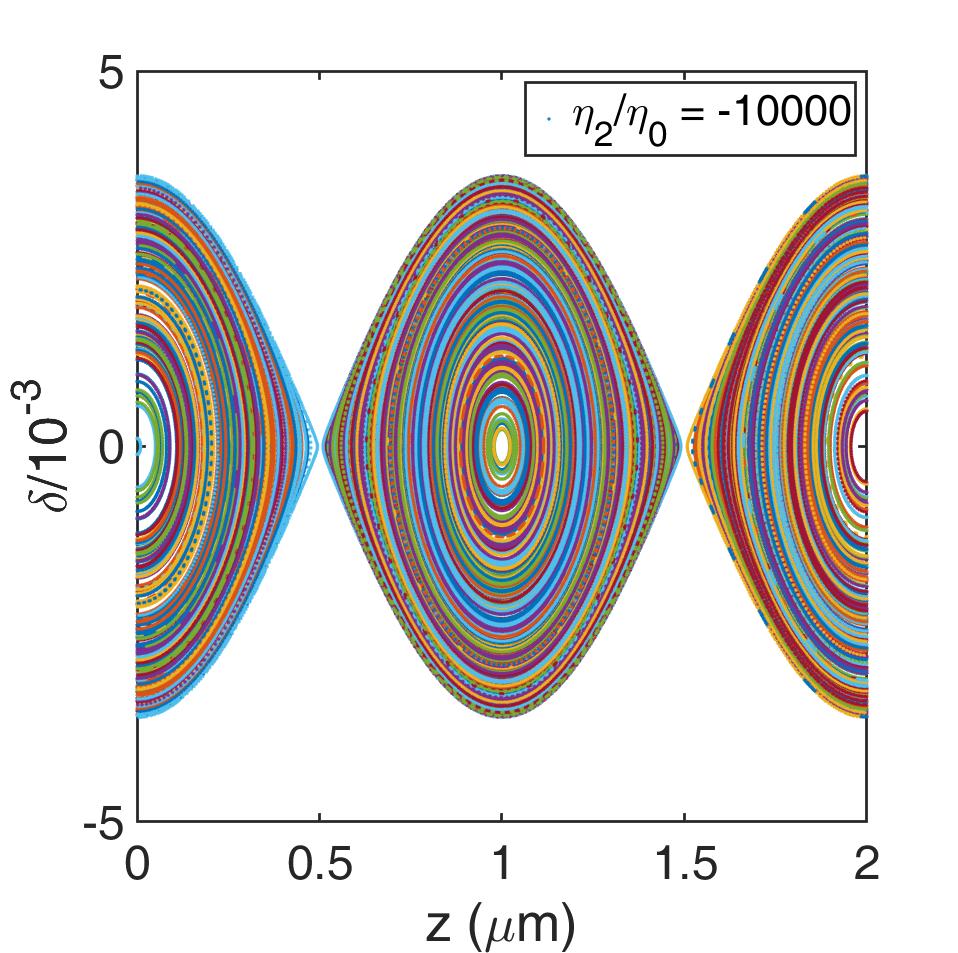}	
	\includegraphics[width=0.32\textwidth]{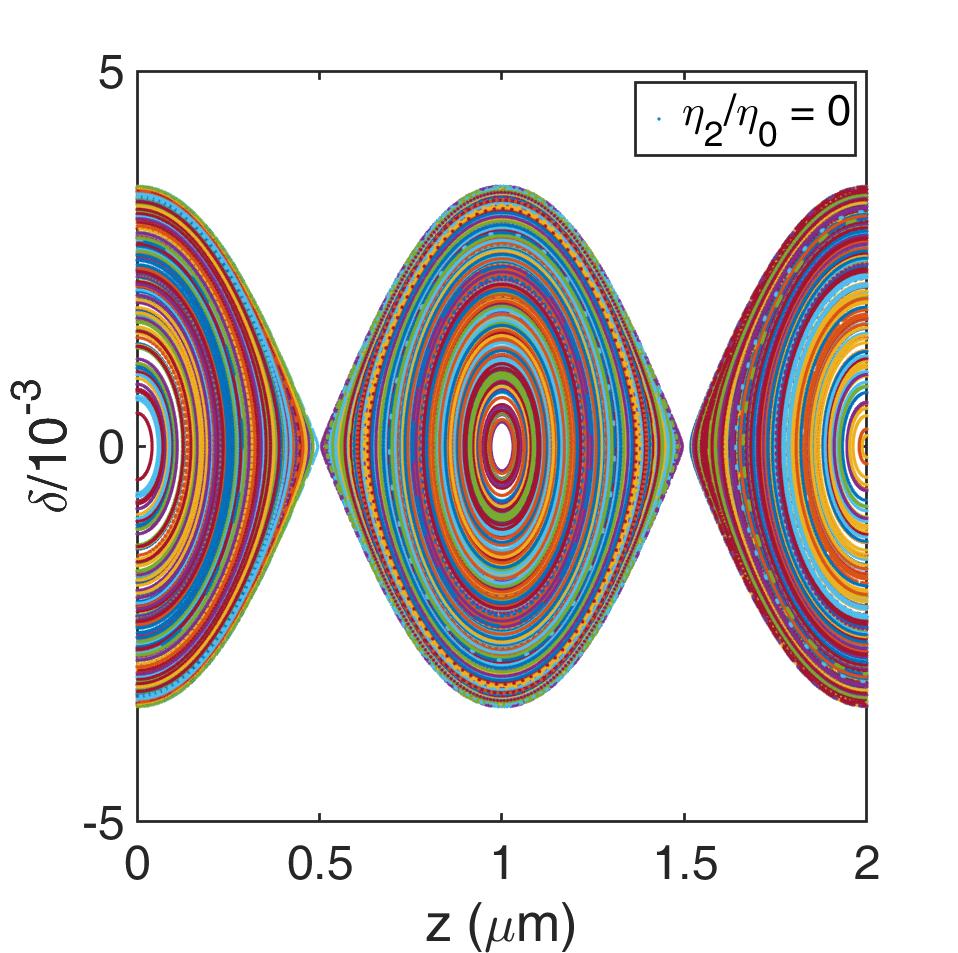}
	\includegraphics[width=0.32\textwidth]{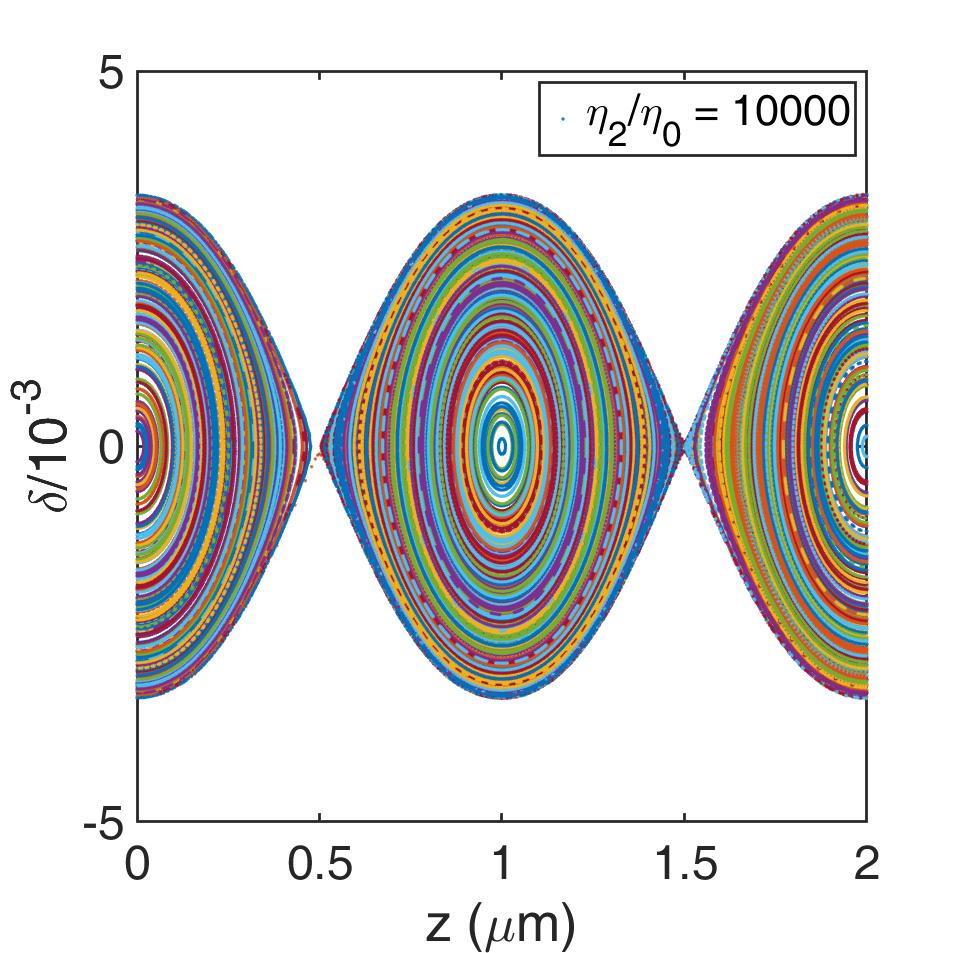}\\
	\includegraphics[width=0.32\textwidth]{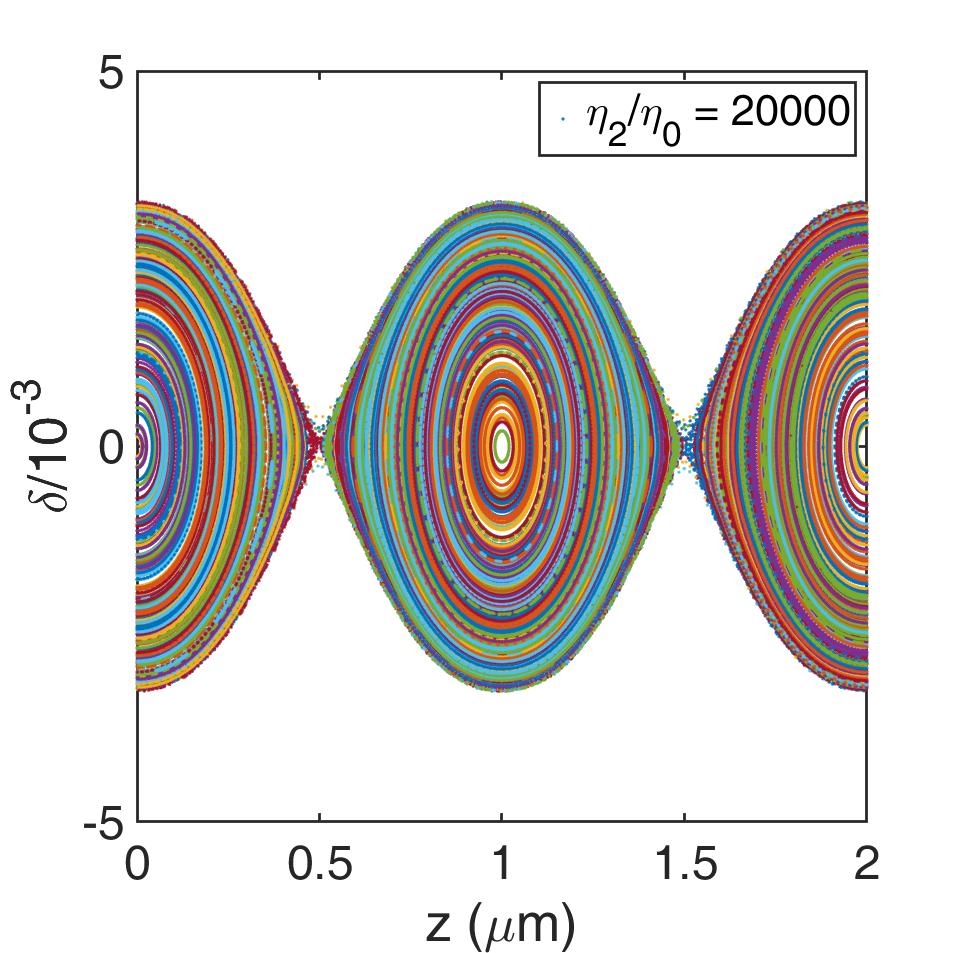}
	\includegraphics[width=0.32\textwidth]{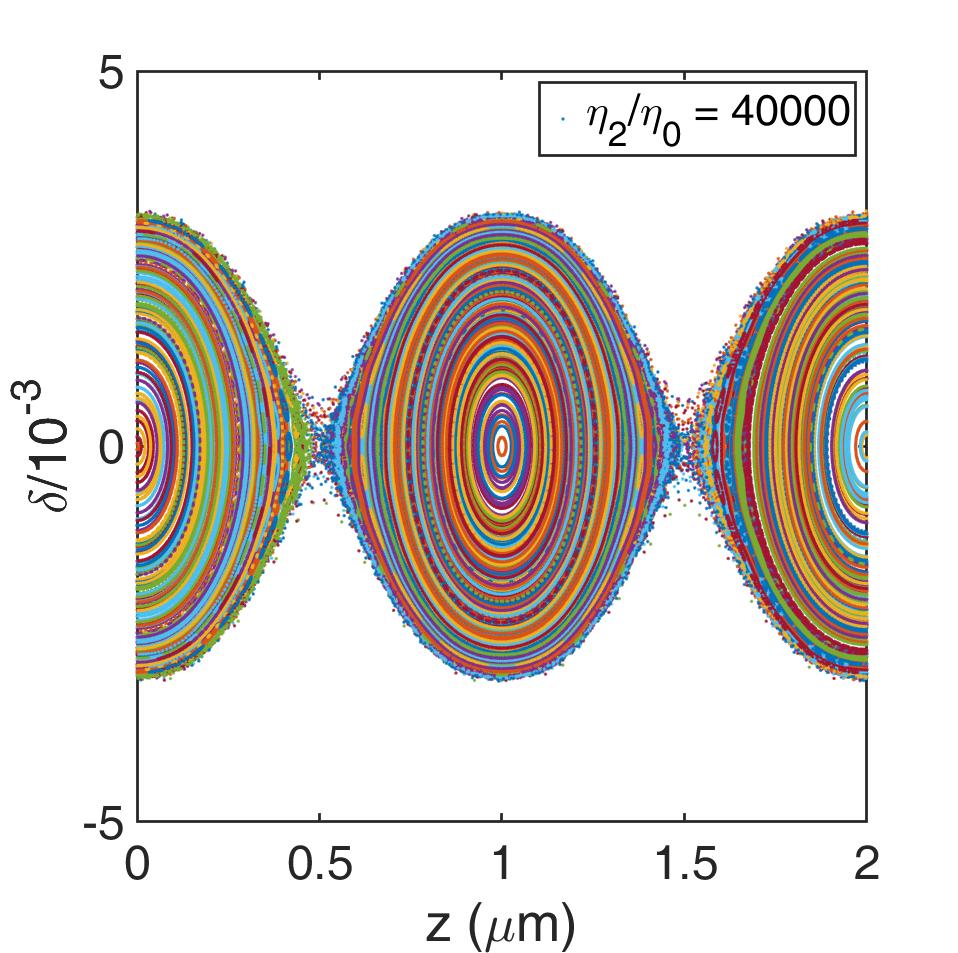}
	\includegraphics[width=0.32\textwidth]{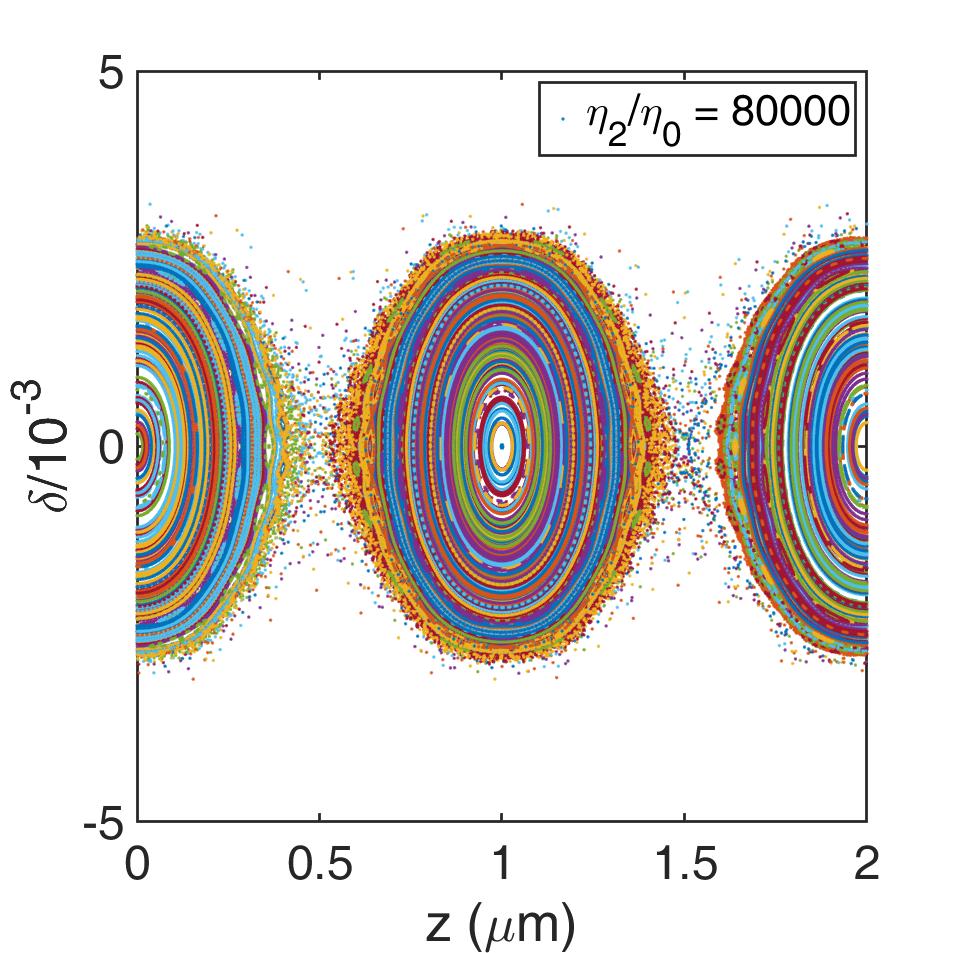}
	
	\caption{
		\label{fig:Chap2-LWFNonLinearAlpha2} 
		The impact of $\eta_{2}$ on the longitudinal phase space bucket in the longitudinal weak focusing regime. Simulation parameters: $\lambda_{\text{RF}}=1\ \mu$m, $A=1\times10^{-3}$,  $C_{0}=100$ m, $\eta_{0}=5\times10^{-7}$. 
	}
\end{figure} 

To see the impact of $\eta_{1}$ and $\eta_{2}$ on the longitudinal phase space bucket, some numerical simulations are conducted. We choose the observation point at the middle of the RF, where $\alpha_{z}=0$ and the beam distribution in the longitudinal phase space is upright. The results of the impact of $\eta_{1}$ and $\eta_{2}$ on longitudinal dynamical aperture are shown in Figs.~\ref{fig:Chap2-LWFNonLinearAlpha1} and \ref{fig:Chap2-LWFNonLinearAlpha2}, respectively. Note that in the plots, we have used the longitudinal coordinate $z$ rather than the phase $\phi$.

As we can see in Fig.~\ref{fig:Chap2-LWFNonLinearAlpha1}, the emergence of $\eta_{1}$ will make the bucket asymmetric in $\delta$, which is as expected as the $\eta=\eta_{0}+\eta_{1}\delta$ is asymmetric in $\delta$. In both directions (positive or negative), the bucket size shrinks with the increase of $\eta_{1}$ and the bucket becomes like an upright $\alpha$-shape, so they are usually referred to as $\alpha$-buckets. Note that we can also classify the bucket to be an RF-bucket or $\alpha$-bucket according to $\delta=0$ or $\eta(\delta)=0$ at the bucket center, respectively. Such classification is more reasonable from beam dynamics consideration. $\alpha$-bucket is also a method to generate short bunch and there are many interesting beam dynamics issues related to such buckets~\cite{ries2014nonlinear}.

However, the impact of $\eta_{2}$ is different. As can be seen from the simulation results presented in Fig.~\ref{fig:Chap2-LWFNonLinearAlpha2}, first, the bucket is still symmetric in $\delta$ as expected. Besides, when $\frac{\eta_{2}}{\eta_{0}}<0$, the stable region of the bucket can be even larger than the case without $\eta_{2}$. We will see later that in the case of longitudinal strong focusing, such observation can be even more notable. Therefore, we can tailor the phase slippage factor as a function of energy in the case of longitudinal strong focusing to enlarge the longitudinal dynamic aperture. This is very helpful as usually the longitudinal dynamics aperture in a longitudinal strong focusing ring is not a trivial issue and needs to be optimized to guarantee a sufficient quantum lifetime for example.

\subsubsection{Longitudinal Strong Focusing}
For a longitudinal strong focusing ring, the motion is strongly chaotic and the motion is not integrable and we cannot refer to the Halmitonian for analysis anymore. Therefore, we use numerical simulations to study the dynamics directly. Instead of a comprehensive investigations, here in this section we aim to give some qualitative remarks on the role of the nonlinear phase slippage, i.e., a proper tailoring of the nonlinear phase slippage can enlarge the longitudinal dynamic aperture significantly.

\begin{figure}[tb] 
	\centering
	\includegraphics[width=0.32\textwidth]{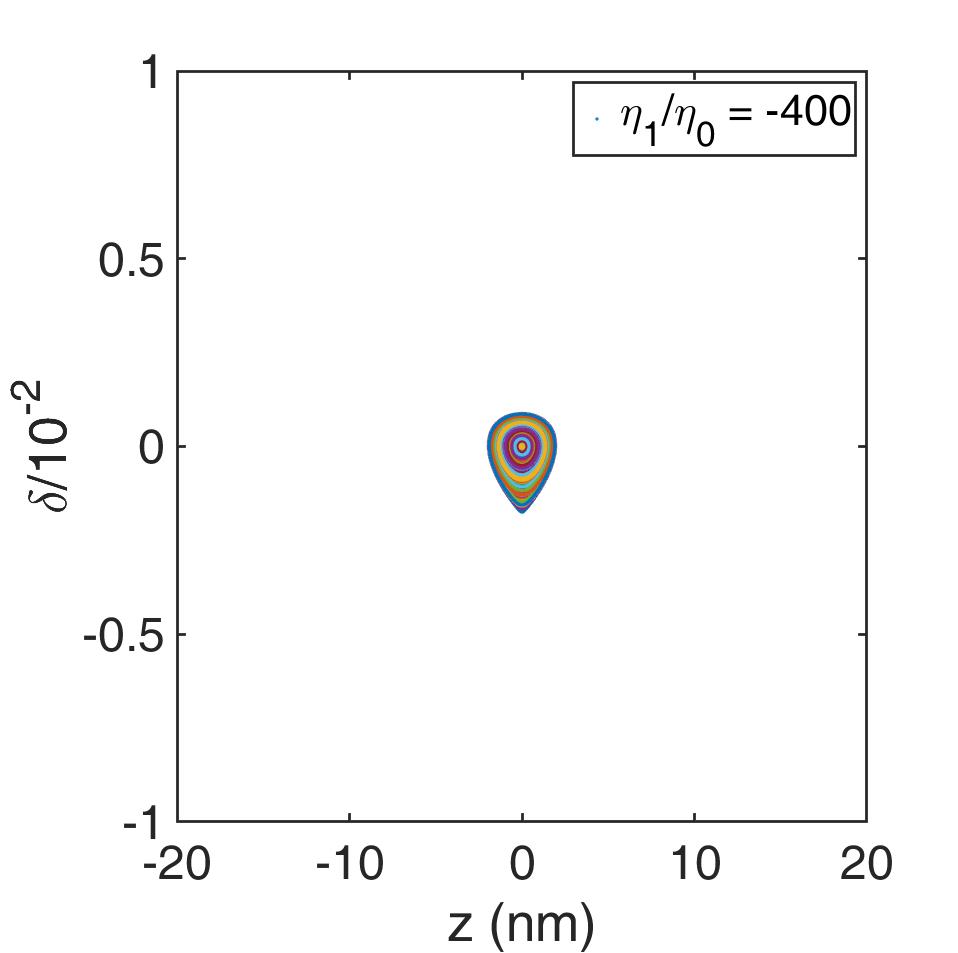}
	\includegraphics[width=0.32\textwidth]{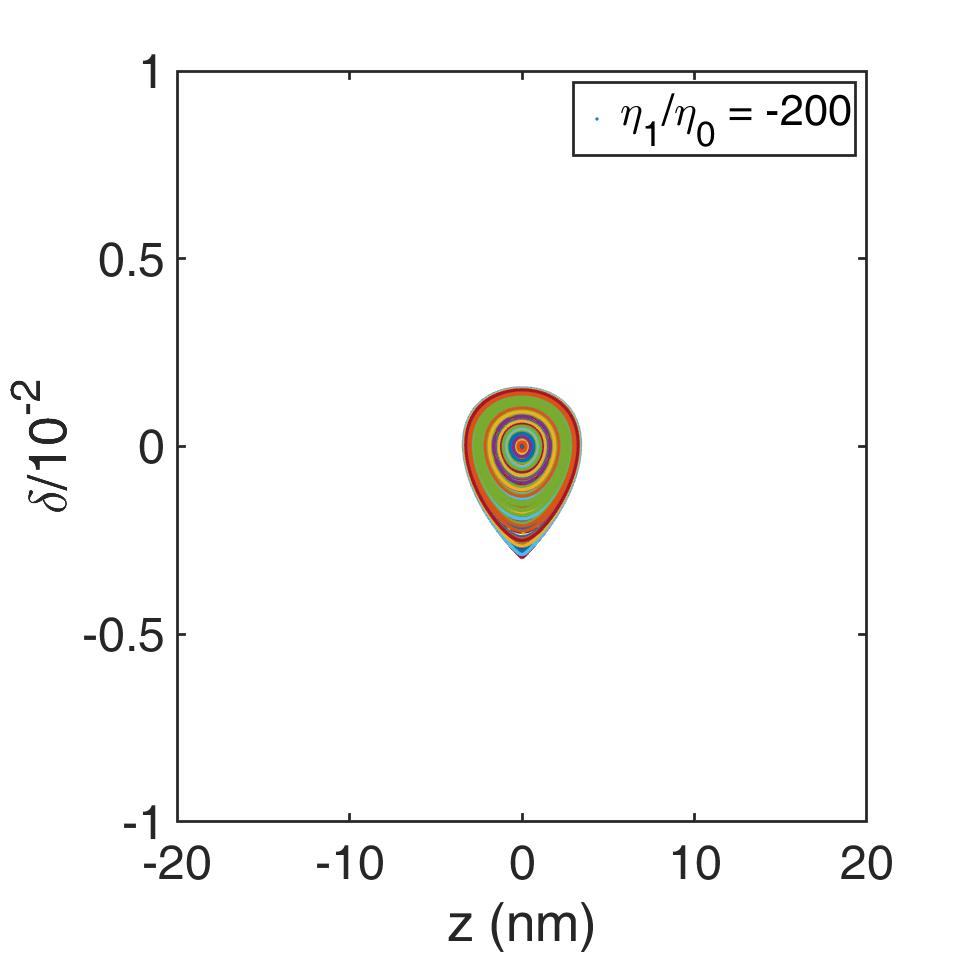}
	\includegraphics[width=0.32\textwidth]{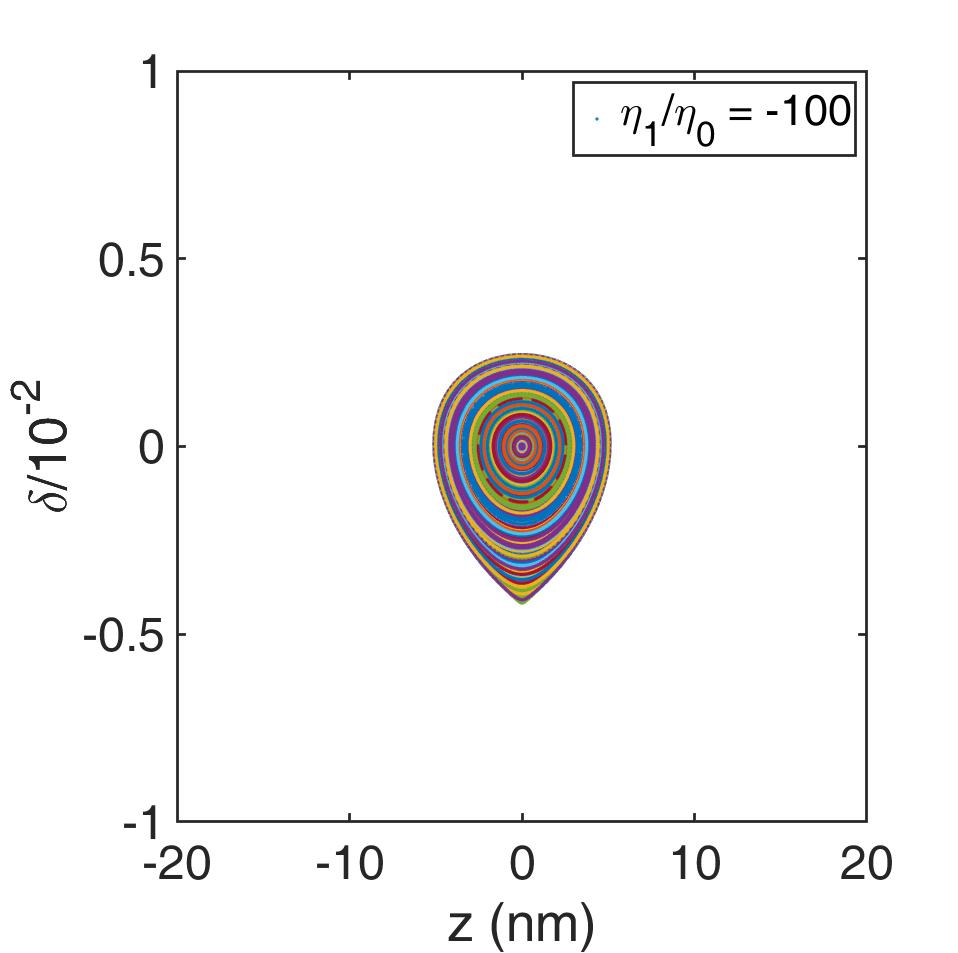}\\	
	\includegraphics[width=0.32\textwidth]{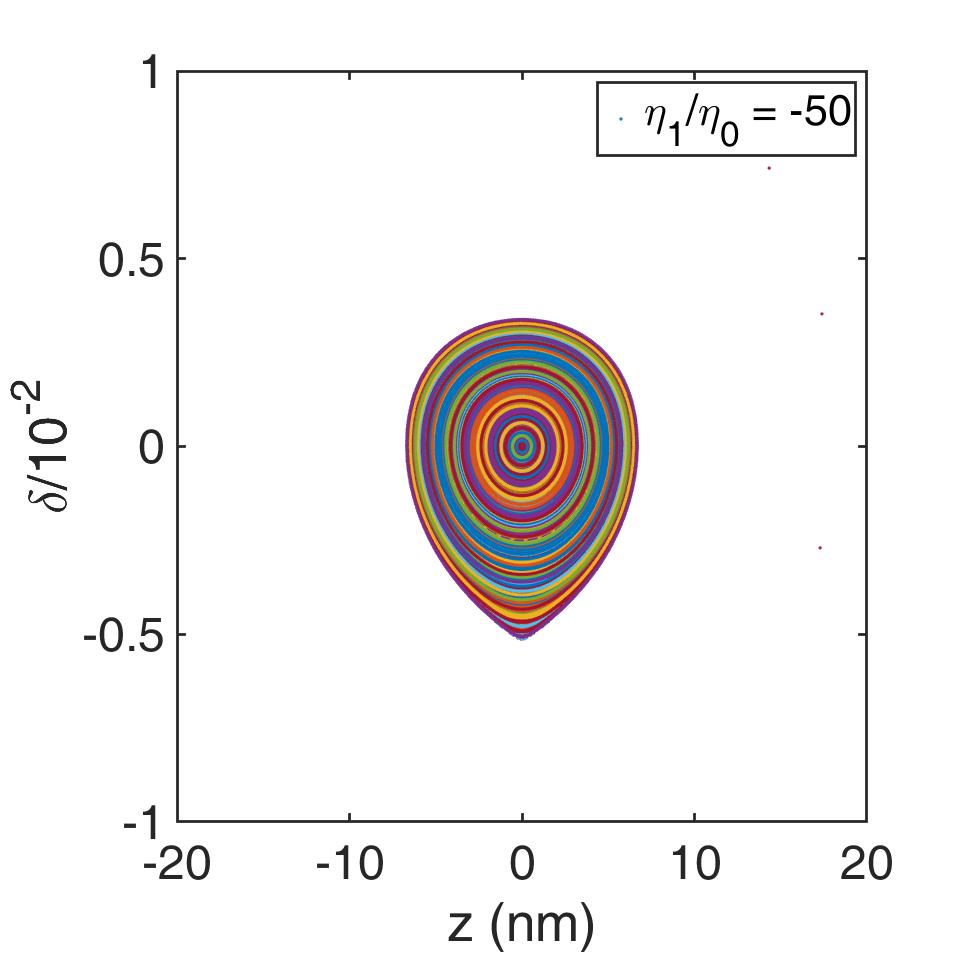}	
	\includegraphics[width=0.32\textwidth]{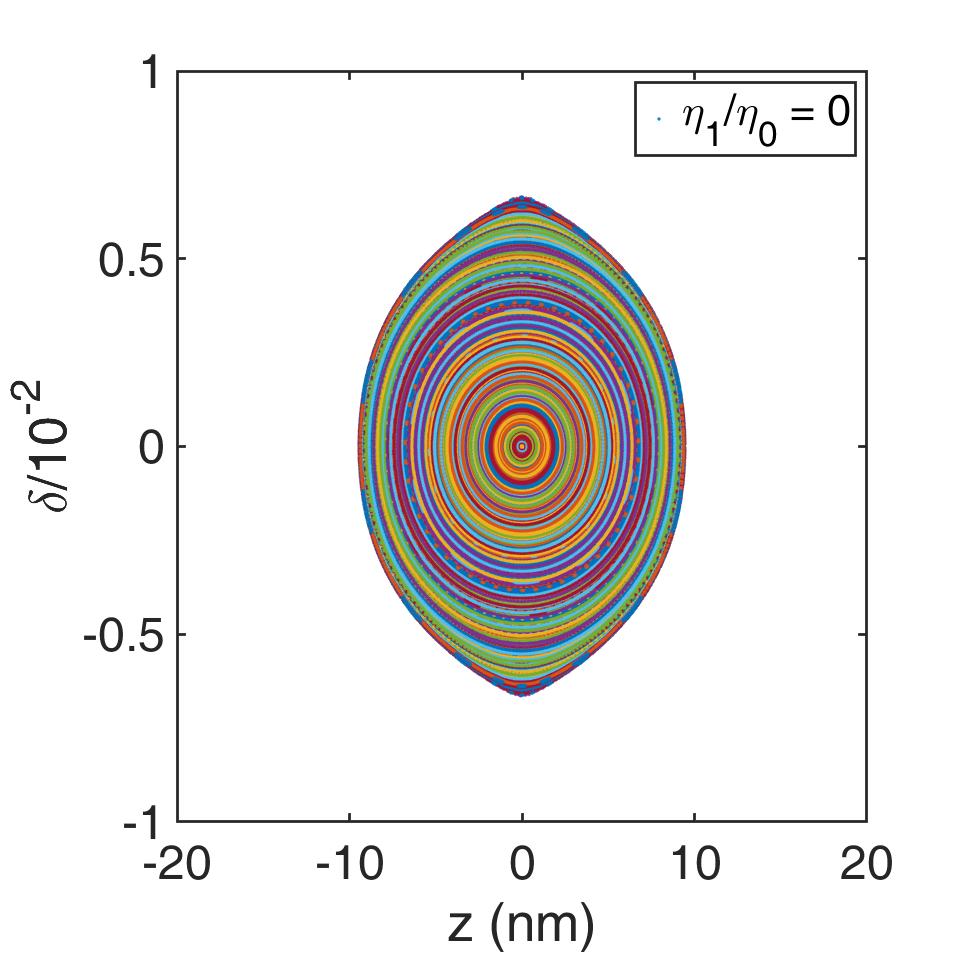}
	\includegraphics[width=0.32\textwidth]{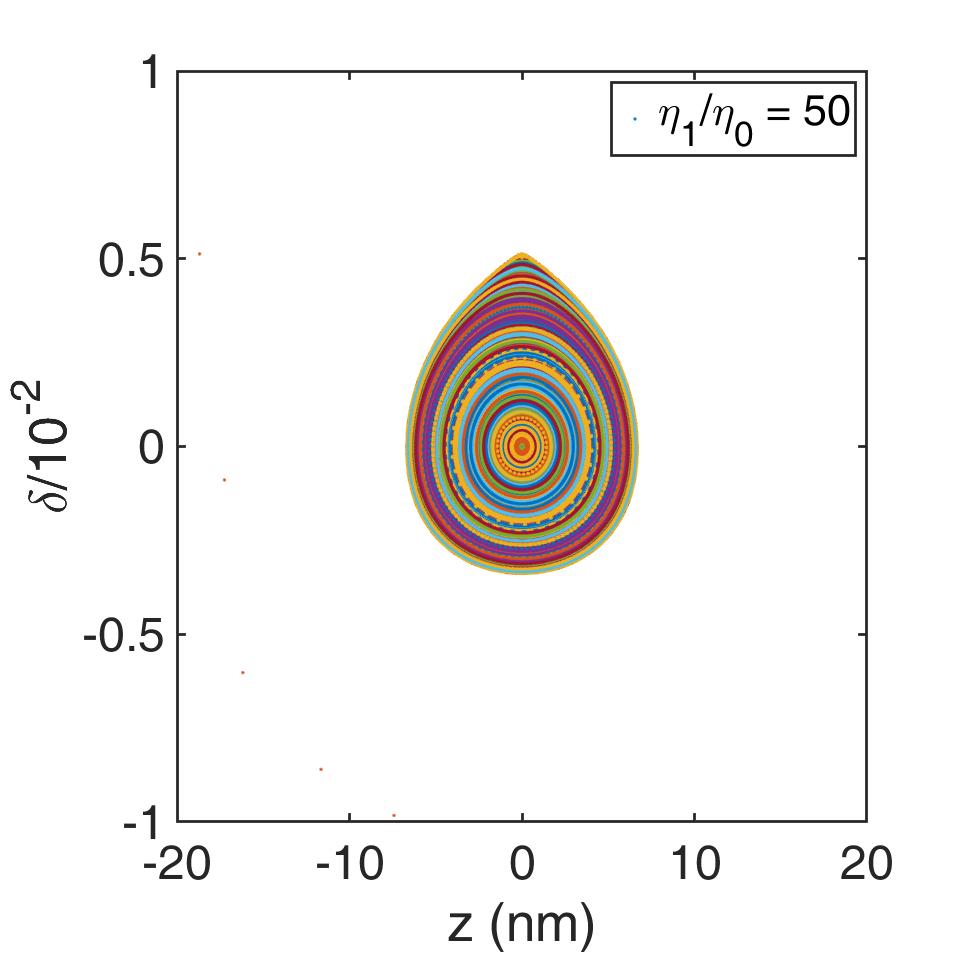}\\
	\includegraphics[width=0.32\textwidth]{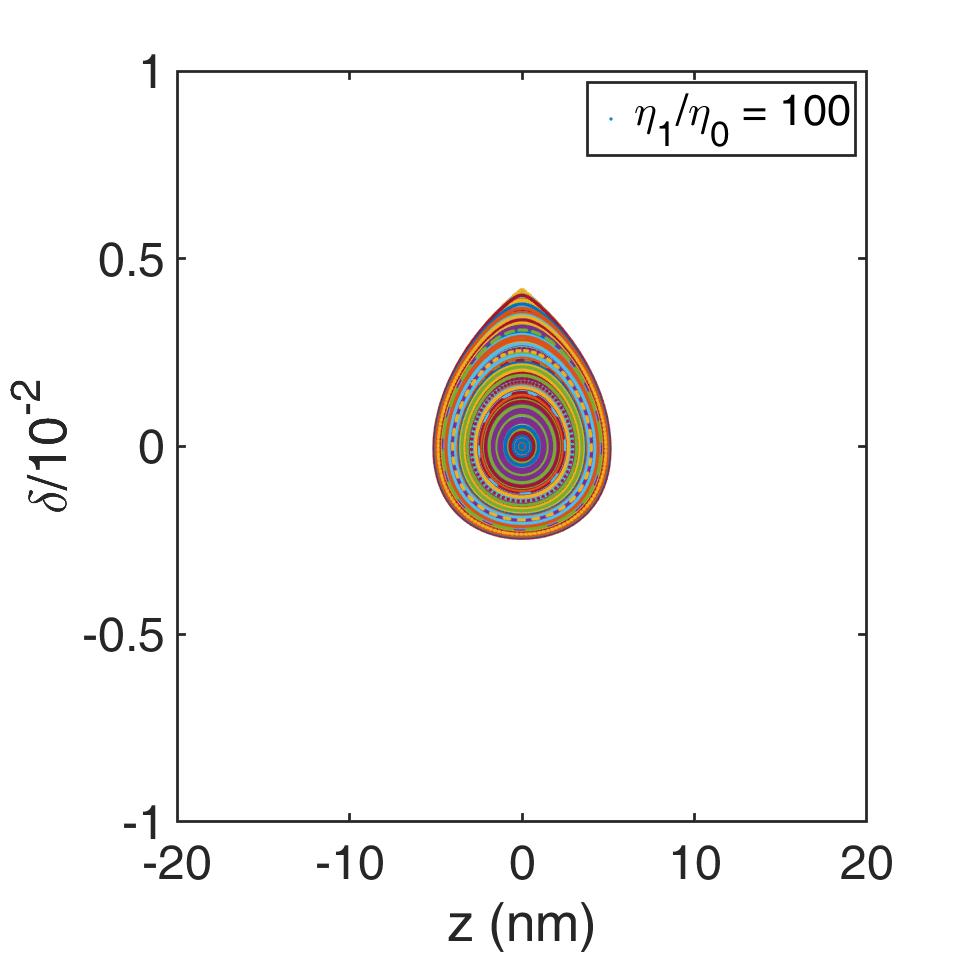}
	\includegraphics[width=0.32\textwidth]{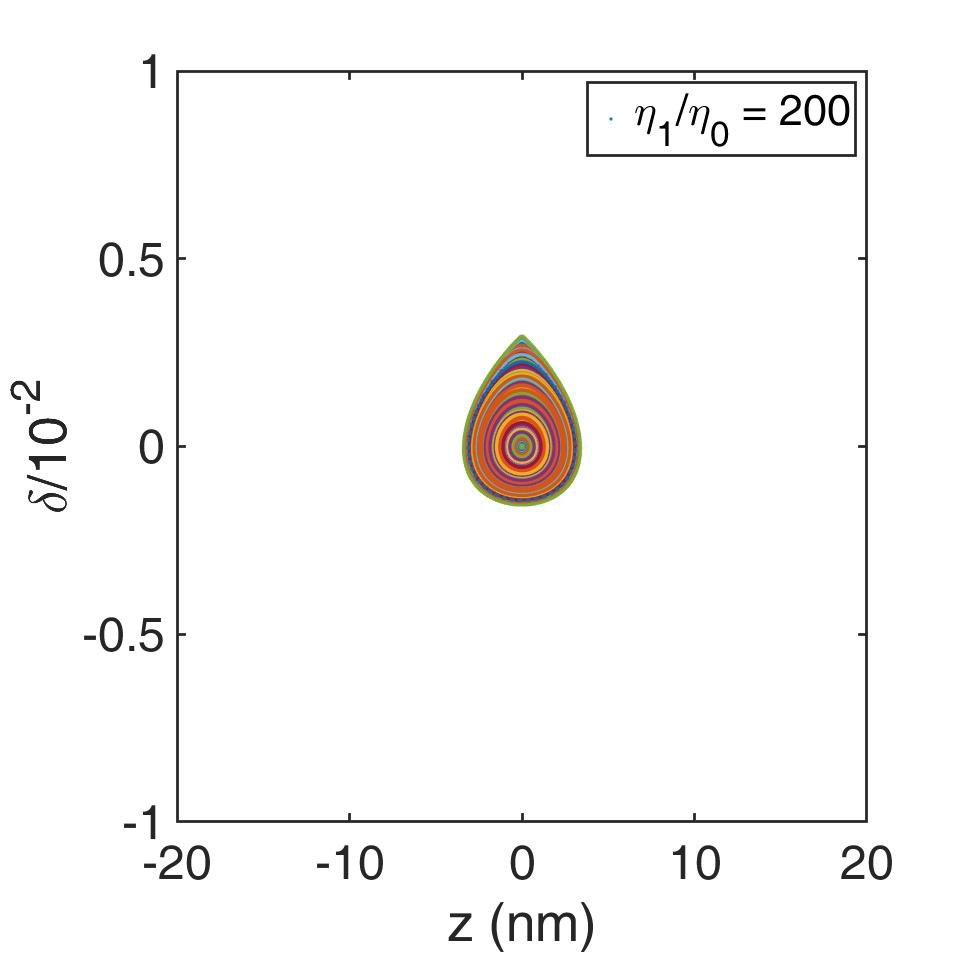}
	\includegraphics[width=0.32\textwidth]{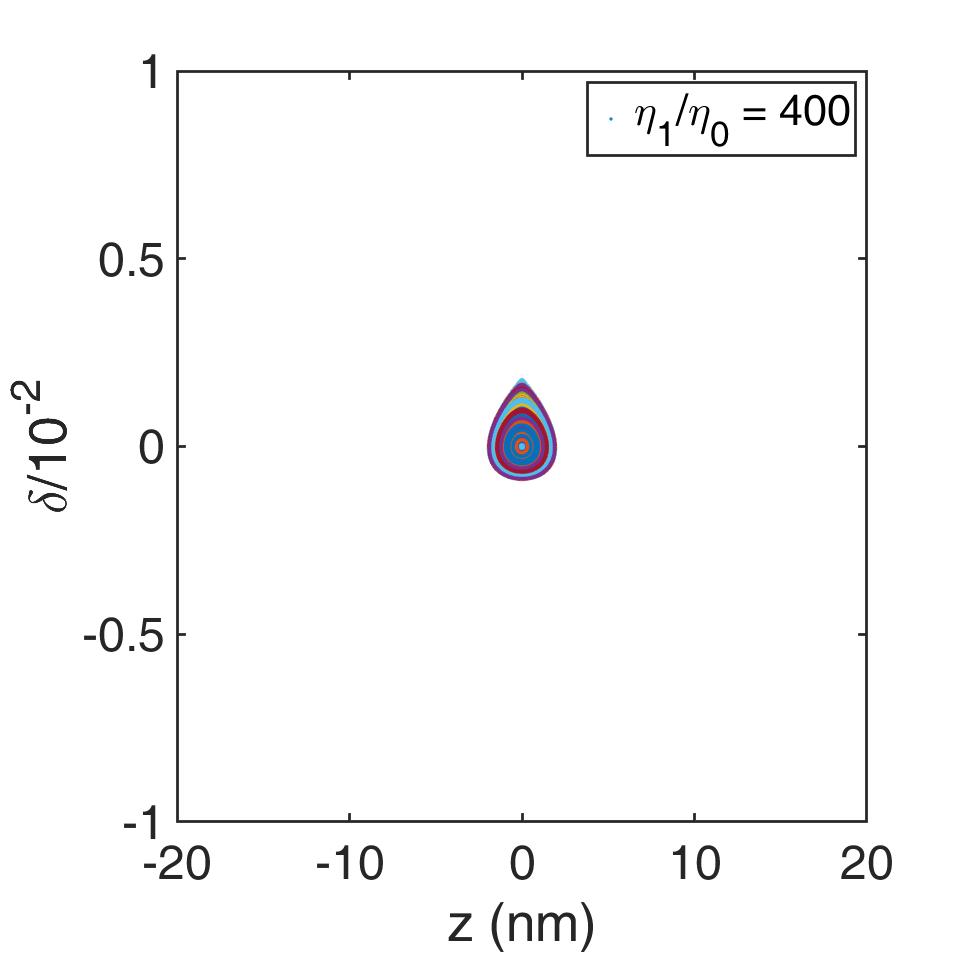}
	\caption{
		\label{fig:Chap2-LSFNonLinearAlpha1} 
		The impact of $\eta_{1}$ on the longitudinal phase space bucket in the longitudinal strong focusing regime. Simulation parameters: $\lambda_{\text{RF}}=1\ \mu$m, $h=-50000\ \text{m}^{-1}$, $R_{56}^{(1)}=15\ \mu$m, $C_{0}=100$~m, $\eta_{0}=1\times10^{-6}$. 
	}
\end{figure} 

\begin{figure}[tb] 
	\centering
	\includegraphics[width=0.32\textwidth]{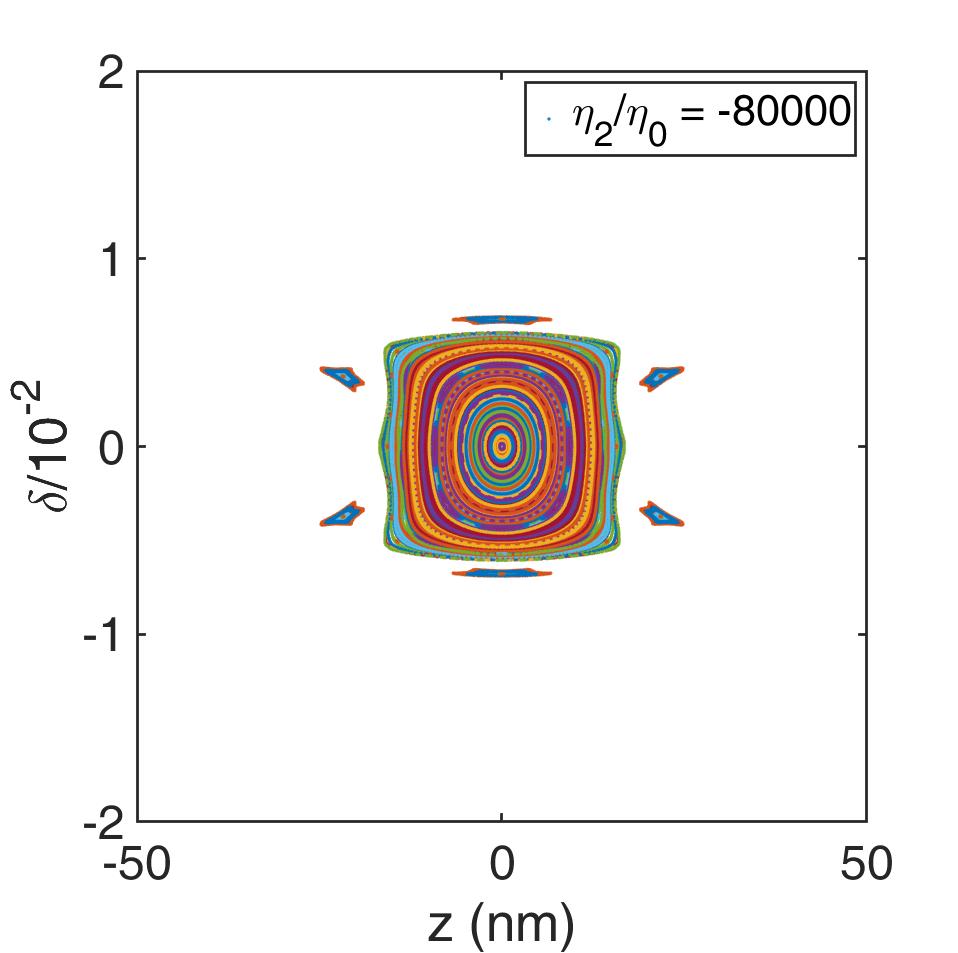}
	\includegraphics[width=0.32\textwidth]{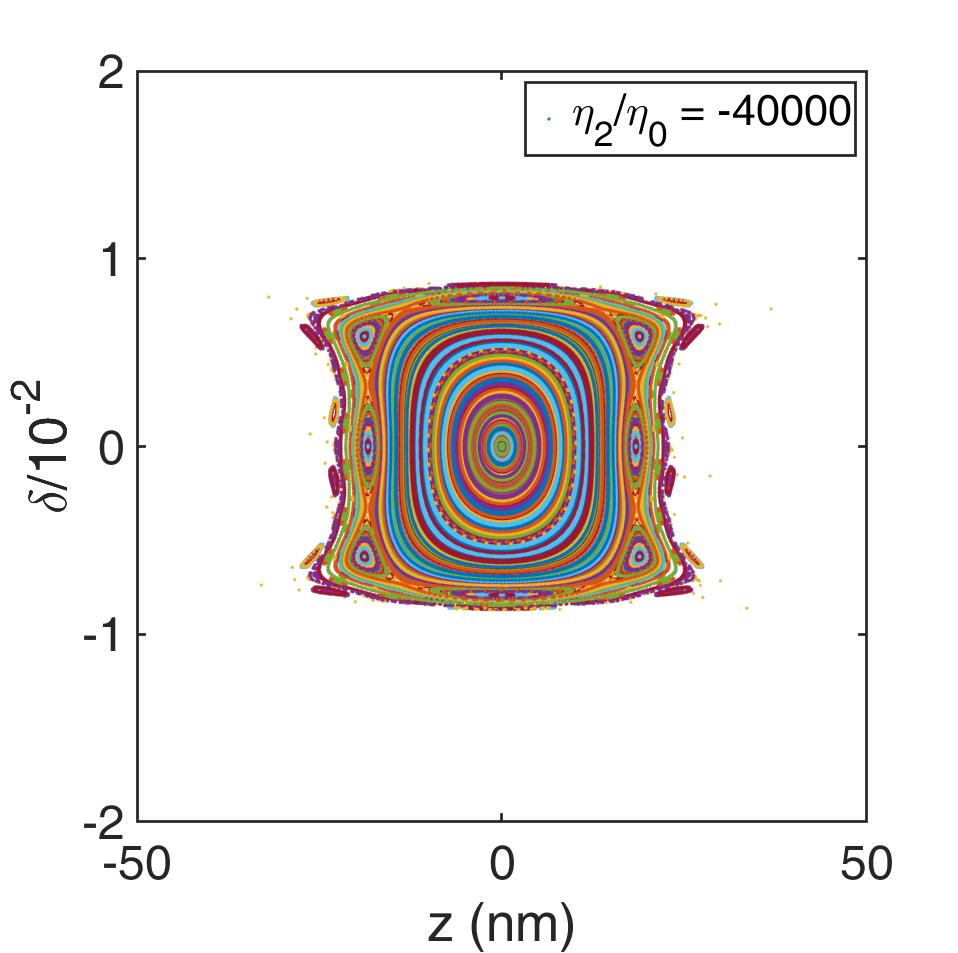}
	\includegraphics[width=0.32\textwidth]{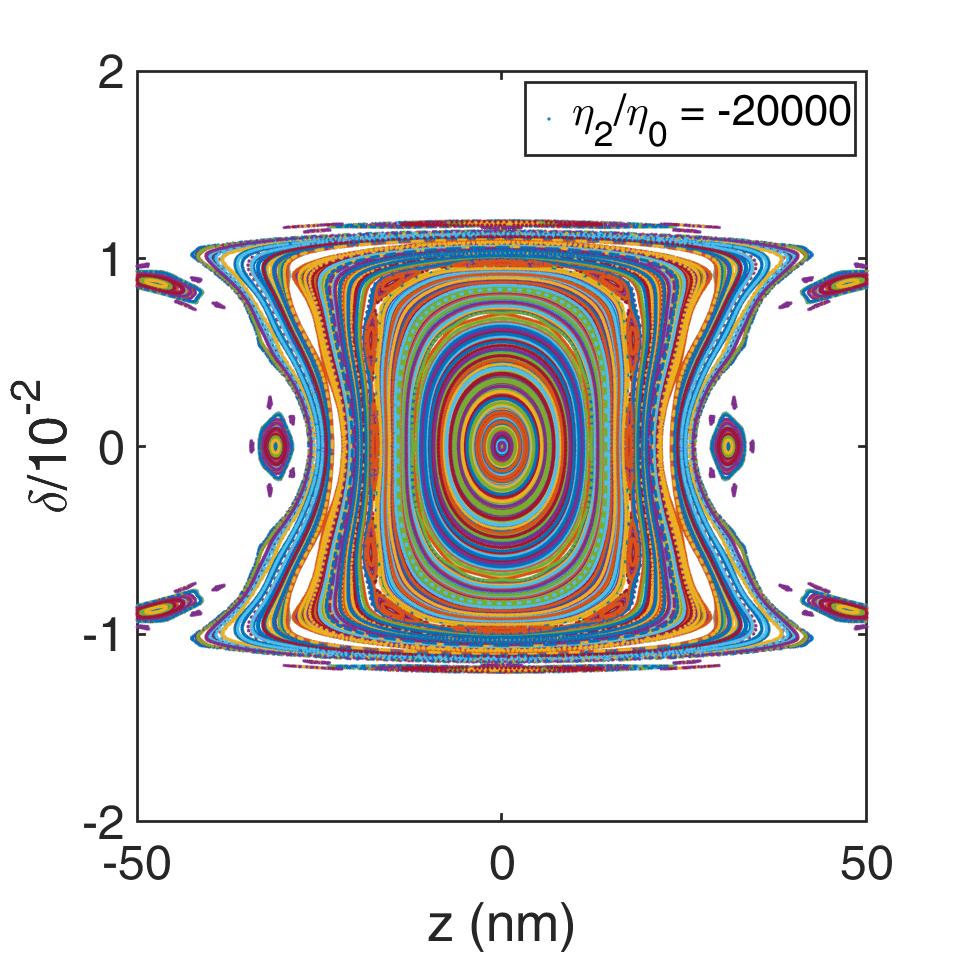}\\	
	\includegraphics[width=0.32\textwidth]{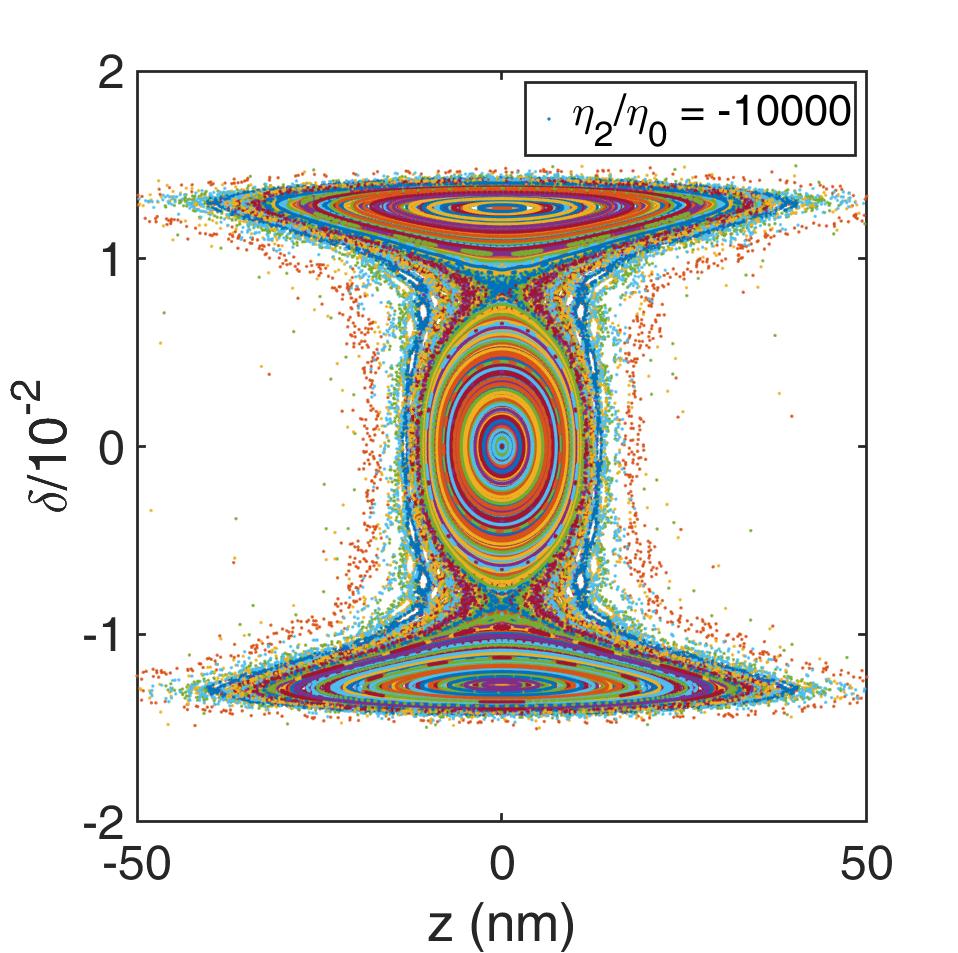}	
	\includegraphics[width=0.32\textwidth]{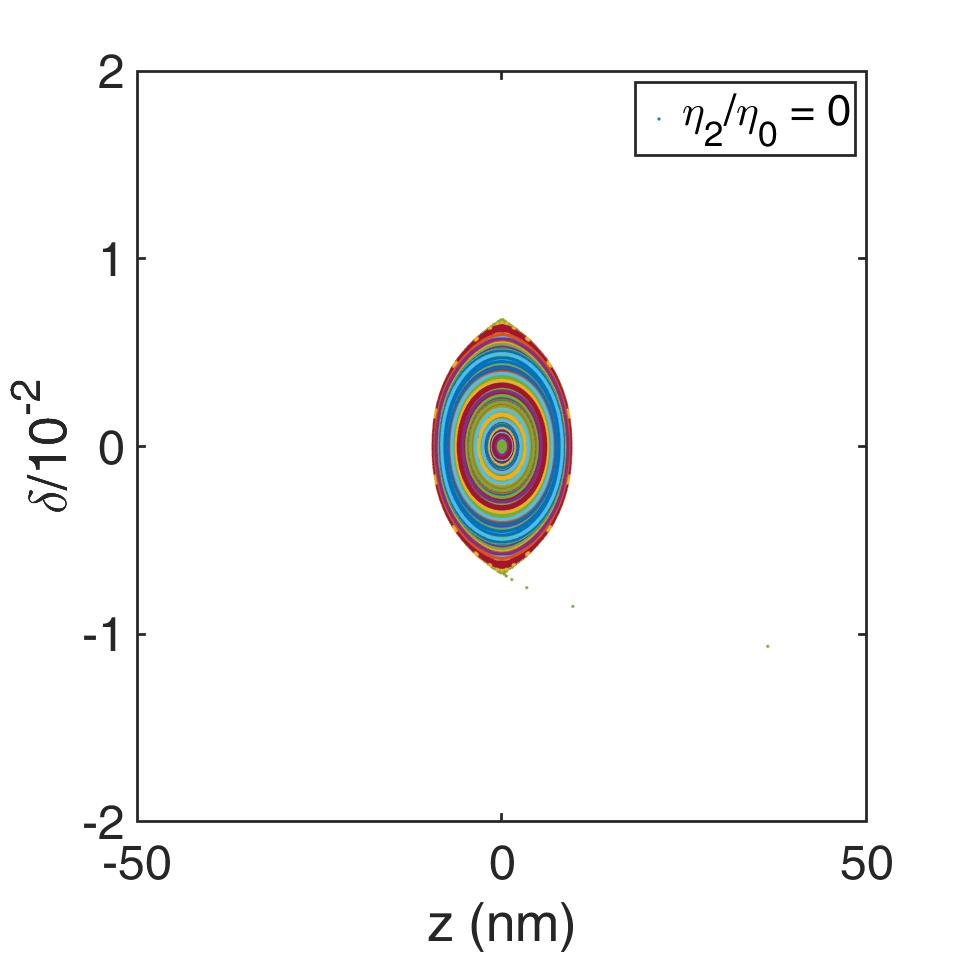}
	\includegraphics[width=0.32\textwidth]{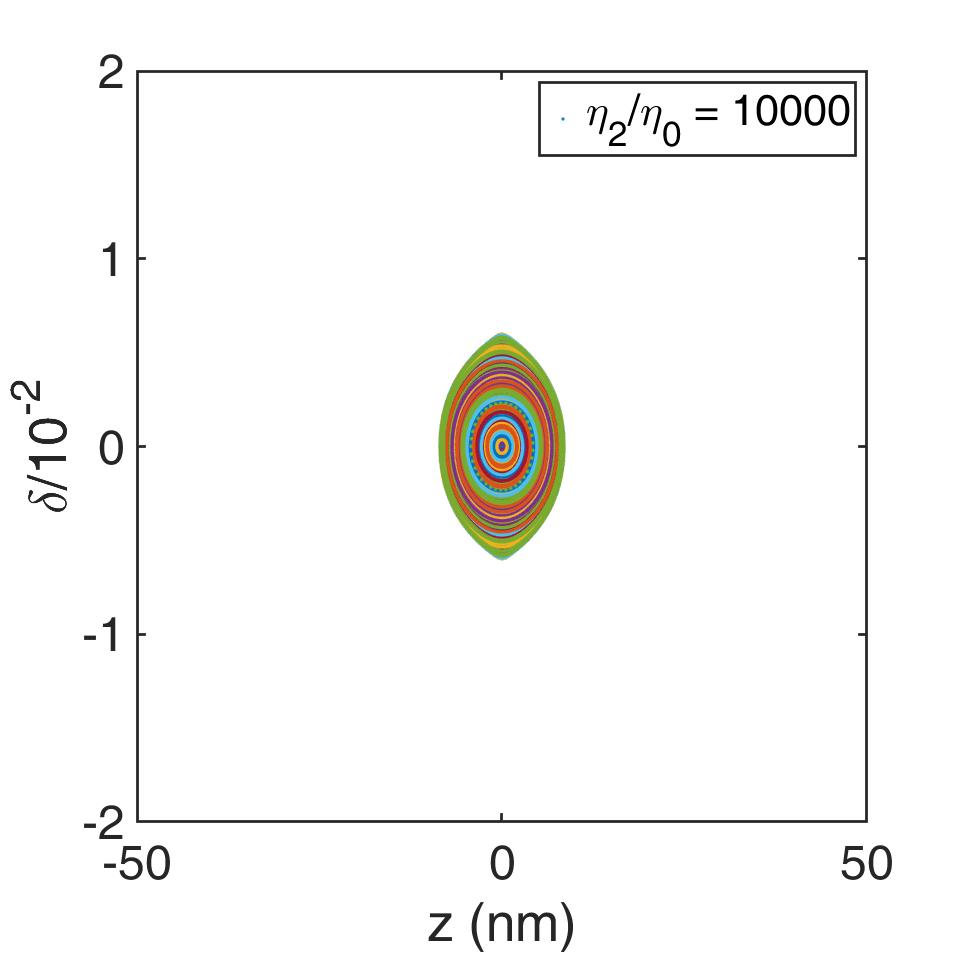}\\
	\includegraphics[width=0.32\textwidth]{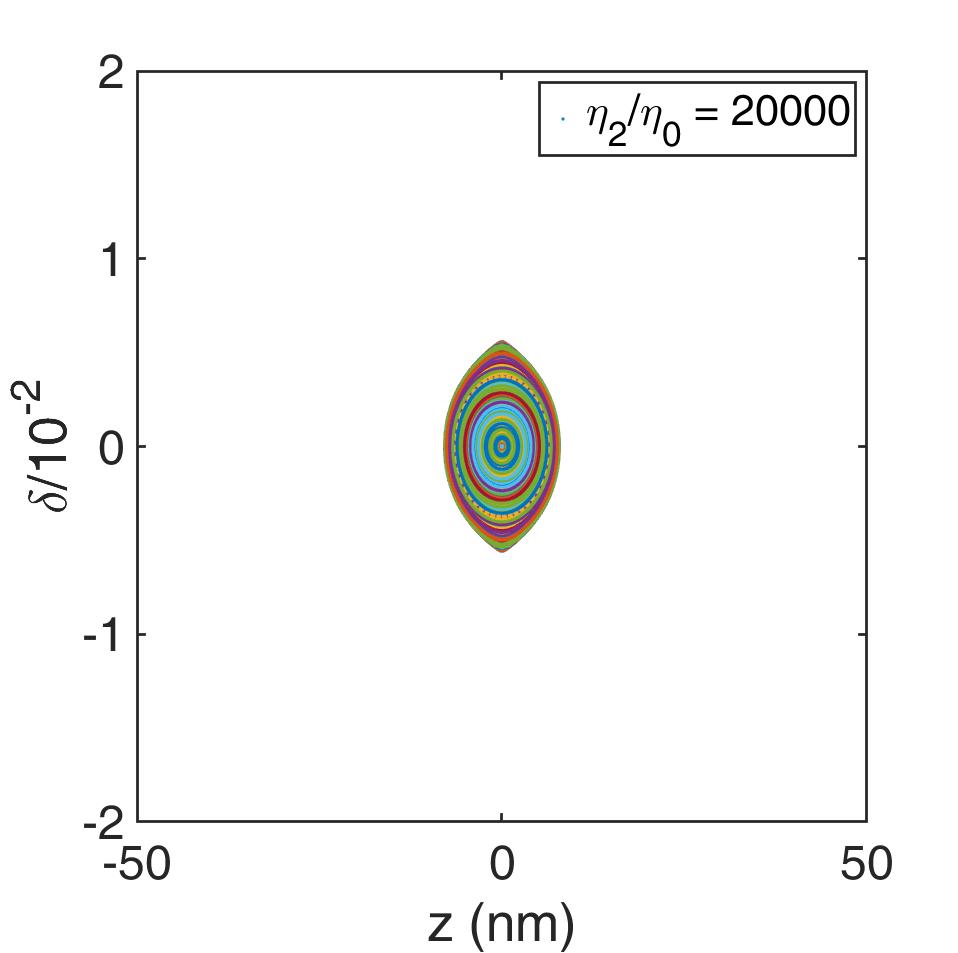}
	\includegraphics[width=0.32\textwidth]{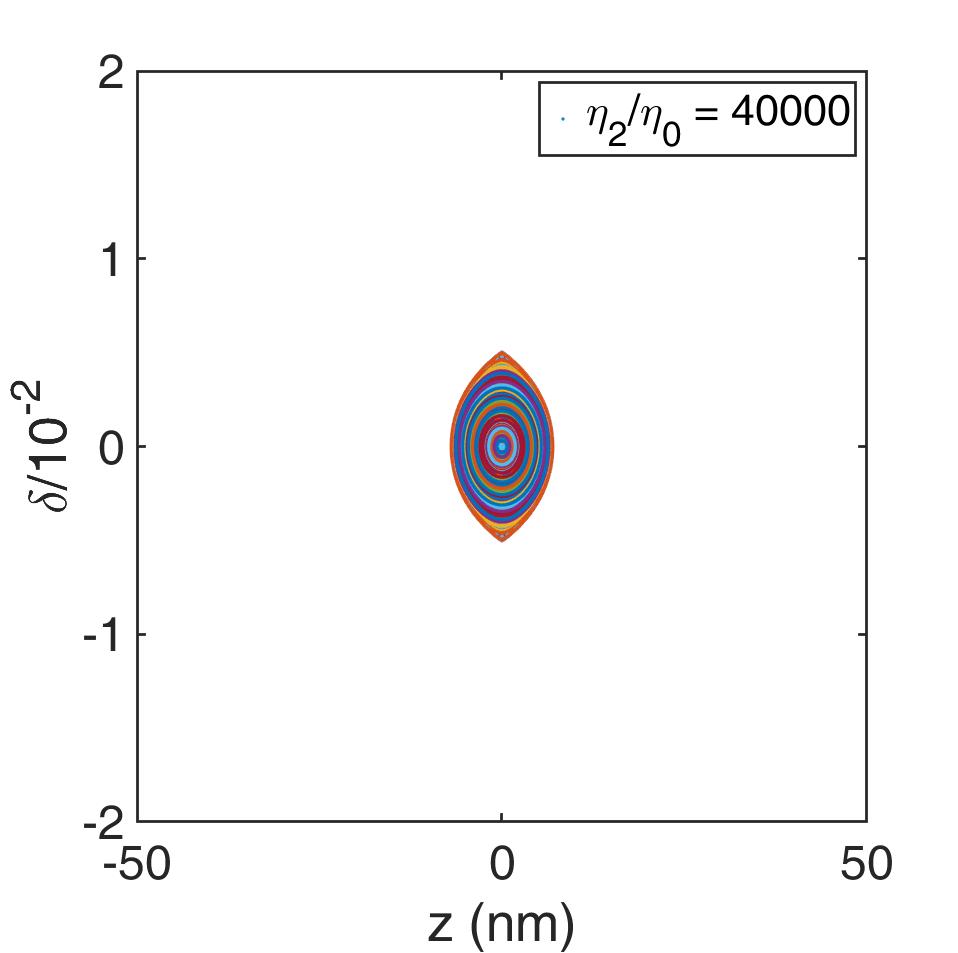}
	\includegraphics[width=0.32\textwidth]{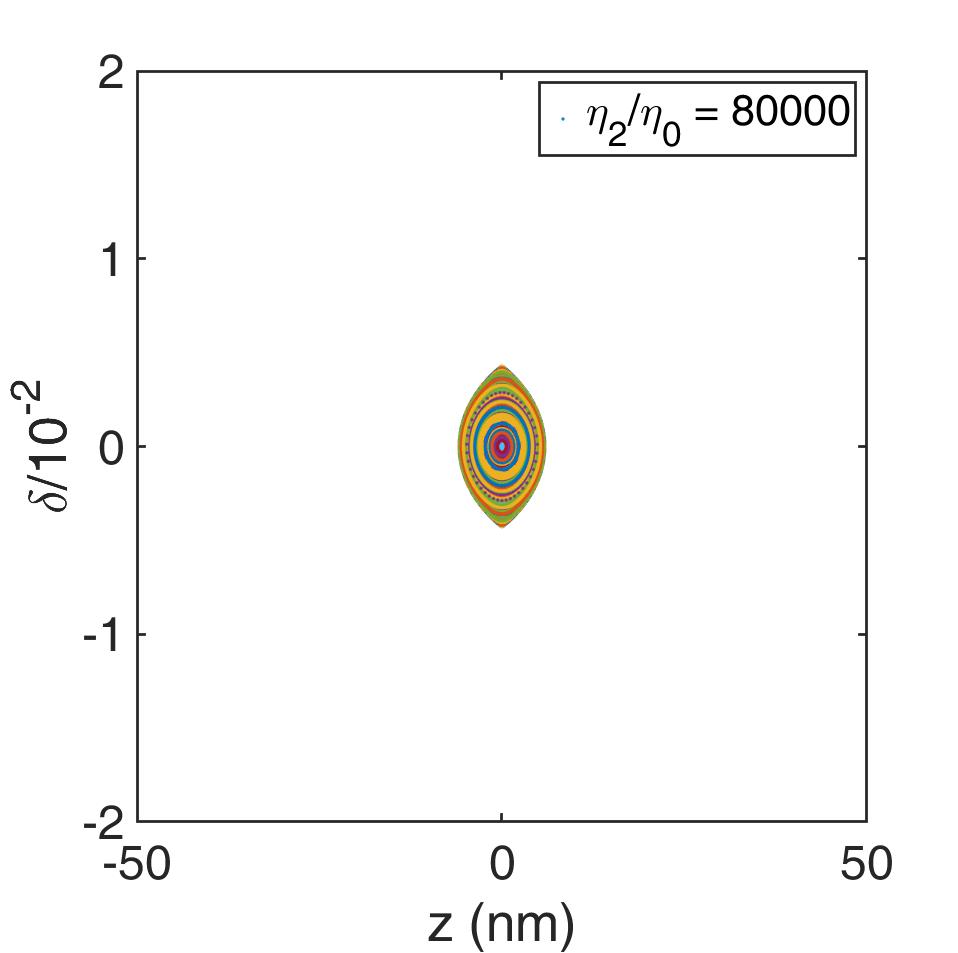}
	\caption{
		\label{fig:Chap2-LSFNonLinearAlpha2} 
		The impact of $\eta_{2}$ on the longitudinal phase space bucket in the longitudinal strong focusing regime. Simulation parameters: $\lambda_{\text{RF}}=1\ \mu$m, $h=-50000\ \text{m}^{-1}$, $R_{56}^{(1)}=15\ \mu$m, $C_{0}=100$~m, $\eta_{0}=1\times10^{-6}$. 
	}
\end{figure}

Let us use the schematic layout shown in Fig.~\ref{fig:Chap2-TwoRFs} and parameters choices given in Tab.~\ref{tab:tab2} as an example for illustration. We choose to observe the beam at the radiator center, and suppose the ring is symmetric with respect to the radiator. The one-turn kick map is then
\begin{equation}\label{eq:one-turn}
\begin{aligned}
z&=z+R_{56}^{(1)}\delta,\\
\delta&=\delta+h/k_{\text{RF}}\sin(k_{\text{RF}}z),\\
z&=z+R_{56}^{(2)}\delta,\\
\delta&=\delta+h/k_{\text{RF}}\sin(k_{\text{RF}}z),\\
z&=z+R_{56}^{(1)}\delta.
\end{aligned}
\end{equation}
As we aim to present the main physical picture, here we only consider the nonlinearity of the main ring first, i.e., $R_{56}^{(2)}(\delta)=-C_{0}(\eta_{0}+\eta_{1}\delta+\eta_{2}\delta^{2})$. $R_{56}^{(1)}$ in principle can also be a function of $\delta$. The simulation results are shown in Figs.~\ref{fig:Chap2-LSFNonLinearAlpha1} and \ref{fig:Chap2-LSFNonLinearAlpha2}.

From Fig.~\ref{fig:Chap2-LSFNonLinearAlpha1}, we know that, like that in the weak focusing case, $\eta_{1}$ makes the bucket asymmetric in $\delta$ and shrink the bucket size whether $\eta_{1}$ is positive or negative. From Fig.~\ref{fig:Chap2-LSFNonLinearAlpha2}, we can see that when $\frac{\eta_{2}}{\eta_{0}}<0$, a proper $\eta_{2}$ can help to merge the island buckets with the main bucket and broaden the stable region of the phase space, i.e., the longitudinal dynamics aperture, significantly. Intuitively this is because $\eta_{2}$ will generate stable fixed points and island buckets near the main bucket in a symmetric way. A proper $\eta_{2}$ makes the amplitude dependent tune shift favorable for the motion to be stable. Note that the fixed points of the island buckets may not have period-1 but period-$n$ stability. 

%
%

%% file: data/chap03.tex
\chapter{Transverse-Longitudinal Coupling Dynamics}
\label{cha:TLC}
After dedicated efforts devoted to the longitudinal dynamics to realize an ultra-small longitudinal emittance or ultra-short bunch length for coherent radiation generation, we need to make sure that the coupling arising from transverse dynamics does not degrade or even destroy the longitudinal fine structures.  The reason is based on the fact that the transverse beam size in an SSMB ring can be orders of magnitude larger than the microbunch length.  This is the motivation for us to investigate the transverse-longitudinal coupling (TLC) dynamics. In this chapter, we start from the linear TLC and then investigate the nonlinear TLC dynamics. For the linear dynamics, first we analyze the passive bunch lengthening induced by bending magnets and discuss the method to preserve microbunching with beam deflection. We then emphasize the fact that TLC can actually be actively applied for efficient bunch compression and high harmonic generation when the transverse emittance is small. We present two theorems and their proofs, concerning the application of such TLC schemes, with their implications discussed. Further, we have analyzed the contribution of modulator to the vertical emittance from quantum excitation,  to obtain a self-consistent evaluation of the required modulation laser power. The two theorems and related analysis provide the theoretical basis for the application of TLC in SSMB to lower the requirement on the modulation laser power, by taking advantage of the fact that the vertical emittance in a planar ring is rather small. Based on the investigations, we have presented an example parameters set for the envisioned SSMB storage ring to generate high-power EUV radiation.  The relation between our TLC analysis and
the transverse-longitudinal emittance exchange is also
discussed.   For the nonlinear dynamics, we present the analysis and the first experiment proof of the second-order TLC effect on the equilibrium parameters, which can help to improve the stable beam current and coherent radiation power of a ring working in quasi-isochronous regime.  

\section{Linear Coupling}\label{sec:LinearCoupling}
\subsection{Passive Bunch Lengthening}\label{sec:passiveLengthening}
\begin{figure}[tb] 
	\centering 
	\includegraphics[width=1\textwidth]{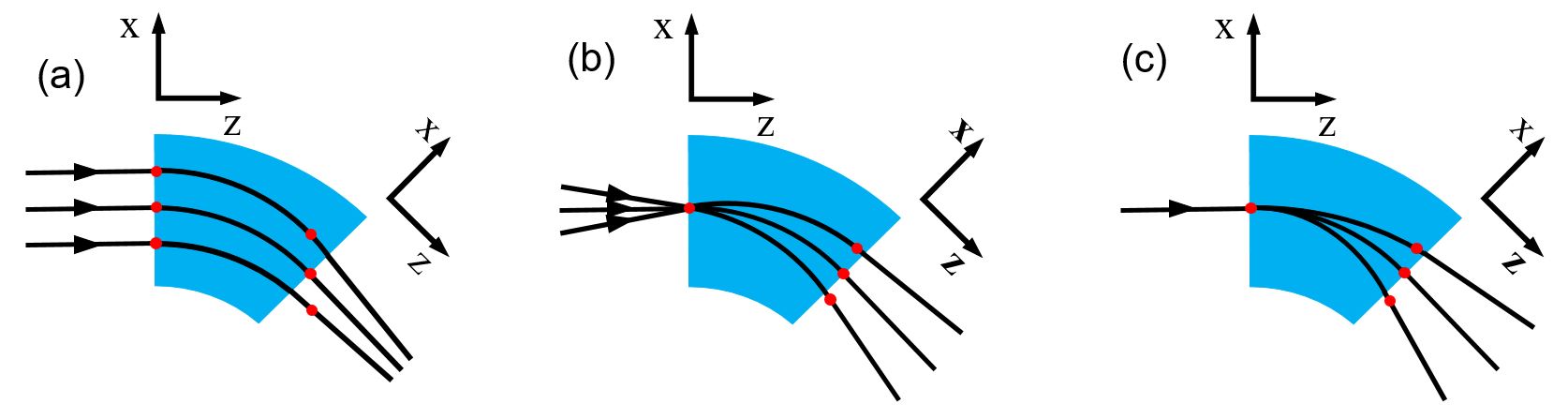}
	\caption{
		\label{fig:Chap3-TLCPysicalPicture} 
		Linear transverse-longitudinal coupling induced by a bending magnet. Particles with different horizontal positions (a) and angles (b) pass the horizontal bending magnet along different paths, resulting in longitudinal coordinate differences. Particles with different energies (c) also pass the horizontal bending magnet along different paths, resulting in horizontal position and angle differences.
	}
\end{figure}

In a linear transport line without bending magnets, the transverse and longitudinal motions are decoupled in a first-order approximation. However, the situation changes when there are bending magnets. Particles with different horizontal (vertical) positions and angles will pass through the horizontal (vertical) bending magnets along different paths, resulting in differences in the longitudinal coordinate. The transverse motion can thus be coupled to the longitudinal dimension. When traversing the bending magnets, particles with different energies will also pass along different paths and exit with different horizontal (vertical) positions and angles. The longitudinal motion can thus also be coupled to the transverse dimension. The physical pictures of the linear transverse-longitudinal coupling introduced by the bending magnets are shown in Fig.~\ref{fig:Chap3-TLCPysicalPicture}. Although this passive TLC is a well-understood effect~\cite{shoji2004bunch,wustefeld2005horizontal, huang2007matrix,shimada2009transverse,shoji2011transient}, here we present a concise analysis of this effect with an emphasis on its vital role in microbunching formation and transportation for both the transient and steady-state cases. 

We start with a planar $x$-$y$ uncoupled lattice and assume that the RF cavities are placed at dispersion-free locations. We temporarily ignoring the vertical dimension, and use the state vector ${\bf X}=(x,x',z,\delta)^{T}$. The subscripts $_{5,\ 6}$ are used for $z,\ \delta$ for consistency with the literature. Hereafter, the subscript $_x$ in this section is omitted unless necessary. 

As introduced in Sec.~\ref{sec:CS}, the betatron coordinate, defined by ${\bf X}_{\beta}={\bf B}{\bf X}$, is first used to parametrize the transport matrix in diagonal form. 
The transport matrix for ${\bf X}_{\beta}$ of a lattice from $s_{1}$ to $s_{2}$ is then
\begin{equation}
{\bf M}_{\beta}(s_{1},s_{2})=\left(\begin{matrix}
{\bf M}_{x_\beta}(s_{1},s_{2})&{\bf 0}\\
{\bf 0}&{\bf M}_{z_\beta}(s_{1},s_{2})
\end{matrix}\right),
\end{equation}
with
\begin{equation}\label{eq:Betatron}
{\bf X}_{\beta}(s_{2})={\bf M}_{\beta}(s_{1},s_{2}){\bf X}_{\beta}(s_{1}).
\end{equation}
Following Courant and Snyder \cite{courant1958theory}, we write ${\bf M}_{x_\beta}(s_{1},s_{2})$ as
\begin{equation}
{\bf M}_{x_\beta}(s_{1},s_{2})={\bf A}^{-1}(s_{2}){\bf T}(s_{1},s_{2}){\bf A}(s_{1}),
\end{equation}
with 
\begin{equation}
{\bf A}(s_{i})=\left(\begin{matrix}
\frac{1}{\sqrt{\beta(s_{i})}}&0\\
\frac{\alpha(s_{i})}{\sqrt{\beta(s_{i})}}&\sqrt{\beta(s_{i})}
\end{matrix}\right)
\end{equation}
and
\begin{equation}
{\bf T}(s_{1},s_{2})=\left(\begin{matrix}
\cos{\psi_{12}}&\sin{\psi_{12}}\\
-\sin{\psi_{12}}&\cos{\psi_{12}}
\end{matrix}\right),
\end{equation}
where 
\begin{equation}\label{eq:betaPhaseAdvance}
\psi_{12}=\psi_{2}-\psi_{1}=\int_{s_{1}}^{s_{2}}\frac{1}{\beta(s)}ds
\end{equation}
is the betatron phase advance from $s_{1}$ to $s_{2}$. 
For ${\bf M}_{z_\beta}(s_{1},s_{2})$, the expression is similar to ${\bf M}_{x_\beta}(s_{1},s_{2})$, but note that if we want to calculate the synchrotron phase advance similar to Eq.~(\ref{eq:betaPhaseAdvance}), the distance $s$ should be replaced by the effective longitudinal drift space, i.e., the $R_{56}$ or more accurately $F$ defined in Eq.~(\ref{eq:Fdefinition}). If there is no RF cavity between $s_{1}$ and $s_{2}$, we have
\begin{equation}
{\bf M}_{z_\beta}(s_{1},s_{2})=\left(\begin{matrix}
1&F(s_{1},s_{2})\\
0&1
\end{matrix}\right).
\end{equation}
In a storage ring, $F=-\tilde{\eta}(s_{1},s_{2})C_{0}$ as shown Eq.~(\ref{eq:Fdefinition}). Note that for a given lattice, $F$ is a function of the initial dispersion $({D}_{1},{D}_{1}')$ at $s_{1}$, although the transfer map ${\bf M}(s_{1},s_{2})$ for the state vector ${\bf X}$ is not, as the transport matrix is fixed once the lattice is given. This dependence is actually a result of the Courant-Snyder parametrization.

From the definition of ${\bf X}_{\beta}$ and Eq.~(\ref{eq:Betatron}), we have
\begin{equation}
{\bf M}(s_{1},s_{2})={\bf B}^{-1}(s_{2}){\bf M}_{\beta}(s_{1},s_{2}){\bf B}(s_{1}).
\end{equation}
After some straightforward algebra, the transport matrix ${\bf M}(s_{1},s_{2})$ can be expressed as
\begin{equation}\label{eq:TransportMatrix}
\begin{aligned}
&{\bf M}(s_{1},s_{2})=\left(\begin{matrix}
R_{11}&R_{12}&0&{D}_{2}-R_{11}{D}_{1}-R_{12}{D}_{1}'\\
R_{21}&R_{22}&0&{D}_{2}'-R_{21}{D}_{1}-R_{22}{D}_{1}'\\
R_{51}&R_{52}&1&F-R_{51}{D}_{1}-R_{52}{D}_{1}'\\
0&0&0&1\\
\end{matrix}\right),\\
&R_{11}=\sqrt{\frac{\beta_{2}}{\beta_{1}}}[\cos{\psi_{12}}+\alpha_{1}\sin{\psi_{12}}],\\
&R_{12}=\sqrt{\beta_{1}\beta_{2}}\sin{\psi_{12}},\\
&R_{21}=-\frac{1}{\sqrt{\beta_{1}\beta_{2}}}[(1+\alpha_{1}\alpha_{2})\sin{\psi_{12}}-(\alpha_{1}-\alpha_{2})\cos{\psi_{12}}],\\
&R_{22}=\sqrt{\frac{\beta_{1}}{\beta_{2}}}[\cos{\psi_{12}}-\alpha_{2}\sin{\psi_{12}}],\\
&R_{51}=R_{21}{D}_{2}-R_{11}{D}_{2}'+{D}_{1}',\\
&R_{52}=-R_{12}{D}_{2}'+R_{22}{D}_{2}-{D}_{1}.
\end{aligned}
\end{equation}	
This matrix can then be used to analyze both the transient and steady-state cases of TLC. 

We consider first the influence of the betatron oscillation on the longitudinal coordinate. With the help of the Courant-Snyder parametrization, the betatron oscillation position and angle at the starting point $s_{1}$ can be expressed as
\begin{gather}
\begin{align}
x_{1}&=\sqrt{2J\beta_{1}}\cos{\psi_{1}}\nonumber\\
x'_{1}&=-\sqrt{2J/\beta_{1}}(\alpha_{1}\cos{\psi_{1}+\sin{\psi_{1}}}),\
\end{align}
\end{gather}
where 
\begin{equation}
J=\frac{1}{2}\left(\gamma x^{2}+2\alpha xx'+\beta x'^{2}\right)
\end{equation}
is the betatron invariant. The longitudinal coordinate displacement relative to the ideal particle due to the betatron oscillation from $s_{1}$ to $s_{2}$ is then
\begin{align}\label{eq:tlcnodelta}
\Delta{z}&=R_{51}x_{1}+R_{52}x'_{1}\notag\\
&=\sqrt{2J\mathcal{H}_{1}}\sin(\psi_{1}-\chi_{1})-\sqrt{2J\mathcal{H}_{2}}\sin(\psi_{2}-\chi_{2}),
\end{align}
where to obtain the final concise result, ${D}$ and ${D}'$ have been expressed in terms of the chromatic $\mathcal{H}$-function and the chromatic phase $\chi$, defined as
\begin{gather}
\begin{align}
{D}&=\sqrt{\mathcal{H}\beta}\cos{\chi},\nonumber\\
{D}'&=-\sqrt{\mathcal{H}/\beta}\left(\alpha\cos{\chi}+\sin{\chi}\right),\nonumber\\
\mathcal{H}&=\gamma{D}^{2}+2\alpha{D}{D}'+\beta{D}'^{2}.
\end{align}
\end{gather}
If there is no dipole kick between point 1 and point 2, $\mathcal{H}$ stays constant and 
$
\chi_{2}-\chi_{1}=\psi_{2}-\psi_{1},
$
which means $\Delta z=0$. Actually, longitudinal coordinate differences of relativistic particles are purely from bending magnets in first-order approximation.

If the particle starts with a relative energy deviation of $\delta$, then
\begin{align}\label{eq:tlcwithdelta}
\Delta{z}=&R_{51}x_{1}+R_{52}x'_{1}+(F-R_{51}{D}_{1}-R_{52}{D}_{1}')\delta\nonumber\\
=&\sqrt{2J\mathcal{H}_{1}}\sin(\psi_{1}-\chi_{1})-\sqrt{2J\mathcal{H}_{2}}\sin(\psi_{2}-\chi_{2})+F\delta.
\end{align}
Note that in Eq.~(\ref{eq:tlcwithdelta}), the betatron invariant and phase should be calculated according to
\begin{gather}
\begin{align}
x_{\beta}&=x-{D}\delta=\sqrt{2J\beta}\cos{\psi}\nonumber\\
x_{\beta}'&=x'-{D}'\delta=-\sqrt{2J/\beta}(\alpha\cos{\psi+\sin{\psi}})\nonumber\\
J&=\frac{1}{2}\left(\gamma x_{\beta}^{2}+2\alpha x_{\beta}x_{\beta}'+\beta x_{\beta}'^{2}\right).
\end{align}
\end{gather}
For a periodic system, if we observe the particle at the same place $n$ periods later, then 
\begin{equation} \label{eq:tlcwithdeltaperiodic}
\Delta z=\sqrt{2J\mathcal{H}}\left[\sin(\psi-\chi)-\sin(\psi+2n\pi\nu_{x}-\chi)\right]-n\eta C_{0}\delta,
\end{equation}
where $\nu_{x}$ is the horizontal betatron tune. From Eq.~(\ref{eq:tlcnodelta}), the RMS value of the transient bunch lengthening of a longitudinal slice from $s_{1}$ to $s_{2}$ caused by this linear TLC can be calculated to be
\begin{equation}\label{eq:TLCBunchlengtheningTransient}
\sigma_{\Delta z}=\sqrt{\epsilon_{x}\left[\mathcal{H}_{1}+\mathcal{H}_{2}-2\sqrt{\mathcal{H}_{1}\mathcal{H}_{2}}\cos\left(\Delta\psi_{21}-\Delta\chi_{21}\right)\right]}.
\end{equation}
The RMS bunch lengthening of an electron beam longitudinal
slice after $n$ complete revolutions in a ring, due to betatron oscillation, is then
\begin{equation}\label{eq:TLCBunchlengtheningmRevolution}
\sigma_{\Delta z}=2\sqrt{\epsilon_{x}\mathcal{H}_{x}}\left|\sin(n\pi\nu_{x})\right|.
\end{equation}

\begin{figure}[tb]
	\centering
	\includegraphics[width=1\textwidth]{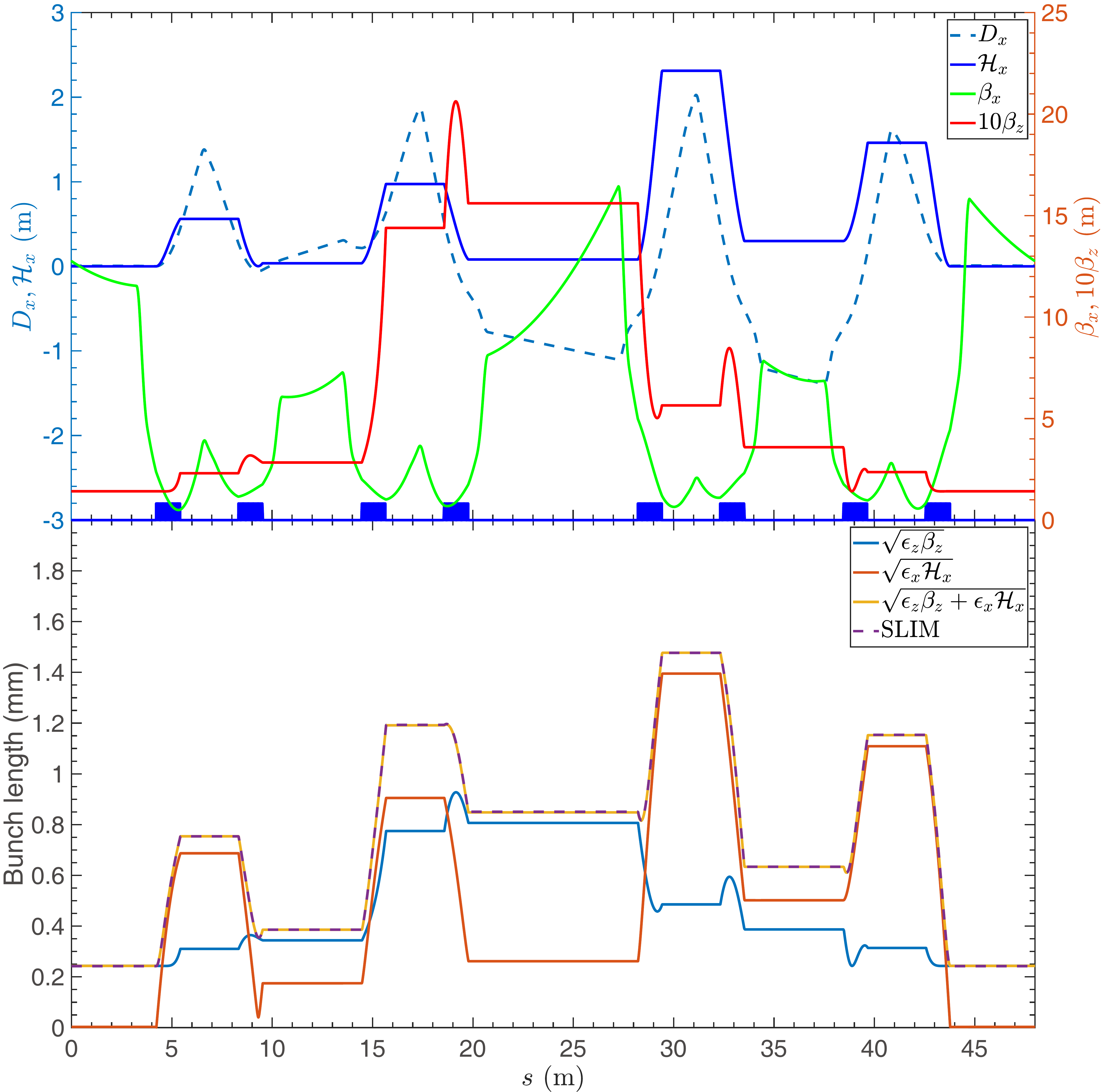}
	\caption{
		\label{fig:Chap3-TLCBunchLength} 
		Bunch length evolution around the ring with lattice optics the same as that in Fig.~\ref{fig:Chap2-LongitudinalBeta}. Other related parameters can be found in Tab.~\ref{tab:tab1}. In getting this plot, the beam energy is 1.2 GeV, and the RF voltage applied is 80 MV.  $\nu_{x}=3.17$, $\nu_{s}=0.016$, $\epsilon_{x}=842$ nm, $\epsilon_{z}=417$ nm. The dipoles are shown at the bottom as blue rectangles. Each dipole has a length of 1.2 m and bends the electron trajectory for an angle of $\pi/4$. 
	}
\end{figure} 

The above equations can be used to explain the dependence of the CSR repetition rate on the betatron tune in the bunch slicing experiment reported in Ref.~\cite{shimada2009transverse}  A similar approach will be applied in the following section to analyze microbunching preservation with beam deflection, for example in FEL multiplexing. These equations are also useful for evaluating the influence of the coupling effect in the SSMB proof-of-principle experiment \cite{deng2021experimental}, which is to be presented in Chap.~\ref{cha:pop}. Note that when Eqs.~(\ref{eq:tlcwithdelta}) and (\ref{eq:tlcwithdeltaperiodic}) are applied for beam analysis, it is assumed that the initial dispersion $(D_{1},D_{1}')$ of the lattice matches that of the beam. If this is not the case, then this mismatch should also be taken into account.

We can also obtain the equilibrium second moments in a storage ring by following the Courant-Snyder parametrization one step further. The result is the same with Eq.~(\ref{eq:2ndmoments}) obtained by SLIM. As can be seen from Eqs.~(\ref{eq:2ndmoments}) and (\ref{eq:2ndmomentsSeparate}), if there is only passive transverse-longitudinal coupling introduced by bending magnet, the transverse emittance always lengthen the bunch at places where $\mathcal{H}_{x}\neq0$,
\begin{equation}\label{eq:steadyBunchLengthening}
\sigma_{z}=\sqrt{\epsilon_{z}\beta_{z}+\epsilon_{x}\mathcal{H}_{x}}.
\end{equation}
Similarly that energy spread always broaden the beam width at places where $D\neq0$,
\begin{equation}
\sigma_{x}=\sqrt{\epsilon_{x}\beta_{x}+\epsilon_{z}\gamma_{z}D^{2}}=\sqrt{\epsilon_{x}\beta_{x}+\sigma_{\delta}^{2}D^{2}}.
\end{equation}

\begin{figure}[tb] 
	\centering 
	\includegraphics[width=0.5\columnwidth]{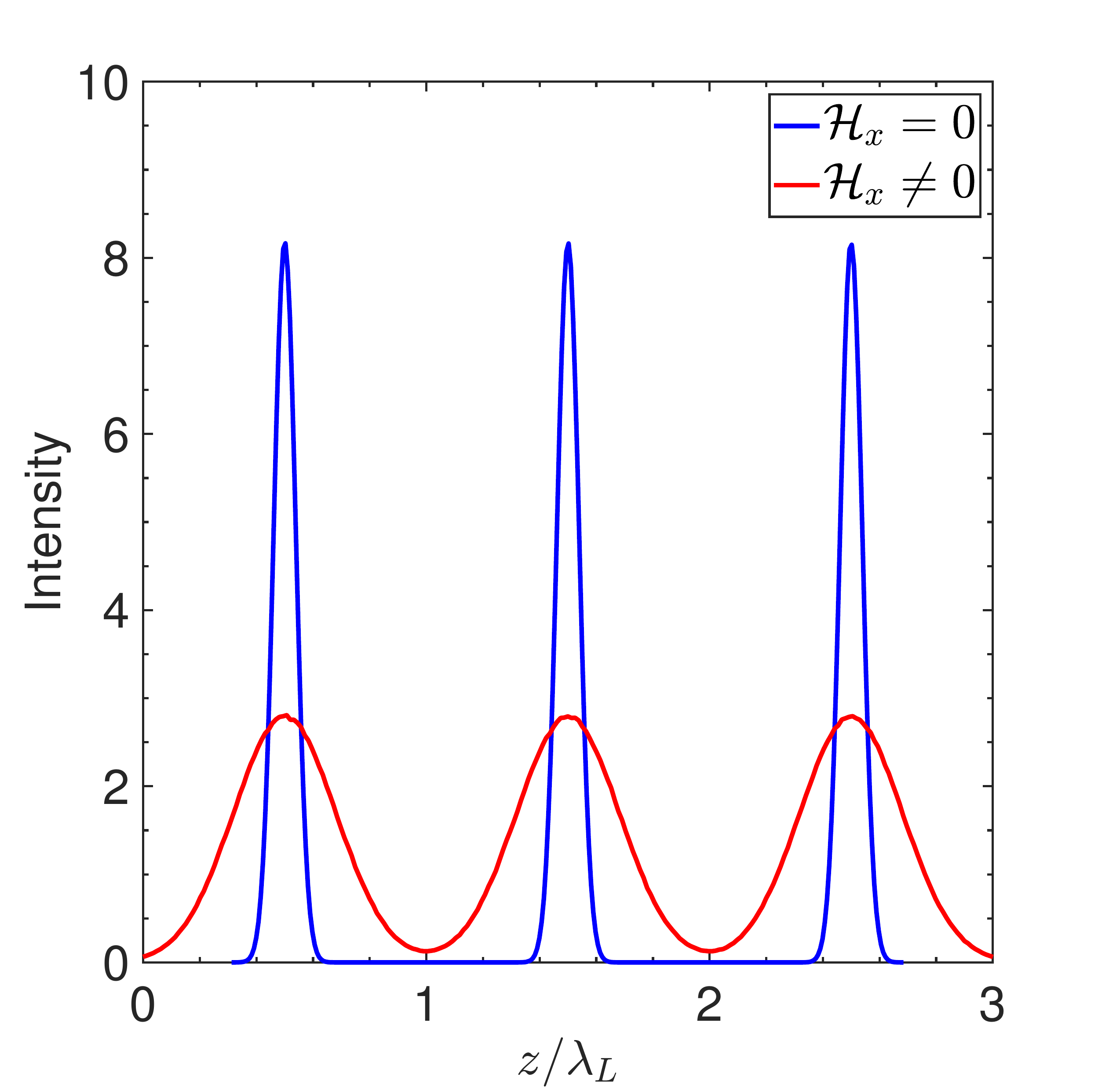}
	\caption{
		\label{fig:Chap3-CurrentOverlap} 
		Beam current distributions at places with different $\mathcal{H}_{x}$. Bunch length in an SSMB ring can easily be dominated by the transverse emittance in places where $\mathcal{H}_{x}\neq0$.
	}
\end{figure}

To give the readers a more concrete feeling about the bunch lengthening from this passive TLC, we present in Fig.~\ref{fig:Chap3-TLCBunchLength} some calculations based on the MLS lattice, whose optics is the same as that used in Sec.~\ref{sec:PartialAlpha} for particle tracking confirming the partial phase slippage effect. As can be seen, indeed that the coupling can contribute significantly, or even dominant the bunch length at places where $\mathcal{H}_{x}$ is large. This observation will especially be true in an SSMB ring, where the transverse size is much larger than the microbunch length. Therefore, the dispersion and dispersion angle should be controlled in precision at places where ultra-short bunch is desired, for example at the radiator. The bunch lengthening from the transverse emittance will make the current distribution less sharp and more like a coasting beam as places where $\mathcal{H}_{x}\neq0$, as shown in Fig.~\ref{fig:Chap3-CurrentOverlap}. Here we make a remark that this coupling effect may be helpful for suppressing unwanted CSR and may mitigate the IBS in SSMB or other applications, as extremely short bunches emerge only at dispersion-free locations. Besides, we will see in the following sections that transverse-longitudinal coupling can actually be actively applied for efficient bunch compression and high harmonic generation. 

We remind the readers again that the chromatic $\mathcal{H}$ function changes only inside dipoles. Between the dipoles, the betatron phase advance $\Delta\psi$ equals to the chromatic phase advance $\Delta\chi$. Physically, this means the contribution of transverse emittance to the bunch length does not change during drifting or experiencing quadrupole kicks, as these manipulations only affect the beam distribution in the transverse phase space.

\subsection{Microbunching Preservation with Deflection}

As mentioned just now, for a microbunch, its transverse size (tens of $\mu$m) is typically orders of magnitude larger than the desired microbunch length (sub-$\mu$m to nm), the longitudinal fine structures can be easily smeared by the coupling from transverse emittance. We know that this passive coupling originates from the bending magnets, but bends are necessary for example in the multiplexing of an X-ray free-electron laser (XFEL)~\cite{macarthur2018microbunch,margraf2019microbunch}, and of course also in a storage ring. The question is then can we preserve the micro-structure after the deflection by the bending magnet? The answer is yes. It can be seen from Eqs. (\ref{eq:tlcwithdelta}) and (\ref{eq:tlcwithdeltaperiodic}) and (\ref{eq:TLCBunchlengtheningTransient}) that the most straightforward way to accomplish this is to let $\mathcal{H}_{1}=\mathcal{H}_{2}=0$, which corresponds to $R_{51}=R_{52}=0$. An achromatic lattice should thus be envisioned. 

\begin{figure}[tb]
	\centering 
	\includegraphics[width=0.6\textwidth]{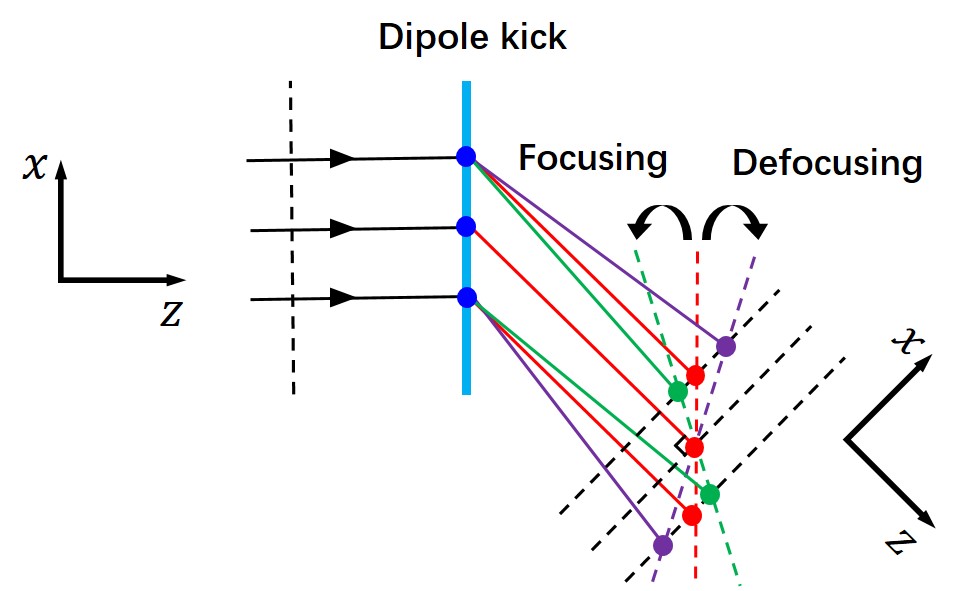}
	\caption{
		\label{fig:Chap3-MicrobunchRotation} 
		Microbunch orientation rotation of after traversing a dipole with focusing (green),  nonfocusing (red) and defocusing (purple) quadrupole component, respectively.  
	}
\end{figure} 

The matrix formalism can also be applied for a more quantitative analysis of the beam evolution with deflection. Here we use the microbunch rotation experiment conducted at the Linac Coherent Light Source (LCLS) \cite{macarthur2018microbunch,margraf2019microbunch} as an example for the analysis. In the experiment, a shifted quadrupole is used to deflect the microbunching for multiplexing. A shifted quadrupole is like a dipole plus a quadrupole.  The schematic of the microbunching rotation after the kick of a shifted quadrupole of the FODO lattice is illustrated in Fig. \ref{fig:Chap3-MicrobunchRotation}.  We can see that the orientation evolution of microbunching after a dipole kick depends on whether there is a focusing or defocusing quadrupole component in the transverse dimension.  


To simplify the discussion, we limit ourselves in the $x$-$z$ dimension and assume that the quadrupole shift is in the horizontal direction. The analysis is the same when the shift is in the vertical dimension. To simplify the calculation further, we treat the kick as a thin-lens one and consider the slice as shown in Fig.~\ref{fig:Chap3-MicrobunchRotation}. 
The longitudinal coordinate of a relativistic particle to first-order approximation stays constant after the shifted quadrupole kick,
\begin{equation}
z(s)=-\theta x_{0},
\end{equation} 
in which $\theta$ is the bending angle, $x_{0}$ the initial horizontal coordinate and $s$ the distance of drifting after the dipole kick. The horizontal coordinate evolves however according to 
\begin{equation}
x(s)=x_{0}-\frac{x_{0}}{2f}s
\end{equation} 
with $f$ the signed focal length of the quadrupole.  The second term is caused by the (de)focusing from the quadrupole middle point to the exit. This is based on the fact that the beam distribution in the transverse phase space is upright at the quadrupole center of a FODO lattice. The tilt angle $t_{xz}$ of the $x$-$z$ beam profile relative to the horizontal axis can be calculated to be  
\begin{equation}\label{eq:pureX}
t_{xz}(s)\approx\frac{z(s)}{x(s)}=-\frac{\theta}{1-\frac{s}{2f}}.
\end{equation}
From the formula above we know that
$
t_{xz}(0)=-\theta
$
and the angle $t_{xz}$ grows (decreases) with $s$ when $f$ is negative (positive), which means the microbunch rotates towards (away from) the new direction of travel when the quadrupole is defocused (focused) as shown in Fig.~\ref{fig:Chap3-MicrobunchRotation}.  This is the microbunching rotation analyzed in Ref.~\cite{macarthur2018microbunch}, and we can see that it is purely a geometric effect.
The above simplified analysis aims to present the main physical picture of this effect and  for simplicity we have ignored the influence from the initial beam divergence. Now we calculate the tilt angle $t_{xz}$ more rigorously from the beam envelope (second moments) matrix. We use ${\bf X}=(x,x',z,\delta)^{T}$ as the state vector. When the quadrupole length $l$ is short and $\theta$ is small, we can approximate the transfer matrix of the shifted quadrupole by letting $l\rightarrow0$ and keeping terms only up to first order in $\theta$ to be
\begin{equation}\label{eq:shiftedQuad}
{\bf R}_{\text{shifted quad}}=
\left(
\begin{matrix}
1 & 0 & 0 & 0\\
-\frac{1}{2f} & 1 & 0 & \theta\\
-\theta & 0 & 1 & 0\\
0 & 0 & 0 & 1
\end{matrix}
\right).
\end{equation}
The total transfer matrix of the shifted quadrupole followed by a drift space of length $s$ is then 
\begin{equation}\label{eq:microbunchrotationMatrix}
{\bf M}={\bf R}_{\text{drift}}{\bf R}_{\text{shifted quad}}=
\left(
\begin{matrix}
1-\frac{s}{2f} & s & 0 & s\theta\\
-\frac{1}{2f} & 1 & 0 & \theta \\
-\theta & 0 & 1 & 0 \\
0 & 0 & 0 & 1
\end{matrix}
\right).
\end{equation}
We further simplify the discussion by assuming that the microbunch is Gaussian and the beam distribution in both the transverse and longitudinal phase spaces are upright at the beginning and the second moments matrix is
\begin{equation}\label{eq:sigma0}
\Sigma_{0}=
\left(
\begin{matrix}
\epsilon_{x}\beta_{x0} & 0 & 0 & 0\\
0 & \epsilon_{x}/\beta_{x0} & 0 & 0 \\
0 & 0 & \epsilon_{z}\beta_{z0} & 0 \\
0 & 0 & 0 & \epsilon_{z}/\beta_{z0}
\end{matrix}
\right).
\end{equation}
with $\beta_{x0}$ and $\beta_{z0}$ the initial horizontal and longitudinal beta function, respectively.
The beam envelope matrix evolves according to
\begin{equation}\label{eq:envelope2}
\Sigma={\bf M}\Sigma_{0}{\bf M}^{T}.
\end{equation}
Actually it is also very straightforward to get the second moments under the assumption of $\alpha_{x0}=\alpha_{z0}=0$ by purely geometric method, but here we use Eq.~(\ref{eq:envelope2}) as it is more general. The tilt angle $t_{xz}$ of the $x$-$z$ beam profile relative to the horizontal axis can be found from 
\begin{equation}
\tan2 t_{xz}=\frac{2\langle xz\rangle}{\langle x^{2}\rangle-\langle z^{2}\rangle}.
\end{equation} 
For a microbunch and when $\theta\ll1$, we have $\langle z^{2}\rangle\ll\langle x^{2}\rangle$ and $\langle xz\rangle\ll\langle x^{2}\rangle$, and therefore $t_{xz}$ is a small quantity and we can ignore the contribution of $\epsilon_{z}$ on the second moments and arrive at
\begin{equation}\label{eq:angle}
t_{xz}(s)\approx\frac{\langle xz\rangle}{\langle x^{2}\rangle-\langle z^{2}\rangle}=\frac{\Sigma_{15}}{\Sigma_{11}-\Sigma_{55}}\approx\frac{-(1-\frac{s}{2f})\theta}{\left(1-\frac{s}{2f}\right)^{2}+s^{2}/\beta_{x0}^{2}-\theta^{2}}.
\end{equation} 
It can be seen from the formula above that
$
t_{xz}(0+)\approx-\theta
$
and
\begin{gather}\label{eq:change}
\frac{\partial{t_{xz}}}{\partial{s}}\bigg|_{s=0^{+}}=-\frac{\theta}{f}.
\end{gather}
It is clear that the microbunch rotation direction right after kick is determined by the sign of $f$. 

For a mathched beam to the FODO lattice, the horizontal beta function at the quadrupole center is
\begin{equation}\label{eq:FODOBeta}
\beta_{x0}=2|f|\sqrt{\frac{2f+L}{2f-L}},
\end{equation}
with the FODO half-period $L<2|f|$. Putting Eq.~(\ref{eq:FODOBeta}) in Eq.~(\ref{eq:angle}) we have 
\begin{equation}
t_{xz}(L)=-(1+\frac{L}{2f})\theta.
\end{equation}
The rotation angle of the microbunch orientation with a drifting length of $L$ is then
\begin{eqnarray}
\Delta t_{xz}=t_{xz}(L)-t_{xz}(0)=-\frac{L\theta}{2f}.
\end{eqnarray}

The same analysis can also be performed using the Courant-Snyder parameterization.  Ignoring the contribution of $\epsilon_{z}$ on second moments, and according to Eq.~(\ref{eq:2ndmomentsSeparate}), we have
\begin{gather}\label{eq:titltGeneral}
\begin{align}
\tan{2t_{xz}}&=\frac{2\langle xz\rangle}{\langle x^{2}\rangle-\langle z^{2}\rangle}=\frac{-2\epsilon_{x}\left(\alpha_{x}D_{x}+\beta_{x}D_{x}'\right)}{\epsilon_{x}\beta_{x}-\epsilon_{x}\mathcal{H}_{x}}=\frac{2\sqrt{\beta_{x}(s)\mathcal{H}_{x}(s)}}{\beta_{x}(s)-\mathcal{H}_{x}(s)}\sin{\chi(s)}.
\end{align}
\end{gather}
Now we can use Eq.~(\ref{eq:titltGeneral}) for the analysis of the above microbunch rotation. Before the shifted quadrupole kick, $\mathcal{H}_{x}=0$ m ($D_{x}=0$ m, $D_{x}'=0$), and $t_{xz}(0{-})=0$. After the shifted quadrupole kick, under the above thin-lens approximation and assume that the initial beam distribution in the transverse phase space is upright, for $s>0$ we have
\begin{equation}\label{eq:TwissEvolution}
\begin{aligned}
\mathcal{H}_{x}(s)&=\beta_{x0}\theta^{2},\\
\beta_{x}(s)&=\left(1-\frac{s}{2f}\right)^{2}\beta_{x0}+\frac{s^{2}}{\beta_{x0}},\\
\chi(s)&=\arccos{\frac{s}{\sqrt{\beta_{x0}\beta_{x}(s)}}}-\pi.
\end{aligned}
\end{equation}
Considering that $\theta\ll1$, we can ignore $\mathcal{H}_{x}$ in the denominator of Eq.~(\ref{eq:titltGeneral}), and arrive at
\begin{gather}\label{eq:newAngle}
\begin{align}
t_{xz}(s)\approx\frac{\langle xz\rangle}{\langle x^{2}\rangle}&=\sqrt{\frac{\mathcal{H}(s)}{\beta(s)}}\sin{\chi(s)}.
\end{align}
\end{gather}
Putting Eq.~(\ref{eq:TwissEvolution}) in Eq.~(\ref{eq:newAngle}), we arrive at the same formulas of Eq.~(\ref{eq:angle}).

Between the dipole kicks, $\mathcal{H}_{x}$ stays constant and $\Delta\chi=\Delta\psi=\int_{s_{1}}^{s_{2}}\frac{1}{\beta_{x}(s)}ds$,  the evolution of $t_{xz}$ depends then purely on the evolution of $\beta_{x}(s)$. If we denote the position right after the shifted quadrupole kick to be $s=0{+}$, then
\begin{equation}
t_{xz}(s)\approx\sqrt{\frac{\mathcal{H}(0{+})}{\beta(s)}}\sin\left[{\chi(0{+})+\int_{0^{+}}^{s}\frac{1}{\beta(s)}ds}\right]
\end{equation}
and
\begin{gather}\label{eq:changenew}
\begin{align}
\frac{\partial{t_{xz}}}{\partial{s}}
&=\sqrt{\mathcal{H}(0^{+})}\frac{\cos{\chi(s)}+\alpha(s)\sin{\chi(s)}}{\sqrt{\beta(s)^{3}}}.
\end{align}
\end{gather}
So the rotation direction of the microbunch orientation is determined by the sign of $\left[\cos{\chi(s)}+\alpha(s)\sin{\chi(s)}\right]$.

We have analyzed the microbunch rotation using matrix formalism directly and also the Courant-Snyder parameterization, here we remind the readers one more point. In the case of microbunch rotation after a shifted quadrupole kick analyzed above, an increase of $\beta_{x}(s)$ (defocusing) means an increase of $t_{xz}$, but we should keep in mind that this correspondence between the signs of $\frac{\partial{t_{xz}}}{\partial{s}}$ and $\beta'(s)$ depends on the chromatic phase $\chi(s)$ and there is not a fixed correspondence between the signs of $\frac{\partial{t_{xz}}}{\partial{s}}$ and $\beta'(s)$. For example, if there is a very long transport line with adequate quadrupole focusing following the dipole kick,  $t_{xz}$ will swing back and forth due to the advance of betatron phase $\psi(s)$ and thus also the chromatic phase $\chi(s)$. The physical reason is that $z$ of a particle stays constant during drifting, while $x$ of a particle alternates between positive and negative values due to betatron oscillation.

As mentioned, for the purpose of deflecting the beam while preserving the microbunch for FEL multiplexing, the most natural way is to correct $\mathcal{H}_{x}$ to zero before entering the radiator after the deflection. An achromatic lattice needs to be invoked for this purpose. For an XFEL and if we aim at mrad level deflection, this may result in a too large $R_{56}$ which degrades microbunching, so it would be better to invoke an isochronous achromat lattice.


\subsection{Active Coupling for Harmonic Generation and Bunch Compression}

The analysis  in previous sections may lead us to conclude that transverse-longitudinal coupling always lengthens the bunch and degrades the microbunching. This, however, is not true, as the above analysis is based on the assumption of a planar $x$-$y$ uncoupled lattice with only the passive coupling induced by the bending magnets. In addition to this passive coupling, an RF cavity (laser modulator) placed at a dispersive location, a transverse deflecting RF cavity, etc., are other sources of coupling that can be used for subtle manipulation of particle beam in 6D phase space. In fact, one can take advantage of the transverse-longitudinal coupling for efficient harmonic generation or bunch compression when the transverse emittance is small. The reason is that there is some flexibility in the projection of the three eigenemittances of a beam into different physical dimensions, although their values cannot be changed in a linear symplectic lattice~\cite{dragt1992general}. Here in this and the following sections we investigate the active application of transverse-longitudinal coupling (TLC) for harmonic generation and bunch compression by taking advantage of the fact that the vertical emittance of an electron beam in a planar $x$-$y$ uncoupled ring is rather small.  

Currently, the laser manipulations of electrons are widely used in the accelerator-based light sources \cite{hemsing2014beam}, especially in the fully coherent free-electron lasers (FELs) \cite{madey1971stimulated,kroll1978stimulated,kondratenko1980generating,bonifacio1984collective,emma2010first,huang2007review,pellegrini2016physics} aiming at short wavelength radiation, e.g., VUV and soft X-ray.  Compared to lasing from Self-Amplified Spontaneous Emission (SASE) \cite{kondratenko1980generating,bonifacio1984collective}, laser-induced harmonic generation improves the longitudinal coherence and stability of the FEL radiation and shortens the radiator undulator length.  The well-known examples of such harmonic generation schemes are high-gain harmonic generation (HGHG) \cite{yu1991generation,yu2000high} and echo-enabled harmonic generation (EEHG) \cite{stupakov2009using,xiang2009echo,zhao2012first,hemsing2016echo,ribivc2019coherent}, where the beam dynamics focus on the longitudinal dimension. One of the key issues in harmonic generation is the efficient frequency up-conversion when using the traditional seed laser.  In HGHG, the harmonic number reachable is approximately proportional to the laser-induced energy modulation strength. Therefore, a high harmonic number requires a large energy modulation, which may hurt the following lasing process. EEHG can reach a much higher harmonic number with a given energy modulation strength, although its implementation is more complex than HGHG. 

Besides the beam manipulations in longitudinal dimension alone, TLC schemes can also be applied for high harmonic generation when the transverse emittance is small. To our knowledge, the first approach of applying such TLC schemes for microbunching is the angular-modulation scheme proposed in Ref.~\cite{xiang2010generating}. 
After that, another TLC scheme based on emittance-exchange (EEX) has also been proposed for high harmonic generation \cite{jiang2011emittance}. Besides the above schemes which invoke transverse deflecting RF cavities or transverse electromagnetic (TEM)01 mode lasers, there are also TLC schemes based on the application of normal RFs or TEM00 mode lasers. The representative scheme is phase-merging enhanced harmonic generation (PEHG) \cite{deng2013using,feng2014phase}, which is then followed by angular dispersion-induced microbunching (ADM) \cite{feng2017storage}. There are also other novel ideas being proposed \cite{feng2014three,wang2020transverse}.

Viewing the harmonic generation schemes from another perspective, they can be applied for bunch length compression if the duration of the whole bunch is shorter than the wavelength of the modulation system. These schemes can therefore be applied in linacs or storage rings to generate short bunches, which are highly desired for coherent radiation generation. Here we investigate the application of these bunch compression schemes in SSMB. 

As mentioned in Chap.~\ref{cha:intro}, the goal of the Tsinghua SSMB task force is to develop an EUV SSMB storage ring.  To generate temporally coherent 13.5 nm EUV radiation, the microbunch length at the radiator should be as short as $3$ nm. And as analyzed in Sec.~\ref{sec:miniLonemit}, it is a challenging task to realize 3 nm directly using a longitudinal weak focusing quasi-isochronous lattice alone, as the global phase slippage required is too small. To relax the burden on the storage ring lattice, the bunch compression schemes are therefore invoked. For example, in the envisioned EUV SSMB storage ring, the microbunch length is a couple of 10 nm in the main sections of the ring, and is compressed to around $3$ nm only at the radiator where 13.5 nm coherent radiation is generated. After the radiator, the microbunch then recovers its longer length. We remind the readers that an SSMB storage ring is different from a single-pass device, and we need to make sure such compression can repeat turn after turn, i.e., the ultra-short bunch length at the radiator is the eigen state of this multi-pass device. 

The first natural idea for such microbunch compression is to apply the HGHG scheme, or more accurately the longitudinal strong focusing regime for a multi-pass device like a storage ring as analyzed in Sec.~\ref{sec:multiRFs}, which however requires a high modulation laser power due to the large compression ratio of the bunch length, and at the same time results in a large electron energy spread at the radiator and can affect the radiation considering the $R_{56}$ of the radiator itself as analyzed in Sec.~\ref{eq:secLM}. By taking advantage of the fact that the vertical emittance is rather small in a planar $x$-$y$ uncoupled storage ring, TLC schemes like PEHG and ADM are thus in favored to lower the requirement on modulation laser power and reduce the final energy spared at the radiator. This is the motivation for us to investigate the applications of TLC schemes in SSMB. 

The contents of this section and the following are organized as follows. First, we give a brief review of the PEHG and ADM schemes to appreciate their physics and link. Then we generalize the analysis and prove a theorem concerning the application of such TLC schemes for harmonic generation or bunch compression. This theorem relates the energy modulation or chirp strength with the chromatic $\mathcal{H}$ functions at the modulator and radiator, respectively. As the $\mathcal{H}$ function quantifies the bunch lengthening from the transverse emittance, the theorem therefore dedicates the relation between the energy chirp strength and the bunch length at the modulator and radiator, respectively. For a pre-microbunched beam, like that in some SSMB scenarios, this bunch lengthening at the modulator can degrade the bunching factor at the radiator since the sinusoidal modulation waveform is nonlinear. This bunching factor degradation is analyzed and the addition of laser harmonics to mitigate such degradation is also discussed. After that, a corresponding theorem of applying TLC with a transverse deflecting RF or TEM01 mode laser is also presented, where the energy chirp strength is replaced by the angular chirp strength and the $\mathcal{H}$ function at the modulator is replaced by the $\beta$ function there.  Further, we have analyzed the contribution of modulator to the vertical emittance by means of quantum excitation,  to obtain a self-consistent evaluation of the required modulation laser power. The two theorems and related analysis provide the theoretical basis for the application of TLC in SSMB to lower the requirement on the modulation laser power, by taking advantage of the fact that the vertical emittance in a planar ring is rather small. Based on the analysis, we have presented an example parameters set for the envisioned SSMB storage ring to generate high-power EUV radiation.   The relation between our TLC analysis and the transverse-longitudinal emittance exchange (EEX) is also briefly discussed.

As a preparation for the theorem to be presented in the next section, here we give a brief review of the mechanism of PEHG and ADM. In this section, we use $y$-$z$ coupling for the analysis and state vector $(y,y',z,\delta)^{T}$ is used. The conclusion is the same for $x$-$z$ coupling. The subscripts of $_{3,4,5,6}$ of the matrix elements are used for the $y$-$z$ dimensions for consistency with literature.

\subsubsection{PEHG}

\begin{figure}[tb]
	\centering 
	\includegraphics[width=0.7\textwidth]{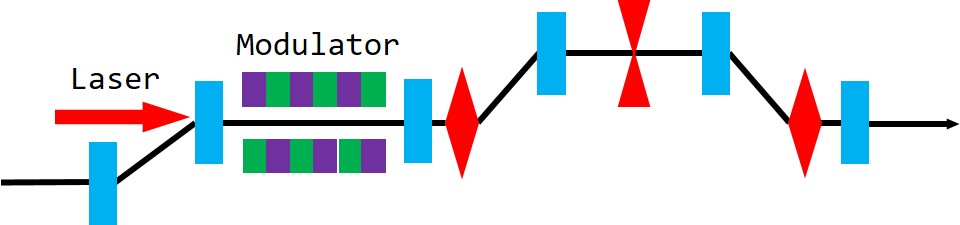}
	\caption{
		\label{fig:Chap3-PEHG} 
		Schematic layout of a PEHG unit.}
\end{figure}

PEHG was proposed in Ref.~\cite{deng2013using,feng2014phase} for an efficient frequency up-conversion in FEL seeding.  As shown in Fig.~\ref{fig:Chap3-PEHG}, a PEHG unit can be divided into four key functional components: (i) a dispersion generation ($R_{36}=d$) device, for example, a dogleg, to correlate $y$ with $\delta$; (ii) a laser modulator to generate an energy chirp ($R_{65}=h$) to correlate $\delta$ with $z$; (iii) a dispersive section ($R_{56}$) for longitudinal phase-space shearing to correlate $z$ with $\delta$ to form microbunching; and (iv) a phase-merging unit ($R_{53}=D'$) to correlate $z$ with $y$. The trick for phase merging is to achieve cancellation between the longitudinal displacement contribution from $\delta$ (through $R_{56}$) and that from $y$ introduced in the first step (through $dD'$), thus causing the longitudinal phases of different particles to merge together. This process can be understood with the help of phase space evolution as shown Fig~\ref{fig:Chap3-Summary}.

To better illustrate the mechanism, we have divided the PEHG process into four functions. Note that there is some flexibility in their order. Additionally, these functions can be realized one by one, or with two or three of them simultaneously. In the original publication on PEHG  \cite{deng2013using}, a transverse-gradient undulator was applied along with the modulation laser to realize energy chirping and phase merging at the same time. There are PEHG variants proposed for example in Ref.~\cite{deng2020single,zhao2017phase} where a normal undulator is used and phase merging is realized elsewhere. 

Now we conduct a matrix analysis to present the physical principle of PEHG as explained above. By using matrix, it means that we have linearized the {\it sine} waveform around the zero-crossing phase. Therefore, here we actually treat PEHG as a bunch compression scheme. The thin-lens transfer matrices of the four functions mentioned above are
\begin{equation}
	\begin{aligned}
	&{\bf M}_\text{1}=\left(\begin{matrix}
	1&0&0&d\\
	0&1&0&0\\
	0&d&1&0\\
	0&0&0&1\\
	\end{matrix}\right),\
	{\bf M}_\text{2}=\left(\begin{matrix}
	1&0&0&0\\
	0&1&0&0\\
	0&0&1&0\\
	0&0&h&1\\
	\end{matrix}\right),\\
	&{\bf M}_\text{3}=\left(\begin{matrix}
	1&0&0&0\\
	0&1&0&0\\
	0&0&1&R_{56}\\
	0&0&0&1\\
	\end{matrix}\right),\
	{\bf M}_\text{4}=\left(\begin{matrix}
	1&0&0&0\\
	0&1&0&-D'\\
	D'&0&1&0\\
	0&0&0&1\\
	\end{matrix}\right).
	\end{aligned}
	\end{equation}	
Note that here we focus on the key matrix elements of each function, with only fundamental physical principle of symplecticity taken into account. With such simplification, the total transfer matrix of a PEHG unit is
\begin{equation}
	{\bf T}={\bf M}_\text{4}{\bf M}_\text{3}{\bf M}_\text{2}{\bf M}_\text{1}=\left(
	\begin{matrix}
	1 & 0 & 0 & d \\
	0 & 1- dh D' & -h D' & -D' \\
	D' &  d\left(h R_{56}+1\right) & h R_{56}+1 & d D'+R_{56} \\
	0 & h D' & h & 1 \\
	\end{matrix}
	\right).
	\end{equation}
To achieve optimal bunch compression, we need
$h R_{56}+1=0$ and $ d D'+R_{56}=0$,
then 
\begin{equation}
	\sigma_{zf}=|D'|\sigma_{yi}=\bigg|\frac{R_{56}}{d}\bigg|\sigma_{yi},
	\end{equation}
where the subscripts $_i$ and $_f$ represent `initial' and `final', respectively. Note that $\sigma_{z}$ here means the microbunch length, and is different from the usual bunch length in an FEL which is much longer than the modulation laser wavelength.  Therefore, the final bunch length in PEHG depends solely on the initial vertical beam size. As a comparison, in HGHG, $\sigma_{zf}=|R_{56}|\sigma_{\delta i}$. In correspondence to that of HGHG, the effective energy spread of PEHG is $\sigma_{\delta,\text{eff}}=\frac{\sigma_{yi}}{|d|}$. So PEHG is favorable for harmonic generation or bunch compression when $\frac{\sigma_{yi}}{|d|}<\sigma_{\delta i}$.  

To quantify the bunching behavior of the particles, a parameter named bunching factor is defined as that in Eq.~(\ref{eq:BF}).
The coherent radiation power at wavelength $\lambda=2\pi/k$ is proportional to $|b(k)|^{2}$. For a Gaussian beam with an RMS bunch length of $\sigma_{z}$,
\begin{equation}
	b(k)=\text{exp}\left[-{\left(k\sigma_{z}\right)^{2}}/{2}\right].
	\end{equation}
Considering that the modulation waveform is actually a nonlinear {\it sine}, the final line density distribution of the electron beam is non-Gaussian. For an electron beam much longer than the laser wavelength, i.e., a coasting beam, the final bunching factor at the $n^{\text{th}}$ laser harmonic, i.e., $k=nk_{\text{L}}$ with $k_{\text{L}}=2\pi/\lambda_{\text{L}}$ being the wavenumber of the modulation laser, in PEHG is therefore \cite{deng2013using,feng2014phase}
\begin{equation}\label{eq:PEHGBF}
	b_{n,\ \text{coasting}}=|J_{n}(n)|\text{exp}\left[-{\left(nk_{\text{L}}D'\sigma_{yi}\right)^{2}}/{2}\right],
	\end{equation}
with $J_{n}$ being the first kind  Bessel function of the $n$-th order. Note that considering the nonlinear {\it sine} modulation waveform, the optimal microbunching condition for a specific harmonic is slightly different from our simplified linear analysis, and $J_{n}(n)$ in Eq.~(\ref{eq:PEHGBF}) should be replaced by $J_{n}(nR_{56}h)$. In the following discussions, we will use the simplified optimal bunch compression conditions analyzed here, as the main physics is the same.


\subsubsection{ADM}

\begin{figure}[tb]
	\centering 
	\includegraphics[width=0.5\textwidth]{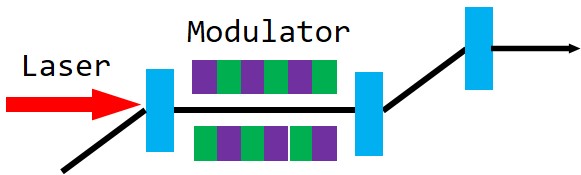}
	\caption{
		\label{fig:Chap3-ADM} 
		Schematic layout of an ADM unit.}
\end{figure}

ADM was first proposed in Ref.~\cite{feng2017storage} for high coherent harmonic generation in a storage ring FEL, by making use of the fact that the vertical emittance in a planar $x$-$y$ uncoupled ring is rather small. Different from PEHG, which uses the dispersion at the modulator, ADM invokes the dispersion angle, based on which the name angular dispersion-induced microbunching is coined. As shown in Fig.~\ref{fig:Chap3-ADM}, an ADM unit consists of (i) a dipole to generate the dispersion angle ($R_{46}=-d'$); (ii) a laser modulator to energy modulate the electron beam ($R_{65}=h$); (iii) a dogleg to generate micorbunching ($R_{56}$ and $R_{54}=D$). Here we also conduct a matrix analysis of ADM like that for PEHG. The thin-lens transfer matrices of the three parts of an ADM unit are
\begin{equation}
	\begin{aligned}
	&{\bf M}_\text{1}=\left(
	\begin{array}{cccc}
	1 & 0 & 0 & 0 \\
	0 & 1 & 0 & -d'  \\
	d'  & 0 & 1 & 0 \\
	0 & 0 & 0 & 1 \\
	\end{array}
	\right),\
	{\bf M}_\text{2}=\left(\begin{matrix}
	1&0&0&0\\
	0&1&0&0\\
	0&0&1&0\\
	0&0&h&1\\
	\end{matrix}\right),\ {\bf M}_\text{3}=\left(
	\begin{array}{cccc}
	1 & 0 & 0 & D \\
	0 & 1 & 0 & 0 \\
	0 & D & 1 & R_{56} \\
	0 & 0 & 0 & 1 \\
	\end{array}
	\right).
	\end{aligned}
	\end{equation}
Then the total transfer matrix is
\begin{equation}
	\begin{aligned}
	&{\bf T}={\bf M}_\text{3}{\bf M}_\text{2}{\bf M}_\text{1}=\left(
	\begin{array}{cccc}
	d'h  D+1 & 0 & h D & D \\
	0 & 1 & 0 & -d'  \\
	d'\left(h R_{56}+1\right) & D & h R_{56}+1 & -d'  D+R_{56} \\
	d'h   & 0 & h & 1 \\
	\end{array}
	\right).
	\end{aligned}
	\end{equation}
To achieve optimal bunch compression, we need
$
h R_{56}+1=0,\ -d'  D+R_{56}=0,
$
then 
\begin{equation}
	\sigma_{zf}=|D|\sigma_{y'i}=\bigg|\frac{R_{56}}{d'}\bigg|\sigma_{y'i}.
	\end{equation}
So the final bunch length depends solely on the initial vertical beam divergence. In correspondence to that of HGHG, the effective energy spread of ADM is  $\sigma_{\delta,\text{eff}}=\frac{\sigma_{y'i}}{|d'|}$. Therefore, ADM is favorable for harmonic generation or bunch compression when $\frac{\sigma_{y'i}}{|d'|}<\sigma_{\delta i}$. Considering the nonlinear {\it sine} modulation waveform, and for a coasting beam, the final bunching factor at the $n^{\text{th}}$ laser harmonic in ADM is \cite{feng2017storage}
\begin{equation}\label{eq:ADMBF}
	b_{n,\  \text{coasting}}=|J_{n}(n)|\text{exp}\left[-{\left(nk_{\text{L}}D\sigma_{y'i}\right)^{2}}/{2}\right].
\end{equation}

We appreciate the link in the analysis of PEHG and ADM before introducing the theorem in next section. Although both applying TLC, PEHG requires dispersion $d$ while ADM requires dispersion angle $d'$ at the modulator. Depending on the different specific situations, one might be in favored compared to the other. For example in a storage ring, it seems to be easier to generate a small $\sigma_{y'}$ with a large $d_{y}'$ than a small $\sigma_{y}$ with a large $d_{y}$, so ADM might be in favored.

\subsection{Theorem on Normal RF or TEM00 Laser for Coupling}
\subsubsection{Theorem and the Proof}
In the previous section, we have reviewed two examples, i.e., PEHG and ADM, of applying TLC for harmonic generation or bunch length compression. As the dispersion $d$ in PEHG, or dispersion angle $d'$ in ADM, is an adjustable parameter, the effective energy spread can be lowered, thus also the required energy modulation strength to realize the same bunch compression ratio or harmonic generation number compared to that in HGHG. The readers may wonder where does the power or flexibility of such TLC schemes come from? Is there any sacrifice to pay for such flexibility? In this section we generalize the analysis and prove a theorem about applying TLC with a normal RF or TEM00 mode laser modulator for harmonic generation or bunch compression. The theorem originates from the sympleticity of Hamiltonian dynamics. According to the theorem, a reduction of energy chirp strength means an increase of bunch lengthening from the transverse emittance at the place of energy chirping, i.e, the RF cavity or laser modulator. For a pre-microbunched beam, this bunch lengthening at the modulator will degrade the bunching factor at the radiator if we account for the fact that the modulation waveform is actually a nonlinear {\it sine}. Besides, this bunch lengthening at the modulator will have crucial impacts on the nonlinear dynamics, for example the transverse and longitudinal dynamic aperture, in the application of TLC in SSMB. 

\begin{figure}[tb]
	\centering 
	\includegraphics[width=1\textwidth]{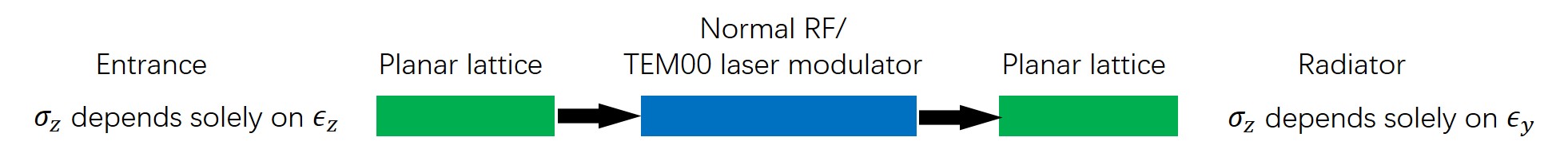}
	\caption{
		\label{fig:Chap3-TLCSetup} 
		Schematic layout of applying $y$-$z$ coupling for bunch compression, using a normal RF or TEM00 mode laser modulator.}
\end{figure}

Before presenting the theorem, let us first define the problem we are trying to solve.  Suppose the beam at the entrance is $y$-$z$ decoupled and Gaussian in both planes, i.e., its second moments matrix is
\begin{equation}
	\Sigma_{i}=\left(\begin{matrix}
	\epsilon_{y}\beta_{y}&-\epsilon_{y}\alpha_{y}&0&0\\
	-\epsilon_{y}\alpha_{y}&\epsilon_{y}{\gamma_{y}}&0&0\\
	0&0&\epsilon_{z}\beta_{z}&-\epsilon_{z}\alpha_{z}\\
	0&0&-\epsilon_{z}\alpha_{z}&\epsilon_{z}{\gamma_{z}}\\
	\end{matrix}\right),
	\end{equation}
where $\alpha$, $\beta$ and $\gamma$ are the Courant-Snyder functions \cite{courant1958theory} at the entrance and $\epsilon_{y,z}$ are the eigen emittances of the beam which are invariants in a linear symplectic lattice \cite{dragt1992general}. For the application of TLC for bunch compression, it means that the final bunch length at the radiator depends only on the vertical emittance $\epsilon_{y}$ and not on the longitudinal one $\epsilon_{z}$.  The magnet lattices are all planar and $x$-$y$ decoupled. The schematic layout of the bunch compression section is shown in Fig.~\ref{fig:Chap3-TLCSetup} .

With the above definition of the problem, the theorem then states
\begin{equation}\label{eq:theorem}
	h^2(\text{Mod})\mathcal{H}_{y}(\text{Mod})\mathcal{H}_{y}(\text{Rad})\geq1,
	\end{equation}
where $h(\text{Mod})$ is the transfer matrix term $R_{65}$ of the RF or laser modulator which quantifies the energy chirp strength, and  $\mathcal{H}_{y}$ is a parameter quantifying the bunch length contribution from the vertical emittance $\epsilon_{y}$. The more specific definition of $\mathcal{H}_{y}$ will be clear soon. The `Mod' and `Rad' in the brackets represent the places of modulator and radiator, respectively. 
The theorem in Eq.~(\ref{eq:theorem}) can also be expressed as
\begin{equation}
	|h(\text{Mod})|\geq\frac{\sqrt{\epsilon_{y}/\mathcal{H}_{y}(\text{Mod})}}{\sigma_{z}(\text{Rad})}=\frac{\epsilon_{y}}{\sigma_{zy}(\text{Mod})\sigma_{z}(\text{Rad})}.
	\end{equation}
Note that in the above formula, the bunch length at the modulator $\sigma_{zy}(\text{Mod})$ contains only that contributed from the vertical emittance $\epsilon_{y}$. So given a fixed $\epsilon_{y}$ and $\sigma_{z}(\text{Rad})$, a smaller $h(\text{Mod})$, i.e., a smaller RF gradient or modulation laser power ($P_{\text{laser}}\propto|h(\text{Mod})|^2$), means a larger $\mathcal{H}_{y}(\text{Mod})$, thus a longer $\sigma_{zy}(\text{Mod})$, is needed.  As $|h(\text{Mod})|\sigma_{z}(\text{Mod})$ quantifies the energy spread introduced by the modulator, we thus also have 
\begin{equation}\label{eq:ES}
	\sigma_{z}(\text{Rad})\sigma_{\delta}(\text{Rad})\geq\epsilon_{y}.
	\end{equation}
Note that Eq.~(\ref{eq:ES}) applies when the modulation waveform is linearized, or the bunch length at the modulator is much shorter than the modulation wavelength. For a coasting beam, the energy spread induced by the nonlinear {\it sine} energy modulation is solely proportional to $|h(\text{Mod})|$ and is independent of the bunch length.

Now we present the proof of the theorem, which invokes the symplectic condition and Cauchy-Schwarz inequality. We divide the schematic layout as shown in Fig.~\ref{fig:Chap3-TLCSetup} into three parts and write the transfer matrices of the magnet lattices in their general thick-lens forms
\begin{equation}
	\begin{aligned}
	&{\bf M}_{1}=\left(\begin{matrix}
	r_{33}&r_{34}&0&d\\
	r_{43}&r_{44}&0&d'\\
	r_{53}&r_{54}&1&r_{56}\\
	0&0&0&1\\
	\end{matrix}\right),\ {\bf M}_{2}=\left(\begin{matrix}
	1&0&0&0\\
	0&1&0&0\\
	0&0&1&0\\
	0&0&h&1\\
	\end{matrix}\right),\ {\bf M}_{3}=\left(\begin{matrix}
	R_{33}&R_{34}&0&D\\
	R_{43}&R_{44}&0&D'\\
	R_{53}&R_{54}&1&R_{56}\\
	0&0&0&1\\
	\end{matrix}\right),\\
	&r_{53}=r_{43}d-r_{33}d',\ r_{54}=-r_{34}d'+r_{44}d,\ r_{33}r_{44}-r_{34}r_{43}=1,\\
	&R_{53}=R_{43}D-R_{33}D',\ R_{54}=-R_{34}D'+R_{44}D,\ R_{33}R_{44}-R_{34}R_{43}=1.
	\end{aligned}
	\end{equation}	
with ${\bf M}_{1}$ representing ``from entrance to RF or laser modulator", ${\bf M}_{2}$ representing ``RF or laser modulator kick" and ${\bf M}_{3}$ representing ``from RF or laser modulator to radiator". 
The transfer matrix from the entrance to the radiator is then
\begin{equation}
	{\bf T}={\bf M}_{3}{\bf M}_{2}{\bf M}_{1}.
	\end{equation}	

From the problem definition, for $\sigma_{z}(\text{Rad})$ to be independent of $\epsilon_{z}$, 
we need
\begin{equation}\label{eq:TLCCondition}
	\begin{aligned}
	T_{55}&= h R_{56}+1=0,\\
	T_{56}&= dR_{53} +d'R_{54}+r_{56}(h R_{56}+1)+R_{56}=0.
	\end{aligned}
	\end{equation}
PEHG and ADM are specific examples of the above general relations. Other possible lattice realizations can also be envisioned. 
Under the conditions of Eq.~(\ref{eq:TLCCondition}), we have
\begin{equation}\label{eq:TransportMatrix1}
	\begin{aligned}
	&{\bf T}=\left(
	\begin{matrix}
	{\bf A}&{\bf B}\\
	{\bf C}&{\bf E}
	\end{matrix}
	\right),
	\end{aligned}
	\end{equation}
with ${\bf A}$, ${\bf B}$, ${\bf C}$, ${\bf E}$ being the $2\times2$ sub-matrices of ${\bf T}$ given by 	
\begin{equation}\label{eq:TransportMatrixSub}
	\begin{aligned}
	&{\bf A}=\left(\begin{matrix}
	r_{33} R_{33}+r_{43} R_{34} + r_{53}h   D& r_{34} R_{33}+r_{44} R_{34} + r_{54}h  D\\
	r_{33} R_{43}+r_{43} R_{44} + r_{53}h  D'& r_{34} R_{43}+r_{44} R_{44} + r_{54}h D'\\
	\end{matrix}\right),\\
	&{\bf B}=\left(\begin{matrix}
	hD&  dR_{33}+ d'R_{34} + \left(r_{56}h +1\right)D\\
	h D'& dR_{43}+ d'R_{44} + \left(r_{56}h +1\right) D'\\
	\end{matrix}\right),\\
	&{\bf C}=\left(\begin{matrix}
	r_{33} R_{53}+r_{43} R_{54}&r_{34} R_{53}+r_{44} R_{54}\\
	r_{53}h &r_{54}h\\
	\end{matrix}\right),\\
	&{\bf E}=\left(\begin{matrix}
	0&0\\
	h&r_{56}h +1\\
	\end{matrix}\right).
	\end{aligned}
	\end{equation}
The bunch length squared at the modulator and the radiator are
\begin{equation}\label{eq:BL}
	\begin{aligned}
	\sigma_{z}^{2}(\text{Mod})
	&=\epsilon_{z}\left(\beta_{z}-2\alpha_{z}r_{56}+\gamma_{z}r_{56}^2\right)+\epsilon_{y}\frac{\left(\beta_{y}r_{53}-\alpha_{y}r_{54}\right)^2+r_{54}^2}{\beta_{y}}\\
	&=\epsilon_{z}\beta_{z}(\text{Mod})+\epsilon_{y}\mathcal{H}_{y}(\text{Mod}),\\
	\sigma_{z}^{2}(\text{Rad})
	&=\epsilon_{y}\frac{\left(\beta_{y}T_{53}-\alpha_{y}T_{54}\right)^2+T_{54}^2}{\beta_{y}}
	=\epsilon_{y}\mathcal{H}_{y}(\text{Rad}).
	\end{aligned}
	\end{equation}
According to Cauchy-Schwarz inequality, we have
\begin{equation}\label{eq:proof}
	\begin{aligned}
	h^2(\text{Mod})\mathcal{H}_{y}(\text{Mod})\mathcal{H}_{y}(\text{Rad})&=h^2\frac{\left[\left(\beta_{y}r_{53}-\alpha_{y}r_{54}\right)^2+r_{54}^2\right]}{\beta_{y}}\frac{\left[\left(\beta_{y}T_{53}-\alpha_{y}T_{54}\right)^2+T_{54}^2\right]}{\beta_{y}}\\
	&\geq
	\frac{h^2}{\beta_{y}^{2}}\left[-\left(\beta_{y}r_{53}-\alpha_{y}r_{54}\right)T_{54}+r_{54}\left(\beta_{y}T_{53}-\alpha_{y}T_{54}\right)\right]^{2}\\
	&=\left(T_{53}r_{54}h-T_{54}r_{53}h\right)^{2}=\left(T_{53}T_{64}-T_{54}T_{63}\right)^{2}.
	\end{aligned}
	\end{equation}
The equality holds when
$
\frac{-\left(\beta_{y}r_{53}-\alpha_{y}r_{54}\right)}{T_{54}}=\frac{r_{54}}{\left(\beta_{y}T_{53}-\alpha_{y}T_{54}\right)}.
$
So if $|T_{53}T_{64}-T_{54}T_{63}|\geq1$, then the theorem is proven. 

The symplecticity of ${\bf T}$ requires that ${\bf T}{\bf S}{\bf T}^{T}={\bf S}$, where ${\bf S}=\left(
\begin{matrix}
{\bf J}&0\\
0&{\bf J}
\end{matrix}
\right)$ and $
{\bf J}=\left(
\begin{matrix}
0&1\\
-1&0
\end{matrix}
\right),$ so we have
\begin{equation}
	\left(
	\begin{matrix}
	{\bf A}{\bf J}{\bf A}^{T}+{\bf B}{\bf J}{\bf B}^{T}&{\bf A}{\bf J}{\bf C}^{T}+{\bf B}{\bf J}{\bf E}^{T}\\
	{\bf C}{\bf J}{\bf A}^{T}+{\bf E}{\bf J}{\bf B}^{T}&{\bf C}{\bf J}{\bf C}^{T}+{\bf E}{\bf J}{\bf E}^{T}
	\end{matrix}
	\right)
	=
	\left(
	\begin{matrix}
	{\bf J}&{\bf 0}\\
	{\bf 0}&{\bf J}
	\end{matrix}
	\right).
	\end{equation}
As shown in Eq.~(\ref{eq:TransportMatrixSub}), ${\bf E}=\left(
\begin{matrix}
0&0\\
h&r_{56}h+1
\end{matrix}
\right),$ then ${\bf E}{\bf J}{\bf E}^{T}=\left(
\begin{matrix}
0&0\\
0&0
\end{matrix}
\right)$. Therefore,
\begin{equation}
	{\bf C}{\bf J}{\bf C}^{T}={\bf J},
	\end{equation}
which means ${\bf C}$ is also a symplectic matrix. So we have
\begin{equation}
	T_{53}T_{64}-T_{54}T_{63}=\text{det}({\bf C})=1,
	\end{equation} 
where $\text{det}({\bf C})$ means the determinant of ${\bf C}$. The theorem is thus proven.

\subsubsection{Impact on Bunching Factor}

As mentioned, the bunch lengthening at the modulator will affect the bunching factor at the radiator for a pre-microbunched beam, considering the nonlinear nature of the modulation waveform. Here we use ADM as an example to derive the bunching factor at the radiator. PEHG has a similar result as we have proven in the last section the general theorem of bunch lengthening in this kind of TLC schemes. Thin-lens kick maps in the last section are again used for the analysis, but now with the fact that the modulation waveform is a nonlinear {\it sine} taken into account. 
Putting in the optimized bunch compression conditions for ADM, namely $
hR_{56}+1=0$ and $
-d' D+R_{56}=0
$,
and using the mathematical identity
$
e^{ia\sin(b)}=\sum_{m=-\infty}^{\infty}e^{imb}J_{m}[a],
$ the final bunching factor at the $n^{\text{th}}$ laser harmonic in ADM is
\begin{equation}
	b_{n}=\left|\sum_{m=-\infty}^{\infty}J_{m}\left(n\right)\int_{-\infty}^{\infty}\int_{-\infty}^{\infty}\int_{-\infty}^{\infty}dy_{i}dy_{i}'dz_{i} e^{-ink_{\text{L}}\left[-\frac{y_{i}'}{hd'} +\left(1-\frac{m}{n}\right)\left(d' y_{i}+z_{i}\right)\right]}f_{i}(y_{i},y_{i}',z_{i})\right|.\\
	\end{equation}
For a coasting beam, $\left\langle e^{-ink_{\text{L}}\left[\left(1-\frac{m}{n}\right)\left(d' y_{i}+z_{i}\right)\right]}\right\rangle$ will be non-zero only if $m=n$, where the bracket $\langle\cdot\cdot\cdot\rangle$ means the average over all the particles. Therefore,
\begin{equation}\label{eq:BFCoasting}
	b_{n,\  \text{coasting}}=|J_{n}(n)|\text{exp}\left[-\left(nk_{\text{L}}\sigma_{z}(\text{Rad})\right)^2/2\right].
	\end{equation}
Note that $\sigma_{z}(\text{Rad})=|D|\sigma_{y'i}$ in this section follows the definition in the linear matrix analysis in the previous section, and does not represent the real bunch length at the radiator considering the nonlinear modulation waveform.
For a pre-microbunched beam, $\left\langle e^{-ink_{\text{L}}\left[\left(1-\frac{m}{n}\right)\left(d' y_{i}+z_{i}\right)\right]}\right\rangle$ will be non-zero for all $m$, thus 
\begin{equation}\label{eq:BFADM}
	\begin{aligned}
	b_{n,\ \text{pre-microbunch}}=\ &\left|\sum_{m=-\infty}^{\infty}J_{m}\left(n\right)\text{exp}\left[-\left((n-m)k_{\text{L}}\sigma_{z}(\text{Mod})\right)^2/2\right]\right|\\
	&\text{exp}\left[-\left(nk_{\text{L}}\sigma_{z}(\text{Rad})\right)^2/2\right],
	\end{aligned}
	\end{equation}
with $\sigma_{z}(\text{Mod})=\langle d' y_{i}+z_{i}\rangle$. If the initial beam is $y$-$z$ decoupled, then $\sigma_{z}(\text{Mod})=\sqrt{\left(d'\sigma_{y'i}\right)^2+\sigma_{zi}^{2}}$. Note that the bunch length $\sigma_{z}(\text{Mod})$ here contains contribution from both $\epsilon_{y}$ and $\epsilon_{z}$. 

We start the discussion by investigating two limiting cases. First, if $\sigma_{z}(\text{Mod})=0$, then we have
\begin{equation}
	\begin{aligned}
	b_{n}&=\left|\sum_{m=-\infty}^{\infty}J_{m}\left(n\right)\right|\text{exp}\left[-\left(nk_{\text{L}}\sigma_{z}(\text{Rad})\right)^2/2\right]=\text{exp}\left[-\left(nk_{\text{L}}\sigma_{z}(\text{Rad})\right)^2/2\right].
	\end{aligned}
	\end{equation}
This result is the same as that assuming the modulation waveform is linear. 
Second, if $\sigma_{z}(\text{Mod})$ is much longer than the modulation laser wavelength, i.e., $k_{\text{L}}\sigma_{z}(\text{Mod})\gg1$, then the summation terms in Eq.~(\ref{eq:BFADM}) 
will be nonzero only for $m=n$ and we arrive at  the same result as the coasting beam case Eq.~(\ref{eq:BFCoasting}) as expected.

\begin{figure}[tb]
	\centering 
	\includegraphics[width=0.6\columnwidth]{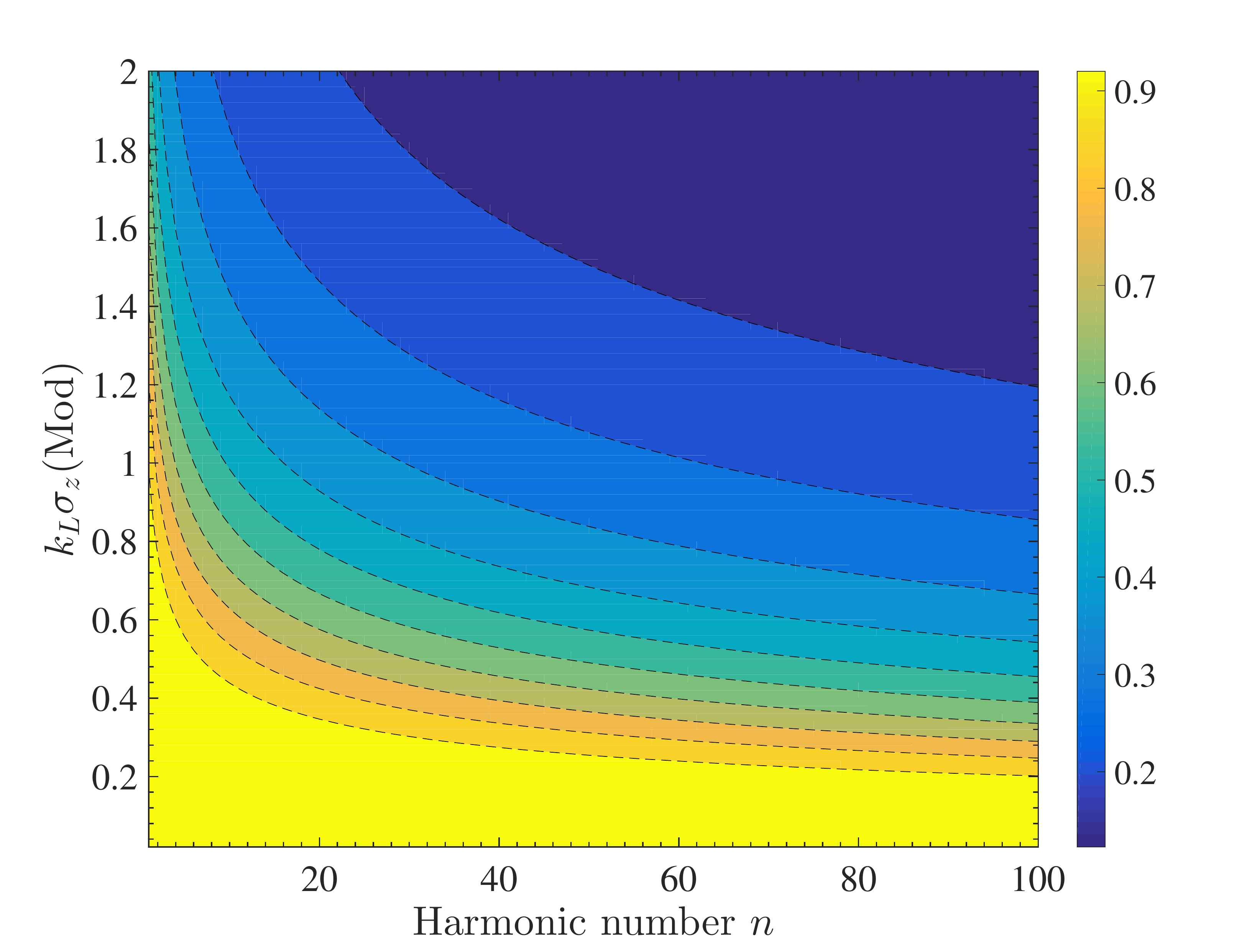}
	\caption{
		\label{fig:Chap3-BFRn} 
		Flat contour plot for the bunching factor reduction factor $R_{n}$ of Eq.~(\ref{eq:BFcorrectfactor1}) as a function of the harmonic number $n$ and the modulation wavelength-normalized bunch length at the modulator $k_{\text{L}}\sigma_{z}(\text{Mod})$.}
\end{figure}

Now we conduct a bit more general discussion. Compared to the linear modulation case, the reduction factor of the bunching factor Eq.~(\ref{eq:BFADM})  is 
\begin{equation}\label{eq:BFcorrectfactor1}
	R_{n}=\left|\sum_{m=-\infty}^{\infty}J_{m}\left(n\right)\right|\text{exp}\left[-\left((n-m)k_{\text{L}}\sigma_{z}(\text{Mod})\right)^2/2\right].
	\end{equation}
Figure~\ref{fig:Chap3-BFRn} shows the flat contour plot for the bunching factor reduction factor $R_{n}$ of Eq.~(\ref{eq:BFcorrectfactor1}) as a function of the harmonic number $n$ and the modulation wavelength-normalized bunch length at the modulator $k_{\text{L}}\sigma_{z}(\text{Mod})$. As can be seen from the figure, the bunch lengthening at the modulator indeed degrades the bunching factor at the radiator, due to the nonlinearity nature of {\it sine} modulation. The longer this bunch lengthening, the more degradation of the bunching factor. The higher the harmonic number, the more significant the impact is. The limit of $R_{n}$ with an infinite long $\sigma_{z}(\text{Mod})$ is $|J_{n}(n)|$. Equation~(\ref{eq:BFcorrectfactor1}) and Fig.~\ref{fig:Chap3-BFRn} are the general result of this bunching factor degradation analysis.

As the decrease of bunching factor originates from the nonlinearity of the {\it sine} modulation, we expect that this reduction will be less if we make the modulation waveform more like linear, for example by adding a third-harmonic RF or laser to broaden the linear zone of the modulation waveform, as also suggested before in Ref.~\cite{ratner2011seeded} and Ref.~\cite{stupakov2011using}. The energy modulation then becomes
$
\delta=\delta+\frac{h_{1}}{k_{\text{L}}}\sin(k_{\text{L}}z)+\frac{h_{3}}{3k_{\text{L}}}\sin(3k_{\text{L}}z).
$
The optimized bunch compression conditions for ADM are now 
$
(h_{1}+h_{3})R_{56}+1=0
$ and $-d' D+R_{56}=0.$
The $n^{\text{th}}$ laser harmonic bunching factor at the radiator is then
\begin{equation}
	b_{n,\  \text{coasting}}=\left|\sum_{m_{1}+3m_{3}=n}J_{m_{1}}\left(\frac{h_{1}}{h_{1}+h_{3}}n\right)J_{m_{3}}\left(\frac{h_{3}}{h_{1}+h_{3}}\frac{n}{3}\right)\right|\text{exp}\left[-\left(nk_{\text{L}}\sigma_{z}(\text{Rad})\right)^2/2\right]
	\end{equation}
for a coasting beam,
and
\begin{equation}\label{eq:BFADM3rd}
	\begin{aligned}
	b_{n,\  \text{pre-microbunch}}=&\Bigg|\sum_{m_{1}=-\infty}^{\infty}\sum_{m_{3}=-\infty}^{\infty}J_{m_{1}}\left(\frac{h_{1}}{h_{1}+h_{3}}n\right) J_{m_{3}}\left(\frac{h_{3}}{h_{1}+h_{3}}\frac{n}{3}\right)\\
	&\ \ \ \text{exp}\left[-\left((n-m_{1}-3m_{3})k_{\text{L}}\sigma_{z}(\text{Mod})\right)^2/2\right]\Bigg|\text{exp}\left[-\left(nk_{\text{L}}\sigma_{z}(\text{Rad})\right)^2/2\right]
	\end{aligned}
	\end{equation}
for a pre-microbunched beam.
Therefore, the reduction factor of the bunching factor Eq.~(\ref{eq:BFADM3rd}), compared to the linear modulation case, is now
\begin{equation}\label{eq:thirdHarmonic}
	\begin{aligned}
	R_{n}&=\Bigg|\sum_{m_{1}=-\infty}^{\infty}\sum_{m_{3}=-\infty}^{\infty}J_{m_{1}}\left(\frac{h_{1}}{h_{1}+h_{3}}n\right) J_{m_{3}}\left(\frac{h_{3}}{h_{1}+h_{3}}\frac{n}{3}\right)\\
	&\ \ \ \ \text{exp}\left[-\left((n-m_{1}-3m_{3})k_{\text{L}}\sigma_{z}(\text{Mod})\right)^2/2\right]\Bigg|.
	\end{aligned}
	\end{equation}
The limit of $R_{n}$ with an infinite long $\sigma_{z}(\text{Mod})$ is $\bigg|\sum_{m_{1}+3m_{3}=n}J_{m_{1}}\left(\frac{h_{1}}{h_{1}+h_{3}}n\right) J_{m_{3}}\left(\frac{h_{3}}{h_{1}+h_{3}}\frac{n}{3}\right)\bigg|$. It is straightforward to generalize the above derivation and result to the case of adding more laser harmonics.

\begin{figure}[tb]
	\centering 
	\includegraphics[width=1\columnwidth]{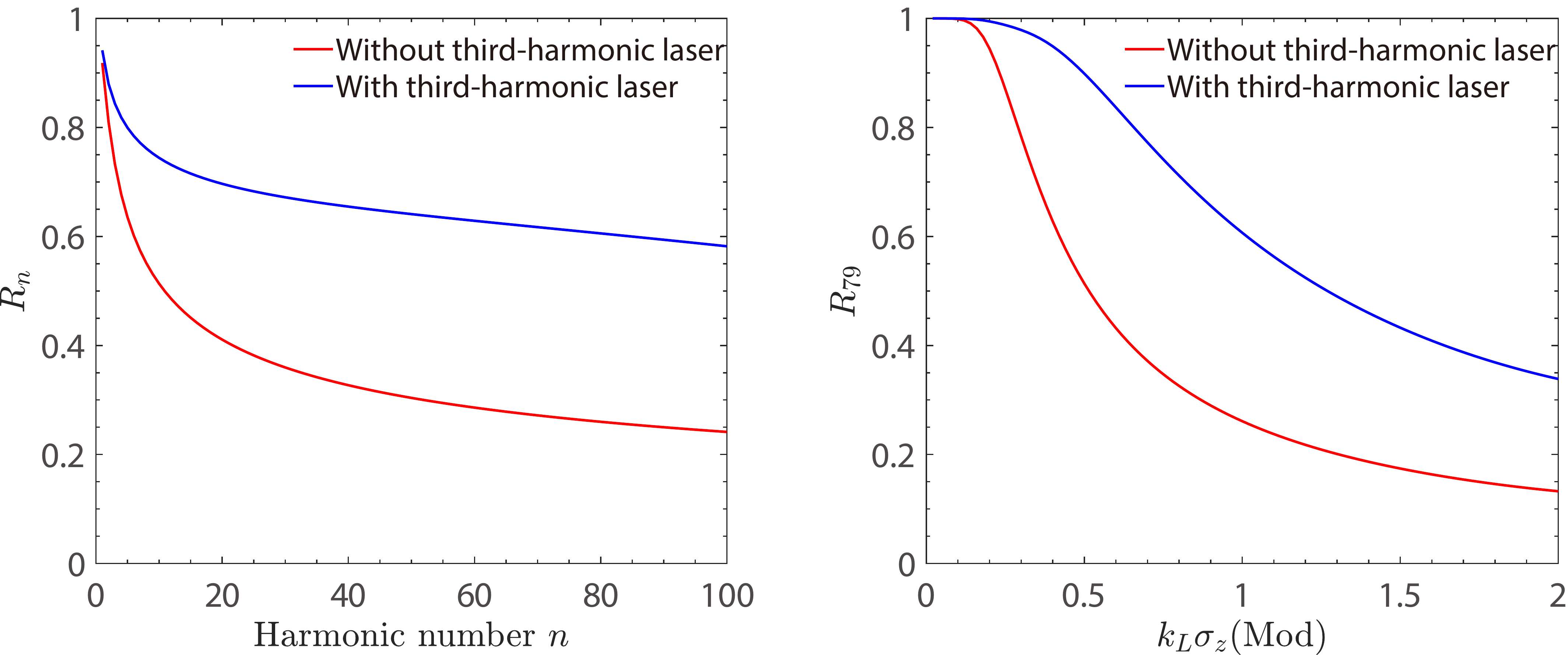}
	\caption{
		\label{fig:Chap3-BFRn3rd} 
		Left: the bunching factor reduction factor $R_{n}$ of Eq.~(\ref{eq:thirdHarmonic}) as a function of the harmonic number $n$ for $k_{\text{L}}\sigma_{z}(\text{Mod})=1$, with $h_{3}=0$ (red) and $h_{3}=-0.15h_{1}$ (blue), respectively. Right: the bunching factor reduction factor $R_{n}$ of Eq.~(\ref{eq:thirdHarmonic}) as a function of the modulation wavelength-normalized bunch length at the modulator $k_{\text{L}}\sigma_{z}(\text{Mod})$ for $n=79$, with $h_{3}=0$ (red) and $h_{3}=-0.15h_{1}$ (blue), respectively. $n=79$ corresponds to the case for example a modulation wavelength of $\lambda_{\text{L}}=1064$ nm and a radiation wavelength of $\lambda_{R}=\lambda_{\text{L}}/79=13.5$~nm. }
\end{figure}

\begin{figure}[tb]
	\centering 
	\includegraphics[width=1\columnwidth]{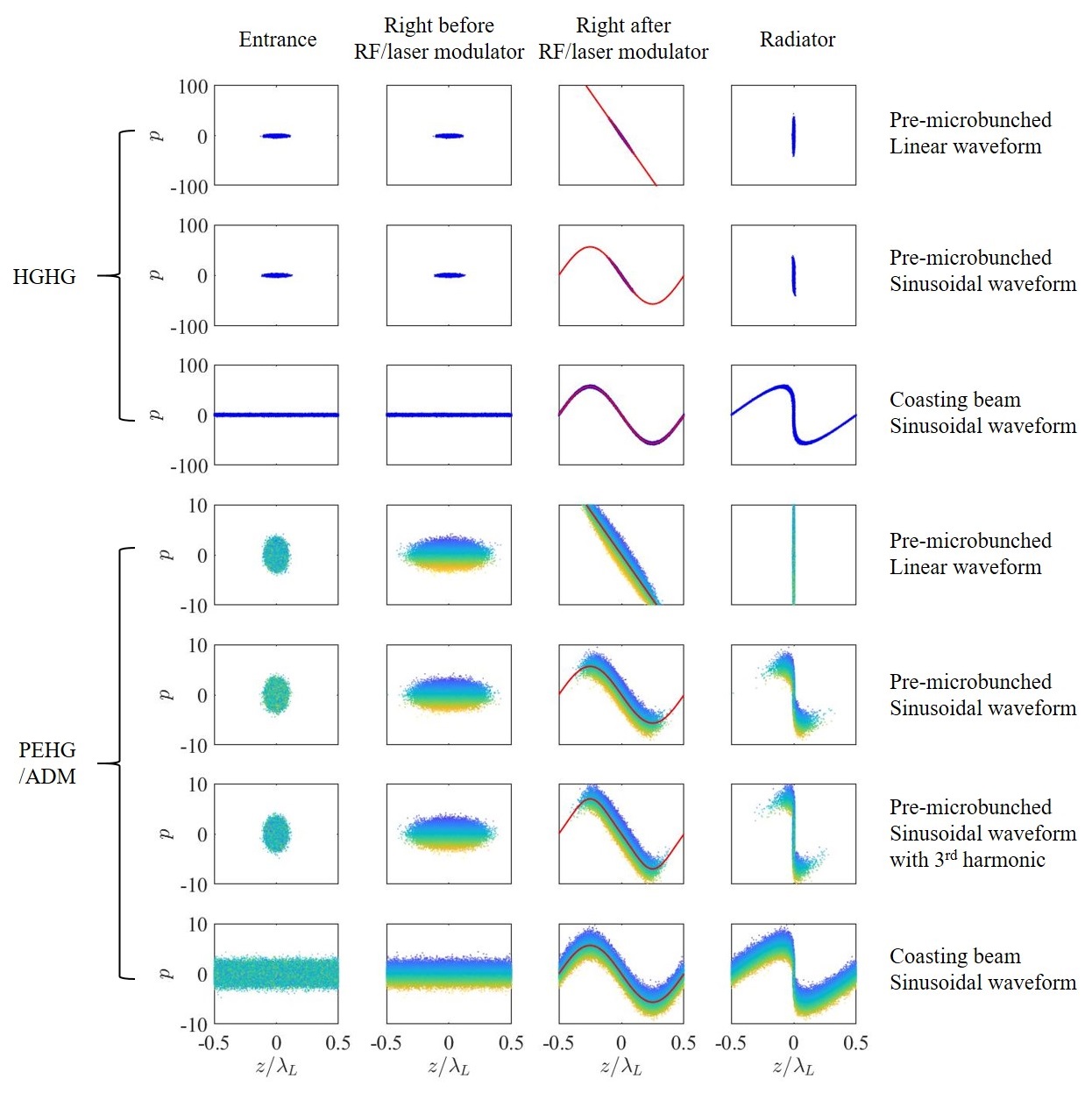}
	\caption{
		\label{fig:Chap3-Summary} 
		Application of HGHG and PEHG/ADM for bunch compression and harmonic generation. Parameters used in this example plot: $\lambda_{\text{L}}=1064$~nm, $\sigma_{zi}=30$ nm and $\sigma_{zf}=3$ nm for the case of a pre-microbunched beam, $\sigma_{\delta i}=3\times10^{-4}$, $\sigma_{yi}=2\ \mu$m, $\sigma_{y'i}=1\ \mu$rad. The figures show the beam distribution evolution in the longitudinal phase space.   The different colors of the particles in the plots of PEHG/ADM correspond to different $y$ for PEHG or $y'$ for ADM. The modulation waveforms are shown in the figure as the red curves.  Note that the range of vertical axis $p=\delta/\sigma_{\delta i}$ in PEHG/ADM is one order of magnitude smaller than that of HGHG in this figure. }
\end{figure}

Now we can use the above formula of $R_{n}$ to do comparison between the cases with and without the third-harmonic laser. If $h_{3}=0$, then Eq.~(\ref{eq:thirdHarmonic}) reduces to Eq.~(\ref{eq:BFcorrectfactor1}). As can be seen in Fig.~\ref{fig:Chap3-BFRn3rd}, indeed addition of a third-harmonic laser is effective in mitigating the bunching factor degradation from the bunch lengthening at the modulator. 

To give the readers a better picture about the above derivations and discussions, here we summarize the main information of the analysis we have presented about the application of TLC with a normal RF or TEM00 mode laser modulator with the help of Fig.~\ref{fig:Chap3-Summary}: (i) Compared to HGHG, TLC schemes like PEHG or ADM can reduce the required energy chirp strength, to realize the same bunch length compression ratio or harmonic generation number,  when the transverse emittance is small. (ii) This lowering of chirp strength is realized through bunch lengthening from transverse emittance at the modulator, which can degrade the bunching factor at the radiator for a pre-microbunched beam due to the nonlinear nature of the {\it sine} modulation. (iii) Addition of the RF or laser harmonics is an effective way to mitigate this bunching factor degradation by broadening the linear zone of the modulation waveform. We emphasize the fact that the discussion of bunching factor degradation is more relevant for a pre-microbunhed beam, like that in some SSMB scenarios, and is generally not an issue for a coasting beam where the bunch duration is much longer than the modulation wavelength, like that in an FEL.

\subsubsection{Contribution of Modulators to Vertical Emittance}

We have stated that the main motivation of applying TLC scheme like PEHG and ADM in SSMB is to lower the requirement on the modulation laser power $P_{\text{L}}$. This is based on the fact that the vertical emittance $\epsilon_{y}$ in a planar $x$-$y$ uncoupled ring is rather small. However, since the modulator in PEHG or ADM is placed at a dispersive location, i.e., $\mathcal{H}_{y}(\text{Mod})\neq0$, therefore the quantum excitation at the modulator will also contribute to $\epsilon_{y}$. 
With this consideration taken into account,  below we try to give a self-consistent analysis of the required modulation laser power $P_{\text{L}}$.


To make sure that the TLC-based bunch compression can repeat turn-by-tun in a ring, usually two laser modulators are placed upstream and downstream of the radiator, respectively, to form a pair. Assuming the modulator undulator is planar, the contribution of these two modulators to $\epsilon_{y}$ is then
\begin{equation}
\begin{aligned}
\Delta\epsilon_{y}(\text{Mod, QE})&=2\times\frac{55}{96\sqrt{3}}\frac{\alpha_{F}{\lambdabar}_{e}^{2}\gamma^{5}}{\alpha_{\text{V}}}\int_{0}^{L_{u}} \frac{\mathcal{H}_{y}(\text{Mod})}{|\rho(s)|^{3}}ds\\
&=2\times\frac{55}{96\sqrt{3}}\frac{\alpha_{F}{\lambdabar}_{e}^{2}\gamma^{5}}{\alpha_{\text{V}}}\frac{\mathcal{H}_{y}(\text{Mod})}{\rho_{0}^{3}}\int_{0}^{L_{u}} |\sin(k_{u}s)|^{3} ds\\
&=2\times\frac{55}{96\sqrt{3}}\frac{\alpha_{F}{\lambdabar}_{e}^{2}\gamma^{5}}{\alpha_{\text{V}}}\frac{\mathcal{H}_{y}(\text{Mod})}{\rho_{u0}^{3}}\frac{4}{3\pi}L_{u},\\
\end{aligned}
\end{equation}
with the vertical damping constant
\begin{equation}
\alpha_{\text{V}}\approx\frac{1}{2}\frac{U_{0}}{E_{0}}=\frac{1}{2}C_{\gamma}\frac{E_{0}^{3}}{\rho_{\text{ring}}}=\frac{1}{2}C_{\gamma}\times0.2998B[\text{T}]E_{0}^{2}[\text{GeV}]
\end{equation}
where $C_{\gamma}=8.85\times10^{-5}\frac{\text{m}}{\text{GeV}^{3}}$. 
Note that $\rho_{\text{ring}}$ is the bending radius of the dipole in the ring, and $\rho_{u0}$ is the minimum being radius corresponding to the peak magnetic field $B_{0}$ of the modulator. 

If we keep $\rho_{\text{ring}}$ in the ring a fixed value, which means $\alpha_{\text{V}}\propto\gamma^{3}$, then
\begin{equation}
\Delta\epsilon_{y}(\text{Mod, QE})\propto\gamma^{2}\frac{L_{u}}{\rho_{u0}^{3}}\mathcal{H}_{y}(\text{Mod})\propto\gamma^{-1}B_{0}^{3}L_{u}\mathcal{H}_{y}(\text{Mod}),
\end{equation}
with $B_{0}$ the peak field of the modulator undulator. Or put it inversely
\begin{equation}\label{eq:HyMod}
\frac{1}{\mathcal{H}_{y}(\text{Mod})}\propto\frac{\gamma^{-1}B_{0}^{3}L_{u}}{\Delta\epsilon_{y}(\text{Mod, QE})}.
\end{equation}
The resonant condition of the laser-electron interaction inside a planar undulator is 
\begin{equation}
\lambda_{\text{L}}=\frac{1+\frac{K^{2}}{2}}{2\gamma^{2}}\lambda_{u},
\end{equation}
with $K=\frac{eB_{0}\lambda_{u}}{2\pi m_{e}c}=0.934\cdot B_{0}[\text{T}]\cdot\lambda_{u}[\text{cm}]$ the undulator parameter.
Assuming that $\frac{K^{2}}{2}\gg1$, then
$
\frac{K^{2}}{4\gamma^{2}}\lambda_{u}\approx\lambda_{\text{L}}.
$
The energy chirp strength induced by a TEM00 mode laser modulator is \cite{chao2022FocusedLaser}
\begin{equation}\label{eq:ChaoChirp}
h=\frac{e[JJ] K}{\gamma^{2}mc^{2}}\sqrt{\frac{4P_{\text{L}}Z_{0}Z_{R}}{\lambda_{\text{L}}}}\tan^{-1}\left(\frac{L_{u}}{2Z_{R}}\right)k_{\text{L}},
\end{equation}
with $[JJ]=J_{0}(\chi)-J_{1}(\chi)$ and $\chi=\frac{K^{2}}{4+2K^{2}}$, $Z_{R}$ the Rayleigh length and $Z_{0}=376.73\ \Omega$ the impedance of free space.
In the optimized case, $\frac{Z_{R}}{L_{u}}\approx\frac{1}{3}$, therefore,
\begin{equation}
h\propto\frac{K}{\gamma^{2}}\sqrt{P_{\text{L}}L_{u}}\lambda_{\text{L}}^{-\frac{3}{2}}
\end{equation}
and 
\begin{equation}\label{eq:PL}
P_{\text{L}}\propto\frac{1}{L_{u}}\left(\frac{h\gamma^{2}}{K}\right)^{2}\lambda_{\text{L}}^{3}\propto\frac{h^{2}\gamma^{2}}{N_{u}}\lambda_{\text{L}}^{2}.
\end{equation}
And according to Eq.~(\ref{eq:theorem}), (\ref{eq:HyMod}) and (\ref{eq:PL}), 
then
\begin{equation}
P_{\text{L}}\propto\frac{h^{2}\gamma^{2}}{N_{u}}\lambda_{\text{L}}^{2}\propto \frac{\gamma^{2}\lambda_{\text{L}}^{2}}{N_{u}\mathcal{H}_{y}(\text{Mod})\mathcal{H}_{y}(\text{Rad})}\propto \frac{\gamma B_{0}^{3}\lambda_{u}\lambda_{\text{L}}^{2}}{\sigma_{z}^{2}(\text{Rad})}\frac{\epsilon_{y}}{\Delta\epsilon_{y}(\text{Mod, QE})}.
\end{equation}
For simplicity, we set $\epsilon_{y}=\Delta\epsilon_{y}(\text{Mod, QE})$, i.e., the vertical emittance is purely from the contribution of these two modulators, then
\begin{equation}\label{eq:scaling}
P_{\text{L}}\propto \frac{\gamma B_{0}^{3}\lambda_{u}\lambda_{\text{L}}^{2}}{\sigma_{z}^{2}(\text{Rad})}\propto\frac{\gamma^{1+\frac{2}{3}}B_{0}^{3-\frac{2}{3}}\lambda_{\text{L}}^{2+\frac{1}{3}}}{\sigma_{z}^{2}(\text{Rad})}.
\end{equation}
This is the scaling law of the modulation laser power, with respect to  beam energy, laser wavelength, modulator undulator field strength, and the linear bunch length at the radiator. Note that $\epsilon_{y}$ is not shown explicitly in the scaling of the laser power, it however affects the bunch length at the modulator and therefore the bunching factor at the radiator as we have explained. In other words, the smaller $\epsilon_{y}$ is, the larger $\sigma_{z}(\text{Rad})$ we can use, thus a lower modulation laser power, to obtain the desired bunching factor.



\begin{table}[tb]
	\caption{\label{tab:exm-EUV-SSMB}
		Example parameters of an SSMB storage ring using TLC-based bunch compression scheme for EUV radiation generation.}
	\centering
	\begin{tabular}{lll}
		\hline
		Parameter & \multicolumn{1}{l}{\textrm{Value}}  & Description \\
		\hline 
		$E_{0}$ & $400$ MeV & Beam energy \\	
		 $\lambda_{\text{L}}$ & 1064 nm & Modulation laser wavelength	\\	
		 $\epsilon_{y}$ & 4 pm & Vertical emittance	 \\ 
		 $\sigma_{zy}(\text{Mod})$ & 505 nm & Bunch lengthening from $\epsilon_{y}$	\\ 
		 $\sigma_{z}(\text{Rad})$ & 2 nm & Linear bunch length at the radiator	\\ 
		 $\lambda_{R}$ &  13.5 nm & Radiation wavelength \\  
		 $b_{79}$ & 0.07  & 13.5 nm bunching factor at radiator\\
		  &   &  (with Bessel suppression)\\ 
		 $h$ &  $h=\frac{\epsilon_{y}}{\sigma_{zy}(\text{Mod})\sigma_{z}(\text{Rad})}=3956\ \text{m}^{-1}$ & Energy chirp strength \\ 
		 $\lambda_{u}$ &  30 cm & $\lambda_{u}$ of modulator undulator\\  
		 $K_{u}$ &  2.6 & $K$ of modulator undulator\\ 
		 $B_{0}$ &  0.092 T & Peak magnetic field of modulator\\ 
		 $N_{u}$ &  10 & $N_{u}$ of modulator undulator \\ 
		 $L_{u}$ &  3 m & Modulator length	\\ 
		 $Z_{R}$ &  1 m & Rayleigh length \\ 
		 $P_{\text{L}}$ &  7.9 MW & Modulation laser power\\ 
		\hline
	\end{tabular}
\end{table}

Based on our analysis, an example parameters set of the envisioned EUV SSMB ring by invoking this TLC scheme is given in Tab.~\ref{tab:exm-EUV-SSMB}. Note that the modulator field $B_{0}$ used is rather weak, which is to control the contribution of the two modulators to $\epsilon_{y}$.

Note also that the energy chirp strength here, and therefore the modulation laser power, is much smaller than that given in Tab.~\ref{tab:tab2}. To further lower the requirement on modulation laser power, we may invoke a helical undulator for laser-electron interaction, which is more effective than a planar one. Besides, we may incorporate multiple laser-electron interaction sections to realize a single-stage energy modulation. To keep contribution to $\epsilon_{y}$ unchanged, then we need to lower $\mathcal{H}_{y}(\text{Mod})$ by a factor of $N$ with $N$ the number of interaction sections, so the benefit in lowering required laser power by applying $N$-times laser-electron interaction is not $N^2$, but $N$. However, this multiple laser-electron interaction approach requires a complex lattice optics.

\subsection{Theorem on Transverse Deflecting RF or TEM01 Laser for Coupling}

\subsubsection{Theorem and the Proof}
\begin{figure}[tb]
	\centering 
	\includegraphics[width=1\textwidth]{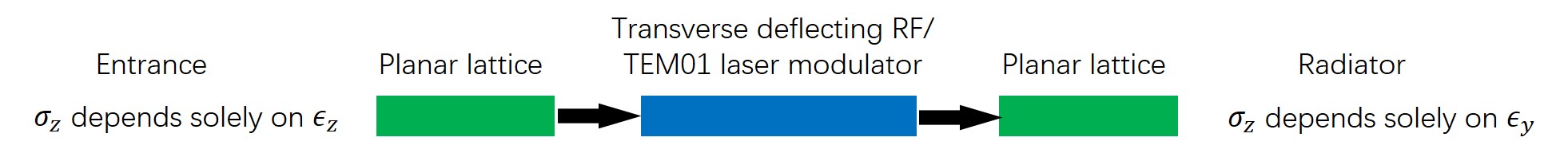}
	\caption{
		\label{fig:Chap3-Transverse} 
		Schematic layout of applying $y$-$z$ coupling  for bunch compression, using a transverse deflecting RF or TEM01 mode laser modulator.}
\end{figure}

In the previous sections, we have focused on TLC schemes using normal RFs or TEM00 mode lasers. Given the same problem definition and schematic layout as shown in Fig.~\ref{fig:Chap3-Transverse}, there is a corresponding theorem for TLC-based harmonic generation and bunch compression schemes using transverse deflecting RFs or TEM01 mode lasers. All we need is to replace $h(\text{Mod})$ by $t(\text{Mod})$  and $\mathcal{H}_{y}(\text{Mod})$ by $\beta_{y}(\text{Mod})$, i.e., the theorem is now
\begin{equation}\label{eq:theorem2}
	t^2(\text{Mod})\beta_{y}(\text{Mod})\mathcal{H}_{y}(\text{Rad})\geq1,
	\end{equation}
where $t(\text{Mod})$ is the transfer matrix term $R_{45}$ of the transverse deflecting RF or TEM01 mode laser modulator which quantifies the angular chirp strength. 
The theorem in Eq.~(\ref{eq:theorem2}) can also be expressed as
\begin{equation}\label{eq:TRFBeta}
	|t(\text{Mod})|\geq\frac{\sqrt{\epsilon_{y}/\beta_{y}(\text{Mod})}}{\sigma_{z}(\text{Rad})}=\frac{\epsilon_{y}}{\sigma_{y\beta}(\text{Mod})\sigma_{z}(\text{Rad})}.
	\end{equation}
Note that in the above formula, the vertical beam size contains only the betatron part, i.e., that from the vertical emittance $\epsilon_{y}$.
So given a fixed $\epsilon_{y}$ and $\sigma_{z}(\text{Rad})$, a smaller $t(\text{Mod})$, i.e., a smaller RF gradient or modulation laser power ($P_{\text{laser}}\propto|t(\text{Mod})|^2$), means a larger $\beta_{y}(\text{Mod})$, thus a larger $\sigma_{y\beta}(\text{Mod})$, is needed. As $|t(\text{Mod})|\sigma_{y}(\text{Mod})$ quantifies the energy spread introduced by the modulator, we thus also have the result of Eq.~(\ref{eq:ES}). 

The proof of this corresponding theorem is basically the same as that for the normal RF or TEM00 mode laser modulator. We just need to replace ${\bf M}_{2}$ with
\begin{equation} 
	\begin{aligned}
	{\bf M}_{2}=\left(\begin{matrix}
	1&0&0&0\\
	0&1&t&0\\
	0&0&1&0\\
	t&0&0&1\\
	\end{matrix}\right).
	\end{aligned}
	\end{equation}	
A transverse deflecting RF cavity here is the counterpart of skew quadrupole in the $y$-$z$ dimension.  For $\sigma_{z}(\text{Rad})$ to be independent of $\epsilon_{z}$, 
we need
\begin{equation}\label{eq:TRFConditions}
	\begin{aligned}
	T_{55}&= t R_{54}+1=0,\\
	T_{56}&= d R_{53}+d'R_{54} +r_{56} \left(t R_{54}+1\right)+(d t+1) R_{56}=0.
	\end{aligned}
	\end{equation} 
Under the conditions of Eq.~(\ref{eq:TRFConditions}), we have
\begin{equation}\label{eq:TransportMatrixSub2}
	\begin{aligned}
	&{\bf A}=\left(\begin{matrix}
	r_{33} \left( tD+R_{33}\right)+R_{34} \left(r_{53}t +r_{43}\right) & r_{34} \left( tD+R_{33}\right)+R_{34} \left(r_{54}t +r_{44}\right)\\
	r_{33} \left( tD'+R_{43}\right)+R_{44} \left(r_{53}t +r_{43}\right) & r_{34} \left( tD'+R_{43}\right)+R_{44} \left(r_{54}t +r_{44}\right)\\
	\end{matrix}\right),\\
	&{\bf B}=\left(\begin{matrix}
	t R_{34} &  dR_{33}+d'R_{34}+ r_{56}t R_{34}+(d t+1)D \\
	t R_{44} &  dR_{43}+d'R_{44} + r_{56}t R_{44}+(d t+1)D'\\
	\end{matrix}\right),\\
	&{\bf C}=\left(\begin{matrix}
	r_{33} R_{53}+r_{43} R_{54}+r_{33}t R_{56} & r_{34} R_{53}+r_{44} R_{54}+r_{34} t R_{56}\\
	r_{33} t  & r_{34} t \\
	\end{matrix}\right),\\
	&{\bf E}=\left(\begin{matrix}
	0&0\\
	0 & d t+1\\
	\end{matrix}\right).
	\end{aligned}
	\end{equation}
The vertical beam size squared at the modulator is
\begin{equation}
	\begin{aligned}
	\sigma_{y}^{2}(\text{Mod})
	=&\epsilon_{y}\frac{\left(\beta_{y}r_{33}-\alpha_{y}r_{34}\right)^2+r_{34}^2}{\beta_{y}}+d^{2}\epsilon_{z}\gamma_{z}=\epsilon_{y}\beta_{y}(\text{Mod})+d^{2}\sigma_{\delta i}^{2}.
	\end{aligned}
	\end{equation}
According to Cauchy-Schwarz inequality, we have
\begin{equation}\label{eq:proof2}
	\begin{aligned}
	t^2(\text{Mod})\beta_{y}(\text{Mod})\mathcal{H}_{y}(\text{Rad})&=t^2\frac{\left[\left(\beta_{y}r_{33}-\alpha_{y}r_{34}\right)^2+r_{34}^2\right]}{\beta_{y}}\frac{\left[\left(\beta_{y}T_{53}-\alpha_{y}T_{54}\right)^2+T_{54}^2\right]}{\beta_{y}}\\
	&\geq
	\frac{t^2}{\beta_{y}^{2}}\left[-\left(\beta_{y}r_{33}-\alpha_{y}r_{34}\right)T_{54}+r_{34}\left(\beta_{y}T_{53}-\alpha_{y}T_{54}\right)\right]^{2}\\
	&=\left(T_{53}r_{34}t-T_{54}r_{33}t\right)^{2}=\left(T_{53}T_{64}-T_{54}T_{63}\right)^{2}.
	\end{aligned}
	\end{equation}
The equality holds when
$
\frac{-\left(\beta_{y}r_{33}-\alpha_{y}r_{34}\right)}{T_{54}}=\frac{r_{34}}{\left(\beta_{y}T_{53}-\alpha_{y}T_{54}\right)}.
$
Similar to the procedures presented in the section about the theorem of normal RF,
we know that ${\bf C}$ is a symplectic matrix and  $|T_{53}T_{64}-T_{54}T_{63}|=\text{det}({\bf C})=1$. The theorem is thus proven.



\subsubsection{Possible Applications}

Here we briefly investigate the possibility of applying the TEM01 mode laser for bunch compression in SSMB. 
A laser modulator implementing a TEM01 mode laser is a transverse deflecting RF cavity in the optical wavelength range, as dictated by Panofsky\--Wenzel theorem~\cite{panofsky1956some}
\begin{equation}
\frac{\partial\Delta y'}{\partial s}=\frac{\partial}{\partial y_{0}}\left(\frac{\Delta\gamma}{\gamma}\right),
\end{equation}
where $\Delta y'$ and $\Delta\gamma$ are the electron angular kick and energy change in the laser modulator. 
For a TEM01 mode laser modulator, the angular chirp strength introduced by a laser with power of $P_{\text{L}}$ is \cite{zholents2008attosecond} 
\begin{equation}
t=\frac{2Kk_{\text{L}}}{\gamma^{2}}\sqrt{\frac{P_{\text{L}}}{P_{0}}}[JJ]f
\end{equation}
where $K$ is the dimensional-less undulator parameter, $P_{0}=I_{\text{A}}m_{e}c^{2}/e$, $I_{A}=\frac{ec}{r_{e}}\approx17$ kA is the Alfv\`en current, $f$ is a function depending on the undulator length and Rayleigh length. For a reasonable choice of parameters, $[JJ]f\sim1$, $K=5$, $E_{0}=400$ MeV, $\lambda_{\text{L}}=1064$~nm,  $P_{\text{L}}=1$~MW, we have $t\sim1\ \text{m}^{-1}$. If $\epsilon_{y}=1$ pm and $\sigma_{z}(\text{Rad})=3$ nm, then according to Eq.~(\ref{eq:TRFBeta}) we have
\begin{equation}
\beta_{y}(\text{Mod})\geq\frac{\epsilon_{y}}{t^{2}(\text{Mod})\sigma_{z}^{2}(\text{Mod})}=1.1\times10^{5}\ \text{m}.
\end{equation}
The application of TLC with a TEM01 mode laser to compress bunch length in SSMB therefore seems to face the issue of a too large $\beta_{y}$ at the modulator, if we desire a final bunch length of nm level, like that demanded in 13.5 nm coherent EUV radiation generation. 

However, if our target wavelength region is THz, then the idea looks appealing. For example, if $\epsilon_{y}=10$ pm, $t\sim1\ \text{m}^{-1}$ and $\sigma_{z}(\text{Rad})=3\ \mu$m (10 fs), then
\begin{equation}
\beta_{y}(\text{Mod})\geq\frac{\epsilon_{y}}{t^{2}(\text{Mod})\sigma_{z}^{2}(\text{Mod})}=1.1\ \text{m},
\end{equation}
which is a reasonable value. Such a short bunch can generate strong coherent radiation with frequency at a couple of 10 THz and smaller. In order to make sure such a bunch compression can repeat turn-by-turn, we need to restore the original beam state. A natural idea is to exchange the $y$-$z$ emittances twice, which will be discussed soon. As we will see, the transverse deflecting RF should then be placed at a dispersive location for a complete $y$-$z$ emittance exchange. Th dispersion at the modulators may lead to the growth of vertical emittance, and thus a vertical emittance $\epsilon_{y}=10$ pm, instead of 1 pm, is used in this example.

\subsubsection{Contribution of Modulators to Vertical Emittance}

We remind the readers that for bunch compression (instead of complete $y$-$z$ emittance exchange), in principle, we can place the modulator in a dispersion-free location. To minimize the contribution of modulators to $\epsilon_{y}$, we choose to place the modulator at dispersion-free location, which means $d=0$ and $d'=0$, then the bunch compression condition is
\begin{equation}\label{eq:TRFConditions2}
\begin{aligned}
T_{55}&= t R_{54}+1=0,\\
T_{56}&= R_{56}=0.
\end{aligned}	
\end{equation}	

Although we have placed the modulator at a dispersion-free location, there is still some residual contribution to $\epsilon_{y}$ since the transfer matrix of a TEM01 mode laser modulator is intrinsic transverse-longitudinal coupled. Now let us evaluate such contribution. Here for simplicity we consider only the linear transfer matrix. We need to consider the physical length $L_{u}$ and $r_{56}=2N_{u}\lambda_{\text{L}}$ of the modulator. The thick-lens matrix of the laser modulator can be obtained by slicing the laser modulator to slices and use the kick and drift method to get the total map. 
The exact map can be obtained by increasing the slice number to infinity, and the slice precision to 0. If we consider only terms to first order of $t$, $r_{56}$ and $L_{u}$, then the thick-lens matrix of the laser modulator for angular modulation is
\begin{equation}\label{eq:thickLens}
{\bf M}_{2}\approx\left(
\begin{array}{cccc}
1 & L_u & \frac{t L_u}{2} & \frac{r_{56} t L_u}{6}  \\
0 & 1 & t & \frac{r_{56} t}{2} \\
\frac{r_{56} t}{2} & \frac{r_{56} t L_u}{6}  &1 & r_{56} \\
t & \frac{t L_u}{2} & 0 & 1 \\
\end{array}
\right).
\end{equation}
Note that 
\begin{equation}
{\bf M}_{2}{\bf S}{\bf M}_{2}^{T}=\left(
\begin{array}{cccc}
0 & \frac{1}{12} r_{56} t^2 L_u+1 & 0 & 0 \\
-\frac{1}{12} r_{56} t^2 L_u-1 & 0 & 0 & 0 \\
0 & 0 & 0 & \frac{1}{12} r_{56} t^2 L_u+1 \\
0 & 0 & -\frac{1}{12} r_{56} t^2 L_u-1 & 0 \\
\end{array}
\right).
\end{equation}
So ${\bf M}_{2}$ is symplectic to first order of $t$.


Assuming that the one-turn map observed at the bunch compression section entrance is 
\begin{equation}
{\bf T}(0)=\left(
\begin{array}{cccc}
\cos\Phi_{y}+\alpha_{y}\sin\Phi_{y}&\beta_{y}\sin\Phi_{y} & 0 & 0\\
-\gamma_{y}\sin\Phi_{y}&\cos\Phi_{y}-\alpha_{y}\sin\Phi_{y} & 0 & 0\\
0 & 0 & \cos\Phi_{z}+\alpha_{z}\sin\Phi_{z}&\beta_{z}\sin\Phi_{z}\\
0 & 0 & -\gamma_{z}\sin\Phi_{z}&\cos\Phi_{z}-\alpha_{z}\sin\Phi_{z}
\end{array}
\right),
\end{equation}
in which $\Phi_{y}=2\pi\nu_{y}$ and $\Phi_{z}=2\pi\nu_{s}$. 
The eigen vectors of the one-turn map at the undulator entrance is
\begin{equation}
{\bf v}_{y}(0)=\frac{1}{\sqrt{2}}\left(\begin{matrix}
\sqrt{\beta_{y}}\\
\frac{i-\alpha_{y}}{\sqrt{\beta_{y}}}\\
0\\
0\\
\end{matrix}\right),\
{\bf v}_{z}(0)=\frac{1}{\sqrt{2}}\left(\begin{matrix}
0\\
0\\
\sqrt{\beta_{z}}\\
\frac{i-\alpha_{z}}{\sqrt{\beta_{z}}}\\
\end{matrix}\right).
\end{equation}
Then the one-turn map observed at the position $s$ of the first modulator is
\begin{equation}
{\bf T}(s)={\bf M}_{2s}{\bf T}(0){\bf M}_{2s}^{-1},
\end{equation}
in which
%
\begin{equation}
{\bf M}_{2s}=\left(\begin{matrix}
1&\left(\frac{s}{L_{u}}\right)L_{u}&\left(\frac{s}{L_{u}}\right)^{2}\frac{t L_u}{2}&\left(\frac{s}{L_{u}}\right)^{3}\frac{r_{56} t L_u}{6}\\
0&1&\left(\frac{s}{L_{u}}\right)t&\left(\frac{s}{L_{u}}\right)^{2}\frac{r_{56} t}{2}\\
\left(\frac{s}{L_{u}}\right)^{2}\frac{r_{56} t}{2}&\left(\frac{s}{L_{u}}\right)^{3}\frac{r_{56} t L_u}{6}&1&\left(\frac{s}{L_{u}}\right)r_{56}\\
\left(\frac{s}{L_{u}}\right)t&0&0&1\\
\end{matrix}\right).
\end{equation}
Note that for simplicity, here we have assumed that the laser is a plane wave such that the induced angular modulation strength is proportional to the distance traveled inside the modulator.
The eigen vectors of the one-turn map at the position $s$ of the first modulator is then 
\begin{equation}
\begin{aligned}
{\bf v}_{y}(s)&={\bf M}_{2s}{\bf v}_{y}(0)=\frac{1}{\sqrt{2}}\left(\begin{matrix}
\sqrt{\beta_{y}}+\left(\frac{s}{L_{u}}\right)L_{u}\frac{i-\alpha_{y}}{\sqrt{\beta_{y}}}\\
\frac{i-\alpha_{y}}{\sqrt{\beta_{y}}}\\
\left(\frac{s}{L_{u}}\right)^{2}\frac{r_{56} t}{2}\sqrt{\beta_{y}}+\left(\frac{s}{L_{u}}\right)^{3}\frac{r_{56} t L_u}{6}\frac{i-\alpha_{y}}{\sqrt{\beta_{y}}}\\
\left(\frac{s}{L_{u}}\right)t\sqrt{\beta_{y}}\\
\end{matrix}\right),\\
{\bf v}_{z}(s)&={\bf M}_{2s}{\bf v}_{z}(0)=\frac{1}{\sqrt{2}}\left(\begin{matrix}
\left(\frac{s}{L_{u}}\right)^{2}\frac{t L_u}{2}\sqrt{\beta_{z}}+\left(\frac{s}{L_{u}}\right)^{3}\frac{r_{56} t L_u}{6}\frac{i-\alpha_{z}}{\sqrt{\beta_{z}}}\\
\left(\frac{s}{L_{u}}\right)t\sqrt{\beta_{z}}+\left(\frac{s}{L_{u}}\right)^{2}\frac{r_{56} t}{2}\frac{i-\alpha_{z}}{\sqrt{\beta_{z}}}\\
\sqrt{\beta_{z}}+\left(\frac{s}{L_{u}}\right)r_{56}\frac{i-\alpha_{z}}{\sqrt{\beta_{z}}}\\
\frac{i-\alpha_{z}}{\sqrt{\beta_{z}}}\\
\end{matrix}\right).
\end{aligned}
\end{equation}
According to Chao's SLIM formalism~\cite{chao1979evaluation}, we know that the contribution of the two modulators to $\epsilon_{y}$ is
\begin{equation}\label{eq:verticalEmittance}
\begin{aligned}
\Delta\epsilon_{y}(\text{Mod, QE})&=2\times\frac{55}{48\sqrt{3}}\frac{\alpha_{F}{\lambdabar}_{e}^{2}\gamma^{5}}{\alpha_{\text{V}}}\int_{0}^{L_{u}} \frac{|{\bf E}_{k5}(s)|^{2}}{|\rho(s)|^{3}}ds\\
&=2\times\frac{55}{96\sqrt{3}}\frac{\alpha_{F}{\lambdabar}_{e}^{2}\gamma^{5}}{\alpha_{\text{V}}}\int_{0}^{L_{u}} \frac{\bigg|\left(\frac{s}{L_{u}}\right)^{2}\frac{r_{56} t}{2}\sqrt{\beta_{y}}+\left(\frac{s}{L_{u}}\right)^{3}\frac{r_{56} t L_u}{6}\frac{i-\alpha_{y}}{\sqrt{\beta_{y}}}\bigg|^{2}}{|\rho(s)|^{3}}ds\\
&=2\times\frac{55}{96\sqrt{3}}\frac{\alpha_{F}{\lambdabar}_{e}^{2}\gamma^{5}}{\alpha_{\text{V}}}\frac{1}{\rho_{0}^{3}}\int_{0}^{L_{u}} \bigg|\sin\left[2N_{u}\pi\left(\frac{s}{L_{u}}\right)\right]\bigg|^{3}\\
&\ \ \ \ \ \ \ \ \ \ \ \ \times\bigg|\left(\frac{s}{L_{u}}\right)^{2}\frac{r_{56} t}{2}\sqrt{\beta_{y}}+\left(\frac{s}{L_{u}}\right)^{3}\frac{r_{56} t L_u}{6}\frac{i-\alpha_{y}}{\sqrt{\beta_{y}}}\bigg|^{2} ds.
\end{aligned}
\end{equation}
Similar method can be invoked to calculate their contribution to $\epsilon_{z}$. Note that to ensure the TLC-based bunch compression can repeat turn-by-tun in a ring, usually two laser modulators are placed upstream and downstream of the radiator, respectively, to form a pair. This is the reason why we times a factor of two in the above calculation of modulator contribution to $\epsilon_{y}$.

When $N_{u}\gg1$, due to the fast oscillating behaviour of $\sin\left[2N_{u}\pi\left(\frac{s}{L_{u}}\right)\right]$, we can adopt the approximation:
\begin{equation}\label{eq:verticalEmittance2}
\begin{aligned}
\Delta\epsilon_{y}(\text{Mod, QE})&\approx2\times\frac{55}{96\sqrt{3}}\frac{\alpha_{F}{\lambdabar}_{e}^{2}\gamma^{5}}{\alpha_{\text{V}}}\frac{1}{\rho_{0}^{3}}\frac{1}{L_{u}}\int_{0}^{L_{u}} \bigg|\sin\left[2N_{u}\pi\left(\frac{s}{L_{u}}\right)\right]\bigg|^{3}\\
&\ \ \ \ \ \ \ \ \ \ \ \ \times\int_{0}^{L_{u}}\bigg|\left(\frac{s}{L_{u}}\right)^{2}\frac{r_{56} t}{2}\sqrt{\beta_{y}}+\left(\frac{s}{L_{u}}\right)^{3}\frac{r_{56} t L_u}{6}\frac{i-\alpha_{y}}{\sqrt{\beta_{y}}}\bigg|^{2} ds\\
&=2\times\frac{55}{96\sqrt{3}}\frac{\alpha_{F}{\lambdabar}_{e}^{2}\gamma^{5}}{\alpha_{\text{V}}}\frac{1}{\rho_{0}^{3}}\frac{4}{3\pi}\frac{r_{56}^2 t^2 \left[-35 L_u \alpha _y \beta _y+5 L_u^2 \left(\alpha _y^2+1\right)+63 \beta _y^2\right]}{1260 \beta _y}L_{u}.\\
\end{aligned}
\end{equation}
For simplicity, we assume that $\alpha_{y}=0$ at the modulator entrance, then
\begin{equation}\label{eq:verticalEmittance3}
\begin{aligned}
\Delta\epsilon_{y}(\text{Mod, QE})&\approx2\times\frac{55}{96\sqrt{3}}\frac{\alpha_{F}{\lambdabar}_{e}^{2}\gamma^{5}}{\alpha_{\text{V}}}\frac{1}{\rho_{0}^{3}}\frac{4}{3\pi}\frac{r_{56}^2 t^2 \left(5 L_u^2+63 \beta _y^2\right)}{1260 \beta _y}L_{u}.\\
\end{aligned}
\end{equation}
To lower the requirement on modulation laser power, we will let $\beta_{y}$ at the modulator to be a large value, for example 100 meter level, so $\beta_{y}(\text{Mod})\gg L_{u}$, therefore,
\begin{equation}\label{eq:verticalEmittance4}
\begin{aligned}
\Delta\epsilon_{y}(\text{Mod, QE})&\approx2\times\frac{55}{96\sqrt{3}}\frac{\alpha_{F}{\lambdabar}_{e}^{2}\gamma^{5}}{\alpha_{\text{V}}}\frac{1}{\rho_{u0}^{3}}\frac{4}{3\pi}\frac{r_{56}^2 t^2 \beta _y}{20 }L_{u}.\\
\end{aligned}
\end{equation}
If we keep $\rho_{\text{ring}}$ in the ring a fixed value, which means $\alpha_{\text{V}}\propto\gamma^{3}$, then
\begin{equation}
\Delta\epsilon_{y}(\text{Mod, QE})\propto\gamma^{2}\frac{L_{u}}{\rho_{u0}^{3}}r_{56}^2 t^2 \beta _y(\text{Mod})\propto\gamma^{-1}B_{0}^{3}L_{u}r_{56}^2\frac{1}{\mathcal{H}_{y}(\text{Rad})},
\end{equation}
with $B_{0}$ the peak field of the modulator.

Now let us put in some numbers to get a more concrete feeling. For example, if $E_{0}=400$ MeV, $\rho_{\text{ring}}=2$ m ($B_{\text{ring}}=0.67$ T) for the ring dipole, $\lambda_{\text{L}}=1064$ nm, $\lambda_{u}=0.06$ m, $K=6.4$, $B_{0}=1.15$ T, $N_{u}=10$, $r_{56}=2N_{u}\lambda_{\text{L}}=21.28\ \mu$m, $L_{u}=0.6$~m, $\epsilon_{y}=4$ pm, $\beta_{y}(\text{Mod})=100$ m,  $\sigma_{y}(\text{Mod})=\sqrt{\epsilon_{y}\beta_y(\text{Mod})}=20\ \mu$m, $\sigma_{z}(\text{Rad})=2$ nm, $t=\frac{1}{\sqrt{\beta_{y}(\text{Mod})\mathcal{H}_{y}(\text{Rad})}}=\frac{\epsilon_{y}}{\sigma_{y}(\text{Mod})\sigma_{z}(\text{Rad})}=100\ \text{m}^{-1}$, then the contribution of the two modulators to $\epsilon_{y}$ is
\begin{equation}\label{eq:verticalEmittance5}
\begin{aligned}
\Delta\epsilon_{y}(\text{Mod, QE})&\approx2\times\frac{55}{96\sqrt{3}}\frac{\frac{1}{137}\left(386\times10^{-15}\right)^{2}\left(\frac{400}{0.511}\right)^{5}}{\frac{1}{2}\times8.85\times10^{-5}\times\frac{0.4^3}{2}}\left(0.2998\times\frac{1.15}{0.4}\right)^{3}\frac{4}{3\pi}\\
&\ \ \ \ \times\frac{(21.28\times10^{-6})^2 \times100^2\times100}{20 }\times0.6\ \text{m}\\
&=0.55\ \text{pm}.
\end{aligned}
\end{equation}
The more accurate numerical integration of Eq.~(\ref{eq:verticalEmittance2}) gives the same result, which confirms that our approximation adopted in Eq.~(\ref{eq:verticalEmittance2}) is rather accurate. So generally, the contribution of the two modulators to $\epsilon_{y}$ is non-negligible. But on the other hand, their contribution seems okay as the value is not so large.

\subsection{Dragt's Minimum Emittance Theorem}\label{sec:minimum}

In the last two sections, we have discussed the applications of TLC schemes with a normal or a transverse deflecting RF (corresponding TEM00 and TEM01 mode laser modulator, respectively) for bunch compression, with the presentation and proof of two related theorems which can guide us in the application of such coupling schemes. A sharp reader may question that our definition of the problem is still not general enough, and wonder if there could be some clever manipulation of the beam by applying the general 4D or even 6D coupling schemes, which may result in an ultra-small emittance in a specific dimension. For completeness of such a discussion, it might be helpful here to introduce Dragt's minimum emittance theorem~\cite{dragt2011lie}. This theorem states that a 6D linear symplectic manipulation of a particle beam cannot make any of the projected emittance smaller than the minimum one of the three eigen emittances. Therefore, this theorem tells us the best we can hope for by utilizing this kind of linear coupling schemes. Here we present the proof of this theorem following the approach of Dragt~\cite{dragt2011lie}. 


\subsubsection{Williamson’s Symplectic Normal Form}



According to Chao's SLIM formalism~\cite{chao1979evaluation}
\begin{equation}\label{eq:Sigma}
\Sigma_{ij}=\langle X_{i}X_{j}\rangle=2\sum_{K=I,II,III}\epsilon_{k}\text{Re}\left(E_{ki}E_{kj}^{*}\right).
\end{equation}
The eigenvectors satisfy ${\bf E}_{k}^{*}={\bf E}_{-k}$, ${\bf E}_{j}^{T}{\bf S}{\bf E}_{i}=0$ unless $j=-i$, and the normalization condition
\begin{equation}
\left({\bf E}_{k}^{*}\right)^{T}{\bf S}{\bf E}_{k}=\begin{cases}
&i,\ k=I,II,III,\\
&-i,\ k=-I,-II,-III.
\end{cases}
\end{equation}
The SLIM formalism was originally proposed for storage ring analysis, nevertheless its method of parametrization can also be applied for the particle beam. This is similar to the Courant-Snyder parametrization, which can be applied for the lattice as well as for the particle beam. The SLIM formalism can be viewed as a generalization of the Courant-Snyder parametrization to higher dimension. Note that when applying the SLIM parametrization in the single-pass case, the artificial symplectic one-turn map ${\bf M}$ should satisfy ${\bf M}\Sigma{\bf M}^{T}=\Sigma$, or ${\bf M}(i\Sigma{\bf S}){\bf M}^{-1}=i\Sigma{\bf S}$. 

Equation~(\ref{eq:Sigma}) can be cast into a form of matrix equation
\begin{equation}\label{eq:Williamson}
\Sigma={\bf A}\text{diag}\left\{\epsilon_{I},\epsilon_{I},\epsilon_{II},\epsilon_{II},\epsilon_{III},\epsilon_{III}\right\}{\bf A}^{T},
\end{equation}
in which 
\begin{equation}
\begin{aligned}
{\bf A}&=\frac{1}{\sqrt{2}}\left[{\bf E}_{I}+{\bf E}_{-I},i({\bf E}_{-I}-{\bf E}_{I}),{\bf E}_{II}+{\bf E}_{-II},\right.\\
&\left. \ \ \ \ \ \ \ \ \ \ \ \ i({\bf E}_{-II}-{\bf E}_{II}),{\bf E}_{III}+{\bf E}_{-III},i({\bf E}_{-III}-{\bf E}_{III})\right].
\end{aligned}
\end{equation}
It can be shown that ${\bf A}$ is a real symplectic matrix. Equation~(\ref{eq:Williamson}) can be written in the Williamson’s symplectic normal form~\cite{Williamson}
\begin{equation}\label{eq:Williamson2}
{\bf N}\Sigma{\bf N}^{T}=\text{diag}\left\{\epsilon_{I},\epsilon_{I},\epsilon_{II},\epsilon_{II},\epsilon_{III},\epsilon_{III}\right\},
\end{equation}
where ${\bf N}={\bf A}^{-1}$. 
So here we have presented an explicit way of obtaining the Williamson’s symplectic normal form of the second moment matrix based on SLIM. 
For example if the three dimensions are decoupled, then ${\bf N}$ is a familiar result that can be expressed using Courant-Snyder functions 
\begin{equation}
{\bf N}=\left(
\begin{array}{cccccc}
\frac{1}{\sqrt{\beta _x}} & 0 & 0 & 0 & 0 & 0 \\
\frac{\alpha _x}{\sqrt{\beta _x}} & \sqrt{\beta _x} & 0 & 0& 0 & 0  \\
0 & 0 & \frac{1}{\sqrt{\beta _y}} & 0 & 0 & 0 \\
0 & 0 & \frac{\alpha _y}{\sqrt{\beta _y}} & \sqrt{\beta _y}& 0 & 0  \\
0 & 0 & 0 & 0 & \frac{1}{\sqrt{\beta _z}} &  0 \\
0 & 0 & 0 & 0 & \frac{\alpha _z}{\sqrt{\beta _z}} & \sqrt{\beta _z}  \\
\end{array}
\right).
\end{equation}


\subsubsection{Classical Uncertainty Principle}\label{sec:uncertainity}


With the help of Williamson’s symplectic normal form, now we first prove the classical uncertainty principle, i.e., 
\begin{equation}
\begin{aligned}
\Sigma_{11}\Sigma_{22}&\geq\epsilon_{\text{min}}^{2}\\
\Sigma_{33}\Sigma_{44}&\geq\epsilon_{\text{min}}^{2}\\
\Sigma_{55}\Sigma_{66}&\geq\epsilon_{\text{min}}^{2},
\end{aligned}
\end{equation}
in which $\epsilon_{\text{min}}$ is the minimum one among the three eigen emittances $\epsilon_{I,II,III}$.
We remind the readers that the proofs of the classical uncertainty principle and the minimum emittance theoreom presented in this subsection and below are from Dragt~\cite{dragt2011lie}. 
What we add here is to give an explicit form of the normal form based on Chao's SLIM formalism, and supplement some details of the proof.
 
According to Eq.~(\ref{eq:Williamson}), we have
\begin{equation}
\begin{aligned}
\Sigma_{11}&=\left({\bf A}\text{diag}\left\{\epsilon_{I},\epsilon_{I},\epsilon_{II},\epsilon_{II},\epsilon_{III},\epsilon_{III}\right\}{\bf A}^{T}\right)_{11}\geq\epsilon_{\text{min}}\sum_{i} A_{1i}^{2}.
\end{aligned}
\end{equation}
Similarly
\begin{equation}
\begin{aligned}
\Sigma_{22}&\geq\epsilon_{\text{min}}\sum_{j} A_{2j}^{2}.
\end{aligned}
\end{equation}
Therefore, 
\begin{equation}
\begin{aligned}
\Sigma_{11}\Sigma_{22}&\geq\epsilon_{\text{min}}^{2}\left(\sum_{i} A_{1i}^{2}\right)\left(\sum_{j} A_{2j}^{2}\right)=\epsilon_{\text{min}}^{2}||{\bf A}_{1}||^{2}\cdot||{\bf A}_{2}||^{2},
\end{aligned}
\end{equation}
where $||{\bf A}_{1,2}||$ mean the Euclidean norms of ${\bf A}_{1,2}$, and ${\bf A}_{1,2}$ are vectors whose elements are $A_{1i}$ and $A_{2j}$, respectively. 
Note that ${\bf A}$ is a real sympletic matrix, according to the symplectic condition ${\bf A}{\bf S}{\bf A}^{T}={\bf S}$, we have 
\begin{equation}
({\bf A}_{1},{\bf S}{\bf A}_{2})=1,
\end{equation}
According to Cauchy-Schwarz inequality, we have
\begin{equation}
1=({\bf A}_{1},{\bf S}{\bf A}_{2})\leq||{\bf A}_{1}||\cdot||{\bf S}{\bf A}_{2}||
\end{equation}
As ${\bf S}{\bf S}^{T}=-I$, the spectral norm of ${\bf S}$ is 1, then
\begin{equation}
||{\bf S}{\bf A}_{2}||\leq||{\bf A}_{2}||.
\end{equation}
Therefore,
\begin{equation}
||{\bf A}_{1}||\cdot||{\bf A}_{2}||\geq1.
\end{equation}
QED.

Actually our result Eq.~(\ref{eq:ES}) can be viewed as a special example of the classical uncertainty principle.


\subsubsection{Minimum Emittance Theorem}
Actually there is a stronger inequality compared to the classical uncertainty principle, i.e, the minimum emittance theorem,
\begin{equation}
\begin{aligned}
\epsilon_{x,\text{pro}}^{2}=\Sigma_{11}\Sigma_{22}-\Sigma_{12}^{2}&\geq\epsilon_{\text{min}}^{2}\\
\epsilon_{y,\text{pro}}^{2}=\Sigma_{33}\Sigma_{44}-\Sigma_{34}^{2}&\geq\epsilon_{\text{min}}^{2}\\
\epsilon_{z,\text{pro}}^{2}=\Sigma_{55}\Sigma_{66}-\Sigma_{56}^{2}&\geq\epsilon_{\text{min}}^{2}.
\end{aligned}
\end{equation}
The proof is based on the fact that any $6\times6$ $\Sigma$ matrix can be transformed to be a $2\times2$ sub-diagonal form using a $6\times6$ symplectic matrix. For example, if we want to make the upper-left $2\times2$ matrix of $\Sigma$ to be diagonal, then we just need
\begin{equation}
\Sigma_{\text{new}}=
\left(
\begin{array}{cccccc}
\frac{1}{\sqrt{\beta _1}} & 0 & 0 & 0 & 0 & 0 \\
\frac{\alpha _1}{\sqrt{\beta _1}} & \sqrt{\beta _1} & 0 & 0& 0 & 0  \\
0 & 0 & 1 & 0 & 0 & 0 \\
0 & 0 & 0 & 1& 0 & 0  \\
0 & 0 & 0 & 0 & 1 &  0 \\
0 & 0 & 0 & 0 & 0 & 1  \\
\end{array}
\right)\Sigma_{\text{old}}\left(
\begin{array}{cccccc}
\frac{1}{\sqrt{\beta _1}} & 0 & 0 & 0 & 0 & 0 \\
\frac{\alpha _1}{\sqrt{\beta _1}} & \sqrt{\beta _1} & 0 & 0& 0 & 0  \\
0 & 0 & 1 & 0 & 0 & 0 \\
0 & 0 & 0 & 1& 0 & 0  \\
0 & 0 & 0 & 0 & 1 &  0 \\
0 & 0 & 0 & 0 & 0 & 1  \\
\end{array}
\right)^{T},
\end{equation}
with
\begin{equation}
\begin{aligned}
\alpha_1&=-\frac{\Sigma_{12}}{\epsilon_{x,\text{pro}}},\\
\beta_1&=\frac{\Sigma_{11}}{\epsilon_{x,\text{pro}}},\\
\gamma_1&=\frac{\Sigma_{22}}{\epsilon_{x,\text{pro}}}.\\
\end{aligned}
\end{equation}
The upper-left $2\times2$ block of $\Sigma_{\text{new}}$ is then
\begin{equation}
\left(
\begin{array}{cccccc}
\epsilon_{x,\text{pro}} & 0  \\
0 & \epsilon_{x,\text{pro}}   \\
\end{array}
\right)
\end{equation}
As shown in Sec.~\ref{sec:CS}, a linear symplectic transport does not change the eigen-emittances of a Gaussian beam. Then according to the classical uncertainty principle, we thus have the minimum emittance theorem.

We remind the readers that the minimum emittance theorem is for the emittances, and does not give any limit for the size in a specific phase space coordinate, for example $z$. The bunch length in principle can be any desired short value, except that the optics manipulation would be demanding. The larger the minimum eigen-emittance, the more demanding the optics manipulation is. So in order to realize a short bunch length with a reasonable lattice optics, we need at least one of the three eigen emittances small. As mentioned, our application of TLC for bunch compression in SSMB is based on the fact that the vertical emittance in a planar $x$-$y$ uncoupled storage ring is rather small.

\subsection{Emittance Exchange}
\subsubsection{Lattice Condition}
For completeness of the investigation, it might also be helpful to make a short discussion on the relation between our TLC analysis and the transverse-longitudinal EEX \cite{cornacchia2002transverse,emma2006transverse,fliller2006general,xiang2011emittance,jiang2011emittance,wang2020transverse}. For a complete EEX, we need the transfer matrix of the form
\begin{equation}
	{\bf T}=\left(
	\begin{matrix}
	{\bf 0}&{\bf B}\\
	{\bf C}&{\bf 0}
	\end{matrix}
	\right).
	\end{equation}
Therefore, EEX is a special case in the context of our problem definition, i.e., the final beam is then also $y$-$z$ decoupled. As can be seen from Eq.~(\ref{eq:TransportMatrixSub}), the application of a normal RF cannot accomplish a complete EEX as $T_{65}=h\neq0$. PEHG and ADM can thus be viewed as partial EEXs. In contrast, as can bee seen from Eq.~(\ref{eq:TransportMatrixSub2}), a transverse deflecting RF as shown in Fig.~\ref{fig:Chap3-Transverse} can accomplish a complete EEX. All we need is to add another condition to Eq.~(\ref{eq:TRFConditions}), i.e.,
\begin{equation}\label{eq:EEX}
	dt+1=0.
	\end{equation}
After some straightforward algebra, the relations in Eqs.~(\ref{eq:TRFConditions}) and (\ref{eq:EEX}) can be summarized in an elegant form as follows
\begin{equation}\label{eq:EEXCondition}
	\begin{aligned}
	t&=-\frac{1}{d},\\
	D&=R_{34} d'+R_{33}d,\\
	D'&=R_{44} d'+R_{43}d.
	\end{aligned}
	\end{equation}
We notice that the relations in Eq.~(\ref{eq:EEXCondition}) for EEX have been obtained earlier in Ref.~\cite{fliller2006general}.  Note that the above relations mean that the lattices upstream and downstream the transverse deflecting RF are not mirror symmetry with respect to each other~\cite{xiang2011emittance}.
Under the conditions given in Eq.~(\ref{eq:EEXCondition}), we have
\begin{equation}\label{eq:EEXMatrix}
	\begin{aligned}
	{\bf T}&=\left(\begin{matrix}
	0&0&-\frac{R_{34}}{d}& d R_{33}-R_{34}\frac{r_{56}-dd'}{d}\\
	0&0&-\frac{R_{44}}{d}& d R_{43}-R_{44}\frac{r_{56}-dd'}{d}\\
	dr_{43}-r_{33}\frac{R_{56}+dd'}{d}&dr_{44}-r_{34}\frac{R_{56}+dd'}{d}&0&0\\
	-\frac{r_{33}}{d}&-\frac{r_{34}}{d}&0&0\\
	\end{matrix}\right)\\
	&=\left(
	\begin{array}{cccc}
	0 & 0 & T_{35} & T_{36} \\
	0 & 0 & T_{45} & T_{46} \\
	T_{53} & T_{54} & 0 & 0 \\
	T_{63} & T_{64} & 0 & 0 \\
	\end{array}
	\right).
	\end{aligned}
	\end{equation}

\subsubsection{Applications}
\begin{figure}[tb] 
	\centering 
	\includegraphics[width=1\columnwidth]{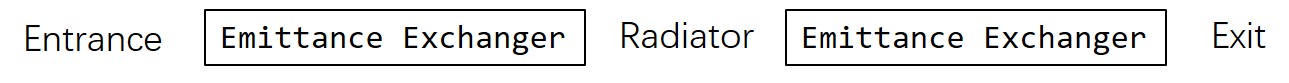}
	\caption{
		\label{fig:Chap3-2EEX} 
		Application of two transverse-longitudinal emittance exchangers to manipulate the bunch length in a storage ring.
	}
\end{figure}

In order to apply EEX to generate short bunch in a storage ring on a turn-by-turn basis, another EEX might be needed following the radiator to swap back the $\epsilon_{y}$ and $\epsilon_{z}$ to maintain the ultra-small vertical emittance $\epsilon_{y}$. Therefore, the application of two $y$-$z$ EEXs for bunch length manipulation in a
storage ring can be envisioned as shown in Fig.~\ref{fig:Chap3-2EEX}. As mentioned, the potential issue of applying transverse deflecting RFs or TEM01 mode lasers for bunch compression is the large $\beta_{y}$ at the modulator, especially if the target bunch length is as short as nm level like that in EUV SSMB. But the scheme looks appealing in the THz wavelength region. Besides, there can be other innovative angular modulation scheme invented such that the issue of large $\beta_{y}$ at the modulator can solved in the future, even if our target radiation wavelength is still in EUV. If there is only one transverse-longitudinal EEX, the ring will then be a transverse-longitudinal M\"obius accelerator \cite{talman1995proposed}, which is also an interesting topic we are not going into in this dissertation.

Now we consider the application of two $y$-$z$ EEXs for bunch length manipulation in a storage ring as shown in Fig.~\ref{fig:Chap3-2EEX}. The motivation is still to make use of the fact that the vertical emittance $\epsilon_{y}$ is rather small in a planar $x$-$y$ uncoupled ring. We hope that the total insertion does not introduce $y$-$z$ coupling, i.e.,
\begin{equation}
\begin{aligned}
{\bf M}&={\bf T}_{\text{downstream}}{\bf T}_{\text{upstream}}=\left(
\begin{array}{cccc}
{\bf A'} & {\bf 0} \\
{\bf 0} & {\bf E'} \\
\end{array}
\right),
\end{aligned}
\end{equation}
such that the small $\epsilon_{y}$ can be kept.
The first natural idea is to add an inverse EEX unit following the EEX, 
\begin{equation}
{\bf T}^{-1}=\left(
\begin{array}{cccc}
0 & 0 & T_{64} & -T_{54} \\
0 & 0 & -T_{63} & T_{53} \\
T_{46} & -T_{36} & 0 & 0 \\
-T_{45} & T_{35} & 0 & 0 \\
\end{array}
\right),
\end{equation}
then the total insertion will be an identity matrix and be transparent to the ring.
The issue of this approach, however, is that we need to design the downstream beamline with an $R_{56}$ having opposite sign to the upstream beamline, which might be a challenging task if we aim at a compact lattice.

The second natural idea is to implement the mirror symmetry of the upstream beamline as the downstream beamline, which is straightforward for the lattice design. 
A beamline matrix $T$ and its mirror image $T_{\text{mirror}}$ is related to each other according to~\cite{berz2015introduction,chao2002lecture}
\begin{equation}
{\bf T}_{\text{mirror}}{\bf U}{\bf T}={\bf U},\ {\bf U}=\left(
\begin{array}{cccc}
1 & 0 & 0 & 0 \\
0 & -1 & 0 & 0 \\
0 & 0 & -1 & 0 \\
0 & 0 & 0 & 1 \\
\end{array}
\right).
\end{equation}
Therefore,
\begin{equation}
\begin{aligned}
{\bf T}_{\text{mirror}}&={\bf U}{\bf T}^{-1}{\bf U}^{-1}=\left(
\begin{array}{cccc}
0 & 0 & -T_{64} & -T_{54} \\
0 & 0 & -T_{63} & -T_{53} \\
-T_{46} & -T_{36} & 0 & 0 \\
-T_{45} & -T_{35} & 0 & 0 \\
\end{array}
\right).
\end{aligned}
\end{equation}
Note, however, the transfer matrix of the total insertion in this case is generally not an identity matrix,
\begin{equation}
\begin{aligned}
{\bf M}&={\bf T}_{\text{mirror}}{\bf T}=\left(
\begin{array}{cccc}
{\bf A'} & {\bf 0} \\
{\bf 0} & {\bf E'} \\
\end{array}
\right),
\end{aligned}
\end{equation}
with
\begin{equation}
\begin{aligned}
{\bf A'}&=\left(
\begin{matrix}
-(T_{53} T_{64}+T_{54} T_{63}) & -2 T_{54} T_{64}\\
-2 T_{53} T_{63} & -(T_{53} T_{64}+T_{54} T_{63})
\end{matrix}
\right),\\
{\bf E'}&=\left(
\begin{matrix}
-(T_{35} T_{46}+T_{36} T_{45}) & -2 T_{36} T_{46}\\
-2 T_{35} T_{45} & -(T_{35} T_{46}+T_{36} T_{45})
\end{matrix}
\right).
\end{aligned}
\end{equation}
A matching cell may be needed to tailor the impact of the insertion on the optics of the ring.

A special case of EEX is the phase space exchange (PSX), i.e., the exchange happens in the phase space variables apart from a magnification factor. In this case, a PSX followed by its mirror can form an identity or a negative identity matrix.\\ 
Case one:
\begin{equation}
\begin{aligned}
{\bf T}=
\left(
\begin{array}{cccc}
0 & 0 & 0 & m_1 \\
0 & 0 & -\frac{1}{m_1} & 0 \\
0 & m_2 & 0 & 0 \\
-\frac{1}{m_2} & 0 & 0 & 0 \\
\end{array}
\right),\ {\bf T}_{\text{mirror}}=
\left(
\begin{array}{cccc}
0 & 0 & 0 & -m_2 \\
0 & 0 & \frac{1}{m_2} & 0 \\
0 & -m_1 & 0 & 0 \\
\frac{1}{m_1} & 0 & 0 & 0 \\
\end{array}
\right),\ {\bf M}={\bf T}_{\text{mirror}}{\bf T}={\bf I},
\end{aligned}
\end{equation}
Case two:
\begin{equation}
\begin{aligned}
{\bf T}=
\left(
\begin{array}{cccc}
0 & 0 & m_1 & 0 \\
0 & 0 & 0 & \frac{1}{m_1} \\
m_2 & 0 & 0 & 0 \\
0 & \frac{1}{m_2} & 0 & 0 \\
\end{array}
\right),\ {\bf T}_{\text{mirror}}=
\left(
\begin{array}{cccc}
0 & 0 & -\frac{1}{m_2} & 0 \\
0 & 0 & 0 & -m_2 \\
-\frac{1}{m_1} & 0 & 0 & 0 \\
0 & -m_1 & 0 & 0 \\
\end{array}
\right),\ {\bf M}={\bf T}_{\text{mirror}}{\bf T}=-{\bf I}.
\end{aligned}
\end{equation}
According to Eq.~(\ref{eq:EEXMatrix}), for case one, we need 
\begin{equation}
\begin{aligned}
-\frac{R_{34}}{d}&=0,\\
d R_{43}+d'R_{44}-\frac{r_{56} R_{44}}{d}&=0,\\
dr_{43}-d'r_{33}-\frac{r_{33}R_{56}}{d}&=0,\\
-\frac{r_{34}}{d}&=0,
\end{aligned}
\end{equation}
and 
\begin{equation}
{\bf T}=\left(\begin{matrix}
0&0&0& \frac{d}{R_{44}}\\
0&0&-\frac{R_{44}}{d}& 0\\
0&\frac{d}{r_{33}}&0&0\\
-\frac{r_{33}}{d}&0&0&0\\
\end{matrix}\right).
\end{equation}
For case two, we need 
\begin{equation}
\begin{aligned}
d R_{33}+d'R_{34}-\frac{r_{56} R_{34}}{d}&=0,\\
-\frac{R_{44}}{d}&=0,\\
dr_{44}-d'r_{34}-\frac{r_{34}R_{56}}{d}&=0,\\
-\frac{r_{33}}{d}&=0,
\end{aligned}
\end{equation}
and
\begin{equation}
{\bf T}=\left(\begin{matrix}
0&0&-\frac{R_{34}}{d}& 0\\
0&0&0& -\frac{d}{R_{34}}\\
-\frac{d}{r_{34}}&0&0&0\\
0&-\frac{r_{34}}{d}&0&0\\
\end{matrix}\right).
\end{equation}

\section{Nonlinear Coupling}\label{sec:NonlinearCoupling}

\subsection{Average Path Length Dependence on Betatron Amplitudes}

After investigating the linear transverse-longitudinal coupling, we will now examine the case of nonlinear coupling. However, here we consider only the second-order path lengthening or shortening from betatron oscillations. A general discussion of the nonlinear dynamics is beyond the scope of this dissertation. The second-order transverse-longitudinal coupling considered here originates from a dependence of the average path length on the betatron oscillation amplitudes, which can be expressed by a concise formula
\begin{equation}\label{eq:deltaC}
\Delta C=-2\pi(\xi_{x}J_{x}+\xi_{y}J_{y}),
\end{equation}
where $\Delta C$ is the average path-length deviation relative to the ideal particle, and $\xi_{x,y}=\frac{d\nu_{x,y}}{d\delta}$ and $J_{x,y}$ are the horizontal (vertical) chromaticity and betatron invariant, respectively. This simple relation is a result of the symplecticity of the Hamiltonian dynamics, as pointed out by Forest \cite{forest1998beam} and other authors \cite{chao2002lecture,berg2007amplitude}. It is called a second-order coupling because the betatron invariant is a second-order term with respect to the transverse position and angle. Note that Eq.~(\ref{eq:deltaC}) is accurate only for the cases of multiple passes or multiple betatron oscillations as it is a betatron-phase-averaged result. For the case of a single pass with only a few betatron oscillations, there will be an extra term, depending on the betatron phase advance, in the path length formula. 

This path length effect has previously been theoretically analyzed by several authors in different contexts \cite{artamonov1989possibility,emery1993proceedings,forest1998beam,chao2002lecture,shoji2005dependence,berg2007amplitude}.  Due to this effect, particles with different betatron amplitudes lose synchronization with each other when traversing a lattice with nonzero chromaticity. This leads to a stringent requirement on the beam emittance for FELs in the X-ray regime (XFELs), as microbunching can be smeared out by this effect when the beam is traveling through the undulator \cite{sessler1992radio}. This effect is also crucial in non-scaling fixed-field alternating-gradient (FFAG) accelerators for muon acceleration \cite{berg2007amplitude}, as a muon beam typically has a large emittance. Furthermore, the natural chromaticities of a linear non-scaling FFAG accelerator are usually not corrected to achieve a large transverse acceptance. This effect may also have an impact on the momentum and dynamic aperture in a storage ring \cite{takao2008impact,hoummi2019beam}, for example, due to the Touschek scattering-induced large betatron amplitude or the large natural chromaticity in a low-emittance lattice. It is also relevant in other applications demanding particle synchronization. In this dissertation, we wish to investigate and emphasize the importance of this nonlinear transverse-longitudinal coupling in precision longitudinal dynamics in a storage ring, such as SSMB. 

\subsection{Energy Widening and Distortion}

As mentioned, the second-order transverse-longitudinal coupling effect can disperse microbunching in XFELs. Methods of overcoming this influence are referred to as ``beam conditioning". Several such methods have been proposed since the first publication of Ref.~\cite{sessler1992radio}. The basic idea of these proposals is to compensate the difference in path length through a difference in velocity by establishing a correlation between the betatron amplitude and the particle energy. In a storage ring, unlike in a single-pass device, the RF cavity will ``condition" the beam automatically, causing all particles to synchronize with it in an average sense through phase stabilization (bunching). This is accomplished by introducing a betatron-amplitude-dependent energy shift to compensate for the path-length difference arising from the betatron oscillations,
\begin{equation}\label{eq:momentumshift}
\Delta\delta = -\frac{\Delta C}{\alpha C_{0}},
\end{equation}
where $\alpha $ is the momentum compaction factor of the ring. This shift will result in the beam energy widening in a quasi-isochronous ring with nonzero chromaticity, because different particles have different betatron invariants~\cite{shoji2005dependence}. This widening will become more significant with the decreasing of the momentum compaction.

\begin{figure}[tb] 
	\centering 
	\includegraphics[width=1\textwidth]{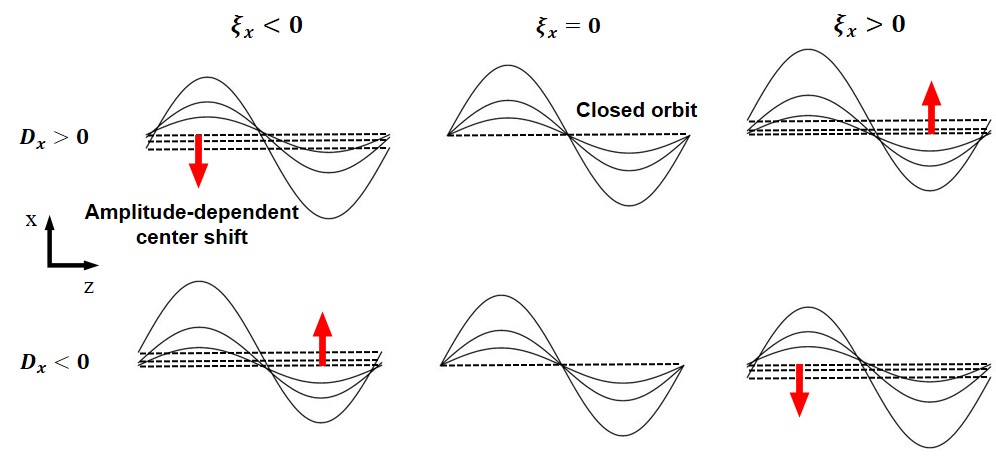}
	\caption{
		\label{fig:ADCS} 
		Physical picture of the amplitude-dependent shift of betatron oscillation center in the case of a positive momentum compaction. Only a horizontal betatron oscillation is considered in this illustration. 
	}
\end{figure}

Due to the energy shift, there will also be an amplitude-dependent shift in the betatron oscillation center at dispersive locations. The shift direction is determined by the signs of $\alpha $, $\xi_{x,y}$ and $D_{x,y}$, and the magnitude of the shift is determined by the magnitudes of $J_{x,y}$, $\alpha $, $\xi_{x,y}$ and $D_{x,y}$. The physical pictures of the betatron center shift resulting from this effect are shown in Fig.~\ref{fig:ADCS}.

When quantum excitation is also taken into account, the total relative energy deviation of a particle with respect to the ideal particle is
\begin{equation}
\delta=\Delta\delta+\delta_{\text{qe}},
\end{equation}
where $\delta_{\text{qe}}$ represents the quantum excitation contribution.  
Finding a general analytical formula for the steady-state distribution of the particles is a complex task and, at the same time, not very useful. Simpler expressions can be obtained based on reasonable approximations. Since the vertical emittance is usually much smaller than the horizontal emittance in a planar uncoupled ring, here, we consider only the contribution from the horizontal emittance. When the coupling is not very strong, the distributions of $J_{x}$ and $\delta_{\text{qe}}$ are still approximately exponential and Gaussian, respectively, and are independent of each other,
\begin{equation}\label{eq:twoDistribution}
\begin{aligned}
\psi(J_{x})=\frac{1}{2\pi\epsilon_{x0}}e^{-\frac{J_{x}}{\epsilon_{x0}}},\ \psi(\delta_{\text{qe}})=\frac{1}{\sqrt{2\pi}\sigma_{\delta0}}e^{-\frac{\delta_{\text{qe}}^{2}}{2\sigma_{\delta0}^{2}}},
\end{aligned}
\end{equation}
where $\epsilon_{x0}$ and $\sigma_{\delta0}$ are the natural horizontal emittance and energy spread. The distribution of $\delta$ is thus an exponentially modified Gaussian because it is the sum of an exponential and a normal random variable,
\begin{equation}\label{eq:deltaDistributin}
\psi(\delta)=\frac{|\lambda|}{2}e^{\frac{\lambda\left(\lambda\sigma_{\delta0}^{2}-2\delta\right)}{2}}\text{erfc}\left[\frac{\text{sgn}\left(\lambda)(\lambda\sigma_{\delta0}^{2}-\delta\right)}{\sqrt{2}\sigma_{\delta0}}\right],
\end{equation} 
where $\lambda=\frac{\alpha C_{0}}{2\pi\xi_{x}\epsilon_{x0}}$, sgn(x) is the sign function and erfc(x) is the complementary error function, defined as
$
\text{erfc}(x)=1-\text{erf}(x)=\frac{2}{\sqrt{\pi}}\int_{x}^{\infty}e^{-t^{2}}dt.
$	The direction of long non-Gaussian tail of the energy distribution is determined by the signs of $\alpha$ and $\xi_{x}$. 

Because of the dispersion and dispersion angle, the non-Gaussian particle energy distribution can also be reflected in the transverse dimension.  When this nonlinear coupling is considered, the horizontal position and angle of a particle in the storage ring are
\begin{gather}\label{eq:xx}
\begin{aligned}
x&=\sqrt{2J_{x}\beta_{x}}\cos{\varphi_{x}} + D_{x}\left(\delta_{\text{qe}}+\frac{2\pi\xi_{x}J_{x}}{\alpha C_{0}}\right),\\
x'&=-\sqrt{2J/\beta_{x}}(\alpha_{x}\cos{\psi_{x}+\sin{\psi_{x}}}) + D_{x}'\left(\delta_{\text{qe}}+\frac{2\pi\xi_{x}J_{x}}{\alpha C_{0}}\right).
\end{aligned}
\end{gather}
It is assumed that the concept of the Courant-Snyder functions is still approximately valid in Eq.~(\ref{eq:xx}).

With these approximations, the variance of $\delta$ is then
\begin{equation}\label{eq:energWidening2nd}
\sigma_{\delta}^{2}=\sigma_{\delta0}^{2}+\left(\frac{2\pi\epsilon_{x0}\xi_{x}}{\alpha C_{0}}\right)^{2}.
\end{equation}
By assuming the MLS parameters shown in Tab.~\ref{tab:MLS_para2} and applying $\alpha=1\times10^{-4}$ and $\xi_{x}=2$, one can find that the energy spread contributed by this effect can be as significant as its natural value.

As discussed in Ref.~\cite{shoji2005dependence}, a shift in the energy center corresponds to a shift in the synchronous RF phase $\phi_{s}$, 
\begin{equation}
\Delta \phi_{s}\approx J_{s}\tan\phi_{s}\Delta\delta,
\end{equation}
where $J_{s}$ is the longitudinal damping partition number and nominally $J_{s}\approx2$. Therefore, particles with different betatron amplitudes will oscillate around different fixed points in the longitudinal dimension, thus lengthening the bunch. The change in the synchronous RF phase in a unit of the longitudinal coordinate, $\Delta z_{s}$, is related to the relative change in energy, $\Delta\delta$, according to
\begin{equation}\label{eq:DeltaCompare}
\frac{\Delta z_{s}}{\sigma_{z0}}=\frac{\nu_{s}J_{s}\tan\phi_{s}}{f_{\text{RF}}/f_{\text{rev}}|\alpha|}\frac{\Delta \delta}{\sigma_{\delta0}}\propto\frac{1}{\sqrt{|\alpha|}}\frac{\Delta \delta}{\sigma_{\delta0}},
\end{equation}
where $\nu_{s}$ is the synchrotron tune and $\sigma_{z0}$ and $\sigma_{\delta0}$ are the natural bunch length and energy spread, respectively. 

\begin{figure}[tb] 
	\centering 
	\includegraphics[width=0.245\textwidth]{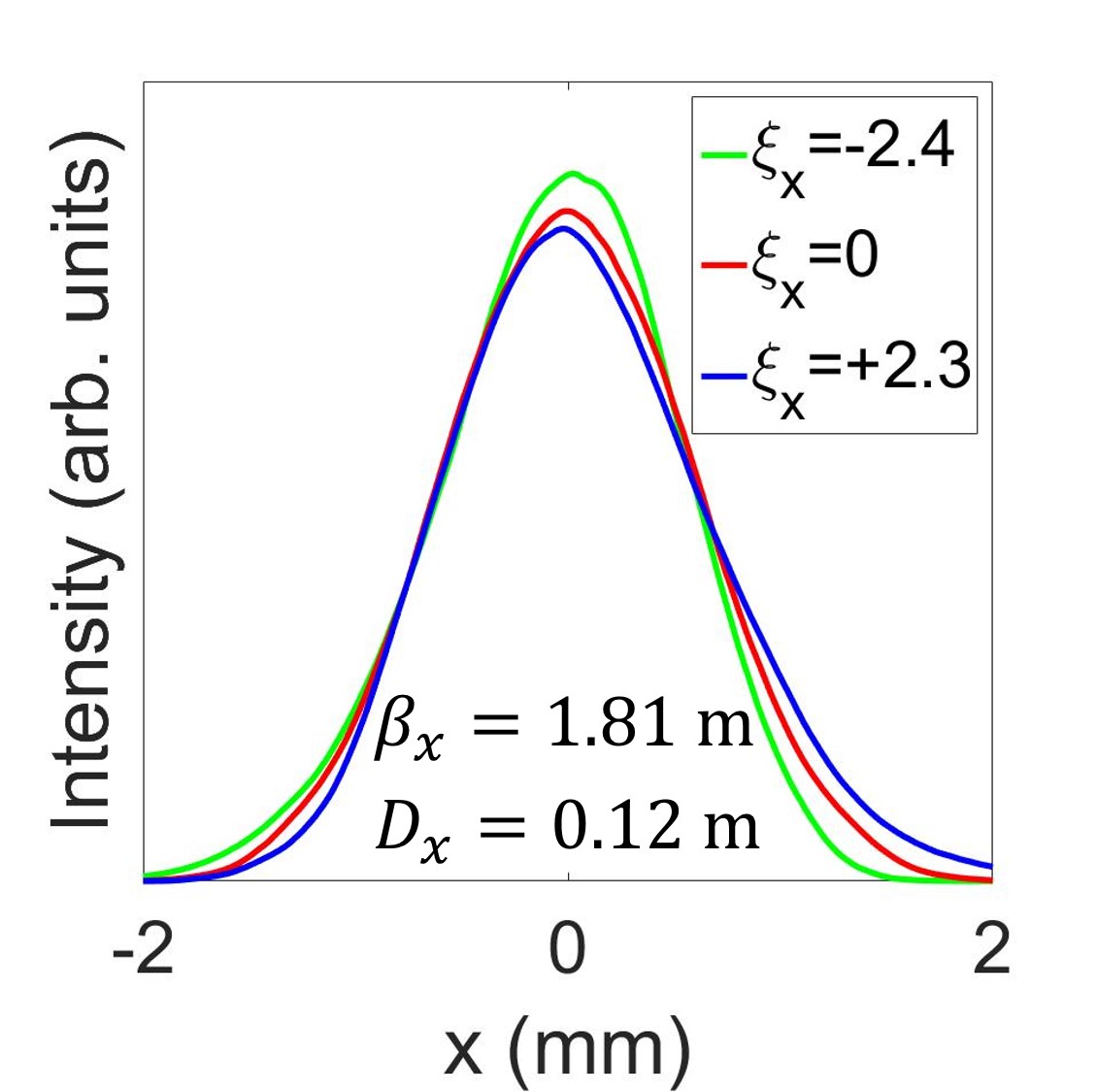}
	\includegraphics[width=0.245\textwidth]{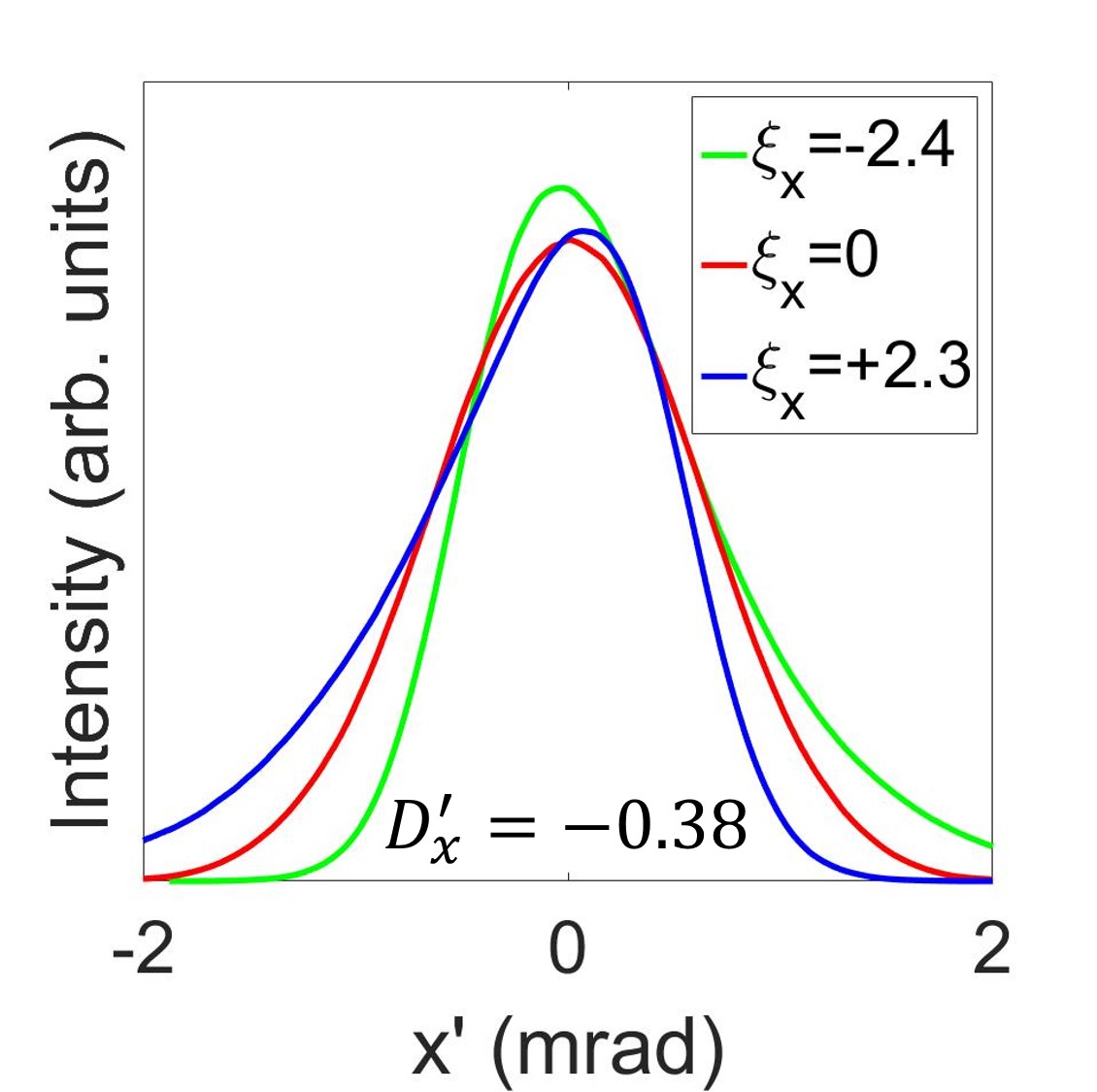}
	\includegraphics[width=0.245\textwidth]{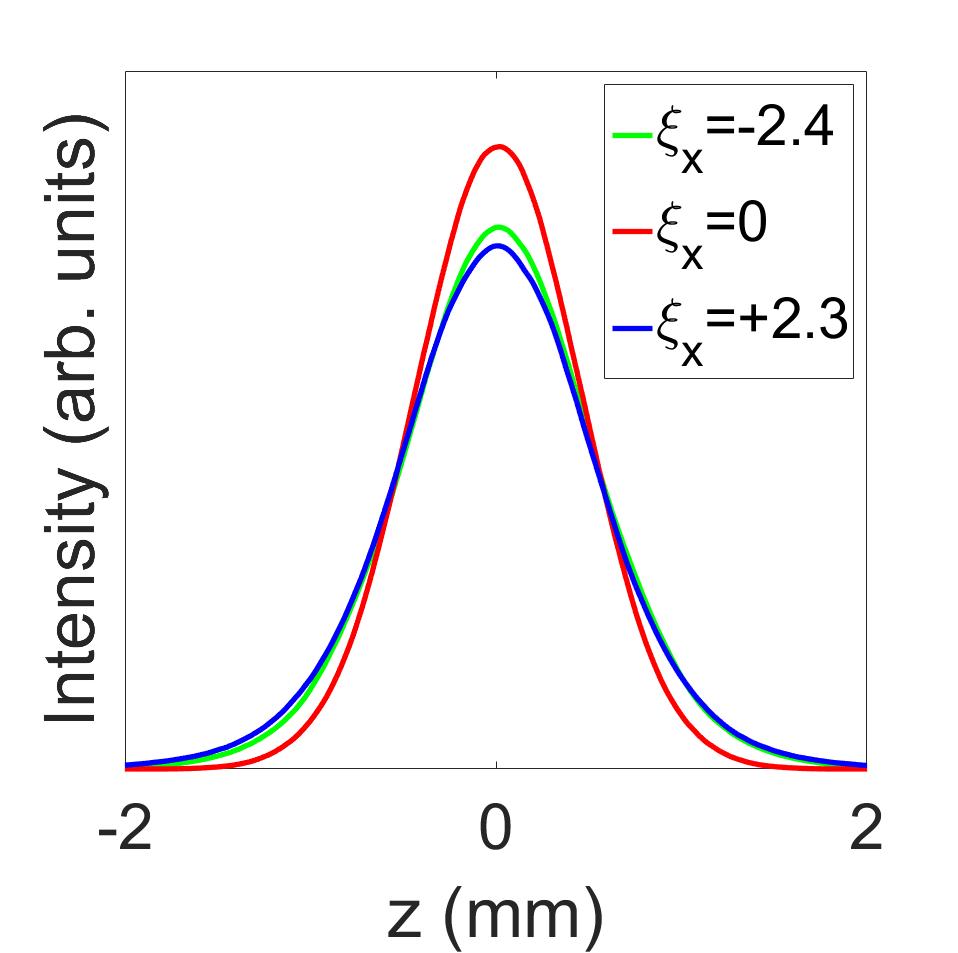}
	\includegraphics[width=0.245\textwidth]{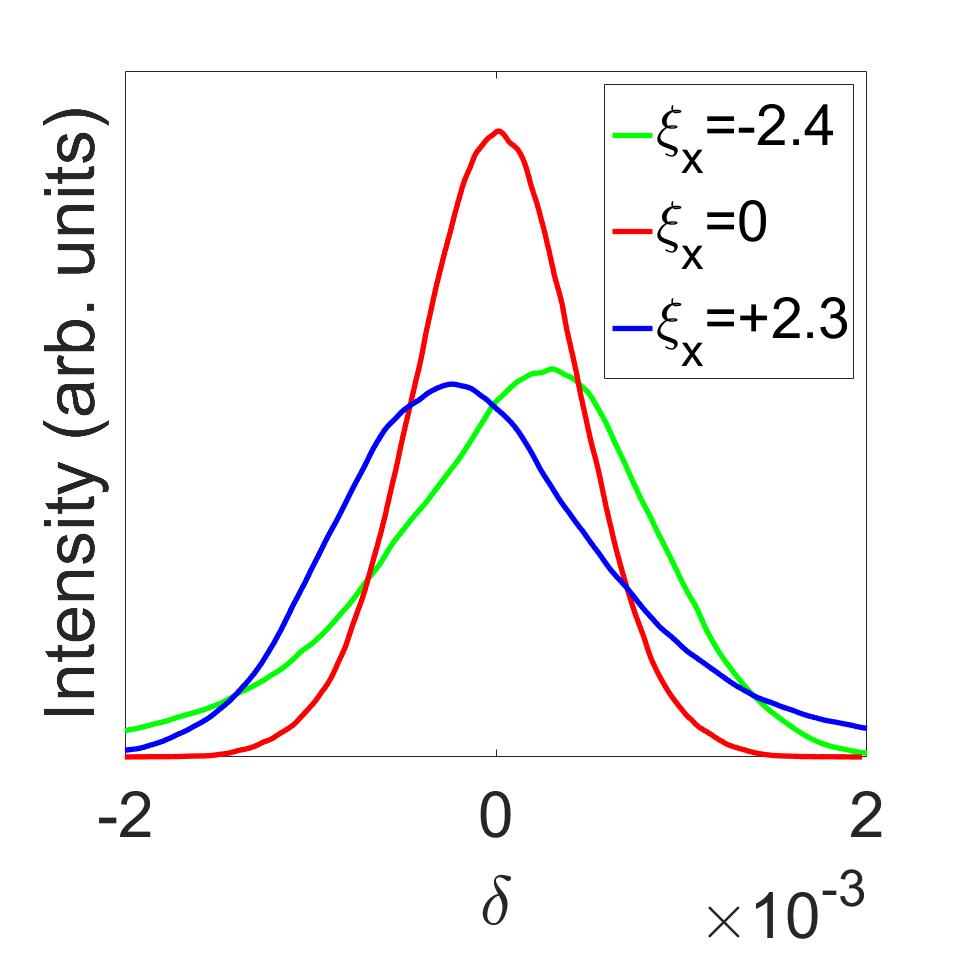}\\
	\includegraphics[width=0.245\textwidth]{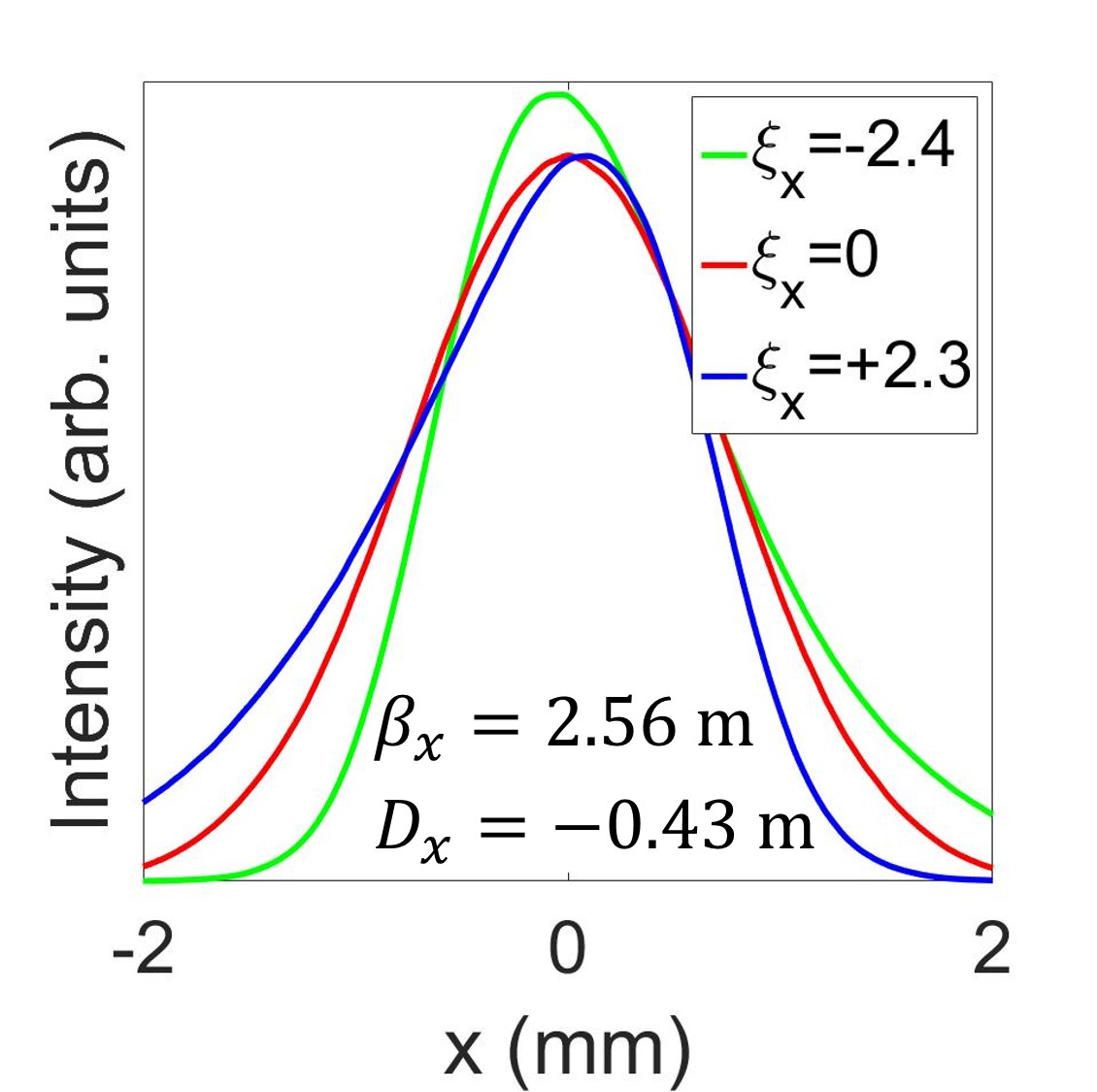}
	\includegraphics[width=0.245\textwidth]{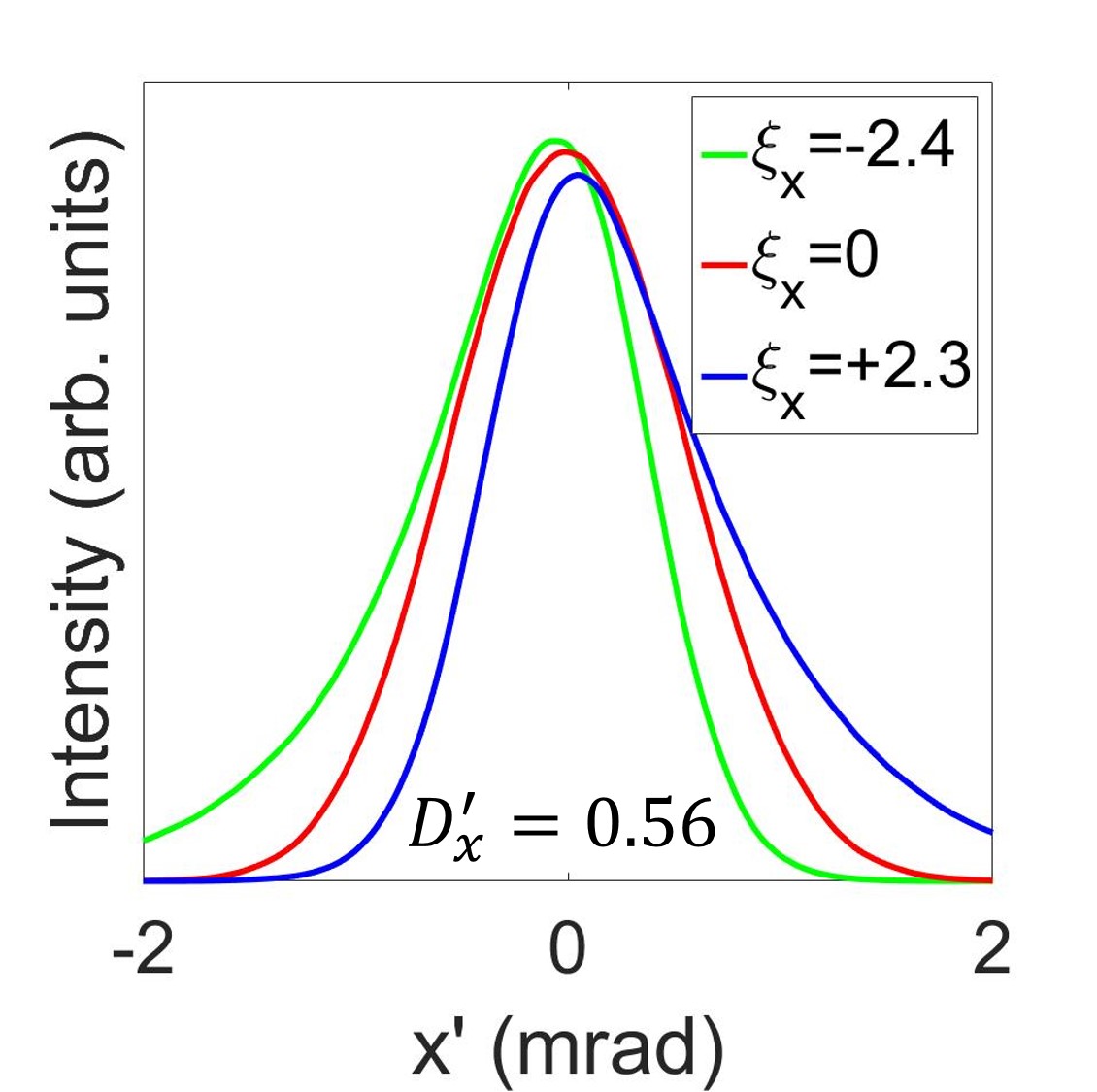}
	\includegraphics[width=0.245\textwidth]{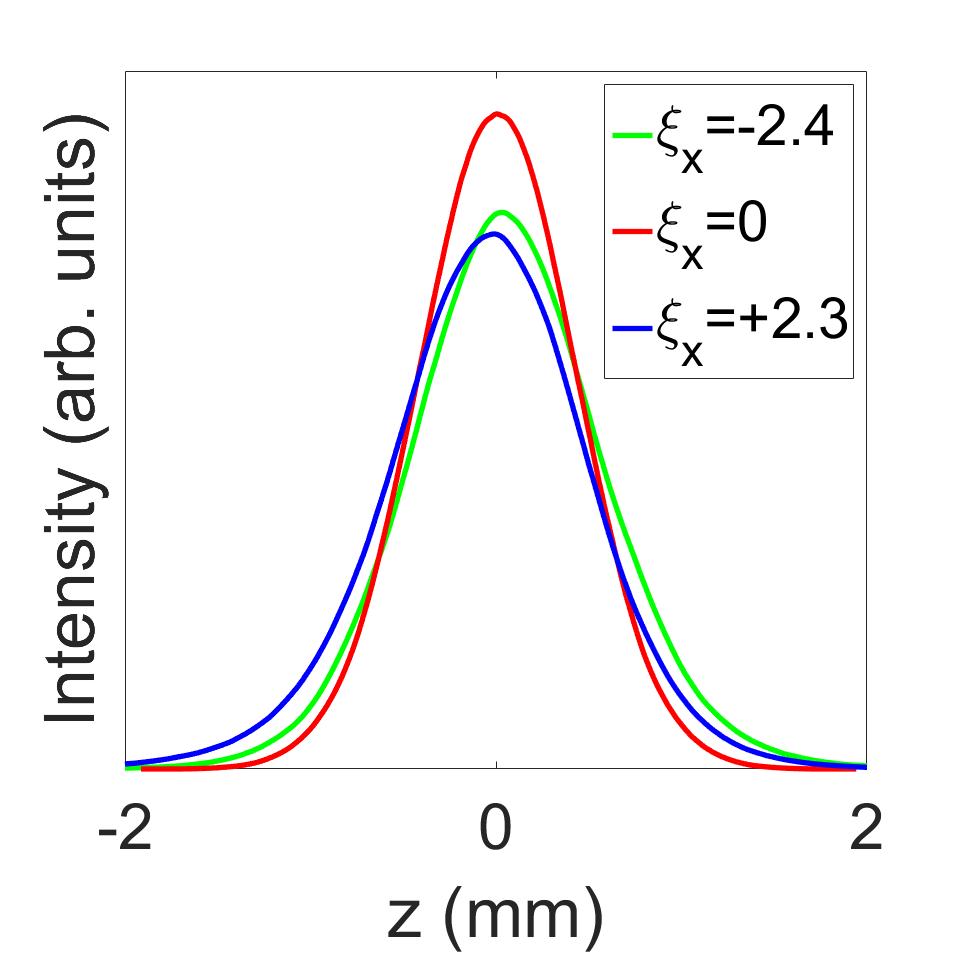}
	\includegraphics[width=0.245\textwidth]{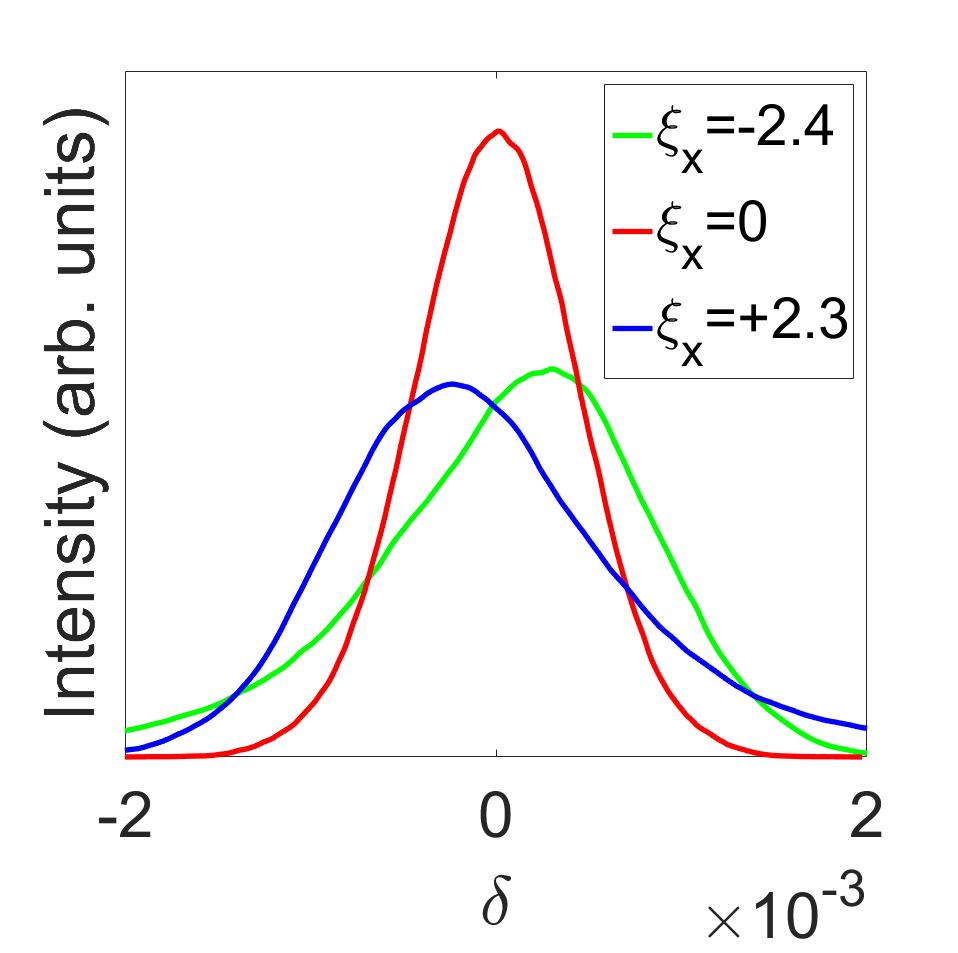}
	\caption{
		\label{fig:second} 
		Energy widening, bunch lengthening and distortion from a Gaussian distribution induced by a nonvanishing horizontal chromaticity. From up to bottom, the particle tracking results for distributions of $x$, $x'$, $z$ and $\delta$ are shown at two dispersive locations in the MLS under three different horizontal chromaticities $\xi_{x}$. The direction of the long non-Gaussian tail for $\delta$ is related to the signs of $\alpha$ and $\xi_{x}$, while for $x$ and $x'$ they are also dependent on $D_{x}$ and $D_{x}'$, respectively. The simulation was conducted using the code ELEGANT \cite{borland2000elegant} with a beam energy of 630 MeV, an RF voltage of 500 kV and the application of $\alpha =1\times10^{-4}$. In each simulation, eight particles were tracked for $5\times10^{6}$ turns, corresponding to approximately 73 longitudinal radiation damping times.
	}
\end{figure}

The critical value of alpha, $\alpha_{\text{c}}$, when the relative change of bunch length and energy spread are the same can be calculated to be 
\begin{equation}\label{eq:criticalAlpha}
\frac{\nu_{s}J_{s}\tan\phi_{s}}{f_{\text{RF}}/f_{\text{rev}}|\alpha_{\text{c}}|}=1\Rightarrow|\alpha_{\text{c}}|=\frac{J_{s}T_{0}\tan\phi_{s}}{\pi f_{\text{RF}}/f_{\text{rev}}\tau_{\delta}},
\end{equation}
where $\tau_{\delta}=1/\alpha_{\text{L}}=\frac{2E_{0}}{J_{s}U_{0}}T_{0}$ is the longitudinal damping time. As an example, we use the MLS parameters given in Tab.~\ref{tab:MLS_para2} and consider the application of an RF voltage of $500$~kV, which corresponds to a synchronous RF phase of $\phi_{s}=0.018$~rad. The critical value of alpha is then $|\alpha_{\text{c}}|\approx2.1\times10^{-9}$, which is about four orders of magnitude smaller than the alpha value reachable at the present MLS. Therefore, the relative bunch lengthening resulting from this effect is much less significant than the corresponding energy widening at the MLS.

Several particle tracking simulations were conducted using the MLS lattice using the parameters presented in Tab.~\ref{tab:MLS_para2} to confirm the analysis. Two dispersive locations, with different signs and magnitudes of $D_{x}$, were selected as the observation points in the simulations. The simulation results are shown in Fig.~\ref{fig:second}. The energy widening and distortion from Gaussian behaviors are as expected and, indeed, are more significant than the bunch lengthening when $\alpha=1\times10^{-4}$. At the two observation points, widening and distortion of the particle energy distribution also manifest in the transverse dimension through $D_{x}$ and $D_{x}'$. The related optic functions at the two observation points are also shown in the profiles of $x$ and $x'$. Note that the directions of the long non-Gaussian tails of the profiles and their relations to the signs of $\xi_{x}$, $D_{x}$ and $D_{x}'$. We conclude that the simulation results agree well with the analysis and physical pictures presented above.

\begin{table}[tb]
	\centering
	\caption{\label{tab:MLS_para2}
		Parameters of the MLS in the experiment.}
	\begin{tabular}{lll}
		\hline
		Parameter & \multicolumn{1}{c}{\textrm{Value}}  & Description \\
		\hline
		$C_0$ & 48  m & Ring circumference \\
		$E_0$ & 630  MeV & Beam energy \\
		$f_{\text{rf}}$  & 500  MHz & RF frequency \\
		$V_{\text{rf}}$  & $\leq600$  kV & RF voltage \\
		$U_{0}$  & 9.1  keV & Radiation loss per turn \\
		$J_{s}$  & 1.95 & Longitudinal damping partition \\	
		$\tau_{\delta}$ & 11.4 ms & Longitudinal radiation damping time \\
		$\epsilon_{x0}$ & 250 nm &Horizontal emittance \\
		$\sigma_{\delta0}$ & $4.4\times10^{-4}$  & Natural energy spread \\
		\hline
	\end{tabular}
\end{table}

\subsection{Experimental Verification}

Here we report the first experimental verification of the energy widening and particle distribution distortion from Gaussian due to this second-oder transverse-longitudinal coupling effect as analyzed above. At the MLS, the Compton-backscattering (CBS) method is applied to measure the electron energy~\cite{klein2008operation}. Nevertheless, within certain limitations, the electron beam energy spread can also be evaluated from the CBS photon spectra~\cite{klein1997beam}. The non-Gaussian momentum distribution makes the evaluation a bit more involved, but
we can assume a Gaussian distribution with an equivalent mean energy spread.
This is a good approximation as long as $\xi_{x}$ is not too large. 

The experiment is conducted with all 80 RF buckets equally filled. To exclude a severe impact from energy widening collective effects, the beam current is decayed till the horizontal beam size is not sensitively dependent on it. The average single-bunch current is below 12.5~$\mu$A ($1$ electron / $1$ pA) while doing the CBS measurements. To mitigate the influence of a nonlinear momentum compaction, the longitudinal chromaticity has been corrected close to zero. The other parameters of the ring in the experiment are presented in Tab.~\ref{tab:MLS_para2}.

\begin{figure}[tb]
	\centering
	\includegraphics[width=1\columnwidth]{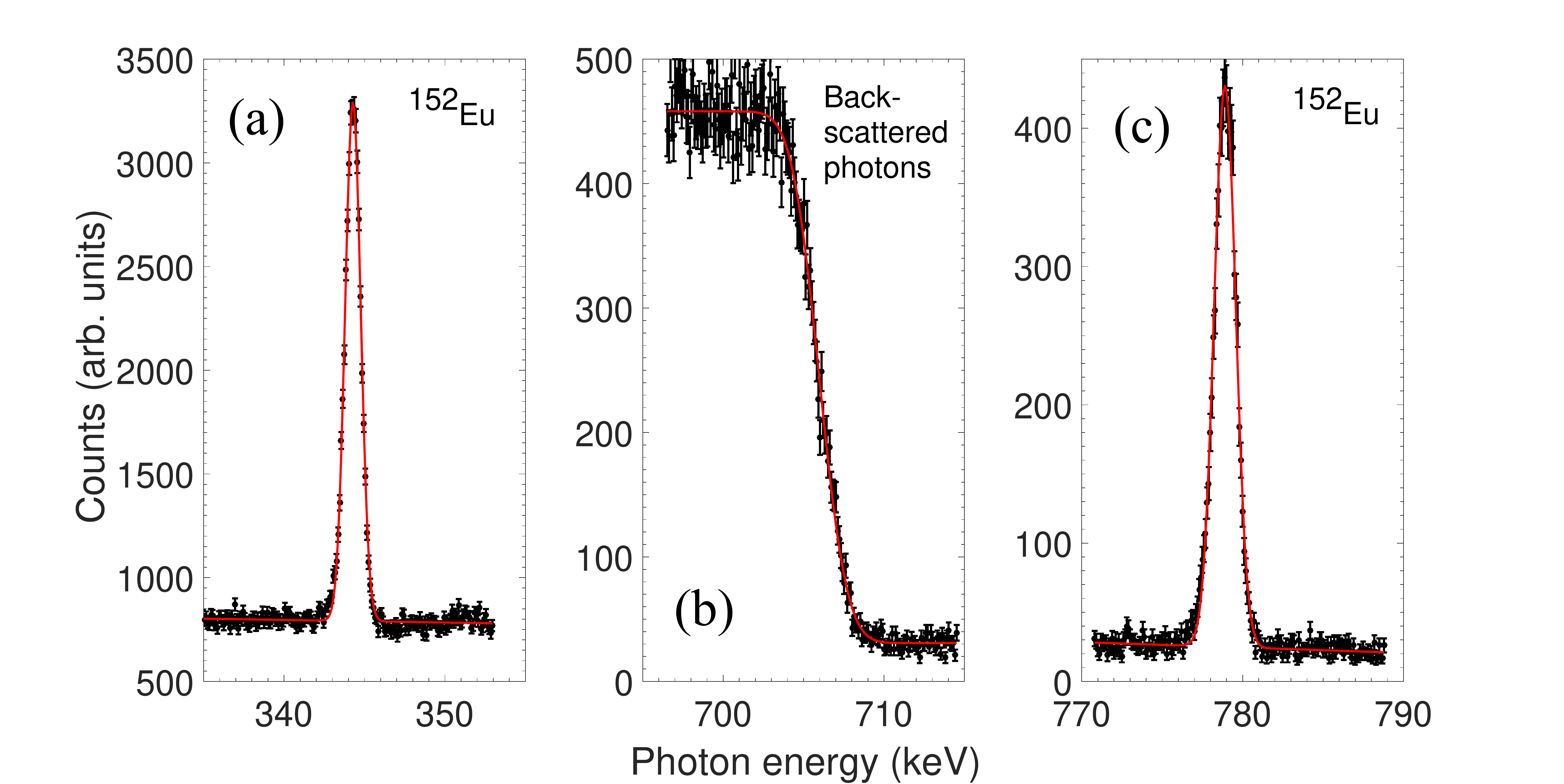}
	\caption{\label{fig:Chap3-CBSCal} Measurements of the CBS photons spectra at the 344.28 keV (a) and 778.90 keV (c) emission lines of $^{152}$Eu radionuclide to calibrate the channel numbers in terms of keV and the HPGe-detector resolution at the CBS cutoff edge (b) in the CBS method of measuring beam energy spread. } 
\end{figure}

To get the energy spread based on the CBS method with precision, the HPGe-detector used in the measurement should be calibrated in terms of photon energy per channel. This is realized by recording the emission lines from a $^{152}$Eu radionuclide simultaneously during the measurement of the CBS photons. 
Moreover, the width of the fitted $^{152}$Eu lines that are close to the edge of the CBS photons have been used to determine the detector resolution $\sigma_{\text{det}}$ at the photon energy of the CBS cutoff edge, $E_{\text{edge}}$, in our case 707~keV. This is done by a linear interpolation of the width of the $^{152}$Eu lines at 344.28 keV and 778.90~keV. The detector resolution $\sigma_{\text{det}}$ at 707 keV is thus determined to be 0.64(4) keV. The calibration scheme and result is shown in Fig.~\ref{fig:Chap3-CBSCal}.

\begin{figure}[tb]
	\centering
	\includegraphics[width=0.8\columnwidth]{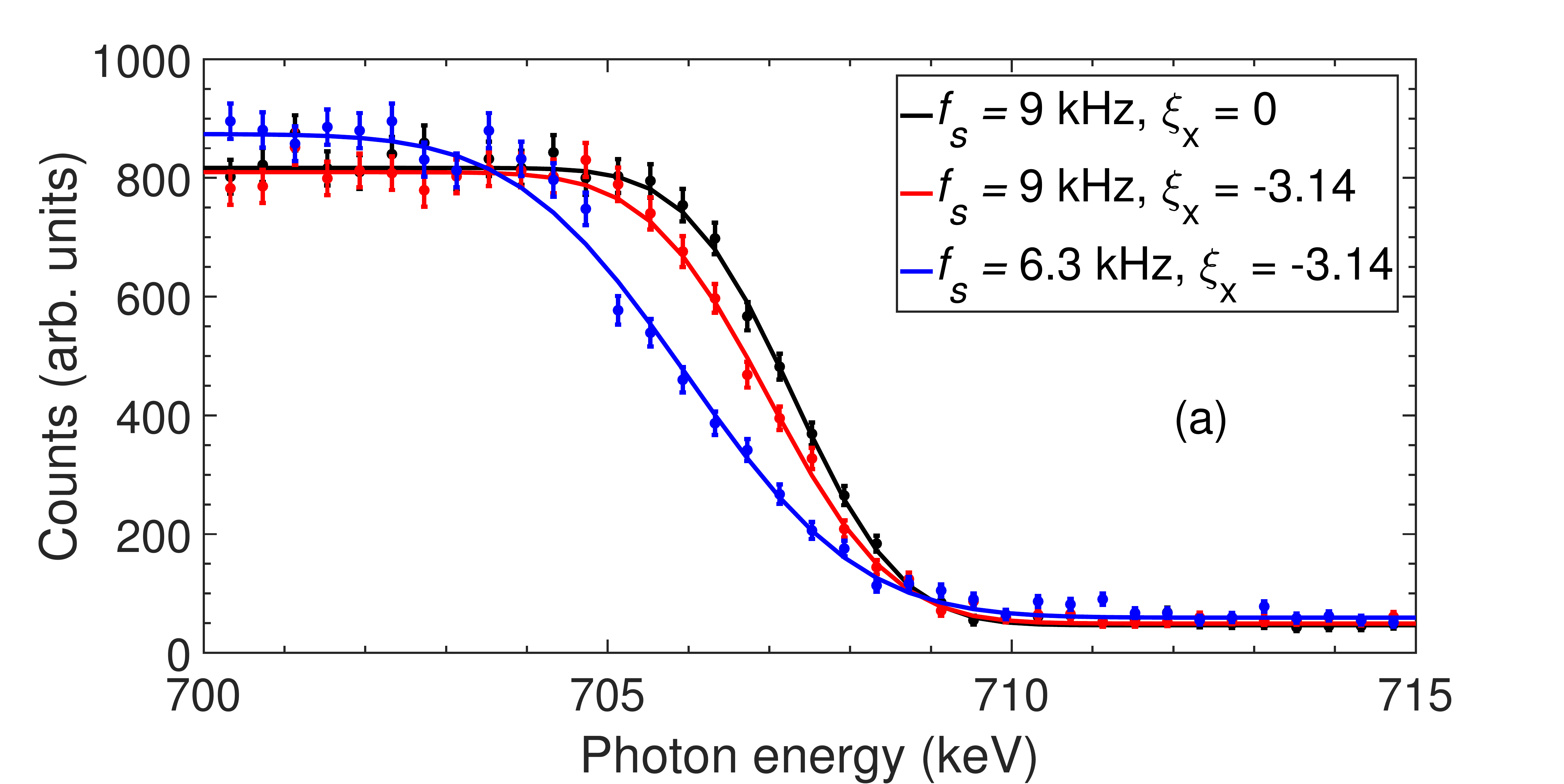}\\
	\includegraphics[width=0.8\columnwidth]{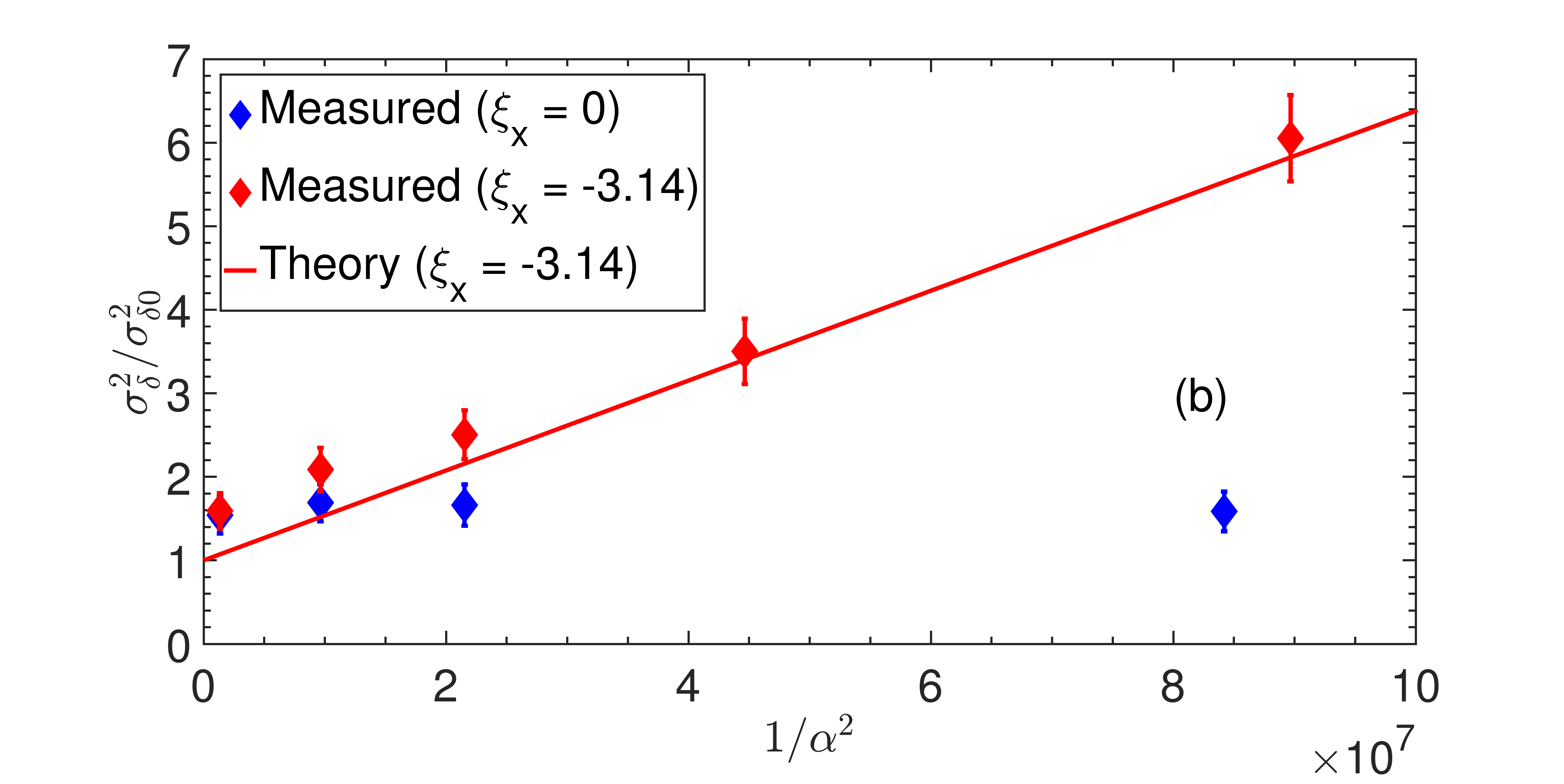}
	\caption{\label{fig:EnergySpreadBroadening} Measurement of the electron beam energy widening brought by the horizontal chromaticity using the CBS method. (a) The cutoff edges of CBS photon spectra under different $\xi_{x}$ and $f_{s}$ (therefore $\alpha$). (b) Quantitative evaluation of the cut-off edges revealing the energy spread and its comparison with theory Eq.~(\ref{eq:energWidening2nd}). The error bars in both figures are the one sigma uncertainties of the measurements and are due to calibration errors and counting statistics. The data acquisition of each spectrum takes 15 minutes. } 
\end{figure}

Figure~\ref{fig:EnergySpreadBroadening} (a) shows the typical CBS photon spectra close to the cutoff edge under the cases of different $\xi_{x}$ and $\alpha$. The adjustment of $\xi_{x}$ is accomplished by the implementation of different chromatic sextupole strengths and the $\alpha$ by slightly tuning quadrupole strengths. In the experiment the RF voltage is kept constant and the synchrotron frequency $f_{s}$ is proportional to the square root of the magnitude of $\alpha$. The edge in the figure is a convolution of a step function representing the CBS cutoff edge with a Gaussian function which attributes to the finite HPGe-detector energy resolution and the electron beam energy spread. The fitted line is basically an error function from which the energy width of the CBS photons at the edge $\sigma_{\text{edge}}$, and therefore the electron beam energy spread $\sigma_{\delta}$, can be deduced. It is assumed  $\sigma_{\text{edge}}$ is given by $\sigma_{\text{edge}}=\sqrt{\sigma_{\text{det}}^2+(2E_{\text{edge}}\sigma_{\delta})^2}$. The second term in the square root is due to the electron beam energy spread and is based on the fact that the energy of the backscattered photon is proportional to the electron energy squared.

It can be seen from Fig.~\ref{fig:EnergySpreadBroadening} (a) that the edge slope decreases with the magnitude increasing of $\xi_{x}$ and lowering of $\alpha$ when $\xi_{x}\neq0$, which indicates that there is an energy widening in the process. Quantitative evaluation of the edges revealing the energy spread and its comparison with theory of Eq.~(\ref{eq:energWidening2nd}) are shown in Fig.~\ref{fig:EnergySpreadBroadening} (b). The energy spread grows significantly with the magnitude decrease of $\alpha$ when $\xi_{x}=-3.14$ while it stays almost constant in the $\xi_{x}=0$ case. The agreement between measurements and theory is quite satisfactory. This is the first direct experimental proof of the impact of this effect on the equilibrium beam parameters in a storage ring.


As analyzed before, the bunch lengthening due to this second-order coupling is much less notable and also due to the limited resolution of the present streak camera, we do not measure the bunch lengthening in the experiment. Nevertheless, a more comprehensive investigation of this effect can be conducted on the other beam characteristics like the transverse intensity distribution. As can be seen from Eq.~(\ref{eq:xx}), particles with different betatron amplitudes oscillate around different closed orbits, which is the amplitude dependent center shift \cite{shoji2014amplitude}. 
Because of the dispersion, the non-Gaussian particle momentum distribution can also be reflected to the transverse dimension, which can be observed by the beam imaging systems installed at the MLS~\cite{koschitzki2010highly}. 

\begin{figure}[tb]
	\includegraphics[width=1\textwidth]{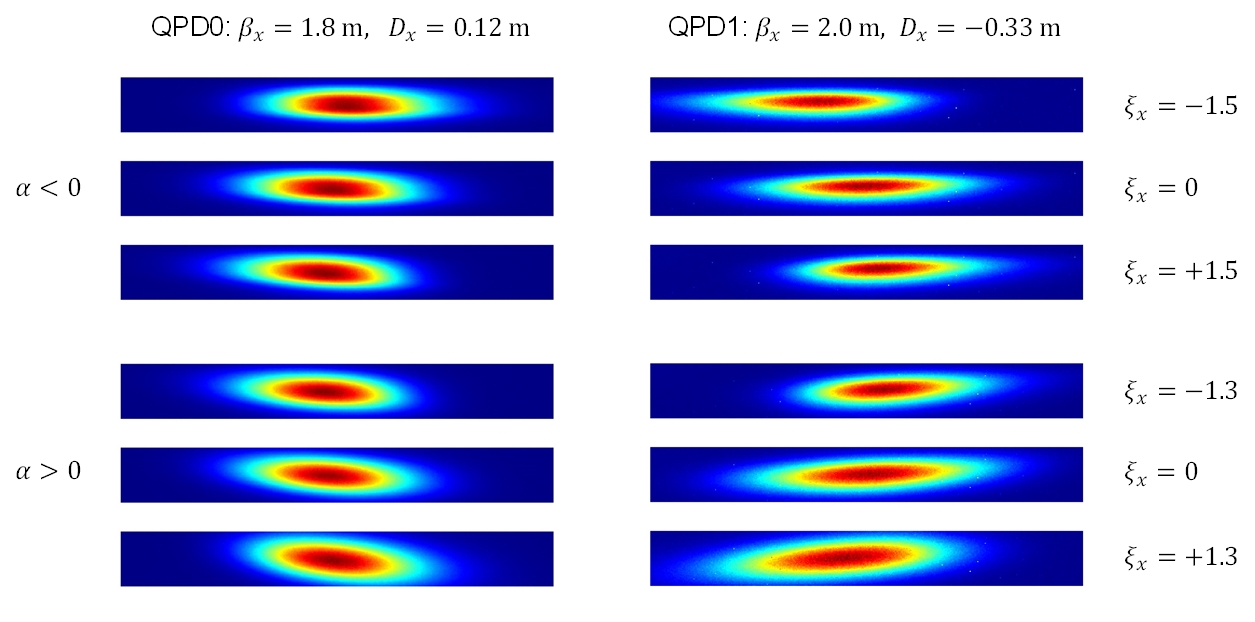}
	\caption{\label{fig:profile} Transverse beam intensity distortion from Gaussian at dispersive locations due to a non-vanishing horizontal chromaticity measured by the imaging systems installed at QPD0 and QPD1. Three different $\xi_{x}$ are applied in both the positive and negative momentum compaction modes with $\alpha=\pm8.4\times10^{-5}$. There is some residual horizontal-vertical coupling in the positive momentum compaction case, which do not influence the principle observation of the non-Gaussian behavior.}
\end{figure}

Figure \ref{fig:profile} shows the typical transverse beam intensity distribution measured by the imaging systems at two dispersive locations, QPD0 and QPD1, with different values of $\xi_{x}$ in both the negative and positive momentum compaction modes. The relevant optics functions, $\beta_{x}$ and $D_{x}$, at the two observation points are also shown in the figure. Note that $D_{x}$ have different signs and magnitudes at QPD0 and QPD1.  It can be seen that the horizontal beam distribution at these dispersive locations becomes asymmetric when $\xi_{x}\neq0$. The long tail direction and the magnitude of deviation from Gaussian are determined by the signs and magnitudes of $\alpha$, $D_{x}$, $\xi_{x}$ and also the value of $\epsilon_{x0}$ and $\beta_{x}$, which fits with the expectations.

\begin{figure}[tb]
	\centering
	\includegraphics[width=0.32\textwidth]{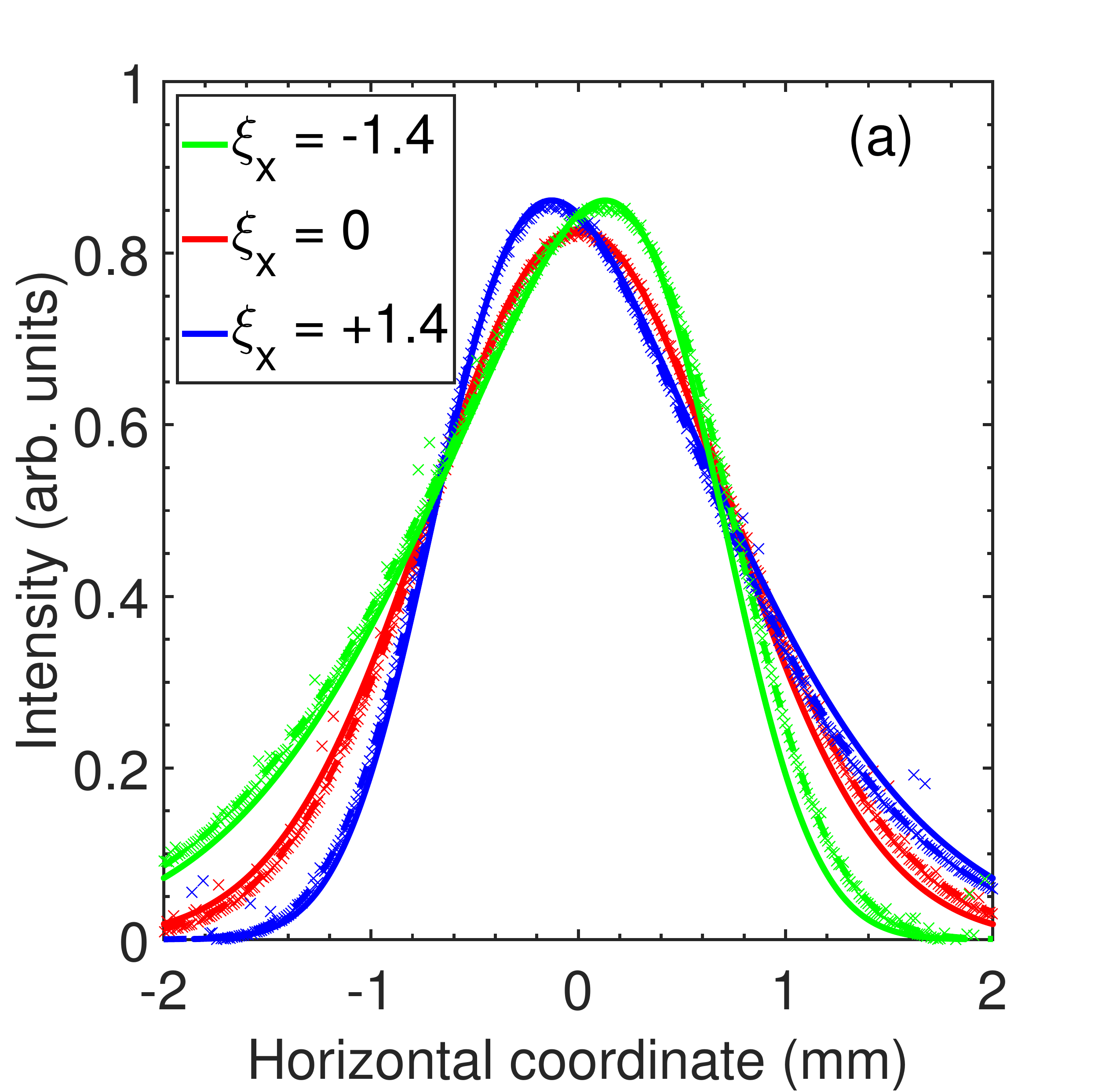}
	\includegraphics[width=0.32\textwidth]{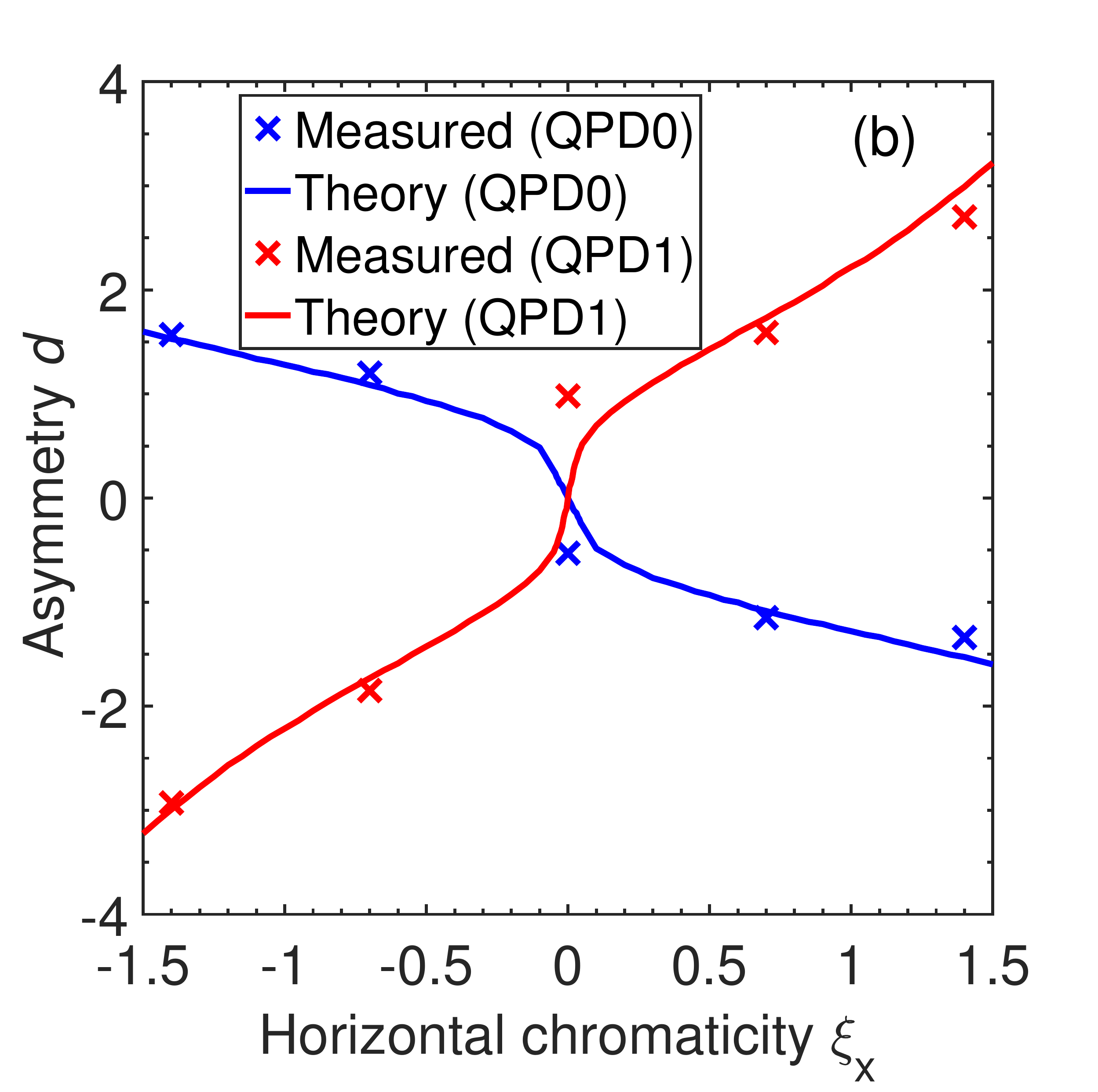}
	\includegraphics[width=0.32\textwidth]{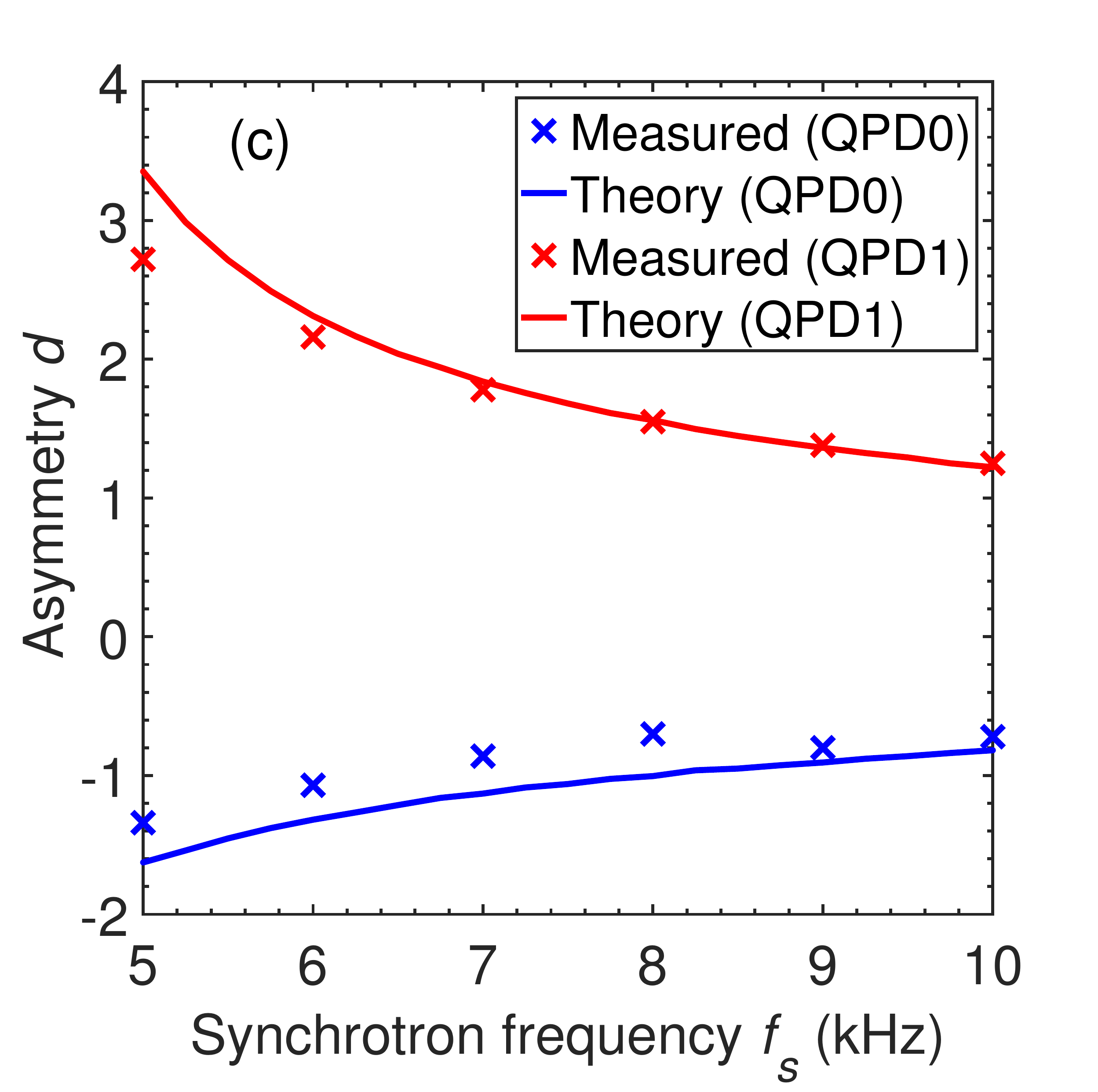}
	\caption{\label{fig:dvsfs} Horizontal beam profile distortion from Gaussian by horizontal chromaticty. (a) Typical horizontal beam profile at QPD1 with $\alpha=-7\times10^{-5}$ under three different $\xi_{x}$. The closed orbit movements of the ideal particle due to the sextupole strengths changes when adjusting $\xi_{x}$ have been compensated in the plot.  Cross: beam imaging system measurement results. Dashed line: fit of the measurement data by an exponentially modified Gaussian function Eq.~(\ref{eq:profileFit}). Solid line: theoretical prediction.  (b) Measured and theoretical asymmetry parameter $d$ versus $\xi_{x}$ at QPD0 and QPD1 with $f_{s}=5$ kHz ($\alpha=-7\times10^{-5}$); (c) Measured and theoretical asymmetry parameter $d$ versus $f_{s}$ at QPD0 and QPD1 with $\xi_{x}=1.4$. All the theoretical curves are obtained based on Eq.~(\ref{eq:twoDistribution}) and Eq.~(\ref{eq:xx}).}
\end{figure}

Figure~\ref{fig:dvsfs}~(a) demonstrates the typical horizontal beam profiles measured at QPD1 in the negative momentum compaction mode under three different $\xi_{x}$ and their good agreements with theory. It turns out that both the theoretical and experimental measured horizontal coordinate distribution $\psi(x)$ can be excellently fitted by an exponentially modified Gaussian function
\begin{equation}\label{eq:profileFit}
\psi(x)=\frac{1}{\sqrt{2\pi}\sigma}\cdot e^{-\frac{(x-b)^2}{2\sigma^2}}\cdot\left(1+\text{erf}\left[d\cdot\frac{x-b}{\sqrt{2}\sigma}\right]\right).
\end{equation}
The reason is that the distribution of particle momentum is approximately an exponentially modified Gaussian as we have analyzed. The asymmetry parameter $d$ in Eq.~(\ref{eq:profileFit}) is used to quantitatively  describe the deviation from Gaussian and as a criterion to do comparison between measurements and theory. 
Figure~\ref{fig:dvsfs}~(b) and (c) show the asymmetry parameter $d$ versus  $\xi_{x}$ and $f_{s}$, therefore $\alpha$, from measurements and theory at QPD0 and QPD1. It can be seen that the larger the $\xi_{x}$ and the smaller the $\alpha$, the more asymmetric the distribution is. Also the asymmetry at QPD1 is more significant than that at QPD0 as the magnitude of $D_{x}$ at QPD1 is larger while the $\beta_{x}$ difference at two places is not much. The agreement between measurements and theory confirms that this effect distorts the beam from Gaussian in both the longitudinal and transverse dimensions. 

While the energy widening and beam distortion could be a detrimental outcome for some applications, it may actually also be beneficial as it can help to stabilize collective instabilities. The bunch lengthening on the other hand is much less notable compared to the energy widening. So quasi-isochronous ring-based coherent radiation schemes, like some of the SSMB scenarios, may boost the stable coherent radiation power by taking advantage of this effect. For example, the stable single-bunch current at the MLS can grow for more than one order of magnitude by increasing the absolute value of the horizontal chromaticity from zero to a value larger than three, with the head-tail and the other collective effects like the longitudinal microwave instability properly suppressed. It has been proved at the MLS that the increase of THz power due to a higher stable beam current overcompensates the decrease due to the slight bunch lengthening of this effect. Therefore, this is now the standard low momentum compaction mode at the MLS for the application of Fourier Transform Spectroscopy.

This effect may also be relevant for the momentum and dynamic aperture
in storage rings \cite{takao2008impact,hoummi2019beam}, for instance due to the large betatron amplitude induced by Touschek scattering or the large natural chromaticity in the low emittance lattice design. Another example where this effect could be crucial is the non-scaling fixed-field alternate gradient (FFAG) accelerators for muon acceleration~\cite{berg2007amplitude}. A linear non-scaling FFAG has its large natural chromaticity uncorrected to achieve a large transverse acceptance as the emittance of the muon beam is usually very large. The energy spread grows significantly after traversing the acceleration lattice. Solutions similar to the beam conditioning for free-electron lasers in the X-ray regime can be invoked. Investigations presented in this dissertation can give clues for new beam conditioning ideas. For example, an upstream transport lattice with an opposite chromaticity to the natural chromaticity of the FEL undulator or non-scalling FFAG can mitigate the influence of this effect.    

This nonlinear transverse-longitudinal coupling may also be useful in some more applications. For example, it can be used for the real-time emittance evaluation in storage rings if the chromaticities, beta function and dispersion are known, which are usually easier to get than measuring the emittance directly. The amplitude dependent center shift can be applied to detect beam instabilities which blow up the transverse emittance~\cite{kobayashi2011amplitude}. A strongly asymmetric particle momentum distribution due to this effect cooperating with a large momentum compaction lattice can generate a strongly asymmetric distributed current, which is favored in some applications such as beam-driven wakefield acceleration~\cite{chen1986energy}.

%% file: data/chap04.tex
\chapter{Microbunching Radiation}
\label{cha:Radiation}

Having discussed the methods to form and preserve microbunching in the last two chapters, now let us see what radiation can we obtained from the formed microbunching. In this chapter, we present theoretical and numerical studies of the average and statistical property of the coherent radiation from SSMB.  Our results show that 1 kW average power of 13.5 nm wavelength EUV radiation can be obtain from an SSMB ring, provided that an average current of 1 A and bunch length of 3~nm microbunch train can be formed at the radiator. Such a high-power EUV source is a promising candidate to fulfill the urgent need of semiconductor industry for EUV lithography. Together with the narrow-band feature, the EUV photon flux can reach $6\times10^{15}$ phs/s within a 0.1 meV energy bandwidth, which is appealing for fundamental condensed matter physics research. In the theoretical investigation, we have generalized the definition and derivation of the transverse form factor of an electron beam which can quantify the impact of its transverse size on the coherent radiation. In particular, we have shown that the narrow-band feature of SSMB radiation is strongly correlated with the finite transverse electron beam size. Considering the pointlike nature of electrons and quantum nature of radiation, the coherent radiation fluctuates from microbunch to microbunch, or for a single microbunch from turn to turn. Some important results concerning the statistical property of SSMB radiation have been presented, with a brief discussion on its potential applications for example the beam diagnostics. The presented work is of value for the development of SSMB and better serve the potential synchrotron radiation users. In addition, it also sheds light on understanding the radiation characteristics of FELs, CHG, etc.

\section{Formulation of Radiation from a Rigid Beam}\label{sec:generalformulation}

For simplicity, in this dissertation we adopt the rigid beam approximation. More specifically, we consider only the impacts of particle position $x$, $y$ and $z$, but ignore the particle angular divergence $x'$, $y'$ and energy deviation $\delta$, on the radiation.  Under this approximation, concise and useful analytical formulas of the coherent radiation can be obtained. This approximation is accurate when the beam size does not change much inside the radiator, i.e., $\beta_{x,y}\gtrsim L_{r}$ and $\beta_{z}\gtrsim R_{56,r}$, where $\beta_{x,y,z}$ are the Courant-Snyder functions of the beam, $L_{r}$ and $R_{56,r}$ are the length and momentum compaction of the radiator, respectively. When this is not the case, applying the average beam size inside the radiator still gives a reasonable result.

\begin{figure}[tb]
\centering
\includegraphics[width=0.5\columnwidth]{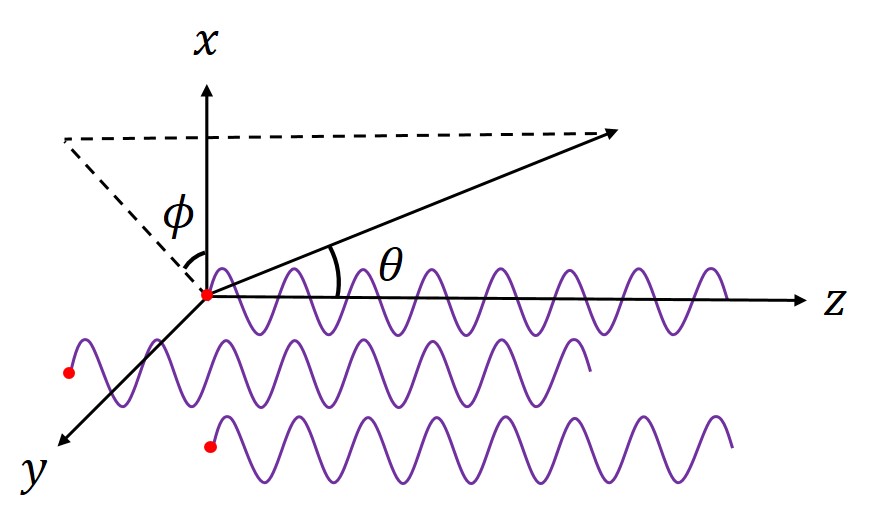}
\caption{\label{fig:RadiationPattern} 
Assumption adopted in this work: the particles (red dots) with different positions $(x,y,z)$ have the same radiation pattern (purple `noses').
}
\end{figure}

Assuming that the radiation vector potential of the reference particle at the observation location is $\vec{A}_{\text{point}}(\theta,\varphi,t)$, with $\theta$ and $\varphi$ being the polar and azimuthal angles in a spherical coordinate system, respectively, as shown in Fig.~\ref{fig:RadiationPattern}.  Then under far-field approximation, the vector potential of a three-dimensional (3D) rigid electron beam containing $N_{e}$ electrons is
\begin{equation}
\begin{aligned}
\vec{A}_{\text{beam}}(\theta,\varphi,t)&=N_{e}\int_{-\infty}^{\infty}\int_{-\infty}^{\infty}\int_{-\infty}^{\infty}\\
&\vec{A}_{\text{point}}\left(\theta,\varphi,t+\frac{x\sin\theta\cos\varphi+y\sin\theta\sin\varphi}{c}+\frac{z}{\beta c}\right)\\
&\rho(x,y,z)dxdydz,\\
\end{aligned}
\end{equation}
in which $\beta$ is the particle velocity normalized by the speed of light in vacuum $c$, and
$
\int_{-\infty}^{\infty}\int_{-\infty}^{\infty}\int_{-\infty}^{\infty}\rho(x,y,z)dxdydz=1.
$
Note that we have assumed that the particle motion pattern, therefore also the radiation pattern of a single electron, does not depend on $x$, $y$ and $z$ of the particle, as shown in Fig.~\ref{fig:RadiationPattern}.  In other words, $x,y,z$ of a particle only influence the arrival time of the radiation at the observation. This is the reason why their impacts can be treated within a single framework. The impacts of $x'$, $y'$ and $\delta$ are different.  Generally, their impacts are two-fold. First, they affect the radiation of the single particle itself, i.e., the radiation pattern. Second, they affect the beam distribution, therefore the coherence of different particles, during the radiation process. In this dissertation, we focus on the case of a `3D rigid' beam.

According to the convolution theorem, for a 3D rigid beam, we now have
\begin{equation}
\vec{A}_{\text{beam}}(\theta,\varphi,\omega)=N_{e}\vec{A}_{\text{point}}(\theta,\varphi,\omega)b(\theta,\varphi,\omega),
\end{equation}
where
\begin{equation}\label{eq:BF3}
\begin{aligned}
b(\theta,\varphi,\omega)&=\int_{-\infty}^{\infty}\int_{-\infty}^{\infty}\int_{-\infty}^{\infty}\rho(x,y,z)\\
&e^{-i\omega\left(\frac{x\sin\theta\cos\varphi+y\sin\theta\sin\varphi}{c}+\frac{z}{\beta c}\right)}dxdydz.
\end{aligned}
\end{equation}
Since $\vec{A}(\theta,\varphi,t)$ is real, then $\vec{A}(\theta,\varphi,-\omega)=\vec{A}^{*}(\theta,\varphi,\omega)$. The energy radiated per unit solid angle per unit frequency interval is then~\cite{jackson1999classical}
\begin{equation}\label{eq:dWdwdO}
\frac{d^{2}W}{d\omega d\Omega}(\theta,\varphi,\omega)=2|\vec{A}(\theta,\varphi,\omega)|^{2}.
\end{equation}
Therefore, we have
\begin{equation}\label{eq:coherent3D}
\frac{d^{2}W}{d\omega d\Omega}(\theta,\varphi,\omega)\Bigg|_{\text{beam}}=N_{e}^{2}{|b(\theta,\varphi,\omega)|^{2}\frac{d^{2}W}{d\omega d\Omega}}(\theta,\varphi,\omega)\Bigg|_{\text{point}}.
\end{equation}
The total radiation energy spectrum of a beam can be obtained by the integration with respect to the solid angle
\begin{equation}\label{eq:totalEnergySpectrum}
\frac{dW}{d\omega}\Bigg|_{\text{beam}}=\int_{0}^{\pi}\sin\theta d\theta \int_{0}^{2\pi}d\varphi\frac{d^{2}W}{d\omega d\Omega}(\theta,\varphi,\omega)\Bigg|_{\text{beam}},
\end{equation}
and the total radiation energy of the beam can be calculated by the further integration with respect to the frequency
\begin{equation}\label{eq:totalEnergy}
W_{\text{beam}}=\int_{0}^{+\infty}d\omega\frac{dW}{d\omega}\Bigg|_{\text{beam}}.
\end{equation}
The reason why the lower limit in the above integration is $0$, instead of $-\infty$, is because that there is a factor of 2 in the definition of Eq.~(\ref{eq:dWdwdO}).
The above formulas can be used to numerically calculate the characteristics  of radiation from an electron beam, once its 3D distribution is given. Note that in the relativistic case, we only need to account for $\theta$ several times of $\frac{1}{\gamma}$, as the radiation is very collimated in the forward direction.

\section{Form Factors}\label{sec:FF}

When the longitudinal and transverse dimensions of the electron beam are decoupled, we can factorize $b(\theta,\varphi,\omega)$ as
\begin{equation}\label{eq:BFFactor}
\begin{aligned}
b(\theta,\varphi,\omega)&=b_{\bot}(\theta,\varphi,\omega)\times b_{z}(\omega),
\end{aligned}
\end{equation}
where
\begin{equation}\label{eq:3DBF}
\begin{aligned}
b_{\bot}(\theta,\varphi,\omega)&=\int_{-\infty}^{\infty}\int_{-\infty}^{\infty}\rho(x,y)e^{-i\omega\left(\frac{x\sin\theta\cos\varphi+y\sin\theta\sin\varphi}{c}\right)}dxdy,
\end{aligned}
\end{equation}
and
\begin{equation}\label{eq:LFF}
b_{z}(\omega)=\int_{-\infty}^{\infty}\rho(z)e^{-i\omega\frac{z}{\beta c}}dz.
\end{equation}
Note that $\rho(x,y)$ and $\rho(z)$ are then the projected charge density. $b_{z}(\omega)$ is the usual bunching factor find in literature and is independent of the observation angle, while $b_{\bot}(\theta,\varphi,\omega)$ generally does.
For example, in the case of a 3D Gaussian $x$-$y$-$z$ decoupled beam
\begin{equation}\label{eq:bunchingfactor}
\begin{aligned}
|b_{\bot}(\theta,\varphi,\omega)|^{2}&=\text{exp}\left\{-\left(\frac{\omega}{c}\right)^{2}\left[\left(\sigma_{x}\sin\theta\cos\varphi\right)^{2}\right.\right.\\
&\left.\left.\ \ \ \ +\left(\sigma_{y}\sin\theta\sin\varphi\right)^{2}\right]\right\},\\
|b_{z}(\omega)|^{2}&=\text{exp}\left[-\left(\frac{\omega}{\beta c}\right)^{2}\sigma_{z}^{2}\right].
\end{aligned}
\end{equation}

In order to efficiently quantify the impact of transverse and longitudinal distributions of an electron beam on the overall radiation energy spectrum, here we define the transverse and longitudinal form factors of an electron beam as
\begin{equation}\label{eq:TFFDefinition}
\begin{aligned}
FF_{\bot}(\omega)&=\frac{\int_{0}^{\pi}\sin\theta d\theta \int_{0}^{2\pi}d\varphi|b_{\bot}(\theta,\varphi,\omega)|^{2}\frac{d^{2}W}{d\omega d\Omega}(\theta,\varphi,\omega)\bigg|_{\text{point}}}{\int_{0}^{\pi}\sin\theta d\theta \int_{0}^{2\pi}d\varphi\frac{d^{2}W}{d\omega d\Omega}(\theta,\varphi,\omega)\bigg|_{\text{point}}},
\end{aligned}
\end{equation}
and
\begin{equation}\label{eq:LFFDef}
FF_{z}(\omega)=|b_{z}(\omega)|^{2},
\end{equation}
respectively. Note that the longitudinal form factor is independent of the radiation processes, while the transverse form factor does.  The overall form factor is then
\begin{equation}
FF(\omega)=FF_{\bot}(\omega)FF_{z}(\omega).
\end{equation}
The total radiation energy spectrum of a beam is related to that of a single electron by
\begin{equation}\label{eq:totalEnergySpectrum2}
\frac{dW}{d\omega}\Bigg|_{\text{beam}}=N_{e}^{2}FF_{\bot}(\omega)FF_{z}(\omega)\frac{dW}{d\omega}\Bigg|_{\text{point}}.
\end{equation}

\subsection{Longitudinal Form Factor}\label{sec:longitudinalFF}
The longitudinal form factor as mentioned is the usual bunching factor squared, and have been discussed by many authors before. Here we use CHG~\cite{girard1984optical}, or HGHG~\cite{yu1991generation,yu2000high} as an example for the derivation of the longitudinal form factor. Of our particular interest in relativistic cases, in what follows we adopt the approximation $\beta=1$ in the calculation of longitudinal form factor.  The microbunching dynamics of an electron beam in CHG or HGHG is modeled by
\begin{equation}
\begin{aligned}
\delta_{\text{f}}&=\delta_{\text{i}}+A\sin(k_{\text{L}}z_{\text{i}}),\\
z_{\text{f}}&=z_{\text{i}}+R_{56}\delta_{\text{f}},
\end{aligned}
\end{equation}
where $k_{\text{L}}=2\pi/\lambda_{\text{L}}$ is the wavenumber of the modulation laser, and the subscripts `i' and `f' mean the initial and final, respectively.
The final bunching factor can be calculated according to
\begin{equation}
\begin{aligned}\label{eq:bessel}
b_{z}(\omega)
=\int_{-\infty}^{\infty}\int_{-\infty}^{\infty}dzd\delta\ f_{\text{i}}(z,\delta)e^{-i\frac{\omega}{c}\left[z+R_{56}\delta+R_{56}A\sin(k_{\text{L}}z)\right]},
\end{aligned}
\end{equation}
where $f_{\text{i}}(z,\delta)$ is the initial distribution of the beam in the longitudinal phase space.

We start from the case that the initial beam is an infinite-long direct current (DC) with a Gaussian energy spread, i.e., $f_{\text{i}}(\delta)=\frac{1}{\sqrt{2\pi}\sigma_{\delta}}\text{exp}\left(-\frac{\delta^{2}}{2\sigma_{\delta}^{2}}\right)$. Using the mathematical identity $e^{ix\sin(k_{\text{L}}z)}=\sum_{n=-\infty}^{\infty}e^{ink_{\text{L}}z}J_{n}[x],$ then \cite{xiang2009echo}
\begin{equation}
\begin{aligned}
b_{z,\text{coasting}}(\omega)
&=\sum_{n=-\infty}^{\infty}J_{n}\left[-\frac{\omega}{c}R_{56}A\right]\lim_{L\rightarrow\infty}\frac{1}{2L}\int_{-L}^{L}dze^{-i\left(\frac{\omega}{c}-nk_{\text{L}}\right)z}\\
&\ \ \times\frac{1}{\sqrt{2\pi}\sigma_{\delta}}\int_{-\infty}^{\infty}d\delta \text{exp}\left(-\frac{\delta^{2}}{2\sigma_{\delta}^{2}}-i\frac{\omega}{c}R_{56}\delta\right)\\
&=\sum_{n=0}^{\infty}\delta\left(\frac{\omega}{c}-nk_{\text{L}}\right)J_{n}\left[-\frac{\omega}{c}R_{56}A\right]\text{exp}\left[-\frac{1}{2}\left(\frac{\omega}{c}R_{56}\sigma_{\delta}\right)^{2}\right],
\end{aligned}
\end{equation} 
where
\begin{equation}
\delta(x)=\begin{cases}
1,& x=0,\\
0,& \text{else}.
\end{cases}
\end{equation}
Here we have used the subscript `coasting' to denote the DC beam-based microbunching we are treating. As can be seen, for an infinite-long bunch length, only bunching at the laser harmonics are non-zero. In other words, a long bunch results a narrow radiation  bandwidth, which is the Fourier uncertainty principle. This result originates from the definite phase relation of the emitting electrons imprinted by the modulation laser. The non-zero energy spread results in an exponentially decaying envelope of the bunching factor with respect to $\omega$.  Putting $\omega=nk_{\text{L}}c$ in the above formula, we arrive at the usual bunching factor at the $n^{\text{th}}$ laser harmonic for CHG and HGHG~\cite{yu1991generation}, 
\begin{equation}
b_{z,n}=J_{n}\left[nk_{\text{L}}R_{56}A\right]\text{exp}\left[-\frac{1}{2}\left(nk_{\text{L}}R_{56}\sigma_{\delta}\right)^{2}\right].
\end{equation}

Now let us consider the more-often confronted case of a finite bunch length. We assume that the initial beam is not energy chirped, i.e.,
\begin{equation}
f_{\text{i}}(z,\delta)=\frac{1}{\sqrt{2\pi}\sigma_{z}}\text{exp}\left(-\frac{z^{2}}{2\sigma_{z}^{2}}\right)\frac{1}{\sqrt{2\pi}\sigma_{\delta}}\text{exp}\left(-\frac{\delta^{2}}{2\sigma_{\delta}^{2}}\right).
\end{equation}
The beam current in the time domain is then a DC multiplied by a Gaussian distribution. According to the convolution theorem, then
\begin{equation}
\begin{aligned}
b_{z,\text{bunched}}(\omega)
&=b_{z,\text{coasting}}(\omega)\otimes b_{z,\text{Gaussian}}(\omega),
\end{aligned}
\end{equation} 
where $\otimes$ means convolution and 
\begin{equation}
\begin{aligned}
b_{z,\text{Gaussian}}(\omega)&=\int_{-\infty}^{\infty}dz\frac{1}{\sqrt{2\pi}\sigma_{z}}\text{exp}\left(-\frac{z^{2}}{2\sigma_{z}^{2}}\right)e^{-i\frac{\omega}{c}z}\\
&=\text{exp}\left[-\frac{1}{2}\left(\frac{\omega}{c}\sigma_{z}\right)^{2}\right].
\end{aligned}
\end{equation}
Therefore, the longitudinal form factor of the beam is now 
\begin{equation}
FF_{z,\text{bunched}}(\omega)=|b_{z,\text{bunched}}(\omega)|^2=|b_{z,\text{coasting}}(\omega)\otimes b_{z,\text{Gaussian}}(\omega)|^{2}.
\end{equation}
The convolution with $b_{z,\text{Gaussian}}(\omega)$ results in a non-zero bandwidth of each laser harmonic line in the longitudinal form factor spectrum
\begin{equation}
({\Delta\omega})_{\text{FWHM}}=2\sqrt{2\ln2}\frac{c/\sigma_{z}}{\sqrt{2}}=\frac{4\sqrt{2}\ln{2}}{(\Delta t)_{\text{FWHM}}},
\end{equation}
where FWHM means full width at half maximum, and $\Delta t$ is the electron bunch length in unit of time. Then the relative bandwidth of the longitudinal form factor at the $H^{\text{th}}$ harmonic can be expressed as 
\begin{equation}\label{eq:bandwidth}
\begin{aligned}
\left(\frac{{\Delta\omega}}{\omega}\right)_{\text{FWHM}}&=\frac{2\sqrt{2}\ln{2}}{\pi}\frac{1}{(c\Delta t)_{\text{FWHM}}/\lambda}\\
&=\frac{2\sqrt{2}\ln{2}}{\pi}\frac{\lambda_{\text{L}}}{H(c\Delta t)_{\text{FWHM}}}.
\end{aligned}
\end{equation}
Note that the coherent radiation pulse length is $\frac{1}{\sqrt{2}}$ of the electron bunch length due to the scaling of $P_{\text{coh}}\propto N_{e}^{2}$, and the above formula means that the coherent radiation is Fourier-transform limited. Note also that the absolute width $({\Delta\omega})_{\text{FWHM}}$ is independent of $H$, while the relative bandwidth $\left(\frac{{\Delta\omega}}{\omega}\right)_{\text{FWHM}}\propto\frac{1}{H}$. 

\begin{figure}[tb]
	\centering
	\includegraphics[width=1\columnwidth]{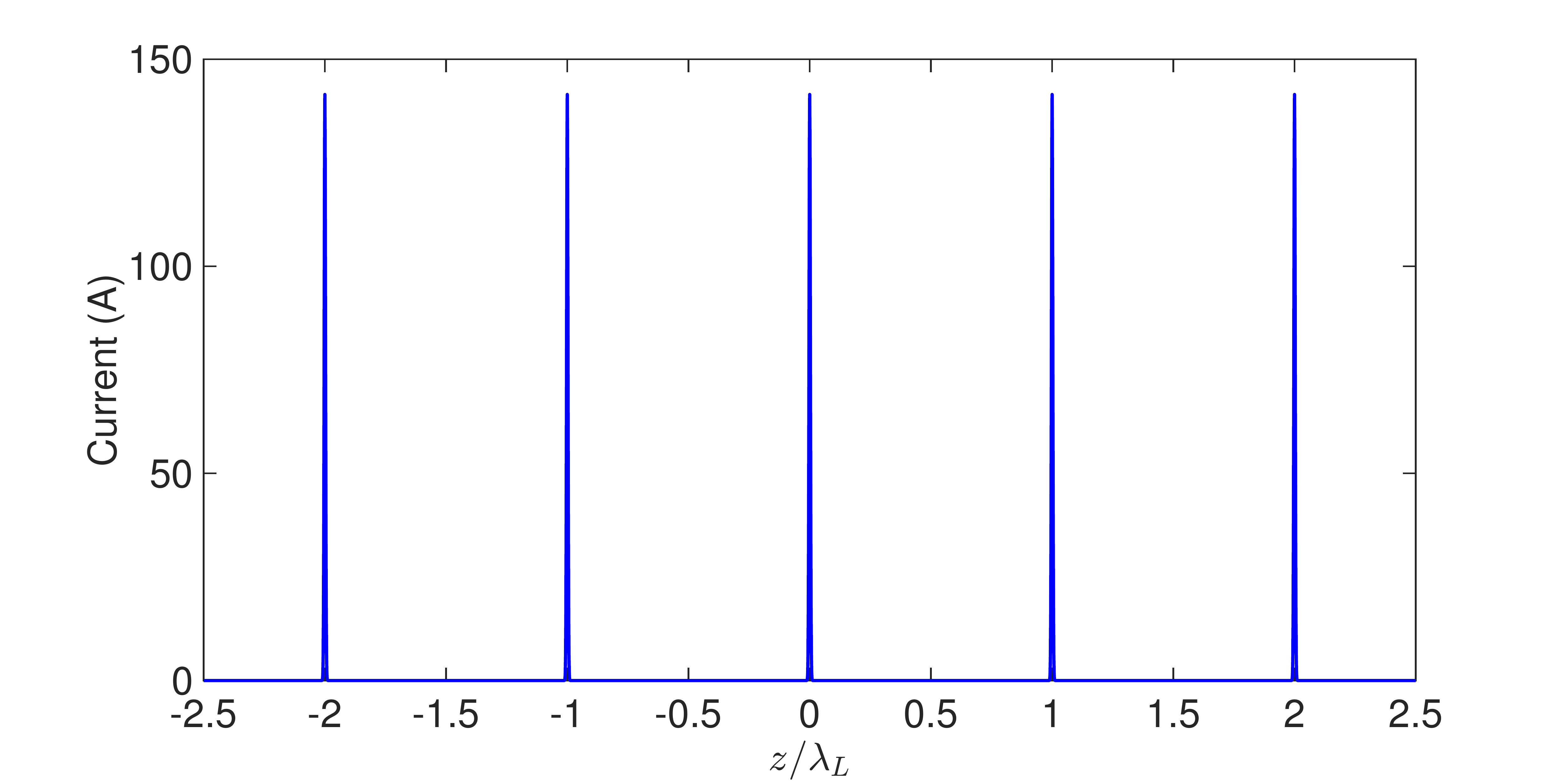}\\
	\includegraphics[width=1\columnwidth]{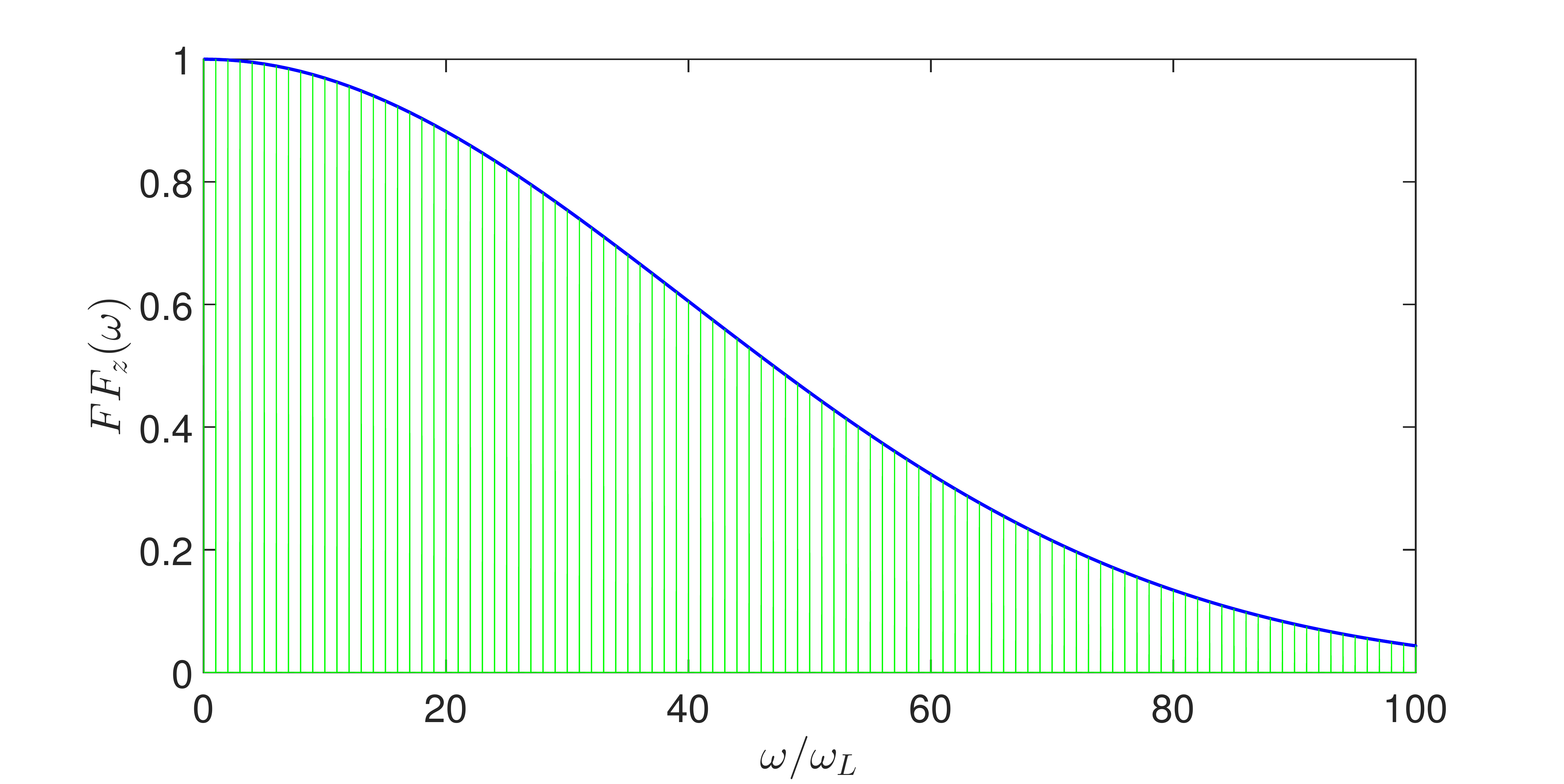}
	\caption{
		\label{fig:MT} 
		An example plot of the beam current and longitudinal form factor of the microbunch train at the radiator in the envisioned EUV SSMB. Up: beam current of the 3~nm microbunch train separated by the modulation laser wavelength $\lambda_{\text{L}}=1064$ nm. Bottom: longitudinal form factor $FF_{z}(\omega)$. The exponential decaying envelope correspond to that of a single 3 nm Gaussian microbunch, and the green periodic $\delta$-function lines correspond to the periodic microbunch train in time domain. The desired radiation wavelength is $\lambda_{0}=\frac{\lambda_{\text{L}}}{79}=13.5$ nm.
	}
\end{figure}

The above derivation considers the case of only a single bunch, which is usually much longer than the modulation laser wavelength. The result applies in FELs and CHGs. In SSMB, the bunch length is shorter than the modulation laser wavelength, and there is one microbunch in each microbucket, i.e., in the time domain the current is a sharp Gaussian distribution  convoluted by a period $\delta$-function. Therefore in the frequency domain, it is a broad Gaussian envelope multiplying a  period $\delta$-function, whose period is the frequency of the modulation laser, 
\begin{equation}
FF_{z,\text{bunch train}}(\omega)=\left|b_{z,\text{bunched}}(\omega)\sum_{n=0}^{\infty}\delta(\omega-nk_{\text{L}}c)\right|^{2}.
\end{equation}
Figure~\ref{fig:MT} is an example plot of the beam current and longitudinal form factor of the envisioned EUV SSMB. Noe that a Gaussian-distributed current at the radiator is assumed here. This is the case corresponding the usual longitudinal strong focusing SSMB. We remind the readers that the current at the radiator in the TLC-based bunch compression scheme (generalized longitudinal strong focusing) is actually non-Gaussian considering the nonlinear modulation waveform, as shown in Fig.~\ref{fig:Chap3-Summary}.

\subsection{Transverse From Factor}
Now let us focus on the transverse form factor. Since the transverse form factor depends on the radiation process, there is not a universal formula involving only the beam distribution. Here for our interest, we investigate the case of undulator radiation. We use a planar undulator as an example. The formulation for a helical undulator is similar.



As well established in literature, the planar undulator radiation of a point charge in the $H^{\text{th}}$ harmonic is \cite{chao2020lectures}
\begin{equation}\label{eq:undulatorRadiation}
\begin{aligned}
&\frac{d^{2}W_{H}}{d\omega d\Omega}(\theta,\varphi,\omega)\Bigg|_{\text{point}}=\frac{2e^2\gamma^2}{\pi\epsilon_{0} c}G(\theta,\varphi)F(\epsilon),\\
&F(\epsilon)=\left(\frac{\sin(\pi N_{u}\epsilon)}{\pi\epsilon}\right)^{2},\\ &\epsilon=\frac{\omega}{\omega_{r}(\theta)}-H=\frac{\omega}{2c\gamma k_{u}^{2}}\left(1+K^{2}/2+\gamma^{2}\theta^{2}\right)-H,\\
&G(\theta,\varphi)=G_{\sigma}(\theta,\varphi)+G_{\pi}(\theta,\varphi),\\
&G_{\sigma}(\theta,\varphi)=\left[\frac{H\left(\frac{K}{\sqrt{2}}\mathcal{D}_{1}+\frac{\gamma\theta}{\sqrt{2}}\mathcal{D}_{2}\cos\varphi\right)}{1+K^{2}/2+\gamma^{2}\theta^{2}}\right]^{2},\\ &G_{\pi}(\theta,\varphi)=\frac{1}{2}\left(\frac{H\gamma\theta\mathcal{D}_{2}\sin\varphi}{1+K^{2}/2+\gamma^{2}\theta^{2}}\right)^{2},\\
&\mathcal{D}_{1}=-\frac{1}{2}\sum_{m=-\infty}^{\infty}J_{H+2m-1}(H\alpha)\left[J_{m}(H\zeta)+J_{m-1}(H\zeta)\right],\\
&\mathcal{D}_{2}=-\frac{1}{2}\sum_{m=-\infty}^{\infty}J_{H+2m}(H\alpha)J_{m}(H\zeta),\\
&\alpha=\frac{2K\gamma\theta\cos\varphi}{1+K^{2}/2+\gamma^{2}\theta^{2}},\ \zeta=\frac{K^{2}/4}{1+K^{2}/2+\gamma^{2}\theta^{2}},
\end{aligned}
\end{equation}
in which $e$ is the elementary charge, $\gamma$ is the Lorentz factor,  $\epsilon_{0}$ is the permittivity of free space, $\omega_{r}(\theta)$ is the fundamental resonant angular frequency at the observation with a polar angle of $\theta$, $k_{u}=\lambda_{u}/2\pi$ is the wavenumber of the undulator, $K=\frac{eB_{0}}{m_{e}ck_{u}}=0.934\cdot B_{0}[\text{T}]\cdot\lambda_{u}[\text{cm}]$ is the undulator parameter, with $B_{0}$ being the peak magnetic field strength of the undulator and $m_{e}$ being the mass of an electron, $J$ means the Bessel function.

Now we try to get some analytical results for the transverse form factor. The motivation is that these analytical results can help us better understand the radiation physics and give us an efficient evaluation of the radiation characteristics.
Instead of a general discussion, here we only consider the simplest case of a transverse Gaussian round beam, i.e,
\begin{equation}\label{eq:BFRound}
\begin{aligned}
|b_{\bot}(\theta,\varphi,\omega)|^{2}&=\text{exp}\left[-\left(\frac{\omega}{c}\sigma_{\bot}\sin\theta\right)^{2}\right].
\end{aligned}
\end{equation}
As the radiation is dominantly in the forward direction in the relativistic case, and $e^{-(\frac{\omega}{c}\sigma_{\bot}\sin\theta)^2}$ approaches zero with the increase of $\theta$, therefore in Eq.~(\ref{eq:TFFDefinition}), the upper limit of $\theta$ in the integration can be effectively replaced by infinity, and $\sin\theta$ can be replaced by $\theta$ in $e^{-(\frac{\omega}{c}\sigma_{\bot}\sin\theta)^2}$.  Then
\begin{equation}
\begin{aligned}
&\int_{0}^{2\pi}d\varphi\int_{0}^{\pi}\sin\theta d\theta e^{-(\frac{\omega}{c}\sigma_{\bot}\sin\theta)^2}\frac{d^2W_{H}}{d\omega d\Omega}(\theta,\varphi,\omega)\Bigg|_{\text{point}}\\
&\approx \int_{0}^{2\pi}d\varphi\int_{0}^{\infty}\theta d\theta e^{-\left(\frac{\omega}{c}\sigma_{\bot}\theta\right)^2}\frac{d^2W_{H}}{d\omega d\Omega}(\theta,\varphi,\omega)\Bigg|_{\text{point}}\\
&\approx \frac{e^{2}}{\pi\epsilon_{0}c}\int_{0}^{2\pi}d\varphi G(\theta_{1},\varphi)\int_{0}^{\infty}d(\gamma\theta)^{2}e^{-\left(\frac{\omega}{c}\sigma_{\bot}\theta\right)^2}F(\epsilon),\\
\end{aligned}
\end{equation}
where
\begin{equation}
\begin{aligned}
\theta_{1}&=\frac{1}{\gamma}\sqrt{\left(1+K^{2}/2\right)\left(\frac{H\omega_{0}}{\omega}-1\right)},\\
\omega_{0}&=\omega_{r}(\theta=0)=\frac{2\gamma^{2}}{1+K^{2}/2}\omega_{u}.
\end{aligned}
\end{equation}
Here we have made use of the fact that there is only one value of $\theta$, i.e., $\theta_{1}$, that contributes significantly to the integration over the solid angle $\Omega$ due to the sharpness of $F(\epsilon)$ when the undulator period number $N_{u}\gg1$, as the spectral width of $F(\epsilon)$ is $1/N_{u}$.

The transverse form factor corresponding to the $H^{\text{th}}$ harmonic can thus be defined as
\begin{equation}\label{eq:TFF}
\begin{aligned}
FF_{\bot}(H,\sigma_{\bot},\omega)&=\frac{\int_{0}^{\infty}d(\gamma\theta)^{2}e^{-\left(\frac{\omega}{c}\sigma_{\bot}\theta\right)^2}\text{sinc}^{2}(N_{u}\pi\epsilon)}{\int_{0}^{\infty}d(\gamma\theta)^{2}\text{sinc}^{2}(N_{u}\pi\epsilon)}.
\end{aligned}
\end{equation}
The radiation spectrum of the $H^{\text{th}}$ harmonic is then
\begin{equation}
\frac{dW_{H}}{d\omega}\Bigg|_{\text{beam}}=N_{e}^{2}FF_{\bot}(H,\sigma_{\bot},\omega)FF_{z}(\omega)\frac{dW_{H}}{d\omega}\Bigg|_{\text{point}},
\end{equation}
and the total radiation spectrum of an electron beam is
\begin{equation}
\frac{dW}{d\omega}\Bigg|_{\text{beam}}=\sum_{H=1}^{\infty}\frac{dW_{H}}{d\omega}\Bigg|_{\text{beam}}.
\end{equation}

Denote
\begin{equation}
\begin{aligned}
\kappa_{1}&=N_{u}\pi\left(\frac{\omega}{\omega_{0}}-H\right),\\ \kappa_{2}&=N_{u}\pi\frac{\omega}{\omega_{0}}\frac{1}{1+K^{2}/2},\\ \kappa_{3}&=\left(\frac{\omega\sigma_{\bot}}{ c\gamma}\right)^2,
\end{aligned}
\end{equation}
then the denominator in Eq.~(\ref{eq:TFF}) is
\begin{equation}\label{eq:TransverseFFD}
\begin{aligned}
\mathcal{D}(\omega)&=\int_{0}^{\infty}d(\gamma\theta)^{2}\text{sinc}^{2}(N_{u}\pi\epsilon)\\
&=\int_{0}^{\infty}dx\text{sinc}^{2}\left(\kappa_{1}+\kappa_{2}x\right)\\
&=\frac{\frac{\pi}{2}-\text{Si}\left(2 \kappa_1\right)+\frac{\sin ^2\left(\kappa_1\right)}{\kappa_1}}{\kappa_2},
\end{aligned}
\end{equation}
where $\text{Si}(x)=\int_{0}^{x}\frac{\sin(t)}{t}dt$ is the sine integral, and the numerator in Eq.~(\ref{eq:TFF}) is
\begin{equation}\label{eq:TransverseFFN}
\begin{aligned}
\mathcal{N}(\omega)&=\int_{0}^{\infty}d(\gamma\theta)^{2}e^{-\left(\frac{\omega}{c}\sigma_{\bot}\theta\right)^2}\text{sinc}^{2}(N_{u}\pi\epsilon)\\
&=\int_{0}^{\infty}dxe^{-\kappa_{3}x}\text{sinc}^{2}\left(\kappa_{1}+\kappa_{2}x\right)\\
&=\frac{e^{\frac{\kappa_1 \kappa_3}{\kappa_2}}}{4 \kappa_1 \kappa_2^2}\left\{-4\kappa_2\sin^{2} (\kappa_{1})e^{-\frac{\kappa_1\kappa_3}{\kappa_2}}\right.\\
&\left. \ \ \ -2 \kappa_1 \kappa_2 i\left[\text{Ei} \left(2\kappa_1 i-\frac{\kappa_1\kappa_3}{\kappa_2}\right)-\text{Ei} \left(-2\kappa_1 i-\frac{\kappa_1\kappa_3}{\kappa_2}\right)\right]\right.\\
&\left. \ \ \ + \kappa_1 \kappa_3\left[\text{Ei} \left(2\kappa_1 i-\frac{\kappa_1\kappa_3}{\kappa_2}\right)- 2\text{Ei}\left(-\frac{\kappa_1 \kappa_3}{\kappa_2}\right)\right.\right.\\
&\left.\left.\ \ \ \ \ \ \ +\text{Ei} \left(-2\kappa_1 i-\frac{\kappa_1\kappa_3}{\kappa_2}\right)\right]\right\},\\
\end{aligned}
\end{equation}
where $\text{Ei}(x)=\int_{-\infty}^{x}\frac{e^{t}}{t}dt$ is the exponential integral.
The transverse form factor is then
\begin{equation}\label{eq:TransverseFF}
FF_{\bot}(H,\sigma_{\bot},\omega)=\frac{\mathcal{N}(\omega)}{\mathcal{D}(\omega)}.
\end{equation}

\begin{figure}[tb]
\centering
\includegraphics[width=0.6\columnwidth]{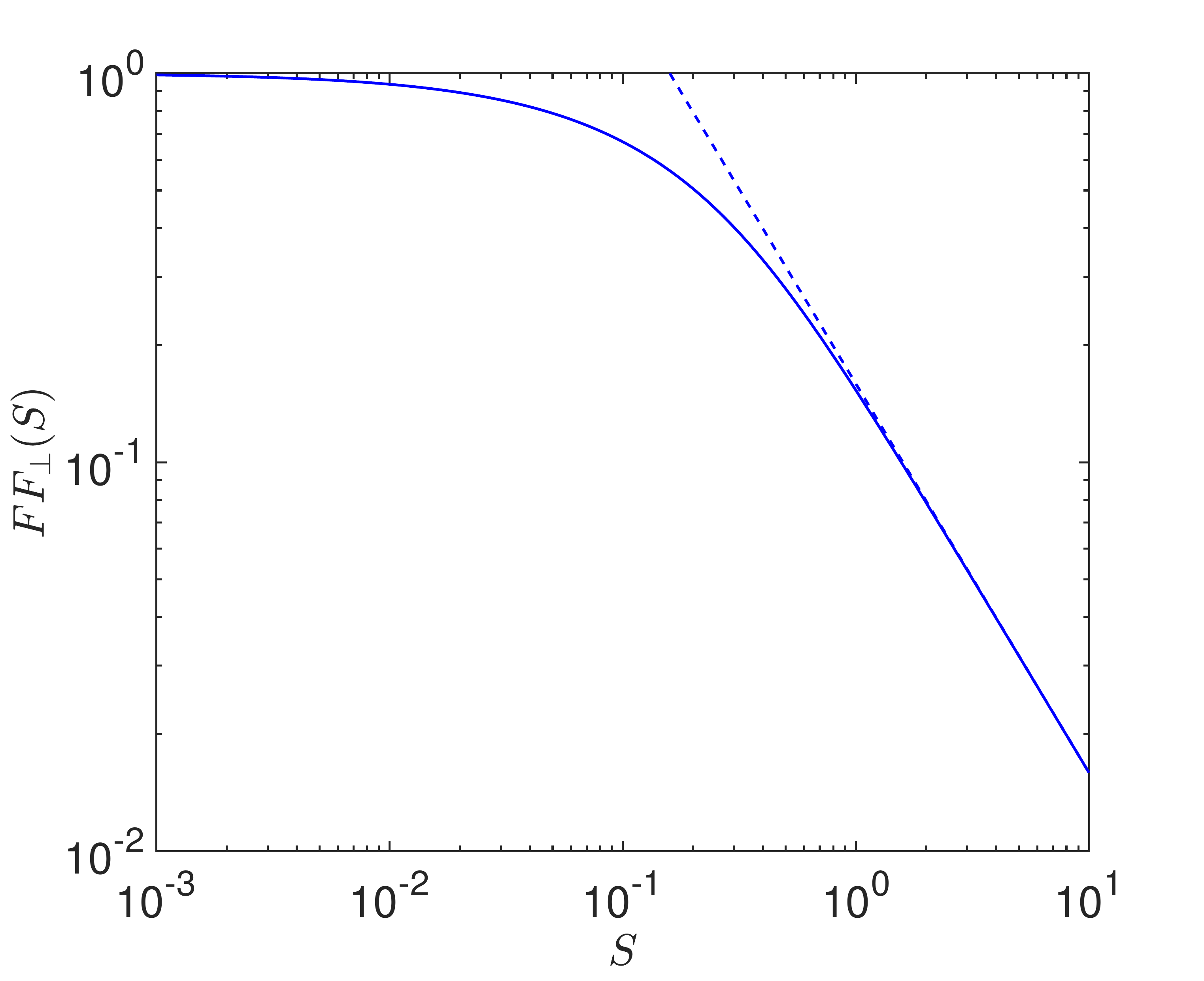}
\caption{\label{fig:Chap4-TFFCurve} 
The universal function $FF_{\bot}(S)$ and its asymptotic value above diffraction limit. The solid line comes from Eq.~(\ref{eq:FFS}), the dashed line from the asymptotic relation above diffraction limit Eq.~(\ref{eq:TFFasm}).
}
\end{figure}

When $\omega= H\omega_{0}$, then $\kappa_{1}=0$, we have
\begin{equation}\label{eq:FFS}
\begin{aligned}
&FF_{\bot}(H,\sigma_{\bot},\omega=H\omega_{0})=\frac{\int_{0}^{\infty}dxe^{-\kappa_{3}x}\text{sinc}^{2}\left(\kappa_{2}x\right)}{\int_{0}^{\infty}dx\text{sinc}^{2}\left(\kappa_{2}x\right)}\\
&=\frac{2}{\pi}\left[\tan^{-1}\left(\frac{1}{2S}\right)+S\ln\left(\frac{(2S)^{2}}{(2S)^{2}+1}\right)\right]\\
&=FF_{\bot}(S),
\end{aligned}
\end{equation}
where
\begin{equation}
S(\sigma_{\bot},L_{u},\omega)=\frac{\kappa_{3}}{4\kappa_{2}}=\frac{\sigma_{\bot}^2k_{u}\frac{\omega}{c}}{ 2N_{u}\pi}=\frac{\sigma_{\bot}^2\frac{\omega}{c}}{ L_{u}}
\end{equation}
is the diffraction parameter. This form factor $FF_{\bot}(S)$ is a universal function and has been obtained before in Ref.~ \cite{saldin2005simple}. Here we have reproduced the result following the general definition of the transverse form factor. The variable $S$ is a parameter used to classify the diffraction limit of the beam
\begin{equation}\label{eq:TFFasm}
FF_{\bot}(S)=\begin{cases}
&1,\ S\ll1,\ \text{below diffraction limit,}\\
&\frac{1}{2\pi S},\ S\gg1,\ \text{above diffraction limit.}
\end{cases}
\end{equation}
This function along with its asymptotic result above diffraction limit are shown in Fig.~\ref{fig:Chap4-TFFCurve}.

Note that the decrease of $FF_{\bot}(S)$ with the increase of $\sigma_{\bot}$ (as $S\propto\sigma_{\bot}^{2}$) means that the coherent radiation at the frequency $\omega=H\omega_{0}$ becomes less when the transverse electron beam size becomes larger. This reflects the fact that for a given radiation frequency, there is a range of polar angle $\theta$ that can contribute. For $\omega=H\omega_{0}$, not only $\theta=0$, but also $\theta$ very close to $0$ contribute. With the increase of $\sigma_{\bot}$, the effective bunching factor $b(\theta,\phi,\omega)$ at $H\omega_{0}$ drops for these non-zero $\theta$ due to the projected bunch lengthening, therefore the coherent radiation becomes less. Another way to appreciate the drop of $FF_{\bot}(S)$ with the increase of $\sigma_{\bot}$ is that there is a transverse coherence area whose radius is proportional to $\sqrt{L_{u}\lambda_{0}/H}$ with $\lambda_{0}=2\pi\frac{c}{\omega_{0}}$, and less particles can cohere with each other when the transverse size of the electron beam increases.

Note that our definition Eq.~(\ref{eq:TFFDefinition}) and derivation of the transverse form factor Eq.~(\ref{eq:TransverseFF}) are more general than that given in Ref.~\cite{saldin2005simple}, as it covers other frequencies in addition to a singe frequency $\omega_{0}$. Therefore, it contains more information than Eq.~(\ref{eq:FFS}) as will be presented soon. The issue of Eq.~(\ref{eq:TransverseFF}) is that it is still not simple enough for analytical evaluation to provide physical insight. A further approximation is thus introduced,
\begin{equation}\label{eq:TransverseFF2}
\begin{aligned}
FF_{\bot}(H,\sigma_{\bot},\omega)
&=\frac{\int_{0}^{\infty}dxe^{-\kappa_{3}x}\text{sinc}^{2}\left(\kappa_{1}+\kappa_{2}x\right)}{\int_{0}^{\infty}dx\text{sinc}^{2}\left(\kappa_{1}+\kappa_{2}x\right)}\\
&=e^{\frac{\kappa_1 \kappa_3}{\kappa_2}}\frac{\int_{\kappa_{1}}^{\infty}dye^{-\frac{\kappa_{3}}{\kappa_{2}}y}\text{sinc}^{2}(y)}{\int_{\kappa_{1}}^{\infty}dy\text{sinc}^{2}(y)}\\
&\approx e^{\frac{\kappa_1 \kappa_3}{\kappa_2}}\frac{\int_{0}^{\infty}dye^{-\frac{\kappa_{3}}{\kappa_{2}}y}\text{sinc}^{2}(y)}{\int_{0}^{\infty}dy\text{sinc}^{2}(y)}\\
&=e^{-4N_{u}\pi S\left(H-\frac{\omega}{\omega_{0}}\right)}FF_{\bot}(S).\\
\end{aligned}
\end{equation}
The condition of applying such simplification is
\begin{equation}
\frac{\kappa_{3}}{\kappa_{2}}\left(\omega=H\omega_{0}\right)\ll1\ \text{or}\ S\left(\omega=H\omega_{0}\right)\ll1,
\end{equation}
i.e., the beam is below diffraction limit for the on-axis radiation $\omega=H\omega_{0}$.  Therefore, the conditions of applying Eq.~(\ref{eq:TransverseFF2}) are
\begin{equation}\label{eq:conditions}
\begin{aligned}
&N_{u}\gg1\ \text{and}\ \sigma_{\bot}\ll\frac{1}{\sqrt{ H}}\sqrt{\frac{L_{u}\lambda_{0}}{2\pi}}.
\end{aligned}
\end{equation}
If the second condition is not satisfied, the more accurate result Eq.~(\ref{eq:TransverseFF}) should be referred.

\begin{figure}[tb]
\centering
\includegraphics[width=0.49\columnwidth]{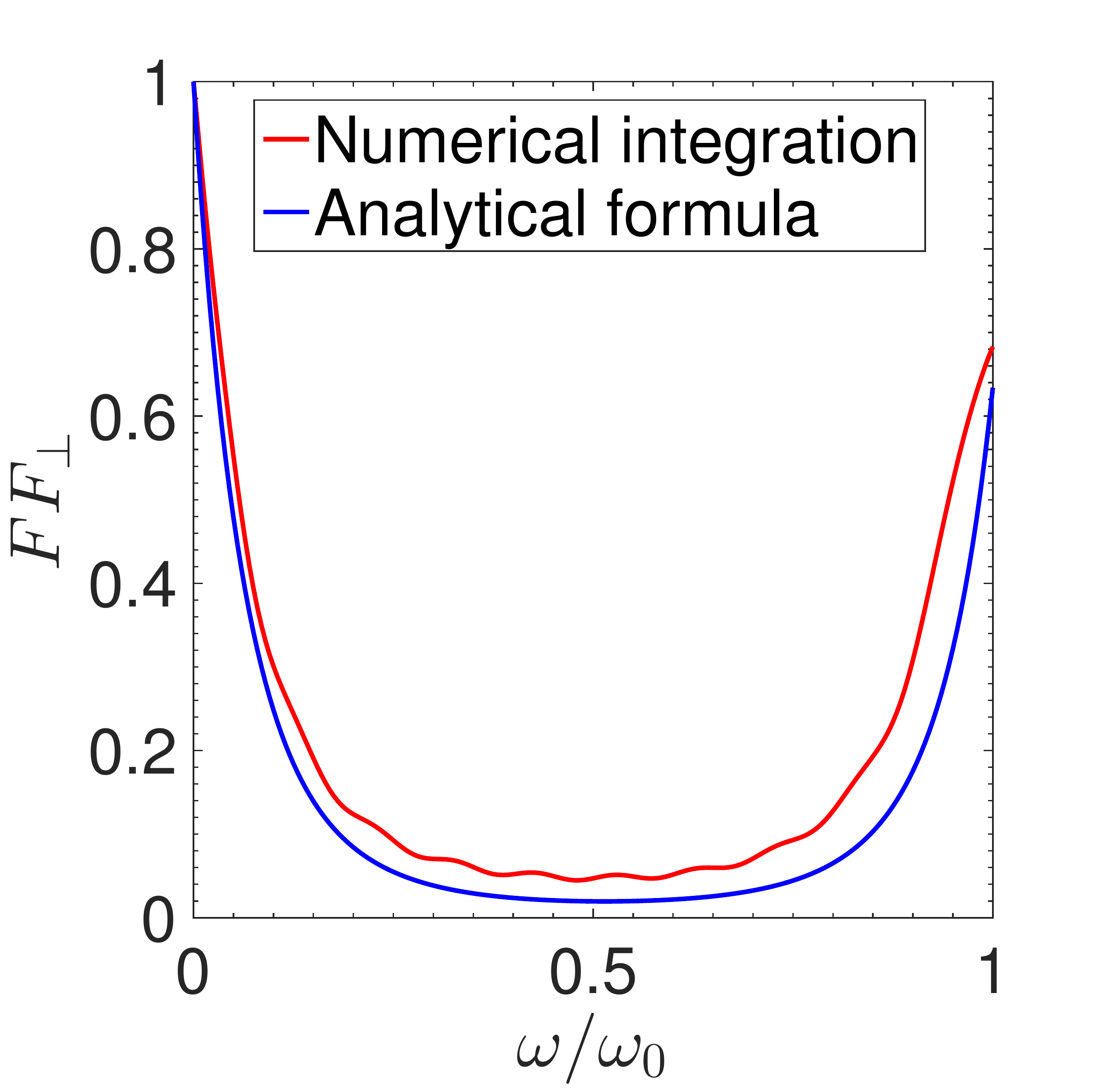}
\includegraphics[width=0.49\columnwidth]{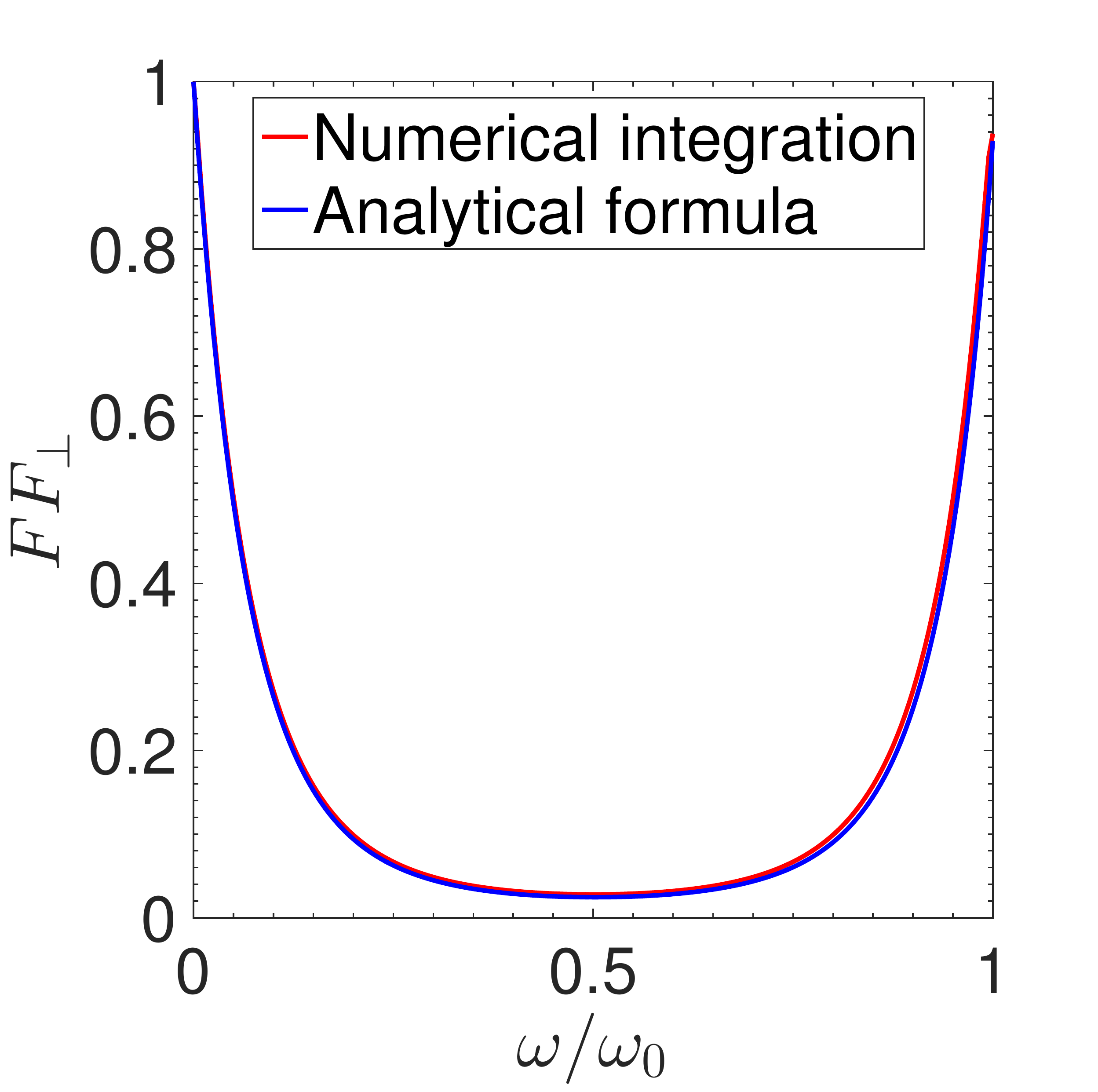}
\caption{\label{fig:Chap4-CompareTFF} 
The comparison of the transverse form factor, between that calculated from our simplified analytical formula Eq.~(\ref{eq:TransverseFF2}) and that from the direct numerical integration of Eq.~(\ref{eq:TFFDefinition}) for the case of $H=1$, with $N_{u}=10$ (left) and $N_{u}=100$ (right), respectively. Other related parameters used in the calculation: $E_{0}=400$~MeV, $\lambda_{0}=13.5$~nm, $\lambda_{u}=1$ cm, $K=1.14$, $\sigma_{\bot}=5\ \mu$m.
}
\end{figure}

As a benchmark of our derivation, here we conduct some calculations of the transverse form factor based on direct numerical integration of Eq.~(\ref{eq:TFFDefinition}) and compare them with our simplified analytical formula Eq.~(\ref{eq:TransverseFF2}). The parameters used is for the envisioned EUV SSMB to be presented in Sec.~\ref{sec:EUVSSMB}. As can be seen in Fig.~\ref{fig:Chap4-CompareTFF}, their agreement when $N_{u}=100$ is remarkably well. Even in the case of $N_{u}=10$, the agreement is still satisfactory. There are two reasons why the agreement is better in the case of a large $N_{u}$. The first is that in the derivation we have made use of the sharpness of $F(\epsilon)$, whose width is $1/N_{u}$. The second is that $S\propto\frac{1}{N_{u}}$ with a given transverse beam size, and our simplified analytical formula Eq.~(\ref{eq:TransverseFF2}) applies when $S\left(\omega=H\omega_{0}\right)\ll1$.

\begin{figure}[tb]
\centering
\includegraphics[width=0.49\columnwidth]{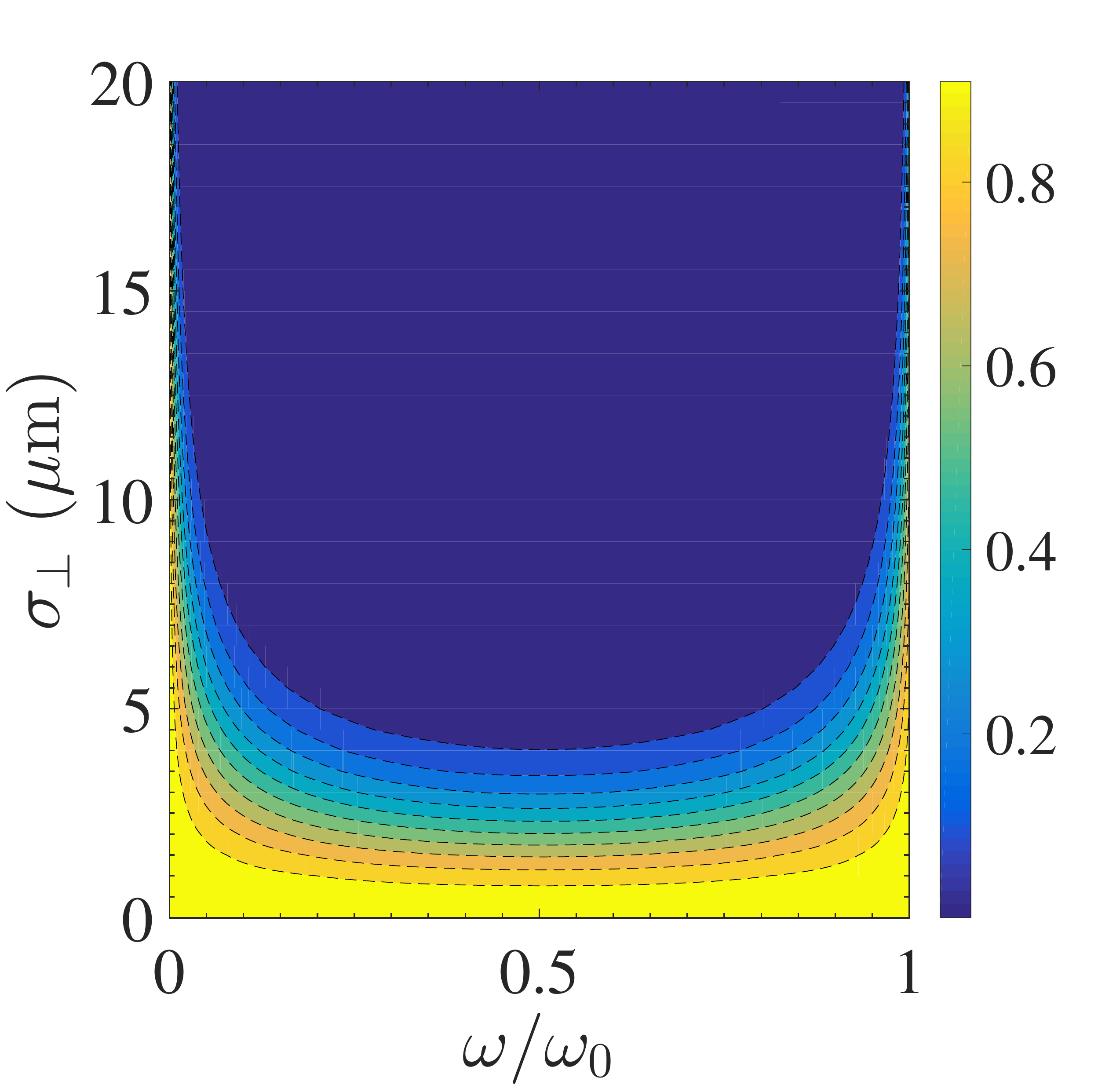}
\caption{\label{fig:Chap4-TFFFlatCoutor} 
Flat contour plot of the transverse form factor $FF_{\bot}(H,\sigma_{\bot},\omega)$ for $H=1$, as a function of the radiation frequency $\omega$ and transverse electron beam size $\sigma_{\bot}$, calculated using our simplified analytical formula Eq.~(\ref{eq:TransverseFF2}). Parameters used in the calculation: $E_{0}=400$ MeV, $\lambda_{0}=13.5$~nm, $\lambda_{u}=1$ cm, $K=1.14$, $N_{u}=79$.
}
\end{figure}

To appreciate the implication of the generalized transverse form factor further,  an example flat contour plot of the transverse form factor as a function of the radiation frequency $\omega$ and transverse electron beam size $\sigma_{\bot}$ is shown in Fig.~\ref{fig:Chap4-TFFFlatCoutor}. As can be seen, a large transverse electron beam size will suppress the off-axis red-shifted radiation due to the projected bunch lengthening from the transverse size, thus the effective bunching factor degradation, when observed off-axis. Therefore, a large transverse electron beam size will make the coherent radiation more collimated in the forward direction, and more narrow-banded around the harmonic lines. But note that not only the red-shifted radiation is suppressed, the radiation strength of each harmonic line $\omega=H\omega_{0}$ actually also decreases with the increase of transverse size, the reason of which we have just explained.

Now we evaluate the bandwidth and opening angle of the radiation at different harmonics due to the transverse form factor. In particular, we are interested in the case where the off-axis red-shifted radiation is significantly suppressed by the transverse size of the electron beam, which requires that
\begin{equation}
N_{u}\pi S(\omega=H\omega_{0})\gg1,
\end{equation}
i.e., $\sigma_{\bot}\gg\sqrt{\frac{H}{2}}\frac{\sqrt{\lambda_{u}\lambda_{0}}}{2\pi}$.  Note that  to apply Eq.~(\ref{eq:TransverseFF2}), we still need the conditions in Eq.~(\ref{eq:conditions}). For example,  to apply the analytical estimation for the case of the example calculation to be presented in Sec.~\ref{sec:EUVSSMB} in which $\lambda_{u}=1$ cm, $\lambda_{0}=13.5$ nm and $N_{u}=79$, we need $1.3\ \mu\text{m}\ll\sigma_{\bot}\ll41\ \mu\text{m}$. The typical transverse beam size in an EUV SSMB ring is in such range.

With these conditions satisfied, the value of the exponential factor in Eq.~(\ref{eq:TransverseFF2}) is more sensitive to the change of $\omega$, compared to the universal function $FF_{\bot}(S)$. Therefore, here we consider only the exponential term when $\omega$ is close to $H\omega_{0}$. We want to know the $\omega$, at which the exponential term gives
$
e^{-4N_{u}\pi S\left(H-\frac{\omega}{\omega_{0}}\right)}=e^{-1}.
$
Then
\begin{equation}
\begin{aligned}
\Delta\omega_{e^{-1}}\bigg|_{\bot}&=H\omega_{0} - \omega_{e^{-1}}=\frac{1-\sqrt{1-\frac{2}{H^{2}\sigma_{\bot}^{2}k_{u}k_{0}}}}{2}H\omega_{0}.
\end{aligned}
\end{equation}
As $\sigma_{\bot}\gg\sqrt{\frac{H}{2}}\frac{\sqrt{\lambda_{u}\lambda_{0}}}{2\pi}$, therefore $\frac{2}{H^{2}\sigma_{\bot}^{2}k_{u}k_{0}}\ll1$,  we have
\begin{equation}\label{eq:bandwidthTrans}
\frac{\Delta\omega_{e^{-1}}}{H\omega_{0}}\bigg|_{\bot}\approx \frac{1}{2H^{2}\sigma_{\bot}^{2}k_{u}k_{0}}.
\end{equation}
Correspondingly, the opening angle of the coherent radiation due to the transverse form factor is
\begin{equation}\label{eq:thetaTrans}
\begin{aligned}
&\frac{\gamma^{2}\theta_{e^{-1}}^{2}}{1+K^{2}/2}\approx\frac{\Delta\omega_{e^{-1}}}{H\omega_{0}}\bigg|_{\bot}\Rightarrow \theta_{e^{-1}}\bigg|_{\bot} \approx \frac{\sqrt{2+K^{2}}}{2H\gamma\sigma_{\bot}\sqrt{k_{u}k_{0}}}.
\end{aligned}
\end{equation}

It is interesting to note that
\begin{equation}
\frac{\Delta\omega_{e^{-1}}}{H\omega_{0}}\bigg|_{\bot}\propto\frac{1}{H^{2}}.
\end{equation}
As a comparison, the relative bandwidth at the harmonics due to the longitudinal form factor is
\begin{equation}
\frac{\Delta\omega}{H\omega_{0}}\bigg|_{z}\propto\frac{1}{H}.
\end{equation}
Note also that $\frac{\Delta\omega_{e^{-1}}}{H\omega_{0}}\bigg|_{\bot}$ and $\theta_{e^{-1}}\bigg|_{\bot}$ are independent of $N_{u}$, although the approximations applied actually involve conditions related to $N_{u}$.

\section{Radiation Power and Spectral Flux at Harmonics}\label{sec:radiationpower}

In many cases, the microbunching is formed based on a long (much longer than radiation wavelength) continuous electron beam, for example in an FEL or coherent harmonic generation (CHG). In these cases, the linewidth of the longitudinal form factor at the harmonics are  usually even narrower than that given by the transverse form factor. This also means that the coherent radiation of a long continuous electron bunch-based microbunching will be dominantly in the forward direction, as the bunching factor of the off-axis red-shifted frequency is suppressed very fast compared to the on-axis resonant ones. For a more practical application, here we derive the coherent radiation power at the undulator radiation harmonics in these cases. As we will see, it can be viewed as a useful reference of the radiation power from SSMB.


We assume that the long electron bunch, before microbunching, is Gaussain. Here we assume that the transverse form factors around the harmonics do not change much, i.e., we assume $e^{-4N_{u}\pi S\left(H-\frac{\omega}{\omega_{0}}\right)}\approx1$ when $\omega$ is close to $H\omega_{0}$. Therefore, we only need to take into account the Gaussian shape of the longitudinal form factor at the harmonics. The bandwidth of longitudinal form factor for a Gaussian bunch with a length of $\sigma_{z}$ is $\Delta\omega|_{z}=\frac{c}{\sigma_{z}/\sqrt{2}}$. Therefore, the coherent radiation energy at the $H^{\text{th}}$ harmonic is
\begin{equation}
\begin{aligned}
W_{H}&=\left[N_{e}^{2}FF_{\bot}(\omega)FF_{z}(\omega)\frac{dW}{d\omega}\Bigg|_{\text{point}}\right]\left(\omega=H\omega_{0}\right)\\
&\ \ \ \ \times\int_{-\infty}^{\infty}\text{exp}\left(-\frac{(\omega-H\omega_{0})^2}{(c/\sigma_{z})^{2}}\right)d\omega\\
&= N_{e}^{2}FF_{\bot}(S)|b_{z,H}|^{2}\frac{2e^{2}}{\epsilon_{0}c}{G}(\theta=0)\frac{1+K^2/2}{H}\frac{N_{u}}{2}\times\sqrt{\pi}c/\sigma_{z}.\\
\end{aligned}
\end{equation}
For a planar undulator, the $\sigma$-mode radiation dominates and from Eq.~(\ref{eq:undulatorRadiation}) we have
\begin{equation}
G_{\sigma}(\theta=0)=\left[\frac{HK/\sqrt{2}}{2(1+K^{2}/2)}\right]^{2}[JJ]_{H}^{2}.
\end{equation}
in which the denotation $[JJ]_{H}^{2}=\left[J_{\frac{H-1}{2}}\left(H\chi\right)-J_{\frac{H+1}{2}}\left(H\chi\right)\right]^{2}$ with $\chi=\frac{K^{2}}{4+2K^{2}}$ is used.
Note, however, the above expression is meaningful only for an odd $H$, as the on-axis even harmonic radiation is rather weak.
The peak power of the  odd-$H^{\text{th}}$ harmonic coherent radiation is then
\begin{equation}\label{eq:RPH}
\begin{aligned}
P_{H,\text{peak}}&=\frac{W_{H}}{\sqrt{2\pi}\frac{\sigma_{z}/c}{\sqrt{2}}}=\frac{\pi}{\epsilon_{0}c}N_{u}H\chi[JJ]_{H}^{2}I_{\text{P}}^{2}FF_{\bot}(S)|b_{z,H}|^{2},
\end{aligned}
\end{equation}
where $I_{\text{P}}=\frac{N_{e}e}{\sqrt{2\pi}\sigma_{z}/c}$ is the peak current of the Gaussian bunch before microbunching.  For a more practical application of the formula, we put in the numerical value of the constants,
\begin{equation}\label{eq:RPHNumerical}
\begin{aligned}
P_{H,\text{peak}}[\text{kW}]&=1.183N_{u}H\chi[JJ]_{H}^{2}I_{\text{P}}^{2}[\text{A}]FF_{\bot}(S)|b_{z,H}|^{2}.
\end{aligned}
\end{equation}
Our derivation of the coherent radiation power above is for a Gaussian bunch-based microbunching. For a coasting or DC beam, we just need to replace $I_{\text{P}}$ in Eq.~(\ref{eq:RPH}) by the average current $I_{\text{A}}$, and the peak power is then the average power. For a helical undulator, we need to replace $K_{\text{planar}}/\sqrt{2}$ with $K_{\text{helical}}$, and $[JJ]_{1}^{2}$ with 1, in the evaluation of the radiation power at fundamental frequency.

The spectral flux of the odd-$H^{\text{th}}$ harmonic coherent radiation is
\begin{equation}\label{eq:flux}
\begin{aligned}
\mathcal{F}_{H}&=\frac{W_{H}/(\hbar H\omega_{0})}{\sqrt{2\pi}\frac{c/\sigma_{z}}{\sqrt{2}}/(H\omega_{0})}=\frac{P_{H,\text{peak}}}{\hbar(c/\sigma_{z})^{2}}\\
&=\frac{1}{1000}\frac{e^{2}}{2\epsilon_{0}c\hbar }N_{u}H\chi[JJ]_{H}^{2}N_{e}^{2}FF_{\bot}(S)|b_{z,H}|^{2}\ \text{(photons/pass/0.1\% b.w.)},
\end{aligned}
\end{equation}
where $\hbar$ is reduced Planck's constant. For a more practical application of the formula, again we put in the numerical value of the constants and arrive at
\begin{equation}\label{eq:fluxnumerical}
\begin{aligned}
\mathcal{F}_{H}&=4.573\times10^{-5}N_{u}H\chi[JJ]_{H}^{2}N_{e}^{2}FF_{\bot}(S)|b_{z,H}|^{2}\ \text{(photons/pass/0.1\% b.w.)}.
\end{aligned}
\end{equation}

At a first glance of Eq.~(\ref{eq:RPH}), the coherent radiation power $P_{\text{coh}}$ seems to be proportional to $N_{u}$, while an intuitive picture of the longitudinal coherence length $l_{\text{coh}}\propto N_{u}$ says that the scaling should be $ P_{\text{coh}}\propto N_{u}^{2}$. This is actually because that $FF_{\bot}(S)$ is also a function of $N_{u}$. It is interesting to note that
\begin{equation}
P_{\text{coh}}=\begin{cases}
&\propto N_{u},\ \text{below diffraction limit,}\\
&\propto N_{u}^{2},\ \text{above diffraction limit,}
\end{cases}
\end{equation}
which can be obtained from the asymptotic expressions of $FF_{\bot}(S)$ as shown in Eq.~(\ref{eq:TFFasm}).
So for a given transverse beam size, $P_{\text{coh}}\propto N_{u}^{2}$ at first when $N_{u}$ is small. When $N_{u}$ is large enough such that the electron beam is below diffraction limit, then $P_{\text{coh}}\propto N_{u}$. Physically this is because with the increase of $N_{u}$, the diffraction of the radiation will make the radiation from one particle cannot so effectively affect the particles far in front of it, as the on-axis field from this particle becomes weaker with the increase of the slippage length of the radiation.

We remind the readers that Eq.~(\ref{eq:RPH}) is for the case of a long continuous bunch-based microbunching, for example in FELs and CHGs. In the case of SSMB, the microbunches are cleanly separated from each other according to the modulation laser wavelength as shown in FIg.~\ref{fig:MT}, and usually the radiation wavelength is at a high harmonic of the modulation laser. Therefore, there actually could be significant red-shifted radiation generated in SSMB. So if we plug in the average current of SSMB in Eq.~(\ref{eq:RPH}), what it evaluates is the radiation power whose frequency content is close to the on-axis harmonics and will underestimate the overall radiation power. We will see this argument clearer in Sec.~\ref{sec:EUVSSMB}.

\section{Statistical Property of Radiation}\label{sec:statistic}

In the previous sections, we have ignored the quantum discrete nature of the radiation. Besides, we have derived the coherent radiation using a smooth distributed charge, i.e., we have treated the charge as a continuum fluid. The number of photons radiated from a charged particle beam actually fluctuates from turn to turn or bunch to bunch if the quantum nature of the radiation and the pointlike nature of the electron are taken into account \cite{goodman2015statistical}. The first mechanism exists even if there is only one electron, and the second mechanism is related to the interference of fields radiated by different electrons \cite{lobach2020statistical}. Using the classical language, the second fluctuation mechanism is from the fluctuation of the form factor of the electron beam.

There have been some study on the statistical property of the radiation in FELs \cite{saldin1998statistical} and also the storage ring-based synchrotron radiation sources, for example the recent seminal work of Lobach et al. \cite{lobach2021transverse,lobach2021measurements,lobach2020statistical} Rich information about the electron beam is embedded in the radiation fluctuations, or more generally the statistical property of the radiation. For example, the turn-by-turn fluctuation of the incoherent undulator radiation can be used to measure the transverse emittance of the beam \cite{lobach2021transverse}.  The previous treatment, however, usually cares about the cases where the  bunch length is much longer than the radiation wavelength, i.e., the  radiation is temporally incoherent (in SASE FEL, incoherent at the beginning).  In SSMB, the bunch length is comparable or shorter than the desired radiation wavelength, and the dominant radiation is temporally coherent. The statistical property of radiation in this parameter space deserves special investigation.  Although numerical calculation is doable following the theoretical formulation of Lobach et al. \cite{lobach2020statistical}, an analytical formula for the fluctuation in this temporally coherent radiation dominant regime is still of value for a better understanding the physics and investigation of its potential applications.

\subsection{Pointlike Nature of Electrons}

Here for SSMB we consider first the second mechanism of fluctuation, i.e., the fluctuation of the form factor arising from the pointlike nature of the radiating electrons. First we investigate the statistical property of the radiation at a specific observation angle $(\theta,\varphi)$ and a specific frequency $\omega$. Corresponding to Eq.~(\ref{eq:BF3}), the bunching factor  of an electron beam at an observation angle $(\theta,\varphi)$ and frequency $\omega$ with the pointlike nature of electrons taken into account is
\begin{equation}
b(\theta,\varphi,\omega)=\frac{1}{N_{e}}\sum_{i}^{N_{e}} e^{i\frac{\omega}{c}\left(x_{i}\sin\theta\cos\varphi+y_{i}\sin\theta\sin\varphi+z_{i}\right)}.
\end{equation}
We start from the simple case with zero transverse size,
\begin{equation}
\begin{aligned}
FF_{z}(\omega)&=|b_{z}(\omega)|^{2}=\frac{1}{N_{e}^{2}}\left[N_{e}+\sum_{i\neq j} e^{i\frac{\omega}{c}(z_{i}-z_{j})}\right].\\
\end{aligned}
\end{equation}
Then we have
\begin{equation}\label{eq:FFZExp}
\begin{aligned}
\langle FF_{z}(\omega)\rangle&=\frac{1}{N_{e}}+\left(1-\frac{1}{N_{e}}\right)\overline{FF}_{z}(\omega),
\end{aligned}
\end{equation}
where $\langle\rangle$ means expectation and
\begin{equation}
\overline{FF}_{z}(\omega)=|\overline{b}_{z}(\omega)|^{2},
\end{equation}
with
\begin{equation}
\overline{b}_{z}(\omega)=\int_{-\infty}^{\infty}\rho(z)e^{i\frac{\omega}{c}z}dz
\end{equation}
the bunching factor we calculated before using a smoothed charge distribution $\rho(z)$. In the language of statistics, $\overline{b}_{z}(\omega)$ is the characteristic function of the random variable $z$. As can be seen from Eq.~(\ref{eq:FFZExp}), when $N_{e}=1$, which corresponds to the case of a single point charge, then $\langle FF_{z}(\omega)\rangle=1$. When $N_{e}\gg1$ and $N_{e}\overline{FF}_{z}(\omega)\ll1$, which corresponds to the case of incoherent radiation dominance, then $\langle FF_{z}(\omega)\rangle=\frac{1}{N_{e}}$. When $N_{e}\gg1$ and $N_{e}\overline{FF}_{z}(\omega)\gg1$, which corresponds to the case of coherent radiation dominance,  then $\langle FF_{z}(\omega)\rangle=\overline{FF}_{z}(\omega)$. These results are as expected.

\begin{table}\caption{\label{tab:15cases}
		The $N_{e}^{4}$ terms in the quadruple sum of Eq.~(\ref{eq:quadruplesum}) can be placed in 15 different classes, as shown in Ref.~\cite{goodman2015statistical}.}
	\centering
	\begin{tabular}{lll}
		\hline
		Item Number & Index Relations  & Number of Terms \\
		\hline
		(1) & $n=m=p=q$  & $N_{e}$ \\
		(2) & $n=m, p=q, n\neq p$  & $N_{e}(N_{e}-1$) \\
		(3) & $n=m, p\neq q\neq n$  & $N_{e}(N_{e}-1)(N_{e}-2)$ \\
		(4) & $n=p, m=q, n\neq m$  & $N_{e}(N_{e}-1)$ \\
		(5) & $n=p, m\neq q\neq n$  & $N_{e}(N_{e}-1)(N_{e}-2)$ \\
		(6) & $n=q, m=p, n\neq m$  & $N_{e}(N_{e}-1)$ \\
		(7) & $n=q, m\neq p\neq n$  & $N_{e}(N_{e}-1)(N_{e}-2)$ \\
		(8) & $n=m=p, n\neq q$  & $N_{e}(N_{e}-1)$ \\
		(9) & $n=m=q, n\neq p$  & $N_{e}(N_{e}-1)$ \\
		(10) & $n=p=q, n\neq m$  & $N_{e}(N_{e}-1)$ \\
		(11) & $p=q=m, n\neq m$  & $N_{e}(N_{e}-1)$ \\
		(12) & $n\neq m\neq p\neq q$  & $N_{e}(N_{e}-1)(N_{e}-2)(N_{e}-3)$ \\
		(13) & $p=q, n\neq m\neq p$  & $N_{e}(N_{e}-1)(N_{e}-2)$ \\
		(14) & $m=q, n\neq m \neq q$  & $N_{e}(N_{e}-1)(N_{e}-2)$ \\
		(15) & $m=p, n\neq m\neq q$  & $N_{e}(N_{e}-1)(N_{e}-2)$ \\
		\hline
	\end{tabular}
\end{table}

The calculation of $\langle FF_{z}^{2}(\omega)\rangle$ is more involved. More specifically
\begin{equation}\label{eq:quadruplesum}
\begin{aligned}
\langle FF_{z}^{2}(\omega)\rangle&=\left\langle\frac{1}{N_{e}^{4}}\sum_{n=1}^{N_{e}}\sum_{m=1}^{N_{e}}\sum_{p=1}^{N_{e}}\sum_{q=1}^{N_{e}}e^{i\frac{\omega}{c}\left(z_{n}-z_{m}+z_{p}-z_{q}\right)}\right\rangle.
\end{aligned}
\end{equation}
The $N_{e}^{4}$ terms in this summation can be placed in 15 different cases, as shown in Tab.~\ref{tab:15cases}.
Corresponding to the 15 cases in Tab.~\ref{tab:15cases}, we have
\begin{equation}\label{eq:FFZ2}
\begin{aligned}
\langle FF_{z}^{2}(\omega)\rangle=&\frac{1}{N_{e}^{4}}\left[N_{e}\right.\\
&\left.\ \ \ \ +N_{e}(N_{e}-1)\right.\\
&\left.\ \ \ \ +N_{e}(N_{e}-1)(N_{e}-2)\overline{FF}_{z}(\omega)\right.\\
&\left.\ \ \ \ +N_{e}(N_{e}-1)\overline{FF}_{z}(2\omega)\right.\\
&\left.\ \ \ \ +N_{e}(N_{e}-1)(N_{e}-2)\sqrt{\overline{FF}_{z}(2\omega)}\overline{FF}_{z}(\omega)\right.\\
&\left.\ \ \ \ +N_{e}(N_{e}-1)\right.\\
&\left.\ \ \ \ +N_{e}(N_{e}-1)(N_{e}-2)\overline{FF}_{z}(\omega)\right.\\
&\left.\ \ \ \ +N_{e}(N_{e}-1)\overline{FF}_{z}(\omega)\right.\\
&\left.\ \ \ \ +N_{e}(N_{e}-1)\overline{FF}_{z}(\omega)\right.\\
&\left.\ \ \ \ +N_{e}(N_{e}-1)\overline{FF}_{z}(\omega)\right.\\
&\left.\ \ \ \ +N_{e}(N_{e}-1)\overline{FF}_{z}(\omega)\right.\\
&\left.\ \ \ \ +N_{e}(N_{e}-1)(N_{e}-2)(N_{e}-3)\overline{FF}_{z}^2(\omega)\right.\\
&\left.\ \ \ \ +N_{e}(N_{e}-1)(N_{e}-2)\overline{FF}_{z}(\omega)\right.\\
&\left.\ \ \ \ +N_{e}(N_{e}-1)(N_{e}-2)\sqrt{\overline{FF}_{z}(2\omega)}\overline{FF}_{z}(\omega)\right.\\
&\left.\ \ \ \ +N_{e}(N_{e}-1)(N_{e}-2)\overline{FF}_{z}(\omega)\right].
\end{aligned}
\end{equation}
When $N_{e}\gg1$ and $N_{e}\overline{FF}_{z}(\omega)\gg1$, which is the case for SSMB, then there are one term of $N_{e}(N_{e}-1)(N_{e}-2)(N_{e}-3)$, i.e., item (12), and six terms of $N_{e}(N_{e}-1)(N_{e}-2)(N_{e}-3)$, i.e., items (3), (5), (7), (13), (14), (15),  which have the main contribution. As the radiation power at frequency $\omega$ is proportional to $N_{e}^{2}FF_{z}(\omega)$, therefore we have
\begin{equation}\label{eq:powerFluctuation}
\begin{aligned}
\frac{\text{Var}\left[P(\omega)\right]}{\langle P(\omega)\rangle^{2}}&=\frac{\text{Var}\left[FF_{z}(\omega)\right]}{\langle FF_{z}(\omega)\rangle^{2}}=\frac{\langle FF_{z}^{2}(\omega)\rangle-\langle FF_{z}(\omega)\rangle^{2}}{\langle FF_{z}(\omega)\rangle^{2}}\\
&=\frac{2}{N_{e}}\left(\frac{1+\sqrt{\overline{FF}_{z}(2\omega)}}{\overline{FF}_{z}(\omega)}-2\right)+\mathcal{O}\left(\frac{1}{N_{e}^{2}}\right).
\end{aligned}
\end{equation}
where Var[] means the variance of.
The above equation is the main result of our analysis of coherent radiation power fluctuation at a specific frequency and observation angle, and to our knowledge is new.
The result for the case of non-zero transverse electron beam size is similar, we only need to replace $\overline{FF}_{z}(\omega)$ with the effective bunching factor squared $|\bar{b}(\theta,\varphi,\omega)|^{2}$.

\begin{figure}[tb]	
\centering
\includegraphics[width=0.49\columnwidth]{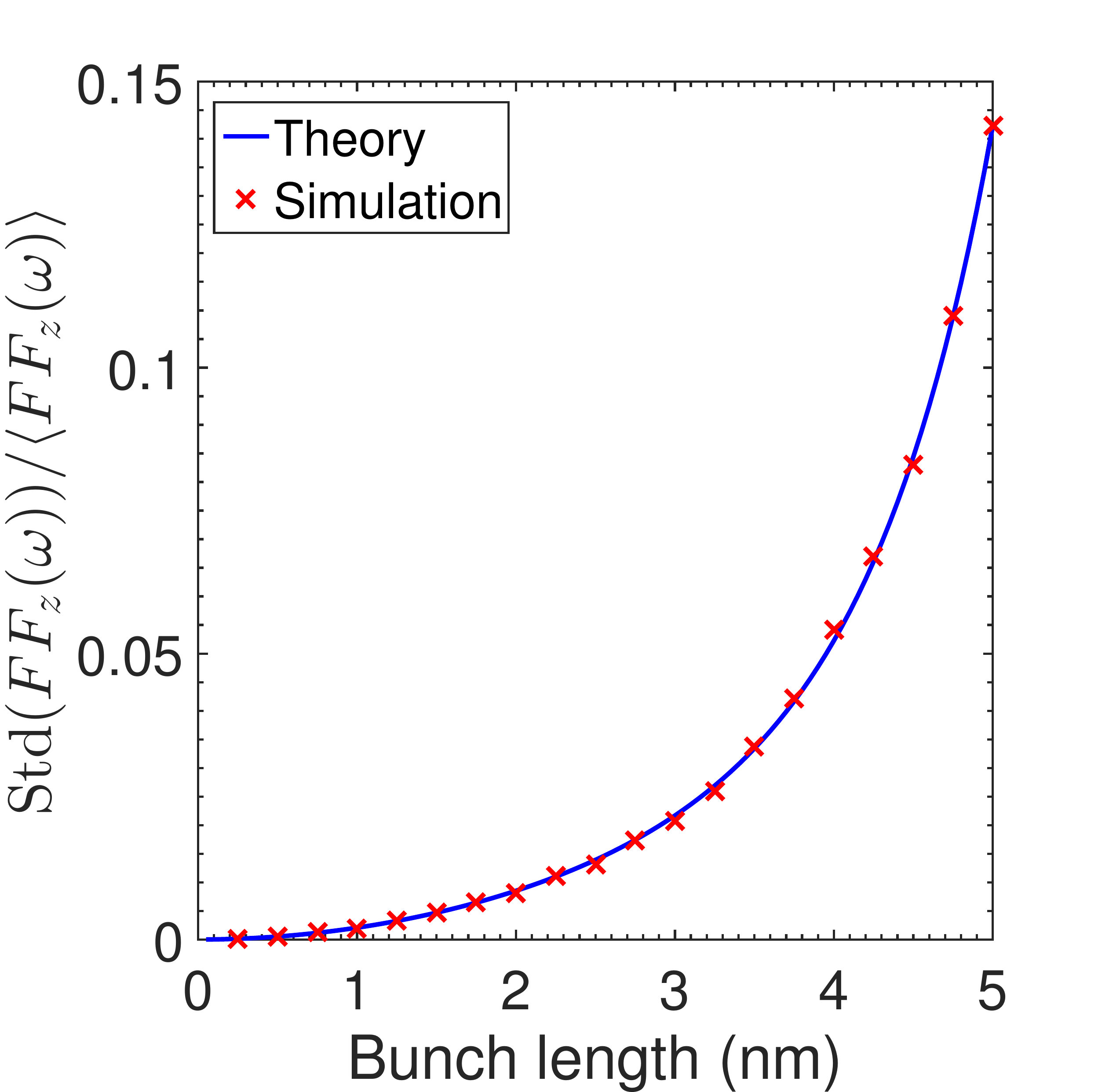}
\includegraphics[width=0.49\columnwidth]{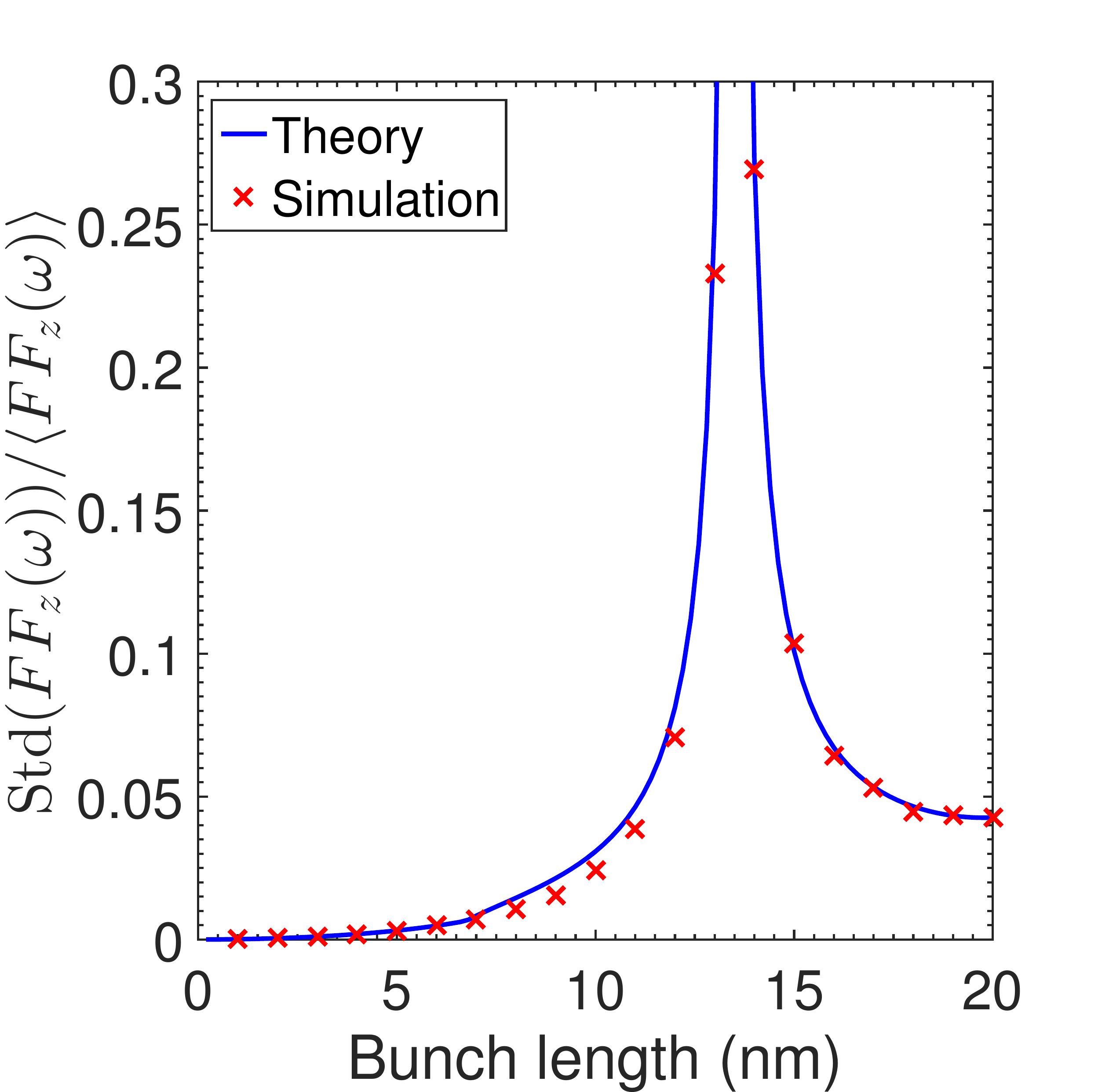}
\caption{\label{fig:PowerFluc} 
Fluctuation of 13.5 nm coherent radiation power v.s. bunch length with $N_{e}=2.2\times10^{4}$. The bunch is assumed to be Gaussian in the left and uniformly distributed in the right, and the theoretical fluctuation is calculated according to Eq.~(\ref{eq:powerFluctuation}), omitting the term $\mathcal{O}\left(\frac{1}{N_{e}^{2}}\right)$. For each parameters choice, $1\times10^{3}$ simulations are conducted to obtain the fluctuation.
}
\end{figure}

Now we conduct some numerical simulations to confirm our analysis. As can be seen from Fig.~\ref{fig:PowerFluc}, which correspond to the cases of a Gaussian and an uniform distributed bunch, respectively, the simulation results agree well with our theoretical prediction based on Eq.~(\ref{eq:powerFluctuation}).

After investigating the expectation  and variance of $FF_{z}(\omega)$, one may be curious about the more detailed distribution of $FF_{z}(\omega)$.  It can be shown that when $N_{e}\overline{FF}_{z}(\omega)\gg1$, the distribution of $FF_{z}(\omega)$ tends asymptotically toward Gaussian.


Now we investigate the more general case of the radiation within at a finite frequency bandwidth gathered at a finite angle acceptance. We use a filter function of $FT(\theta,\varphi,\omega)$ to account for the general case of frequency filter, angle acceptance, and detector efficiency. The expectation of the photon energy and photon energy squared gathered  are
\begin{equation}
\begin{aligned}
\langle W\rangle=& N_{e}^{2}\int_{0}^{\pi}\sin\theta d\theta \int_{0}^{2\pi}d\varphi\int_{0}^{\infty}d\omega FT(\theta,\varphi,\omega)\\
&\frac{d^{2}W}{d\omega d\Omega}(\theta,\varphi,\omega)\Bigg|_{\text{point}}\left\langle|b(\theta,\varphi,\omega)|^{2}\right\rangle.
\end{aligned}
\end{equation}
and
\begin{equation}
\begin{aligned}
\langle W^{2}\rangle=&N_{e}^{4}\int_{0}^{\pi}\sin\theta d\theta \int_{0}^{2\pi}d\varphi\int_{0}^{\infty}d\omega \int_{0}^{\pi}\sin\theta' d\theta' \\
&\int_{0}^{2\pi}d\varphi'\int_{0}^{\infty}d\omega'  FT(\theta,\varphi,\omega)FT(\theta',\varphi',\omega')\\
&\frac{d^{2}W}{d\omega d\Omega}(\theta,\varphi,\omega)\Bigg|_{\text{point}}\frac{d^{2}W}{d\omega d\Omega}(\theta',\varphi',\omega')\Bigg|_{\text{point}}\\
& \left\langle|b(\theta,\varphi,\omega)|^{2}\right\rangle\left\langle|b(\theta',\varphi',\omega')|^{2}\right\rangle g_{2}(\theta,\theta',\varphi,\varphi',\omega,\omega').
\end{aligned}
\end{equation}
where
\begin{equation}
g_{2}(\theta,\theta',\varphi,\varphi',\omega,\omega')=\frac{\left\langle|b(\theta,\varphi,\omega)|^{2}|b(\theta',\varphi',\omega')|^{2}\right\rangle}{\left\langle|b(\theta,\varphi,\omega)|^{2}\right\rangle\left\langle|b(\theta',\varphi',\omega')|^{2}\right\rangle},
\end{equation}
whose calculation can follow the similar approach of calculating $\langle FF_{z}^{2}(\omega)\rangle$ in Eq.~(\ref{eq:FFZ2}). And the relative fluctuation of the gathered photon energy is
\begin{equation}
\sigma^{2}_{W}=\frac{\langle W^{2}\rangle}{\langle W\rangle^{2}}-1.
\end{equation}

\subsection{Quantum Nature of Radiation}
As mentioned, there is another source of fluctuation, i.e., the quantum discrete nature of the radiation. As a result of the Campbell's theorem \cite{campbell1909study}, we know that for a Poisson photon statistics, the variance of photon number arising from this equals its expectation value. With both contribution from pointlike nature of electrons and quantum nature of radiation taken into account, the relative fluctuation of the radiation power or energy at frequency $\omega$ is
\begin{equation}\label{eq:powerFluctuationQuantum}
\begin{aligned}
\frac{\text{Var}\left[P(\omega)\right]}{\langle P(\omega)\rangle^{2}}=&\frac{1}{\langle \mathcal{N}_{\text{ph}}(\omega)\rangle|_{\text{beam}}}+\frac{2}{N_{e}}\left(\frac{1+\sqrt{\overline{FF}_{z}(2\omega)}}{\overline{FF}_{z}(\omega)}-2\right)\\
&+\mathcal{O}\left(\frac{1}{N_{e}^{2}}\right),
\end{aligned}
\end{equation}
in which
\begin{equation}
\begin{aligned}
\langle\mathcal{N}_{\text{ph}}(\omega)\rangle|_{\text{beam}}&=\left[N_{e}+N_{e}(N_{e}-1)\overline{FF}_{z}(\omega)\right]\langle \mathcal{N}_{\text{ph}}(\omega)\rangle|_{\text{point}}\\
&\approx N_{e}^{2}\overline{FF}_{z}(\omega)\langle \mathcal{N}_{\text{ph}}(\omega)\rangle|_{\text{point}}
\end{aligned}
\end{equation}
is the expected radiated photon number from the electron beam, and $\langle \mathcal{N}_{\text{ph}}(\omega)\rangle|_{\text{point}}$ is the expected radiated photon number from a single electron.  For a beam with non-zero transverse size,  we only need to replace $\overline{FF}_{z}(\omega)$ in the above formula with the effective bunching factor squared $|\bar{b}(\theta,\varphi,\omega)|^{2}$.

Note that to obtain a nonzero expected photon number $\langle\mathcal{N}_{\text{ph}}(\omega)\rangle|_{\text{beam}}$, a finite frequency bandwidth is needed. Therefore, the above Eq.~(\ref{eq:powerFluctuationQuantum}) actually applies to a finite frequency bandwidth close to $\omega$ where $\overline{FF}_{z}(\omega)$ does not change much.

From Eq.~(\ref{eq:powerFluctuationQuantum}), it is interesting to note that with the narrowing of the energy bandwidth acceptance, i.e., the decrease of $\langle \mathcal{N}_{\text{ph}}(\omega)\rangle|_{\text{beam}}$, the  contribution to the relative fluctuation from the quantum nature of radiation increases, while the contribution from the pointlike nature of the electron does not change. This reflects the fact that one fluctuation is quantum, while the other is classical.

Note that in our interested  case of SSMB, $N_{e}\langle \mathcal{N}_{\text{ph}}(\omega)\rangle|_{\text{point}}$ is usually much larger than 1, then the second  term in Eq.~(\ref{eq:powerFluctuationQuantum}) dominants. In other words, the fluctuation due to the pointlike nature of electrons dominants. Only when $N_{e}\langle \mathcal{N}_{\text{ph}}(\omega)\rangle|_{\text{point}}$ is close to 1, will the first term be significant compared to the second term.

\subsection{Potential Applications}

As the statistical property of the radiation embeds rich information about the electron beam, innovative beam diagnostics method can be envisioned by making use of this fact. Here we propose an experiment to measure the sub-ps bunch length accurately at a quasi-isochronous storage ring, for example the MLS, at a low beam current, by measuring and analyzing the fluctuation of the coherent THz radiation generated from the electron bunch. Equation~(\ref{eq:powerFluctuationQuantum}) or some numerical code based on the analysis presented in this section will be the theoretical basis for the experimental proposal. In principle, we can also deduce the transverse distribution of the electron beam by measuring the two-dimensional distribution of the radiation fluctuation.  More novel beam diagnostics methods may be invented for SSMB and future light sources by making use of the statistical property of radiation.

\section{Example Calculation for Envisioned EUV SSMB}\label{sec:EUVSSMB}

To summarize our investigations on the average and statistical property of SSMB radiation, here we present an example calculation for the envisioned EUV SSMB. In the envisioned SSMB-based EUV light source, the microbunch length is $\sigma_{z}\approx3$~nm at the radiator where 13.5~nm coherent EUV radiation is generated, and these 3~nm microbunch train are separated from each other with a distance of $\lambda_{\text{L}}=1064\ \text{nm}=79\times13.5\ \text{nm}$, which is the modulation laser wavelength. The beam at the radiator can be round or flat depending on the lattice scheme, and its transverse size can range from a couple of $\mu$m to a couple of 10 $\mu$m.
As our goal is to give the readers a picture of the radiation characteristics, here for simplicity we consider the case of a round beam. We remind the readers that the parameters used in this example EUV SSMB radiation calculation is for an illustration and is not optimized.

\subsection{Average Property}
\begin{figure}[tb]
\centering
\includegraphics[width=1\columnwidth]{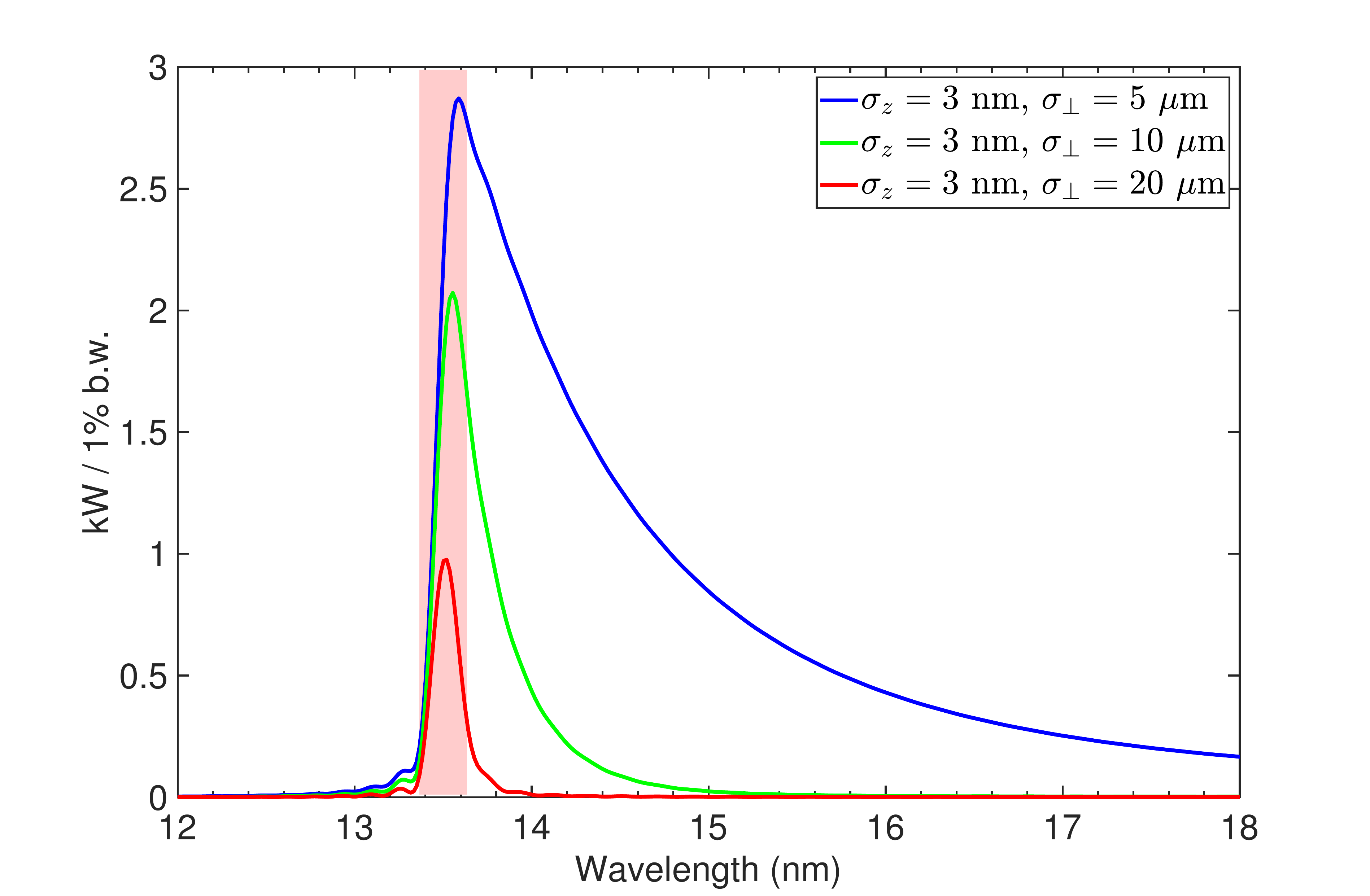}
\includegraphics[width=1\columnwidth]{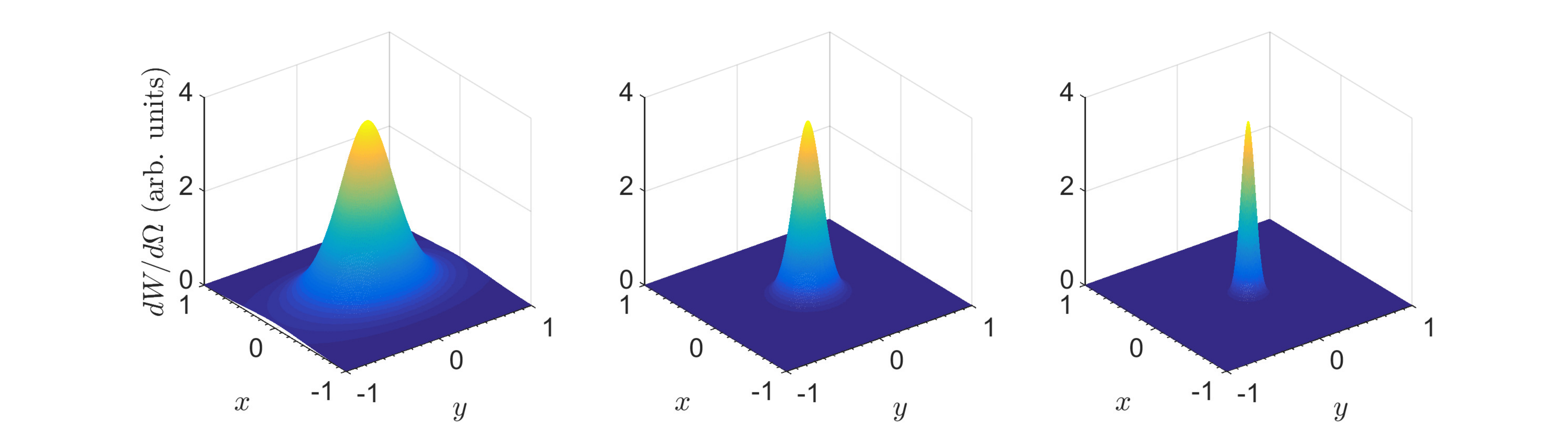}
\caption{\label{fig:EUVSSMB2} 
An example EUV SSMB radiation calculation with a microbunch length of $\sigma_{z}=3$ nm and different transverse sizes $\sigma_{\bot}$. The upper part shows the energy spectrum. The total radiation power are 39~kW, 7~kW, 1.7~kW, corresponding to $\sigma_{\bot}=5,10,20\ \mu$m, respectively. The shaded area corresponds to wavelength of $13.5\pm\frac{13.5}{100}$ nm. The bottom part shows spatial distribution of radiation energy plotted in 3D view in the ($x$, $y$) space with $x=\gamma\theta\cos\varphi, y=\gamma\theta\sin\varphi$. From left to right: $\sigma_{\bot}=5,10,20\ \mu$m. Parameters used for the calculation: $E_{0}=400$ MeV, $I_{\text{avg}}=1$ A, $\lambda_{\text{L}}=1064$ nm, $\lambda_{\text{r}}=\frac{\lambda_{\text{L}}}{79}=13.5$~nm, $\lambda_{u}=1$ cm, $K=1.14$, $N_{u}=79$.
}
\end{figure}

First we present the result for the average property of the EUV radiation.
The calculation is based on Eq. (\ref{eq:totalEnergy}), (\ref{eq:bunchingfactor}) and (\ref{eq:undulatorRadiation}), and the result is shown in Fig.~\ref{fig:EUVSSMB2}.
The total radiation power is calculated according to
\begin{equation}
P=\frac{W}{\lambda_{\text{L}}/c},
\end{equation}
where $W$ is the total radiation energy loss of each microbunch. The upper part of the figure shows the energy spectrum. The lower part shows the spatial distribution of the radiation energy. The total radiation power are 39~kW, 7~kW, 1.7~kW, corresponding to $\sigma_{\bot}=5,10,20\ \mu$m, respectively. As a reference, the radiation power calculated based on Eq.~(\ref{eq:RPH}) for these three transverse beam sizes are 1.8 kW, 1.5 kW and 0.93 kW, respectively. The reason Eq.~(\ref{eq:RPH}) gives a smaller power as explained is that it does not take into account the red-shifted part of the radiation. Therefore, Eq.~(\ref{eq:RPH}) can be used to evaluate the lower bound of the radiation power from SSMB, once a parameters set of electron beam and radiator undulator is given.  It can be seen that generally, 1 kW EUV radiation power can be straightforwardly anticipated from a 3 nm microbunch train with an average beam current of 1 A.

Another important observation is that the spectral and spatial distribution of SSMB radiation depends strongly on the transverse size of the electron beam.  A large transverse size results in a decrease of the overall radiation  power, and will also make the radiation more narrow-banded and collimated in the forward direction. This is an important conclusion drawn from our investigation on the generalized transverse form factor. Using the example parameters, i.e, $E_{0}=400$ MeV, $\lambda_{\text{L}}=1064$ nm, $\lambda_{0}=\frac{\lambda_{\text{L}}}{79}=13.5$ nm, $\lambda_{u}=1$~cm, $N_{u}=79$, $K=1.14$, if $\sigma_{\bot}=10\ \mu$m, then according to Eqs.~(\ref{eq:bandwidthTrans}) and (\ref{eq:thetaTrans}), we can calculate the relative bandwidth and opening angle due to the transverse form factor
\begin{equation}
\begin{aligned}
\frac{\Delta\omega_{e^{-1}}}{\omega_{0}}\bigg|_{\bot}&\approx1.7\%,\\ \theta_{e^{-1}}\bigg|_{\bot} &\approx 0.21\ \text{mrad}.
\end{aligned}
\end{equation}
which is in agreement with the result presented in Fig.~\ref{fig:EUVSSMB2}.

The energy spectrum and spatial distribution presented in Fig.~\ref{fig:EUVSSMB2} is for that of a single microbunch. For the energy spectrum of radiation from a periodic microbunch train, as shown in Fig.~\ref{fig:MT}, we just need to multiply it with periodic delta function in frequency domain whose frequency separation is the modulation laser frequency, the situation of which is similar to that of conventional synchrotron radiation in a storage ring. Corresponding to these delta function lines in the energy spectrum, there will ring-shaped peaks in the spatial distribution of the coherent radiation as a result of the interference of radiation from different microbunches. The polar angles of these rings, corresponding to the delta function lines in energy spectrum, are determined by the off-axis resonant condition. Note, however, the beam energy spread and angular divergence will make the linewidth of these delta function lines non-zero. For example, the relative bandwidth of the radiation caused by an energy spread of $\sigma_{\delta}$ is $2\sigma_{\delta}$.

As a result of the high-power and narrow-band feature of the SSMB radiation, a high EUV photon flux of $6\times10^{15}$ phs / s within a 0.1 meV bandwidth can be obtained, if we can realize 1 kW / 1\% b.w. as shown in Fig.~\ref{fig:EUVSSMB2}. Such a high photon flux is highly desired in condensed matter physics study, for example in the application of ultra-high-energy-resolution angle-resolved photoemission spectroscopy (ARPES) to study the energy gap distribution and electron states of magic-angle graphene in superconductivity \cite{cao2018unconventional}.

\begin{figure}[tb]
	\centering
	\includegraphics[width=0.8\columnwidth]{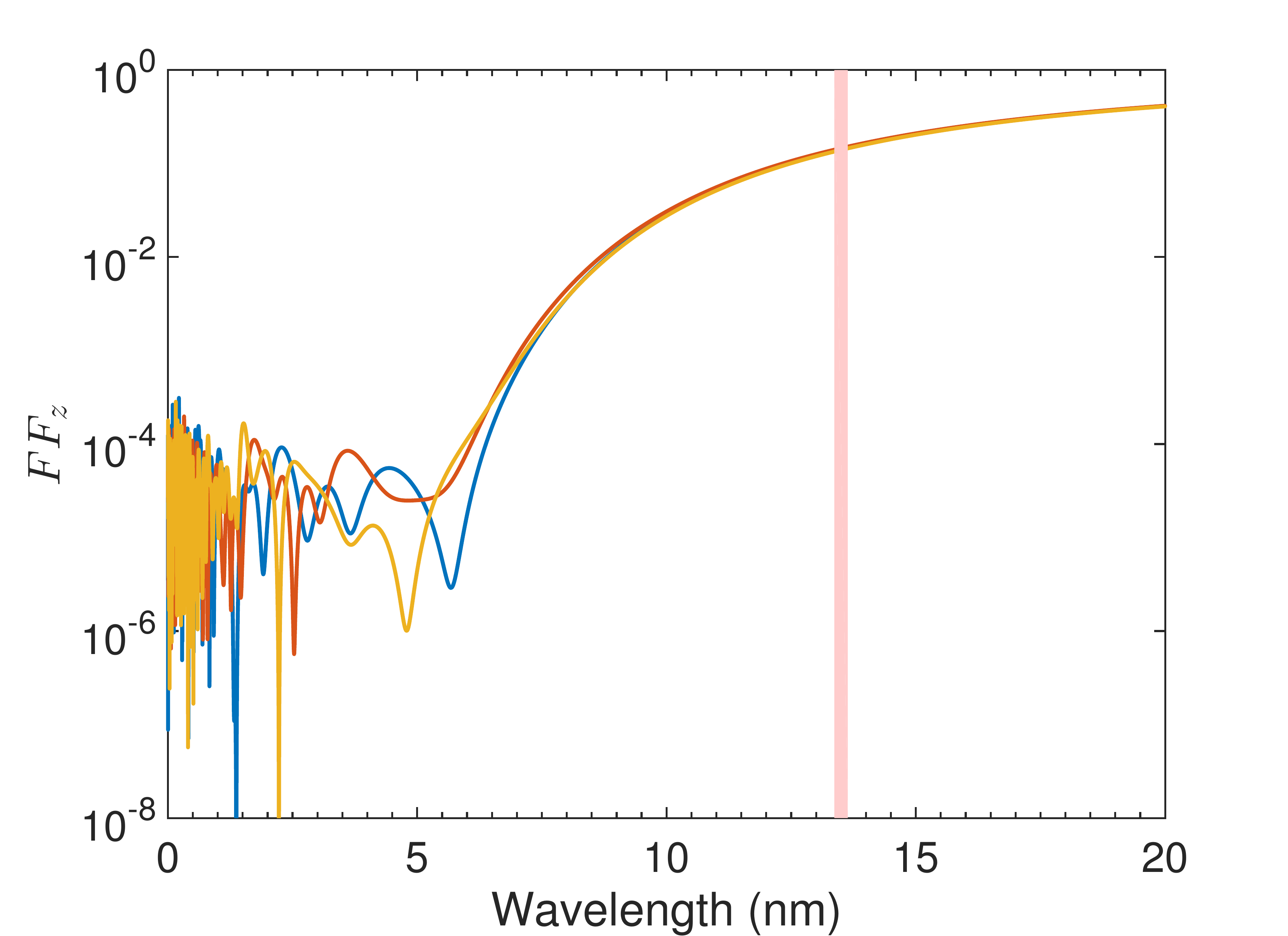}
	\caption{\label{fig:EUVSSMBFluc} 
		The spectrum of the longitudinal form factor of three possible realizations of a Gaussian microbunch length of $\sigma_{z}=3$ nm and $N_{e}=2.2\times10^{4}$. The shaded area corresponds to wavelength of $13.5\pm\frac{13.5}{100}$ nm.
	}
\end{figure}

\subsection{Statistical Property}

Now we present the result for the statistical property of the radiation. For the case of a Gaussian bunch with $\sigma_{z}=3$ nm and $N_{e}=2.2\times10^{4}$ which corresponds to 1 A average current if the modulation laser wavelength is $\lambda_{\text{L}}=1064$ nm, from Eq.~(\ref{eq:powerFluctuation}) we know that the relative fluctuation of the turn-by-turn or microbunch-by-microbunch on-axis 13.5 nm coherent radiation power will be around $2\%$.

Figure~\ref{fig:EUVSSMBFluc} gives an example plot for the longitudinal form factor spectrum of three possible realizations of such a Gaussian microbunch, with an expectation rms bunch length of 3 nm and $N_{e}=2.2\times10^{4}$. As can be seen, the spectrum is noisy mainly in the high-frequency, or short wavelength range. Since our EUV radiation is mainly at the wavelength close to 13.5 nm, and the form factor close to this frequency fluctuates together from turn to turn, or bunch to bunch. So the overall radiation power fluctuation is also about 2\% as analyzed above.  This fluctuation is also the fluctuation of the center of motion of the microbunch induced by the coherent radiation.

Note that this 2\% fluctuation of radiation power should have negligible impact for the application in EUV lithography, since the revolution frequency of the microbunch in the ring is rather high (MHz), let alone if we consider that there is actually a microbunch each modulation laser wavelength.

%% file: data/chap05.tex
\chapter{Proof-of-Principle Experiments}
\label{cha:pop}

\section{Importance and Strategy of the PoP Experiments}
Since the first publication of the SSMB concept~\cite{ratner2010steady}, there is continuous development on its theoretical concepts and beam physics study \cite{ratner2011reversible,jiao2011terahertz,chao2016high,khan2017ultrashort,tang2018overview,chao2018SSMB,rui2018strong,pan2019storage,pan2020research,li2019lattice,deng2020single,deng2020widening,zhang2021ultralow,deng2021courant,deng2021theorem,tsai2021coherent,tsai2021longitudinal,tsai2021theoretical,tang2022steady,deng2022electron,deng2022average}. To make SSMB a real choice for future photon source facility, a crucial step is to experimentally demonstrate its working mechanism. Therefore, the SSMB proof-of-principle (PoP) experiment is launched as one of the key tasks since the SSMB task force was established~\cite{tang2018overview,chao2018SSMB}.  

\begin{figure}[tb] 
	\centering 
	\includegraphics[width=1\textwidth]{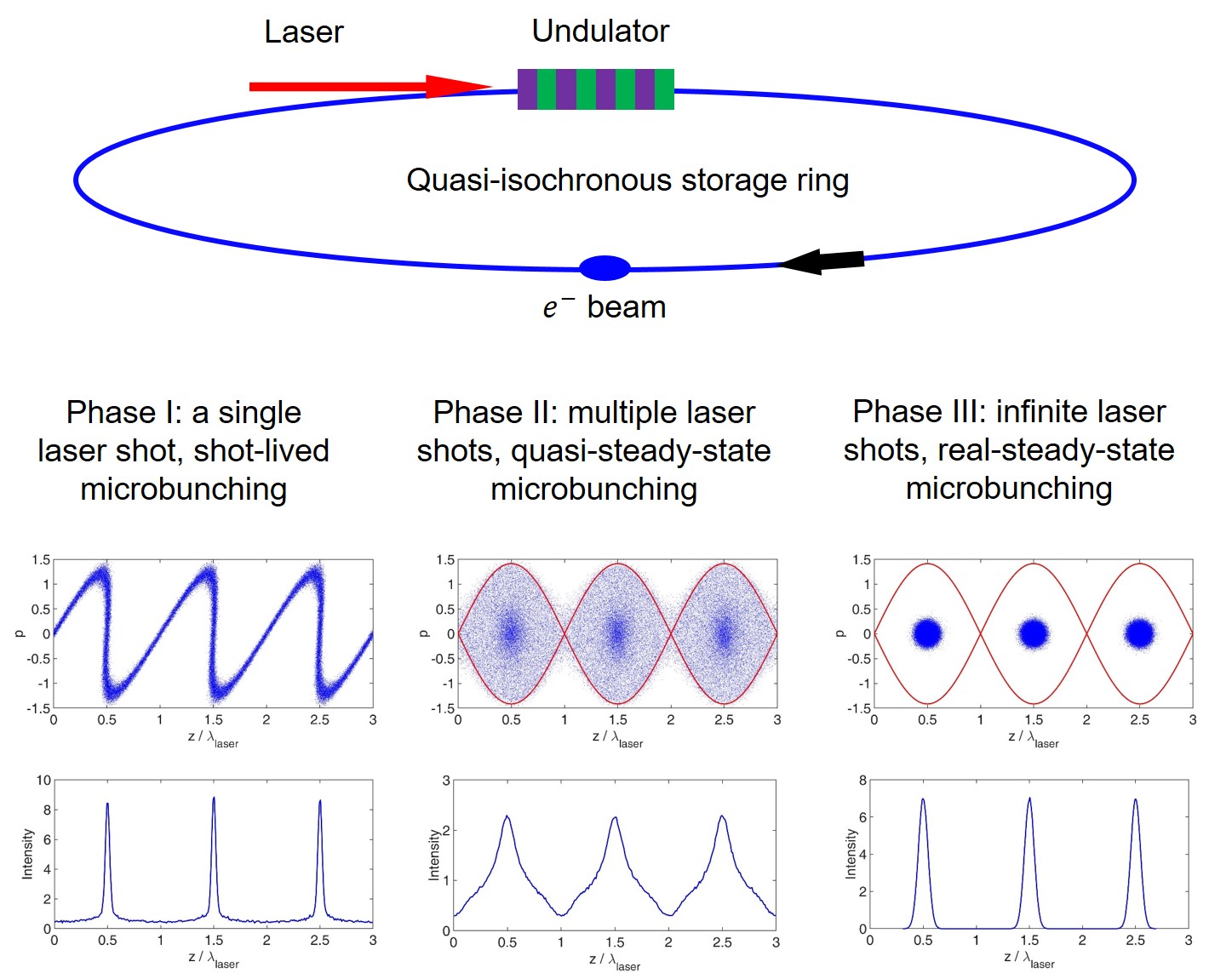}
	\caption{
		\label{fig:Chap5-SSMBPoPThreeSTages} 
		Three stages of SSMB PoP experiments: from single-shot to multiple shots to infinite shots laser pulse; from short-lived to quasi-steady-state to real steady-state microbunching. 
	}
\end{figure}

\subsection{Three Stages of PoP Experiments}\label{sec:threePoP}
Considering the fact that it is a demanding task to realize SSMB directly in an existing machine, part of the reasons we have analyzed in previous chapters, among them the most fundamental one is the large quantum diffusion of bunch length arising from the large partial phase slippage, the SSMB PoP experiment has been divided into three stages. The three stages of the SSMB proof-of-principle experiment are shown in Fig.~\ref{fig:Chap5-SSMBPoPThreeSTages}. Some brief descriptions of the three stages are as follows.
\begin{itemize} 
	\item Phase I: a single-shot laser is fired to interact with the electron beam stored in a quasi-isochronous ring. The modulated electron beam becomes microbunched at the same place of modulation after one complete revolution in the ring and this microbunching can last several revolutions. By doing this experiment, we want to confirm that the optical phases, i.e., the longitudinal coordinates, of electrons can be correlated turn-by-turn in a sub-laser-wavelength precision. The realization of SSMB relies on this precise turn-by-turn phase correlation;
	\item Phase II: on the basis of Phase I, we replace the single-shot laser with a high-repetition phase-locked one to interact with the electrons turn after turn. By doing this, we want to establish stable microbuckets and sustain the microbunching in the microbuckets to reach a quasi-steady state;
	\item Phase III: Phase II is very close to the final SSMB, but is still not, as a true SSMB means the balance of quantum excitation and radiation damping. If the magnet lattice is appropriate, following Phase II, the electron beam will be damped to the microbucket center and form real steady-state microbunches. However, the requirement of a true SSMB on the magnet lattice is demanding, especially the quantum diffusion of longitudinal coordinate $z$ due to the partial phase slippage variance. Therefore, this final stage is more likely to be realized in a dedicated ring designed for SSMB, which is also one of the key ongoing tasks of the SSMB task force~\cite{tang2018overview,rui2018strong,pan2019storage,pan2020research}.
\end{itemize}
Below, we use PoP I, II, III to represent the three stages of the experiment. The key words of the three stage experiments are summarized as follows.
\begin{itemize}
	\item PoP I: stored electron bunch, microbunching, turn-by-turn phase correlation;
	
	\item PoP II: quasi steady state, bounded motion, microbucket;
	
	\item PoP III: real steady state, balance of diffusion and damping.
\end{itemize}

These three stages each have their own significance and are all important for the SSMB development. Recently, we have successfully performed the PoP I and demonstrated the mechanism of SSMB at the Metrology Light Source (MLS) of Physikalisch-Technische Bundesanstalt (PTB) in Berlin~\cite{deng2021experimental,tang2020first,feikes2021progress}. The experiment is a collaboration work of Tsinghua, Helmholtz-Zentrum Berlin (HZB) and PTB. 

\subsection{Metrology Light Source Storage Ring}
\begin{figure}[tb]
	\centering 
	\includegraphics[width=1\textwidth]{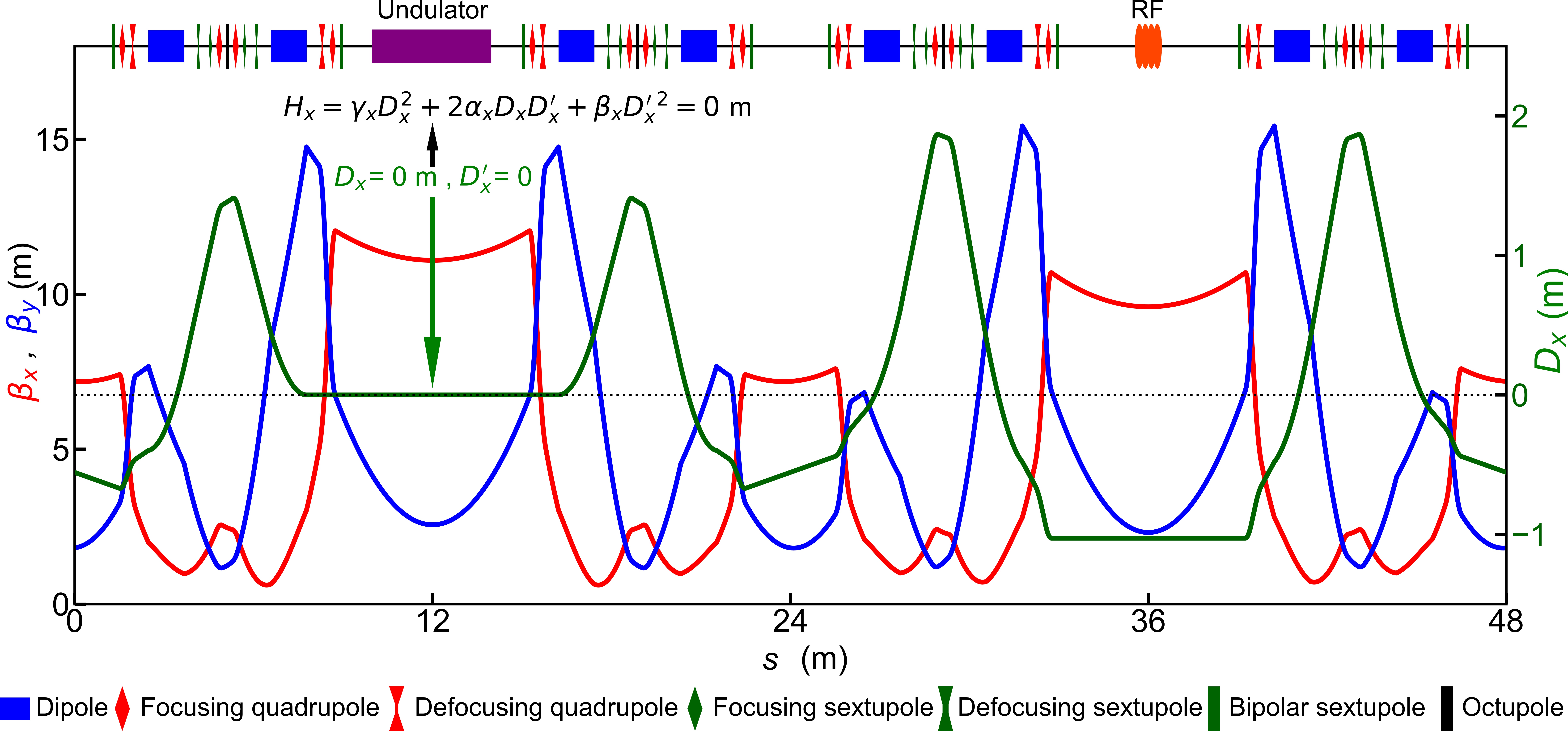}
	\caption{\label{fig:Chap5-SSMBPoPIExtFig1}The MLS quasi-isochronous magnet lattice used to generate microbunching. The magnet lattice and the key are shown at the top. The curves are the model horizontal (red) and vertical (blue) $\beta$-functions and the horizontal dispersion $D_{x}$ (green). Operating parameters of the ring: beam energy, $E_{0}$ = 250 MeV; relative energy spread, $\sigma_{\delta}=1.8\times10^{-4}$ (model); horizontal emittance, $\epsilon_{x}$ = 31 nm (model); horizontal betatron tune, $\nu_{x}=3.18$ (model and measured); vertical betatron tune, $\nu_{y}=2.23$ (model and measured); horizontal chromaticity, $\xi_{x}=-0.5$ (measured). Note that this optics is different from that used in Sec.~\ref{sec:singleRF} for the simulation of partial phase slippage effect.}
\end{figure}

The MLS is a storage ring dedicated optimized for quasi-isochronous operation~\cite{klein2008operation,feikes2011metrology,ries2014nonlinear}, thus an ideal testbed for SSMB physics investigation and PoP experiments. However, the partial phase slippage of the MLS is large as the bending angle of each dipole is large ($\pi/4$) and the dispersion magnitude inside the dipoles is also large, so it is not feasible to realize true SSMB, i.e., PoP III, directly at the MLS. Therefore, the SSMB PoP experiment has been divided into three stages as introduced just now, and PoP I and II are what we have performed and plan to conduct at the MLS. Some basic parameters of the MLS are shown in Tab.~\ref{tab:PoP}. The lattice optics of the MLS used in the SSMB PoP experiments are shown in Fig.~\ref{fig:Chap5-SSMBPoPIExtFig1}.

\begin{table}[tb]
	\caption{\label{tab:PoP}
		Basic parameters of the MLS lattice.}
	\centering
	\begin{tabular}{lll}
		\hline
		Parameter & \multicolumn{1}{l}{\textrm{Value}}  & Description \\
		\hline
		$E_0$ & 50 - 630  MeV & Beam energy \\
		$C_0$ & 48  m & Ring circumference \\
		$f_{\text{rf}}$  & 500  MHz & RF frequency \\
		$V_{\text{rf}}$  & $\leq 600$  kV & RF voltage \\	
		$\eta$  & $3\times10^{-2}$  & Phase slippage factor (standard user) \\
		$\eta$  & $-2\times10^{-5}$  & Phase slippage factor (SSMB experiment) \\
		$U_{0}$  & 226  eV@250 MeV & Radiation loss per turn \\
		$J_{s}$  & 1.95 & Longitudinal damping partition \\	
		$\tau_{\delta}$ & 180 ms@250 MeV & Longitudinal radiation damping time \\
		$\sigma_{\delta}$ & $1.8\times10^{-4}$@250 MeV & Natural energy spread \\
		$\sigma_{z}$ & $36\ \mu$m (120 fs)@250 MeV & Zero-current bunch length (SSMB experiment) \\
		$\nu_{x}$ & 3.18 & Horizontal betatron tune \\
		$\nu_{y}$ & 2.23 & Vertical betatron tune \\	
		$\epsilon_{x}$ & 31 nm@250 MeV & Horizontal emittance \\
		$\lambda_{\text{L}}$ & 1064 nm & Modulation laser wavelength \\
		$\lambda_{u}$ & 125 mm & Undulator period length \\
		$N_{u}$ & 32 & Number of undulator periods\\
		$L_{u}$ & 4 m & Undulator length\\
		$K$ & 2.5 & Undulator parameter \\
		\hline	
	\end{tabular}
\end{table}

\section{PoP I: Turn-by-Turn Laser-Electron Phase Correlation}

\subsection{Experimental Setup}

\begin{figure}[tb]
	\centering 
	\includegraphics[width=1\textwidth]{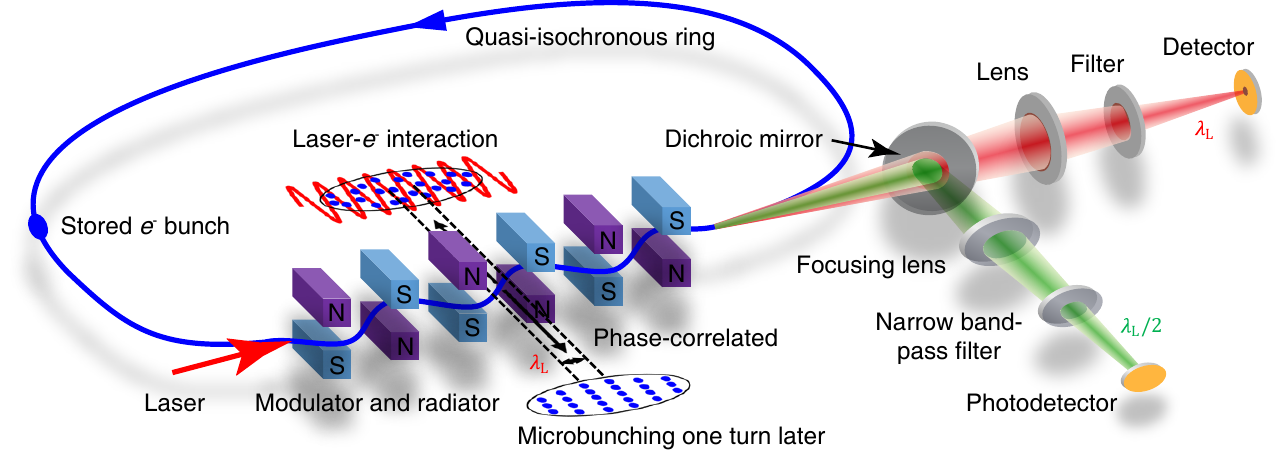}
	\caption{\label{fig:Chap5-SSMBPoP}Schematic of the experimental setup. The stored 250 MeV electron bunches are energy-modulated by a 1064 nm wavelength laser in an undulator, and become microbunched after one complete revolution in the 48 m circumference quasi-isochronous storage ring. This formed microbunching can then preserve for multiple turns in the ring. Each time the microbunching going through the undulator, narrow-band coherent radiation will be generated. The undulator radiation is separated into the fundamental and second harmonics by dichroic mirrors, and sensitive photodiodes are used as the detectors. Narrow band-pass filters can be inserted in front of the photodetectors to pick out the narrow-band coherent radiation generated from the microbunching. (Figure from Ref.~\cite{deng2021experimental})}
\end{figure}

Figure~\ref{fig:Chap5-SSMBPoP} shows the schematic setup of the SSMB PoP I experiment.  A horizontally polarized laser pulse (wavelength, $\lambda_{\text{L}}$ = 1064 nm; full-width at half-maximum, FWHM $\approx$ 10 ns; pulse energy, $\approx$ 50 mJ) is sent into a planar undulator (period, $\lambda_{u}$ = 0.125 m; total length, $L_{u}=4$ m) to co-propagate with the electron bunches (energy, $E_{0}$ = 250 MeV) stored in the MLS storage ring (circumference, $C_{0}$ = 48 m). To maximize the laser-electron energy exchange, the undulator gap is chosen to satisfy the resonance condition $\lambda_{s}=\lambda_{\text{L}}$, where $\lambda_{s}=\frac{1+K^{2}/2}{2\gamma^{2}}\lambda_{u}$ is the central wavelength of the spontaneous undulator radiation, with $\gamma\propto E_{0}$ being the Lorentz factor and $K=\frac{eB_{0}}{m_{e}ck_{u}}=0.934\cdot B_{0}[\text{T}]\cdot\lambda_{u}[\text{cm}]$ being the dimensionless undulator parameter, determined by the undulator period and magnetic field strength. This laser-electron interaction induces a sinusoidal energy modulation pattern on the electron beam with a period of the laser wavelength. Because particles with different energies have different revolution periods, after one revolution in the ring, the energy-modulated electrons shift longitudinally with respect to each other, clumping towards synchronous phases and forming microbunches. The formed microbunches can last several revolutions in the ring. The coherent undulator radiation generated from the microbunches, detected by a high-speed photodetector with a photodiode, confirms microbunching.

The symplectic longitudinal dynamics of the the above experiment processes can be modeled by
\begin{equation}\label{eq:PoPEquation1}
\begin{cases}
&\delta_{1}=\delta_{0}+A\sin(k_{L}z_{0}),\\
&z_{1}=z_{0}-\eta C_{0}\delta_{1},
\end{cases}
\end{equation}
for the first revolution with laser modulation, and 
\begin{equation}\label{eq:PoPEquation2}
\begin{cases}
&\delta_{m+1}=\delta_{m},\\
&z_{m+1}=z_{m}-\eta C_{0}\delta_{m+1},
\end{cases}
\end{equation} 
for the later revolutions. This demonstration proves that the longitudinal dynamics described by the {\it one-turn map} Eq.~(\ref{eq:PoPEquation1}) can be extrapolated from the RF wavelength (metre scale) to laser wavelength (micrometre scale) for a {\it stored} electron beam, thus validating the SSMB microbunching mechanism.

\subsection{Physical Analysis of Microbunching Formation}

\subsubsection{Storage Ring}

{\bf Operation energy} 
The above models Eqs.~(\ref{eq:PoPEquation1}) and (\ref{eq:PoPEquation2}), however, do not consider the non-symplectic, transverse-longitudinal coupling and nonlinear dynamics, which all could lead to degradation of the microbunching. It turns out that the first non-symplectic dynamics we need to account for is the synchrotron radiation itself. As we know, when a relativistic electron is subjected to an acceleration normal to its velocity exerted by a bending magnet, it radiates electromagnetic
energy~\cite{elder1947radiation,tzu1948,schwinger1949classical}. This radiation is characterized by the quantum nature of the photon emission process. The photon energy and place or time of emission are
both stochastic, giving rise to changes on particle energy~\cite{sands1970physics} (instantly)
and the longitudinal coordinate~\cite{shoji1996longitudinal,deng2020single,deng2021courant} $z$ (non-instantly), as analyzed before in Sec.~\ref{sec:PartialAlpha}. Of special interest in the experiment is the RMS quantum diffusion of $z$ in one turn $d_{z}$. According to Eq.~(\ref{eq:zdiffusion}), we have  
\begin{equation}
d_{z}=\sqrt{\langle z^2\rangle-\langle z\rangle^2}=\sqrt{\langle F^{2}\rangle\langle \mathcal{N}\rangle\left\langle \frac{u^{2}}{E_{0}^{2}}\right\rangle}.
\end{equation}
For the MLS quasi-isochronous magnet lattice used in the PoP experiment as shown in Fig.~\ref{fig:Chap5-SSMBPoPIExtFig1}, $d_{z}$ is as large as 260 nm at its standard operation energy of 630 MeV, deteriorating
the sub-micrometre microbunching considerably. Therefore, the beam energy needs to be lowered to mitigate this diffusion, as $\sqrt{\langle \mathcal{N}\rangle \left\langle \frac{u^{2}}{E_{0}^{2}}\right\rangle}\propto\gamma^{2.5}$. At the same time, a lower beam energy gives a smaller energy spread and is also beneficial for microbunching, as the smearing from the natural uncorrelated energy spread becomes smaller. Nevertheless, the beam energy cannot be too low, otherwise the beam parameters and lifetime could be profoundly affected by scattering among particles~\cite{piwinski1974intra,bjorken1982intrabeam,bernardini1963lifetime}. An electron beam energy of 250 MeV is adopted in the experiment to balance these issues. At $E_{0}$ = 250 MeV, we have $d_{z}=26$ nm and $\sigma_{\delta}=1.8\times10^{-4}$.

{\bf Phase slippage factor} Because the laser wavelength is much smaller than that of an RF wave, the phase slippage factor $\eta$ needs to be ultra small. That is, the ring
should be quasi-isochronous to allow turn-by-turn stabilization of the electron optical phases, i.e., the longitudinal coordinates, for particles with different energies.
More quantitatively, the RMS spread
of $z$ in one turn that arises from the uncorrelated electron energy spread
should be adequately smaller than the laser wavelength, 
\begin{equation}
\Delta z_{\text{ES}}=|\eta C_{0}\sigma_{\delta}|\leq\lambda_{\text{L}}/2\pi.
\end{equation}
To fulfill this requirement, the phase slippage factor of the MLS was
lowered to $\eta\approx-2\times10^{-5}$, which is three orders of magnitude smaller than its standard value of $3\times10^{-2}$. By implementing these parameters,
$\Delta z_{\text{ES}}=|\eta C_{0}\sigma_{\delta}|\approx0.17\ \mu$m (about 0.6 fs), enabling the formation and preservation of sub-micrometre microbunching. 

Such a quasi-isochronous magnet lattice is
achieved by tailoring the horizontal dispersion functions $D_{x}$ 
around the ring so that a particle with non-ideal energy travels part of
the ring inwards and part of the ring outwards compared to the reference
orbit, thus having a revolution period nearly the same as that of
the ideal particle. The tailoring of $D_{x}$ is accomplished by adjusting the
(de)focusing strengths of the quadrupole magnet. The operation of the
MLS as a quasi-isochronous ring also benefits from the optimization
of the sextupole and octupole nonlinear magnet schemes to control
the higher-order terms of the phase slippage~\cite{feikes2011metrology,ries2014nonlinear}, which affect both the equilibrium beam distribution in the longitudinal phase space before the laser modulation and
the succeeding microbunching evolution as analyzed in Sec.~\ref{sec:NonlinearAlpha}.

In the experiment, the value of the small phase slippage factor is
quantified by measuring the electron orbit offsets while slightly adjusting
the RF frequency up and down at a beam position monitor (BPM),
where $D_{x}$ is large. From the offsets and $D_{x}$, the phase-slippage-dependent
electron energy shifts caused by the RF frequency adjustment can be
derived, as well as the phase slippage factor. The $D_{x}$ value at the BPM
is acquired using the same method in the opposite way, that is, based
on a known phase slippage factor. This is done at a larger phase slippage
factor, at which its value can be determined from its relation to
the synchrotron oscillation frequency of the electron beam as well-known in storage ring physics~\cite{sands1970physics}. The
synchrotron oscillation frequency can be measured accurately at a
phase slippage factor such as $-5\times10^{-4}$, and the model confirmed that
the relative change of $D_{x}$ at the highly dispersive BPM is small ($<4\%$)
when the phase slippage factor is reduced from $-5\times10^{-4}$ to the desired
$-2\times10^{-5}$ by marginally changing the quadrupole magnet strengths.

\begin{figure}[tb]
	\centering
	\includegraphics[width=0.49\textwidth]{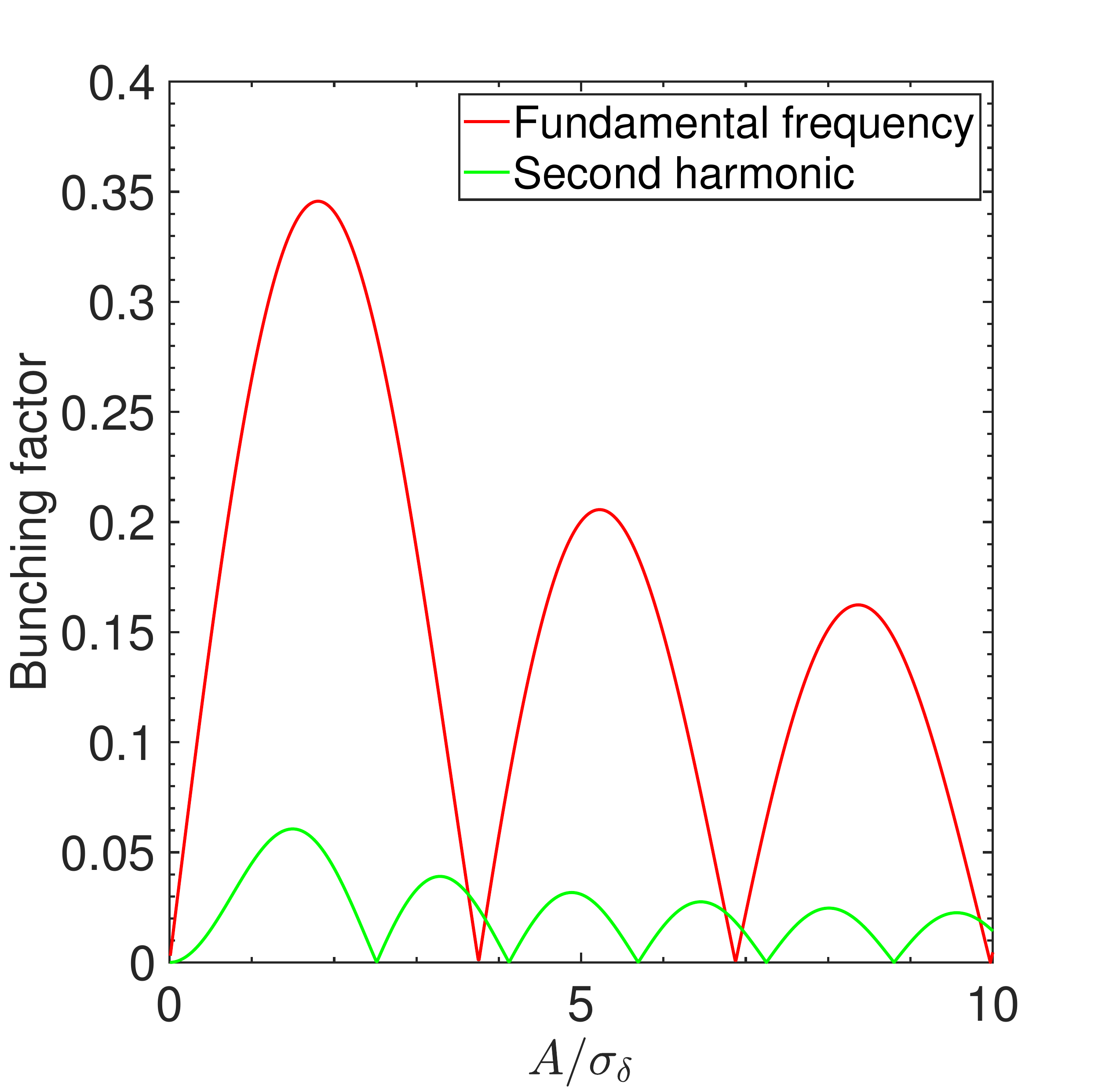}
	\includegraphics[width=0.49\textwidth]{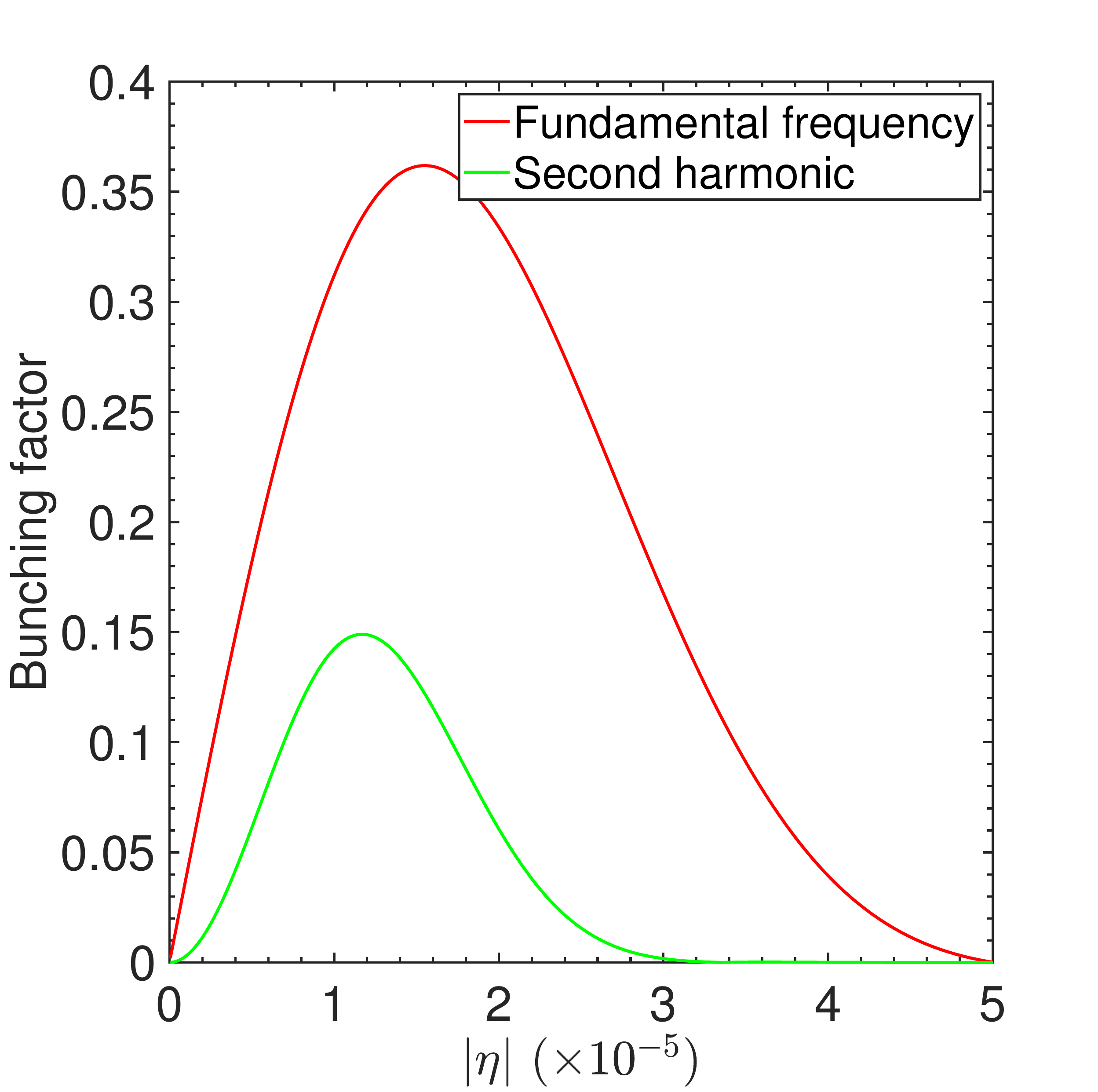}
	
	\caption{
		\label{fig:Chap5-BFEta} 
		Left: impact of the energy modulation strength ($A\propto\sqrt{P_{\text{L}}}$) on the bunching factor at the fundamental and second harmonic, with $\eta=-2\times10^{-5}$.  
		Right: impact of the phase slippage factor $\eta$ on the bunching factor at the fundamental and second harmonic, with $A=1.5\sigma_{\delta}$. 
	}
\end{figure}

\begin{figure}[tb]
	\centering
	\includegraphics[width=0.49\textwidth]{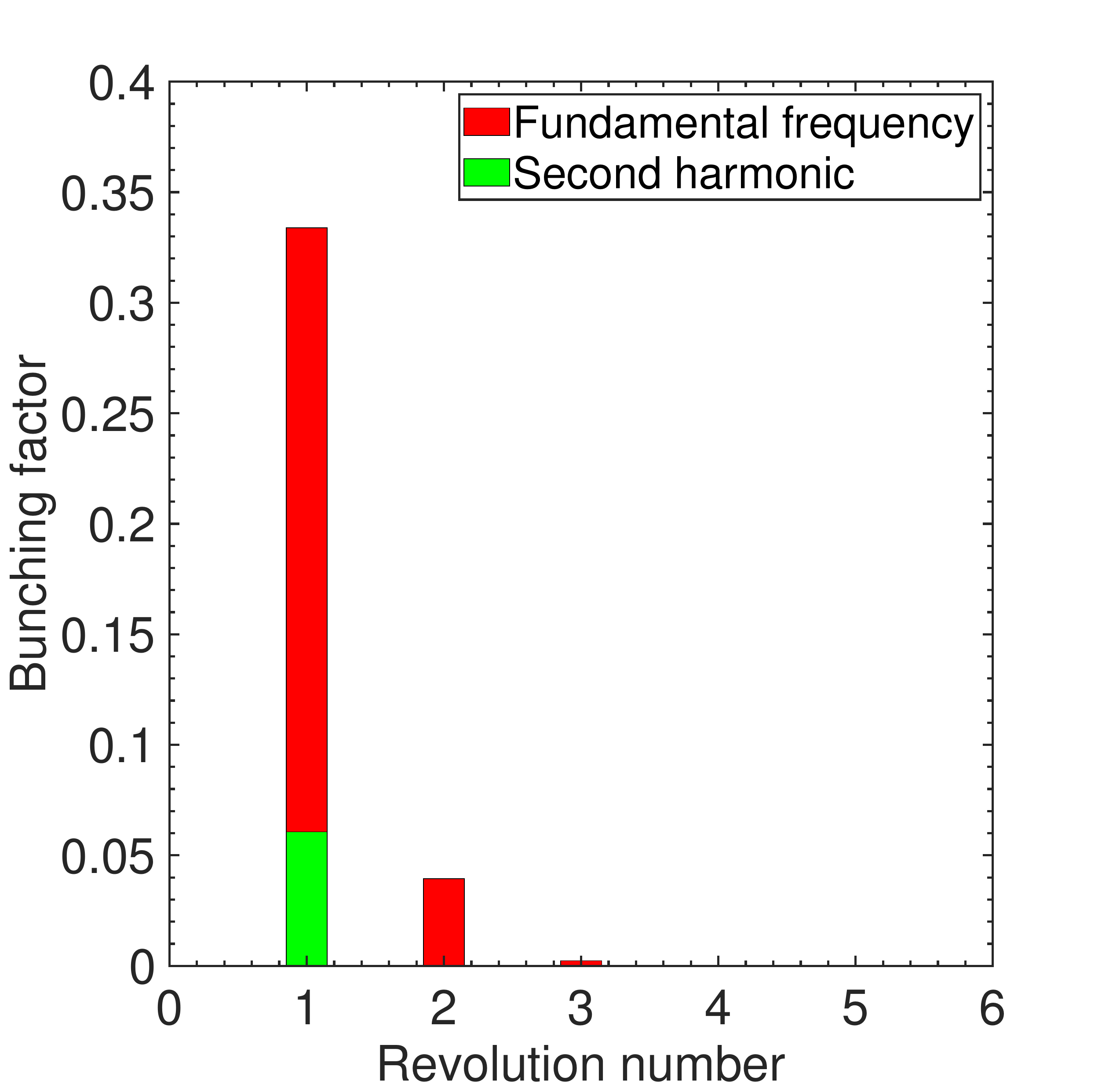}

	\caption{
		\label{fig:Chap5-BFTurnNo} 
		The evolution of the bunching factor at the fundamental and second harmonic with respect to the revolution number, with $A=1.5\sigma_{\delta}$ and $|\eta|=2\times10^{-5}$.
	}
\end{figure}

{\bf Bunching factor} With the electron beam evolved according to Eq.~(\ref{eq:PoPEquation1}) for one revolution
in the ring, the bunching factor at the $n^{\text{th}}$ laser harmonic
as analyzed in Sec.~\ref{sec:NonlinearAlpha} is
\begin{equation}\label{eq:BFPoPI}
b_{n}=\left|J_{n}(nk_{L}\eta C_{0} A)\right|\text{exp}\left[-\frac{(nk_{L}\eta C_{0}\sigma_{\delta})^{2}}{2}\right],
\end{equation}
where $J_{n}$ is the $n$-th order Bessel function of the first kind.  The coherent radiation
power at the $n^{\text{th}}$ harmonic is proportional to the bunching factor
squared, $P_{n,\text{coh}}\propto|b_{n}|^{2}$. With the dynamics in the following turns modeled by Eq.~(\ref{eq:PoPEquation2}), for the $m^{\text{th}}$ revolution, we just need to replace the $\eta C_{0}$ in Eq.~(\ref{eq:BFPoPI}) with $m\eta C_{0}$.

The maximum reachable bunching factor becomes larger with the decrease of $\eta$, given that the optimal $A$ can always be realized. $\eta=-2\times10^{-5}$ is approximately the present lowest reachable value at the MLS, so here below we use this $\eta$ for the analysis. As we will explain soon, our signal detection at first focuses on the second harmonic.   As shown in Fig.~\ref{fig:Chap5-BFEta}, given $\eta=-2\times10^{-5}$, a modulation strength of $A\approx1.5\sigma_{\delta}$ results in the maximum bunching at the second harmonic. Correspondingly, given $A=1.5\sigma_{\delta}$, the optimal $\eta$ for the fundamental frequency and second harmonic bunching is a bit smaller than $2\times10^{-5}$. Note that the optimized conditions for the fundamental frequency and second harmonic bunching are different. Figure~\ref{fig:Chap5-BFTurnNo} shows the bunching factor evolution with respect to the revolution number, with $A=1.5\sigma_{\delta}$ and $|\eta|=2\times10^{-5}$. As can be seen, the second harmonic bunching can last only one turn, while the fundamental frequency bunching can last about three turns or even more if $\eta$ becomes smaller, as also can be seen in the right part of Fig.~\ref{fig:Chap5-BFEta}. These expectations have also been confirmed in the experiment as will be presented soon. 


{\bf Chromatic $\mathcal{H}_{x}$ function and chromaticity $\xi_{x}$} 
Apart from the longitudinal beam dynamics, the coupling of the
particle betatron oscillation to the longitudinal dimension is also critical.
The reason is based on the fact that the horizontal beam width at the
undulator is about $600\ \mu$m (model value), three orders of magnitude larger than
the sub-micrometre longitudinal structures that we aim to produce.
Because the vertical emittance is much smaller than the horizontal one
in a planar $x$-$y$ uncoupled storage ring, in the following we consider only the impact
of the horizontal betatron oscillation.

According to Eq.~(\ref{eq:TLCBunchlengtheningmRevolution}), for a periodic system, 
the RMS bunch lengthening of an electron beam longitudinal slice after $m$ complete revolutions in the ring, due to betatron
oscillation, is
\begin{equation}
\Delta z_{B,m}=2\sqrt{\epsilon_{x}\mathcal{H}_{x}}\left|\sin(m\pi\nu_{x})\right|.
\end{equation}
With this bunch lengthening taken into account, the bunching factor at the $n^{\text{th}}$ laser harmonic after $m$ revolutions will be
\begin{equation}\label{eq:BFPoPIWithHx}
b_{n,m}=\left|J_{n}(nk_{L}m\eta C_{0} A)\right|\text{exp}\left[-\frac{(nk_{L}m\eta C_{0}\sigma_{\delta})^{2}+\left(nk_{L}2\sqrt{\epsilon_{x}\mathcal{H}_{x}}\left|\sin(m\pi\nu_{x})\right|\right)^{2}}{2}\right].
\end{equation}
The relative bunching factor reduction due to the non-zero $\mathcal{H}_{x}$ can thus be defined as 
\begin{equation}
R_{n,m}(\mathcal{H}_{x})=\text{exp}\left[-\frac{\left(nk_{L}2\sqrt{\epsilon_{x}\mathcal{H}_{x}}\left|\sin(m\pi\nu_{x})\right|\right)^{2}}{2}\right].
\end{equation}

\begin{figure}[tb] 
	\centering 
	\includegraphics[width=0.49\columnwidth]{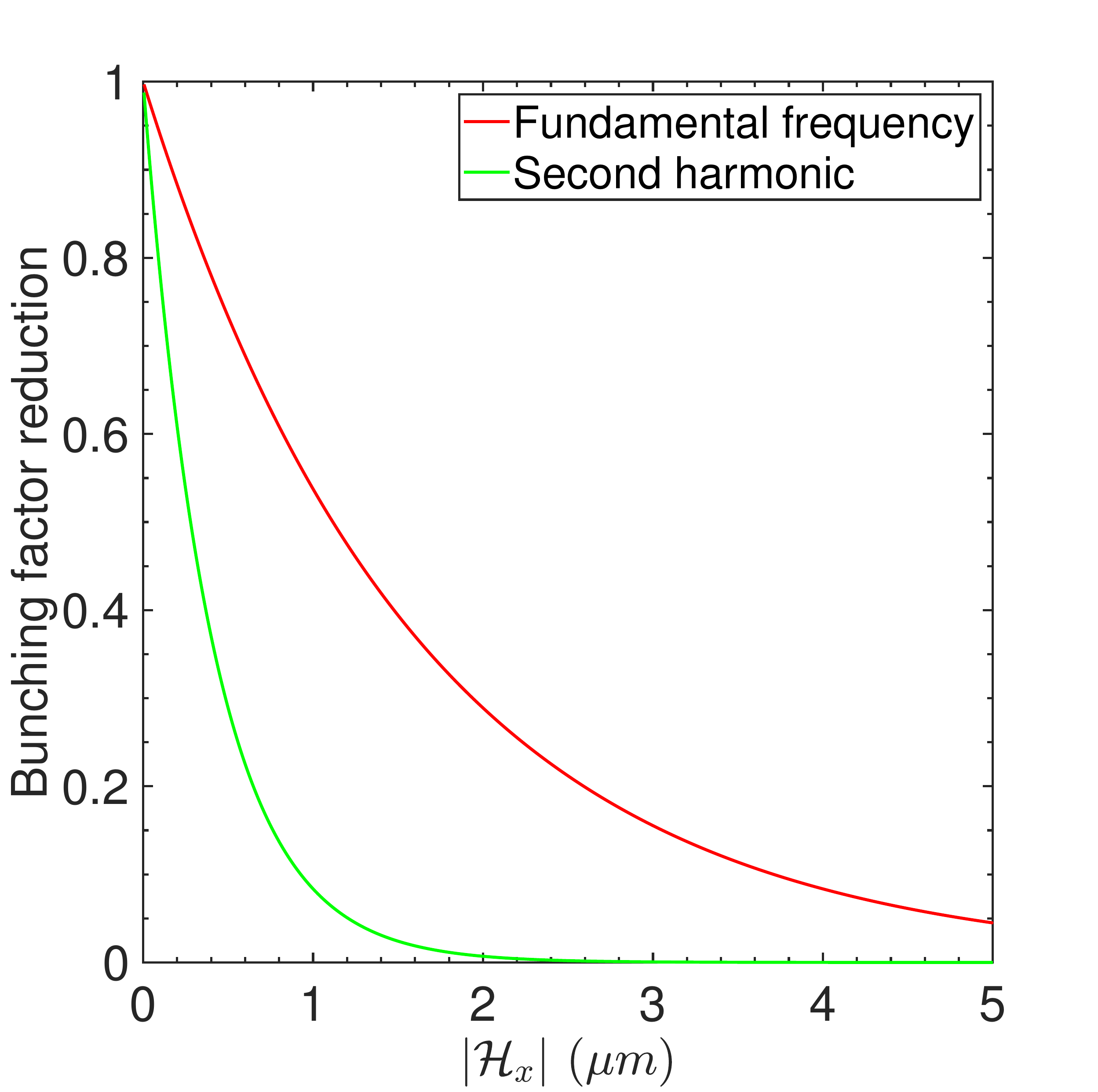}

	\caption{
		\label{fig:Chap5-BFHx} 
		Influence of $\mathcal{H}_{x}$ at the undulator on the bunching factor one turn after laser modulation, with $\epsilon_{x}$ = 31 nm (model) and $\nu_{x}=3.18$. 
	}
\end{figure} 

Putting in $\epsilon_{x}$ = 31 nm (model) and $\nu_{x}=3.18$
(model and measured), we need $\mathcal{H}_{x}\leq0.8\ \mu$m at the undulator to have
$\Delta z_{B,1}\leq\lambda_{\text{L}}/2\pi$. Figure~\ref{fig:Chap5-BFHx} shows the bunching factor reduction at the fundamental frequency and the second harmonic one turn after the laser modulation as a function of the $\mathcal{H}_{x}$ at the undulator. As can be seen, the second-harmonic bunching is even more sensitive to the $\mathcal{H}_{x}$ at the undulator.  This stringent condition on $\mathcal{H}_{x}$ (note that $\mathcal{H}_{x}$ at other places of the ring is
typically $\geq0.1$ m) is satisfied by fine-tuning the quadrupole magnet
(de)focusing strengths to correct the dispersion $D_{x}$ and dispersion
angle $D_{x}'$ at the undulator to the level of millimetre and 0.1 mrad, respectively
(see Fig.~\ref{fig:Chap5-SSMBPoPIExtFig1}). 

In addition to the linear-order oscillating bunch lengthening, as discussed in Sec.~\ref{sec:NonlinearCoupling}, the
betatron oscillation also produces an average path lengthening or
shortening (second-order effect) described by the formula
\begin{equation}
\Delta C_{B}=-2\pi J_{x}\xi_{x}
\end{equation}
with $\Delta C_{B}$ being the average change of the particle recirculation path
length, and $\xi_{x}=d\nu_{x}/d\delta$ being the horizontal chromaticity of the ring.
Because different particles have different betatron oscillation amplitudes
(actions), this effect results in a loss of synchronization between
particles and degrades microbunching. Moreover, it broadens the equilibrium
energy spread and distorts the beam from the Gaussian form~\cite{deng2020widening}
before the laser modulation as investigated in Sec.~\ref{sec:NonlinearCoupling}, which also affects the microbunching. 
Therefore, the horizontal chromaticity should be small, to moderate its detrimental outcome, and simultaneously sufficient to suppress
collective effects such as the head-tail instability~\cite{chao1993physics}. As a consequence, a small negative chromaticity
is used in the experiment.

\subsubsection{Modulation Laser}

\begin{figure}[tb]
	\centering 
	\includegraphics[width=1\textwidth]{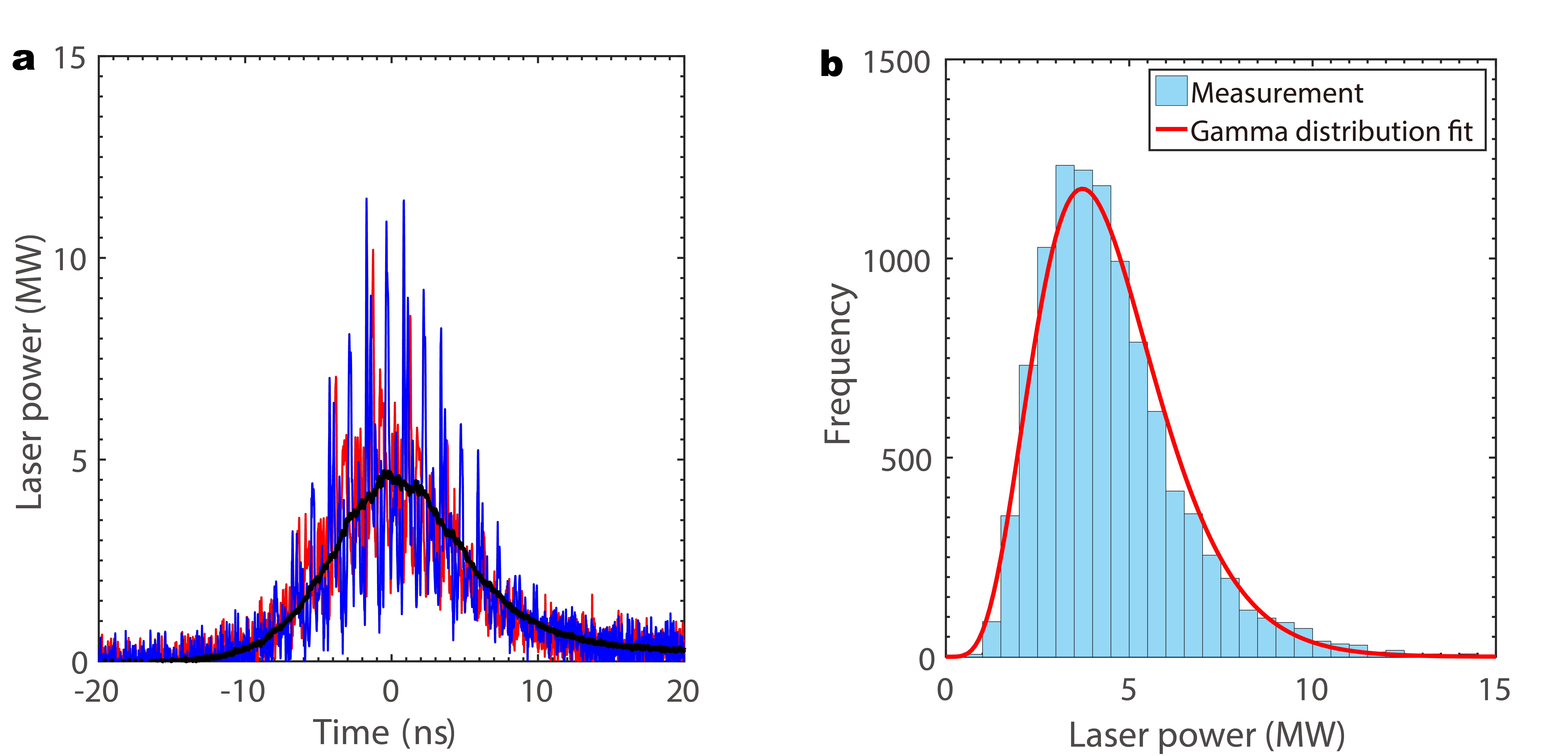}
	\caption{\label{fig:Chap5-SSMBPoPIExtFig2}Fluctuating temporal profiles of the multilongitudinal-mode laser. {\bf a}, Temporal profiles of two example consecutive
		laser shots (red and blue) and the averaged waveform of 200 consecutive laser
		shots (black). {\bf b}, Statistical distribution of the laser power at $t$ = 0 ns in a for
		$10^{4}$ consecutive laser shots, where the red curve is a gamma distribution fit.
		Laser: compact Nd:YAG Q-switched laser (Beamtech Optronics Dawa-200).
		Detector: ultrafast photodetectors (Alphas UPS-40-UVIR-D; rise time $<$ 40 ps).
		Measurement system: digital oscilloscope (Teledyne LeCroy WM825Zi-B;
		bandwidth 25 GHz; sample rate 80 billion samples per second).  (Figure from Ref.~\cite{deng2021experimental})}
\end{figure}

{\bf Long-pulse laser} To simplify the experiment by avoiding a dedicated laser-electron synchronization system, a long-pulse laser (FWHM $\approx$ 10 ns) is used, as the shot-to-shot laser timing jitter is $t_{\text{jitter}}\leq$ 1 ns (RMS). According to Eq.~(\ref{eq:BFPoPI}) and Fig.~\ref{fig:Chap5-BFEta}, for a given phase slippage factor and harmonic number,
there is an optimal laser-induced energy modulation amplitude $A$ ($A\propto\sqrt{P_{\text{L}}}$ with $P_{\text{L}}$ the laser power) that gives the maximum bunching
factor. 
The laser used in the experiment (Beamtech Optronics Dawa-200) has multiple longitudinal modes, and its temporal profile has several peaks and fluctuates considerably from shot to shot (see Fig.~\ref{fig:Chap5-SSMBPoPIExtFig2}). Therefore, the laser-induced electron energy modulation amplitudes are different from shot to shot and from bunch to bunch. When the modulation amplitude matches the phase slippage factor, the energy-modulated electrons are properly focused at synchronous phases, which gives optimal microbunching. For some of the shots, the laser intensity is higher or lower than the optimal value, and the electrons are then over-focused or under-focused, giving weaker microbunching and less coherent radiation. As we will see soon, this explains the shot-to-shot fluctuation of the coherent amplified signals shown in Fig.~\ref{Chap5-fig:SSMBPoPIFig2}c, e.

{\bf Power and Rayleigh length}
The electric field of a TEM00 mode Gaussian laser beam is \cite{chao2022FocusedLaser}
\begin{equation}\label{eq:GaussianLaser}
\begin{aligned}
E_{x}&=E_{x0}e^{ik_{\text{L}}z-i\omega t+i\phi_{0}}\frac{1}{1+i\frac{z}{Z_{R}}}\text{exp}\left[i\frac{k_{\text{L}}Q}{2}(x^2+y^2)\right],\\
E_{z}&\approx-E_{x}x,
\end{aligned}
\end{equation}
with $Z_{R}=\frac{\pi w_{0}^{2}}{\lambda_{\text{L}}}$ the Rayleigh length, $w_{0}$ the beam waist radius, and $Q=\frac{i}{Z_{R}(1+\frac{z}{Z_{R}})}$.
The relation between $E_{x0}$ and the laser peak power is given by
\begin{equation}
P_{\text{L}}=\frac{E_{x0}^{2}Z_{R}\lambda_{\text{L}}}{4Z_{0}},
\end{equation}
in which $Z_{0}=376.73\ \Omega$ is the impedance of free space. 
The electron wiggles in an undulator according to 
\begin{equation}
x(z)=\frac{K}{\gamma k_{u}}\sin(k_{u}z),
\end{equation}
and the laser-electron  exchange energy according to
\begin{equation}
\frac{dW}{dt}=\vec{v}\cdot\vec{E}=v_{x}E_{x}+v_{z}E_{z}.
\end{equation}
Assuming that the laser beam waist is in the middle of the undulator, and when $x,y\ll w(z)$, which is the case for SSMB PoP I, we drop the $\text{exp}\left[i\frac{k_{\text{L}}Q}{2}(x^2+y^2)\right]$ in the laser electric field and also drop the contribution from $E_{z}$, then the integrated modulation voltage induced by the laser in the undulator normalized by the electron beam energy  is \cite{chao2022FocusedLaser}
\begin{equation}
\begin{aligned}
\frac{eV_{\text{mod}}}{E_{0}}&=\frac{e[JJ] K}{\gamma^{2}mc^{2}}\sqrt{\frac{4P_{\text{L}}Z_{0}Z_{R}}{\lambda_{\text{L}}}}\tan^{-1}\left(\frac{L_{u}}{2Z_{R}}\right),
\end{aligned}
\end{equation}
in which $[JJ]=J_{0}(\chi)-J_{1}(\chi)$ and $\chi=\frac{K^{2}}{4+2K^{2}}$.

\begin{figure}[tb] 
	\centering 
	\includegraphics[width=0.49\columnwidth]{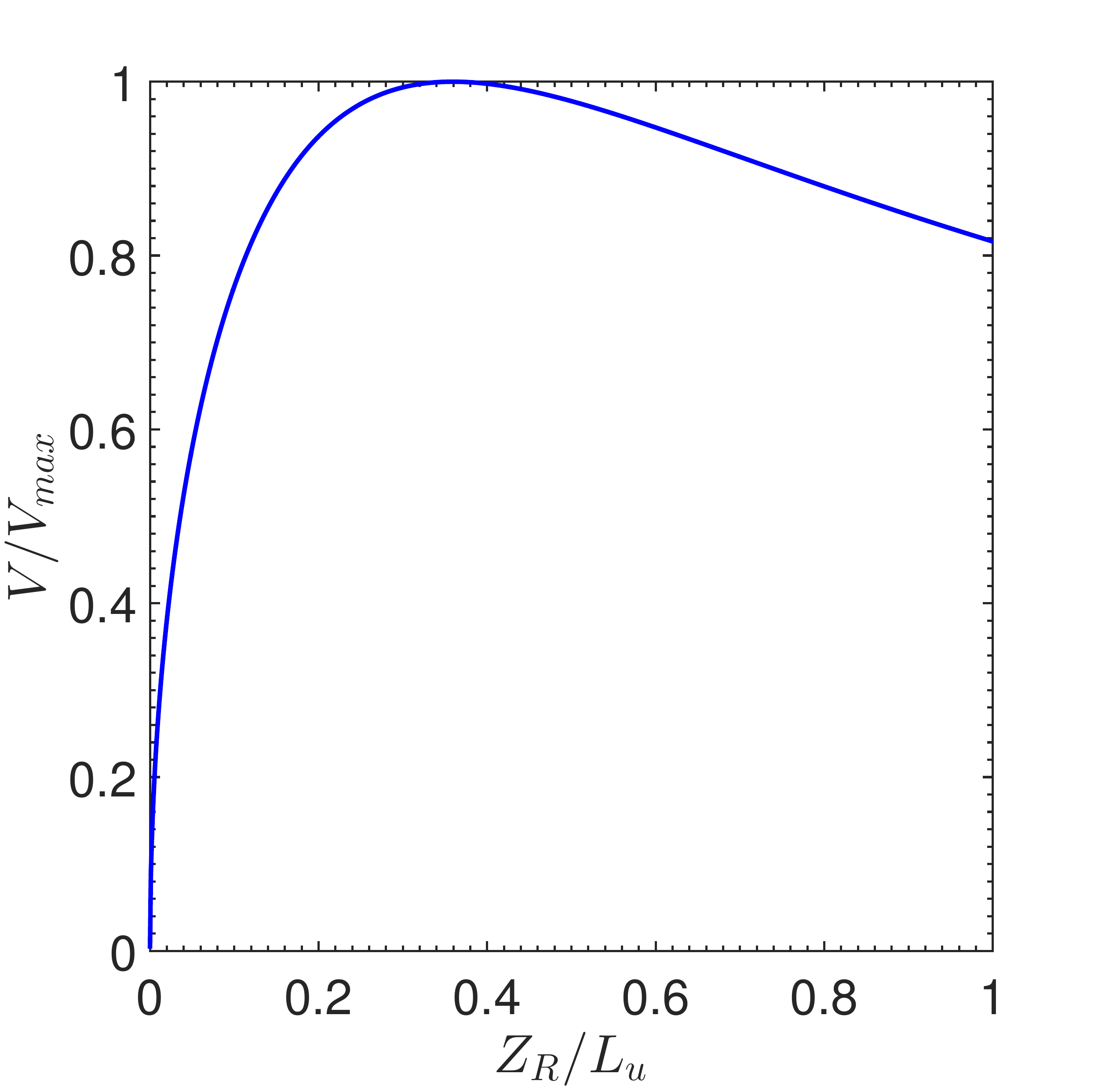}
	\caption{
		\label{fig:Chap5-VvsRy} 
		Integrated modulation voltage versus the Rayleigh length of the laser, with the undulator length $L_{u}$ kept fixed.
	}
\end{figure} 

As can be seen from the above formula, when $\frac{L_{u}}{Z_{R}}$ is kept constant, then $A=V_{\text{mod}}\propto\sqrt{Z_{R}}\propto \sqrt{L_{u}}$. In our case, $L_{u}$ is fixed, then as shown in Fig.~\ref{fig:Chap5-VvsRy}, to maximize $V_{\text{mod}}$ we need $\frac{Z_{R}}{L_{u}}\approx0.359$.  On the other hand, the modulation strength does not depend on $Z_{R}$ sensitively when $Z_{R}$ is larger than this optimal value. To make the laser beam waist larger than the electron beam and thus induce the same energy modulation on different electrons, a larger Rayleigh length might be in favored in our case. For example if $Z_{R}=2L_{u}$, then in order to induce an energy modulation depth of $A=1.5\sigma_{\delta}$, we can calculate that the laser power required is 430 kW. Considering the non-ideal conditions and the fact that the laser used contains higher Gaussian modes, one order of magnitude higher laser power might be required in the actual case.

\begin{figure}[tb]
	\centering
	\includegraphics[width=1\columnwidth]{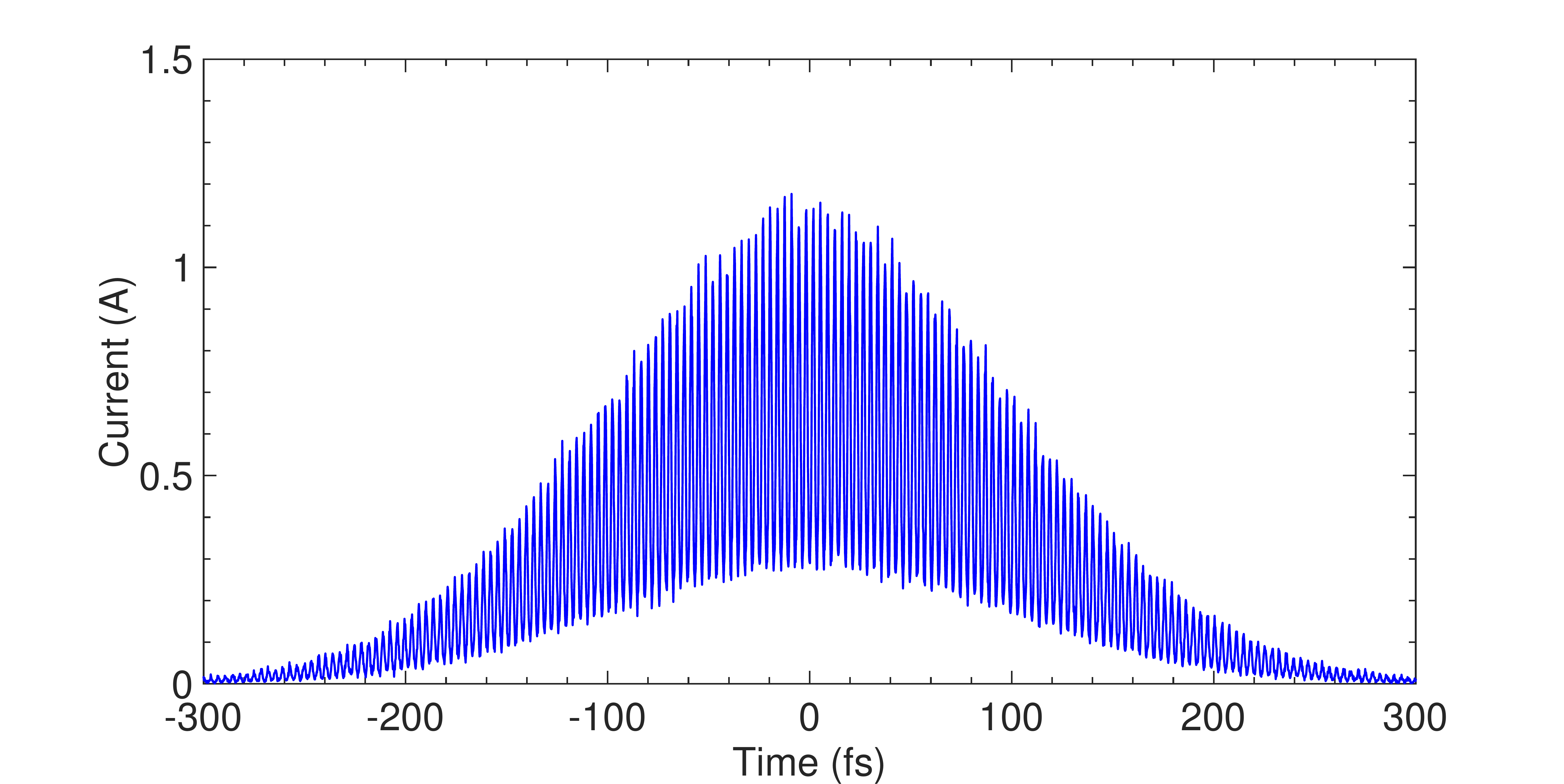}
	\includegraphics[width=1\columnwidth]{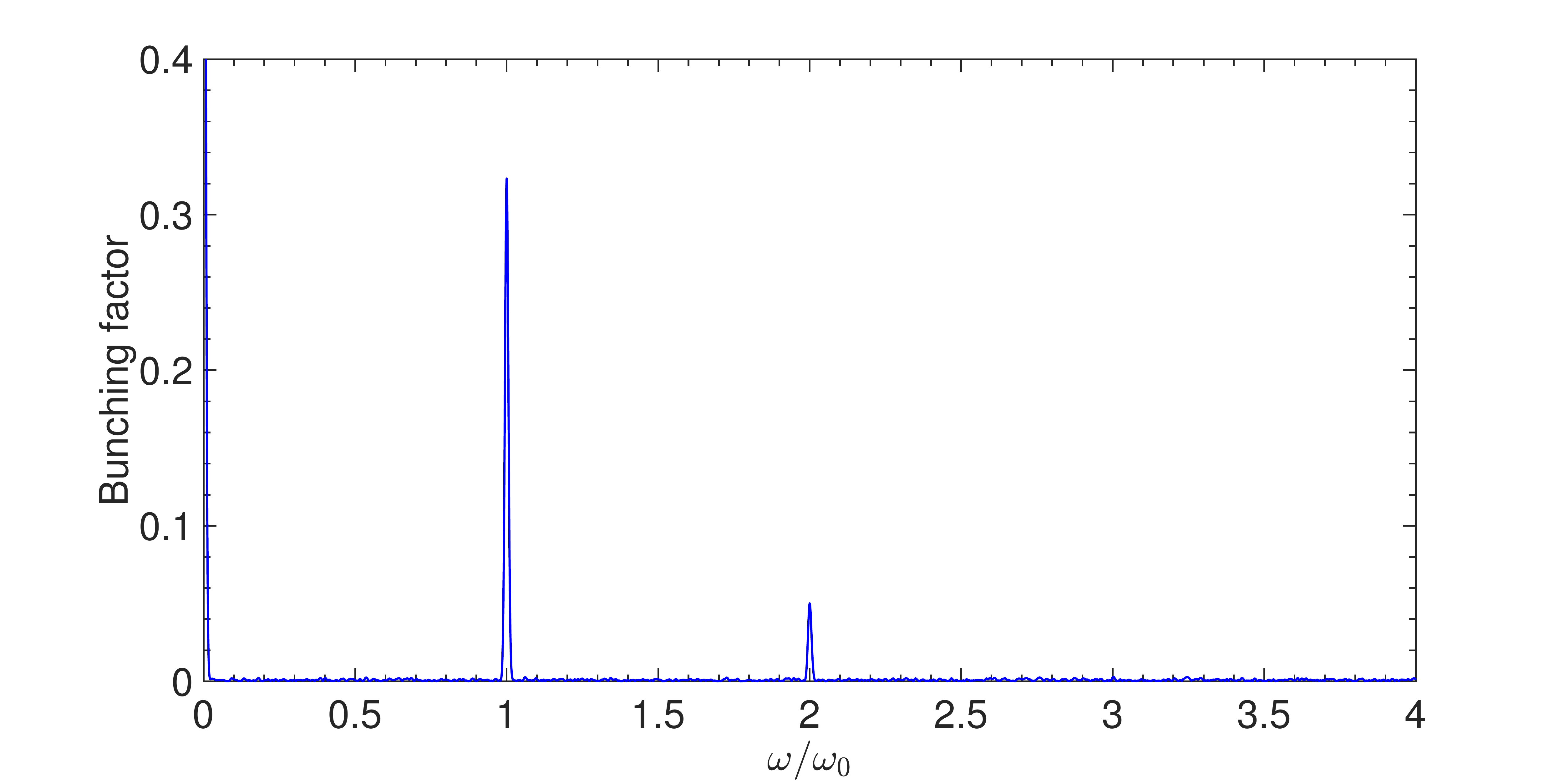}
	\caption{
		\label{fig:Chap5-Current} 
		Example current profile and bunching factor spectrum one turn after the laser modulation in SSMB PoP I, obtained from ELEGANT~\cite{borland2000elegant} tracking.   Parameters used: $\sigma_{t}=100$ fs, $\epsilon_{x}=31$ nm,  $\epsilon_{y}=\frac{1}{10}\epsilon_{x}$.  $1\times10^{6}$ particles are simulated, meaning 160 fC for a one-to-one correspondence.
	}
\end{figure}


\subsubsection{Microbunching Simulation}

With the above parameters, the beam current and bunching factor one turn after the laser modulation is shown in Fig.~\ref{fig:Chap5-Current}.  
The simulation is conducted using ELEGANT \cite{borland2000elegant}. As can be seen from the comparison with Fig.~\ref{fig:Chap5-BFTurnNo}, the simulation agrees with theory well.

\subsection{Microbunching Radiation Calculation}


Now we conduct some numerical calculations to confirm our above analysis of the radiation power and flux.  The numerical calculation of incoherent and coherent undulation radiation shown in this section are obtained using {\it SPECTRA}~\cite{tanaka2001spectra}. The beam energy and undulator parameters used are those in our SSMB proof-of-principle experiment, i.e., $E_{0}=250$ MeV, $\lambda_{\text{L}}=\lambda_{0}=1064$~nm, $\lambda_{u}=125$ mm, $K=2.5$, $N_{u}=32$. The results are also used to compare with the theoretical formulas presented in in Chap.~\ref{cha:Radiation}. Note that the numerical calculation of coherent radiation with a 3D charge distribution is usually time-consuming. This is also one of the motivations for us to develop the simplified analytical formulas with the main physics accounted for, in Chap.~\ref{cha:Radiation}.

\begin{figure}[tb]
	\centering
	\includegraphics[width=0.49\columnwidth]{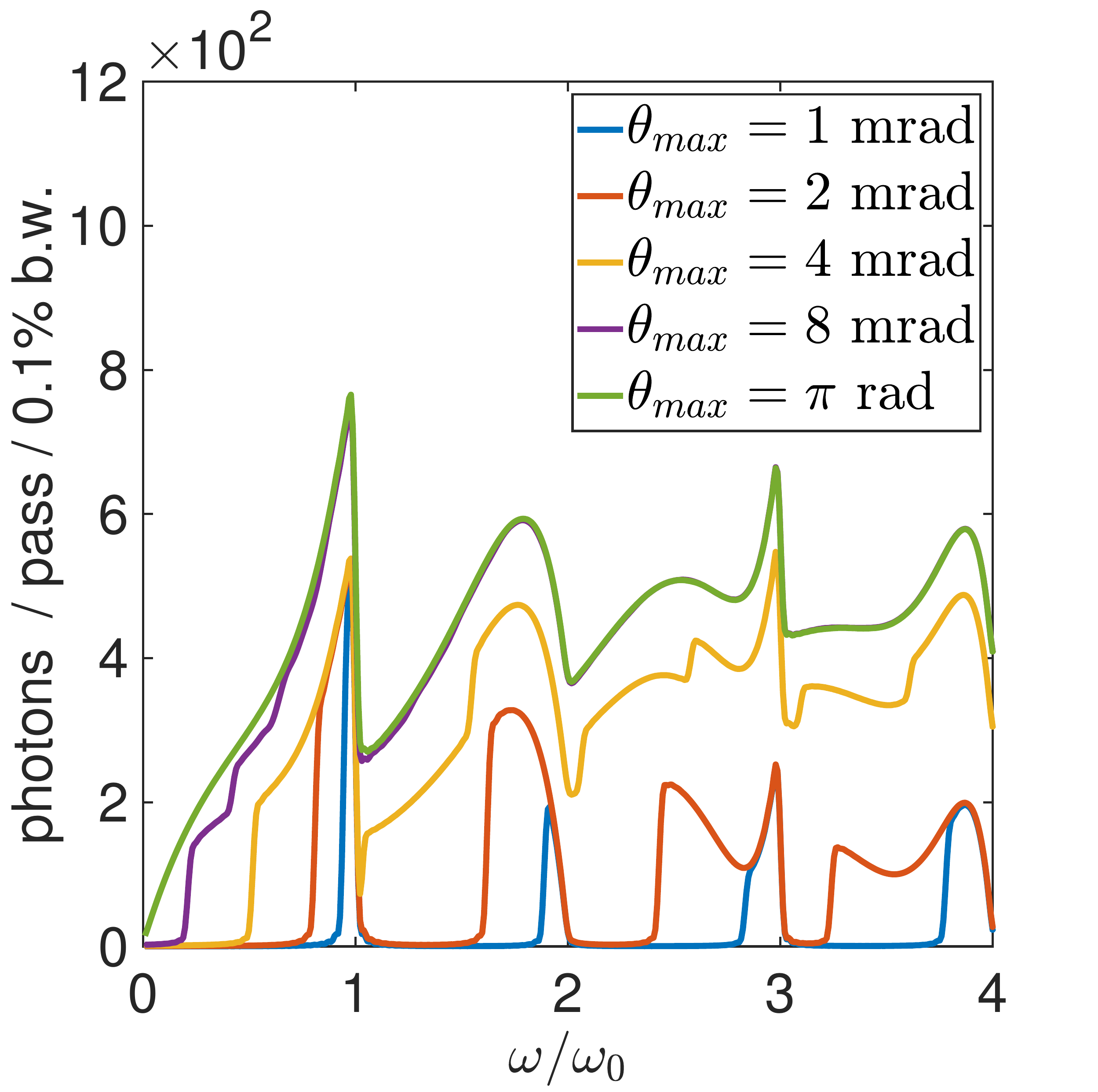}
	\includegraphics[width=0.49\columnwidth]{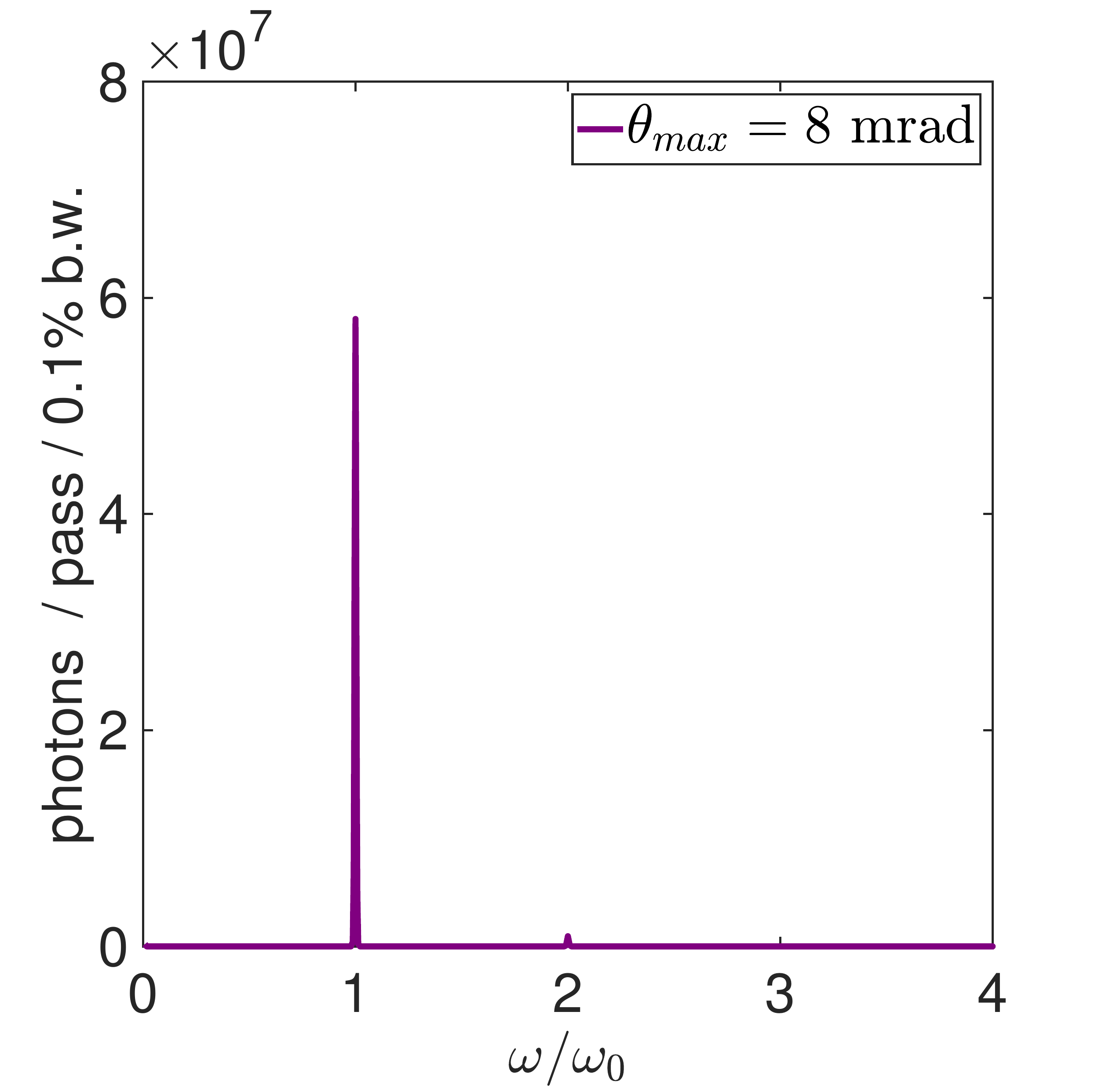}
	\caption{
		\label{fig:FluxISRVSR} 
		Left: incoherent radiation photon flux of $10^{6}$ electrons within a circular aperture placed in the forward direction, whose largest opening angles are $\theta_{\text{max}}$. 
		Right: coherent radiation photon flux of $10^{6}$ electrons within the circular aperture whose largest opening angle $\theta_{\text{max}}=8$ mrad, with the beam current and bunching factor spectrum shown in Fig.~\ref{fig:Chap5-Current}. Other related parameters: $E_{0}=250$~MeV, $\lambda_{0}=1064$ nm, $\lambda_{u}=125$ mm, $K=2.5$, $N_{u}=32$. 
	}
\end{figure}

The left part of Fig.~\ref{fig:FluxISRVSR} shows the incoherent undulator radiation flux  of $10^{6}$ electrons (0.16 pC)  versus the opening angle of a circular aperture placed in the forward direction of electron traveling.  As can be seen, with the increase of the aperture opening angle, the red-shifted part of the radiation grows. For the total flux, there are sharp spikes near the odd harmonics and no clear spikes near the even harmonics. This is due to the fact that there is no on-axis radiation at the even harmonics. Also note that with the increase of the aperture opening angle, there are jumps in the flux at a specific frequency $n\omega_{0}$. This is due to the fact that the red-shifted radiation of higher harmonics $m>n$ can contribute to the flux at $\omega=n\omega_{0}$ when the aperture is large enough.

Now we calculate the coherent radiation of the laser modulation-induced microbunched beam, using an RMS bunch length of 100 fs ( $\sigma_{z}=30\ \mu$m).  An example beam current and bunching factor spectrum of the laser modulation-induced microbunched beam are shown in Fig.~\ref{fig:Chap5-Current}. We remind the readers that the bunch length in the real machine is typically longer than that used in the calculation here. For example, the theoretical zero-current bunch length in the experiment is 120 fs. Therefore, the coherent radiation is even more narrow-banded than presented in the example calculation in this section.

\begin{figure}[tb]
	\centering
	\includegraphics[width=0.4\columnwidth]{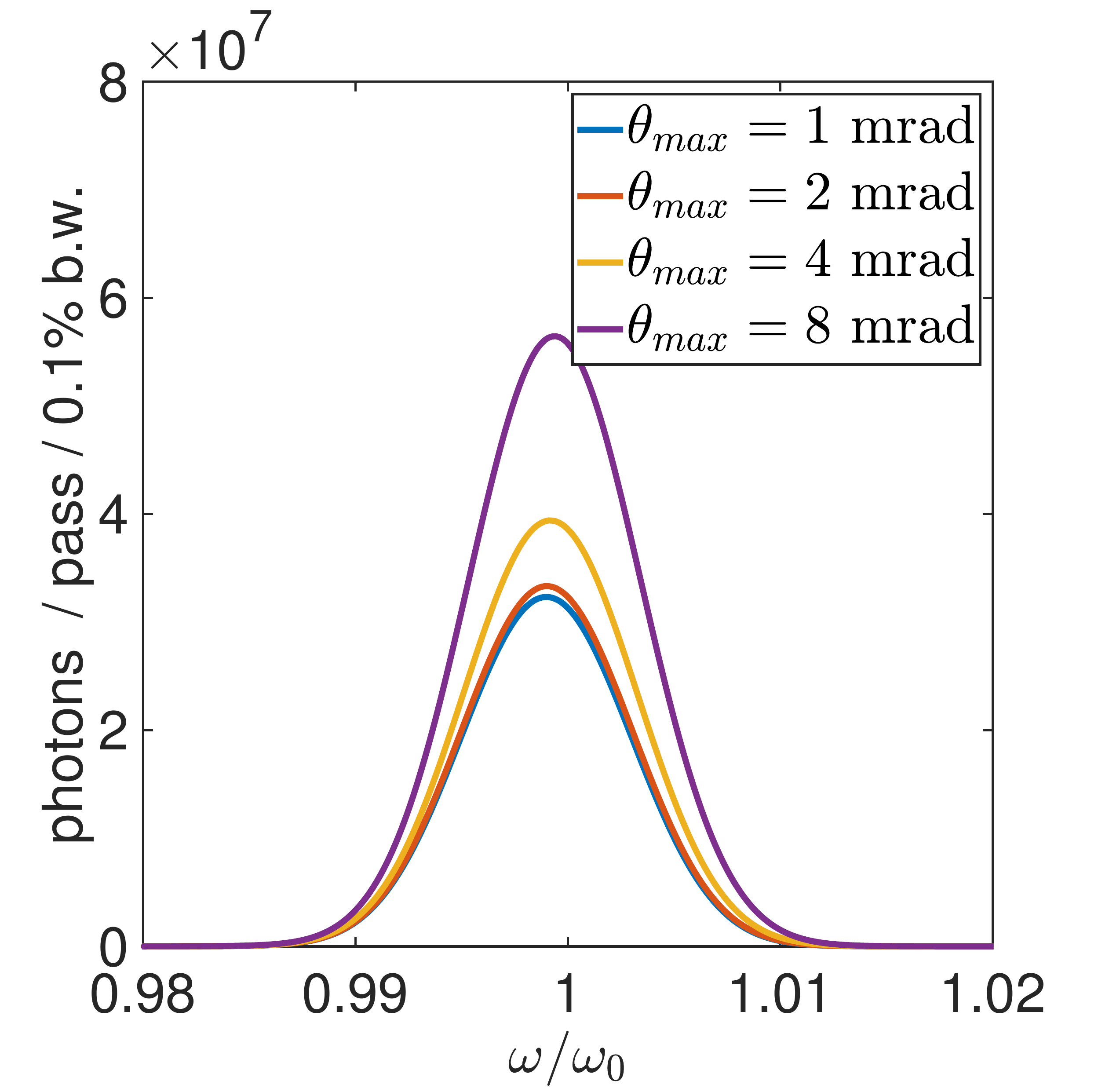}
	\includegraphics[width=0.4\columnwidth]{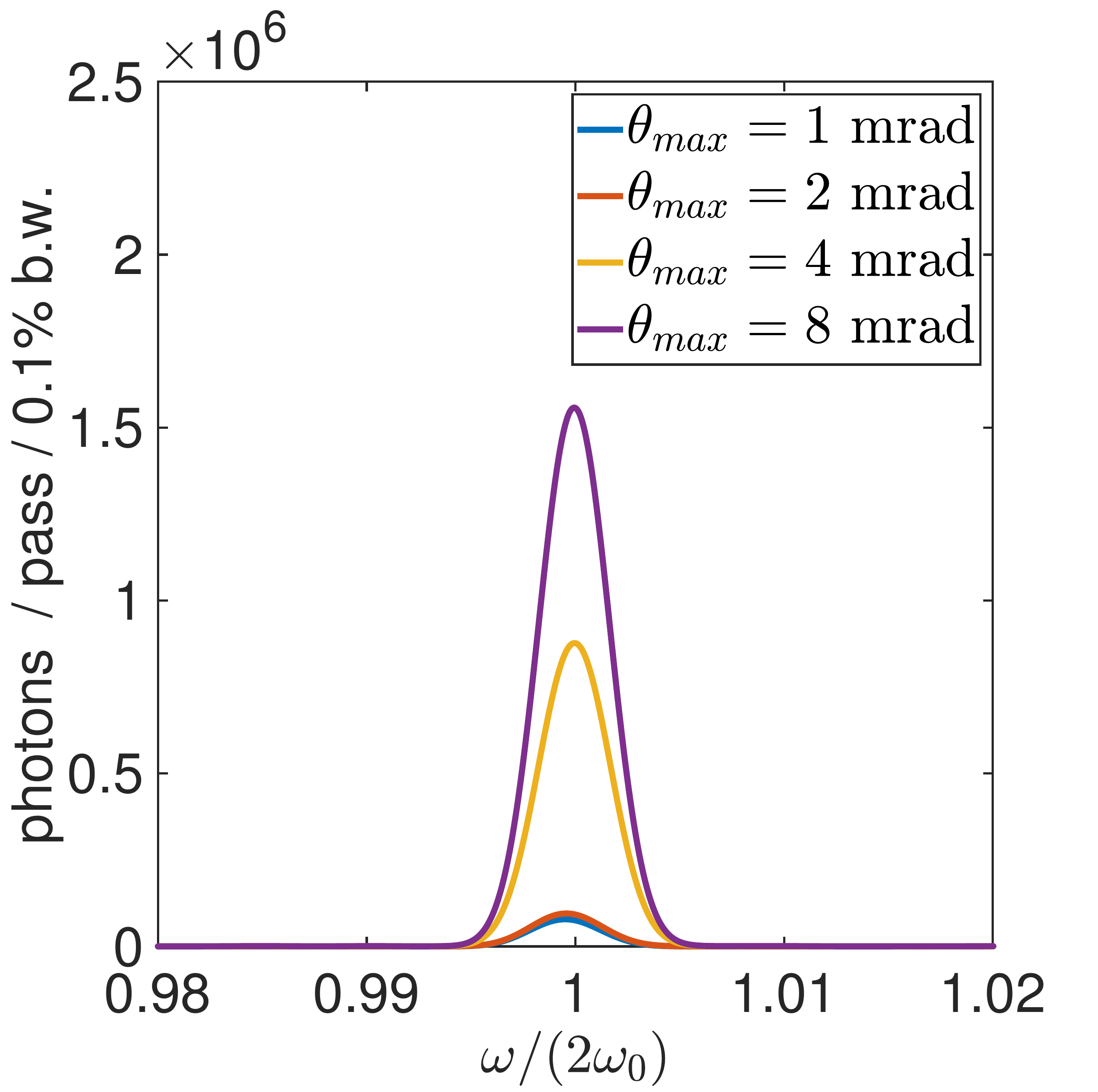}\\
	\includegraphics[width=0.4\columnwidth]{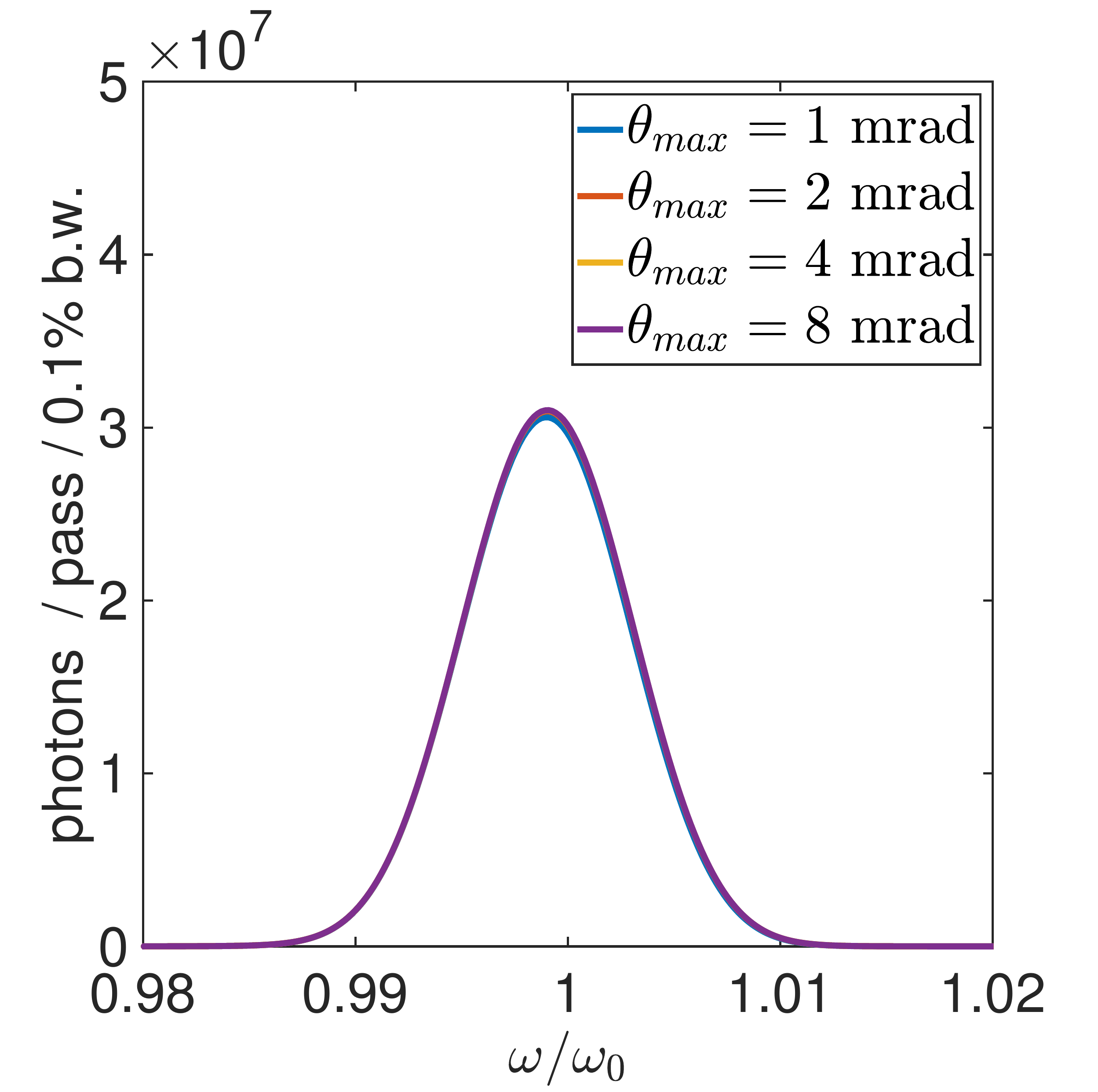}
	\includegraphics[width=0.4\columnwidth]{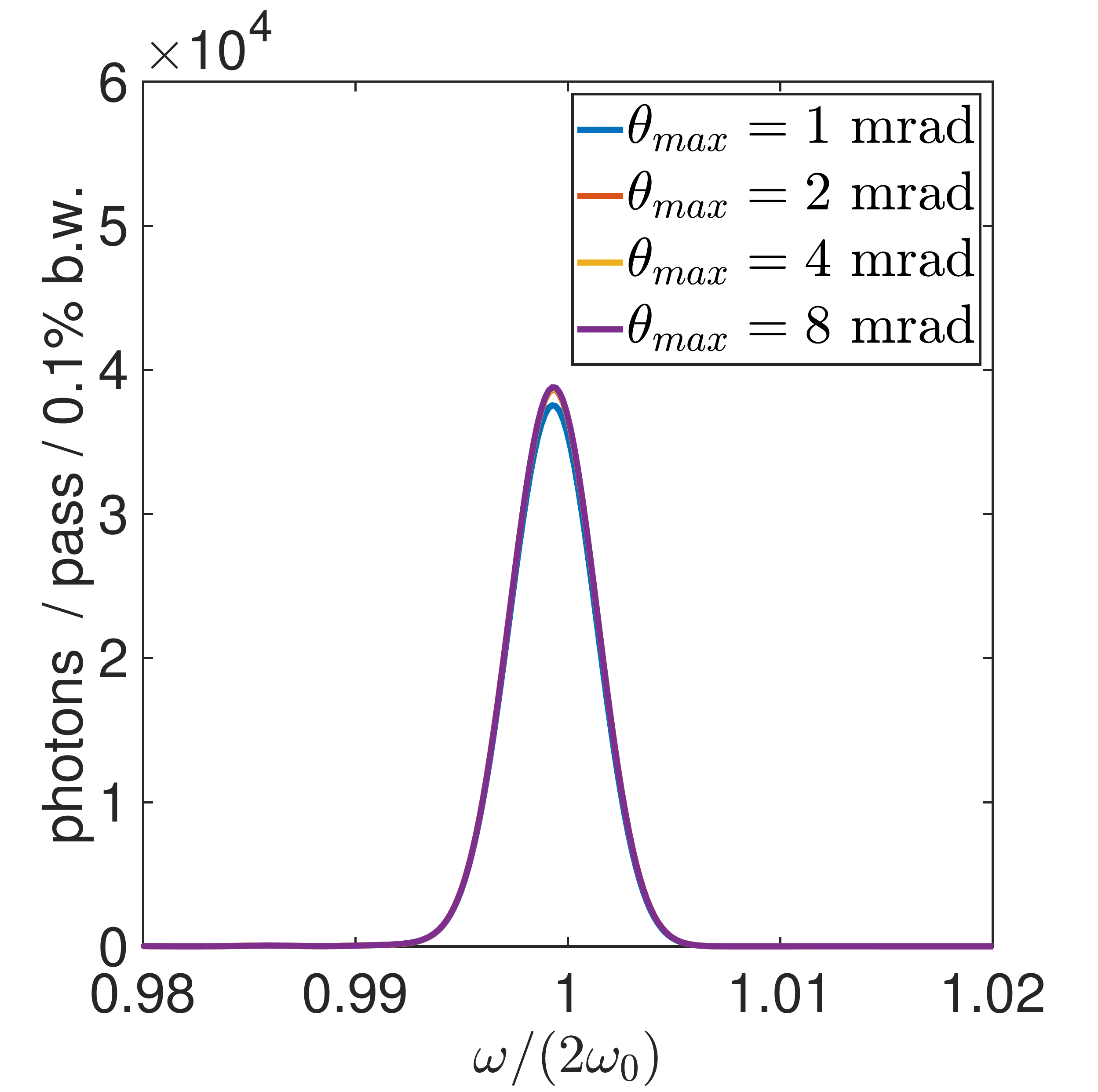}\\
	\includegraphics[width=0.4\columnwidth]{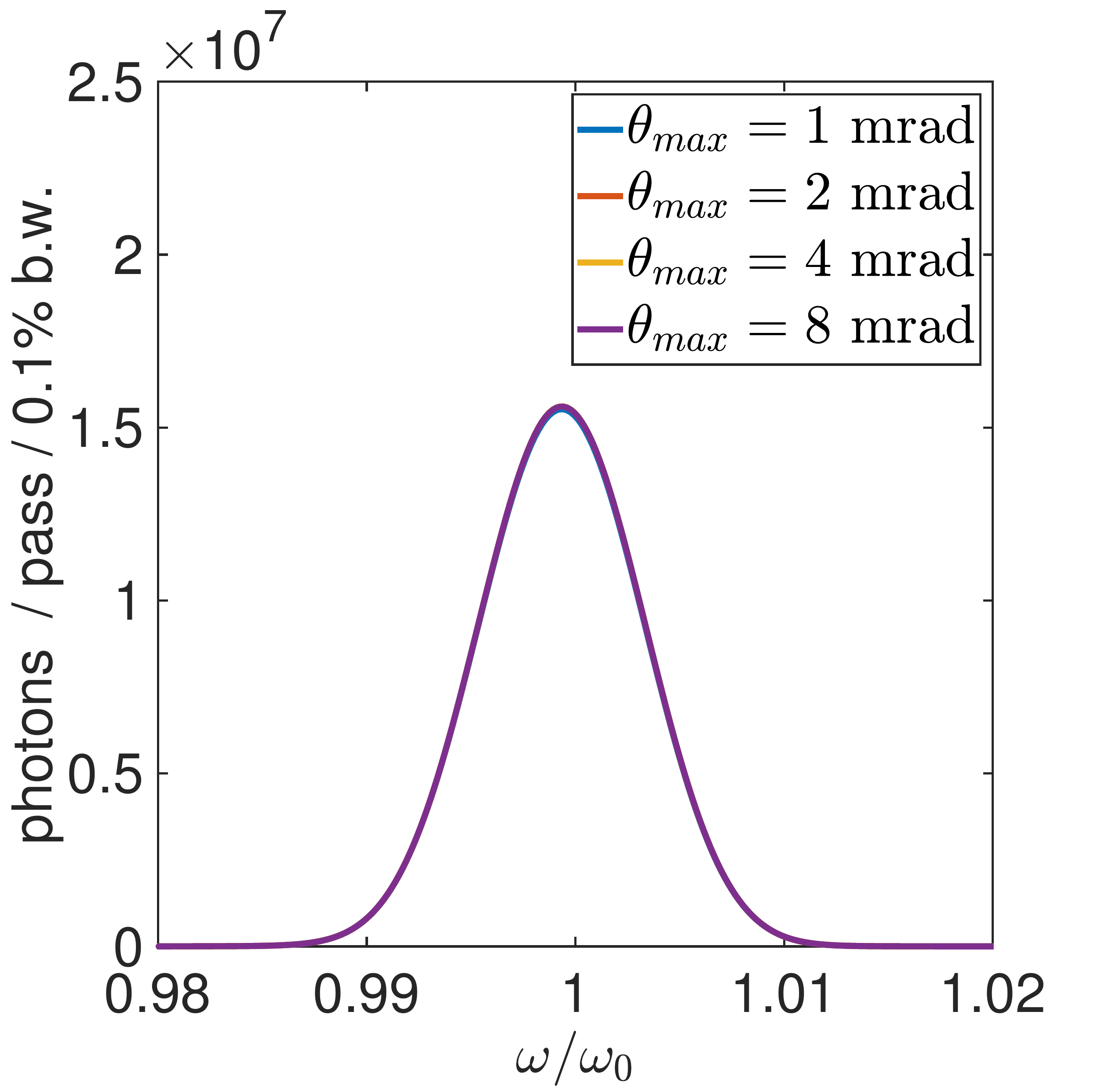}
	\includegraphics[width=0.4\columnwidth]{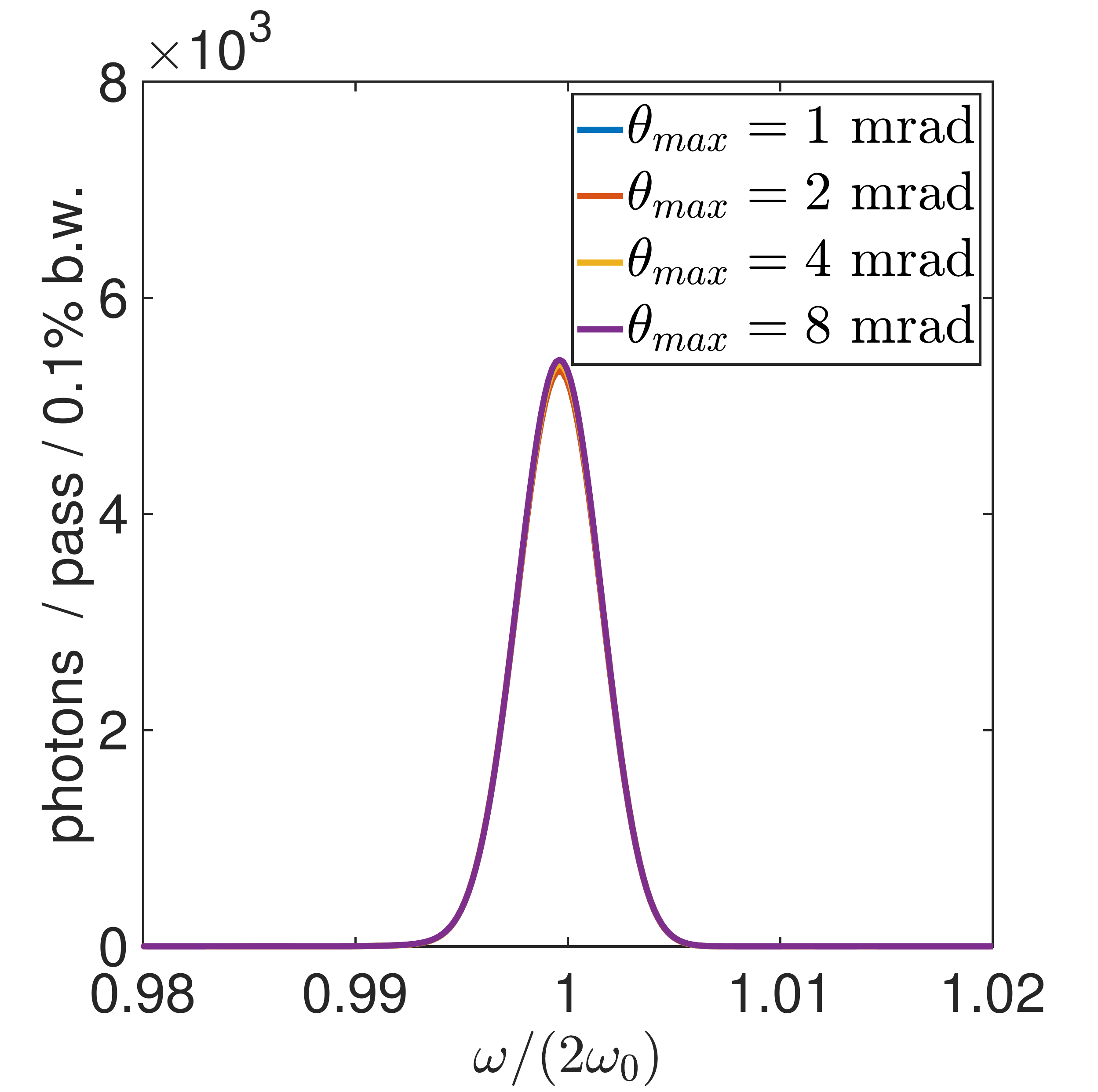}
	\caption{
		\label{fig:Chap4-FluxCSRZeroEmottance} 
		Coherent radiation photon flux of $10^{6}$ electrons versus opening angle $\theta_{\text{max}}$ of the circular aperture for the first two harmonics, with $\sigma_{\bot}=0\ \mu$m (up),  $\sigma_{\bot}=100\ \mu$m (middle) and $\sigma_{\bot}=400\ \mu$m (bottom), respectively. The beam current and bunching factor spectrum are shown in Fig.~\ref{fig:Chap5-Current}. Other related parameters: $E_{0}=250$~MeV, $\lambda_{0}=1064$ nm, $\lambda_{u}=125$ mm, $K=2.5$, $N_{u}=32$.  
	}
\end{figure}

\begin{figure}[tb]
	\centering
	\includegraphics[width=0.7\columnwidth]{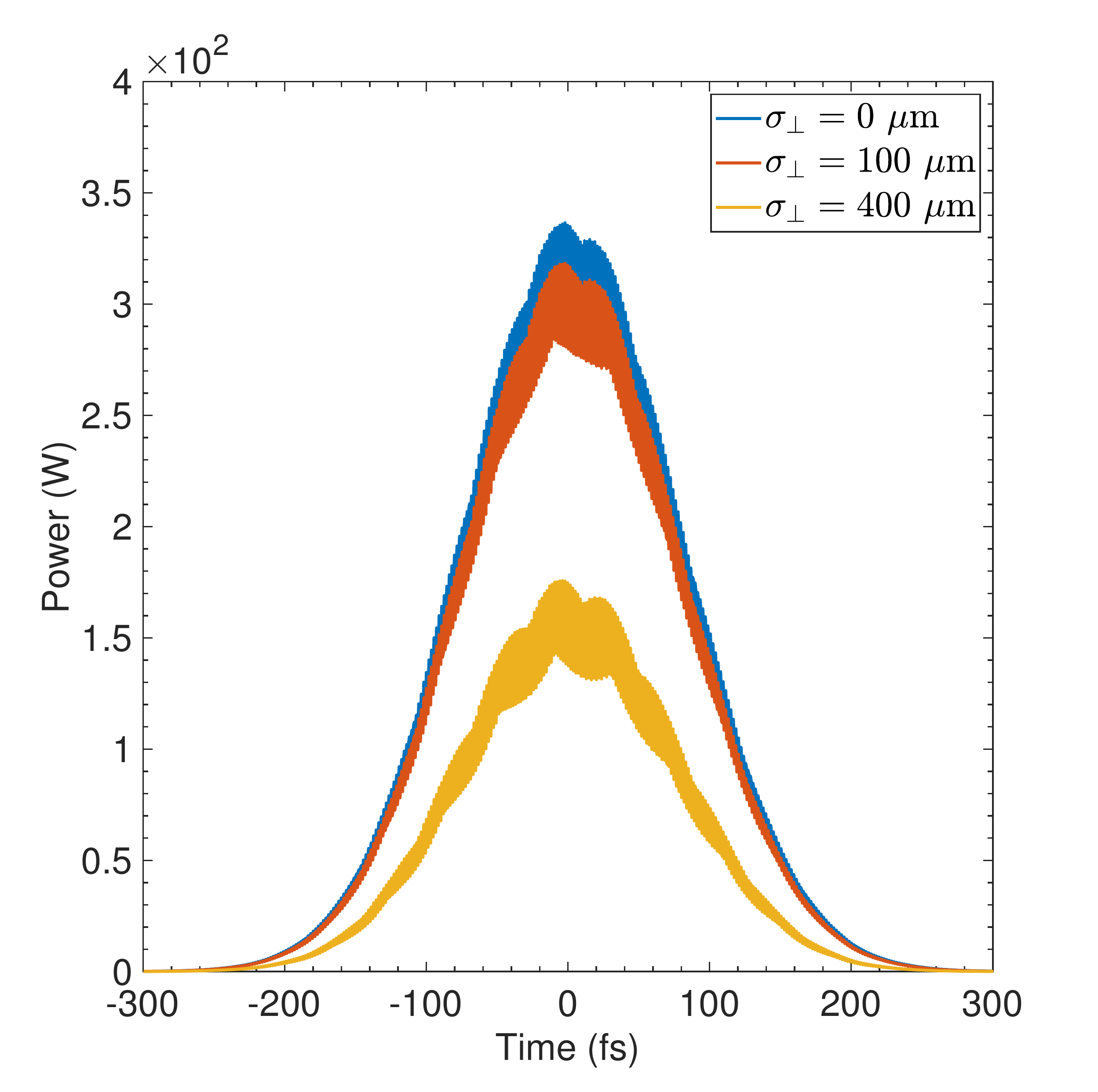}
	\caption{
		\label{fig:PowerVSTime} 
		Total radiation power as a function of the observation time gathered within a circular aperture with $\theta_{\text{max}}=1$~mrad, for $\sigma_{\bot}=0,100,400\ \mu$m, respectively. The beam current and bunching factor spectrum are shown in Fig.~\ref{fig:Chap5-Current}. Other related parameters: $E_{0}=250$~MeV, $\lambda_{0}=1064$~nm, $\lambda_{u}=125$~mm, $K=2.5$, $N_{u}=32$.  
	}
\end{figure}

At first, we ignore the influence of transverse dimension, i.e., a thread beam is assumed. The coherent radiation spectrum of $10^{6}$ electrons is shown in the right part of Fig.~\ref{fig:FluxISRVSR}. As can be seen, there is only clear narrow-band coherent radiation at the modulation laser harmonics, which fits with expectation due to the fact that there is only notable bunching factor at the laser harmonics. More closer look of the first two harmonics versus the aperture opening angle are shown in the upper part of Fig.~\ref{fig:Chap4-FluxCSRZeroEmottance}. As can be seen, indeed the radiation bandwidth of the coherent radiation at the fundamental frequency and second harmonic are $1\%$ and $0.5\%$, respectively, agreeing with the values calculated from Eq.~(\ref{eq:bandwidth}). In addition, the flux of the fundamental mode $H=1$ at the fundamental frequency agrees reasonably well with that according to Eq.~(\ref{eq:flux}), i.e., $\mathcal{F}_{1}(\omega=\omega_{0},\sigma_{\bot}=0\ \mu{\text{m}})=3.4\times10^{7}\ \text{(photons/pass/0.1\% b.w.)}$. In other words, the amplification factor of the flux at $\omega=H\omega_{0}$ is indeed $N_{e}^{2}|b_{z,H}|^{2}$ when $\sigma_{\bot}=0\ \mu$m. Also note that the jumps of the flux with the change of aperture opening angle as we commented just now. 

Now we investigate the impact of transverse electron beam sizes on the coherent radiation. As can be seen from the middle and bottom parts of Fig.~\ref{fig:Chap4-FluxCSRZeroEmottance} and the comparison with the upper part, the transverse sizes of the electron beam suppress the coherent radiation. And the calculated fluxes at the fundamental frequency agrees well with those predicted by Eq.~(\ref{eq:fluxnumerical}), i.e., $\mathcal{F}_{1}(\omega=\omega_{0},\sigma_{\bot}=100\ \mu{\text{m}})=3.1\times10^{7}\ \text{(photons/pass/0.1\% b.w.)}$ and $\mathcal{F}_{1}(\omega=\omega_{0},\sigma_{\bot}=400\ \mu{\text{m}})=1.6\times10^{7}\ \text{(photons/pass/0.1\% b.w.)}$. The suppression is more significant at the higher harmonics and the suppression factors agree with those predicted according to the transverse form factor Eq.~(\ref{eq:FFS}). Also note that different from that of incoherent radiation, when $\sigma_{\bot}=100\ \mu$m or $400\ \mu$m, there is no visible jump of the flux with the aperture opening angle $\theta_{\text{max}}$ grown from 1 mrad to 8 mrad. This is because that the off-axis red-shifted coherent radiation of higher modes are suppressed now.

We have also confirmed our derivation of the coherent radiation power by comparing it with simulation. As shown in Fig.~\ref{fig:PowerVSTime}, the calculated peak powers of coherent radiation with different transverse electron beam sizes also agree well with the theoretical predictions from Eq.~(\ref{eq:RPHNumerical}), i.e., $P_{1,\text{peak}}(\sigma_{\bot}=0\ \mu\text{m})=363$~W, $P_{1,\text{peak}}(\sigma_{\bot}=100\ \mu\text{m})=332$~W, $P_{1,\text{peak}}(\sigma_{\bot}=400\ \mu\text{m})=168$~W.

From the calculation and analysis, we know that the coherent radiation from the formed microbunching is mainly in the fundamental frequency and second  harmonic of the modulation laser, and in the forward direction. The coherent radiation is is narrow-banded, and stronger than the incoherent radiation. These calculations and observation are the basis for our signal detection scheme. 

\begin{figure}[tb]
	\centering 
	\includegraphics[width=1\textwidth]{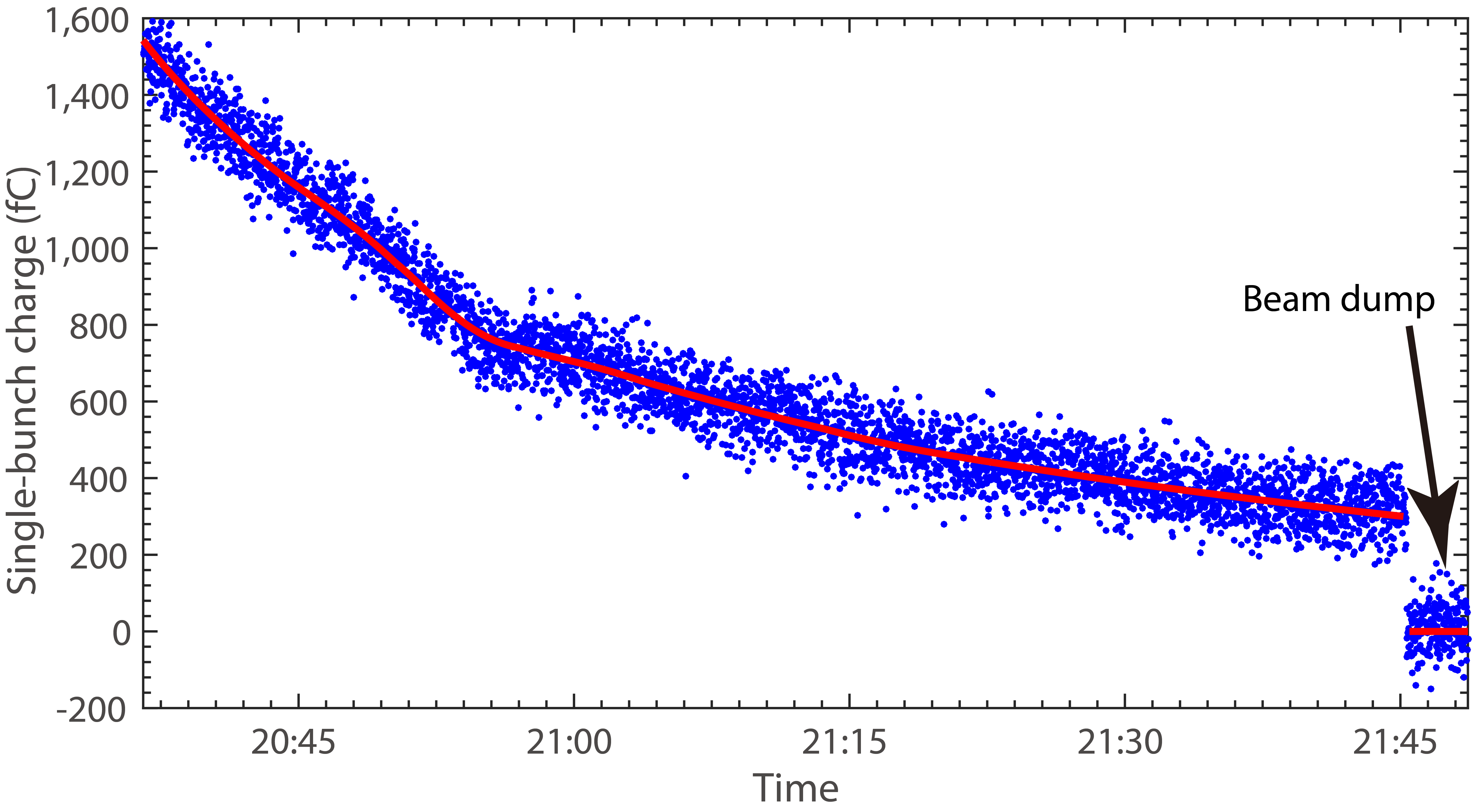}
	\caption{\label{fig:Chap5-SSMBPoPIExtFig3}Evaluation of bunch charge based on the stripe line measurement.
		Blue dots are the measurement results with the systematic offset subtracted
		and the red curve is a fit by the sum of two exponential functions,
		$Q(t)=Q_{1}\text{exp}(-t/\tau_{1})+Q_{2}\text{exp}(-t/\tau_{2})$, performed at different time intervals, with the
		fit results connected by a smoothed line.  (Figure from Ref.~\cite{deng2021experimental})}
\end{figure}

\subsection{Signal Detection and Evaluation}
After investigating the microbunching formation beam dynamics and radiation characteristics of the formed microbunching in the above sections, now we consider how we can measure and evaluate the signals.

\begin{figure}[tb]
	\centering 
	\includegraphics[width=1\textwidth]{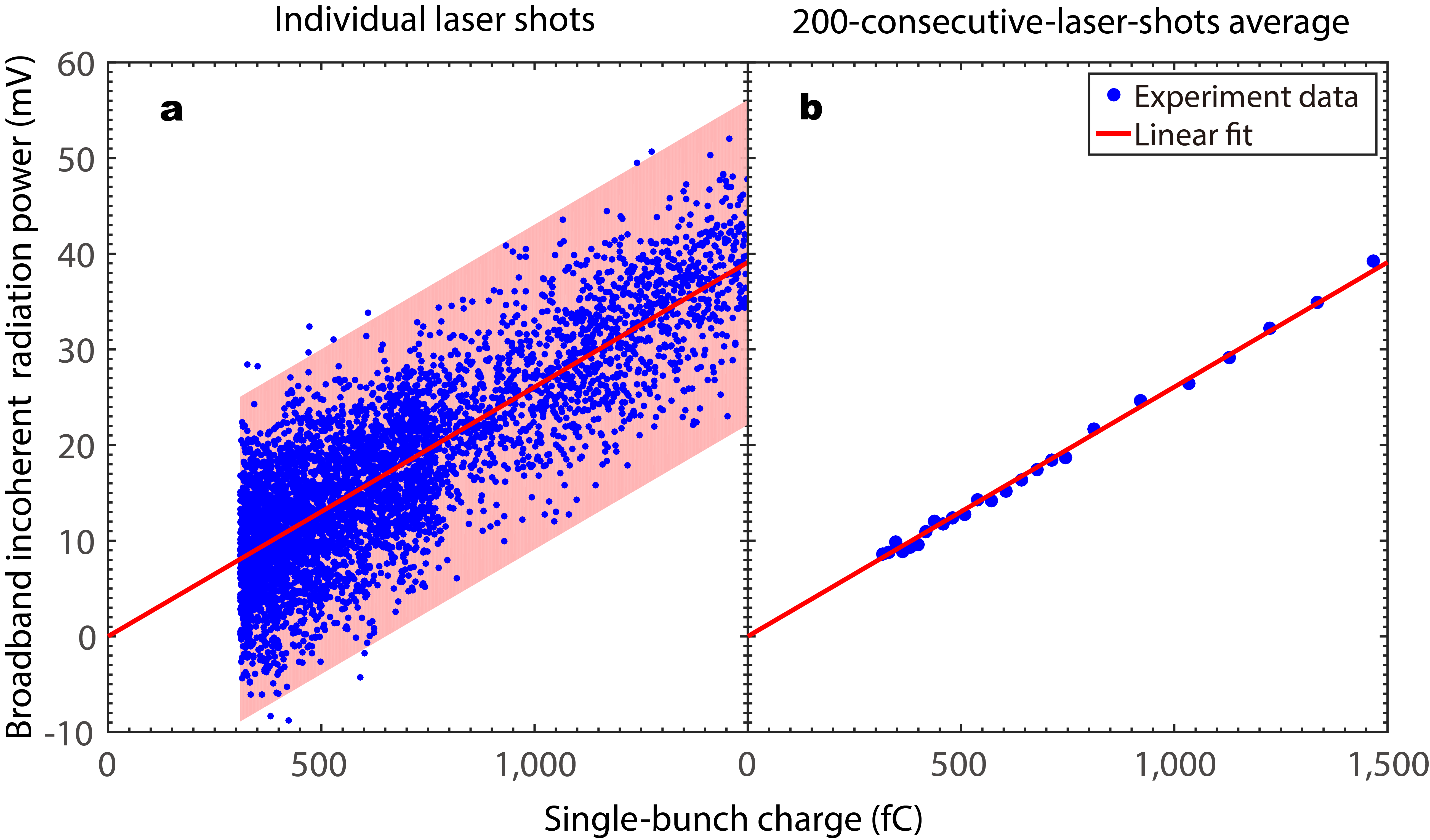}
	\caption{\label{fig:Chap5-SSMBPoPIExtFig4}Linear dependence of the broadband incoherent
		undulator radiation on the bunch charge. {\bf a}, Results corresponding to
		individual laser shots; the shading (light red) represents $3\sigma$ of the detection
		noise. {\bf b}, The result after 200-consecutive-laser-shot averaging. The blue dots
		are the experimental data of a bunch not modulated by the laser and the red
		curves are linear fits.  (Figure from Ref.~\cite{deng2021experimental})}
\end{figure}

{\bf Measurement and evaluation of bunch charge} 
The bunch-by-bunch charge (current) in the experiment is measured
by a single-bunch current monitor~\cite{falkenstern2009bunch}, which analyses the electron
beam-induced RF signals from a set of four stripline electrodes (3 GHz
bandwidth). To minimize the influence of neighbouring bunches on the
signal, the pulse response of the electrodes is reshaped by a 500-MHz
low-pass filter. The current calibration of the monitor is conducted
using a parametric current transformer~\cite{klein2008operation} at higher current, and the
linearity of the system at lower current is checked with the signal of the
photodiode illuminated by synchrotron radiation. During the current
decay in the experiment, one data point of the result given by the monitor
is saved every second for each individual bunch. The averaged measurement
of ten unfilled bunches preceding the homogeneous filled
bunches (10 ns time gap in between) is used as the systematic offset. To
smooth the measurement noise and at the same time account for the
change of the beam lifetime, the time evolution of the offset-removed
data points is then fitted by the sum of two exponential functions,
$Q(t)=Q_{1}\text{exp}(-t/\tau_{1})+Q_{2}\text{exp}(-t/\tau_{2})$, at different time intervals, with the fit results connected smoothly. One example evaluation of the bunch
charge measurement result is presented in Fig.~\ref{fig:Chap5-SSMBPoPIExtFig3}. Based
on the evaluated data, we obtain a linear bunch-charge dependence of
the broadband incoherent signal that is detected by the photodetector
without the 3-nm-bandwidth band-pass filter, as shown in Fig.~\ref{fig:Chap5-SSMBPoPIExtFig4}, confirming the reliability of the bunch-charge measurement
and evaluation method.

\begin{figure}[tb]
	\centering 
	\includegraphics[width=1\textwidth]{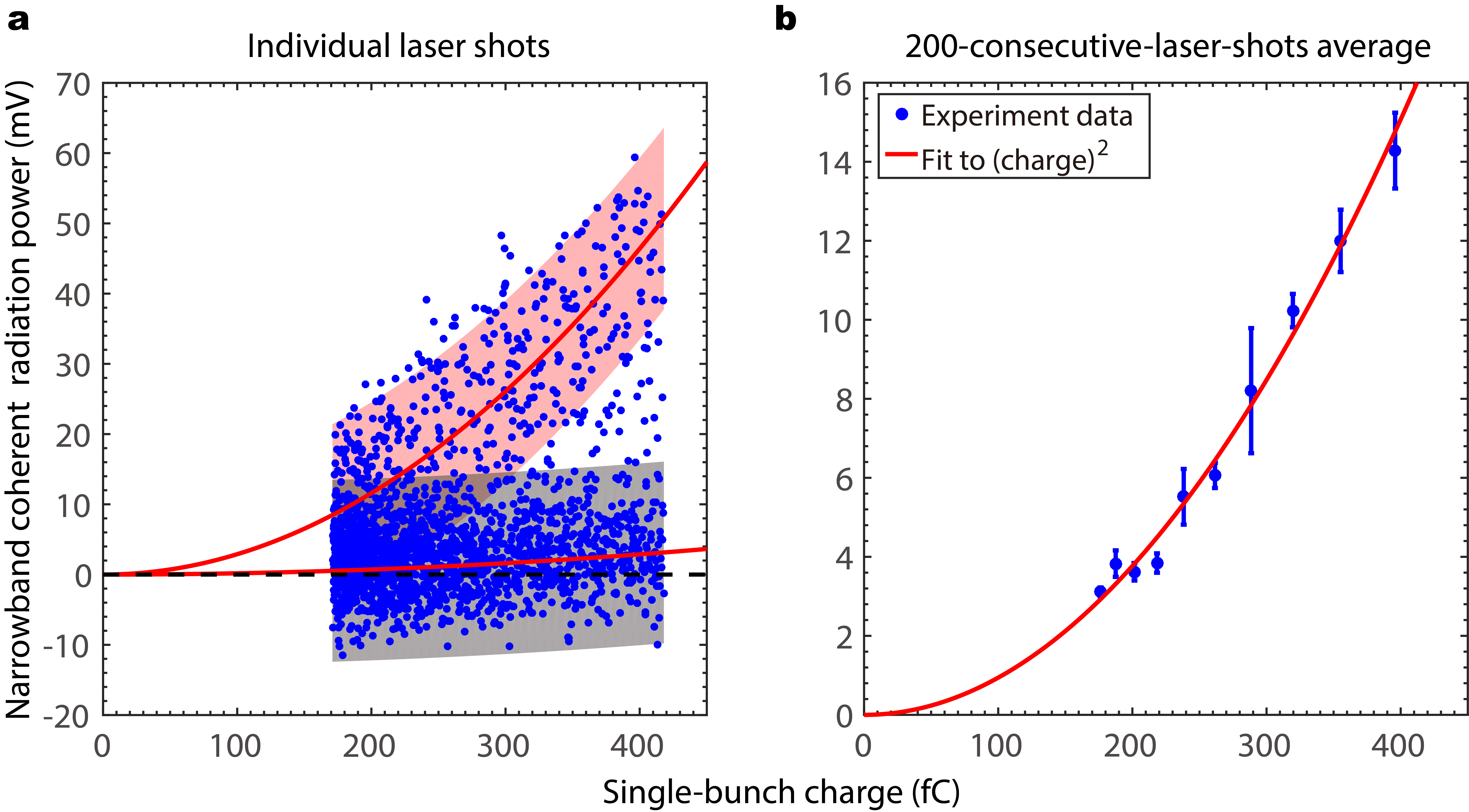}
	\caption{\label{fig:Chap5-SSMBPoPIExtFig5}Quadratic dependence of the narrowband coherent
		undulator radiation generated from microbunching on the bunch
		charge. {\bf a}, Results corresponding to individual laser shots; the shading
		(light red and grey) represents $3\sigma$ of the detection noise. {\bf b}, The result after
		200-consecutive-laser-shot averaging; the plot is the same as Fig.~\ref{fig:Chap5-SSMBPoPIFig3} and is
		presented again here for comparison with a and with the incoherent signal in
		Fig.~\ref{fig:Chap5-SSMBPoPIExtFig4}. The blue dots represent the experimental data and the
		red curves are quadratic fits.  (Figure from Ref.~\cite{deng2021experimental})}
\end{figure}

If the ring works in single-bunch mode, there is a more direct and accurate method of bunch charge measurement based on the measurement of synchrotron radiation strength using photodiodes. Both method have been used in our experiments.

{\bf Detection and evaluation of undulator radiation} The long-pulse laser (FWHM $\approx$ 10 ns) is used to simplify the experiment by avoiding a dedicated laser–electron synchronization system, given that the shot-to-shot laser timing jitter is $t_{\text{jitter}}\leq$ 1 ns (RMS). However, the photodetector (Femto HSPR-X-I-1G4-SI; rise/fall time, 250 ps)
becomes saturated and even damaged by the powerful laser (Beamtech Optronics Dawa-200) if it is placed in the path of the laser. To address this issue, the undulator radiation is separated into the fundamental and second harmonics by appropriate dichroic mirrors (Thorlabs Harmonic
Beamsplitters HBSY21/22), as shown in Fig.~\ref{fig:Chap5-SSMBPoP}, and the signal detection at first focuses on the second harmonic with the wavelength centred at 532 nm. The photodetector output voltage, which is proportional to the radiation power, is then measured by a digital oscilloscope (Tektronix MSO64:6-BW-4000; bandwidth, 4 GHz; sample rate, 25 billion samples per second). Later we will also present the result of the 1064 nm radiation by implementing the Pockels cells to block the modulation laser and let pass the radiation in the following turns.

An example radiation waveform of the second harmonic is shown in Fig.~\ref{Chap5-fig:SSMBPoPIFig2}. To avoid the impact of the signal waveform offset caused by stray laser
light, the data analysis takes the peak-to-peak value of the photodetector output voltage as a measure of the radiation power.
The coherent radiation power that corresponds to each individual
laser shot, obtained during a time interval with a decaying beam current,
is presented in Fig.~\ref{fig:Chap5-SSMBPoPIExtFig5}a, where the modest contribution
on the measured quantity from the small amount of incoherent
radiation transmitted through the 3-nm-bandwidth band-pass filter
has been eliminated. As can be seen, the coherent signal fluctuates
considerably from shot to shot. This is attributable to the shot-to-shot
fluctuation of the laser intensity profile (see Fig.~\ref{fig:Chap5-SSMBPoPIExtFig2}) and
the measurement noise. Despite the fluctuation, quadratic functions
fit reasonably to the lower and upper bounds of the data points, which
correspond to the cases of minimum and maximum bunching factors
induced by the fluctuating laser, respectively. When performing the fits,
we took into account that the measured quantity is the real radiation
signal convoluted with the detection noise. The impact of this noise,
obtained by analysing the measurement result of the unfilled bunches,
on the bounds of the measured data points is visualized as shading in
Fig.~\ref{fig:Chap5-SSMBPoPIExtFig5}a for the coherent signal and in Fig.~\ref{fig:Chap5-SSMBPoPIExtFig4}a for the incoherent signal. To smooth this shot-to-shot fluctuation,
a 200-consecutive-laser-shot averaging is conducted and the
results are presented in Fig.~\ref{fig:Chap5-SSMBPoPIExtFig5}b and Fig.~\ref{fig:Chap5-SSMBPoPIExtFig4}b for the narrowband
coherent and broadband incoherent signals, where a quadratic and a
linear fit have been performed, respectively.

\subsection{Experimental Results}
\subsubsection{Second Harmonic Radiation}

\begin{figure}[tb]
	\centering 
	\includegraphics[width=1\textwidth]{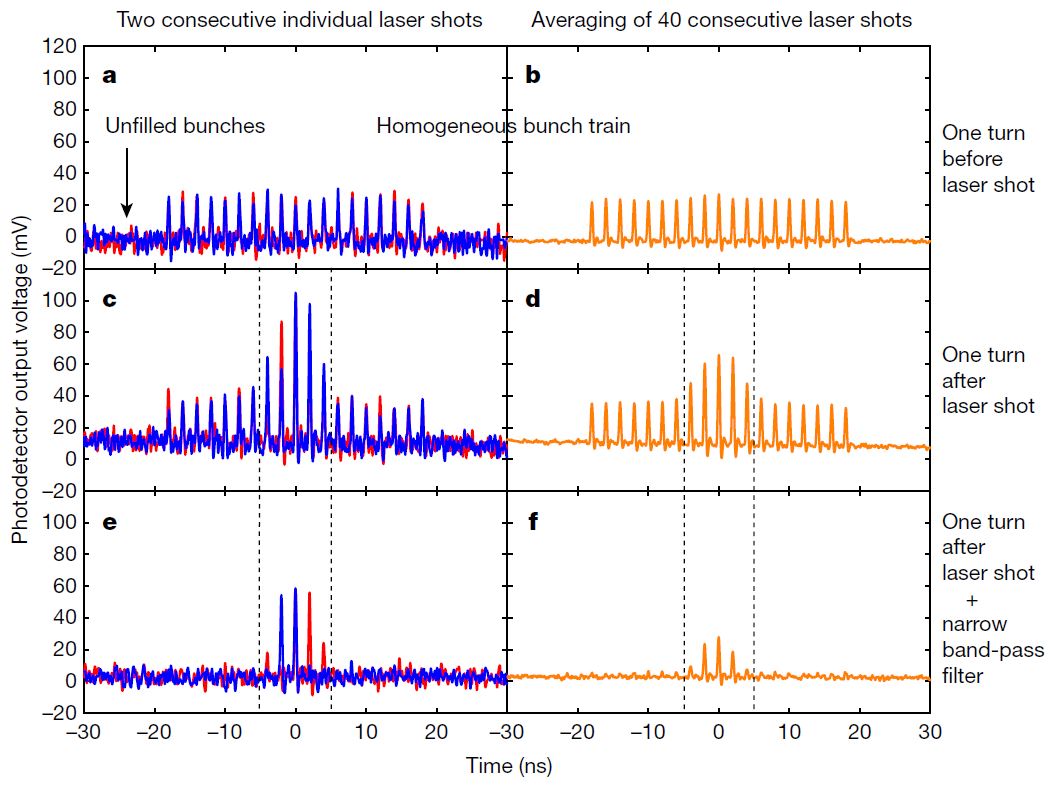}
	\caption{\label{Chap5-fig:SSMBPoPIFig2}Waveforms of the undulator radiation produced from a homogeneous stored bunch train. {\bf a}, {\bf b}, Radiation one turn before the laser shot. The photodetector output voltage is proportional to the radiation power. {\bf c}, {\bf d}, Radiation one turn after the laser shot, from the same bunches as those of {\bf a} and {\bf b}, where the central five bunches are modulated by the laser pulse. The offset and general slight decreasing trend of the waveforms are due to the photodetector being saturated by stray light from the modulation laser one revolution before and not having completely recovered. {\bf e}, {\bf f}, Radiation one turn after the laser shot, obtained with a narrow band-pass filter (centre wavelength, 532 nm; bandwidth, 3 nm FWHM) placed in front of the photodetector, with bunch filling and charge similar to those in {\bf a} to {\bf d}.  (Figure from Ref.~\cite{deng2021experimental})}
\end{figure}

Figure~\ref{Chap5-fig:SSMBPoPIFig2} shows typical measurement results of the second-harmonic undulator radiation emitted from a homogeneous stored bunch train, with a charge of about 1 pC per bunch and a time spacing of 2 ns, supplied by the 500 MHz RF cavity at the MLS. The spikes in the waveforms are the signals of different bunches. The left and right panels show the results corresponding to 2 and 40 consecutive laser shots, respectively. To smooth the measurement noise and signal fluctuation, the waveforms in the right panels have been averaged. Figure~\ref{Chap5-fig:SSMBPoPIFig2}a, b shows the signals one turn before the laser shot, which
correspond to the incoherent radiation and reflect the homogeneous
bunch filling pattern. Figure~\ref{Chap5-fig:SSMBPoPIFig2}c, d shows the radiation one turn after the laser shot, from the same bunches as those in Fig.~\ref{Chap5-fig:SSMBPoPIFig2}a, b. The five larger spikes at the centre correspond to the bunches modulated by the laser. The enhanced signals of these five spikes indicate the formation of microbunching and the generation of coherent radiation from the laser-modulated bunches.

\begin{figure}[tb]
	\centering 
	\includegraphics[width=0.7\textwidth]{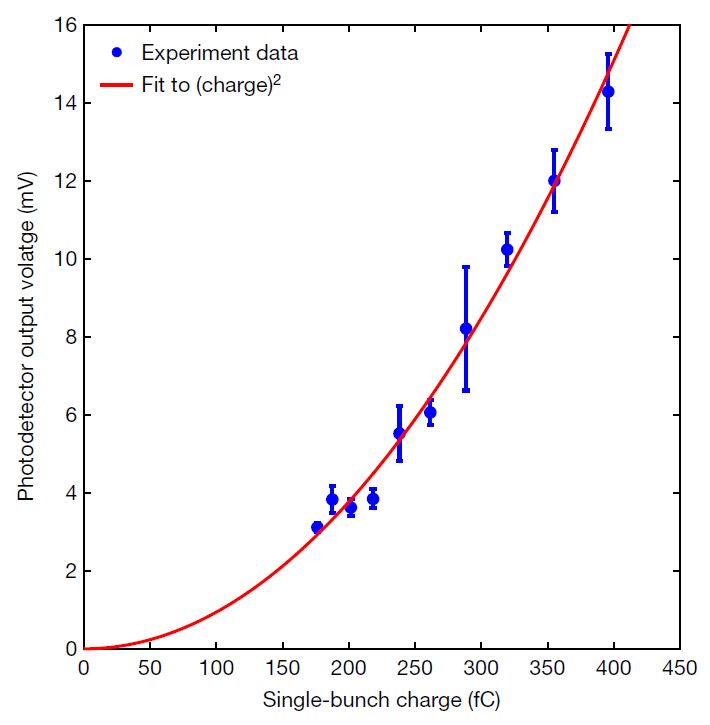}
	\caption{\label{fig:Chap5-SSMBPoPIFig3}Quadratic dependence of the coherent undulator radiation
		generated from microbunching on the bunch charge. The blue dots
		represent the experimental data and the red curve is a quadratic fit. Each data
		point represents the averaged result of 200 consecutive laser shots. The error
		bars denote the standard deviation of the averaged results when the averaging
		time window shifts for ±100 consecutive laser shots from the corresponding
		data point.  (Figure from Ref.~\cite{deng2021experimental})}
\end{figure}

The laser used in the experiment has multiple longitudinal modes, and its temporal profile has several peaks and fluctuates considerably from shot to shot (see Fig.~\ref{fig:Chap5-SSMBPoPIExtFig2}). Therefore, the laser-induced electron energy modulation amplitudes are different from shot to shot and from bunch to bunch. When the modulation amplitude matches the phase slippage factor, the energy-modulated electrons are properly focused at synchronous phases, which gives optimal microbunching. For some of the shots, the laser intensity is higher or lower than the optimal value, and the electrons are then over-focused or under-focused, giving weaker microbunching and less coherent radiation. This explains the shot-to-shot fluctuation of the coherent amplified signals shown in Fig.~\ref{Chap5-fig:SSMBPoPIFig2}c, e.

As analyzed before, the microbunching coherent radiation is much narrowbanded compared to the incoherent radiation. 
To confirm that the amplified radiation is due to microbunching, we tested this narrowband feature of the coherent radiation. A band-pass filter (Thorlabs FL532-3; centre wavelength, 532 nm; bandwidth, 3 nm FWHM) was inserted in front of the detector. The radiation one turn after the laser shot is shown in Fig.~\ref{Chap5-fig:SSMBPoPIFig2}e, f, which was obtained with a bunch filling and charge similar to that of Fig.~\ref{Chap5-fig:SSMBPoPIFig2}c, d. From the comparison between Fig.~\ref{Chap5-fig:SSMBPoPIFig2}e and Fig.~\ref{Chap5-fig:SSMBPoPIFig2}c (Fig.~\ref{Chap5-fig:SSMBPoPIFig2}f and Fig.~\ref{Chap5-fig:SSMBPoPIFig2}d), we can see that the broadband incoherent signals are nearly completely blocked by the filter, whereas the amplified part is not affected much, confirming that the amplification is the narrowband coherent radiation generated by the microbunches.

Finally, we investigated the dependence of the coherent radiation on the bunch charge. To mitigate collective effects such as intrabeam scattering~\cite{piwinski1974intra,bjorken1982intrabeam} and head–tail instability~\cite{chao1993physics}, which could change the electron beam parameters, this investigation was conducted at low beam current, and the coherent signal was optimized by fine-tuning the machine to ensure a sufficient signal-to-noise ratio. Because the longitudinal radiation damping time in the experiment was 180 ms, we operated the laser at 1.25 Hz repetition rate to ensure that the electron bunches had time to recover their equilibrium parameters before each laser shot. The 3-nm-bandwidth band-pass filter was inserted to block the incoherent radiation, and the coherent signal corresponding to each individual laser shot was saved, with the beam current decaying naturally until the signal was at the detection noise
level. The measurement results of the bunch closest to the laser temporal centre ($t = 0$ ns in Fig.~\ref{Chap5-fig:SSMBPoPIFig2}) are used for quantitative analysis as introduced above. To lessen the impact of the laser temporal profile fluctuation and measurement noise, a 200-consecutive-laser-shot averaging is performed to obtain the data point for each bunch charge. The coherent undulator radiation power versus the single-bunch charge is shown in
Fig.~\ref{fig:Chap5-SSMBPoPIFig3}, where a quadratic function fits well to the experiment data. The quadratic bunch charge dependence, together with the narrowband feature of the coherent radiation, demonstrates unequivocally the formation of microbunching.

\subsubsection{Fundamental Frequency Radiation}

The experimental results shown for the second harmonic are obtained in the year of 2020~\cite{deng2021experimental,tang2020first}. After that, there are two main upgrades on the experimental setup. First, the multi-longitudinal-mode laser has been replaced by a single-longitudinal-mode one (Amplitude Surelite I-10). Second, Pockel Cells have been installed along the optical path to block the modulation laser and let pass the radiation in later revolutions, thus allowing the detection of the fundamental frequency radiation~\cite{kruschinski2021fundamental}. As shown in our analysis, we expect that the coherent radiation at the fundamental frequency is much stronger than that at the second harmonic and the microbunching can last multiple turns. 
Recently in April of 2021, we have relaunched the SSMB experiments at the MLS and these expectations have been confirmed \cite{feikes2021progress}.

\begin{figure}[tb]
	\centering 
	\includegraphics[width=1\textwidth]{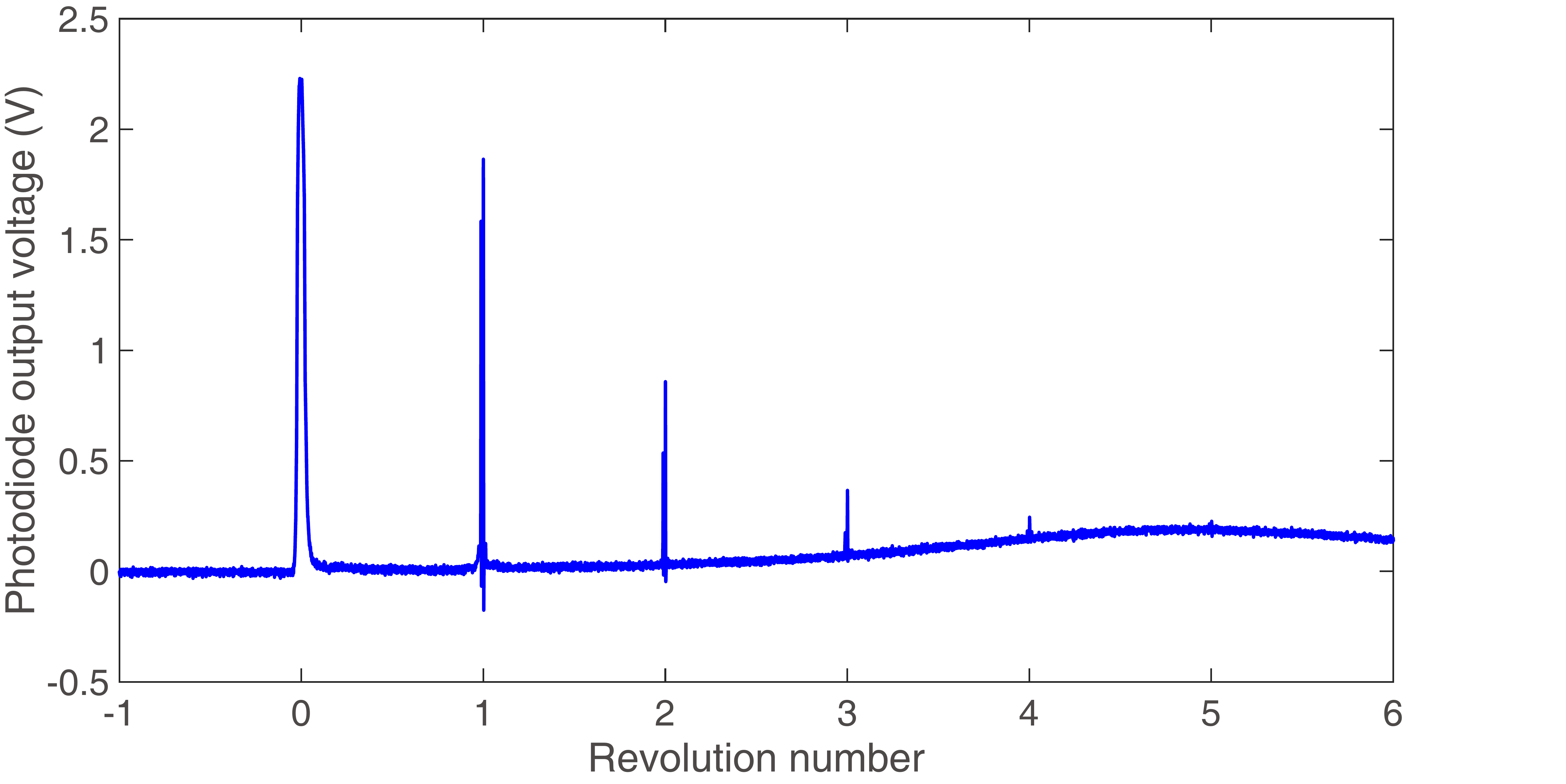}
	\caption{\label{fig:Chap5-MultiTurnMB}Raw data of the multi-turn microbunching preservation experiment result. The signal is for the fundamental freuqency radiation, i.e., 1064 nm. (Refer to J. Feikes' talk in IPAC2021~\cite{feikes2021progress} for more details and courtesy of MLS colleagues)}
\end{figure}

\begin{figure}[tb]
	\centering 
	\includegraphics[width=1\textwidth]{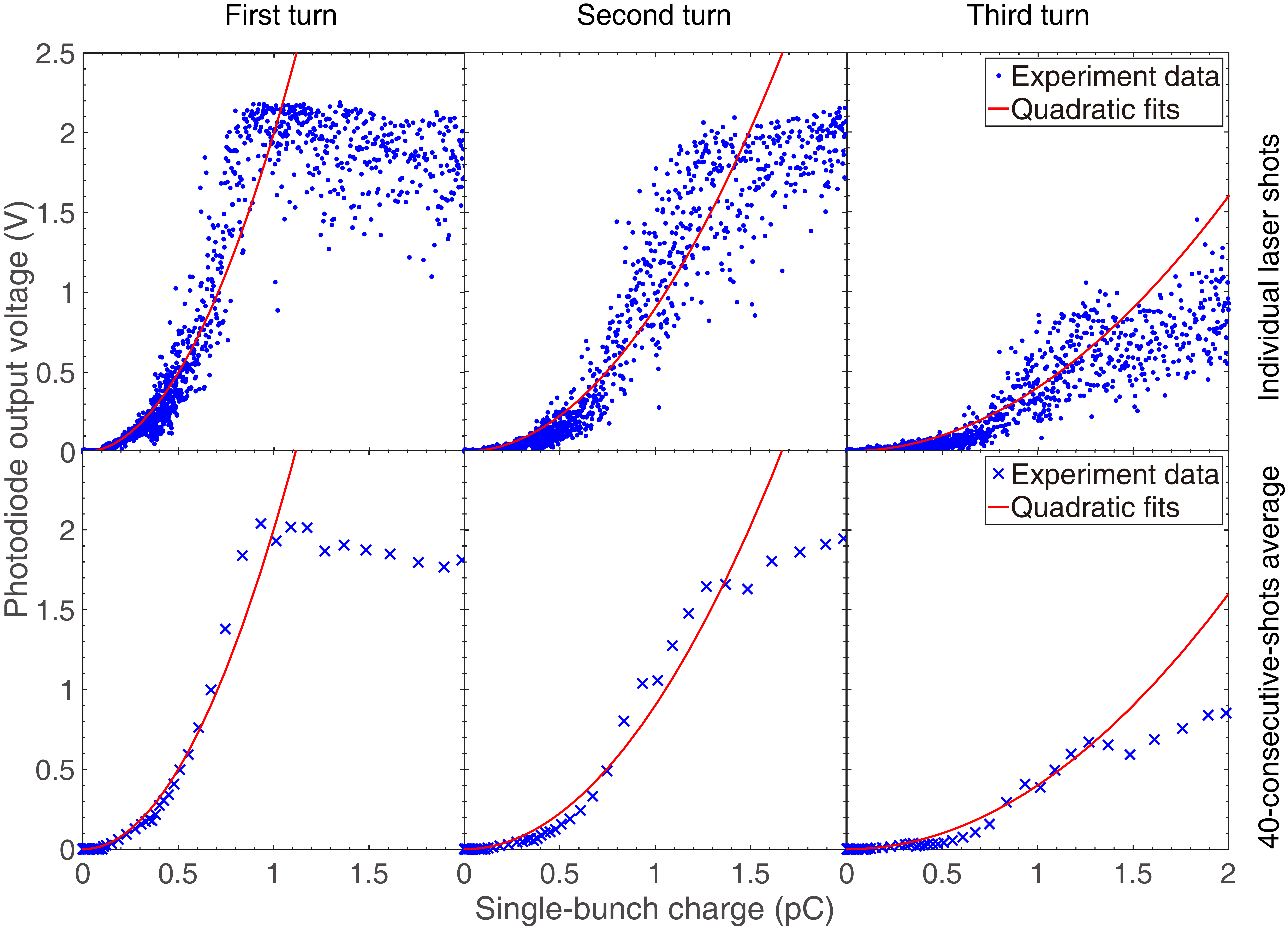}
	\caption{\label{fig:Chap5-SSMBPoPNewComparison}The bunch-charge scaling of the coherent undulator radiation signals of the first three turns after the laser shots. The saturation level of the detector is about 2 V. (Refer to J.~Feikes' talk in IPAC2021\cite{feikes2021progress} for more details and courtesy of MLS colleagues)}
\end{figure}

Figure~\ref{fig:Chap5-MultiTurnMB} shows the preliminary experimental results of the multi-turn coherent radiation at the fundamental frequency. A narrow-band-pass filter has been inserted to select the narrowband coherent radiation. Figure.~\ref{fig:Chap5-SSMBPoPNewComparison} is the more quantitative data analysis of the signal of the first three turns after each laser shots. The bunch charge has now also been obtained in a more accurate way by using the incoherent synchrotron radiation signal measured by a photodiode. There are several important observations from the experiment results: 
\begin{itemize}
	\item First, signals of all three turns have shown nice quadratic bunch charge fits, confirming again the formation of microbunching and coherent radiation generation from it; 
	\item Second, we mentioned in the above section and also in Ref.~\cite{deng2021experimental} that the huge shot-to-shot signal fluctuation is mainly due to the laser profile fluctuation due to its multi-longitudinal-mode nature. This argument has also been confirmed by the results shown in  Fig.~\ref{fig:Chap5-SSMBPoPNewComparison}. From the comparison of Fig.~\ref{fig:Chap5-SSMBPoPNewComparison} and Fig.~\ref{fig:Chap5-SSMBPoPIExtFig5}a, we can see that the shot-to-shot coherent signal with the present single-longitudinal-mode laser is much more stable;
	\item Third, the signal deviates from the quadratic scaling at high current and starts to saturates about 1.3 pC, which we believe is due to the influence of collective effects.
\end{itemize}
More in-depth investigations on the multi-turn microbunching is still ongoing and will be reported in the future.



\subsubsection{Short Summary}
In conclusion, we have demonstrated the mechanism of SSMB in an electron storage ring. This demonstration represents the first milestone towards the implementation of an SSMB-based high-repetition, high-power photon source.  We emphasize that here we do not report an actual demonstration of SSMB, but rather a demonstration of the mechanism by which SSMB will eventually be attained. First, the formation of microbunching after one {\it complete} revolution of a laser-modulated bunch in a quasi-isochronous ring and the maintenance of microbunching for multiple turns demonstrate the viability of a turn-by-turn electron optical phase correlation with a precision of sub-laser wavelength. Second, this microbunching is produced on the {\it stored} electron bunch, the equilibrium parameters and distribution of which before the laser modulation are defined by the same storage ring as a whole. The combination of these two crucial factors establishes a closed loop to support the realization of SSMB, provided that a phase-locked laser interacts with the electrons turn by turn.


\section{PoP II: Quasi-steady-state Microbunching}

On the basis of the PoP I, the next step is to replace the single-shot laser by a high-repetition phase-locked one to interact with the electrons turn-by-turn. By doing so, we want to form stable microbucket to constrain the microbunching in it to reach a quasi steady state. This is the SSMB PoP II as introduced in Sec.~\ref{sec:threePoP}. 

\subsection{Phase-mixing in Buckets}

To reach a quasi steady state, the particles need to do synchrotron oscillations to reach phase mixing in the microbuckets, as a result of longitudinal amplitude dependent tune spread of the electron beam. Here we present a remarkable feature of phase mixing (filamentation) in RF or optical buckets. As we will see soon, given an initial DC mono-energetic beam, there will be an equilibrium phase space distribution after phase mixing in the bucket. We find that in this final steady state, the current distribution has little dependence on the bucket height. This feature is favorable for the SSMB PoP II, as the requirement on the modulation laser power can then be much relaxed compared to PoP I. This effect is also of relevance to the injection process of the final real SSMB storage ring.

\begin{figure}[tb] 
	\centering 
	\includegraphics[width=1\columnwidth]{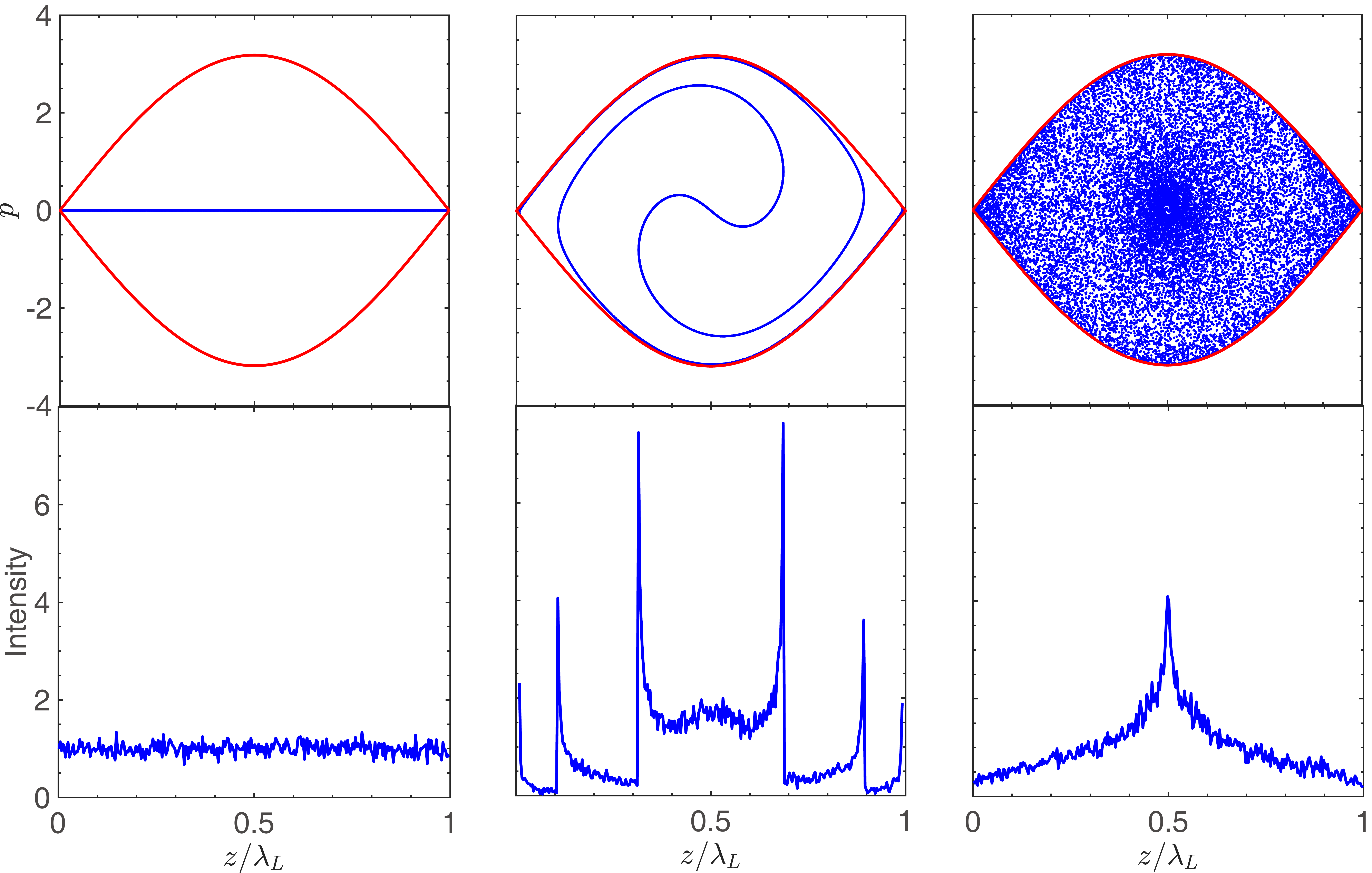}
	\caption{
		\label{fig:Chap5-PhaseMixing} 
		Phase mixing (filamentation or decoherence) in optical bucket with $K=0.01$. Red line: separatrix.
	}
\end{figure} 

Now we start the analysis. As the phase mixing is a rather fast process compared to radiation damping, we consider only the symplectic dynamics in this section for simplicity. The symplectic longitudinal dynamics of a particle in a storage ring with a single RF system can be modeled by the well-known ``standard map''~\cite{chirikov1979universal}
\begin{equation}\label{eq:standardmap}
\begin{cases}
&I_{n+1}=I_{n}+K\sin{\theta_{n}},\\
&\theta_{n+1}=\theta_{n}+I_{n+1},
\end{cases}
\end{equation}
in which 
\begin{equation}
\theta=k_{\text{RF}}z,\ I=R_{56}k_{\text{RF}}\delta,\ K=V_{\text{RF}}R_{56}k_{\text{RF}},
\end{equation}
with $R_{56}=-\eta C_{0}$.
Note that $K$ in this section is not the undulator parameter. Equation~(\ref{eq:standardmap}) can be described with the pendulum Hamiltonian driven by a periodic perturbation 
\begin{equation}\label{eq:HamiltonianStandardMap}
H(I,\theta,t)=\frac{1}{2}I^{2}+K\cos{\theta}\sum_{n=-\infty}^{\infty}\cos(2\pi nt).
\end{equation}
The dynamics is given by a sequence of free propagations interleaved with periodic kicks. For $K\neq0$, the dynamics is non-integrable and chaotic. 
But for a $K$ much smaller than 1, which is the case for usual storage rings working in the longitudinal weak focusing scheme, the motion is close to integrable and the differences in Eq.~(\ref{eq:standardmap}) can be approximately replaced by differentiation,
and the Hamiltonian Eq.~(\ref{eq:HamiltonianStandardMap}) can be replaced by a pendulum Hamiltonain 
\begin{equation}
H=\frac{1}{2}I^{2}+K\cos{\theta}.
\end{equation}
The separatrix of the pendulum bucket is $H=K$ with a bucket half-height of $2\sqrt{K}$. When $K$ is large, the strongly chaotic dynamics can also be used for interesting applications, for example applying the bucket burfication to generate short bunches as proposed in Ref.~\cite{jiao2011terahertz}.

Figure~\ref{fig:Chap5-PhaseMixing} shows a simulation result of the evolution in the longitudinal phase space evolution of a mono-energetic DC beam after injection into RF or optical bucket described by Eq.~(\ref{eq:standardmap}). We have chosen to observe the beam in the middle of the RF kick so the beam distribution in the longitudinal phase space is upright. As can be seen, there is a steady-state beam distribution due to phase mixing in the bucket. 

\begin{figure}[tb] 
	\centering 
	\includegraphics[width=0.7\columnwidth]{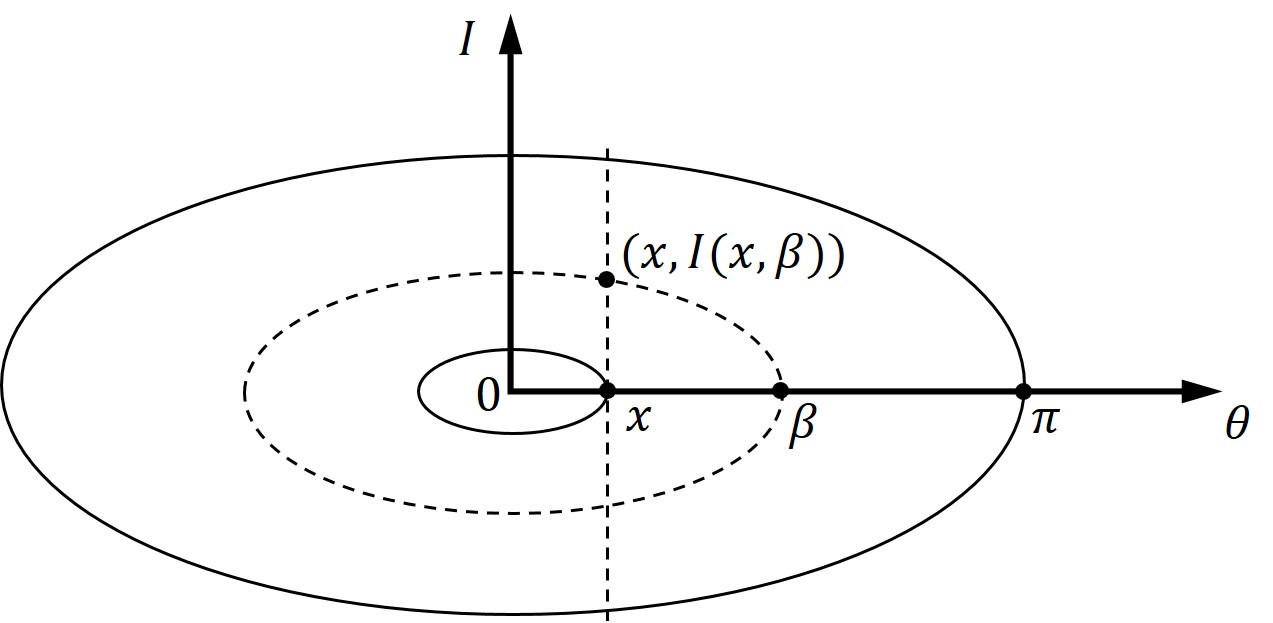}
	\caption{
		\label{fig:Chap5-DistbutionCalculation} 
		A plot to help better understand Eq.~(\ref{eq:RFBucket}).}
\end{figure} 

As the longitudinal form factor, thus the coherent radiation, depends more directly on the beam current (namely the longitudinal coordinate $z$ of the electrons) rather than the longitudinal phase space distribution, now we try to get an analytical formula for the steady-state beam current. For convenience, we shift the bucket center to the origin, which means $\theta-\pi\rightarrow\theta$. What we want to know is the steady-state distribution of $\theta$, i.e., $f(\theta,t\rightarrow\infty)$. In action-angle $(\phi,J)$ phase space, the distribution function evolves according to
\begin{equation}
f(\phi,J,t)=f(\phi-\omega(J)t,J,0).
\end{equation}
In the limit of $t\rightarrow\infty$, the steady-state distribution depends only on the initial distribution of action $J$, when there is a tune dependence $\omega(J)$ on $J$, as a result of phase mixing 
\begin{equation}\label{eq:phasemixing}
f(\phi,J,t\rightarrow\infty)=\frac{1}{2\pi}\int_{0}^{2\pi}f(\phi,J,t=0)d\phi=\frac{1}{2\pi}f(J,t=0).
\end{equation}
As shown in Fig~\ref{fig:Chap5-DistbutionCalculation}, after reaching the steady state, the percentage of the particles with $x\geq\theta$ for $0<x<\pi$ is
\begin{equation}\label{eq:RFBucket}
P(x\geq\theta)=\frac{x}{\pi}+\int_{x}^{\pi}\left(1-\frac{\phi(x,I(x,\beta))}{\pi}\right)\frac{1}{\pi}d\beta,
\end{equation}
in which $I(x,\beta)$ represents the $I$-coordinate of a point on the $(\phi,J)$ phase space trajectory traversing $(\beta,0)$ with a $\theta$-coordinate of $x$. 

\begin{figure}[tb]
	\centering 
	\includegraphics[width=0.5\columnwidth]{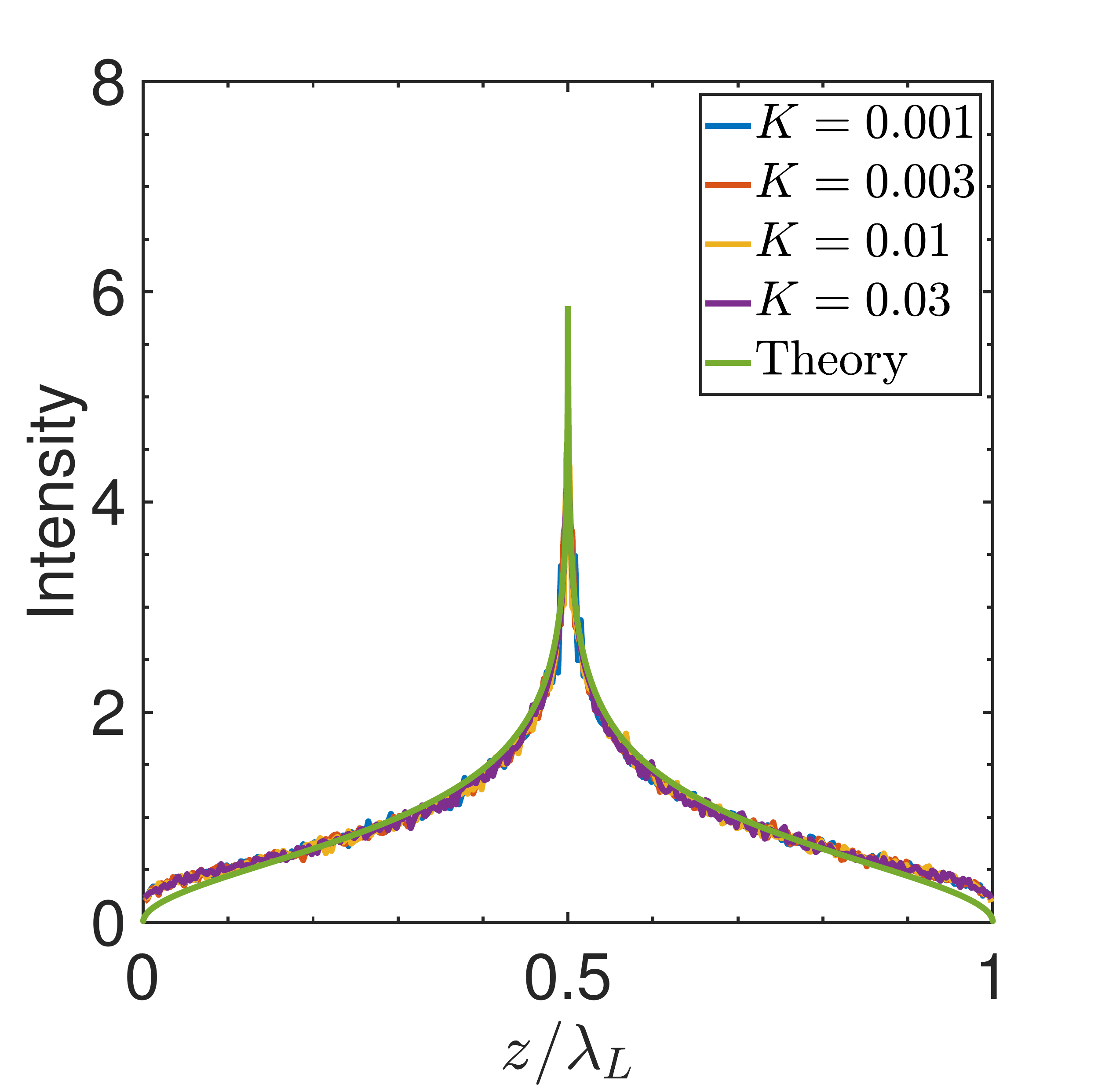}
	\caption{
		\label{fig:Chap5-PhaseMixingCurrent} 
		The steady-state current distribution after phase mixing in RF or optical bucket, with different $K$. In each simulation, $2\times10^{5}$ particles are tracked $2\times10^{5}$ turns. Also presented in the figure is the theoretical prediction of Eq.~(\ref{eq:theoreticalDis}).  
	}
\end{figure}

After getting $P(x\geq\theta)$, the current distribution can then be calculated according to
\begin{equation}\label{eq:distribution}
f(\theta)=\frac{\partial{P}}{\partial{x}}\bigg|_{x=\theta}.
\end{equation}
However, $\phi(x,I(x,\beta))$ has a complex form, and it is hard to get a simple analytical expression for $f(\theta)$.  Here we simplify the discussion by approximating all the phase space trajectories in the bucket by ellipses to arrive at an analytical formula for $f(\theta)$.  For an ellipse phase space trajectory, we have
\begin{equation}\label{eq:ellipse}
\phi(x,I(x,\beta))=\arccos{\frac{x}{\beta}}.
\end{equation}
Note that the result in Eq.~(\ref{eq:ellipse}) has no dependence on $K$. For a real RF or optical bucket, there is a dependence of $\phi(x,I(x,\beta))$ on $K$, but the dependence is weak, especially for trajectories close to the origin. So we expect our approximated Eq.~(\ref{eq:ellipse}) is valid to a large extent. Substituting Eq.~(\ref{eq:ellipse}) into Eqs.~(\ref{eq:RFBucket}) and~(\ref{eq:distribution}), we have 
\begin{equation}
\begin{aligned}\label{eq:theoreticalDis}
f(\theta)=\frac{\partial{P}}{\partial{x}}\bigg|_{x=\theta}
=\int_{\theta}^{\pi}\left(\frac{1}{\pi\sqrt{1-\left(\frac{\theta}{\beta}\right)^{2}}}\frac{1}{\beta}\right)\frac{1}{\pi}d\beta
=\frac{1}{\pi^{2}}\ln\bigg|\frac{\pi+\sqrt{\pi^{2}-\theta^{2}}}{\theta}\bigg|.
\end{aligned}
\end{equation}
Note that our simplified theoretically current distribution $f(\theta)$ is independent of $K$, which means the steady-state current distribution is independent of the bucket height.

\begin{figure}[tb]
	\centering 
	\includegraphics[width=0.5\columnwidth]{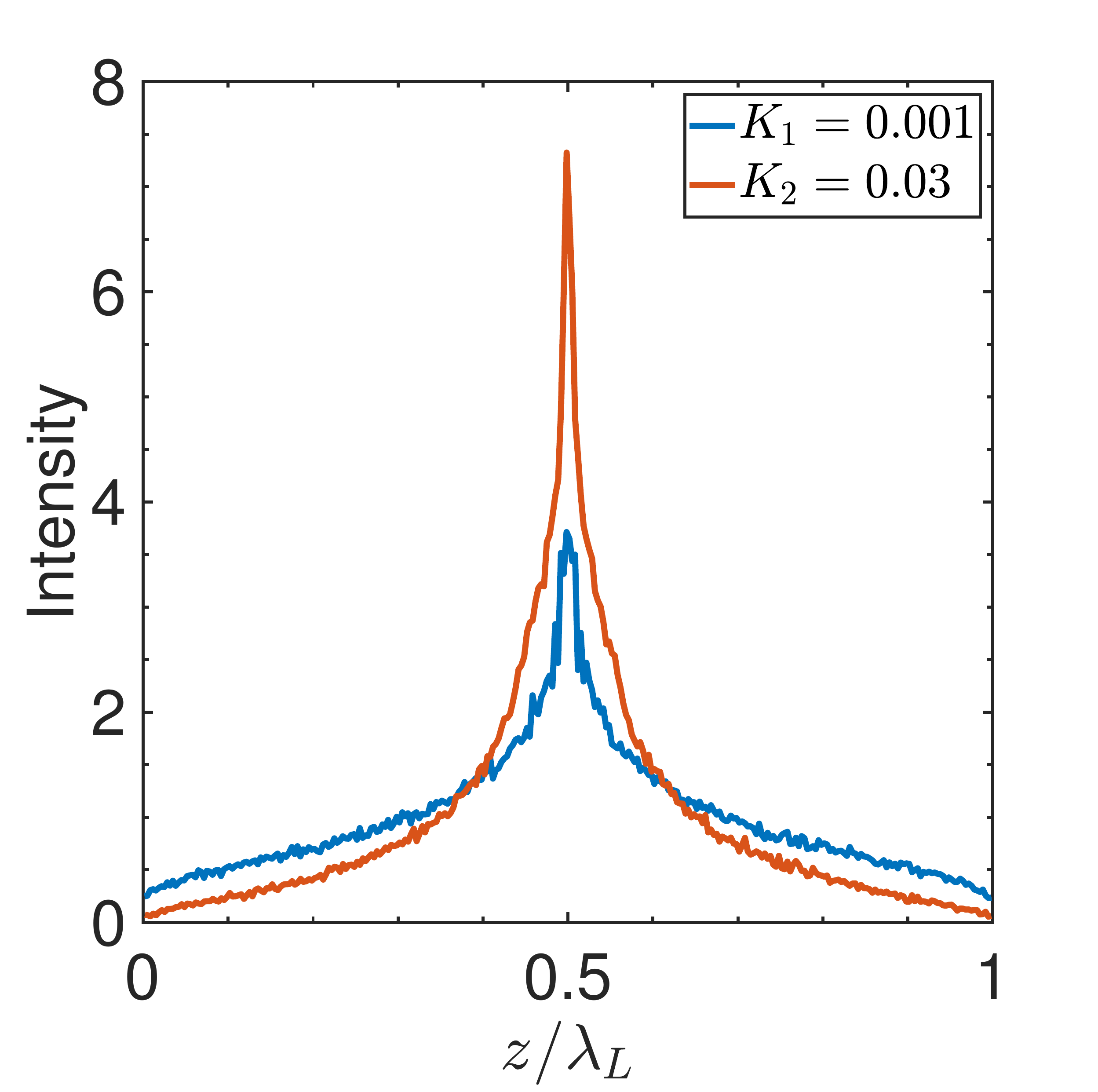}
	\caption{
		\label{fig:Chap5-PhaseMixingCurrent2} 
		The steady-state current distribution after phase mixing in RF or optical bucket, with an increase of $K$ from 0.001 to 0.03 in two consecutive steps. In each step, $2\times10^{5}$ particles are tracked $2\times10^{5}$ turns. 
	}
\end{figure}

Figure~\ref{fig:Chap5-PhaseMixingCurrent} shows the simulation result of the steady-state current distribution under different $K$,  i.e., different bucket heights, and simultaneously our simplified theoretical distribution Eq.~(\ref{eq:theoreticalDis}). As can be seen that indeed the steady-state current distribution has little dependence on the bucket height, and our simplified analysis is quite accurate. Note that the origin is not shifted in the plot.

The analysis reveals a remarkable feature of phase mixing in RF or optical bucket, i.e., the final steady-state current distribution after a mono-energetic beam getting trapped by RF or optical bucket has little dependence on the bucket height. This is helpful for our Quasi-SSMB experiment since it means the requirement on the modulation laser power is not that demanding. A bucket height which is several times higher than the natural energy spread is sufficient. 


The above result is based on a constant RF voltage in the phase mixing process, it can be anticipated that more particles will be bunched closer to the bucket center phase when we increase $K$ after the beam reach its steady-state distribution after phase mixing. Similar step to the above section can be invoked for calculating the new steady-state current distribution. A transformation of the action when $K$ changes is all that needed. Figure~\ref{fig:Chap5-PhaseMixingCurrent2} shows the simulation result of the steady-state current distribution by increasing $K$ in two consecutive steps from 0.001 to 0.03. As can be seen, the current are more concentrated to the center after the increase of $K$.

\begin{figure}[tb]
	\centering
	\includegraphics[width=0.32\textwidth]{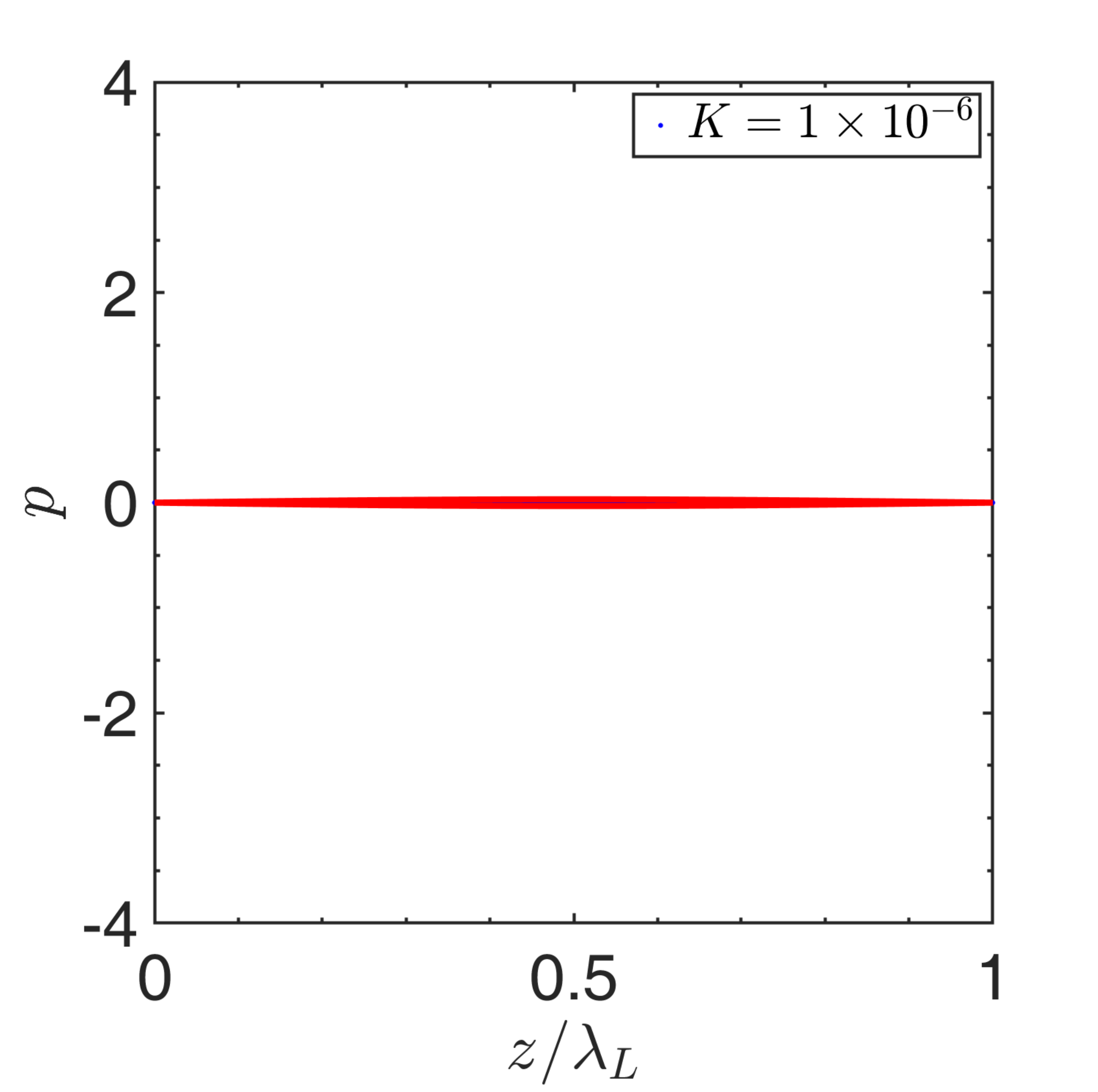}
	\includegraphics[width=0.32\textwidth]{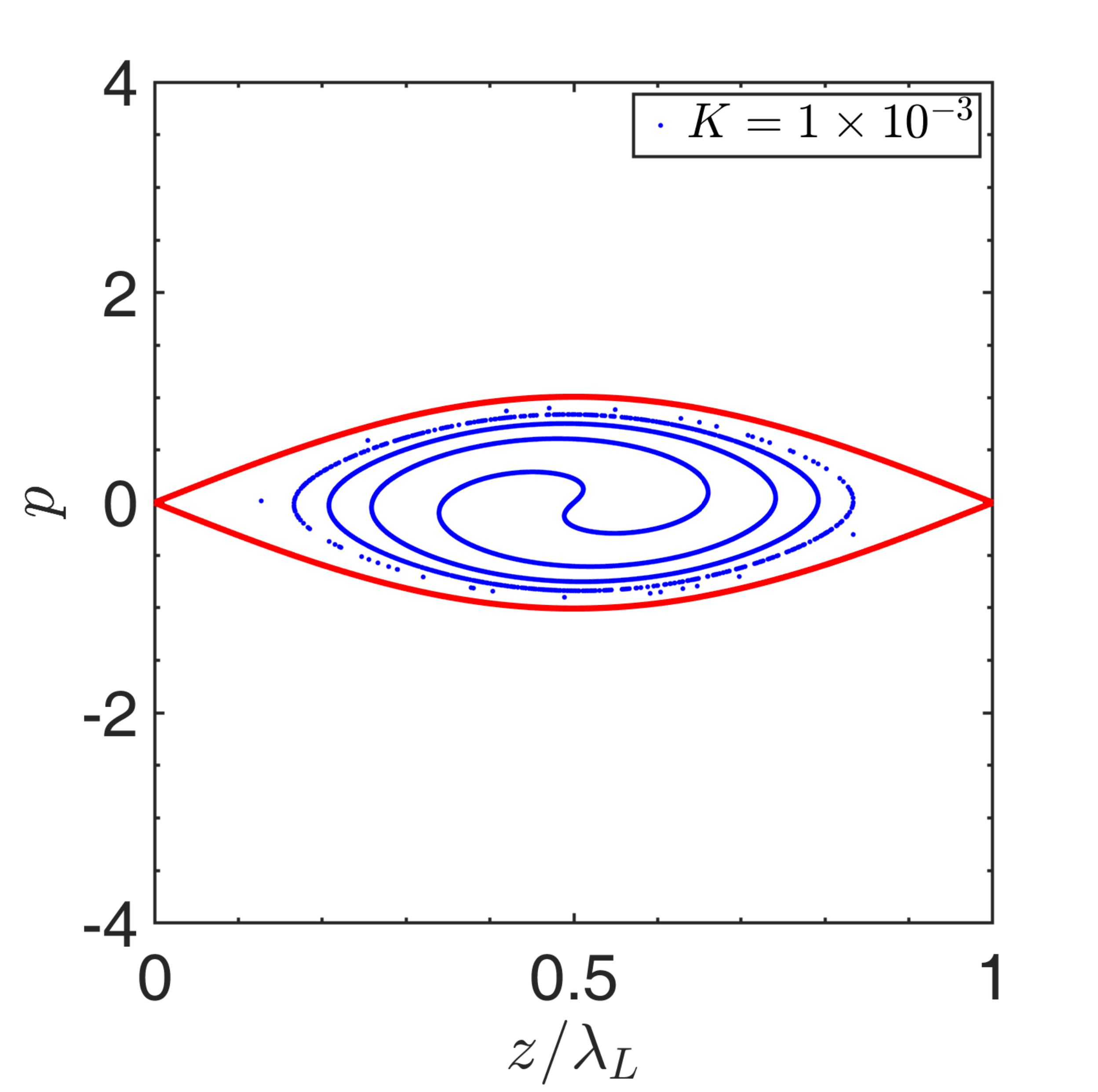}
	\includegraphics[width=0.32\textwidth]{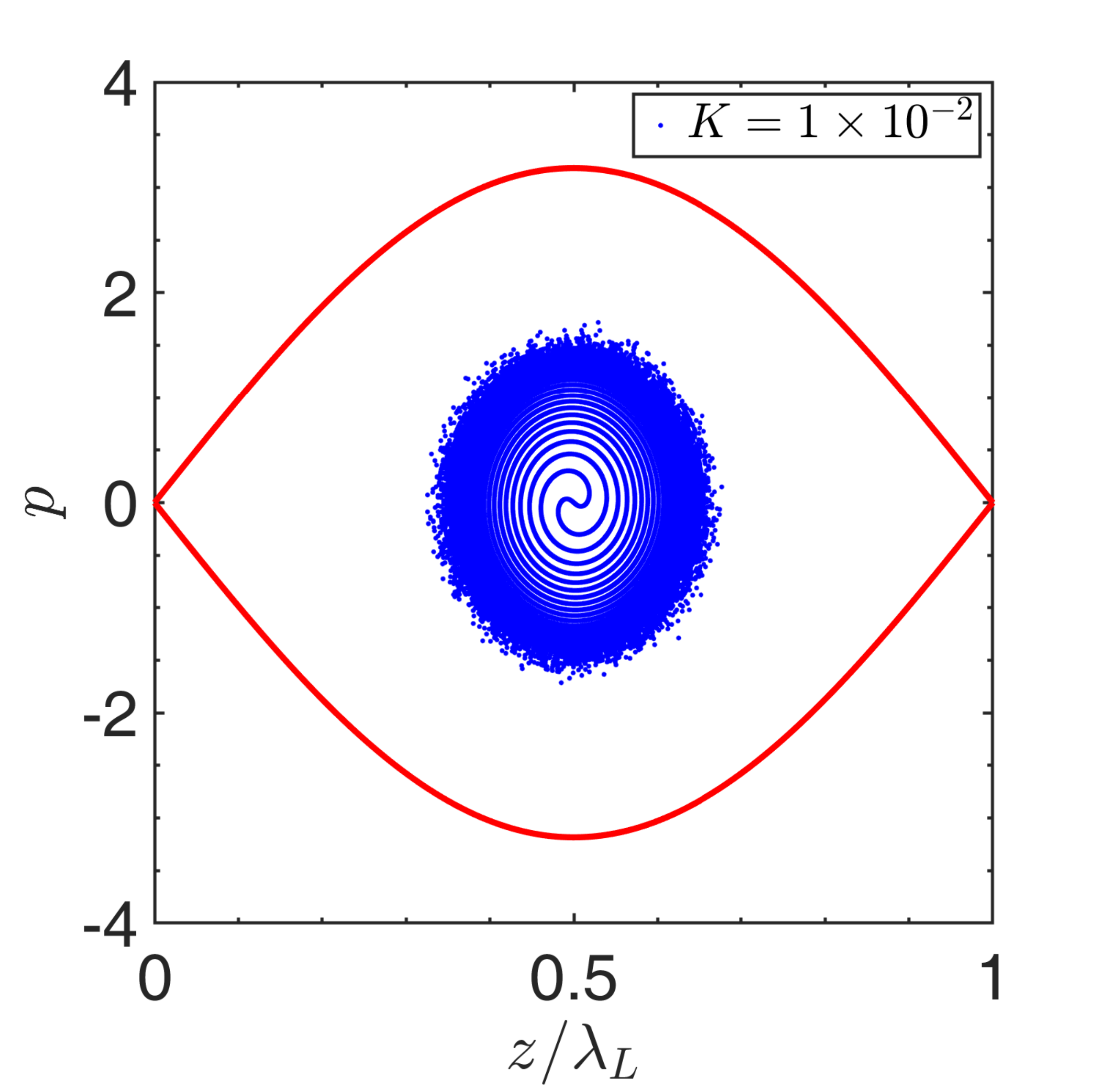}\\
	\includegraphics[width=0.32\textwidth]{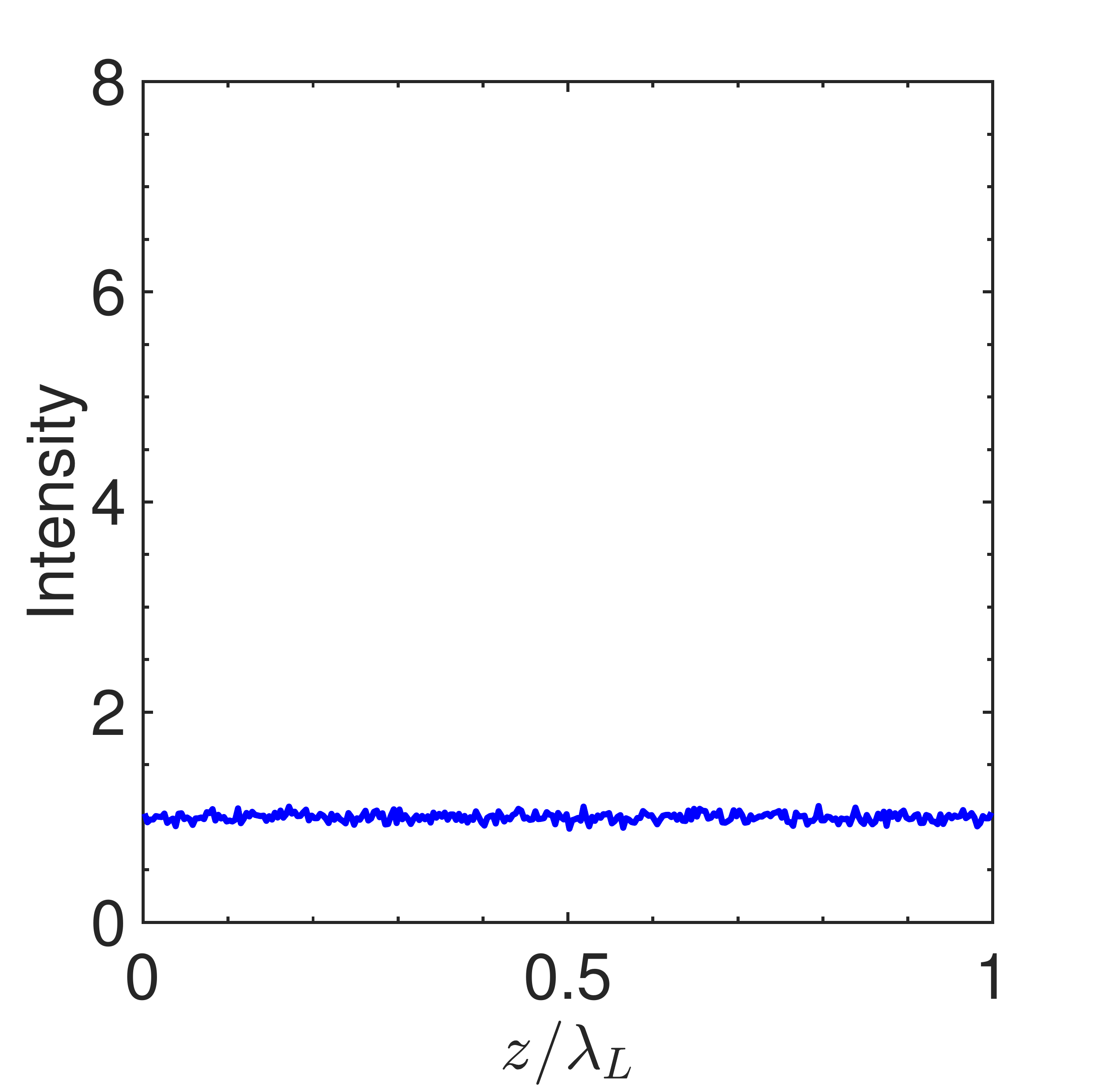}
	\includegraphics[width=0.32\textwidth]{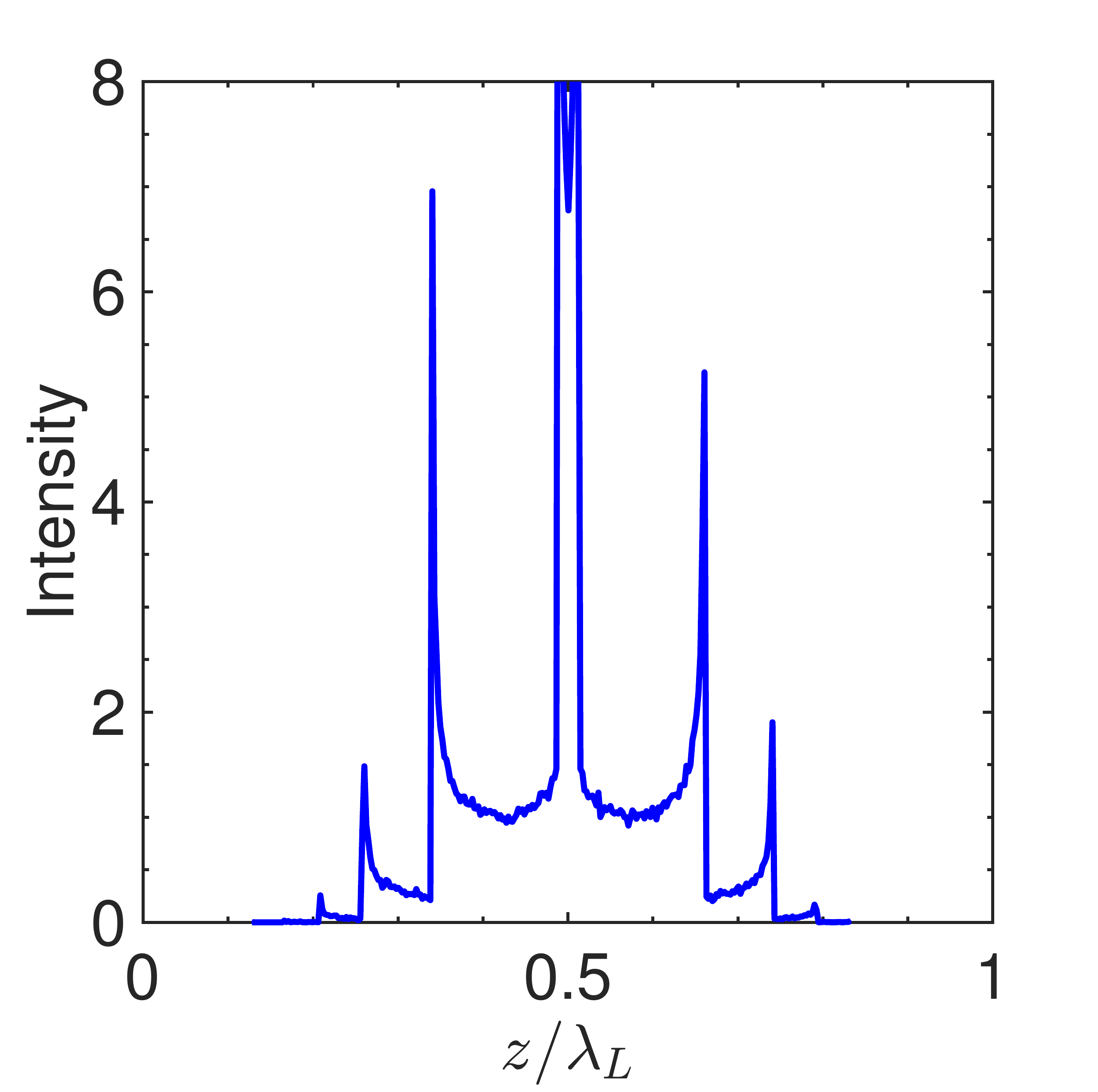}
	\includegraphics[width=0.32\textwidth]{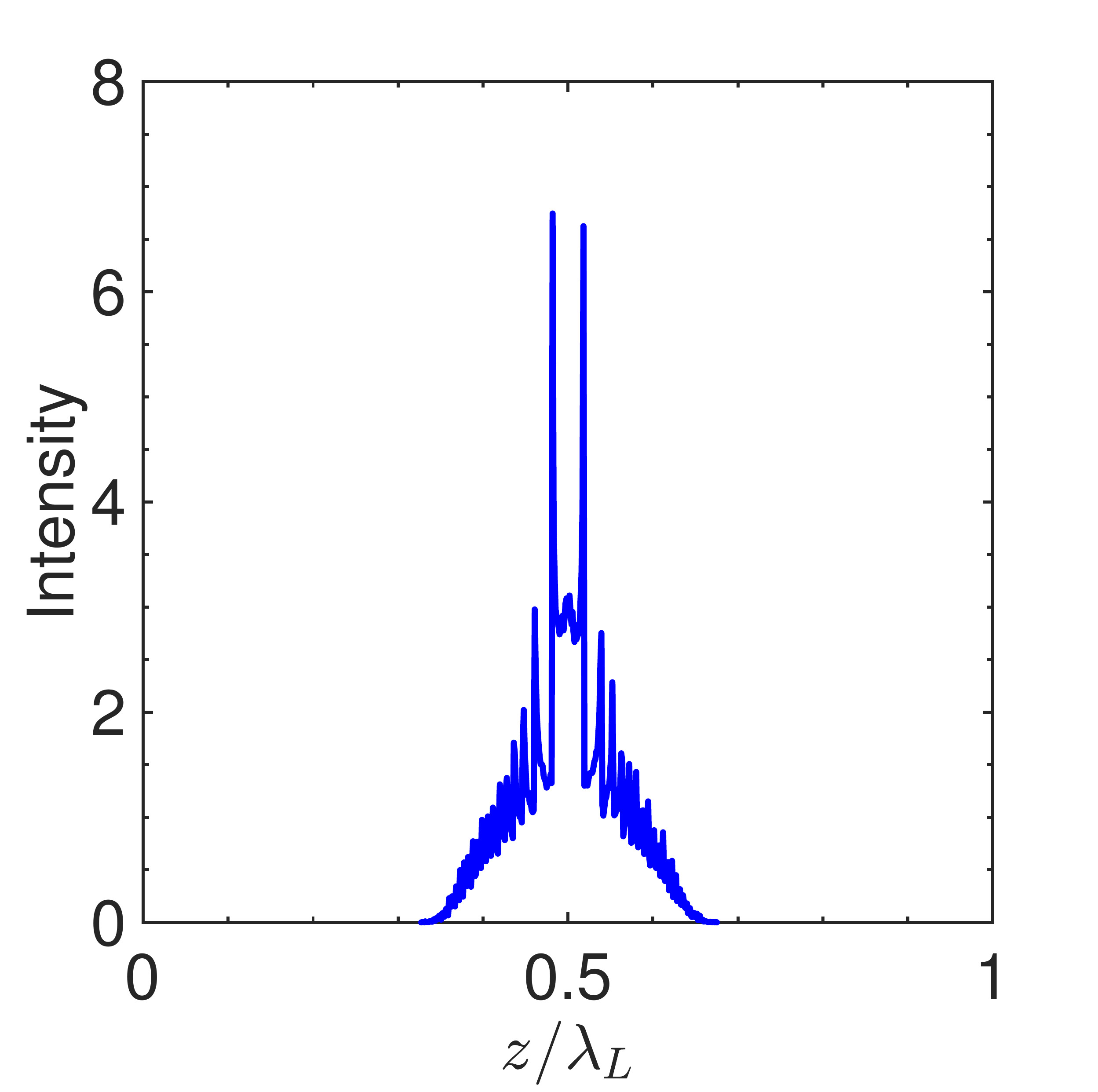}
	\caption{Trapping of particles with a linear increase of $K$ from $1\times10^{-6}$ to $1\times10^{-2}$ in $10^{4}$ turns. Red curves: separatrices.}
	\label{fig:Chap5-AdiabaticTrapping}
\end{figure}

A discrete change of $K$ can boost bunching as shown in Fig.~\ref{fig:Chap5-PhaseMixingCurrent2}. However, it is not without sacrifice, as the filamentation process will result in beam emittance growth. This emittance increase is unwanted in some cases. As well-studied in RF gymnastics \cite{garoby2011rf}, an adiabatic change of RF voltage or lattice parameters can manipulate the bunch length while preserving the longitudinal emittance. Similar ideas can also be applied to boost microbunching with little emittance growth. A simulation of trapping of microbunch with $K$ linearly ramped from $1\times10^{-6}$ to $1\times10^{-2}$ is shown in Fig.~\ref{fig:Chap5-AdiabaticTrapping}. Note the drastic difference between Fig.~\ref{fig:Chap5-AdiabaticTrapping} and Fig.~\ref{fig:Chap5-PhaseMixing}. The spirit of adiabatic buncher \cite{hemsing2013cascaded,sudar2018demonstration} is the same with adiabatic trapping introduced here, for enhancing microbunching while preserving longitudinal emittance which is useful for FEL and inverse FEL. The adiabatic trapping mechanism can also be applied in the beam injection of the SSMB or other storage rings whose momentum aperture is of concern. It is interesting to note the connection of adiabatic trapping with the microbunching process in a high-gain FEL \cite{kroll1978stimulated,kondratenko1980generating,bonifacio1984collective}.

\subsection{Experimental Parameters Choice}
As the quantum diffusion of $z$ is large for the MLS lattice (26 nm RMS per turn at 250~MeV corresponding to optics in Fig.~\ref{fig:Chap5-SSMBPoPIExtFig1}), it is not feasible to realize true SSMB inside a 1064 nm wavelength microbucket at the MLS. Therefore, the goal of SSMB PoP II is to accomplish microbunching for 100 to 1000 consecutive turns to reach a quasi steady state.  Based on the beam physics and noises analysis, the experimental parameters choice is tentatively as shown in Tab.~\ref{tab:Chap5-Quasi-SSMB_para}. Based on the parameters choice we have conducted numerical simulations, from which we can see that the typical evolution of electrons in PoP II experiment can be divided into several stages.
\begin{itemize}
	\item I: with the modulation laser turned on, the bunching factor reaches the maximum after about one quarter of the synchrotron oscillation period; 
	\item II: after several synchrotron oscillation periods, the whole microbuckets are filled with particles like shown in the right part of Fig.~\ref{fig:Chap5-PhaseMixing} as a result of phase mixing. Bunching factor after reaching this quasi-steady state will be stable if there is no quantum excitation or other diffusion effects; 
	\item III: due to quantum excitation and various diffusion effects, energy spread starts to increase and particles continue to leak out the microbuckets and begin to hit on the vacuum pipe and become lost. Bunching factor in this stage decreases;
	\item IV: after a while, all the particles are lost in the end. 
\end{itemize}
To make the experiment more realistic, each time we fire the laser, we only want to accomplish the above stages I and II, but avoid III and IV, i.e., to avoid particle loss, otherwise it will be too time-consuming to do the experiment. Since preparing the beam and storage ring state is time-consuming, while the particle can be lost with milli seconds with the laser keep firing. This is the reason why we aim for preserving micorbunching for $10^{3}$ turns, instead of $10^{6}$ turns or a longer time. The experiment is under preparation and more progress will be reported in the future. One thing worth mentioning is that the second-harmonic bunching in the quasi-steady state is negligible at the MLS, therefore, fundamental frequency radiation detection is needed in SSMB PoP II.

\begin{table}[tb]
	\caption{\label{tab:Chap5-Quasi-SSMB_para}
		Tentative parameters of the Quasi-SSMB experiment to be conducted at the MLS.}
	\centering
	\begin{tabular}{lll}
		\hline
		Parameter & \multicolumn{1}{l}{\textrm{Value}}  & Description \\
		\hline
		$E_0$ & 250  MeV & Beam energy \\
		$C_0$ & 48  m & Ring circumference \\
		$f_{\text{rf}}$  & 500  MHz & RF frequency \\
		$\eta$  & $|\eta|\leq2\times10^{-5}$  & Phase slippage factor \\
		$\tau_{\delta}$ & 180 ms@250 MeV & Longitudinal radiation damping time \\
		$\sigma_{\delta}$ & $1.8\times10^{-4}$@250 MeV & Natural energy spread \\	
		$\epsilon_{x}$ & 31 nm@250 MeV & Horizontal emittance \\
		$\lambda_{u}$ & 125 mm & Undulator period length \\
		$N_{u}$ & 32 & Number of undulator periods\\
		$L_{u}$ & 4 m & Undulator length\\
		$K$ & 2.5 & Undulator parameter \\
		$\lambda_{\text{L}}$ & 1064 nm & Modulation laser wavelength \\
		$Z_{R}$ & $\sim\frac{L_{u}}{3}$ & Rayleigh length\\
		 $P_{\text{peak}}$ & $\geq10$ kW & Modulation laser peak power \\
		 $\delta_{\frac{1}{2}}$ & $\geq1.5\sigma_{\delta}$ & Half bucket height\\
		  $\lambda_{r}$ & 1064 nm & 	Radiation wavelength\\
		 $b_{1}$ & $\geq0.01$ &  Bunching factor (1064 nm) in quasi-steady state\\
		\hline	
	\end{tabular}
\end{table}

%% file: data/chap06.tex
\chapter{Summary and Outlook}
\label{cha:summary}

This dissertation is devoted to the theoretical and experimental studies of SSMB, with important results achieved. The work presented can be summarized as: first, how to realize SSMB; second, what radiation characteristics can we obtain from the formed SSMB; and third, experimentally demonstrate the working mechanism of SSMB in a real machine. In this chapter, we give a brief review of the dissertation work, and some outlooks for future research.

\section{Summary}

In Chap.~\ref{cha:Longitudinal}, to account for the impact of local phase slippage factors in a quasi-isochronous electron storage ring, we have developed and applied the Courant-Snyder formalism in the longitudinal dimension to derive new formulae of bunch length, energy spread and longitudinal emittance beyond the classical scaling laws. Such new derivations, without adiabatic approximation, are necessary to accurately describe the dynamics of the SSMB mechanism. The method of optimizing the global and local phase slippages simultaneously to tailor the longitudinal $\beta$ function around ring has been proposed based on the derivations and in-depth analysis to generate an ultra-short bunch length and ultra-small longitudinal emittance in an electron storage ring, as required by SSMB. Based on the developed method, we have derived the scaling law of the theoretical minimum bunch length and longitudinal emittance with respect to the bending radius and angle of the bending magnet. The applications of transverse and longitudinal gradient bends for minimizing the longitudinal emittance have also been investigated.  The application of multiple RF cavities, or laser modulators in an SSMB storage ring, for longitudinal strong focusing has been discussed using the same formalism with important observations made. For a more accurate modeling of beam dynamics in a laser modulator, its thick-lens linear and nonlinear maps have been derived and simulated. We have also studied the applications of the higher-order terms of the phase slippage for high-harmonic bunching and optimization of the longitudinal dynamic aperture. 

In Chap.~\ref{cha:TLC}, we we have presented a concise analysis of the bending magnet-induced passive bunch lengthening from transverse emittance. Based on the analysis, we then discussed the method, i.e., incorporating an isochronous achromat lattice, to preserve longitudinal fine structure with beam deflection, which can be used for example in FEL beam multiplexing. After that, we have generalized the analysis and proved two theorems concerning the active applications of transverse-longitudinal coupling (TLC) for efficient harmonic
generation or bunch length compression. These two theorems dictate the relation between the energy chirp strength and the lattice optics functions at the modulator and radiator, respectively. Further, we have analyzed the contribution of modulator to the vertical emittance from quantum excitation,  to obtain a self-consistent evaluation of the required modulation laser power.  The two theorems and related analysis provide the theoretical basis for the application of TLC in SSMB to lower the requirement on the modulation laser power, by taking advantage of the fact that the vertical emittance in a planar ring is rather small. Based on the investigations, we have presented an example parameters set for the envisioned SSMB storage ring to generate high-power EUV radiation. The relation between our TLC analysis and
the transverse-longitudinal emittance exchange is also
discussed.  In addition to the investigation on linear TLC dynamics, we have also reported the first experimental validation of particle energy widening and distortion 
by nonlinear TLC in a quasi-isochronous ring, which originates from an average path-length dependence on the betatron oscillation amplitudes. The results could be important for quasi-isochronous rings, SSMB, nonscaling fixed-field alternate gradient accelerators, etc., where very small phase slippage factor or large chromaticity is required.

In Chap.~\ref{cha:Radiation}, we have presented theoretical and numerical studies of the average and statistical property of the coherent radiation from SSMB.  Our results show that 1 kW average power of 13.5 nm wavelength EUV radiation can be obtained from an SSMB ring, provided that an average current of 1 A and bunch length of 3~nm microbunch train can be formed at the radiator. Such a high-power EUV source is a promising candidate to fulfill the urgent need of semiconductor industry for EUV lithography. Together with the narrow-band feature, the EUV photon flux can reach $6\times10^{15}$ phs/s within a 0.1 meV energy bandwidth, which is appealing for fundamental condensed matter physics research. In the theoretical investigation, we have generalized the definition and derivation of the transverse form factor of an electron beam which can quantify the impact of its transverse size on the coherent radiation. In particular, we have shown that the narrow-band feature of SSMB radiation is strongly correlated with the finite transverse electron beam size. Considering the pointlike nature of electrons and quantum nature of radiation, the coherent radiation fluctuates from microbunch to microbunch, or for a single microbunch from turn to turn. Some important results concerning the statistical property of SSMB radiation have been presented, with a brief discussion on its potential applications for example the beam diagnostics. The presented work is of value for the development of SSMB and better serve the potential synchrotron radiation users. In addition, it also sheds light on understanding the radiation characteristics of FELs, CHG, etc.

In Chap.~\ref{cha:pop}, we have reported the first demonstration of the mechanism of SSMB at the Metrology Light Source (MLS) in Berlin. We have shown that electron bunches stored in a quasi-isochronous ring
can yield sub-micrometre microbunching and coherent radiation, one complete
revolution after energy modulation induced by a 1064 nm wavelength laser, and this microbunching can preserve for multiple turns.
These results verify that the optical phases, i.e, the longitudinal coordinates, of electrons can be correlated turn by turn at a precision of sub-laser wavelengths. On the basis of this phase correlation, we expect
that SSMB will be realized by applying a phase-locked laser that interacts with the
electrons turn by turn. This demonstration represents the first milestone towards the
implementation of an SSMB-based high-power, high-repetition photon source.

\section{Future Work}

Important progresses have been achieved in this dissertation as summarized above, there are still much to investigate about the SSMB physics.  Below we give a personal perspective with a hope to stimulate more interesting work.

{\bf Quasi-SSMB experiment} For SSMB PoP II, we want to establish the stable microbuckets and sustain the microbunching for multiple turns, such that the electron beam can reach a quasi steady state in the microbuckets. Once realized, this experiment will verify that stable microbucket can really be formed in a real machine and is a crucial step in the realization of real SSMB. The challenges of this experiment may include the development of the high-repetition high-power phase-locked laser, the synchronization between electron and laser, the coherent radiation signal detection, etc. The laser development is nearly completed and the experiment is planned to be conducted at the MLS soon. We will report the experiment results in the future.

{\bf Nonlinear dynamics} For the nonlinear dynamics, further quantitative analysis and optimization are needed. The goal is to obtain an EUV SSMB lattice which gives large enough transverse and longitudinal dynamic apertures, or a 6D dynamic aperture, and at the same time have a reasonable requirement on the modulation laser power.   Lie algebra-based analysis and multi-objective genetic optimization can be invoked in optimizing the nonlinear dynamics in SSMB.  Tentatively, our goal is to operate the envisioned EUV SSMB storage ring using a CW mode optical enhancement cavity (OEC), such that the filling factor of the electron beam in the ring can be very large, for example larger than $50\%$, as shown in Fig.~\ref{fig:Chap1-SSMBSchematic}. The average stored laser power of the OEC is set to be in MW level, which is a demanding but reachable value~\cite{liu2018r,wang2020study}.

{\bf Collective effects} For the collective effects, as mentioned in the previous chapters, there are several unique characteristics of SSMB which make the study of its collective effects a multifaceted work. New instabilities could arise and corresponding formalisms may need to be developed. The possible instabilities and methods to stabilize or mitigate them need to be investigated, such that a high enough stable beam current can be achieved in an SSMB ring to generate high-power radiation.  Especially, the microbunch train coherent synchrotron radiation-induced collective instability and intra-beam scattering in a general coupled lattice with longitudinal strong focusing need in-depth study in SSMB. 

CSR is the reason why SSMB can provide powerful coherent radiation. On the other hand, CSR is also the effect which sets the upper limit of the stable beam current. We have presented some important results on the investigations of average and statistical property of the SSMB radiation in Chap.~\ref{cha:Radiation}. The later work may include the more quantitative study of these radiation characteristics with 6D phase space distribution of the electron beam taken into account, and also the radiation acting back on the beam.  Recently, study on the coherent radiation induced single-pass and multi-pass instabilities in the SSMB laser modulator based on the macroparticle model are reported in Ref.~\cite{tsai2021coherent,tsai2021longitudinal,tsai2021theoretical}. The results emphasize the potential importance of CSR in SSMB. The following study may generalize the analysis from the macroparticle model to the more practical beam distribution with the finite longitudinal and transverse sizes taken into account, as we have shown in Chap.~\ref{cha:Radiation} that the longitudinal and transverse size of the electron beam can affect the coherent radiation significantly. Vlasov and Vlasov Fokker-Planck  formalism can be invoked for the analysis. And numerical code may be developed to study such effects. In addition, the CSR generated inside bending magnets can also induce beam instability\cite{bane2010threshold}. As analyzed in Chap.~\ref{cha:TLC}, the transverse dimension can have a significant impact on the longitudinal dimension in SSMB, therefore the study of such collective effects may need to be done in 4D or even 6D phase space to account for such impact self-consistently. 

The intrabeam scattering (IBS) is the effect determining the lower limit of the operation energy.  IBS in some sense is a well-understood effect. The most well-known formalisms are that of Piwinski~\cite{piwinski1974intra}  and that of Bjorken and Mtingwa (BM)~\cite{bjorken1982intrabeam}. However, ths IBS study in SSMB is different from that in a conventional ring, as the lattice can be 6D general coupled and the ring can  work in the longitudinal strong focusing regime. According to our knowledge, the IBS treatment of Kubo and Oide, which is a generalization of BM model in a 6D general coupled lattice and is implemented in the code SAD~\cite{kubo2001intrabeam}, and that of Nash~\cite{nash2006analytical} are ready for IBS study in SSMB, as they are $6\times6$ general transport matrix based. An IBS formalism can also be developed based on Chao's SLIM formalism \cite{chao1979evaluation}, in which eigen analysis has been invoked and applies to 3D general coupled lattice with longitudinal strong focusing.  Optimization of the lattice based on these formalisms can be invoked to mitigate the IBS in SSMB. 

{\bf Noise analysis and error tolerance} To make SSMB a real photon source, we need to confirm that the anticipated SSMB ring is robust enough for practical operation, considering various noises and imperfections in engineering. The first natural noise come to us is the modulation laser phase and amplitude (power) noise. Besides the laser noises, the magnet field and alignment errors, etc., also deserve careful look. Here we present some of our preliminary investigation and thoughts about the laser phase and amplitude noise.

The impact of laser noise to the electron beam dynamics in an SSMB storage ring is similar to the RF noise in a conventional ring. Note that RF noise is different from quantum excitation as it shakes the beam as a whole, while quantum excitation of different electrons are un-correlated. In this respect, the RF noise is similar to the noised induced by a stripline kicker, although the stripline  usually kicks in the transverse dimension. Our preliminary study indicates that the impact of laser noises can be divided into two parts depending on the frequency: the low frequency part and the high frequency part. Synchrotron oscillation frequency is the value to classify a frequency to be `high' or `low'. 


A criterion on the low frequency part of the phase and amplitude noises can be given to guarantee the change of the laser phase and amplitude is adiabatic enough such that the electrons can catch up and the emittance is kept invariant. If the laser noise does not fulfill such criterion, then feedback system is needed to ensure the phase locking of laser and electron. For the high frequency noise, there have been some study on the RF noise-induced diffusion in a proton storage ring based on Fokker-Planck equation \cite{dome1984theory,krinsky1982bunch}. We need to generalize such analysis to the electron storage ring with quantum excitation and radiation damping taken into account. In addition, we need to generalize the analysis to the case of longitudinal strong focusing, general coupled lattice, and multiple RFs or laser modulators. Therefore, there are many interesting issues to be investigated. In the meanwhile, we also recognize that such analysis in a longitudinal strong focusing ring is more involved than that in a longitudinal weak focusing ring, as the motion is non-integrable and the canonical perturbation theory can not be applied so straightforwardly. 

On the other hand, once the spectral density of the noise is given or measured, a straightforward approach is to implement a numerical simulation to study its impacts on beam dynamics. In the simulation, we can generate the time series of the noise according to the spectral density of the noise (see Appendix~\ref{app:Rice}). Note that one spectral density can correspond to infinite possible realizations of noise time series, so the simulation should be performed several times using different realizations of the noise to obtain a reliable conclusion.

More in-depth work on the above mentioned respects is ongoing and will be reported in the future.

%



%

%% file: data/appendix.tex
\chapter{Simplified Derivation Beyond Classical Scaling}\label{app:Fokker-Planck}
In Sec.~\ref{sec:CS}, we have present the derivation of the bunch length beyond the classical $\sqrt{|\eta|}$ scaling using SLIM. We can also derive the accurate bunch length and energy spread formulas using the Fokker-Planck equation~\cite{chandrasekhar1943stochastic,wang1945theory}. To simplify the analysis, we choose the observation point in the middle of RF cavity, where $\alpha_{z}=0$. 
With radiation damping and quantum excitation considered, the equation of motion is
\begin{equation}\label{eq:eom}
\begin{aligned}
\left(\begin{matrix}
z_{n}\\ \delta_{n}
\end{matrix}\right)=&\left(\begin{matrix}
1-\frac{h}{2}\eta C_{0}&-\eta C_{0}\\
h-\left(\frac{h}{2}\right)^{2}\eta C_{0}&1-\frac{h}{2}\eta C_{0}
\end{matrix}\right)
\left(\begin{matrix}
z_{n-1}\\ \delta_{n-1}
\end{matrix}\right)+\left(\begin{matrix}
(\Delta z)_{n-1}-\langle\Delta z\rangle\\
\left((\Delta\delta)_{n-1}-\langle\Delta\delta\rangle\right)
\end{matrix}\right),
\end{aligned}
\end{equation}
with $\Delta X$ being the change of corresponding variable $X$ in one turn. 


Using smooth approximation and the statistical properties shown in Eqs.~(\ref{eq:zdiffusion}) and (\ref{eq:CampbellStorageRing}), we can get the Fokker-Planck equation of the particle distribution function~\cite{chao1976particle}
\begin{equation}\label{eq:Fokker-Planck}
\frac{\partial\psi}{\partial t}=-\sum_{i}C_{ii}\psi-\sum_{i}\sum_{j}C_{ij}x_{j}\frac{\partial\psi}{\partial x_{i}}+\sum_{i}\sum_{j}D_{ij}\frac{\partial^{2}\psi}{\partial x_{i}\partial x_{j}}
\end{equation}
in which 
\begin{equation}\label{eq:statisticalProperty}
\begin{aligned}
{\bf C}&=\left(
\begin{array}{cc}
-\alpha _{\text{L}} & \omega _s \beta _{\text{zS}} \\
-\frac{\omega _s}{\beta _{\text{zS}}} & -\alpha _{\text{L}} \\
\end{array}
\right),\\
D_{11}&=\frac{1}{2}\langle F(s_{i},s_{\text{RF}})^{2}\rangle\dot{\mathcal{N}}\left\langle \frac{u^{2}}{E_{0}^{2}}\right\rangle, D_{12}=D_{21}=0,\\
 D_{22}&=\frac{1}{2}\dot{\mathcal{N}}\left\langle \frac{u^{2}}{E_{0}^{2}}\right\rangle\left[1+\frac{\eta C_{0}\left\langle F(s_{i},s_{\text{RF}})\right\rangle }{\beta_{z\text{S}}^{2}}\right].
\end{aligned}
\end{equation}
Note that the result of $D_{22}$ is modified to effectively account for the contribution of $\left\langle (\Delta z-\langle\Delta z\rangle)(\Delta\delta-\langle\Delta\delta\rangle)\right\rangle$ from quantum excitation before the RF kick, and we have used the relation 
\begin{equation}
\begin{aligned}
\left\langle (\Delta z-\langle\Delta z\rangle)(\Delta\delta-\langle\Delta\delta\rangle)\right\rangle&=\left\langle\left(-\sum_{i}F(s_{i},s_{\text{RF}})\frac{u_{i}}{E_{0}}-\left\langle -\sum_{i}F(s_{i},s_{\text{RF}})\frac{u_{i}}{E_{0}}\right\rangle\right)\right.\\
&\left.\ \ \ \ \ \ \ \ \left(-\sum_{i}\frac{u_{i}}{E_{0}}-\left\langle-\sum_{i}\frac{u_{i}}{E_{0}}\right\rangle\right)\right\rangle\\
&=\langle F(s_{i},s_{\text{RF}})\rangle \langle \mathcal{N}\rangle\left\langle \frac{u^{2}}{E_{0}^{2}}\right\rangle.
\end{aligned}
\end{equation}


The Fokker-Planck equation can also be used to study the transition behavior after beam injection. But here we focus on the equilibrium beam distribution which can be solved by letting $\frac{\partial\psi}{\partial t}=0$.
Due to central limit theorem, the equilibrium distribution $\psi$ is always Gaussian in a linear system and is uniquely determined by the second moments matrix $\Sigma$
\begin{equation}
\psi({X})=\frac{1}{2\pi\sqrt{\text{det}\Sigma}}\text{exp}\left(-\frac{1}{2}{X}^{T}\Sigma^{-1}{X}\right)
\end{equation}

Following Wang and Uhlenbeck~\cite{wang1945theory}, and also Chao and Lee~\cite{chao1976particle}, we can get the second moment matrix $\Sigma$ of the longitudinal dimension by the following procedures. Eigenvalues $\lambda$ and eigenvectors ${\bf V}$ of matrix ${\bf C}$ are
\begin{equation}
\lambda_{\pm}=-\alpha_{\text{L}}\pm i\omega_{s},\ {\bf V}=\left(
\begin{array}{cc}
i \beta _{\text{zS}} & -i \beta _{\text{zS}} \\
1 & 1 \\
\end{array}
\right),\ {\bf V}^{-1}=\left(
\begin{array}{cc}
-\frac{i}{2 \beta _{\text{zS}}} & \frac{1}{2} \\
\frac{i}{2 \beta _{\text{zS}}} & \frac{1}{2} \\
\end{array}
\right).
\end{equation}
Equilibrium second moment matrix viewed in the eigenvector space is
\begin{equation}
B_{ij}=-\frac{2}{\lambda_{i}+\lambda_{j}}[{\bf V}^{-1}{\bf D}({\bf V}^{T})^{-1}]_{ij}
\end{equation}
And the $\Sigma$ matrix in real space is
\begin{equation}
\begin{aligned}
\Sigma&={\bf V}{\bf B}{\bf V}^{T}\\
&=\left(
\begin{array}{cc}
\frac{D_{11} \left(2 \alpha_{\text{L}}^2+\omega _s^2\right)+\omega _s \beta _{\text{zS}} \left(2 D_{12} \alpha_{\text{L}}+D_{22} \omega _s \beta _{\text{zS}}\right)}{2 \alpha_{\text{L}} \left(\alpha_{\text{L}}^2+\omega _s^2\right)} & \frac{2 D_{12} \alpha_{\text{L}} \beta _{\text{zS}}+\omega _s \left(D_{22} \beta _{\text{zS}}^2-D_{11}\right)}{2 \beta _{\text{zS}} \left(\alpha_{\text{L}}^2+\omega _s^2\right)} \\
\frac{2 D_{12} \alpha_{\text{L}} \beta _{\text{zS}}+\omega _s \left(D_{22} \beta _{\text{zS}}^2-D_{11}\right)}{2 \beta _{\text{zS}} \left(\alpha_{\text{L}}^2+\omega _s^2\right)} & \frac{D_{22} \beta _{\text{zS}}^2 \left(2 \alpha_{\text{L}}^2+\omega _s^2\right)+\omega _s \left(D_{11} \omega _s-2 D_{12} \alpha_{\text{L}} \beta _{\text{zS}}\right)}{2 \alpha_{\text{L}} \beta _{\text{zS}}^2 \left(\alpha_{\text{L}}^2+\omega _s^2\right)} \\
\end{array}
\right).
\end{aligned}
\end{equation}
The problem is now solved completely. We can use the
$\Sigma$ matrix to get the equilibrium bunch length and energy
spread. When $|\omega_{s}|\gg\alpha_{\text{L}}$, which is usually the case, then
\begin{equation}
\begin{aligned}
\Sigma\approx\left(
\begin{array}{cc}
\frac{\beta_{\text{zS}}^{2}}{2\alpha_{\text{L}}}\left(\frac{D_{11}}{\beta_{\text{zS}}^{2}}+D_{22}\right) & \frac{\beta_{\text{zS}}}{2\omega_{s}}\left(-\frac{D_{11}}{\beta_{\text{zS}}^{2}}+D_{22}\right) \\
\frac{\beta_{\text{zS}}}{2\omega_{s}}\left(-\frac{D_{11}}{\beta_{\text{zS}}^{2}}+D_{22}\right) & \frac{1}{2\alpha_{\text{L}}}\left(\frac{D_{11}}{\beta_{\text{zS}}^{2}}+D_{22}\right) \\
\end{array}
\right).
\end{aligned}
\end{equation}

For the diffusion terms, if only term $D_{22}=\frac{1}{2}\dot{\mathcal{N}}\left\langle \frac{u^{2}}{E_{0}^{2}}\right\rangle$ is
considered, then we arrive at the result of classical Sands analysis~\cite{sands1970physics}.
With $D_{11}$ and the modified $D_{22}=\frac{1}{2}\dot{\mathcal{N}}\left\langle \frac{u^{2}}{E_{0}^{2}}\right\rangle\left[1+\frac{\eta C_{0}\left\langle F(s_{i},s_{\text{RF}})\right\rangle }{\beta_{z\text{S}}^{2}}\right]$ both taken into account, the result will be the same with what we obtained in Sec.~\ref{sec:CS} derived using SLIM, i.e.,
\begin{equation}
\begin{aligned}
\epsilon_{z}&=\epsilon_{z\text{S}}\left(1+\frac{\left\langle F^{2}(s_{i},s_{\text{RF}})\right\rangle+\eta C_{0}\left\langle F(s_{i},s_{\text{RF}})\right\rangle }{\beta_{z\text{S}}^{2}}\right),\\
\sigma_{z}(s_{\text{RF}})&=\sqrt{\epsilon_{z}\beta_{z\text{S}}}=\sigma_{z\text{S}}\sqrt{1+\frac{\left\langle F^{2}(s_{i},s_{\text{RF}})\right\rangle+\eta C_{0}\left\langle F(s_{i},s_{\text{RF}})\right\rangle }{\beta_{z\text{S}}^{2}}},\\
\sigma_{\delta}(s_{\text{RF}})&=\sqrt{\epsilon_{z}/\beta_{z\text{S}}}=\sigma_{\delta\text{S}}\sqrt{1+\frac{\left\langle F^{2}(s_{i},s_{\text{RF}})\right\rangle+\eta C_{0}\left\langle F(s_{i},s_{\text{RF}})\right\rangle }{\beta_{z\text{S}}^{2}}}.
\end{aligned}
\end{equation} 


\chapter{Simulation of Random Processes}\label{app:Rice}

We mention in the summary that a straightforward method to study the impact of various noises in a storage ring is to implement a numerical simulation. Here for the convenience of interested practitioners, we give a brief introductions on generating time series of the noise based on the spectral density of the noise, as the power spectral density (PSD) is usually used to quantify the noise characteristics and is given in the datasheet or can be measured.

\section{One-Dimensional Noise}
The PSD  of a stationary random process $x(t)$ is defined as
\begin{equation}\label{eq:definition}
\begin{aligned}
S_{x}(f)&=\lim_{T\rightarrow\infty}\frac{1}{T}\left|\int_{0}^{T}e^{-i2\pi f t}x(t)dt\right|^{2}\\
&=\left\langle\frac{1}{T}\left|\int_{0}^{T}e^{-i2\pi f t}x(t)dt\right|^{2}\right\rangle\\
&=\int_{-\infty}^{\infty}e^{-i2\pi f \tau}R_{x}(\tau)d\tau,
\end{aligned}
\end{equation}
in which 
$
R_{x}(\tau)=\langle x(t)x(t+\tau)\rangle
$ 
is the auto-correlation function of $x(t)$.  Here the brackets represent an ensemble average of noises of the same kind, i.e., all the possible realizations of a random or stochastic process in a given time periods $[0,T]$. Note that $x(t)$ is assumed to be ergodic. So $S_{x}(f)$ and $R_{x}(\tau)$ forms a Fourier transform pair, which is usually referred to as Wiener-Khinchin theorem. 

With $S_{x}(f)$ known, measured or provided in the datasheet, now we present a method of generating $x(t)$ according to  $S_{x}(f)$, which is from Rice~\cite{rice1944mathematical}:
\begin{equation}\label{eq:rice1}
x(t)=\sum_{k=0}^{N-1}\left\{a_{k}\cos\left[2\pi (k\Delta f+f_{1})t\right]+b_{k}\sin\left[2\pi (k\Delta f+f_{1})t\right]\right\},
\end{equation}
where $\Delta f=(f_{2}-f_{1})/N$ with the interested frequency region $[f_{1},f_{2}]$, and $a_{k}$ and $b_{k}$ are independent random variables which are distributed normally about zero with the standard deviation $\sqrt{S_{x}(k\Delta f+f_{1})\Delta f}$. 
There is another frequently used method which is:
\begin{equation}\label{eq:fn}
x(t)=\sum_{k=0}^{N-1}c_{k}\cos\left[2\pi (k\Delta f+f_{1})t+\phi_{k}\right],
\end{equation}
where $c_{k}=\sqrt{2S_{x}(k\Delta f+f_{1})\Delta f}$ and $\phi_{k}$ is a random phase angle distributed uniformly over the range $[0,2\pi]$.  At an appropriate point in the analysis, $N$ and $\Delta f$ are made to approach infinity and zero, respectively, in such a manner that the entire frequency band is covered by the summations, which then become integrations. That the two representations Eq.~(\ref{eq:rice1}) and (\ref{eq:fn}) lead to the same statistical properties is a consequence of the fact that they are always used in such a way that the ``Central Limit Theorem" may be used in both cases~\cite{rice1944mathematical}. 

Note that as shown in the definition in Eq.~(\ref{eq:definition}), the PSD of a stochastic process is a quantity which has a statistical meaning, therefore the generation of $x(t)$ in  a finite time period from a given PSD is not unique. Therefore, the simulation should be performed several times using different realizations to obtain a reliable conclusion.

Note also that in the above methods of generating time series $x(t)$, the noise is still assumed Gaussian, although colored. The Gaussian here means that the set of probability distributions below are all Gaussian~\cite{wang1945theory}: 
\begin{itemize}
	\item $W_{1}(xt)dt=$ probability of finding $x$ in the range of $(x,x+dx)$ at time $t$;
	\item $W_{2}(x_{1}t_{1};x_{2}t_{2})dx_{1}dx_{2}=$ joint probability of finding $x$ in the range of $(x_{1},x_{1}+dx_{1})$ at time $t_{1}$ and in the range of $(x_{2},x_{2}+dx_{2})$ at time $t_{2}$;
	\item $W_{3}(x_{1}t_{1};x_{2}t_{2};x_{3}t_{3})dx_{1}dx_{2}dx_{3}=$ joint probability of finding a triple of values of $y$ in the range of $dx_{1}$, $dx_{2}$, $dx_{3}$ at times $t_{1}$, $t_{2}$, $t_{3}$.
	\item And so on ...
\end{itemize}

\section{Multi-Dimensional Noises}

Now we consider the case of multi-dimensional noises ${\bf X}(t)=\left(x_{1}(t),...,x_{n}(t)\right)$. If the noises $x_{i}(t)$ are un-correlated, they can be generated independently using Rice method introduced above. But in reality there could be some kind of correlation between the noises. For example, if the laser phase and amplitude noise are not completely independent of each other, how do we generate the time series for these two noises in simulation? First let us introduce the concepts of cross-correlation matrix and cross-spectral density matrix.\\~\\
Cross-correlation matrix:
\begin{equation}\label{eq:correlation}
{\bf R}(\tau)=
\left(
\begin{matrix}
\langle x_{1}^{*}(t)x_{1}(t+\tau)\rangle&...&\langle x_{1}^{*}(t)x_{n}(t+\tau)\rangle\\
.& &.\\
.&...&.\\
.& & .\\
\langle x_{n}^{*}(t)x_{1}(t+\tau)\rangle&...&\langle x_{n}^{*}(t)x_{n}(t+\tau)\rangle\\
\end{matrix}
\right)
\end{equation}
where $^{*}$ means complex conjugate. If $x_{i}(t)$ are real, the complex conjugate can be omitted.\\
Cross-spectral density matrix:
\begin{equation}
\begin{aligned}
S_{ij}(f)&=\int_{-\infty}^{\infty}\left[\lim_{T\rightarrow\infty}\frac{1}{T}\int_{0}^{T}x_{i}^{*}(t)x_{j}(t+\tau)dt\right]e^{-i2\pi f\tau}d\tau=\int_{-\infty}^{\infty}R_{ij}(\tau)e^{-i2\pi f\tau}d\tau\\
S_{ji}(f)&=\int_{-\infty}^{\infty}\left[\lim_{T\rightarrow\infty}\frac{1}{T}\int_{0}^{T}x_{j}^{*}(t)x_{i}(t+\tau)dt\right]e^{-i2\pi f\tau}d\tau=\int_{-\infty}^{\infty}R_{ji}(\tau)e^{-i2\pi f\tau}d\tau
\end{aligned}
\end{equation}
Note that $S_{ii}(f)$ is the $S_{x_{i}}(f)$ defined in the first section. According to the definition we have
\begin{equation}
\begin{aligned}
S_{ij}(f)&=S_{ji}^{*}(f),\\
R_{ij}(\tau)&=R_{ji}^{*}(\tau).
\end{aligned}
\end{equation}

Given the cross-spectral density matrix $S_{ij}(f)$, here we introduce one method of generating the times series of $x_{i}(t)$ for the simulation of multi-dimensional random processes, which is from Ref.~\cite{shinozuka1971simulation}. First we need to find a matrix $H_{ij}(f)$ satisfying
\begin{equation}
S_{ij}(f)=H_{ij}(f)H_{ij}^{*T}(f)
\end{equation}
where $^{T}$ means transpose. As $S_{ij}(f)=S_{ji}^{*}(f)$, one can show that the number of unknown functions $H_{ij}(f)$ is $\frac{n(n+1)}{2}$. In this connection, one interesting method of obtaining $H_{ij}(f)$ is to let 
$
H_{ij}(f)=0\ \text{for}\ j > i,
$
then $H_{ij}(f)$ can be solved according to $S_{ij}(f)$. For example, for the case of $n=2$, then
\begin{equation}
\begin{aligned}
H_{11}(f)&=\sqrt{S_{11}(f)},\\
H_{21}(f)&=\frac{S_{21}(f)}{\sqrt{S_{11}(f)}},\\
H_{22}(f)&=\frac{\sqrt{S_{22}(f)S_{11}(f)-|S_{21}(f)|^{2}}}{\sqrt{S_{11}(f)}}.\\
\end{aligned}
\end{equation}
Define 
\begin{equation}
\gamma_{ij}(f)=\frac{|H_{ij}(f)|}{|H_{jj}(f)|},
\end{equation}
and 
\begin{equation}
\theta_{ij}(f)=\tan^{-1}\left[\frac{\text{Im}H_{ij}(f)}{\text{Re}H_{ij}(f)}\right]
\end{equation}
where $\theta_{ii}(f)=0$.
Note that $\gamma_{ij}(f)=1$ for $i=j$, and $\gamma_{ij}(f)=0$ for $j>i$.  Assume $x_{i}(t)$ are real, therefore $\text{Re}S_{ij}(f)=\text{Re}S_{ij}(-f)$ and $\text{Im}S_{ij}(f)=-\text{Im}S_{ij}(-f)$,  and we have
\begin{equation}
\begin{aligned}
\text{Re}H_{ij}(f)&=\text{Re}H_{ij}(-f),\\
\text{Im}H_{ij}(f)&=-\text{Im}H_{ij}(-f).
\end{aligned}
\end{equation}
It follows then
\begin{equation}
\theta_{ij}(f)=-\theta_{ij}(-f).
\end{equation}

After doing these preparations, $x_{i}(t)$ can now be generated according to
\begin{equation}
x_{i}(t)=\sum_{j=1}^{n}\left\{\sum_{k=0}^{N_{j}-1}\sqrt{2S_{jj}(f_{jk})\Delta f_{j}}\gamma_{ij}(f_{jk})\cos\left[2\pi f_{jk}t+\phi_{jk}+\theta_{ij}(f_{jk})\right]\right\}
\end{equation}
where $\Delta f_{j}=(f_{j2}-f_{j1})/N_{j}$ with the interested frequency region $[f_{j1},f_{j2}]$ for $x_{j}$, and $f_{jk}=(k\Delta f_{j}+f_{j1})$ and $\phi_{jk}$ is a random phase angle distributed uniformly over the range $[0,2\pi]$. 

For example, in the case of $n=2$, we have
\begin{equation}
\begin{aligned}
x_{1}(t)&=\sum_{k=0}^{N_{1}-1}\sqrt{2S_{11}(f_{1k})\Delta f_{1}}\cos\left[2\pi f_{1k}t+\phi_{1k}\right],\\
x_{2}(t)&=\sum_{k=0}^{N_{1}-1}\sqrt{2S_{11}(f_{1k})\Delta f_{1}}\gamma_{21}(f_{1k})\cos\left[2\pi f_{1k}t+\phi_{1k}+\theta_{21}(f_{1k})\right]\\
&+\sum_{k=0}^{N_{2}-1}\sqrt{2S_{22}(f_{1k})\Delta f_{2}}\cos\left[2\pi f_{2k}t+\phi_{2k}\right].\\
\end{aligned}
\end{equation}

\chapter{Statistical Parameter Fluctuations of Random Processes}

This chapter deviates a bit from our main focus of this dissertation on SSMB. Therefore, we choose to present it in the appendix.  The question we are trying to answer is: what is the standard error of using particle tracking result to represent the real beam parameter? Or in other words how much confidence do we have when we say that the tracking gives the correct theoretical prediction?
Particle tracking (numerical simulation) is used by accelerator physicists and also other scientists every day and we believe this question needs to be seriously answered. 
It is mainly a mathematical analysis. But the discussion may provide some insight for the numerical simulation used everyday in accelerator physics study.  


We use electron storage ring as an example for the analysis. The mathematical derivations in this chapter is lengthy, but the conclusion is concise, 
\begin{equation}
\frac{\text{Std}(X)}{\langle X\rangle}\propto\frac{1}{\sqrt{t_{\text{simulation}}/\tau_{\text{damping}}}},
\end{equation} 
in which  $X$ is a statistical parameter of the random process evaluated in each specific simulation, Std() means the standard deviation of different possible simulation realizations, $t_{\text{simulation}}$ and $\tau_{\text{damping}}$ are the simulation time and damping time, respectively.

\section{1D Process}\label{sec:1D}
We start from a one-dimensional model:
\begin{equation}\label{eq:moe1}
x_{n}=x_{n-1}e^{-\alpha}+\mathcal{D}\xi_{n}
\end{equation}
with the damping constant $\alpha$ and the diffusion rate $\mathcal{D}$ are both larger than zero, and $\xi$ the unit white  Gaussian noise
\begin{equation}
\langle\xi_{i}\rangle=0,\
\langle\xi_{i}\xi_{j}\rangle=\delta_{ij}
\end{equation} 
where $\delta_{ij}$ is the Kronecker delta. According to Eq.~(\ref{eq:moe1}), we can write $x_{n}$ in terms of the initial point $x_{0}$ as
\begin{equation}\label{eq:xn}
x_{n}=x_{0}e^{-n\alpha}+\sum_{m=0}^{m=n-1}e^{-m\alpha}\mathcal{D}\xi_{n-m}.
\end{equation}
Then 
\begin{equation}\label{eq:inf}
x_{\infty}=\sum_{m=0}^{\infty}e^{-m\alpha}\mathcal{D}\xi_{m}.
\end{equation}
Here the statistic independence of $\xi_{i}$ is used in replacing $\xi_{n-m}$ by $\xi_{m}$. From Eq.~(\ref{eq:inf}) we know
\begin{equation}
\langle x_{\infty}\rangle=0,\
\langle x_{\infty}^{2}\rangle=\frac{\mathcal{D}^{2}}{1-e^{-2\alpha}}.
\end{equation}
So the theoretical variance of $x$ is
\begin{equation}
\sigma_{x}^{2}=\langle x_{\infty}^{2}\rangle-\langle x_{\infty}\rangle^{2}=\frac{\mathcal{D}^{2}}{1-e^{-2\alpha}}.
\end{equation}
When $\alpha\ll1$, then
\begin{equation}
\sigma_{x}^{2}\approx\frac{\mathcal{D}^{2}}{2\alpha}.
\end{equation}

Now the question is what is the error when we use the variance (could also be other parameters, here we use variance as an example) of $n$ adjacent points $\{x_{1},x_{2},...,x_{n}\}$ in a single particle tracking or numerical simulation to represent the real $\sigma_{x}^{2}$? According to Eq.~(\ref{eq:xn}), we have
\begin{align}\label{eq:var}
\langle x_{\text{track}}\rangle&=x_{0}\frac{e^{-\alpha}-e^{-n\alpha}}{n(1-e^{-\alpha})}+\mathcal{D}\sum_{k=1}^{k=n}\frac{1-e^{-(n-k+1)\alpha}}{n(1-e^{-\alpha})}\xi_{k}\notag\\
\langle x^{2}_{\text{track}}\rangle&=\frac{1}{n}\sum_{k=1}^{k=n}\left(x_{0}e^{-k\alpha}+\sum_{m=0}^{m=k-1}e^{-m\alpha}\mathcal{D}\xi_{k-m}\right)^{2}\notag\\
\sigma_{x_{\text{track}}}^{2}&=\langle x^{2}_{\text{track}}\rangle-\langle x_{\text{track}}\rangle^2\notag\\
&=x_{0}^{2}\left(\frac{e^{-2\alpha}-e^{-2n\alpha}}{n(1-e^{-2\alpha})}-\frac{(e^{-\alpha}-e^{-n\alpha})^{2}}{n^{2}(1-e^{-\alpha})^{2}}\right)\notag\\
&\ \ \ \  +\mathcal{D}^{2}\sum_{k=1}^{k=n}\left(\frac{1-e^{-2(n-k+1)\alpha}}{n(1-e^{-2\alpha})}-\left(\frac{1-e^{-(n-k+1)\alpha}}{n(1-e^{-\alpha})}\right)^{2}\right)\xi_{k}^{2}\notag\\
&\ \ \ \  +2\mathcal{D}^{2}\sum_{i=2}^{i=n}\sum_{j=1}^{j=i-1}\left(\frac{e^{-(i-j)\alpha}\left(1-e^{-2(i-1)\alpha}\right)}{n(1-e^{-2\alpha})}-\frac{\left(1-e^{-(n-i+1)\alpha}\right)\left(1-e^{-(n-j+1)\alpha}\right)}{n^{2}(1-e^{-\alpha})^{2}}\right)\xi_{i}\xi_{j}\notag\\
&\ \ \ \ +2x_{0}\mathcal{D}\sum_{k=1}^{k=n}\left(\frac{e^{-k\alpha}\left(1-e^{-2(n-k+1)\alpha}\right)}{n(1-e^{-2\alpha})}-\frac{(e^{-\alpha}-e^{-n\alpha})(1-e^{-(n-k+1)\alpha})}{n^{2}(1-e^{-\alpha})^{2}}\right)\xi_{k}
\end{align}
Since $\xi$ is a unit white Gaussian noise, we have
\begin{equation}\label{eq:statistics}
\begin{aligned}
&\mathcal{X}=\xi_{i}^{2}\sim\Phi_{z}^{2}(1),\
\text{E}\left(\mathcal{X}\right)=1,\
\text{Var}\left(\mathcal{X}\right)=2;\\
&\mathcal{Y}=\xi_{i}\xi_{j}\ (i\neq j),\
\text{E}\left(\mathcal{Y}\right)=0,\
\text{Var}\left(\mathcal{Y}\right)=1;\\
&\text{Cov}\left(\mathcal{X},\mathcal{Y}\right)=\text{Cov}\left(\mathcal{X},\xi\right)=\text{Cov}\left(\mathcal{Y},\xi\right)=0,
\end{aligned}
\end{equation} 
where $\Phi_{z}^{2}$ means the chi-square distribution.

We assume $x_{0}$ is a known value. Then from Eq.~(\ref{eq:var}) and Eq.~(\ref{eq:statistics}), the expectation of the tracking points' variance is
\begin{equation}
\begin{aligned}
\text{E}\left(\sigma_{x_{\text{track}}}^{2}\right)
&=x_{0}^{2}\left(\frac{e^{-2\alpha}-e^{-2n\alpha}}{n(1-e^{-2\alpha})}-\frac{(e^{-\alpha}-e^{-n\alpha})^{2}}{n^{2}(1-e^{-\alpha})^{2}}\right)\\
&\ \ \ \  +\mathcal{D}^{2}\sum_{k=1}^{k=n}\left(\frac{1-e^{-2(n-k+1)\alpha}}{n(1-e^{-2\alpha})}-\left(\frac{1-e^{-(n-k+1)\alpha}}{n(1-e^{-\alpha})}\right)^{2}\right)
\end{aligned}
\end{equation}
Note $\text{E}\left(\sigma_{x_{\text{track}}}^{2}\right)$ depends both on $x_{0}$ and n. Generally, for a finite $n$,
\begin{equation}
\text{E}\left(\sigma_{x_{\text{track}}}^{2}\right)\neq\sigma_{x}^{2},
\end{equation}  which means particle tracking actually does not give the correct statistic parameter, even in an expectation value sense. This is due to the correlation of $x_{n}$ and $x_{n-1}$, which means the particle has a memory. Only when we track the particle for an infinite-long time will we arrive at the correct prediction
\begin{equation}
\lim_{n\rightarrow\infty}\text{E}\left(\sigma_{x_{\text{track}}}^{2}\right)=\sigma_{x}^{2}
\end{equation}
which is a manifestation of the ergodic theorem. 

Similarly, the variance of the tracking points' variance can be calculated from Eq.~(\ref{eq:var}) and Eq.~(\ref{eq:statistics}) to be 
\begin{equation}
\begin{aligned}
&\text{Var}\left(\sigma_{x_{\text{track}}}^{2}\right)\\
&=\sum_{k=1}^{k=n}\left(\mathcal{D}^{2}\left(\frac{1-e^{-2(n-k+1)\alpha}}{n(1-e^{-2\alpha})}-\left(\frac{1-e^{-(n-k+1)\alpha}}{n(1-e^{-\alpha})}\right)^{2}\right)\right)^{2}\times2\\
&\ \ \ \  +\sum_{i=2}^{i=n}\sum_{j=1}^{j=i-1}\left(2\mathcal{D}^{2}\left(\frac{e^{-(i-j)\alpha}\left(1-e^{-2(i-1)\alpha}\right)}{n(1-e^{-2\alpha})}-\frac{\left(1-e^{-(n-i+1)\alpha}\right)\left(1-e^{-(n-j+1)\alpha}\right)}{n^{2}(1-e^{-\alpha})^{2}}\right)\right)^{2}\times1\\
&\ \ \ \ +\sum_{k=1}^{k=n}\left(2x_{0}\mathcal{D}\left(\frac{e^{-k\alpha}\left(1-e^{-2(n-k+1)\alpha}\right)}{n(1-e^{-2\alpha})}-\frac{(e^{-\alpha}-e^{-n\alpha})(1-e^{-(n-k+1)\alpha})}{n^{2}(1-e^{-\alpha})^{2}}\right)\right)^{2}\times1
\end{aligned}
\end{equation}
So the standard relative error of the particle tracking result is
\begin{equation}
\begin{aligned}
\text{Err}(n,\alpha)=\frac{\text{Std}\left(\sigma_{x_{\text{track}}}^{2}\right)}{\text{E}\left(\sigma_{x_{\text{track}}}^{2}\right)}&=\frac{\sqrt{\text{Var}\left(\sigma_{x_{\text{track}}}^{2}\right)}}{\text{E}\left(\sigma_{x_{\text{track}}}^{2}\right)}
&\approx\frac{\sqrt{\text{Var}\left(\sigma_{x_{\text{track}}}^{2}\right)}}{\sigma_{x}^{2}}\\
\end{aligned}
\end{equation}
When $n$ is large, the terms containing $\frac{1}{n}$ in $\text{Err}(n,\alpha)$ dominate over those with $\frac{1}{n^{2}}$ or higher power index and to simplify the analysis further we let $x_{0}=0$, then
\begin{equation}\label{eq:varerr1D}
\begin{aligned}
&\frac{\text{Var}(\sigma^{2}_{x\ {\text{track}}})}{\langle \sigma^{2}_{x\ {\text{track}}}\rangle^{2}}\\
&\approx\frac{{\sum_{k=1}^{k=n}\left(\mathcal{D}^{2}\left(\frac{1-e^{-2(n-k+1)\alpha}}{n(1-e^{-2\alpha})}\right)\right)^{2}\times2+\sum_{i=2}^{i=n}\sum_{j=1}^{j=i-1}\left(2\mathcal{D}^{2}\frac{e^{-(i-j)\alpha}\left(1-e^{-2(i-1)\alpha}\right)}{n(1-e^{-2\alpha})}\right)^{2}}}{\left(\frac{\mathcal{D}^{2}}{1-e^{-2\alpha}}\right)^{2}}\\
&={\frac{2}{n^{2}}\sum_{k=1}^{k=n}\left(1-e^{-2(n-k+1)\alpha}\right)^{2}+\frac{4}{n^{2}}\sum_{i=2}^{i=n}\sum_{j=1}^{j=i-1}\left(e^{-(i-j)\alpha}\left(1-e^{-2(i-1)\alpha}\right)\right)^{2}}\\
&\approx{\frac{2}{n}+\frac{4}{n(e^{2\alpha}-1)}}
\end{aligned}
\end{equation}
When $\alpha\ll1$, which is the case for usual electron storage rings, we have
\begin{equation}\label{eq:err1D}
\begin{aligned}
\frac{\text{Std}(\sigma_{x_{\text{track}}})}{\langle \sigma_{x_{\text{track}}}\rangle}=\frac{1}{2}\frac{\text{Std}(\sigma^{2}_{x_{\text{track}}})}{\langle \sigma^{2}_{x_{\text{track}}}\rangle}
&\approx\frac{1}{2}\sqrt{\frac{2}{n}+\frac{2}{n\alpha}}\approx\sqrt{\frac{1}{2n\alpha}}
\end{aligned}
\end{equation} 
We remark that when $\alpha$ is small, which means the particle has a long memory or long range of correlation, the main contribution of $\text{Err}(n,\alpha)$ is the fluctuation of $\xi_{i}\xi_{j}$ as there are $C_{n}^{2}=\frac{n(n-1)}{2}$ these terms, while there are only $n$ terms of $\xi_{i}^{2}$.  

\begin{figure}[tb] 
	\centering 
	\includegraphics[width=0.49\columnwidth]{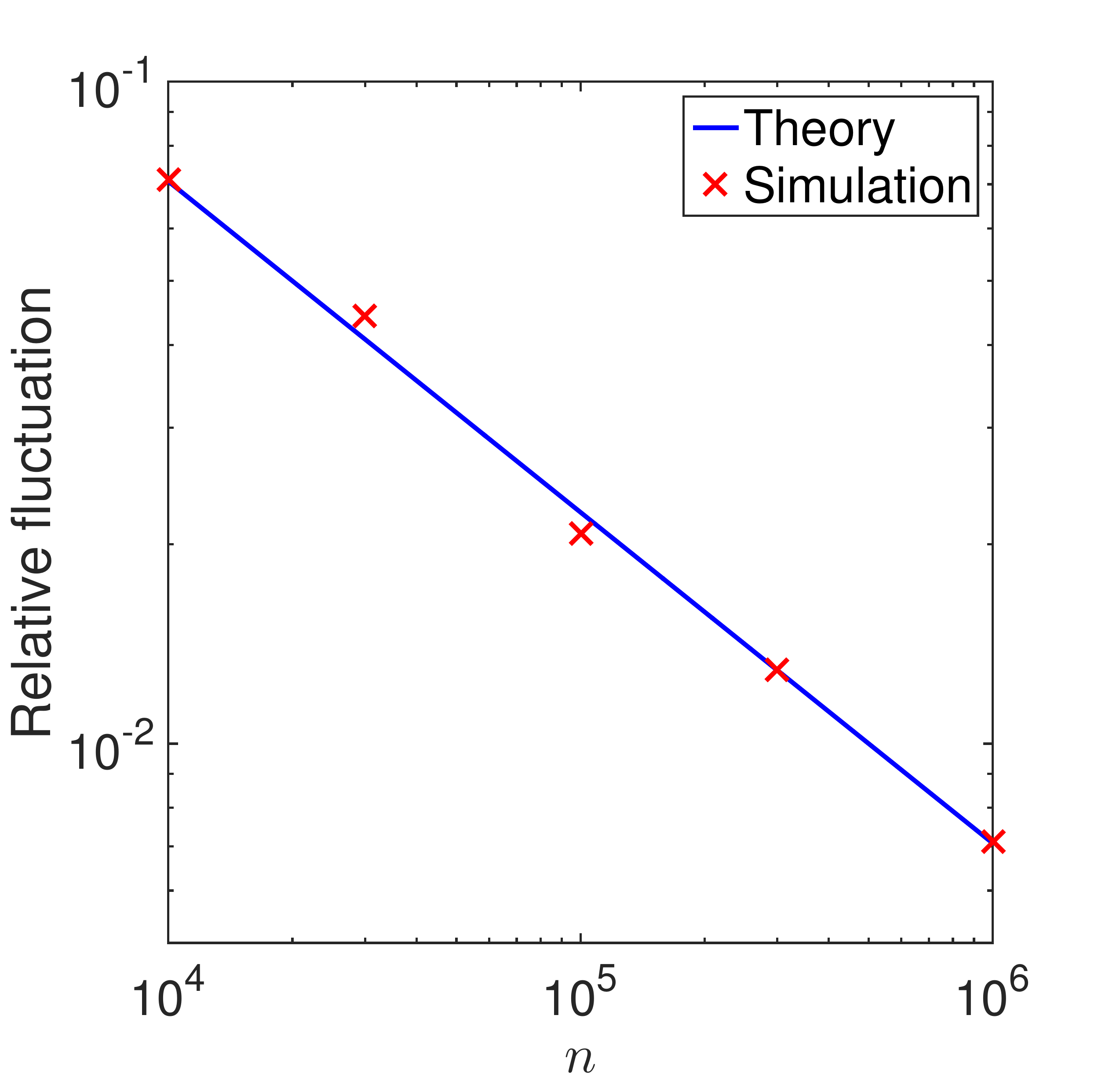}
	\includegraphics[width=0.49\columnwidth]{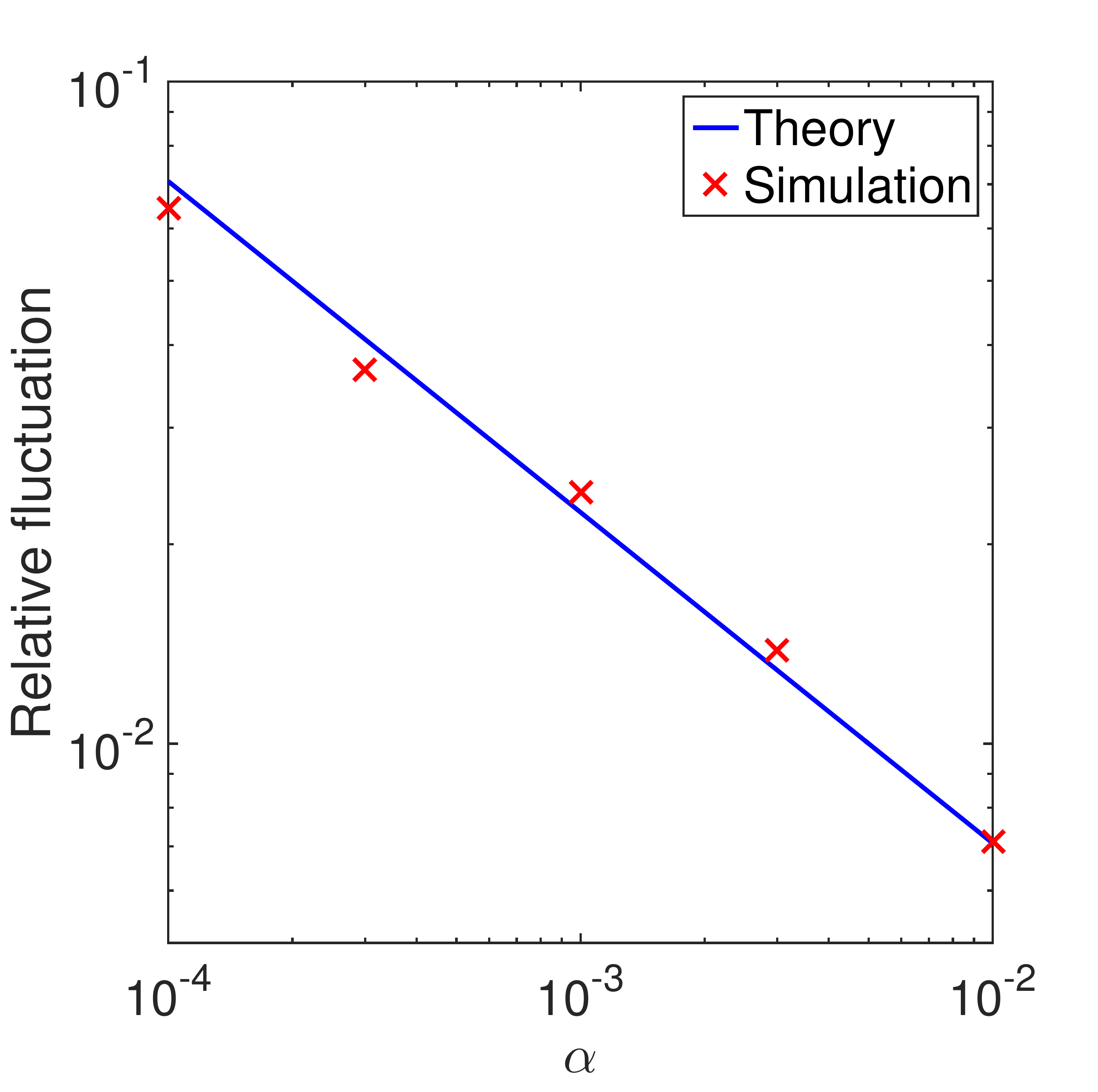}
	\caption{
		\label{fig:Simulation1D} 
		Simulation and theoretical result of relative fluctuation of $\sigma_{x_{\text{track}}}$ v.s. $n$ with $\alpha=10^{-2}$ (left), and $\alpha$ with $n=1\times10^{6}$ (right). For each setup, 100 simulation is conducted to get the data points presented in the figure. The theoretical curve is from Eq.~(\ref{eq:err1D}). }
\end{figure}

Some numerical simulations have been conducted to check validity of the analysis above. The result is shown in Fig.~\ref{fig:Simulation1D}. As can be seen, the tracking result agree well with theory, therefore confirming our analysis.

\section{2D Longitudinal Phase Space}
Here we study the 2D longitudinal phase space case with the state vector ${\bf X}=\left(\begin{matrix}
-\frac{\omega_{s}}{\eta c}z\\ \delta
\end{matrix}\right)$ and the equation of motion
\begin{equation}\label{eq:emoX}
{\bf X}_{n}={\bf A}{\bf X}_{n-1}+\left(\begin{matrix}
\lambda\vartheta\\
\mathcal{D}\xi
\end{matrix}\right)
\end{equation}
$\vartheta$ and $\xi$ the unit white Gaussian noise and the transport matrix ${\bf A}$ with synchrotron oscillation and damping
\begin{equation}
{\bf A}=\left(\begin{matrix}
1 & \omega_{s} T_{0}\\
-\omega_{s} T_{0} & 1-2\alpha_{s}' T_{0}
\end{matrix}\right)
\end{equation}
in which
\begin{equation}
\begin{aligned}
\alpha_{s}'&=\alpha_{s}+\frac{1}{2}\omega_{s}^{2}T_{0}
\end{aligned}
\end{equation}
Note here $\alpha_{s}$ is the (emittance) radiation damping coefficient. As the first step we let $\lambda=0$, which means the noise is purely from $\delta$ (quantum excitation) and ignore other noises on longitudinal coordinate like that from what we have analyzed in Sec. 2.1 (update), then
\begin{equation}
\mathcal{D}=\sqrt{T_{0} \langle\mathcal{N}\rangle\left\langle\frac{u^{2}}{E^{2}}\right\rangle}.
\end{equation}
According to Eq.~(\ref{eq:emoX}), we can write ${\bf X}_{n}$ with respect to the initial state ${\bf X}_{0}$ as
\begin{equation}\label{eq:Xn}
{\bf X}_{n}={\bf A}^{n}{\bf X}_{0}+\sum_{m=0}^{m=n-1}{\bf A}^{m}\mathcal{D}\left(\begin{matrix}
0\\
\xi_{n-m}
\end{matrix}\right).
\end{equation}
in which ${\bf A}^{m}$ can be written as
\begin{equation}
\begin{aligned}
{\bf A}^{m}={\bf V}\left(\begin{matrix}
(1+\lambda_{+}T_{0})^{m} & 0\\
0 & (1+\lambda_{-}T_{0})^{m}
\end{matrix}\right){\bf V}^{-1}
\end{aligned}
\end{equation}
with the eigenvalue of the matrix ${\bf A}-{\bf I}$ being
\begin{equation}\label{eq:lambda}
\begin{aligned}
\lambda_{\pm}&=-\alpha_{s}'\pm i\sqrt{\omega_{s}^{2}-\alpha_{s}'^{2}}
\end{aligned}
\end{equation}
and the corresponding eigenvector 
\begin{equation}\label{eq:V}
{\bf V}=\left(\begin{matrix}
1 & 1\\
\frac{\lambda_{+}}{\omega_{s}} & \frac{\lambda_{-}}{\omega_{s}}
\end{matrix}\right)=\left(\begin{matrix}
1 & 1\\
e^{i\zeta} & e^{-i\zeta}
\end{matrix}\right). 
\end{equation}
For later convenience, we also give the expression of ${\bf V}^{-1}$
\begin{equation}\label{eq:VInverse}
{\bf V}^{-1}=\left(\begin{matrix}
\frac{\lambda_{-}}{\lambda_{-}-\lambda_{+}}&\frac{1}{\frac{\lambda_{+}}{\omega_{s}}-\frac{\lambda_{-}}{\omega_{s}}}\\
\frac{\lambda_{+}}{\lambda_{-}-\lambda_{+}}&-\frac{1}{\frac{\lambda_{+}}{\omega_{s}}-\frac{\lambda_{-}}{\omega_{s}}}
\end{matrix}\right)
=
\left(\begin{matrix}
\frac{1}{1-e^{2i\zeta}}&\frac{1}{e^{i\zeta}-e^{-i\zeta}}\\
\frac{1}{e^{-2i\zeta}-1}&\frac{1}{e^{-i\zeta}-e^{i\zeta}}
\end{matrix}\right)
\end{equation}
${\bf A}^{m}$ can also be expressed in another compact form \cite{schin1996equilibrium} to be
\begin{equation}\label{eq:elegantAm}
\begin{aligned}
{\bf A}^{m}=\frac{I_{0}^{m}}{\sin\zeta}\left(\begin{matrix}
-\sin(m\Phi_{z}-\zeta) & \sin(m\Phi_{z})\\
-\sin(m\Phi_{z}) & \sin(m\Phi_{z}+\zeta)
\end{matrix}\right)
\end{aligned}
\end{equation}
where $I_{0}e^{\pm i\Phi_{z}}=1+\lambda_{\pm}T_{0}$ and $I_{0}=\sqrt{1-2\alpha_{s} T_{0}}<1$ and some of the derivations below can actually become more concise using this representation.

Now we can calculate the theoretical bunch length and energy spread from the statistics of ${\bf X}_{\infty}$. As $I_{0}<1$ and according to Eq.~(\ref{eq:Xn}) we have
\begin{equation}\label{eq:Xinfty}
{\bf X}_{\infty}=\sum_{m=0}^{\infty}{\bf A}^{m}\mathcal{D}\left(\begin{matrix}
0\\
\xi_{m}
\end{matrix}\right).
\end{equation}
Here the statistic independence of $\xi_{i}$ is again used in replacing $\xi_{n-m}$ by $\xi_{m}$.
Therefore,
\begin{equation}
\langle {\bf X}_{\infty}\rangle=\left(\begin{matrix}
0\\
0
\end{matrix}\right),\
\langle \Sigma_{\infty}\rangle=\langle {\bf X}_{\infty}{\bf X}_{\infty}^{T}\rangle=\mathcal{D}^{2}\sum_{m=0}^{m=\infty}{\bf A}^{m}\Sigma[\xi]\left({\bf A}^{m}\right)^{T},
\end{equation}
with the noise correlation matrix
\begin{equation}
\Sigma[\xi]=\left(\begin{matrix}
0 & 0\\
0 & 1
\end{matrix}\right).
\end{equation}
So we have
\begin{equation}
\begin{aligned}
\left\langle \Sigma_{\infty}\right\rangle&=\mathcal{D}^{2}\sum_{m=0}^{m=\infty}{\bf V}\left(\begin{matrix}
(1+\lambda_{+}T_{0})^{m} & 0\\
0 & (1+\lambda_{-}T_{0})^{m}
\end{matrix}\right){\bf V}^{-1}\left(\begin{matrix}
0 & 0\\
0 & 1
\end{matrix}\right)\\
&\ \ \ \ \ \ \ \  \ \ \ \ \left({\bf V}^{-1}\right)^{T}\left(\begin{matrix}
(1+\lambda_{+}T_{0})^{m} & 0\\
0 & (1+\lambda_{-}T_{0})^{m}
\end{matrix}\right){\bf V}^{T}.
\end{aligned}
\end{equation}
And the beam energy spread is
\begin{align}
\langle \sigma_{\delta}^{2}\rangle&=\left\langle \Sigma_{\infty}\right\rangle_{22}\notag\\
&=-\frac{\mathcal{D}^{2}}{4\sin^{2}{\zeta}}\left(\frac{\left(\frac{\lambda_{+}}{\omega_{s}}\right)^{2}}{1-(1+\lambda_{+}T_{0})^{2}}-\frac{2\frac{\lambda_{-}}{\omega_{s}}\frac{\lambda_{+}}{\omega_{s}}}{1-(1+\lambda_{+}T_{0})(1+\lambda_{-}T_{0})}+\frac{\left(\frac{\lambda_{-}}{\omega_{s}}\right)^{2}}{1-(1+\lambda_{-}T_{0})^{2}}\right)\notag\\
&=\frac{\mathcal{D}^{2}}{4\alpha_{s}T_{0}}\frac{1}{\left(1-\alpha_{s}T_{0}-\frac{1}{4}\omega_{s}^{2}T_{0}^{2}\right)}.
\end{align}
When the longitudinal damping constant timing revolution period $\alpha_{s}T_{0}$ and synchrotron tune $\nu_{s}$ are both much smaller than $1$, then
\begin{equation}
\langle \sigma_{\delta}^{2}\rangle\approx\frac{\mathcal{D}^{2}}{4\alpha_{s}T_{0}}=\frac{\langle\mathcal{N}\rangle\left\langle\frac{u^{2}}{E^{2}}\right\rangle}{4\alpha_{s}}
\end{equation}
which is a familiar result from electron storage ring physics. Similarly, the bunch length is
\begin{equation}\label{eq:sigmaTau}
\begin{aligned}
\langle \sigma_{z}^{2}\rangle&=\left(\frac{\eta c}{\omega_{s}}\right)^{2}\left\langle \Sigma_{\infty}\right\rangle_{11}\\
&=\left(\frac{\eta c}{\omega_{s}}\right)^{2}(1-\alpha_{s}T_{0})\left\langle \Sigma_{\infty}\right\rangle_{22}\\
&=\left(\frac{\eta c}{\omega_{s}}\right)^{2}\frac{\mathcal{D}^{2}}{4\alpha_{s}T_{0}}\frac{1-\alpha_{s}T_{0}}{\left(1-\alpha_{s}T_{0}-\frac{1}{4}\omega_{s}^{2}T_{0}^{2}\right)}
\end{aligned}
\end{equation}
As an exercise, here we try to get the above second moments by using the other representation of ${\bf A}^{m}$, i.e. Eq.~(\ref{eq:elegantAm}),
\begin{align}
\left\langle \Sigma_{\infty}\right\rangle
&=\frac{\mathcal{D}^{2}}{\sin^{2}{\zeta}}\sum_{m=0}^{\infty}I_{0}^{2m}\left(\begin{matrix}
\sin^{2} (m\Phi_{z}) & \sin\left(m\Phi_{z}\right)\sin\left(m\Phi_{z}+\zeta\right)\\
\sin\left(m\Phi_{z}+\zeta\right)\sin\left(m\Phi_{z}\right) & \sin^{2}\left(m\Phi_{z}+\zeta\right)
\end{matrix}\right)
\end{align}
We use $\langle \Sigma_{\infty}\rangle_{11}$ as an example to check
\begin{align}
\langle \Sigma_{\infty}\rangle_{11}&=\frac{\mathcal{D}^{2}}{(1-I_{0}^{2})}\frac{(I_{0}^{2}+I_{0}^{4})\sin^{2}(\Phi_{z})}{\sin^{2}{\zeta}\left(1+I_{0}^{4}-2I_{0}^{2}\cos(2\Phi_{z})\right)}\notag\\
&=\frac{\mathcal{D}^{2}}{2\alpha_{s}T_{0}}\frac{\left(1+(1-2\alpha_{s}T_{0})\right)(\omega_{s}^{2}-\alpha_{s}'^{2})T_{0}^{2}}{\frac{\omega_{s}^{2}-\alpha_{s}'^{2}}{\omega_{s}^{2}}\left(1+(1-2\alpha_{s}T_{0})^{2}-2(1-2\alpha_{s}T_{0})+4(\omega_{s}^{2}-\alpha_{s}'^{2})T_{0}^{2}\right)}\notag\\
&=\frac{\mathcal{D}^{2}}{4\alpha_{s}T_{0}}\frac{1-\alpha_{s}T_{0}}{\left(1-\alpha_{s}T_{0}-\frac{1}{4}\omega_{s}^{2}T_{0}^{2}\right)}
\end{align}
which agrees with the previous derivation result Eq.~(\ref{eq:sigmaTau}).

Now we start to analyze the variance of $\langle \delta_\text{track}^{2}\rangle$ of a single particle tracking $\{{\bf X}_{1},{\bf X}_{2},...,{\bf X}_{n}\}$. First we simplify the analysis by letting ${\bf X}_{0}=\left(\begin{matrix}
0\\
0
\end{matrix}\right)$, the second moment matrix of $\{{\bf X}_{1},{\bf X}_{2},...,{\bf X}_{n}\}$ is then	
\begin{align}
\Sigma_{n}&=\frac{1}{n}\sum_{t=1}^{n}{\bf X}_{t}{\bf X}_{t}^{T}\notag\\
&=\frac{\mathcal{D}^{2}}{n}\sum_{t=1}^{n}\left(\sum_{m=0}^{t-1}{\bf A}^{m}\left(\begin{matrix}
0\\
\xi_{t-m}
\end{matrix}\right)\right)
\left(\sum_{k=0}^{t-1}{\bf A}^{k}\left(\begin{matrix}
0\\
\xi_{t-k}
\end{matrix}\right)\right)^{T}\notag\\
&=\frac{\frac{\mathcal{D}^{2}}{n}}{\sin^{2}{\zeta}}\sum_{t=1}^{n}\sum_{m=0}^{t-1}\sum_{k=0}^{t-1}I_{0}^{m+k}\left(\begin{matrix}
\sin m\Phi_{z}\sin k\Phi_{z} & \sin\left(m\Phi_{z}\right)\sin\left(k\Phi_{z}+\zeta\right)\\
\sin\left(m\Phi_{z}+\zeta\right)\sin\left(k\Phi_{z}\right) & \sin\left(m\Phi_{z}+\zeta\right)\sin\left(k\Phi_{z}+\zeta\right)
\end{matrix}\right)\xi_{t-m}\xi_{t-k}
\end{align}
The formula above is similar to that of 1D case in Sec.~\ref{sec:1D}. The difference is now there is a synchrotron oscillation phase factor as shown in the matrix above, which makes the algebra a bit more involved, but the steps are basically the same. 

Below we focus on ${\Sigma_{n}}_{11}$ 
\begin{equation}
\begin{aligned}
&\frac{\sin^{2}\zeta}{\mathcal{D}^{2}}{\Sigma_{n}}_{11}\\
&=\frac{1}{n}\sum_{t=1}^{n}\sum_{m=0}^{t-1}\sum_{k=0}^{t-1}I_{0}^{m+k}\sin\left(m\Phi_{z}\right)\sin\left(k\Phi_{z}\right)\xi_{t-m}\xi_{t-k}\\
&=\frac{1}{2n}\frac{1}{(1-I_{0}^{2})\left(1+I_{0}^{4}-2I_{0}^{2}\cos(2\Phi_{z})\right)}\left\{\sum_{k=1}^{n}\left[\left(I_{0}^{2}+I_{0}^{4}\right)\left(1-\cos(2\Phi_{z})\right)\right.\right.\\
&\left.\ \ \ \ -I_{0}^{2(n-k)}-I_{0}^{2(n-k+2)}+2I_{0}^{2(n-k+1)}\cos(2\Phi_{z})\right.\\
&\left.\ \ \ \ +I_{0}^{2(n-k+1)}\cos(2(n-k+1)\Phi_{z})-I_{0}^{2(n-k+1)}\cos(2(n-k)\Phi_{z})\right.\\
&\left.\ \ \ \ -I_{0}^{2(n-k+1)}\cos(2(n-k+1)\Phi_{z})+I_{0}^{2(n-k+3)}\right]\xi_{k}^{2}\\
&\ \ +\frac{1}{n}\sum_{i=2}^{n}\sum_{j=1}^{i-1}I_{0}^{i-j+2}\left[\cos((i-j)\Phi_{z})-\cos((i-j+2)\Phi_{z})\right.\\
&\left.\ \ \ \ +I_{0}^{2}\left(\cos((i-j)\Phi_{z})-\cos((i-j-2)\Phi_{z})\right)\right.\\
&\left.\ \ \ \ +I_{0}^{2(n-i)}\left(-\cos((i-j)\Phi_{z})+\cos((2n-i-j+2)\Phi_{z})\right)\right.\\
&\left.\ \ \ \
+I_{0}^{2(n-i+1)}\left(\cos((i-j)\Phi_{z})\cos(2\Phi_{z})-\cos((2n-i-j)\Phi_{z})-\cos((2n-i-j+2)\Phi_{z})\right)\right.\\
&\left.\left.\ \ \ \ +I_{0}^{2(n-i+2)}\left(-\cos((i-j)\Phi_{z})+\cos((2n-i-j)\Phi_{z})\right)\right]\xi_{i}\xi_{j}\right\}.
\end{aligned}
\end{equation}
Then the variance of $\frac{\Sigma_{n11}}{\langle\Sigma_{\infty}\rangle_{11}}$ is
\begin{equation}	
\begin{aligned}\label{eq:varsigma112D}
&\text{Var}\left(\frac{\Sigma_{n11}}{\langle\Sigma_{\infty}\rangle_{11}}\right)\\
&=2\times\frac{1}{n^{2}}\frac{1}{2\left(I_{0}^{2}+I_{0}^{4}\right)\left(1-\cos(2\Phi_{z})\right)}\sum_{k=1}^{n}\left(
1-I_{0}^{2(n-k)}-I_{0}^{2(n-k+2)}\right.\\
&\left.\ \ \ \ +2I_{0}^{2(n-k+1)}\cos(2\Phi_{z})+I_{0}^{2(n-k+1)}\cos(2(n-k+1)\Phi_{z})\right.\\
&\left.\ \ \ \ 
{-I_{0}^{2(n-k+1)}\cos(2(n-k)\Phi_{z})-I_{0}^{2(n-k+1)}\cos(2(n-k+1)\Phi_{z})+I_{0}^{2(n-k+3)\Phi_{z}}}
\right)^{2}\\
&\ \ \ \  +1\times\frac{1}{n^{2}}\frac{1}{\left(1+I_{0}^{2}\right)\sin^{2}(\Phi_{z})}\sum_{i=2}^{n}\sum_{j=1}^{i-1}I_{0}^{2(i-j)}\left(
\cos((i-j)\Phi_{z})-\cos((i-j+2)\Phi_{z})\right.\\
&\left.\ \ \ \ +I_{0}^{2}\left(\cos((i-j)\Phi_{z})-\cos((i-j-2)\Phi_{z})\right)\right.\\
&\left.\ \ \ \
+I_{0}^{2(n-i)}\left(-\cos((i-j)\Phi_{z})+\cos((2n-i-j+2)\Phi_{z})\right)\right.\\
&\left.\ \ \ \ +I_{0}^{2(n-i+1)}\left(\cos((i-j)\Phi_{z})\cos(2\Phi_{z})-\cos((2n-i-j)\Phi_{z})-\cos((2n-i-j+2)\Phi_{z})\right)\right.\\
&\left.\ \ \ \ 
+{I_{0}^{2(n-i+2)}\left(-\cos((i-j)\Phi_{z})+\cos((2n-i-j)\Phi_{z})\right)}
\right)^{2}\\
&\approx\frac{2}{n}+\frac{1}{n^{2}}\frac{1}{\left(1+I_{0}^{2}\right)\sin^{2}(\Phi_{z})}\sum_{i=2}^{n}\sum_{j=1}^{i-1}I_{0}^{2(i-j)}\left(\cos((i-j)\Phi_{z})-\cos((i-j+2)\Phi_{z})\right.\\
&\left.\ \ \ \ +I_{0}^{2}\left(\cos((i-j)\Phi_{z})-\cos((i-j-2)\Phi_{z})\right)\right)^{2}\\
&=\frac{2}{n}+\frac{2}{n^{2}}\sum_{i=2}^{n}\left(\left(1+\left(\frac{1-I_{0}^{2}}{1+I_{0}^{2}}\frac{1}{\tan{(\Phi_{z})}}\right)^{2}\right)\frac{I_{0}^{2}-I_{0}^{2i}}{1-I_{0}^{2}} +\left(1-\left(\frac{1-I_{0}^{2}}{1+I_{0}^{2}}\frac{1}{\tan{(\Phi_{z})}}\right)^{2}\right)\right.\\
&\left.\ \ \ \ \left(\frac{1-I_{0}^{2}\cos(2\Phi_{z})-I_{0}^{2(i-1)}\cos(2(i-1)\Phi_{z})+I_{0}^{2i}\cos(2(i-2)\Phi_{z})}{1+I_{0}^{4}-2I_{0}^{2}\cos(2\Phi_{z})}-1\right)\right.\\
&\left.\ \ \ \ 
+ 2\frac{1-I_{0}^{2}}{1+I_{0}^{2}}\frac{1}{\tan{(\Phi_{z})}}\frac{I_{0}^{2}\sin(2\Phi_{z})-I_{0}^{2(i-1)}\sin(2(i-1)\Phi_{z})+I_{0}^{2i}\sin(2(i-2)\Phi_{z})}{1+I_{0}^{4}-2I_{0}^{2}\cos(2\Phi_{z})}\right).
\end{aligned}
\end{equation}
In the second step, we have ignored the terms obviously containing $\frac{1}{n^{2}}$ as we are treating the case when $n$ is large. Continue this approximation, the fluctuation of $\Sigma_{n11}$ can be calculated to be
\begin{equation}
\begin{aligned}
{\text{Var}\left(\frac{\Sigma_{n11}}{\langle\Sigma_{\infty}\rangle_{11}}\right)}&\approx\frac{2}{n}+\frac{2}{n}\left[\left(1+\left(\frac{1-I_{0}^{2}}{1+I_{0}^{2}}\frac{1}{\tan{(\Phi_{z})}}\right)^{2}\right)\frac{I_{0}^{2}}{1-I_{0}^{2}}\right.\\
&\left.\ \ \ \ +\left(1-\left(\frac{1-I_{0}^{2}}{1+I_{0}^{2}}\frac{1}{\tan{(\Phi_{z})}}\right)^{2}\right)\frac{I_{0}^{2}\cos(2\Phi_{z})-I_{0}^{4}}{1+I_{0}^{4}-2I_{0}^{2}\cos(2\Phi_{z})}\right.\\
&\left.\ \ \ \ + 2\frac{1-I_{0}^{2}}{1+I_{0}^{2}}\frac{1}{\tan{(\Phi_{z})}}\frac{I_{0}^{2}\sin(2\Phi_{z})}{1+I_{0}^{4}-2I_{0}^{2}\cos(2\Phi_{z})}\right]\\
&={\frac{2}{n}+\frac{2}{n}\left(\frac{2I_{0}^{4}}{1-I_{0}^{4}}+\frac{8I_{0}^{2}(1-I_{0}^{2})\cos^{2}(\Phi_{z})}{(1+I_{0}^{2})\left(1+I_{0}^{4}-2I_{0}^{2}\cos(2\Phi_{z})\right)}\right)}
\end{aligned}
\end{equation}
When $\alpha_{s}T_{0}\ll\pi\nu_{s}<1$ ($\Phi_{z}\approx2\pi\nu_{s}$), using Taylor expansion and keeping terms up to the linear order of $\alpha_{s}T_{0}$ 
\begin{equation}\label{eq:varerr}
\begin{aligned}
{\text{Var}\left(\frac{\Sigma_{n11}}{\langle\Sigma_{\infty}\rangle_{11}}\right)}&\approx{\frac{2}{n}+\frac{2}{n}\left(\frac{2}{1-(1-4\alpha_{s}T_{0})}+\frac{8\cos^{2}(\Phi_{z})}{2\left(1+1-2\cos(2\Phi_{z})\right)}\right)}\\
&={\frac{2}{n}+\frac{1}{n}\left(\frac{1}{\alpha_{s}T_{0}}+\frac{2}{\tan^{2}(\Phi_{z})}\right)}\\
&\approx{\frac{1}{n\alpha_{s}T_{0}}}=\frac{1}{{N_{\text{track}}/N_{\text{damping}}}}
\end{aligned}
\end{equation}
in which $N_{\text{damping}}$ is the damping time in unit of revolution numbers. Note that Eq.~(\ref{eq:varerr}) is basically the same with the 1D correspondence Eq.~(\ref{eq:varerr1D}). The only difference is there is an extra term $\frac{2}{\tan^{2}(\Phi_{z})}$ from synchrotron oscillation. 

\begin{figure}[tb] 
	\centering 
	
	\includegraphics[width=0.49\columnwidth]{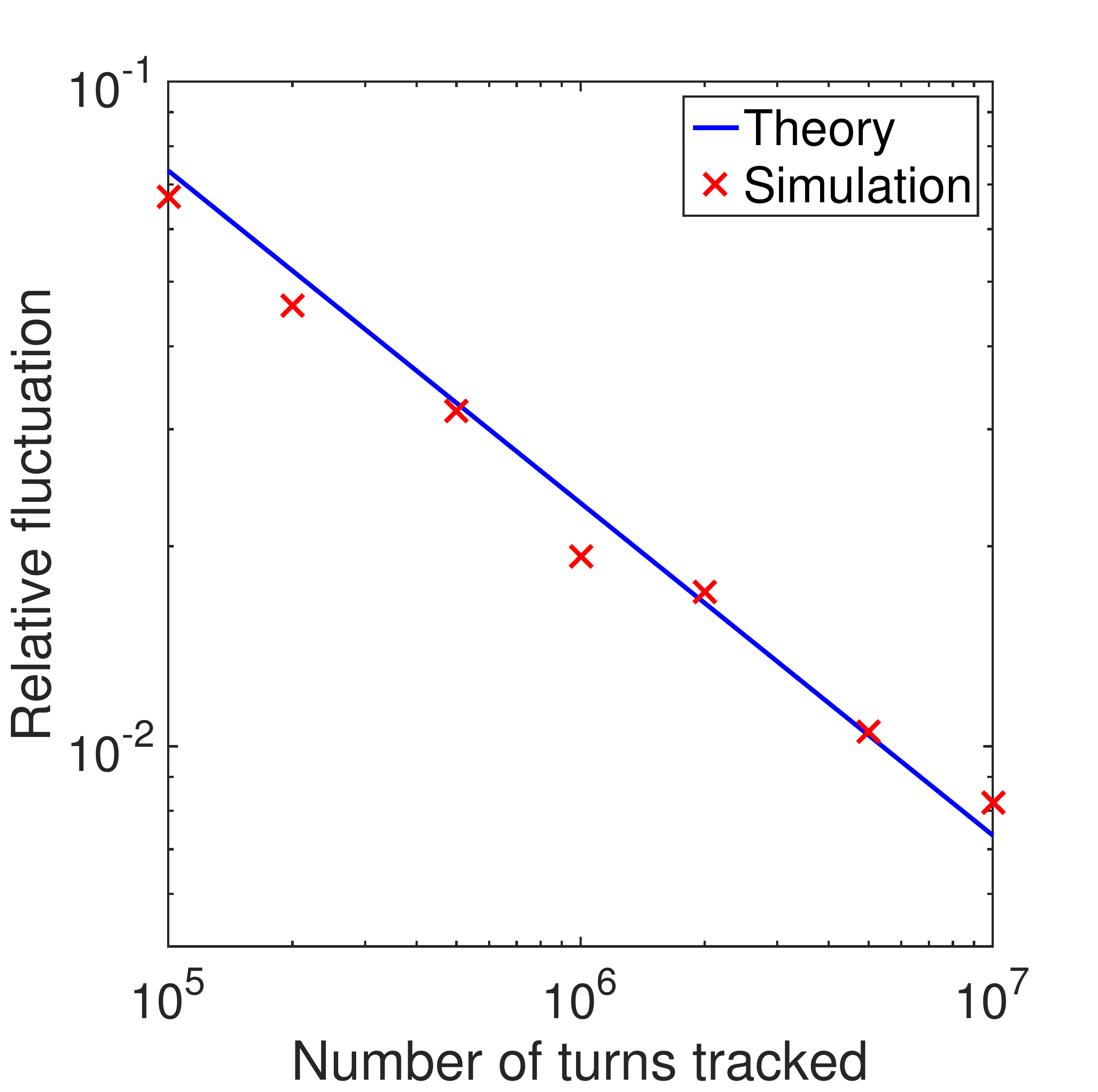}
	\includegraphics[width=0.49\columnwidth]{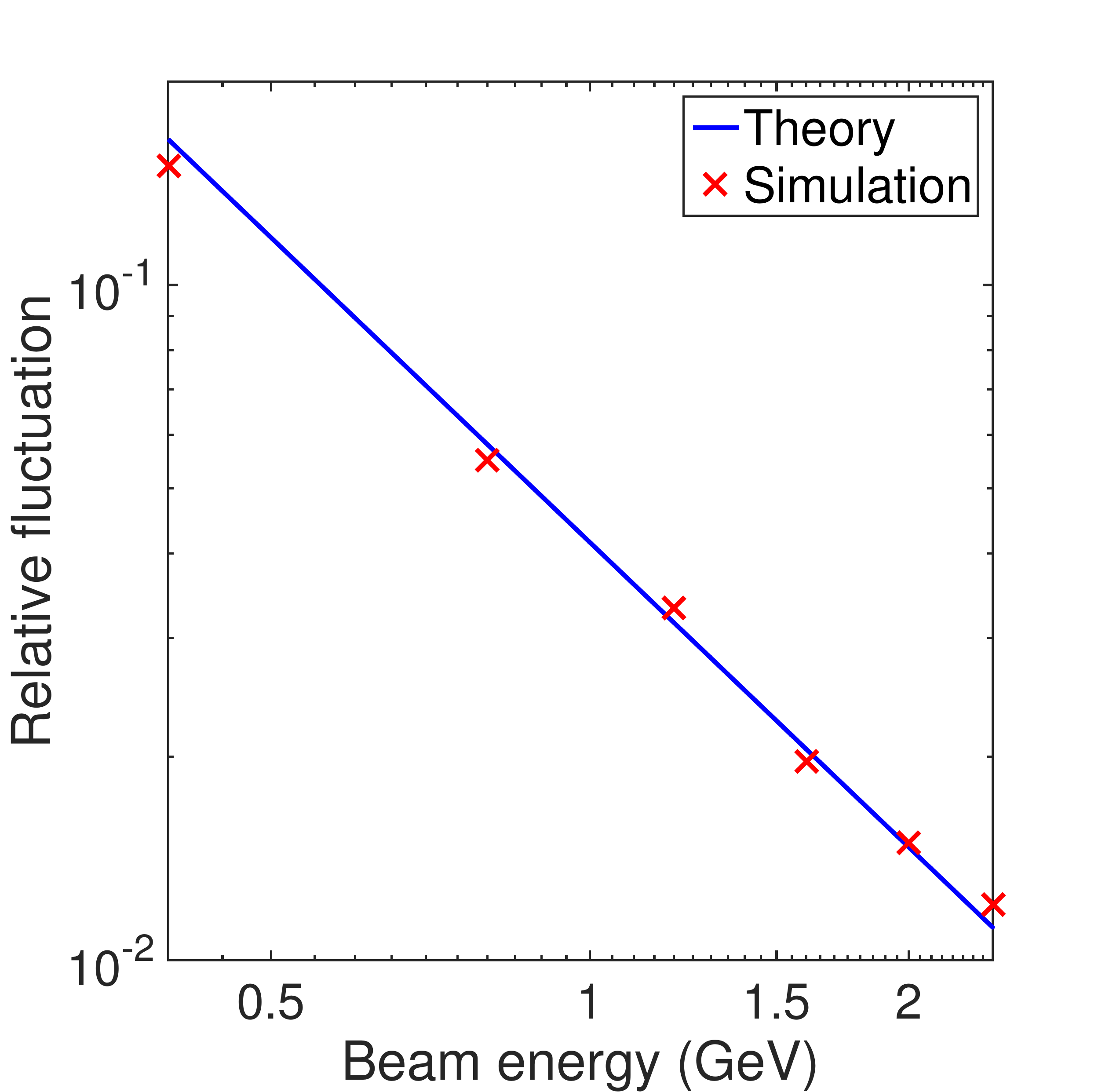}
	\caption{
		\label{fig:Simulation2D} 
		Simulation and theoretical result of relative fluctuation of $\sigma_{\delta_{\text{track}}}$ v.s. $n$ with a beam energy 2 GeV (left), and beam energy with $n=1\times10^{7}$ (right).  For each setup, 100 simulation is conducted to get the data points presented in the figure. The theoretical curve is from Eq.~(\ref{eq:std}).}
\end{figure}

The relative fluctuation of $\sigma_{z,\delta\ {\text{track}}}$ is half of that of the variance (concluded from simulation, check rigorously later), therefore,
\begin{equation}\label{eq:std}
\begin{aligned}
\frac{\text{Std}(\sigma_{z,\delta\ {\text{track}}})}{\langle \sigma_{z,\delta\ {\text{track}}}\rangle}=\frac{1}{2}\frac{\text{Std}(\sigma^{2}_{z,\delta\ {\text{track}}})}{\langle \sigma^{2}_{z,\delta\ {\text{track}}}\rangle}
&\approx\frac{1}{2}\sqrt{\frac{1}{n\alpha_{s}T_{0}}}=\frac{1}{2\sqrt{N_{\text{track}}/N_{\text{damping}}}}
\end{aligned}
\end{equation} 


Some numerical simulations have been conducted to check the analysis above.  The result is shown in Fig.~\ref{fig:Simulation2D}. As can be seen, the tracking result agree well with theory, therefore confirming our analysis.

\section{6D Phase Space}
Following the above analysis of 2D phase space, for the case of 6D phase space simulation of electron storage ring dynamics with quantum excitation and radiation damping, there is a sum rule of the fluctuation of the three eiegn emittance, which corresponds to the Robinson sum rule~\cite{orlov2010robinson,robinson1958radiation,talman2009orlov}
\begin{equation}
\alpha_{I}+\alpha_{II}+\alpha_{III}=\frac{2U_{0}}{E_{0}}.
\end{equation}
Here we present the sum rule of the simulated fluctuation of the three eigen emittances 
\begin{equation}
\frac{1}{\text{Var}\left(\frac{\epsilon_{I,\text{track}}}{\langle\epsilon_{I,\text{track}}\rangle}\right)}+\frac{1}{\text{Var}\left(\frac{\epsilon_{II,\text{track}}}{\langle\epsilon_{II,\text{track}}\rangle}\right)}+\frac{1}{\text{Var}\left(\frac{\epsilon_{III,\text{track}}}{\langle\epsilon_{III,\text{track}}\rangle}\right)}=\frac{2U_{0}}{E_{0}}N_{\text{track}}.
\end{equation}
The derivation is lengthy but straightforward, similar to that presented for the cases of 1D and 2D.